\documentclass[review]{elsarticle}
\pdfoutput=1

\biboptions{numbers,sort&compress}
\usepackage{lineno,hyperref}
\modulolinenumbers[5]

\usepackage{amsmath}
\usepackage{amsfonts}
\usepackage{amssymb}

\usepackage[hang,small,bf]{caption}
\usepackage[labelformat=simple]{subcaption}

\usepackage{enumitem}
\usepackage{booktabs}
\usepackage{slashed}
\usepackage{tikz}
\usepackage{float}
\usetikzlibrary{arrows,shapes.misc,decorations.markings,decorations.pathmorphing,positioning,intersections}

\tikzset{cross/.style={cross out, draw=black, minimum size=2*(#1-\pgflinewidth), inner sep=0pt, outer sep=0pt},
cross/.default={1.5mm}}
\tikzset{mydash/.style={dashed, dash pattern=on 4pt off 5pt}}

\renewcommand{\Im}{\mathrm{Im}}
\newcommand{\MEG}{\mu \rightarrow e \gamma}

\newcommand{\GeV}{\,\mathrm{GeV}}
\newcommand{\TeV}{\,\mathrm{TeV}}
\newcommand{\be}{\begin{equation}}
\newcommand{\ee}{\end{equation}}
\newcommand{\bea}{\begin{eqnarray}}
\newcommand{\eea}{\end{eqnarray}}

\begin{document}

\begin{frontmatter}

\title{A Call for New Physics : The Muon Anomalous Magnetic Moment and Lepton Flavor Violation}


\author{Manfred Lindner\fnref{manfred}}\fntext[manfred]{lindner@mpi-hd.mpg.de}

\author{Moritz Platscher\fnref{moritz}}\fntext[moritz]{moritz.platscher@mpi-hd.mpg.de}

\author{Farinaldo S. Queiroz\fnref{farinaldo}}\fntext[farinaldo]{farinaldo.queiroz@mpi-hd.mpg.de}

\address{Max-Planck-Institut f\"ur Kernphysik\\ Saupfercheckweg 1, 69117 Heidelberg, Germany}

\begin{abstract}
We review how the muon anomalous magnetic moment ($g-2$) and the quest for lepton flavor violation are intimately correlated. Indeed the decay $\mu \to e \gamma$ is induced by the same amplitude for different choices of in- and outgoing leptons. In this work, we try to address some intriguing questions such as:{\it Which hierarchy in the charged lepton sector one should have in order to reconcile possible signals coming simultaneously from $g-2$ and lepton flavor violation? What can we learn if the $g-2$ anomaly is confirmed by the upcoming flagship experiments at FERMILAB and J-PARC, and no signal is seen in the decay $\mu \rightarrow e\gamma$ in the foreseeable future? On the other hand, if the $\mu \rightarrow e\gamma$ decay is seen in the upcoming years, do we need to necessarily observe a signal also in $g-2$?}. In this attempt, we generally study the correlation between these observables in a detailed analysis of simplified models. We derive master integrals and fully analytical and exact expressions for both phenomena, and adress other flavor violating signals. We investigate under which conditions the observations can be made compatible and discuss their implications. Lastly, we discuss in this context several extensions of the SM, such as the Minimal Supersymmetric Standard Model, Left-Right symmetric model, $B-L$ model, scotogenic model, two Higgs doublet model, Zee-Babu model, 331 model, and $L_{\mu}-L_{\tau}$, dark photon, seesaw models type~I, II and III, and also address the interplay with $\mu \to eee$ decay and $\mu-e$ conversion.
\end{abstract}

\end{frontmatter}

\tableofcontents

\newpage

\section{Introduction}
\begin{figure}[t]
\centering
\includegraphics[scale=1]{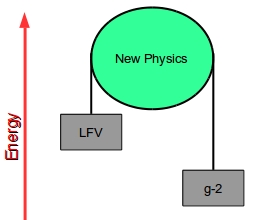}
\caption{Muon anomalous magnetic moment points to new physics at relatively low scale, whereas lepton flavor violating processes typically push new physics to high scales. How are they connected? Can signals in both observables be reconciled? In this review we address these questions.} 
\label{figure3LFV}
\end{figure}

The muon anomalous magnetic moment ($g-2$) is a prime example of the success of quantum field theory~\cite{GellMann:1954kc}. Its precise measurement is paramount to understanding the effects of higher order corrections arising in perturbation theory. Furthermore, it potentially indicates the existence of new physics since there is a long standing deviation between the Standard Model (SM) prediction and the measurement, which raised much interest in the past~\cite{Blum:2013xva}. On the other hand, lepton flavor violation (LFV) has been observed via neutrino oscillations since the late 90's~\cite{Fukuda:1998mi,Ahmad:2002jz}, but has thus far not been detected among charged leptons. Typically, new physics models that accommodate $g-2$ advocate the existence of new particles with masses around or below the TeV scale and face serious problems when confronted with constraints from LFV, which tend to force these particle to be rather heavy as illustrated in Fig.~\ref{figure3LFV}. Possibly, the ongoing $g-2$ experiments may reach a $5\sigma$ deviation from the SM in the foreseeable future, constituting an augury for new physics. Can such a signal be reconciled with limits from LFV? Is there room for signals in both observables? A new era in particle physics may be ahead of us and this review plans to pave the way. We outline what sort of models can accommodate signals and constraints once both data sets are accounted for in a systematic way. To do so, we structured the manuscript as follows:
\begin{enumerate}[label=(\roman*)]
  \item in Sec.~\ref{Secg2} we review the theoretical aspects of the $g-2$;
  \item in Sec.~\ref{Secg2Exp} we discuss the experimental apparatus for $g-2$;
  \item in Sec.~\ref{Secg2new} we review the call for new physics in the $g-2$;
  \item in Sec.~\ref{LFVintro} we give a brief introduction to LFV;
  \item in Sec.~\ref{LFVexp} we provide the experimental status;
  \item in Sec.~\ref{sec:framework} we describe the foundation of the relation between $g-2$ and $\mu \rightarrow e \gamma$;
  \item in Sec.~\ref{sec:numerics}, we investigate how one can reconcile possible signal seen in $g-2$ and $\mu \rightarrow e \gamma$ 		and the implications in several simplified models guarding $SU(2)_L$ invariance;
  \item in Sec.~\ref{sec:UVModels} we discuss the correlation between $g-2$ and $\MEG$, $\mu \rightarrow 3e$, and $\mu-e$ conversion in several UV complete models: the MSSM, 		$U(1)_{B-L}$, Left-Right symmetry, two Higgs doublets, Zee-Babu, scotogenic, the 331 model and $L_{\mu}-L_{\tau}$, dark photon and seesaw models, to show that our findings are applicable to a multitude of popular particle physics models;
  \item  in Sec.~\ref{sec:conclusion} we finally draw our conclusions.
\end{enumerate}

\subsection{\label{Secg2}Muon Anomalous Magnetic Moment}

\begin{figure}[t]
\centering
\includegraphics[scale=0.4]{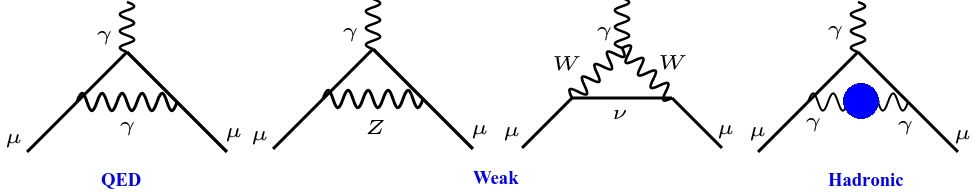}
\caption{Lowest-order SM corrections to $a_{\mu}$. From left to right: QED , weak and hadronic.} 
\label{fig:1}
\end{figure}

In quantum mechanics we have learned that any charged particle has a magnetic dipole
moment ($\overrightarrow{\mu}$) which is aligned with its spin ($\overrightarrow{s}$) and linked through the equation, 
\begin{equation}
\overrightarrow{\mu} = g \left( \frac{q}{2m}\right) \overrightarrow{s},
\end{equation}
where $g$ is the gyromagnetic ratio, $q =\pm e$ is the electric charge of a given charged particle, and m its mass. In classical quantum mechanics, $g=2$. However, loop corrections calculable in quantum field theories, such as the SM, yield small corrections to this number, as shown in Fig.~\ref{fig:1}. These corrections are parametrized in terms of $a_{\mu}=(g_{\mu}-2)/2$, referred as the anomalous magnetic moment which has been calculated since the 1950s~\cite{GellMann:1954kc}. Ever since, a great deal of effort has been put forth to determine the SM prediction including higher orders of perturbation theory~\cite{Garwin:1960zz,Burnett:1967zfb,Kinoshita:1967txv,Terazawa:1968jh,Lautrup:1969fr,Aldins:1970id,Lautrup:1972iw,
Bramon:1972bc,Auberson:1972xd}. Considering SM contributions up to three orders in the electromagnetic constant, one finds:
\begin{equation}
\begin{aligned}
a_{\mu}^{SM}&&= 116591802 (2)(42)(26)  \times 10^{-11}\ \textnormal{\cite{Blum:2013xva}}\,, \\
a_{\mu}^{SM}&&= 116591828 (2)(43)(26)  \times 10^{-11}\ \textnormal{\cite{Hagiwara:2011af}}\,. 
\end{aligned}
\end{equation}
The different central values are due to different results found for the hadronic vacuum polarization contributions. The three errors in parenthesis account for electroweak, lowest-order hadronic, and higher-order hadronic contributions, respectively~\cite{Amsler:2008zzb}. 

Moreover, there is an ongoing aspiration to reduce the theoretical errors~\cite{Ellis:1994qf,Bijnens:1995xf,Alemany:1997tn,Hayakawa:1997rq,Knecht:2001qf,
Blokland:2001pb,Melnikov:2003xd,deTroconiz:2004yzs,Kinoshita:2005sm,Kataev:2006yh,Aoyama:2008gy,Prades:2009tw,
Goecke:2010if,Davier:2010nc,Boughezal:2011vw,Aoyama:2011dy,Goecke:2012qm,Bijnens:2012an,
Dorokhov:2012qa,Aoyama:2012wk,Williams:2013tia,Kurz:2014wya,Colangelo:2014qya,Masjuan:2014rea,Dorokhov:2014iva,Roig:2014dya,Blum:2014oka,
Blum:2015gfa,Eidelman:2015isu,Bijnens:2016hgx,Sanchez-Puertas:2016mmz,Nyffeler:2016xul,Bernard:2016tlj,Dominguez:2016eol,Ananthanarayan:2016mns}; however, calculating the SM contribution to $a_{\mu}$ is still burdensome with large uncertainties arising~\cite{Knecht:2014sea,KNECHT:2014bsa} most prominently from hadronic light-light corrections~\cite{Colangelo:2014pva,Kurz:2015bia,Stoffer:2015fvt,Green:2015mva,Jin:2015bty,Nyffeler:2016gnb,Bijnens:2016hgx,Asmussen:2016lse}. 

Since the SM prediction to the muon anomalous magnetic moment is proportional to the electromagnetic constant, it is also important to measure the latter to a high precision. Up to now the electromagnetic constant is obtained through measurements of the electron magnetic moment~\cite{Kinoshita:1981vs,Kinoshita:1981ww,Kinoshita:1981wm,Odom:2006zz,Hanneke:2008tm,Hanneke:2010au,Haffner:2000zzi,Bouchendira:2010es,Aoyama:2014sxa}. In other words, the electron magnetic moment serves as input to determine the muon anomalous magnetic moment in the SM. Ideally, one could independently measure the electromagnetic constant and then compare with the measured electron magnetic moment. Albeit, the electron magnetic moment is not as sensitive to new physics effects due to the small electron mass. Indeed, its relative sensitivity to new physics is reduced by a factor of $m_{\mu}^2/m_e^2 \sim 40000$. Nevertheless, there are specific cases in which the electron magnetic moment plays a complementary role to probe new physics~\cite{Giudice:2012ms,Aboubrahim:2014hya}. As for the tau magnetic moment~\cite{Eidelman:2007sb,Eidelman:2007fj}, which would, in principle, be an excellent probe for new physics, it is measured with very poor precision, $-0.052 < a_{\tau}^{exp} < 0.013$ at $95\%$~C.L. (also quoted as $a_{\tau}^{exp}=-0.018(17)$~\cite{Abdallah:2003xd}) due to its very short lifetime ($\sim 10^{-13}$~s). The quoted limit is the one adopted by the PDG~\cite{Abdallah:2003xd}, but there are many competing bounds in the literature~\cite{Silverman:1982ft,Domokos:1985rp,Grifols:1990ha,delAguila:1990jg,Escribano:1996wp,Ackerstaff:1998mt}. Some are actually more stringent than the PDG one, lying in the range of $-0.007 < a_{\tau}^{exp} < 0.005$~\cite{Acciarri:1998iv,GonzalezSprinberg:2000mk}, using data on tau lepton production at LEP1, SLC, and LEP2. The SM prediction is 0.0117721(5)~\cite{Eidelman:2007sb,Passera:2007fk}, showing that even using this more restrictive limit we are still several orders of magnitude away from probing new physics using the tau magnetic moment. 

Now that we have understood what is the muon anomalous magnetic moment, discussed the SM prediction to $a_{\mu}$ and the theoretical uncertainties, we shall have a look at the measurement procedure.

\subsection{\label{Secg2Exp}Measuring the Muon Anomalous Magnetic Moment}

\begin{figure}[t]
\centering
\includegraphics[scale=0.5]{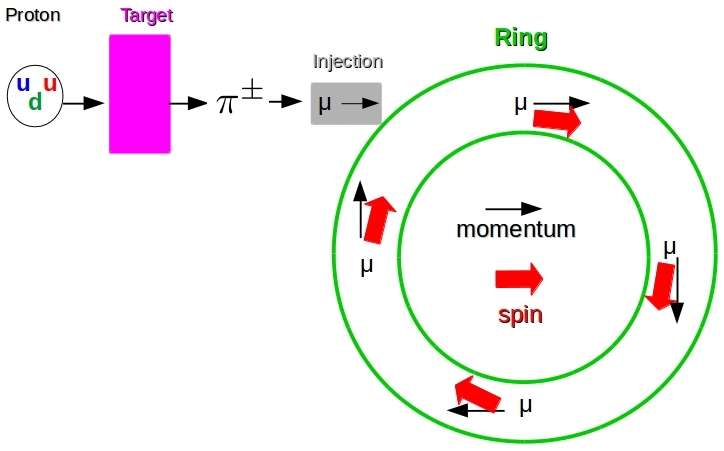}
\caption{Illustrative figure showing how experiments measure the muon anomalous magnetic moment, using a beam of polarized muons and Larmor precession physics.} 
\label{fig:2}
\end{figure} 

\begin{table}[!h]
\begin{center}
\begin{tabular}{|l|c|l|}
\hline
Determination & Beam & $a_\mu$\\ \hline
SM~\cite{Blum:2013xva} & & 0.00116 591 803(1)(42)(26)\\
\hline
\hline
Columbia-Nevis($1957$)~\cite{nevis57} & $\mu^+$ & $0.00 \pm 0.05$ \\
Columbia-Nevis($1959$)~\cite{nevis59} & $\mu^+$ & $0.001\,13^{+(16)}_{-(12)}$ 
 \\ \hline
CERN ($1961$)~\cite{cern1-61,cern2-61} & $\mu^+$ & $0.001\,145(22)$ \\
CERN ($1962$)~\cite{cern1-62} & $\mu^+$ & $0.001\,162(5)$ \\
CERN ($1968$)~\cite{cern2-68} & $\mu^\pm$ & $0.001\,166\,16(31)$ \\
CERN (1975)~\cite{cern3-75} & $\mu^\pm$ & 0.001\,165\,895(27) \\
CERN (1979)~\cite{cern3-79} & $\mu^\pm$ & 0.001\,165\,911(11) \\
\hline
BNL E821 (2000)~\cite{bnl00} & $\mu^+$ & 0.001\,165\,919\,1(59) \\
BNL E821 (2001)~\cite{bnl01} & $\mu^+$ & 0.001\,165\,920\,2(16) \\
BNL E821 (2002)~\cite{bnl02} & $\mu^+$ & 0.001\,165\,920\,3(8) \\
BNL E821 (2004)~\cite{bnl04} & $\mu^-$ & 0.001\,165\,921\,4(8)(3) \\
BNL E821 (2006)~\cite{bnl06} & $\mu^\pm$ & 0.001\,165\,920\,89(63)\\
\hline
\hline
Current Discrepancy~\cite{Agashe:2014kda} & &  $\Delta a_\mu =287(80) \times 10^{-11}$ \\
                                                            & &  $\Delta a_\mu =261 (78) \times 10^{-11}$ \\
\hline
Future sensitivity~\cite{Carey:2009zzb} & &  $\Delta a_\mu = 288(34) \cdot 10^{-11}$\\
\hline
\end{tabular}
\end{center}
\caption{\label{tab:amuOverview}Historic overview of the measurements of the muon anomalous magnetic moment $a_\mu$, along with current and future sensitivity of $\Delta a_\mu = a_\mu^\mathrm{exp} - a_\mu^\mathrm{SM}$. The errors in the SM values are due to electroweak corrections, leading and next-to-leading order hadronic corrections, respectively. Throughout this work we will adopt  $\Delta a_\mu =287(80) \times 10^{-11}$ as used by the PDG~\cite{Beringer:1900zz}. For recent theoretical reviews see e.g.~\cite{Zhang:2008pka,Agashe:2014kda}. }
\end{table}

\begin{figure}[t]
  \centering
  \includegraphics[width=.85\textwidth]{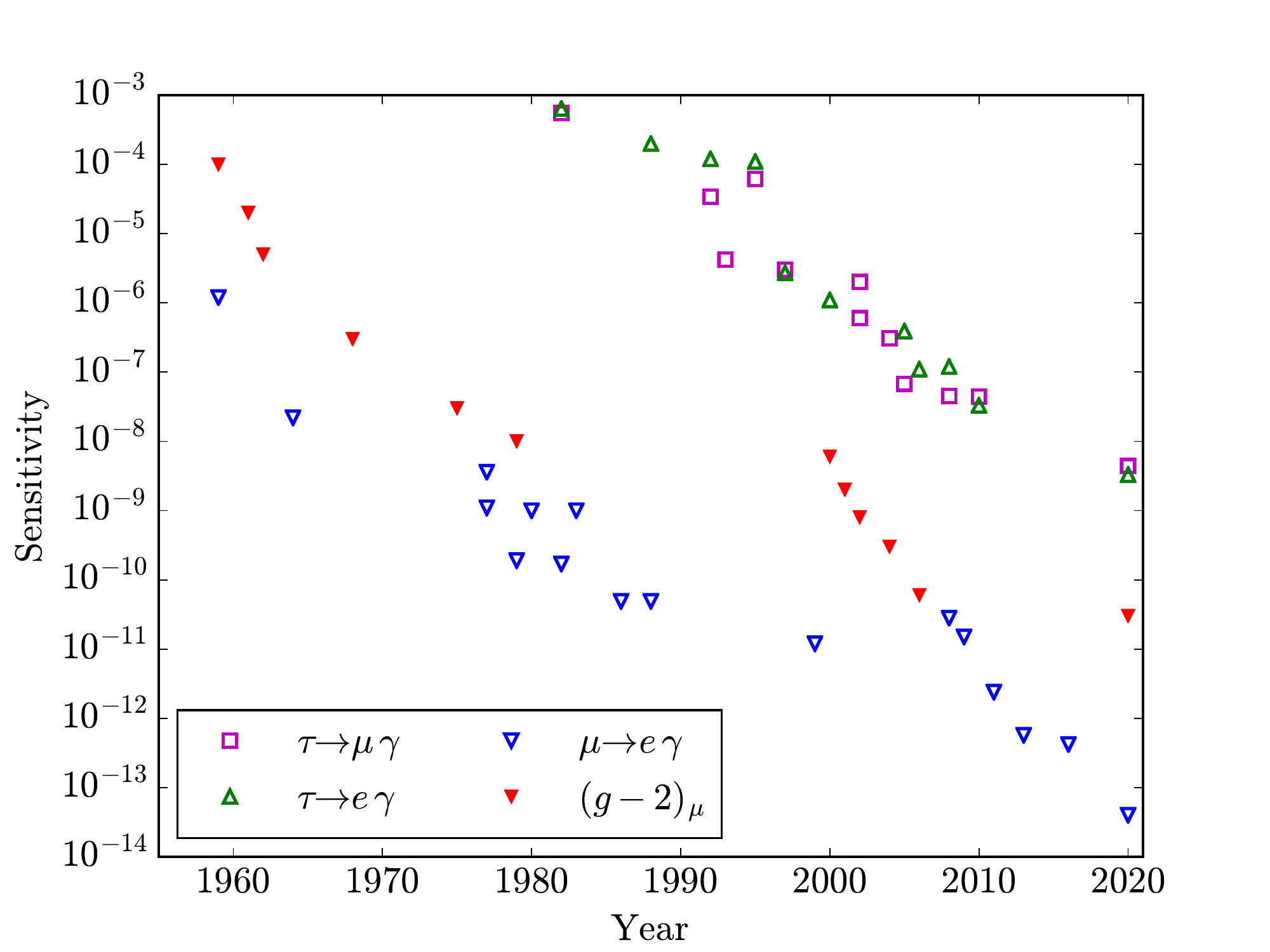}
  \caption{\label{fig:sensitivity}Historic development of the experimental sensitivities.}
\end{figure}

A multitude of experiments have measured the muon anomalous magnetic moment through the principle of Larmor precession, whose frequency is proportional to the magnetic field which the charged particle is subject to. The measurement of the muon anomalous magnetic moment is illustrated in Fig.~\ref{fig:2}. After protons hit a target, charged pions are produced which then decay into polarized muons which are used by an injector which injects muons into the storage ring to which a uniform magnetic field ($\overrightarrow{B}$) perpendicular to muon spin and orbit plane is applied. Using a vertically focused quadrupole electric field $\overrightarrow{E}$, one can find the frequency difference between the spin precession ($\overrightarrow{w_{a_{\mu}}}$) and the  the cyclotron motion~\cite{Zhang:2008pka},
\begin{equation}
\overrightarrow{w_{a_{\mu}}}=\frac{e}{m_{\mu}} \left[ a_{\mu} \overrightarrow{B} - \left(a_{\mu}- \frac{1}{\gamma^2-1}\right) \overrightarrow{v}\times \overrightarrow{E}\right],
\end{equation}
where $\gamma=(1-v^2)^{-1/2}$, with $v$ being the muon velocity. The fundamental idea concerning the measurement of $a_{\mu}$ consist of tuning the muon velocity such that $\gamma=29.3$, removing the dependence on the electric field. This particular value is known as ``magic $\gamma$''~\cite{Bargmann:1959gz}. Next, one needs to measure the frequency $\overrightarrow{w_{a_{\mu}}}$ with high precision and extract $a_{\mu}$. In Tab.~\ref{tab:amuOverview} we present a comprehensive historic perspective of $a_{\mu}$ measurements going back to the first measurement in 1957. Interestingly, two Nobel prize winners (Leon Lederman, 1988 and Georges Charpak, 1992) were at some point involved in the measurement of the muon anomalous magnetic moment. In Fig.~\ref{fig:sensitivity} one can easily see how the sensitivity has improved with time.

The most recent measurement comes from BNL (2006) data which found $a_{\mu}^{exp}=(116592089 \pm 63)\times 10^{-11}$, i.e.~$\delta a_{\mu}^{exp}= 63 \times 10^{-11}$, reaching unprecedented sensitivity. The Muon $g-2$ Experiment at Fermilab (FNAL) aims to improve the statistical error by a factor of four, reaching a precision of $\pm 0.14$~ppm, which translates into $\delta a_{\mu}^{exp}= 16 \times 10^{-11}$~\cite{Venanzoni:2014ixa,Anastasi:2015oea,Korostelev:2016qaf,Gray:2015qna}. The FNAL functioning is similar to that illustrated in Fig.~\ref{fig:2} since the FNAL experiment uses the BNL ring, which was brought to the FERMILAB site. In other words, FNAL is a more sophisticated version of the BNL experiment.

If everything goes smoothly, the first results are expected to be announced around the beginning of 2019~\cite{FNALtalk}, which will be followed by two other publications, in the course of a few years, aiming to reduce the systematic uncertainties by a factor of three and possibly achieve a $\pm 0.1$~ppm statistical precision~\cite{FNALtalk}. An important cross-check will be performed by the J-PARC experiment, located in Tokai, Japan, which initially plans to reach a statistical precision of $0.37$~ppm, and should start taking data around 2020-2022. Its final goal is similar to FNAL, i.e.~to reduce the statistical uncertainty down to a $0.1$~ppm precision, as well as the systematics by a factor of three. We highlight that J-PARC experiment features a different setup  though~\cite{Ishida:2009zz,Mibe:2010zz,Iinuma:2011zz,Saito:2012zz,Eads:2015arb}, because it uses incident muons with much lower energies compared to FNAL~\cite{Ishida:2009zz,Mibe:2011zz}, a stronger magnetic field, and it does not adopt the ``magic $\gamma$" approach. Rather, it will run with with zero electric field. Consequently, its systematic errors are also distinct~\cite{JPARCproposal,Otani:2015jra,Tanida:2016ryv}. In the foreseeable future these two flagship experiments will play an important role in particle physics regardless which direction their measurement will point to, but if indeed the central value remains roughly the same, the significance of the anomaly will be around or over $5\sigma$, 
 constituting a strong call for new physics~\cite{Blum:2013xva}, which is the focus of the following section. 

\subsection{\label{Secg2new}An Augury for New Physics: The Muon Anomalous Magnetic Moment}

Comparing the SM prediction with the recent measurement from Brookhaven National Lab we find two values for the discrepancies depending on the value used for the hadronic vacuum polarization~\cite{Blum:2013xva},

\begin{equation}
\begin{aligned}
\Delta a_{\mu} &= a_{\mu}^{exp} -a_{\mu}^{SM} = (287 \pm 80 ) \times 10^{-11}\, (3.6 \sigma),\\
\Delta a_{\mu} &= a_{\mu}^{exp} -a_{\mu}^{SM} = (261 \pm 78 ) \times 10^{-11}\, (3.3 \sigma),
\end{aligned}
\end{equation}
which stands for the $3.6\sigma$ and $3.3 \sigma$ deviations from the SM predictions, respectively. The significance of the excess can, however, be dwindled with the use of $\tau$ data in the hadronic contributions to $2.4\sigma$~\cite{Beringer:1900zz}. Conversely, using recent results on the lowest-order hadronic evaluation, the difference between the SM prediction and the experimental value becomes larger yielding a significance of over $4\sigma$~\cite{Benayoun:2012wc,Benayoun:2015gxa}.  Unfortunately, the hadronic corrections, which are the main source of error in the SM prediction, result in uncertainties that mask the impact of the deviation.

Anyhow, the muon anomalous magnetic moment has triggered various interpretations in terms of new physics effects. Fortunately, we are currently at a very special moment because both experiment, FNAL and J-PARC are expected to reach unprecedented sensitivity and report results in the upcoming years. If the current anomaly is confirmed, the beginning of a new era might be ahead of us. Therefore, it is timely sensitive to discuss models which could possibly accommodate the deviation and their implications in a broad sense. (see~\cite{Czarnecki:2001pv,McKeen:2009ny,Jegerlehner:2009ry,
Branco:2011iw,Cho:2011rk,Miller:2012opa,Stockinger:2013rna,
Blum:2013xva,Fargnoli:2013zia,Fargnoli:2013fra,Agrawal:2014ufa,
Freitas:2014pua,Chiu:2014oma,Queiroz:2014zfa,Aboubrahim:2014hya} for reviews on $g-2$ with different focus and~\cite{Gomes:2016ixw} for a discussion in the context of Lorentz symmetry breaking). However, in the attempt to address the muon anomalous magnetic moment in the context of new physics, there are often stones encountered on the way, namely constraints stemming from LFV probes. Therefore, in the next section, we put the muon anomalous magnetic moment into perspective with LFV.

\begin{table}[!h]
\begin{center}
Historic overview of constraints on $\mu \rightarrow e \gamma$.
\begin{tabular}{|l|c|l|}
\hline
Determination & Beam & ${\rm BR} (\mu \rightarrow e \gamma) <$\\ \hline
\hline
SM~\cite{Cheng:1977nv} & & $10^{-55}$\\
\hline
\hline
AFFLM ($1959$)~\cite{Meyer:1959zz} & $\mu^+$ & $ 1.2 \times 10^{-6}$ \\
\hline
PAR ($1964$)~\cite{Parker:1964zz} & $\mu^+$ & $2.2 \times 10^{-8}$ \\ 
\hline
NaI  ($1977$)~\cite{Depommier:1977yk} & $\mu^+$ & $3.6 \times 10^{-9}$ \\
\hline
SIN ($1977$)~\cite{Povel:1977sk} & $\mu^+$ & $1.1 \times 10^{-9}$ \\
\hline
Clinton Anderson ($1979$)~\cite{Bowman:1979bj} & $\mu^+$ & $1.9 \times 10^{-10}$ \\
\hline
SIN ($1980$)~\cite{vanderSchaaf:1979hz} & $\mu^+$ & $1 \times 10^{-9}$ \\
\hline
NaI ($1982$)~\cite{Kinnison:1981gc} & $\mu^+$ & $1.7 \times 10^{-10}$\\ 
\hline
TRIUMF ($1983$)~\cite{Azuelos:1983wx} & $\mu^+$ & $1 \times 10^{-9}$ \\
\hline
NaI-2 ($1986$)~\cite{Bolton:1986tv} & $\mu^+$ & $4.9 \times 10^{-11}$\\
\hline
Crystal Box ($1988$)~\cite{Bolton:1988af} & $\mu^+$ & $4.9 \times 10^{-11}$\\
\hline
MEGA ($1999$)~\cite{Brooks:1999pu} & $\mu^+$ & $1.2 \times 10^{-11}$\\
\hline
MEG ($2008$)~\cite{Adam:2009ci} & $\mu^+$ & $2.8 \times 10^{-11}$\\
MEG ($2009$)~\cite{Golden:2011zz} & $\mu^+$ & $1.5 \times 10^{-11}$\\
\hline
MEG ($2011$)~\cite{Adam:2011ch} & $\mu^+$ & $2.4 \times 10^{-12}$\\
\hline
MEG ($2013$)~\cite{Adam:2013mnn} & $\mu^+$ & $5.7 \times 10^{-13}$\\
MEG ($2016$)~\cite{TheMEG:2016wtm} & $\mu^+$ & $4.2 \times 10^{-13}$\\
\hline
\hline
Current Bound~\cite{TheMEG:2016wtm} & &  ${\rm BR (\mu \rightarrow e \gamma) < 4.2\times 10^{-13} } $ \\
\hline
Projected Sensitivity~\cite{Mori:2016vwi} & &  ${\rm BR (\mu \rightarrow e \gamma) < 4\times 10^{-14} } $ \\
\hline
\end{tabular}
\end{center}
\caption{\label{tab:LFVOverviewmu}Comprehensive overview of constraints on the decay $\mu \rightarrow e \gamma$.}
\end{table}

\begin{table}[p]
\begin{center}
Complete overview of constraints on the  $\tau \rightarrow X \gamma$ decay, $X=e,\mu$.
\begin{tabular}{|l||l|}
\hline
Determination &  Bound\\ \hline
\hline
MARK-II (1982)~\cite{Hayes:1981bn,Hayes:1981ta}  & $ {\rm BR}(\tau \rightarrow \mu\gamma) <5.5 \times 10^{-4}$  \\
MARK-II (1982)~\cite{Hayes:1981bn,Hayes:1981ta}  & $ {\rm BR}(\tau \rightarrow e\gamma) <6.4 \times 10^{-4}$  \\
CRYSTAL BALL (1988)~\cite{Keh:1988gs}& ${{\rm BR}(\tau \rightarrow e\gamma) <2 \times 10^{-4}}$\\
\hline
ARGUS (1992)~\cite{Albrecht:1992uba} & $ {\rm BR}(\tau \rightarrow \mu\gamma) <  3.4\times 10^{-5}$  \\
ARGUS (1992)~\cite{Albrecht:1992uba} & $ {\rm BR}(\tau \rightarrow e\gamma) <  1.2\times 10^{-4}$  \\
\hline
CLEO-II (1993)\cite{Bean:1992hh} & $ {\rm BR}(\tau \rightarrow \mu\gamma) < 4.2 \times 10^{-6}$  \\
\hline
DELPHI (1995)\cite{Abreu:1995gs} & $ {\rm BR}(\tau \rightarrow \mu\gamma) < 6.2 \times 10^{-5}$  \\
DELPHI (1995)\cite{Abreu:1995gs} & $ {\rm BR}(\tau \rightarrow e\gamma) < 1.1 \times 10^{-4}$  \\
\hline
CLEO-II (1997)\cite{Edwards:1996te} & $ {\rm BR}(\tau \rightarrow \mu\gamma) < 3 \times 10^{-6}$  \\
CLEO-II (1997)\cite{Edwards:1996te} & $ {\rm BR}(\tau \rightarrow e\gamma) < 2.7 \times 10^{-6}$  \\ 
CLEO-II (2000)\cite{Ahmed:1999gh} & $ {\rm BR}(\tau \rightarrow e\gamma) < 1.1 \times 10^{-6}$  \\ 
\hline
BaBar ($2002$)~\cite{Brown:2002mp} & $ {\rm BR}(\tau \rightarrow \mu\gamma) < 2 \times 10^{-6}$  \\
\hline
Belle ($2002$)~\cite{Inami:2002us} & $ {\rm BR}(\tau \rightarrow \mu\gamma) < 6 \times 10^{-7}$  \\
Belle ($2004$)~\cite{Abe:2003sx} & $ {\rm BR}(\tau \rightarrow \mu\gamma) < 3.1 \times 10^{-7}$  \\
Belle ($2005$)~\cite{Hayasaka:2005xw} & ${\rm BR}(\tau \rightarrow e\gamma) < 3.9 \times 10^{-7}$  \\
\hline
BaBar ($2005$)~\cite{Aubert:2005ye} & $ {\rm BR}(\tau \rightarrow \mu\gamma) < 6.8 \times 10^{-8}$  \\
BaBar ($2006$)~\cite{Aubert:2005wa} & $ {\rm BR}(\tau \rightarrow e\gamma) < 1.1 \times 10^{-7}$  \\
\hline
Belle ($2008$)~\cite{Hayasaka:2007vc} & $ {\rm BR}(\tau \rightarrow \mu\gamma) < 4.5 \times 10^{-8}$  \\
Belle ($2008$)~\cite{Hayasaka:2007vc} & $ {\rm BR}(\tau \rightarrow e\gamma) < 1.2 \times 10^{-7}$  \\
\hline
BaBar ($2010$)~\cite{Aubert:2009ag} & $ {\rm BR}(\tau \rightarrow \mu\gamma) < 4.4 \times 10^{-8}$  \\
BaBar ($2010$)~\cite{Aubert:2009ag} & $ {\rm BR}(\tau \rightarrow e\gamma) < 3.3 \times 10^{-8}$  \\
\hline
Current Bound~\cite{Aubert:2009ag}  &  ${\rm BR (\tau \rightarrow \mu \gamma) < 4.4\times 10^{-8} } $\\  
				    &  ${\rm BR (\tau \rightarrow e \gamma) < 3.3 \times 10^{-8}}$ \\
\hline
Projected Sensitivity~\cite{Hayasaka:2012pj}  &   ${\rm BR (\tau \rightarrow \mu \gamma) < 4.4 \times 10^{-9} } $\\   
					      &  ${\rm BR (\tau \rightarrow e \gamma) < 3.3 \times 10^{-9}}$ \\
\hline
\end{tabular}
\end{center}
\caption{Comprehensive overview of constraints on the decay $\tau \rightarrow e \gamma$. The projected sensitivity comes from  superKEKB/Belle II which is expected to improve existing limits by a factor of 10~\cite{Hayasaka:2012pj}.}
\label{tab:LFVOverviewtau}
\end{table}

\subsection{\label{LFVintro}Lepton Flavor Violation}

Lepton flavor violation is known today as a canonical search for new physics. However, its history goes far beyond simply LFV and has led us to a much better understanding of the fundamental laws of nature. The first search for the radiative decay of the muon was conducted in 1948~\cite{Hincks:1948vr}. Since it was known at the time that muon had a mass of about 200 times of the electron, then one could imagine that the muon could decay into an electron and a neutral particle. This is a landmark event in the search for LFV. Shortly after the search for $\mu-e$ conversion were also conducted.  In particular, the first upper limit obtained in 1955 read ${\rm Br} \mu \to e\gamma < 10^{-5}$~\cite{Bernstein:2013hba}.\footnote{This bound supposedly obtained in 1955 was presented in a talk, but there is no manuscript available~\cite{Bernstein:2013hba} and for this reason this result was removed from Tab.~\ref{tab:LFVOverviewmu}}.As for the $\mu-e$ conversion the first bound obtained in 1955 read ${\rm CR}(\mu\, \text{Cu} \to e\, \text{Cu}) < 5 \times 10^{-4}$~\cite{Steinberger:1955hfk}. 


Therefore, instead of reviewing the muon anomalous magnetic moment individually, it is perhaps more important to review its connection to the quest for LFV, such as the decay $\MEG$~\cite{Marciano:1977wx,Ilakovac:1994kj,Gomez:1998wj,Feng:1999wt,Kuno:1999jp,Petcov:2003zb,Pascoli:2003uh,Cirigliano:2005ck,Marciano:2008zz,
Casas:2010wm,Deppisch:2012vj,Davidson:2016edt}. In the SM, lepton flavor is a conserved quantity since neutrinos are massless. However, we know that neutrinos do have masses and that they experience flavor oscillations~\cite{Fukuda:1998mi,Ahmad:2002jz}. Though we have experimental confirmation of LFV from neutrino oscillations, we have not yet observed such violation in processes involving charged leptons.\footnote{Notice that LFV does not imply that neutrinos oscillate, since there are models where LFV occurs but neutrinos remain massless~\cite{Marciano:2008zz,Davidson:2016utf}.} Nevertheless, if LFV occurs among neutrino flavors, it is arguably natural to expect that this violation also happens among charged leptons. From now on LFV will be regarded in the context of charged leptons.

In Fig.~\ref{figure3LFV} we illustrate in a nutshell the effects that LFV and $g-2$ observables bring to new physics. While the non-observation of LFV typically pushes new physics to high energy scales, as indicated in Fig.~\ref{figure3LFV}, the $g-2$ anomaly favors lower energy scales. Thus, one may wonder whether the same new physics can still plausibly accommodate signals in both fronts.

Keep in mind though that, converse to the muon magnetic moment, an observation of LFV would undoubtedly confirm the existence of new physics with tremendous implications. Many new physics models can accommodate LFV processes. Several LFV processes other than $\MEG$ have been experimentally searched for, such as $\tau \rightarrow e\gamma$,  $\tau \rightarrow \mu\gamma$, $\mu^+ \to e^+e^+e^-$ and $\tau \rightarrow  \mu\mu\mu$, and since no signal was observed stringent limits were derived~\cite{Aubert:2009ag} (See a sensitivity perspective in Fig.~\ref{fig:sensitivity}). In particular the $\tau$ decays, which are intimately connected to our reasoning in the context of $g-2$, yield much weaker limits. For completeness, we provide in Tab.~\ref{tab:LFVOverviewtau} a complete overview of all existing limits on this LFV decay mode. Eventually, we will discuss $SU(2)_L$ invariant simplified models, and when we do we will have the limits on the ${\rm BR} (\tau \rightarrow X \gamma)$ in mind, but it is clear that they will impose no further constraints on the model in the light of the loose constraints stemming from this decay. 

In summary, in this review we will focus on LFV involving the charged leptons. Since we are also investigating the interplay with $g-2$, the $\mu \to e\gamma$ decay will be the primary focus, but we will also approach other LFV observables such as $\mu^+ \to e^+e^+e^-$ and $\mu-e$ conversion, specially when we address UV complete models.  We start with the LFV muon decays.

\subsection{\label{LFVexp}The Decays $\mu \rightarrow e \gamma$ and $\mu \to 3e$}

The well known normal muon decay $\mu \to e\overline{\nu}_{e}\nu_\mu$, also known the Michel decay, where the muon decays into a muon-neutrino plus an electron accompanied by a electron-anti-neutrino is a purely leptonic processes governed by weak interactions and for this reason is subject to precise measurements with high statistics.  In fact, the precise measurements on the muon decay were crucial to our understanding of the Standard Model. For instance, the Fermi coupling constant, which is one of the most precisely measured quantities in nature, is extracted from muon lifetime ($\tau_\mu$) that reads
\begin{equation}
\tau_\mu = \frac{G_F^2 m_\mu^5}{192\pi^3}F\left(\frac{m_e^2}{m_\mu^2}\right)\left(1+\frac{3m_\mu^2}{5 m_W^2}\right) \left[1+ \frac{\alpha_{em}(m_\mu)}{2\pi}\left(\frac{25}{4}-\pi^2\right)\right],
\end{equation}
where $F(x)=1-8x+8x^3-x^4-12x^2\ln(x)$.

By measuring the muon mass using a muonium atom with QED corrections~\cite{Cohen:1987fr} and knowing the electromagnetic fine-structure constant, $\alpha_\text{em}$, at the muon mass scale one can derive $G_F$~\cite{Olive:2016xmw}. The precise determination of $G_F$ is key to perform high precision test to the Standard Model.

The Michel decay of the muon is the dominant muon decay mode and accounts for nearly 100\% of the branching ratio. The second most relevant muon decay is the $\mu \to e\bar{\nu_e} \nu_\mu\gamma$ final state which is around 1\%. Therefore, the muon decay is a great laboratory to test news physics effects, in particular LFV.

\begin{table}[t]
\begin{center}
Historic overview of constraints on $\mu^+ \rightarrow e^+e^+e^-$.
\begin{tabular}{|l|c|l|}
\hline
Determination  & Location (Year) & ${\rm BR (\mu^+ \rightarrow e^+e^+e^-)} $\\ 
\hline
\hline
Ref.~\cite{Korenchenko:1975gf} 		& JINR (1976)  		& $<1.9\cdot 10^{-9}$ \\
\hline
Crystal Box~\cite{Bolton:1984qr}  				& LANR (1984)	 	& $<1.3 \times 10^{-10}$ \\
\hline
SINDRUM~\cite{Bertl:1984wi}				& PSI (1984)		& $<1.6 \times 10^{-10}$ \\
SINDRUM~\cite{Bertl:1985mw}			& PSI (1985)		& $<2.4 \times 10^{-12}$ \\
\hline
Crystal Box~\cite{Bolton:1988af}  		& LANR (1988)	 	& $<3.5 \times 10^{-11}$ \\
\hline
SINDRUM~\cite{Bellgardt:1987du}		& PSI (1988)		& $<1.0 \times 10^{-12}$ \\
\hline
Ref.~\cite{Baranov:1990uh} 				& JINR (1991)  		& $<3.6\times 10^{-11}$ \\
\hline
\hline
Current Bound~\cite{Bellgardt:1987du} & &  $\rm BR (\mu \rightarrow 3 e) < 1.0 \times 10^{-12} $\\
\hline
Projected Sensitivity~\cite{Blondel:2013ia} & &  $\rm BR (\mu \rightarrow 3 e) \lesssim 1 \times 10^{-16} $ \\
\hline
\end{tabular}
\end{center}
\caption{\label{tab:mu3everview}Comprehensive overview of constraints on the decay $\mu^+ \rightarrow e^+e^+e^- \gamma$.}
\end{table}

Several searches for LFV have been performed, mostly based on the positively charged muon decay since it provides a unique opportunity to search for LFV compared with other channels.\footnote{A negative muon it is captured by a nucleus when it is stopped in material~\cite{Kuno:1999jp}.}\footnote{$\mu\rightarrow e$ conversion provides a complementary probe to LFV, and, if a great improvement in the bounds occurs, it may give rise to leading constraints for some new physics models.} This is due to the large number of muons available for experimental searches~\cite{Kuno:1999jp}. A positively charged muon decays at rest producing collinearly a $e^+\gamma$ pair with energies equal to half of the muon mass. The signal is clean, and since no excess events has been observed thus far, stringent limits were placed on the branching ratio for the decay $\MEG$ [${\rm BR}(\MEG)$]. See Tab.~\ref{tab:LFVOverviewmu} for a comprehensive overview of all existing bounds. Nevertheless, we know that in the SM with massless neutrinos this flavor violating decay is non-existent. Moreover, this decay is still extremely suppressed with the neutrino masses and mixing angles in agreement with atmospheric and solar-neutrino data~\cite{Bilenky:1977du}. Even in the popular canonical type~I seesaw mechanism with heavy right-handed (RH) neutrinos, the branching ratio of the decay $\mu \rightarrow e\gamma$ is many orders of magnitude below current sensitivity  limits~\cite{Cheng:1980tp}. Hence, the observation of LFV in the foreseeable future would be a paramount event since it conclusively confirms the presence of new physics, beyond the type~I seesaw mechanism. 

Another important LFV observable is the $\mu^+ \to e^+e^+e^-$ decay, which also naturally appears in models where $\mu \to e$ occurs. The $\mu^+ \to e^+e^+e^-$ decay --~different from the $\mu^+ \to e^+\gamma$ decay --~can receive contributions from off-shell photons, $Z$-penguins, and box diagrams, in addition to the photonic penguin diagram that gives rise to $\mu^+ e^+\gamma$ decay. Hence, the direct connection to $\mu^+ \to e^+\gamma$ and $g-2$ is lost in a general setup. Since one of the primary focus on this review is the muon magnetic moment, we will address the interplay between g-2, $\mu^+ \to e^+e^+e^-$ and $\mu^+ \to e^+\gamma$ only when we have a complete model under investigation.

\subsection{$\mu-e$ Conversion}
While the decay $\MEG$ currently provides the strongest constraints and is particularly interesting due to its relation with the $g-2$, one should have in mind that $\mu^- - e^-$ conversion in nuclei might in some cases give rise to stronger bounds. We will check some concrete realizations when we investigate UV complete models. See \cite{Bartolotta:2017mff} for a recent discussion at next-leading order. 

The $\mu^- - e^-$ process refers to the the process where a negative muon is stopped in some material and trapped, forming a muonic atom. After the muon cascades down to the ground state, the muon either experiences its  Michel decay, $\mu \to e\bar{\nu_e}\nu_\mu$, or it is captured by a nucleus of atomic mass A and atomic number Z, thus,
\begin{equation}
\mu^- + {\rm Nucleus} (A,Z) \to \nu_\mu + {\rm Nucleus} (A,Z-1).
\end{equation}
Albeit, in the context of LFV, we should also expect the $\mu^- - e^-$ transition,
\begin{equation}
\mu^- + {\rm Nucleus} (A,Z) \to e^- + {\rm Nucleus} (A,Z)
\end{equation}
which violates both electron and muon lepton flavors.

\begin{table}[t]
\begin{center}
Historic overview of constraints on $\mu^- - e^- $ conversion.
\begin{tabular}{|l|c|l|}
\hline
Determination & Location (Year) & ${\rm CR (\mu^- + N \rightarrow e^- + N)} $\\
\hline
\hline
$\mu^- + \mathrm{Cu} \to e^- + \mathrm{Cu}$~\cite{Steinberger:1955hfk} 							& (1955) & $<5 \times 10^{- 4}$\\
\hline
$\mu^- + \mathrm{Cu} \to e^- + \mathrm{Co}$~\cite{Bryman:1972rf} 									& SREL (1972) & $<1.6 \times 10^{-8}$\\
\hline
$\mu^-  + {}^{32}\mathrm{S} \to e^- + {}^{32}\mathrm{S}$~\cite{Badertscher:1981ay}		& SIN (1982) & $<7 \times 10^{-11} $ \\
\hline
$\mu^- + \mathrm{Ti} \to e^- + \mathrm{Ti}$~\cite{Bryman:1985fq} 									& TRIUMF (1985) 	& $<1.6 \times 10^{-11}$\\
$\mu^- + \mathrm{Ti} \to e^- + \mathrm{Ti}$~\cite{Ahmad:1988ur} 									& TRIUMF (1988) 	& $<4.6 \times 10^{-12}$\\
$\mu^- + \mathrm{Ti} \to e^- + \mathrm{Ti}$~\cite{Dohmen:1993mp} 									& PSI (1993) 			& $<4.3 \times 10^{-12}$\\
$\mu^- + \mathrm{Ti} \to e^- + \mathrm{Ti}$~\cite{Wintz:1998rp} 										& PSI (1998) 			& $<6.1 \times 10^{-13}$\\
\hline
$\mu^- + \mathrm{Pb} \to e^- + \mathrm{Pb}$~\cite{Ahmad:1988ur} 									& TRIUMF (1988) 	& $<4.9 \times 10^{-10}$\\
$\mu^- + \mathrm{Pb} \to e^- + \mathrm{Pb}$~\cite{Honecker:1996zf} 								& PSI (1996) 			& $<4.6 \times 10^{-11}$\\
\hline
$\mu^- + \mathrm{Au} \to e^- + \mathrm{Au}$~\cite{Bertl:2006up} 									& PSI (2006) 			& $<7.0 \times 10^{-13}$\\
\hline
\hline
Current Bound~\cite{Wintz:1998rp} & &  $\rm CR (\mu^-\, \mathrm{Ti} \rightarrow e^-\,  \mathrm{Ti}) < 6.1 \times 10^{-13} $ \\
\hline
Projected Sensitivity~\cite{Bernstein:2013hba} & &  $\rm CR (\mu^-\, \mathrm{Al} \rightarrow e^-\,  \mathrm{Al}) \lesssim 10^{-17}$ \\
\hline
\end{tabular}
\end{center}
\caption{\label{tab:mu-eOverview}Comprehensive overview of constraints on the $\mu^- -e^- $ conversion.}
\end{table}

In this way, the branching ratio of the $\mu^- -e^-$ conversion is found to be,
\begin{equation}
{\rm CR}(\mu-e)= \frac{\Gamma(\mu^- + {\rm Nucleus} (A,Z) \to e^- + {\rm Nucleus} (A,Z))}{\Gamma(\mu^- + {\rm Nucleus} (A,Z) \to \nu_\mu + {\rm Nucleus} (A,Z-1))}.
\label{defmueconversion}
\end{equation}
The bounds on this branching ratio strongly depends on the nucleus under consideration. For instance, for the Titanium nucleus, this branching ratio should be smaller than $6.1 \times 10^{-13}$. This current bounds is not very restrictive, specially having in mind the bound the $\mu \to e\gamma$ decay. However, there is an experimental effort under way to improve this bound in four to five orders of magnitude (see Table~\ref{tab:mu-eOverview} for a historic perspective) \footnote{We adoted the projected limit of $10^{-17}$ but there are old proposals that feature a sensitivity of $10^{-18}$ \cite{Aysto:2001zs,Bernstein:2013hba,deGouvea:2013zba}.} . In this projected scenario $\mu-e$ conversion becomes a powerful tool in the quest for LFV. Furthermore, we will put the $\mu-e$ observable into perspective with the other observables addressed in this review.

Before we turn to our analysis, we shall now describe the foundation of the complementarity between the $g-2$ and LFV observables in more detail.

\section{\label{sec:framework}General Framework}

It is instructive to first take an effective field theory (EFT) viewpoint since it highlights the intimate relation of LFV decays and the anomalous magnetic moments of the leptons.

Assuming only conservation of charge and Lorentz invariance, the effective operators relevant for the on-shell transition $\MEG$ and the $g-2$ are
\begin{equation}
  \mathcal{L}_\textrm{eff}^\text{photonic, on-shell} = \frac{\mu^M_{ij}}{2} \overline{\ell_i} \sigma^{\mu\nu} \ell_j F_{\mu\nu} + 
    \frac{\mu^E_{ij}}{2} \overline{\ell_i} i \gamma^5 \sigma^{\mu\nu} \ell_j F_{\mu\nu},
\end{equation}
where the diagonal elements in the transition magnetic moment $\mu^M$ generate the anomalous magnetic dipole moments $\Delta a = \frac{1}{2}(g-2)$ of the leptons. Similarly, the flavor-diagonal part of $\mu^E$ gives contributions to the electric dipole moments, which we disregard in the present work. The off-diagonal elements, on the other hand, contribute to LFV decays such as $\MEG$. If $m_i \gg m_j$, it is convenient to define the dipole form factors $A_M$ and $A_E$ such that, neglecting contributions proportional to $m_j$, one may write $\mu^{M/E}_{ij} \equiv e m_i A^{M/E}/2$. Notice that $A_M$ and $A_E$ refer to parity conserving and non-conversing operators, respectively. With this definition, one obtains the following expressions for the anomalous magnetic moment and the branching ratio of the LFV decay:
\begin{subequations}
\begin{align}
  \Delta a_{\ell_i} &= A^M_{ii} m_{i}^2 \textrm{ (no sum)}, \\
  \mathrm{BR}\left(\ell_i\to \ell_j\gamma\right) &= \frac{3(4\pi)^3\alpha_\mathrm{em}}{4 G_F^2} \left(\left|A^M_{ji}\right|^2 + \left|A^E_{ji}\right|^2\right) \underbrace{\mathrm{BR}\left(\ell_i\to \ell_j \nu_i \overline{\nu_j}\right)}_{\approx 1 \textrm{ for $\MEG$}}.
\end{align}
\label{Eq.general}
\end{subequations}
Here, $G_F$ is Fermi's constant of weak interactions and $\alpha_\mathrm{em}$ is the electromagnetic fine-structure constant. 

Naturally, any flavor non-diagonal coupling will activate LFV decays and at the same time yield contributions to the anomalous magnetic moments of the leptons. This interesting observation is the foundation of our present work in which we give fully general and yet compact expressions for the evaluation of LFV decays as well as the contributions to $g-2$. This should enable the readership to easily apply our results to any model. We point out that this approach is complementary to the effective field theory approach of Ref.~\cite{Davidson:2016edt}. Furthermore, using the simplified model description guarding $SU(2)_L$ invariance, we believe that the results become more intuitive, motivating our work from a practical point of view.

Be aware that there is a multitude of collider and electroweak precision limits that can be applied to many of the models we describe here. Since they are rather model dependent and our aim is to describe the correlation between the $g-2$ and $\mu \rightarrow e \gamma$ in a simplified framework, we have decided to leave those out of our discussion. Rather, we emphasize that, depending on the context in which the simplified models are embedded, the region of parameter space where one can accommodate $g-2$ and/or $\mu \rightarrow e \gamma$ might be actually ruled out. See~\cite{Freitas:2014pua} for a recent discussion in reference to LEP bounds.

As discussed in  the previous section, the LFV processes $\mu N \to e N$ and $\mu \to 3e$ also involve off-shell contributions from photon exchange, and it is instructive to consider the form of the amplitude for a generic photon leg $A_\mu(q)$, where $q\equiv p_j - p_i$~\cite{Kuno:1999jp}:
\begin{equation}
  \begin{aligned}
  \mathcal{A}^\text{photonic} = -e A_\mu^*(q) \overline{u}_{\ell_j}(p_j) &\left[ \left( f^{ji}_{E0}(q^2) + \gamma_5  f^{ji}_{M0}(q^2)\right)\left(\gamma^\mu-\frac{\slashed{q} q^\mu}{q^2}\right)\right. +\\
    &\quad+ \left.\left(  f^{ji}_{M1}(q^2) + \gamma_5  f^{ji}_{E1}(q^2)\right) \frac{i \sigma^{\mu\nu}q_\nu}{m_i} \right] u_{\ell_i}(p_i)\,.
  \end{aligned}
\end{equation}
The form factors $f_{E0}$ and $f_{M0}$ account for the off-shell photon and do not correspond to real photon emission. Hence, they should vanish when $q^2 \rightarrow 0$, and in this low momentum limit we can recover the on-shell photon contributions with the identifications $A^E = -i m_i^2 f_{E1}(0)$ and $A^M = m_i^2  f_{M1}(0)$. 

In a scenario where the $f_{E/M1}$ are dominating over the $f_{E/M0}$, one can derive the following relations between $\MEG$ and the other LFV transitions~\cite{Kuno:1999jp}:
\begin{subequations}
\begin{align}
  \mathrm{CR}(\mu\, \text{Ti }\to e\, \text{Ti}) &\simeq \frac{1}{200} \mathrm{BR}(\MEG)\,\label{eq:MuEconvEstimate} \\
  \mathrm{CR}(\mu\, \text{Al}\to e\, \text{Al}) &\simeq \frac{1}{350} \mathrm{BR}(\MEG)\,\\
  \mathrm{BR}(\mu \to 3 e) &\simeq \frac{1}{160} \mathrm{BR}(\MEG)\,.\label{eq:Mu3eEstimate}
\end{align}
\end{subequations}
We emphasize that such relations serve merely to have an idea up to an order of magnitude correction of the correlation between these observables. Some setups that we will discuss further predict branching ratios of the same order of magnitude. Moreover, we highlight the next generation of experiments such as Mu2e and COMET will use Aluminum as target aiming at a sensitivity of $\sim 10^{-17}$ on $\mu-e$ conversion \cite{Bernstein:2013hba}. However, there is an old proposal that uses Titanium instead, potentially being capable of reaching a projected bound of $\sim 10^{-18}$ \cite{Aysto:2001zs}. In this work, we will use the former, since it is up-to-date.

These relations will not hold in general and therefore no such relation exists for a generic set-up. For the former transition, one expects a simple result if the photonic, long-range contributions dominate~\cite{Geib:2015unm}:
\begin{equation}
	\mathrm{CR}(\mu\, N \to e\, N) = \frac{8 \alpha_\mathrm{em}^5 m_\mu Z_\text{eff}^4 Z F_p^2}{\Gamma_\text{capture}} \Xi^2_\text{particle}\,,
\end{equation}
where the effective atomic charge and the nuclear matrix element have to be determined from nuclear physics. The particle physics contribution factorizes and reads
\begin{equation}
	\Xi_\text{particle}^2 = \left| f_{E0}(m_\mu^2) + f_{M1}(m_\mu^2)\right|^2 + \left| f_{E1}(m_\mu^2) + f_{M0}(m_\mu^2)\right|^2\,.
\end{equation}

We now turn to the calculation of the contributions from individual fields, focusing for brevity on the on-shell photonic form factors.

\section{\label{sec:results}New Physics Contributions}
In this section we derive the new physics contributions to $g-2$ and $\mu \rightarrow e\gamma$. We focus for now on individual and multiple field corrections without imposing $SU(2)_L$ invariance to keep the results general and model independent, such that one can apply the results to a broader context. Only in Sec.~\ref{sec:numerics} we will put them into perspective preserving $SU(2)_L$ invariance in the context of simplified SM extensions. 

For now, we consider the most generic couplings that are allowed by the conservation of electric charge and Lorentz invariance. We start by addressing scalar mediators.

\subsection{Scalar Mediators}

\subsubsection{Neutral Scalar}

If additional electrically neutral scalar fields are present in a model, they will induce a shift in the leptonic magnetic moments and mediate LFV decays via the following interactions:
\begin{equation}
  \mathcal{L}_\textrm{int} = g_{s\,1}^{ij}  \phi\, \overline{\ell_i}\, \ell_j + i g_{p\,1}^{ij}  \phi\, \overline{\ell_i} \gamma^5 \ell_j,
  \label{Eq:neutralscalar}
\end{equation}
where both terms are manifestly hermitian. Notice that this sort of Lagrangian arises in many models. For instance, in $U(1)_X$ extensions of the SM, a scalar is usually needed to break the additional gauge symmetry spontaneously. If SM particles are charged under $U(1)_X$, interactions between the SM fermions and the new scalar arise, including the ones above. Moreover, in models where there is an inert scalar or the new neutral scalar mixes with the SM Higgs, this interaction Lagrangian also appears. In the latter case, however, the interaction strength is proportional to the (small) ratio $m_{f}/v$, where $v$ is the vacuum expectation value (VEV) of the Higgs and $m_f$ the fermion mass. 

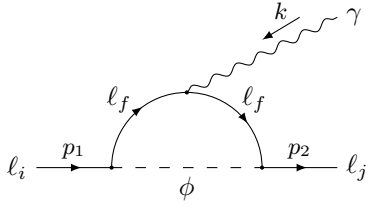
\begin{figure}[t]
  \centering
  \begin{tikzpicture}[x=1mm,y=1mm]
    \node[anchor=east] at (-20,0) (f1) {$\ell_i$};
    \node[anchor=west] at (20,0) (f2) {$\ell_j$};
    \node (V1) at (-10,0) {};
    \node (V2) at (10,0) {};
    \draw[fill=black] (V1) circle(.2);
    \draw[fill=black] (V2) circle(.2);
    \draw[decoration={markings, mark=at position 0.6 with {\arrow{latex}}}, postaction={decorate}] (f1) -- (V1.center) 
      node[midway, above] {\small$p_1$};
    \draw[mydash] (V1.center) -- (V2.center) node[midway, below] {$\phi$};	
    \draw[decoration={markings, mark=at position 0.6 with {\arrow{latex}}}, postaction={decorate}] (V2.center) -- (f2) 
      node[midway, above] {\small$p_2$};
    \draw[decoration={markings, mark=at position 0.6 with {\arrow{latex}}}, postaction={decorate}] (V1.center) arc(180:90:10) 
    node (V3) {};
    \draw[fill] (V3.center) circle(.2);
    \draw[decorate, decoration={snake, segment length=3.3mm, amplitude=.5mm}] (V3.center) -- (20,20) node[anchor=west] {$\gamma$}; 
    \draw[decoration={markings, mark=at position 0.6 with {\arrow{latex}}}, postaction={decorate}] (V3.center) arc(90:0:10);
    \draw[->, >=latex] (15,20) -- (10,17) node[midway, above] {\small$k$};
    \node at (-9,9) {$\ell_f$};
    \node at (9,9) {$\ell_f$};
  \end{tikzpicture}
  \caption{\label{fig:neutralScalar}Diagrammatic representation of amplitude for the process $\ell_i\to \ell_j \gamma$ in the presence of a new neutral scalar $\phi$.}
\end{figure}

Decomposing the amplitude depicted in Fig.~\ref{fig:neutralScalar} as
\begin{equation}
  -i\mathcal{M} = \overline{u}_j(p_2) \left(-i e \Gamma^\mu \right) u_i(p_1) \varepsilon_\mu(k),
\end{equation}
we obtain for the dipole part of the vertex function\footnote{This corresponds to the finite part of the amplitude proportional to $\sigma^{\mu\nu}k_\nu$. Note that all other contributions, especially those where the photon line is attached to the external leptons, yields non-dipole contributions ($\propto \gamma^\mu$). These are irrelevant for both $\MEG$ and $g-2$.}
\begin{align}
  \Gamma^\mu_1 = \frac{i \sigma^{\mu\nu}k_\nu}{8\pi^2} \frac{m_i}{2} \sum_f
    \Big[\underbrace{g_{s\,1}^{fj}g_{s\,1}^{fi} I_{f,\, 1}^{+\,+}+g_{p\,1}^{fj}g_{p\,1}^{fi} I_{f,\, 1}^{+\,-}}_{\equiv (4\pi)^2 A^M_{ji}} + i \gamma^5\underbrace{\left(g_{p\,1}^{fj}g_{s\,1}^{fi} I_{f,\, 1}^{-\,+} - g_{s\,1}^{fj}g_{p\,1}^{fi} I_{f,\, 1}^{-\,-}\right)}_{\equiv (4\pi)^2 A^E_{ji}} \Big]. \label{eq:Gamma1_result}
\end{align}
Note that we have identified the form factors $A_{ji}^M$ and $A_{ji}^E$ in Eq.~\eqref{eq:Gamma1_result}, where 
\begin{equation}\label{eq:dipoleFormFactors_neutralScalar}
\begin{aligned}
  A_{ji}^M=\frac{1}{(4\pi)^2}\left(g_{s\,1}^{fj}g_{s\,1}^{fi} I_{f,\, 1}^{+\,+}+g_{p\,1}^{fj}g_{p\,1}^{fi} I_{f,\, 1}^{+\,-}\right), \\
  A_{ji}^E=\frac{1}{(4\pi)^2}\left(g_{p\,1}^{fj}g_{s\,1}^{fi} I_{f,\, 1}^{-\,+} - g_{s\,1}^{fj}g_{p\,1}^{fi} I_{f,\, 1}^{-\,-}\right).
\end{aligned}
\end{equation}
The exact loop integral $I^{\pm\,\pm}_{f,\,1}$ is given in the appendix and for the special case of $\MEG$ it may be approximated as
\begin{align}
  I^{(\pm)_1\,(\pm)_2}_{f,\, 1} &\simeq \int_0^1 \mathrm{d}x \int_0^{1-x}\hspace{-5mm}\mathrm{d}z \frac{xz +(\pm)_2 (1-x)\frac{m_f}{m_\mu}}{-x z m_i^2+xm_\phi^2 + (1-x)m_f^2}\nonumber\\
  &=  \frac{1}{m_\phi^2}\int_0^1\mathrm{d}x\int_0^1\mathrm{d}y\, x^2 \frac{(1-x)y +(\pm)_2 \epsilon_f}{(1-x)(1-xy\lambda^2)+x\epsilon_f^2\lambda^2}\nonumber\\ 
  &= \frac{1}{m_\phi^2} \left[ \frac{1}{6} + (\mp)_2 \epsilon_f \left(\frac{3}{2} +\log(\epsilon_f^2 \lambda^2) \right)\right] \textrm{, for } m_\phi \to \infty.\label{eq:I1_simplified_mediator}
\end{align}
Here, we work to leading order in $m_j/m_i \ll 1$ and simplify the integration by defining $z \equiv (1-x) y$. Finally, we have used that the resulting integral is invariant under $x \to (1-x)$, and defined $\epsilon_f\equiv \frac{m_f}{m_\mu}$ and $\lambda\equiv \frac{m_\mu}{m_\phi}$.

With the form factors $A_{ji}^M$ and $A_{ji}^E$, we can use Eq.~\eqref{Eq.general} to obtain the corrections to $g-2$ ($i=j$) and $\mu \rightarrow e\gamma$ ($i\neq j$).
\begin{subequations}\label{eq:Delta_a_Neutral_Scalar}
\begin{equation}
  \frac{1}{2}(g-2) \equiv \Delta a_\mu(\phi) = \frac{1}{8\pi^2} \frac{m_\mu^2}{m_\phi^2}\int_0^1 \mathrm{d}x\, \sum_f \frac{\left(g_{s\,1}^{f\mu}\right)^2 P_1^+(x)+\left(g_{p\,1}^{f\mu}\right)^2 P_1^-(x)}{(1-x)(1-x \lambda^2)+x\,\epsilon_f^2\lambda^2}
\end{equation}
where
\begin{equation}
  P_1^\pm(x) = x^2\left(1-x \pm \epsilon_f\right). \label{eq:P1_def}
\end{equation}
\end{subequations}
Note that Eq.~\eqref{eq:P1_def} can be obtained from the second line of Eq.~\eqref{eq:I1_simplified_mediator} by setting $y=1$ in the integrand and omitting the $y$-integration. 
For the case of a heavy mediator, $m_\phi \gg m_\mu,m_f$, the expression for $\Delta a_\mu$ can be approximated as
\begin{equation} \label{eq:Delta_a_Neutral_Scalar_approx}
  \begin{split}
	\Delta a_\mu(\phi) \simeq \frac{1}{4\pi^2}\frac{m_\mu^2}{m_\phi^2} \sum_f \left[\left(g_{s\,1}^{f\mu}\right)^2 \left( \frac{1}{6} - \epsilon_f\left(\frac{3}{4} + \log(\epsilon_f\lambda)\right)\right)\right. + \\
	 + \left.\left(g_{p\,1}^{f\mu}\right)^2 \left( \frac{1}{6} + \epsilon_f\left(\frac{3}{4} + \log(\epsilon_f\lambda)\right) \right)  \right].
  \end{split}
\end{equation}
Our results agree with those found in Refs.~\cite{Queiroz:2014zfa,Freitas:2014pua}. For the related LFV decay we find
\begin{equation}\label{eq:BR1}
    \mathrm{BR}(\MEG) \approx \frac{3(4\pi)^3 \alpha_\mathrm{em}}{4 G_F^2}\left( |A_{e\mu}^M|^2 + |A_{e\mu}^E|^2 \right) ,
\end{equation}
with $A_{e\mu}^M$ and $A_{e\mu}^E$ given in Eq.~\eqref{eq:dipoleFormFactors_neutralScalar}. In the limit where $m_\phi \gg m_\mu, m_f$, we can approximate this expression using the last line in Eq.~\eqref{eq:I1_simplified_mediator}:
\begin{subequations} \label{eq:BR1_approx}
  \begin{equation}
  \begin{split}
    A_{e\mu}^M= \frac{1}{16\pi^2 m_\phi^2 }\sum_f \left\lbrace g_{s\,1}^{fe}g_{s\,1}^{f\mu} \left[ \frac{1}{6}  - \epsilon_f \left(\frac{3}{2} +\log(\epsilon_f^2 \lambda^2) \right)\right]\right. + \\
    \left. \hfill + g_{p\,1}^{fe}g_{p\,1}^{f\mu}  \left[ \frac{1}{6}  + \epsilon_f \left(\frac{3}{2} +\log(\epsilon_f^2 \lambda^2) \right)\right] \right\rbrace,
  \end{split}
  \end{equation}
  \begin{equation}
  \begin{split}
    A_{e\mu}^E= \frac{1}{16\pi^2 m_\phi^2 }\sum_f \left\lbrace g_{p\,1}^{fe}g_{s\,1}^{f\mu} \left[ \frac{1}{6}  - \epsilon_f \left(\frac{3}{2} +\log(\epsilon_f^2 \lambda^2) \right)\right]\right. - \\
    \left. \hfill - g_{s\,1}^{fe}g_{p\,1}^{f\mu}  \left[ \frac{1}{6}  + \epsilon_f \left(\frac{3}{2} +\log(\epsilon_f^2 \lambda^2) \right)\right] \right\rbrace.
  \end{split}
  \end{equation}
\end{subequations}
In summary, Eq.~\eqref{eq:Delta_a_Neutral_Scalar_approx} and Eq.~\eqref{eq:BR1} determine the prediction for $g-2$ and $\MEG$ for a scalar particle which interacts with a charged lepton according to Eq.~\eqref{Eq:neutralscalar}.

\subsubsection{Singly Charged Scalar}

Singly charged scalars are a clear signature of additional scalar doublets. They appear in the Zee-Babu model as well as in models with much richer scalar sectors such as multi-Higgs doublet models or scalar triplet models. For example, Left-Right models usually feature such triplet scalars which encompass singly charged scalars. The relevant interaction terms for the contribution of a scalar with unit charge to the amplitude $\mathcal{M}$ are given by
\begin{equation}
  \mathcal{L}_\textrm{int} = g_{s\,2}^{ij}  \phi^+\, \overline{\nu_i}\, \ell_j + g_{p\,2}^{ij}  \phi^+\, \overline{\nu_i} \gamma^5 \ell_j +\ \mathrm{h.c.}
  \label{eq:singlyscalar}
\end{equation}
In what follows the $\nu_f^i$ do not have to be the SM neutrinos, but they can be any sort of (exotic) neutral leptons with arbitrary mass. Since we will provide a fully general result the reader is welcome to use it. Conversely, we will also give simplified results taking the scalar to be much heavier than all other particles involved in the processes. It is worth emphasizing that if lepton number is explicitly violated in the given model, one might also find operators of the form $\phi^+\overline{\nu^\mathcal{C}}\ell$. For the present analysis this will not change the calculation, hence we do not consider these operators explicitly.

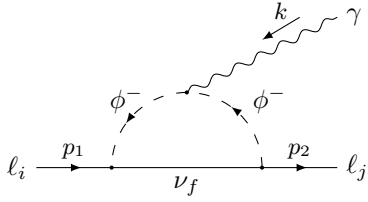
\begin{figure}[t]
  \centering
  \begin{tikzpicture}[x=1mm,y=1mm]
    \node[anchor=east] at (-20,0) (f1) {$\ell_i$};
    \node[anchor=west] at (20,0) (f2) {$\ell_j$};
    \node (V1) at (-10,0) {};
    \node (V2) at (10,0) {};
    \draw[fill=black] (V1) circle(.2);
    \draw[fill=black] (V2) circle(.2);
    \draw[decoration={markings, mark=at position 0.6 with {\arrow{latex}}}, postaction={decorate}] (f1) -- (V1.center) 
      node[midway, above] {\small$p_1$};
    \draw (V1.center) -- (V2.center) node[midway, below] {$\nu_f$};	
    \draw[decoration={markings, mark=at position 0.6 with {\arrow{latex}}}, postaction={decorate}] (V2.center) -- (f2) 
      node[midway, above] {\small$p_2$};
    \draw[mydash, decoration={markings, mark=at position 0.6 with {\arrow{latex}}}, postaction={decorate}] (V2.center) arc(0:90:10) 
    node (V3) {};
    \draw[fill] (V3.center) circle(.2);
    \draw[decorate, decoration={snake, segment length=3.3mm, amplitude=.5mm}] (V3.center) -- (20,20) node[anchor=west] {$\gamma$}; 
    \draw[mydash,decoration={markings, mark=at position 0.6 with {\arrow{latex}}}, postaction={decorate}] (V3.center) arc(90:180:10);
    \draw[->, >=latex] (15,20) -- (10,17) node[midway, above] {\small$k$};
    \node at (-8,9) {$\phi^-$};
    \node at (11,9) {$\phi^-$};
  \end{tikzpicture}
  \caption{\label{fig:SinglychargedScalar}Process $\ell_i\to \ell_j \gamma$ mediated by a charged scalar.}
\end{figure}

From Fig.~\ref{fig:SinglychargedScalar} the vertex function can be computed as
\begin{equation}
  \begin{aligned}
    \Gamma^\mu_2 = - \frac{i \sigma^{\mu\nu}k_\nu}{8\pi^2} \frac{m_i}{2} \sum_f
      \Big[&\underbrace{{g_{s\,2}^{fj}}^*g_{p\,2}^{fi} I_{f,\, 2}^{+\,+} + {g_{p\,2}^{fj}}^* g_{p\,2}^{fi} I_{f,\, 2}^{+\,-}}_{\equiv -(4\pi)^2 A^M_{ji}} -\\
      &- \gamma^5\underbrace{\left({g_{p\,2}^{fj}}^* g_{s\,2}^{fi} I_{f,\, 2}^{-\,+} + {g_{s\,2}^{fj}}^* g_{p\,2}^{fi} I_{f,\, 2}^{-\,-}\right)}_{\equiv  i(4\pi)^2 A^E_{ji}} \Big]. \label{eq:Gamma2_result}
    \end{aligned}
\end{equation}
From the full expression for the integral $I_{f,\, 2}$ given in Eq.~\eqref{eq:I2_def}, we obtain for $\Delta a_\mu$:
\begin{subequations}
  \begin{equation}
    \Delta a_\mu\left(\phi^+\right) = -\frac{1}{8\pi^2} \frac{m_\mu^2}{m_{\phi^+}^2}\int_0^1 \mathrm{d}x\, \sum_f \frac{\left|g_{s\,2}^{f\mu}\right|^2 P_2^+(x)+\left|g_{p\,2}^{f\mu}\right|^2 P_2^-(x)}{\epsilon_f^2\lambda^2(1-x)\left(1-\epsilon_f^{-2}x\right)+x},\label{eq:Delta_a_Singly_Scalar}
  \end{equation}
  where
  \begin{equation}
    P_2^\pm(x) = x(1-x)\left(x \pm \epsilon_f\right),\quad \epsilon_f\equiv \frac{m_{\nu_f}}{m_\mu},\quad \lambda\equiv \frac{m_\mu}{m_{\phi^+}}.\label{eq:P2_def} 
  \end{equation}
\end{subequations}
In the limit of a heavy scalar mediator, i.e.~$\lambda\to 0$, this reduces to
\begin{equation}\label{eq:Delta_a_Singly_Scalar_approx}
	\Delta a_\mu(\phi^+) \simeq \frac{-1}{4\pi^2}\frac{m_\mu^2}{m_{\phi^+}^2} \sum_f \left[ \left|g_{s\,2}^{f\mu}\right|^2 \left( \frac{1}{12} + \frac{\epsilon_f}{4} \right) + \left|g_{p\,2}^{f\mu}\right|^2 \left( \frac{1}{12} - \frac{\epsilon_f}{4} \right) \right].
\end{equation}
Having identified $A_{ji}^M$ and $A_{ji}^E$ in Eq.~\eqref{eq:Gamma2_result}, we can substitute them in Eq.~\eqref{Eq.general} to find   $\mathrm{BR}(\MEG)\approx \frac{3(4\pi)^3 \alpha_\mathrm{em}}{4 G_F^2}\left( |A_{e\mu}^M|^2 + |A_{e\mu}^E|^2 \right)$, where
\begin{subequations}\label{eq:BR2}
  \begin{align}
    A_{e\mu}^M &= \frac{-1}{(4\pi)^2}\sum_f \left( {g_{s\,2}^{fe}}^*g_{s\,2}^{f\mu} I_{f,\, 2}^{+\,+}+{g_{p\,2}^{fe}}^*g_{p\,2}^{f\mu} I_{f,\, 2}^{+\,-}\right)\,,\\ 
    A_{e\mu}^E &= \frac{-i}{(4\pi)^2}\sum_f \left({g_{p\,2}^{fe}}^*g_{s\,2}^{f\mu} I_{f,\, 2}^{-\,+} + {g_{s\,2}^{fe}}^*g_{p\,2}^{f\mu} I_{f,\, 2}^{-\,-}\right)\,,
  \end{align}
\end{subequations}
which in the case of a heavy scalar ($m_{\phi^+} \gg m_{\mu}, m_{\nu_f}$), simplify to
\begin{subequations}
\label{Eq:BRsinglycharged}
\begin{align}
A_{e\mu}^M &\simeq \frac{-1}{16\pi^2 m_{\phi^+}^2}\sum_f \left( {g_{s\,2}^{fe}}^*g_{s\,2}^{f\mu} \left[ \frac{1}{12} + \frac{\epsilon_f}{2} \right] + {g_{p\,2}^{fe}}^*g_{p\,2}^{f\mu} \left[ \frac{1}{12} - \frac{\epsilon_f}{2} \right] \right),\\
A_{e\mu}^E &\simeq \frac{-i}{16\pi^2 m_{\phi^+}^2}\sum_f \left( {g_{p\,2}^{fe}}^*g_{s\,2}^{f\mu} \left[ \frac{1}{12} + \frac{\epsilon_f}{2} \right] + {g_{s\,2}^{fe}}^*g_{p\,2}^{f\mu} \left[ \frac{1}{12} - \frac{\epsilon_f}{2} \right] \right),
\end{align}
\end{subequations}
where we have used the approximate expression for $I_{f,\,2}$ found in the Appendix.

\subsubsection{Doubly Charged Scalar}

Doubly charged scalars are key features of the type~II seesaw model~\cite{Melfo:2011nx}, and they are also predicted in models based on the $SU(3)_C\times SU(3)_L\times U(1)_X$ (331) gauge group. Such particles were vastly used to enhance the signal $H \rightarrow \gamma \gamma$, when a mild excess in the diphoton channel surfaced in the Higgs discovery~\cite{Alves:2011kc,Alves:2012yp,Akeroyd:2012ms,Wang:2012ts}. Typically, such doubly charged scalars are accompanied by a singly charged one, however for now we will be restricted to the doubly charged scalar contribution only. In the case of a doubly charged scalar field, that might e.g.\ be a component of an $SU(2)_L$ triplet, the possible interactions take the form
\begin{equation}
  \mathcal{L}_\mathrm{int} = g_{s\,3}^{ij} \phi^{++} \overline{\ell_i^\mathcal{C}} \ell_j + g_{p\,3}^{ij} \phi^{++} \overline{\ell_i^\mathcal{C}} \gamma^5 \ell_j + \mathrm{h.c.}\,,
\end{equation}
which violate the SM lepton number symmetry explicitly. Note that $g_{s\,3}$ and $g_{p\,3}$ are forced to be symmetric matrices in flavor space.

\begin{figure}[t]
  \centering
  \begin{subfigure}[b]{.45\textwidth}
    \centering
    \begin{tikzpicture}[x=1mm,y=1mm]
      \node[anchor=east] at (-20,0) (f1) {$\ell_i$};
      \node[anchor=west] at (20,0) (f2) {$\ell_j$};
      \node (V1) at (-10,0) {};
      \node (V2) at (10,0) {};
      \draw[fill=black] (V1) circle(.2);
      \draw[fill=black] (V2) circle(.2);
      \draw[decoration={markings, mark=at position 0.6 with {\arrow{latex}}}, postaction={decorate}] (f1) -- (V1.center) 
	node[midway, above] {\small$p_1$};
      \draw[decoration={markings, mark=at position 0.55 with {\arrow{latex}}}, postaction={decorate}] (V2.center) -- (V1.center) node[midway, below] {$\ell_f^\mathcal{C}$};	
      \draw[decoration={markings, mark=at position 0.6 with {\arrow{latex}}}, postaction={decorate}] (V2.center) -- (f2) 
	node[midway, above] {\small$p_2$};
      \draw[mydash, decoration={markings, mark=at position 0.6 with {\arrow{latex}}}, postaction={decorate}] (V2.center) arc(0:90:10) 
      node (V3) {};
      \draw[fill] (V3.center) circle(.2);
      \draw[decorate, decoration={snake, segment length=3.3mm, amplitude=.5mm}] (V3.center) -- (20,20) node[anchor=west] {$\gamma$}; 
      \draw[mydash,decoration={markings, mark=at position 0.6 with {\arrow{latex}}}, postaction={decorate}] (V3.center) arc(90:180:10);
      \draw[->, >=latex] (15,20) -- (10,17) node[midway, above] {\small$k$};
      \node at (-9,9) {$\phi^{++}$};
      \node at (12,9) {$\phi^{++}$};
    \end{tikzpicture}
  \end{subfigure}
  \hspace{1cm}
  \begin{subfigure}[b]{.45\textwidth}
    \centering
    \raisebox{1mm}{
    \begin{tikzpicture}[x=1mm,y=1mm]
      \node[anchor=east] at (-20,0) (f1) {$\ell_i$};
      \node[anchor=west] at (20,0) (f2) {$\ell_j$};
      \node (V1) at (-10,0) {};
      \node (V2) at (10,0) {};
      \draw[fill=black] (V1) circle(.2);
      \draw[fill=black] (V2) circle(.2);
      \draw[decoration={markings, mark=at position 0.6 with {\arrow{latex}}}, postaction={decorate}] (f1) -- (V1.center) 
	node[midway, above] {\small$p_1$};
      \draw[mydash,decoration={markings, mark=at position 0.55 with {\arrow{latex}}}, postaction={decorate}] (V2.center) -- (V1.center) node[midway, below] {$\phi^{++}$};	
      \draw[decoration={markings, mark=at position 0.6 with {\arrow{latex}}}, postaction={decorate}] (V2.center) -- (f2) 
	node[midway, above] {\small$p_2$};
      \draw[decoration={markings, mark=at position 0.6 with {\arrow{latex}}}, postaction={decorate}] (V2.center) arc(0:90:10) 
      node (V3) {};
      \draw[fill] (V3.center) circle(.2);
      \draw[decorate, decoration={snake, segment length=3.3mm, amplitude=.5mm}] (V3.center) -- (20,20) node[anchor=west] {$\gamma$}; 
      \draw[decoration={markings, mark=at position 0.6 with {\arrow{latex}}}, postaction={decorate}] (V3.center) arc(90:180:10);
      \draw[->, >=latex] (15,20) -- (10,17) node[midway, above] {\small$k$};
      \node at (-9,9) {$\ell_f^\mathcal{C}$};
      \node at (9,9) {$\ell_f^\mathcal{C}$};
    \end{tikzpicture}
    }
  \end{subfigure}
  \caption{\label{fig:DoublychargedScalar}Process $\ell_i\to \ell_j \gamma$ mediated by a doubly charged scalar $\phi^{++}$, where two diagrams contribute.}
\end{figure}
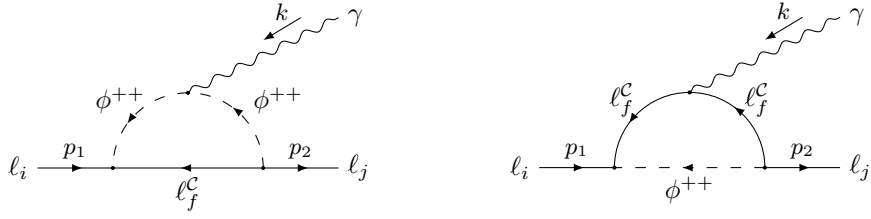

Since all fields in the relevant interactions are electrically charged, there will be two contributions as shown in Fig.~\ref{fig:DoublychargedScalar}. The corresponding expressions can be obtained from the previous subsections with appropriate changes in the parameters. One obtains for the sum of the two diagrams
\begin{equation}
  \begin{aligned}
    \Gamma^\mu_3 =& - 8 \frac{i \sigma^{\mu\nu}k_\nu}{8\pi^2} \frac{m_i}{2} \sum_f \left[ {g_{s\,3}^{fj}}^* g_{s\,3}^{fi} I_{f,\,2}^{+\,+} + {g_{p\,3}^{fj}}^* g_{p\,3}^{fi} I_{f,\,2}^{+\,-} -\right.\\
    &\hspace{35mm} \left.- \gamma^5 \left( {g_{p\,3}^{fj}}^* g_{s\,3}^{fi} I_{f,\,2}^{-\,+} + {g_{s\,3}^{fj}}^* g_{p\,3}^{fi} I_{f,\,2}^{-\,-} \right) \right]
    \\
    & - 4  \frac{i \sigma^{\mu\nu}k_\nu}{8\pi^2} \frac{m_i}{2} \sum_f \left[ {g_{s\,3}^{fj}}^* g_{s\,3}^{fi} I_{f,\,1}^{+\,+} + {g_{p\,3}^{fj}}^* g_{p\,3}^{fi} I_{f,\,1}^{+\,-} -\right.\\
    &\hspace{35mm} \left. - \gamma^5 \left( {g_{p\,3}^{fj}}^* g_{s\,3}^{fi} I_{f,\,1}^{-\,+} + {g_{s\,3}^{fj}}^* g_{p\,3}^{fi} I_{f,\,1}^{-\,-} \right) \right],
  \end{aligned}
  \label{dcharged}
\end{equation}
where it is understood that the replacements $m_{\nu_f}\to m_f,\,m_{\phi, \phi^{+}}\to m_{\phi^{++}}$ are made in the loop functions $I_{f,\,1/2}^{\pm\,\pm}$. The multiple factors of $2$ are due to the double unit charge of the scalar field and symmetry factors. 

Again, we can extract the contributions to $g-2$, which read -- in agreement with Ref.~\cite{Queiroz:2014zfa},
\begin{equation}
  \begin{aligned}
    \Delta a_\mu\left(\phi^{++}\right) =& -\frac{8}{8\pi^2} \frac{m_\mu^2}{m_{\phi^{++}}^2}\int_0^1 \mathrm{d}x\, \sum_f \frac{\left|g_{s\,3}^{f\mu}\right|^2 P_2^+(x)+\left|g_{p\,3}^{f\mu}\right|^2 P_2^-(x)}{\epsilon_f^2\lambda^2(1-x)(1-\epsilon_f^{-2}x)+x} - \\
    & - \frac{4}{8\pi^2} \frac{m_\mu^2}{m_{\phi^{++}}^2}\int_0^1 \mathrm{d}x\, \sum_f \frac{\left|g_{s\,3}^{f\mu}\right|^2 P_1^+(x) + \left|g_{p\,3}^{f\mu}\right|^2 P_1^-(x)}{(1-x)(1-\lambda^2)+x\,\epsilon_f^2\lambda^2}. \label{eq:Delta_a_Doubly_Scalar}
  \end{aligned}
\end{equation}
The functions $P_{1/2}^\pm$ are defined in Eqs.~\eqref{eq:P1_def} and~\eqref{eq:P2_def}, however for $\epsilon_f\equiv\frac{m_f}{m_\mu}$ and $\lambda\equiv \frac{m_\mu}{m_{\phi^{++}}}$ in both functions. Note the relative sign between the second line in Eq.~\eqref{eq:Delta_a_Doubly_Scalar} and Eq.~\eqref{eq:Delta_a_Neutral_Scalar},  which is due to the appearance of a \emph{charge conjugate} lepton coupling to the photon.

For the case of a heavy mediator, one finds the simple expression~\cite{Queiroz:2014zfa}
\begin{equation}
  \Delta a_\mu\left(\phi^{++}\right) =-\frac{1}{4\pi^2} \frac{m_\mu^2}{m_{\phi^{++}}^2} \sum_f \left[ \left|g_{s\,3}^{f\mu}\right|^2 \left(\frac{4}{3}-\epsilon_f\right) + \left|g_{p\,3}^{f\mu}\right|^2 \left(\frac{4}{3}+\epsilon_f\right) \right].
  \label{Eq:doublychargedamu}
\end{equation}

We have seen that the correction to $g-2$ arising from a doubly charged scalar is basically a combination of the neutral and charged scalar contributions with some minor changes. Notice that considering only flavor diagonal couplings, i.e.\ $\epsilon_f=1$ and omitting the sum, the doubly charged scalar contribution to $g-2$ is negative. However, if the $\tau$-lepton contributes significantly then the overall contribution might be positive, but this only true if $g^{\tau \mu}_s \neq  g^{\tau \mu}_p$ for some reason, otherwise the $\epsilon_f$ terms in Eq.~\eqref{Eq:doublychargedamu} cancel out.

As for $\mu \rightarrow e\gamma$ a similar treatment can be applied to find,
\begin{equation}
 \mathrm{BR}(\MEG) \simeq \frac{3(4\pi)^3 \alpha_\mathrm{em}}{4 G_F^2}\left( |A_{e\mu}^M|^2 + |A_{e\mu}^E|^2 \right),
\end{equation}
where
\begin{subequations}\label{eq:BRdoublycharged}\allowdisplaybreaks
  \begin{align}
    A_{e\mu}^M &= \frac{-1}{(4\pi)^2} \sum_f \left( {g_{s\,3}^{fe}}^*g_{s\,3}^{f\mu} I_{f,\, 21}^{+\,+}+{g_{p\,3}^{fe}}^*g_{p\,3}^{f\mu} I_{f,\, 21}^{+\,-}\right)\,,\\
    A_{e\mu}^E &= \frac{-i}{(4\pi)^2} \sum_f \left({g_{p\,3}^{fe}}^*g_{s\,3}^{f\mu} I_{f,\, 21}^{-\,+} + {g_{s\,3}^{fe}}^*g_{p\,3}^{f\mu} I_{f,\, 21}^{-\,-}\right)\,,
  \end{align}
\end{subequations}in agreement with~\cite{Lavoura:2003xp}, and $I_{f,\,21} \equiv 4 (2 I_{f,\,2} + I_{f,\,1})$.

In the limit $m_{\phi^{++}} \gg m_f$ we get
\begin{eqnarray}
A_{e\mu}^M &\simeq & \frac{-1}{8\pi^2 m_{\phi^{++}}^2}\sum_f \left( {g_{s\,3}^{fe}}^*g_{s\,3}^{f\mu} \left[ \frac{2}{3} - \epsilon_f \right] + {g_{p\,3}^{fe}}^*g_{p\,3}^{f\mu} \left[ \frac{2}{3} + \epsilon_f \right] \right),\nonumber\\
A_{e\mu}^E &\simeq & \frac{-i}{8\pi^2 m_{\phi^{++}}^2}\sum_f \left( {g_{p\,3}^{fe}}^*g_{s\,3}^{f\mu} \left[ \frac{2}{3} - \epsilon_f \right] + {g_{s\,3}^{fe}}^*g_{p\,3}^{f\mu} \left[ \frac{2}{3} + \epsilon_f \right] \right)
\end{eqnarray}
For the important special case where we have $y/\sqrt{2} = g_{s\,3} = \pm g_{p\,3}$, this expression reduces to
\begin{equation}
  \mathrm{BR}(\MEG) \simeq \frac{\alpha_\mathrm{em} \left|(y^\dag y)_{e\mu}\right|^2}{3 \pi G_F^2 m_{\phi^{++}}^4}.\
  \label{Eq:doublyBRmutoe}
\end{equation}
We emphasize that Eq.~\eqref{Eq:doublyBRmutoe} accounts only for the doubly charged scalar contribution to ${\rm BR}(\MEG)$. However, as mentioned above, doubly charged scalars usually arise in the context of Higgs triplets, so that a singly charged scalar will also contribute. Combining our results for the singly and doubly charged scalars we may obtain
\begin{equation}
  \mathrm{BR}(\MEG) \simeq \frac{27 \alpha_\mathrm{em} \left|(y^\dag y)_{e\mu}\right|^2}{64 \pi G_F^2 m_{\phi^{++}}^4},
\end{equation}where we assumed $m_{\phi^+} = m_{\phi^{++}}$. This result matches the well-known result for the Higgs triplet contribution derived in ~\cite{Cuypers:1996ia} and quoted in~\cite{Akeroyd:2009nu,Chakrabortty:2012vp,Chakrabortty:2015zpm}. Thus, if one needs a more general assessment of the doubly charged contribution to $g-2$ and $\MEG$, Eq.~\eqref{Eq:doublychargedamu} and Eq.~\eqref{Eq:doublyBRmutoe} should be used, respectively.
  
\subsection{Gauge Boson Mediator}

In this section we will discuss fully general interactions of neutral and charged gauge bosons which arise in many models that augment the SM with a new Abelian symmetry or extended electroweak gauge sectors such as $SU(2)_R$, $SU(3)_L$ etc. Whether we consider new gauge bosons or fermion fields, in many cases the amplitude will involve a gauge boson propagator. In this subsection we list the relevant expressions for $\mathcal{M}$. We start with the charged gauge boson.

\subsubsection{Neutral Fermion -- Charged Gauge Boson}

The introduction of several neutral fermions $N_i$ opens the following vector and axial-vector interaction channels:
\begin{equation}
 \mathcal{L}_\mathrm{int} = g_{v\,1}^{ij} {W^\prime}_\mu^+ \overline{N}_i \gamma^\mu \ell_j + g_{a\,1}^{ij} {W^\prime}_\mu^+ \overline{N}_i \gamma^\mu \gamma^5 \ell_j + \textrm{h.c.}
 \label{Eq:WprimeN}
\end{equation}
This expression may root in several high-energy models. For example, we could have that the $W^\prime$ is actually the SM $SU(2)_L$ gauge boson. In that case we would find that the coupling strength is proportional to some mixing of the $N_i$ with the active neutrinos of the SM. Alternatively, we could have that the $W^\prime$ is due to some extended gauge sector, e.g.\ $SU(2)_R$ of the Left-Right symmetric model.  Since we are not worried about $SU(2)_L$ invariance at this point, we need to make further assumptions to carry out the relevant calculations. We have done so by performing all calculations in unitary gauge and taking the respective propagators for the internal $W^\prime$s. 

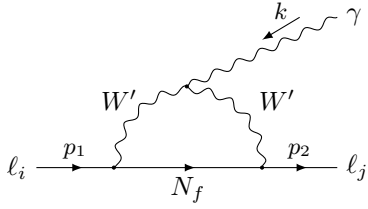
\begin{figure}
  \centering
  \begin{tikzpicture}[x=1mm,y=1mm]
      \node[anchor=east] at (-20,0) (f1) {$\ell_i$};
      \node[anchor=west] at (20,0) (f2) {$\ell_j$};
      \node (V1) at (-9.7,0) {};
      \node (V2) at (10,0) {};
      \draw[fill=black] (V1) circle(.2);
      \draw[fill=black] (V2) circle(.2);
      \draw[decoration={markings, mark=at position 0.6 with {\arrow{latex}}}, postaction={decorate}] (f1) -- (V1.center) 
	node[midway, above] {\small$p_1$};
      \draw[decoration={markings, mark=at position 0.55 with {\arrow{latex}}}, postaction={decorate}] (V1.center) -- (V2.center) node[midway, below] {$N_f$};	
      \draw[decoration={markings, mark=at position 0.6 with {\arrow{latex}}}, postaction={decorate}] (V2.center) -- (f2) 
	node[midway, above] {\small$p_2$};
      \node at(0,10.8) (V3) {};
      \draw[fill] (V3.center) circle(.2);
      \draw[decorate, decoration={snake, segment length=3.3mm, amplitude=.5mm}] (V3.center) -- (20,20) node[anchor=west] {$\gamma$};
      \draw[->, >=latex] (15,20) -- (10,17) node[midway, above] {\small$k$};
      \node at (-9,9) {$W^\prime$};
      \node at (12,9) {$W^\prime$};
      \begin{scope}
      \clip (-12,0) rectangle (12,12);
      \draw[decorate, decoration={snake, segment length=3.1mm, amplitude=-.5mm}] (V2.center) arc(0:200:9.6);
      \end{scope}
  \end{tikzpicture}
  \caption{\label{fig:NeutralFermion}Process $\ell_i\to \ell_j \gamma$ mediated by a neutral fermion $N$ and a $W^\prime$ boson.}
\end{figure}

For the diagram in Fig.~\ref{fig:NeutralFermion}, we obtain, with the aid of the loop function $I_{f,\,3}^{\pm\,\pm}$ defined in Eq.~\eqref{eq:I3_def},
\begin{equation}
  \begin{aligned}
    \Gamma^\mu_4 = \frac{i \sigma^{\mu\nu}k_\nu}{8\pi^2} \frac{m_i}{2} \sum_f &\left[ {g_{v\,1}^{fj}}^* g_{v\,1}^{fi} I_{f,\,3}^{+\,+} + {g_{a\,1}^{fj}}^* g_{a\,1}^{fi} I_{f,\,3}^{+\,-} +\right.\\
     & \left.+ \gamma^5 \left( {g_{a\,1}^{fj}}^* g_{v\,1}^{fi} I_{f,\,3}^{-\,+} + {g_{v\,1}^{fj}}^* g_{a\,1}^{fi} I_{f,\,3}^{-\,-} \right) \right].
  \end{aligned}
\end{equation}
From this we extract the relevant expressions for the $g-2$:
\begin{subequations}\allowdisplaybreaks
  \begin{equation}
  \label{Eq:similartoseesaw}
    \Delta a_\mu\left(N,W^\prime\right) = \frac{-1}{8\pi^2} \frac{m_\mu^2}{m_{W^\prime}^2} \int_0^1 \mathrm{d}x \sum_f \frac{\left|g_{v\,1}^{f\mu}\right|^2 P_3^+(x) + \left|g_{a\,1}^{f\mu}\right|^2 P_3^-(x)}{\epsilon_f^2\lambda^2(1-x)\left(1-\epsilon_f^{-2}x\right)+x},
  \end{equation}
  with
  \begin{equation}\label{eq:P3def}
    P_3^\pm = -2 x^2 (1 + x \mp 2\epsilon_f) + \lambda^2 x (1-x) (1 \mp \epsilon_f)^2\left(x \pm \epsilon_f\right)
  \end{equation}
\end{subequations}
and $\epsilon_f \equiv \frac{m_{N_f}}{m_\mu}$,  $\lambda \equiv \frac{m_\mu}{m_{W^\prime}}$, in agreement with Ref.~\cite{Leveille:1977rc}.

It is interesting to see that the rather lengthy expression for $I_{f,\,3}$ reduces to a much simpler one if the internal boson is assumed to be decoupled, i.e.~$m_{W^\prime}\gg m_N,m_f$:
\begin{equation}
  I_{f,\,3}^{(\pm)_1,(\pm)_2} \simeq \frac{1}{m_{W^\prime}^2} \left[\frac{5}{6} \left(1 + (\pm)_1 m_j/m_i\right) - 2 (\pm)_2 m_{N_f}/m_i\right].
\end{equation}
Thus, if the intermediate $W^\prime$ is heavy, i.e.\ $\lambda \to 0$, we find the useful approximation
\begin{equation}\label{eq:approxamu_W1}
  \Delta a_\mu(N,W^\prime) \simeq \frac{1}{4\pi^2}\frac{m_\mu^2}{m_{W^\prime}^2} \sum_f \left[ \left|g_{v\,1}^{f\mu}\right|^2 \left( \frac{5}{6} - \epsilon_f\right) + \left|g_{a\,1}^{f\mu}\right|^2 \left( \frac{5}{6} + \epsilon_f\right) \right].
\end{equation}
Similarly, taking both $m_N \simeq m_{W^\prime}\gg m_{f}$ such that $\epsilon_f\lambda \simeq 1$, we find that the contribution is independent of $\epsilon_f$ if we allow the vector and axial-vector contributions to cancel, i.e.\  $\left|g_v\right| = \left|g_a\right| = |g|$. We obtain
\begin{equation} \label{eq:approxamu_W2}
	\Delta a_\mu(N,W^\prime) \simeq \frac{17}{48 \pi^2} \frac{m_\mu^2}{m_{W^\prime}^2} \sum_f \left|g^{f\mu}\right|^2.
\end{equation}
As we will see below, this matches a well-known result in the Left-Right model for $\MEG$ where $g=g_R$, with $g_R$ being the gauge coupling from the $SU(2)_R$ group. We emphasize that it is applicable only when $m_N \simeq m_{W^\prime}$ both being much heavier than any charged lepton involved.

As for $\MEG$, we find
\begin{subequations}
\begin{align}
\label{Eq:similartoseesawmutoe}
A_{e\mu}^M =& \frac{-1}{(4\pi)^2}\sum_f \left( {g_{v\,1}^{fe}}^*g_{v\,1}^{f\mu} I_{f,\, 3}^{+\,+}+{g_{a\,1}^{fe}}^*g_{a\,1}^{f\mu} I_{f,\, 3}^{+\,-}\right),\\
A_{e\mu}^E =& \frac{i}{(4\pi)^2}\sum_f \left({g_{a\,1}^{fe}}^*g_{v\,1}^{f\mu} I_{f,\, 3}^{-\,+} + {g_{v\,1}^{fe}}^*g_{a\,1}^{f\mu} I_{f,\, 3}^{-\,-}\right),
\end{align}
\end{subequations}
with $I_{f,\,3}^{\pm\,\pm}$ given in Eq.~\eqref{eq:I3_def} in the Appendix.

Again taking the limit $m_{W^\prime}\gg m_N,m_f$, which is often present in studies involving sterile neutrinos~\cite{Abada:2014kba}, we get
\begin{subequations}\label{eq:BR4}
  \begin{align}
    A_{e\mu}^M \simeq& \frac{-1}{16 \pi^2 m_{W^\prime}^2}\sum_f \left( {g_{v\,1}^{fe}}^*g_{v\,1}^{f\mu} \left[\frac{5}{6} - 2 \epsilon_f \right] +{g_{a\,1}^{fe}}^*g_{a\,1}^{f\mu} \left[\frac{5}{6} + 2 \epsilon_f \right] \right),\\
    A_{e\mu}^E \simeq& \frac{i}{16 \pi^2 m_{W^\prime}^2}\sum_f \left( {g_{a\,1}^{fe}}^*g_{v\,1}^{f\mu} \left[\frac{5}{6} - 2 \epsilon_f \right] +{g_{v\,1}^{fe}}^*g_{a\,1}^{f\mu} \left[\frac{5}{6} + 2 \epsilon_f \right] \right),
  \end{align}
\end{subequations}

However, for the regime in which $m_{N_f} = m_{W^\prime} \gg m_{f}$, this results in ($\left|g_v\right| = \left|g_a\right| = |g|$)
\begin{equation}\label{eq:BR4_1}
  A_{e\mu}^M = i A_{e\mu}^E \simeq \frac{-1}{16 \pi^2 m_{W^\prime}^2} \frac{17}{12} \sum_f {g^{fe}}^*g^{f\mu}.
\end{equation}
and thus
\begin{equation}
 \mathrm{BR}(\MEG) \simeq 6.43 \times 10^{-6} \left(\frac{{\rm 1\, TeV}}{m_{W^\prime}}\right)^4  \left|\sum_f {g^{fe}}^*g^{f\mu}\right|^2.
\end{equation}

\subsubsection{Singly charged Fermion -- Neutral Gauge Boson}
\begin{figure}
  \centering
  \begin{tikzpicture}[x=1mm,y=1mm]
      \node[anchor=east] at (-20,0) (f1) {$\ell_i$};
      \node[anchor=west] at (20,0) (f2) {$\ell_j$};
      \node (V1) at (-9.7,0) {};
      \node (V2) at (10,0) {};
      \draw[fill=black] (V1) circle(.2);
      \draw[fill=black] (V2) circle(.2);
      \draw[decoration={markings, mark=at position 0.6 with {\arrow{latex}}}, postaction={decorate}] (f1) -- (V1.center) 
	node[midway, above] {\small$p_1$};
      \draw[decorate, decoration={snake, segment length=3.5mm, amplitude=.5mm}] (V1.center) -- (V2.center) node[midway, below] {$Z^\prime$};	
      \draw[decoration={markings, mark=at position 0.6 with {\arrow{latex}}}, postaction={decorate}] (V2.center) -- (f2) 
	node[midway, above] {\small$p_2$};
      \node at(0,10) (V3) {};
      \draw[fill] (V3.center) circle(.2);
      \draw[decorate, decoration={snake, segment length=3.3mm, amplitude=.5mm}] (V3.center) -- (20,20) node[anchor=west] {$\gamma$};
      \draw[->, >=latex] (15,20) -- (10,17) node[midway, above] {\small $k$};
      \node at (-9,9) {$E_f$};
      \node at (9,9) {$E_f$};
      \draw[decoration={markings, mark=at position 0.6 with {\arrow{latex}}}, postaction={decorate}] (V1.center) arc(180:90:10);
      \draw[decoration={markings, mark=at position 0.6 with {\arrow{latex}}}, postaction={decorate}] (V3.center) arc(90:0:10);
  \end{tikzpicture}
  \caption{\label{fig:ChargedFermion}Process $\ell_i\to \ell_j \gamma$ mediated by a charged fermion $E$ and a $Z^\prime$ boson.}
\end{figure}
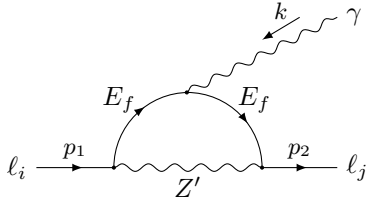

Exotic singly charged fermions commonly arise in 331 models~\cite{Ponce:2006vw,Salazar:2007ym}, in two Higgs doublet models~\cite{DePree:2008st}, and many other theories, cf. Refs.~\cite{Hewett:1988xc,Lee:2012xn,Ari:2013wda}. 
Charged fermions could interact with the SM leptons via both a scalar and/or a neutral vector boson ($Z^\prime$), i.e.~via the Lagrangian
\begin{equation} \label{eq:LagrangianChargedFermion}
  \mathcal{L}_\mathrm{int} = g_{v\,2}^{ij} Z^\prime_\mu \overline{E_i}\gamma^\mu \ell_j + g_{a\,2}^{ij} Z^\prime_\mu \overline{E_i}\gamma^\mu \gamma^5 \ell_j + g_{s\,4}^{ij} h\, \overline{E_i}\, \ell_j + g_{p\,4}^{ij} h\, \overline{E_i} \gamma^5 \ell_j + \textrm{h.c.}
\end{equation}
In the case where the new charged leptons $E_i$ couple only to vector bosons, we set $g_{s/p\,5}^{ij}=0$. As a result we obtain the following expression (cf.\ Fig.~\ref{fig:ChargedFermion}):
\begin{equation}
\begin{aligned}
  \Gamma^\mu_{5,v} = - \frac{i \sigma^{\mu\nu}k_\nu}{8\pi^2} \frac{m_i}{2} \sum_f &\left[ {g_{v\,2}^{fj}}^* g_{v\,2}^{fi} I_{f,\,4}^{+\,+} + {g_{a\,2}^{fj}}^* g_{a\,2}^{fi} I_{f,\,4}^{+\,-} + \right.\\
  &\left. + \gamma^5 \left( {g_{a\,2}^{fj}}^* g_{v\,2}^{fi} I_{f,\,4}^{-\,+} + {g_{v\,2}^{fj}}^* g_{a\,2}^{fi} I_{f,\,4}^{-\,-} \right) \right],
\end{aligned}
\end{equation} 
with $I_{f,\,4}$ given exactly in Eq.~\eqref{eq:I4_def}. 

We can obtain from the exact expression, for the case $m_i = m_j = m_\mu$, the result found in the literature~\cite{Leveille:1977rc, Jegerlehner:2009ry}: 
\begin{subequations}  \label{Eq:doublyEamu}
  \begin{equation}
    \Delta a_\mu\left(E, Z^\prime\right) = \frac{1}{8\pi^2} \frac{m_\mu^2}{m_{Z^\prime}^2} \int_0^1 \mathrm{d}x \sum_f \frac{\left|g_{v\,2}^{f\mu}\right|^2 P_4^+(x) + \left|g_{a\,2}^{f\mu}\right|^2 P_4^-(x)}{(1-x)\left(1-\lambda^2 x\right)+\epsilon_f^2\lambda^2 x},
  \end{equation}
  with
  \begin{equation}
    P_4^\pm = 2x(1-x) (x-2\pm 2\epsilon_f) + \lambda^2x^2(1 \mp \epsilon_f)^2(1-x \pm \epsilon_f)\label{eq:P4def}
  \end{equation}
\end{subequations}
and $\epsilon_f \equiv \frac{m_{E_f}}{m_\mu}$, $\lambda \equiv \frac{m_\mu}{m_{Z^\prime}}$.

Interestingly in the limit of a heavy $Z^{\prime}$, i.e.~when $m_{Z^\prime} \gg m_E ,m_\mu$, $I_{f,\,4}$ simplifies to
\begin{equation}
  I_{f,\,4}^{(\pm)_1,(\pm)_2} \simeq \frac{1}{m_{Z^\prime}^2} \frac{2}{3} \left[1 + (\pm)_1 m_j/m_i - 3 (\pm)_2 m_E/m_i \right],
  \label{Eq:doublyE}
\end{equation}
which gives us an idea of the asymptotic behavior of $g-2$ and $\MEG$. Indeed, 
to leading order for a heavy gauge boson, one finds
\begin{equation} \label{eq:neutralGBapprox}
	\Delta a_\mu\left(E,Z^\prime\right) \simeq \frac{-1}{4\pi^2}\frac{m_\mu^2}{m_{Z^\prime}^2} \sum_f \left[ \left|g_{v\,2}^{f\mu}\right|^2 \left( \frac{2}{3} - \epsilon_f \right) + \left|g_{a\,2}^{f\mu}\right|^2 \left( \frac{2}{3} + \epsilon_f \right) \right].
\end{equation}
As for $\mathrm{BR}(\MEG) \simeq \frac{3(4\pi)^3 \alpha_\mathrm{em}}{4 G_F^2}\left( |A_{e\mu}^M|^2 + |A_{e\mu}^E|^2 \right)$ one obtains:
\begin{subequations}\label{eq:BR5v}
  \begin{align}
    A_{e\mu}^M=& \frac{-1}{(4\pi)^2}\sum_f \left( {g_{v\,2}^{fe}}^*g_{v\,2}^{f\mu} I_{f,\, 4}^{+\,+}+{g_{a\,2}^{fe}}^*g_{a\,2}^{f\mu} I_{f,\, 4}^{+\,-}\right)\\
    \simeq&  \frac{-1}{24\pi^2 m_{Z^\prime}^2}\sum_f \left( {g_{v\,2}^{fe}}^*g_{v\,2}^{f\mu} \left(1 - 3 \epsilon_f \right) + {g_{a\,2}^{fe}}^*g_{a\,2}^{f\mu} \left(1 + 3\epsilon_f \right) \right),\\
    A_{e\mu}^E=& \frac{i}{(4\pi)^2}\sum_f \left({g_{a\,2}^{fe}}^*g_{v\,2}^{f\mu} I_{f,\, 4}^{-\,+} + {g_{v\,2}^{fe}}^*g_{a\,2}^{f\mu} I_{f,\, 4}^{-\,-}\right)\\
    \simeq&  \frac{i}{24\pi^2 m_{Z^\prime}^2}\sum_f \left( {g_{a\,2}^{fe}}^*g_{v\,2}^{f\mu} \left(1 - 3 \epsilon_f \right) + {g_{v\,2}^{fe}}^*g_{a\,2}^{f\mu} \left(1 + 3\epsilon_f \right) \right),
  \end{align}
\end{subequations}
where we have used the approximation $m_i \ll m_j \ll m_{Z^\prime}$ and $m_E \ll m_{Z^\prime}$.

The case where the charged fermions $E_i$ couple to the SM leptons via a scalar only, the result is identical to the case of a neutral scalar. One can readily obtain:
\begin{equation}\label{eq:I5s_def}
  \begin{aligned}
    \Gamma^\mu_{5,s} = \frac{i \sigma^{\mu\nu}k_\nu}{8\pi^2} \frac{m_i}{2} \sum_f & 
    \left[{g_{s\,4}^{fj}}^*g_{s\,4}^{fi} I_{f,\, 5s}^{+\,+} + {g_{p\,4}^{fj}}^* g_{p\,4}^{fi} I_{f,\, 5s}^{+\,-} -\right. \\
    &\left. - \gamma^5 \left({g_{p\,4}^{fj}}^* g_{s\,4}^{fi} I_{f,\, 5s}^{-\,+} + {g_{s\,4}^{fj}}^* g_{p\,4}^{fi} I_{f,\, 5s}^{-\,-}\right) \right].
  \end{aligned}
\end{equation}
with $I_{f,\,5s}^{(\pm)_1,(\pm)_2} \equiv I_{f,\,1}(m_i,(\pm)_1 m_j,(\pm)_2 m_{E_f},m_h)$ as defined in Eq.~\eqref{eq:I1_def}. Similarly, the §$g-2$ is easily generalized from Eq.~\eqref{eq:Delta_a_Neutral_Scalar}:
\begin{equation}
  \Delta a_\mu(E,\phi) = \frac{1}{8\pi^2} \frac{m_\mu^2}{m_{\phi}^2}\int_0^1 \mathrm{d}x\, \sum_f \frac{\left|g_{s\,4}^{f\mu}\right|^2 P_1^+(x)+\left|g_{p\,4}^{f\mu}\right|^2 P_1^-(x)}{(1-x)(1-\lambda^2)+x\,\epsilon_f^2\lambda^2}, \label{eq:Delta_a_charged_Fermion}
\end{equation}
with $P^\pm_1$ defined above and $\epsilon_f \equiv \frac{m_{E_f}}{m_\mu}$, $\lambda \equiv \frac{m_\mu}{m_{\phi}}$. Similarly, the case of a heavy scalar mediator yields Eq.~\eqref{eq:Delta_a_Neutral_Scalar_approx}.

The corresponding $\mathrm{BR}(\MEG)$ can be found by combining the $A_{ji}^M$ and $A_{ji}^E$ functions defined in Eq.~\eqref{Eq.general} and yield an expression identical to Eqs.~\eqref{eq:BR1} and~\eqref{eq:BR1_approx}.

\subsubsection{Doubly Charged Fermion -- Charged Gauge Boson}
\begin{figure}
  \centering
  \begin{subfigure}[b]{.45\textwidth}
    \centering\raisebox{-1.5mm}{
    \begin{tikzpicture}[x=1mm,y=1mm]
	\node[anchor=east] at (-20,0) (f1) {$\ell_i$};
	\node[anchor=west] at (20,0) (f2) {$\ell_j$};
	\node (V1) at (-9.7,0) {};
	\node (V2) at (10,0) {};
	\draw[fill=black] (V1) circle(.2);
	\draw[fill=black] (V2) circle(.2);
	\draw[decoration={markings, mark=at position 0.6 with {\arrow{latex}}}, postaction={decorate}] (f1) -- (V1.center) 
	  node[midway, above] {\small$p_1$};
	\draw[decoration={markings, mark=at position 0.55 with {\arrow{latex}}}, postaction={decorate}] (V1.center) -- (V2.center) node[midway, below] {$\psi_f$};	
	\draw[decoration={markings, mark=at position 0.6 with {\arrow{latex}}}, postaction={decorate}] (V2.center) -- (f2) 
	  node[midway, above] {\small$p_2$};
	\node at(0,10.8) (V3) {};
	\draw[fill] (V3.center) circle(.2);
	\draw[decorate, decoration={snake, segment length=3.3mm, amplitude=.5mm}] (V3.center) -- (20,20) node[anchor=west] {$\gamma$};
	\draw[->, >=latex] (15,20) -- (10,17) node[midway, above] {\small$k$};
	\node at (-9,9) {$X$};
	\node at (11,9) {$X$};
	\begin{scope}
	\clip (-12,0) rectangle (12,12);
	\draw[decorate, decoration={snake, segment length=3.1mm, amplitude=-.5mm}] (V2.center) arc(0:200:9.6);
	\end{scope}
    \end{tikzpicture}
    }
  \end{subfigure}
  \hspace{1cm}
  \begin{subfigure}[b]{.45\textwidth}
    \centering
    \begin{tikzpicture}[x=1mm,y=1mm]
	\node[anchor=east] at (-20,0) (f1) {$\ell_i$};
	\node[anchor=west] at (20,0) (f2) {$\ell_j$};
	\node (V1) at (-9.7,0) {};
	\node (V2) at (10,0) {};
	\draw[fill=black] (V1) circle(.2);
	\draw[fill=black] (V2) circle(.2);
	\draw[decoration={markings, mark=at position 0.6 with {\arrow{latex}}}, postaction={decorate}] (f1) -- (V1.center) 
	  node[midway, above] {\small$p_1$};
	\draw[decorate, decoration={snake, segment length=3.5mm, amplitude=.5mm}] (V1.center) -- (V2.center) node[midway, below] {$X$};	
	\draw[decoration={markings, mark=at position 0.6 with {\arrow{latex}}}, postaction={decorate}] (V2.center) -- (f2) 
	  node[midway, above] {\small$p_2$};
	\node at(0,10) (V3) {};
	\draw[fill] (V3.center) circle(.2);
	\draw[decorate, decoration={snake, segment length=3.3mm, amplitude=.5mm}] (V3.center) -- (20,20) node[anchor=west] {$\gamma$};
	\draw[->, >=latex] (15,20) -- (10,17) node[midway, above] {\small$k$};
	\node at (-9,9) {$\psi_f$};
	\node at (9.5,9) {$\psi_f$};
	\draw[decoration={markings, mark=at position 0.6 with {\arrow{latex}}}, postaction={decorate}] (V1.center) arc(180:90:10);
	\draw[decoration={markings, mark=at position 0.6 with {\arrow{latex}}}, postaction={decorate}] (V3.center) arc(90:0:10);
    \end{tikzpicture}
    
  \end{subfigure}
  \caption{\label{fig:DoublyChargedFermion}Process $\ell_i\to \ell_j \gamma$ mediated by a doubly charged fermion $\psi$ and a $X$ boson. Since both particles are charged, there are two separate contributions.}
\end{figure}
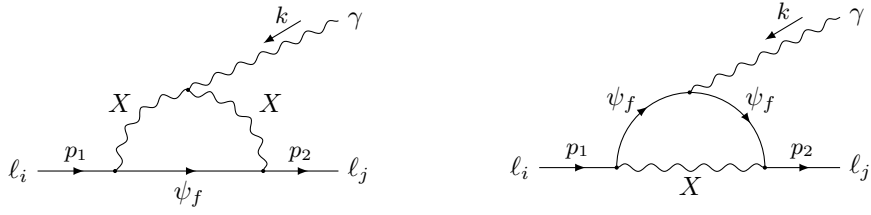

Doubly charged fermions, which appear as components of fermionic $SU(2)_L$ triplet fields~\cite{Ma:2013tda}, exotic doublets~\cite{Ma:2014zda}, as well as in composite Higgs models with extended isospin multiplets~\cite{Altarelli:1977zu,Wilczek:1977wb,Biondini:2012ny,Li:2016ijq}. They interact with SM leptons as follows:
\begin{equation}
  \mathcal{L}_\textrm{int} = g_{v\,3}^{ij} X_\mu^+ \overline{\ell_i}\gamma^\mu \psi_j + g_{a\,3}^{ij} X_\mu^+ \overline{\ell_i}\gamma^\mu \gamma^5 \psi_j + \mathrm{h.c.}\,,
\end{equation}
where the fermion $\psi$ has electric charge $Q_\psi=-2e$. Similar to the case of a doubly charged scalar, the doubly charged fermion will contribute with two diagrams to the dipole form factors as shown in Fig.~\ref{fig:DoublyChargedFermion}. Both contributions are related to the previous fermionic contributions:
\begin{equation}\label{eq:Gamma6}
  \begin{aligned}
    \Gamma^\mu_6 =& - \frac{i \sigma^{\mu\nu}k_\nu}{8\pi^2} \frac{m_i}{2} \sum_f \left[ {g_{v\,3}^{fj}}^* g_{v\,3}^{fi} I_{f,\,3}^{+\,+} + {g_{a\,3}^{fj}}^* g_{a\,3}^{fi} I_{f,\,3}^{+\,-} +\right.\\
    & \hspace{35mm}\left. + \gamma^5 \left( {g_{a\,3}^{fj}}^* g_{v\,3}^{fi} I_{f,\,3}^{-\,+} + {g_{v\,3}^{fj}}^* g_{a\,3}^{fi} I_{f,\,3}^{-\,-} \right) \right] \\
    &- \frac{i \sigma^{\mu\nu}k_\nu}{8 \pi^2} m_i \sum_f \left[ {g_{v\,3}^{fj}}^* g_{v\,3}^{fi} I_{f,\,4}^{+\,+} + {g_{a\,3}^{fj}}^* g_{a\,3}^{fi} I_{f,\,4}^{+\,-} +\right.\\
    & \hspace{35mm}\left. + \gamma^5 \left( {g_{a\,3}^{fj}}^* g_{v\,3}^{fi} I_{f,\,4}^{-\,+} + {g_{v\,3}^{fj}}^* g_{a\,3}^{fi} I_{f,\,4}^{-\,-} \right) \right],
  \end{aligned}
\end{equation}
with the masses $m_E$ and $m_N$ replaced with $m_\psi$ in both loop functions $I_{f,\,3/4}^{\pm\,\pm}$ and the $Z$ mass replaced with $m_X$ in $I_{f,\,4}^{\pm\,\pm}$. This yields
\begin{align}\displaybreak[3]
  \Delta a_\mu \left(\psi,X\right) &= \frac{1}{8\pi^2} \frac{m_\mu^2}{m_X^2} \int_0^1 \mathrm{d}x \sum_f \frac{\left|g_{v\,3}^{f\mu}\right|^2 P_3^+(x) + \left|g_{a\,3}^{f\mu}\right|^2 P_3^-(x)}{\epsilon_f^2\lambda^2(1-x)\left(1-\epsilon_f^{-2}x\right)+x} + \nonumber\\
  &+ \frac{2}{8\pi^2} \frac{m_\mu^2}{m_X^2} \int_0^1 \mathrm{d}x \sum_f \frac{\left|g_{v\,3}^{f\mu}\right|^2 P_4^+(x) + \left|g_{a\,3}^{f\mu}\right|^2 P_4^-(x)}{(1-x)\left(1-\lambda^2 x\right)+\epsilon_f^2\lambda^2 x}, \label{eq:Delta_a_Doubly_Fermion} 
\end{align}
with $\epsilon_f \equiv \frac{m_\psi}{m_\mu}$ and $\lambda \equiv \frac{m_\mu}{m_X}$; $P_{3/4}^\pm$ are defined in Eqs.~\eqref{eq:P3def} and~\eqref{eq:P4def}. For a heavy mediator $(\lambda\to 0)$ we get
\begin{equation}
  \Delta a_\mu\left(\psi,X\right) \simeq \frac{1}{4\pi^2}\frac{m_\mu^2}{m_X^2} \sum_f \left[ \left|g_{v\,1}^{f\mu}\right|^2 \left( 3\epsilon_f- \frac{13}{6}\right) - \left|g_{a\,1}^{f\mu}\right|^2 \left(3 \epsilon_f + \frac{13}{6}\right) \right].
\end{equation}
Notice that if the vector and axial-vector couplings are identical, the contribution to $g-2$ is negative, and there is no dependence on the fermion mass. Keep in mind that this result is valid only in the heavy mediator limit, i.e.~when the mediator is much heavier than all other scales in the problem, so the apparent scaling with fermion mass is a mirage, because very heavy fermions means even heavier mediators and thus the contribution to $g-2$ is further suppressed. In order to compute the contribution to $g-2$ for arbitrary mass hierarchies one needs to solve Eq.~\eqref{eq:Delta_a_Doubly_Fermion} numerically.

Regarding $\MEG$, similarly to the previous subsections we find,
\begin{equation}\label{eq:BR6}
    \mathrm{BR}(\MEG) \approx \frac{3(4\pi)^3 \alpha_\mathrm{em}}{4 G_F^2}\left( |A_{e\mu}^M|^2 + |A_{e\mu}^E|^2 \right), 
\end{equation}
where one should insert $A_{e\mu}^M= \frac{1}{(4\pi)^2}\sum_f \left( {g_{v\,3}^{fe}}^*g_{v\,3}^{f\mu} I_{f,\, 34}^{+\,+}+{g_{a\,3}^{fe}}^*g_{a\,3}^{f\mu} I_{f,\, 34}^{+\,-}\right)$ and\\ \mbox{$A_{e\mu}^E=\frac{-i}{(4\pi)^2}\sum_f \left({g_{a\,3}^{fe}}^*g_{v\,3}^{f\mu} I_{f,\, 34}^{-\,+} + {g_{v\,3}^{fe}}^*g_{a\,3}^{f\mu} I_{f,\, 34}^{-\,-}\right)$}, according to Eq.~\eqref{eq:Gamma6}, and the short-hand notation $I_{f,\,34} = I_{f,\,3} + 2 I_{f,\,4}$ has been used. In the usual approximation $m_\mu \ll m_X $, we obtain
\begin{subequations}
  \begin{align}
    A_{e\mu}^M \simeq & \frac{1}{16\pi^2 m_{W^\prime}^2}\sum_f \left( {g_{v\,3}^{fe}}^*g_{v\,3}^{f\mu} \left(\frac{13}{6} - 6 \epsilon_f \right) + {g_{a\,3}^{fe}}^*g_{a\,3}^{f\mu}  \left(\frac{13}{6} + 6 \epsilon_f \right) \right),\\
    A_{e\mu}^E \simeq & \frac{-i}{16\pi^2 m_{W^\prime}^2}\sum_f \left( {g_{a\,3}^{fe}}^*g_{v\,3}^{f\mu} \left(\frac{13}{6} - 6 \epsilon_f \right) + {g_{v\,3}^{fe}}^*g_{a\,3}^{f\mu}  \left(\frac{13}{6} + 6 \epsilon_f \right) \right).
  \end{align}
\end{subequations}

\subsubsection{Charged Fermion -- Doubly Charged Vector Boson}
Finally, we discuss the case of a doubly charged vector boson, which again contributes through two independent diagrams. Doubly charged gauge bosons are a typical signature of the minimal 331 model~\cite{Pisano:1991ee}. There, the $SU(2)_L$ gauge group is extended to a $SU(3)_L$, with the generations in the fundamental representation of $SU(3)_L$. The third component in the fermion triplet has a charged lepton with opposite electric charge, e.g.~$e^c$, and consequently in the covariant derivative a doubly charged gauge boson arises, with a charged current as shown below, 
\begin{equation}
  \mathcal{L}_\textrm{int} = g^{ij}_{v\,4} U^{++}_\mu\, \overline{\ell_i^\mathcal{C}} \gamma^\mu\, \ell_j + g^{ij}_{a\,4} U^{++}_\mu\, \overline{\ell_i^\mathcal{C}} \gamma^\mu \gamma^5 \ell_j + \mathrm{h.c.}
\end{equation}
Note that, while $g_{a\,4}$ is symmetric in flavor space, $g_{v\,4}$ is anti-symmetric and contains no diagonal entries.\footnote{If the vector current is $j^{\mu}=\overline{\psi_i^c}\gamma^{\mu}\psi_j=-\psi_i^T C^{-1}\gamma^{\mu}CC^{-1}\psi_j=\psi_i^T \gamma^{\mu\, T} C^T\psi_j=\psi_i^T \gamma^{\mu\, T} (\psi_j^T\,C)^T=-\psi_i^T \gamma^{\mu\, T} (\psi_j^T\,C^{-1})^T=(\psi_j^T C^{-1} \gamma^{\mu} \psi_i)^T=-\overline{\psi_j^c}\gamma^{\mu}\psi_i$, so if $i=j$ the vector current must vanish, where we used $\psi^c= C \bar{\psi}^T$ and $\bar{\psi^c}=-\psi^T\, C^{-1}$, with $C$ being the charged conjugation matrix.} Consequently, there will be symmetry factors in the vertex rules due to the appearance of identical fields. The two diagrams that contribute have topologies identical to the ones shown in Fig.~\ref{fig:DoublyChargedFermion} and give
\begin{equation}
  \begin{aligned}
    \Gamma^\mu_9 =& - 8 \frac{i \sigma^{\mu\nu}k_\nu}{8\pi^2} \frac{m_i}{2} \sum_f \left[ {g_{v\,4}^{fj}}^* g_{v\,4}^{fi} I_{f,\,3}^{+\,+} + {g_{a\,4}^{fj}}^* g_{a\,4}^{fi} I_{f,\,3}^{+\,-} +\right.\\
    & \hspace{35mm}\left. + \gamma^5 \left( {g_{a\,4}^{fj}}^* g_{v\,4}^{fi} I_{f,\,3}^{-\,+} + {g_{v\,4}^{fj}}^* g_{a\,4}^{fi} I_{f,\,3}^{-\,-} \right) \right] +
    \\
   & + 4  \frac{i \sigma^{\mu\nu}k_\nu}{8\pi^2} \frac{m_i}{2} \sum_f \left[ {g_{v\,4}^{fj}}^* g_{v\,4}^{fi} I_{f,\,4}^{+\,+} + {g_{a\,4}^{fj}}^* g_{a\,4}^{fi} I_{f,\,4}^{+\,-} +\right.\\
   & \hspace{35mm} \left. + \gamma^5 \left( {g_{a\,4}^{fj}}^* g_{v\,4}^{fi} I_{f,\,4}^{-\,+} + {g_{v\,4}^{fj}}^* g_{a\,4}^{fi} I_{f,\,4}^{-\,-} \right) \right],
  \end{aligned}
\end{equation}
such that
\begin{align}
  \Delta a_\mu\left(U^{++}\right) &= \frac{8}{8\pi^2} \frac{m_\mu^2}{m_U^2} \int_0^1 \mathrm{d}x \sum_f \frac{\left|g_{v\,4}^{f\mu}\right|^2 P_3^+(x) + \left|g_{a\,4}^{f\mu}\right|^2 P_3^-(x)}{\epsilon_f^2\lambda^2(1-x)\left(1-\epsilon_f^{-2}x\right)+x} - \nonumber\\
  &- \frac{4}{8\pi^2} \frac{m_\mu^2}{m_W^2} \int_0^1 \mathrm{d}x \sum_f \frac{\left|g_{v\,4}^{f\mu}\right|^2 P_4^+(x) + \left|g_{a\,4}^{f\mu}\right|^2 P_4^-(x)}{(1-x)\left(1-\lambda^2 x\right)+\epsilon_f^2\lambda^2 x}, \label{eq:Delta_a_Doubly_Vector} 
\end{align}
with $\epsilon_f \equiv \frac{m_f}{m_\mu}$ and $\lambda \equiv \frac{m_\mu}{m_U}$. Note the relative sign between the second term in Eq.~\eqref{eq:Delta_a_Doubly_Vector} and Eq.~\eqref{eq:Delta_a_Doubly_Fermion} due to the charge conjugate lepton coupling to the photon. In the heavy mediator limit ($m_{U^{++}} \gg m_f,m_{\mu}$), one obtains
\begin{equation}
  \Delta a_\mu\left(U^{++}\right) \simeq \frac{1}{\pi^2} \frac{m_\mu^2}{m_U^2} \sum_f \left( \left|g_{v\,4}^{f\mu}\right|^2 \left[ -1 + \epsilon_f\right] - \left|g_{a\,4}^{f\mu}\right|^2 \left[ 1 + \epsilon_f\right]\right),
  \label{Eq:Delta_a_Doubly_Vectorlimit}
\end{equation}
which vanishes for $\epsilon_f=1$ and $g_{a\,4}=0$, in agreement with the above symmetry argument.

Thus, if one considers only the muon in the loop ($\epsilon_f= 1$), the doubly charged gauge boson gives rise to a \emph{negative} contribution to $g-2$. One can use that information to enforce the doubly charged boson correction to $g-2$ to lie below the current and projected error bars to derive tight constraints on the scale of symmetry breaking of the minimal 331 model since the mass of such a boson is directly related to the latter~\cite{Kelso:2014qka}.

Finally, we report the expression for $\mathrm{BR}(\MEG) \approx \frac{3(4\pi)^3 \alpha_\mathrm{em}}{4 G_F^2}\left( |A_{e\mu}^M|^2 + |A_{e\mu}^E|^2 \right)$ with
\begin{equation}
  \begin{aligned}
  A_{e\mu}^M= \frac{-1}{(4\pi)^2} \sum_f\left( {g_{v\,4}^{fe}}^*g_{v\,4}^{f\mu} I_{f,\, 3-4}^{+\,+}+{g_{a\,4}^{fe}}^*g_{a\,4}^{f\mu} I_{f,\, 3-4}^{+\,-}\right),\\
    A_{e\mu}^E= \frac{i}{(4\pi)^2} \sum_f \left({g_{a\,4}^{fe}}^*g_{v\,4}^{f\mu} I_{f,\, 3-4}^{-\,+} + {g_{v\,4}^{fe}}^*g_{a\,4}^{f\mu} I_{f,\, 3-4}^{-\,-}\right),  \end{aligned}
\end{equation}
and $I_{f,\, 3-4} = -4(2I_{f,\, 3} - I_{f,\,4}) \simeq - \left(1 \mp 2 \epsilon_f\right)/m_U^2$ for a heavy mediator and $m_j \ll m_i$.

We have now concluded reviewing individual and multiple field contributions to $g-2$ and $\MEG$ without worrying about $SU(2)_L$ invariance. In what follows, we keep $SU(2)_L$ invariance as a guiding principle and discuss some simplified models.
\section{\label{sec:numerics}SU(2) Invariant Simplified Models}
In this section, we will illustrate our results by numerically analyzing some specific simplified beyond the SM (BSM) scenarios which can be embedded in several extensions of the SM. To this end, we choose to restore $SU(2)_L$ invariance by combining the results of the previous section. For definiteness, we also restrict our attention to coupling matrices with a fixed flavor structure matrix $\Lambda$ multiplied by a universal coupling $g$. Departures from this assumption will be addressed in the following section where we study UV complete models. In any case, the flavor structures we consider are labeled either as {\it strong} hierarchy, in which case we have
\begin{equation}
  \Lambda = \left(
	      \begin{array}[c]{ccc}
		1 & 10^{-5} & 10^{-8} \\
		10^{-5} & 1 & 10^{-5} \\
		10^{-8} & 10^{-5} & 1
	      \end{array}
	    \right).
	    \label{couplingmatrix1}
\end{equation}
For the cases referred to as {\it mildly} hierarchical, we set
\begin{equation}
  \Lambda = \left(
	      \begin{array}[c]{ccc}
		1 & 10^{-3} & 10^{-6} \\
		10^{-3} & 1 & 10^{-3} \\
		10^{-6} & 10^{-3} & 1
	      \end{array}
	    \right).
\label{couplingmatrix2}
\end{equation}
One might question the choices made for these hierarchies, since we know that both CKM and PMNS matrices present much weaker hierarchies. We will briefly motivate this choices below.

In case of weak hierarchies we know from current observations that any new physics giving rise to sizable contribution to $\mu \rightarrow e\gamma$ should usually come from scales much above $1$~TeV. Since new physics contributions do not have to follow either CKM or PMNS patterns, one might wonder: 
\begin{quote}\it
  Which hierarchy in the charged lepton sector should one have in order to reconcile possible signals coming from $g-2$ and LFV? What can we learn if the $g-2$ anomaly is confirmed by the upcoming $g-2$ experiments, and no signal is seen in the $\mu \rightarrow e\gamma$ decay in the foreseeable future? If the $\mu \rightarrow e\gamma$ decay is seen in the upcoming years, do we need to necessarily observe a signal also in $g-2$?
\end{quote}
Those questions motivated the choice for the hierarchies above, in the sense that we choose the hierarchies that give rise to scenarios where one can reconcile possible signals in $g-2$ and $\mu \rightarrow e\gamma$ and/or a signal in one of either observable can be probed within current or future sensitivity of the experiments. In summary, the purpose of this approach is to illustrate three different scenarios:
\begin{enumerate}[label=(\roman*)]
  \item New physics signal in  $a_\mu$:
  
  We will see that null results from LFV places stringent limits on new physics models capable of explaining $g-2$. In order to make both comparable, a strong hierarchy in the coupling is needed.
  
  \item New physics signal in $\MEG$:
  
  Assuming that the $g-2$ anomaly is resolved otherwise, using the $1\sigma$ error bar on $g-2$, we can set strong limits on new physics interpretations to  $\MEG$. We will see that most possible scenarios are already ruled out by $g-2$.
  
  \item New physics signal seen in both $a_\mu$ and $\MEG$:

  Depending on the hierarchy used, some models offer regions of parameter space where both signals can be simultaneously accommodated. 
\end{enumerate}
We discuss those three scenarios for several simplified models that preserve $SU(2)_L$ invariance below.

\subsection{Scalar Contributions}
We begin the discussion of the results with simplified models describing additional scalar fields restoring the previously disregarded $SU(2)_L$ invariance.

\subsubsection{Scalar doublet}\label{sec:Discussion_scalarDoublet}

Scalar doublets, as they occur in two Higgs doublet models, and multi-Higgs doublet models~\cite{Felipe:2014zka}, may also show up in the broken phase of a 331 symmetric model with a scalar $SU(3)_L$ sextet. We consider first the case of a scalar doublet $\phi$ with hypercharge $Y=1/2$, which couples to the SM leptons as does the SM Higgs doublet
\begin{equation} \label{eq:down_type_Yukawa}
  \mathcal{L}_\mathrm{int} = {g_L}_{ij} \overline{e_R^i}\, \phi^\dag \cdot \ell_L^j + \mathrm{h.c.}\, ,
\end{equation}
where `\,$\cdot$\,' denotes the $SU(2)_L$ invariant product. In this case the imposed $SU(2)_L$ invariance dictates that $g_R = {g_L}^\dag$ such that assuming real and symmetric couplings, we find that $g_p = - \Im\, g_L = 0$ and $g_s = g_L$ for the electrically neutral component of $\phi$.\footnote{For the purpose of illustration, we consider here only the CP even part of $\phi^0$. The CP odd scalar would have $g_s=0$ and $g_p = g_L$.} In contrast, for the charged field, we have $g_p = - g_s = -g_L/2$. Since the doublet $\phi$ consists of a neutral and a charged component, the result is a sum of Eqs.~\eqref{eq:Gamma1_result} and~\eqref{eq:Gamma2_result}, with SM leptons and neutrinos running in the loops, respectively.

\begin{figure}[t]
   \centering
   \begin{subfigure}[b]{.45\textwidth}
      \centering
      \includegraphics[width=\textwidth]{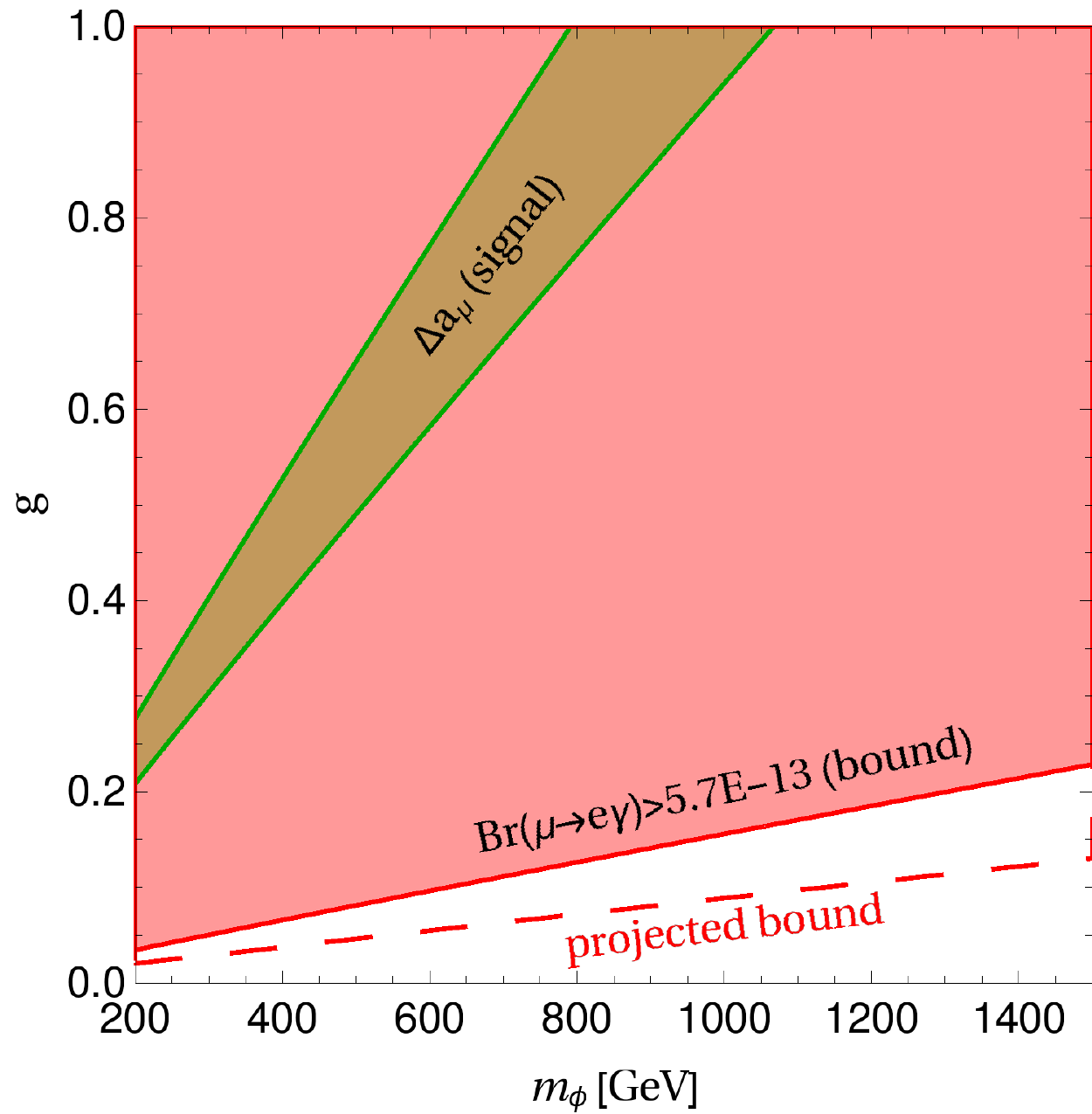}
      \subcaption{mild hierarchy}
   \end{subfigure}
   \hfill
   \begin{subfigure}[b]{.45\textwidth}
      \centering
      \includegraphics[width=\textwidth]{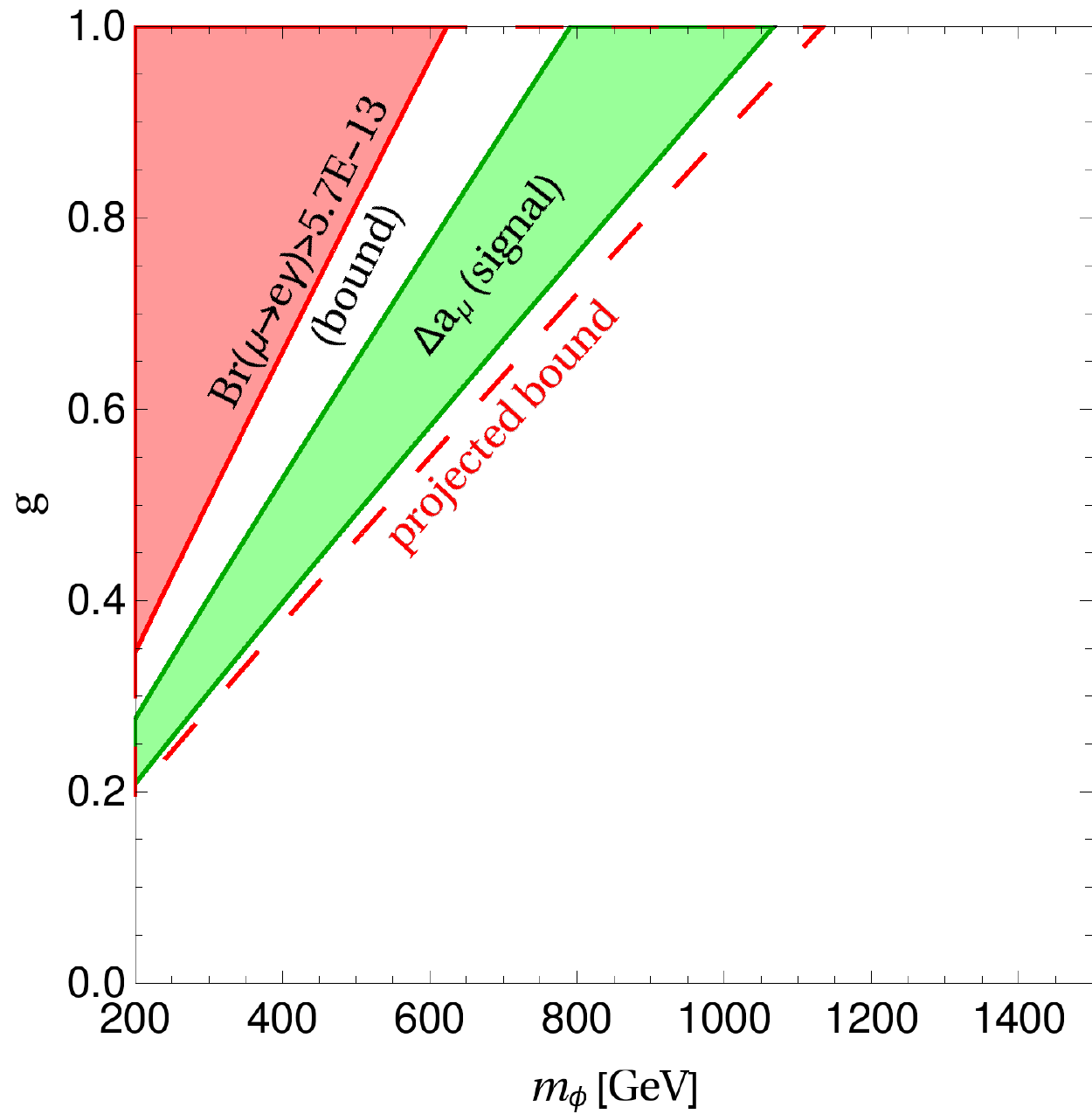}
      \subcaption{strong hierarchy}
   \end{subfigure}
	\caption{\label{fig:results_scalarDoublet1Da}Results for a scalar doublet with hypercharge $Y=1/2$. The green area corresponds to the signal region for $\Delta a_\mu$, while the red region is excluded by $\MEG$ at $1\sigma$. The projected bound is shown as a dashed line.}
\end{figure}

\begin{figure}[t]
   \begin{subfigure}[b]{.45\textwidth}
      \centering
      \includegraphics[width=\textwidth]{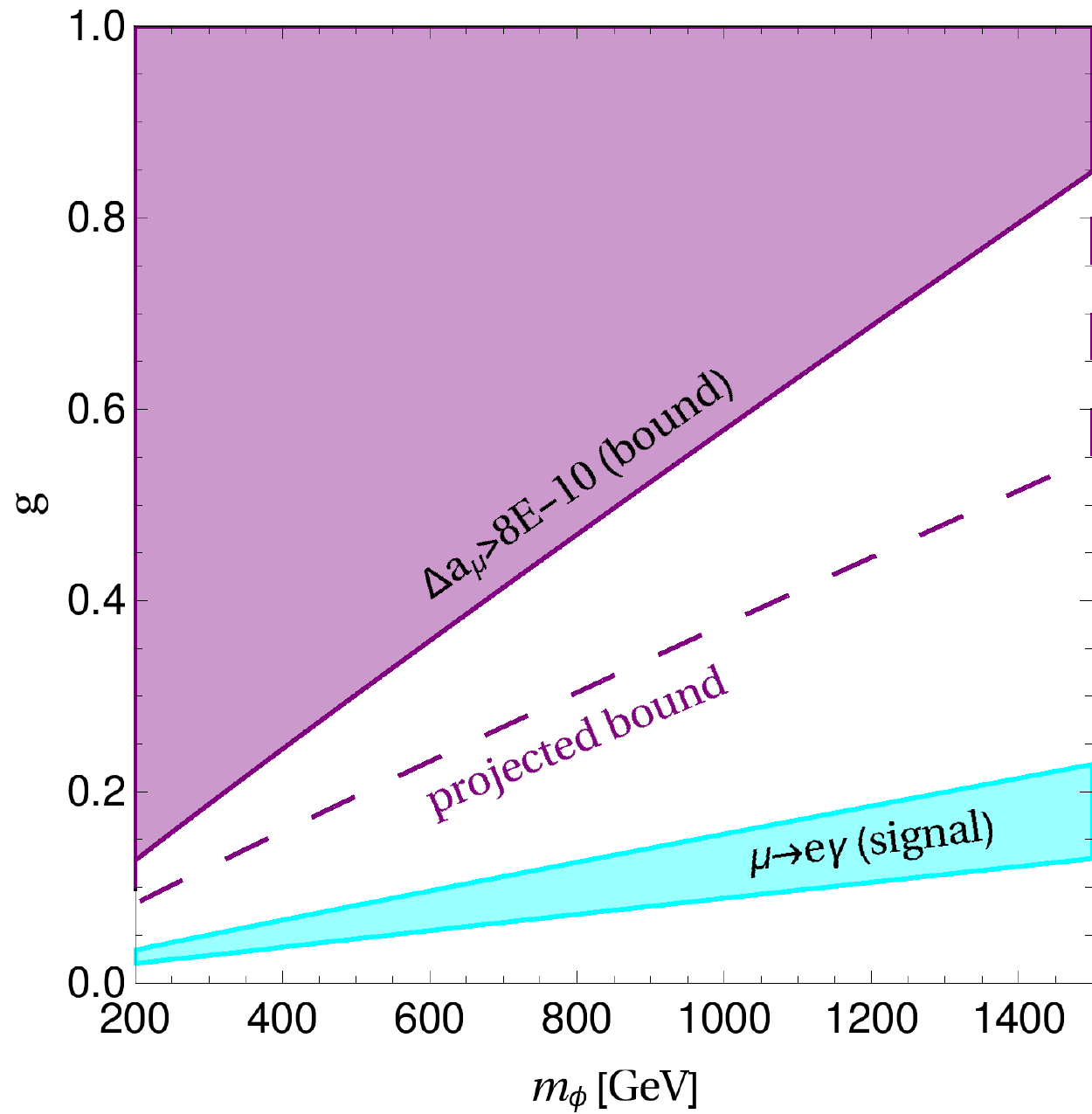}
      \subcaption{\label{fig:scalarDoublet1_Brmild}mild hierarchy}
   \end{subfigure}
   \hfill
   \begin{subfigure}[b]{.45\textwidth}
      \centering
      \includegraphics[width=\textwidth]{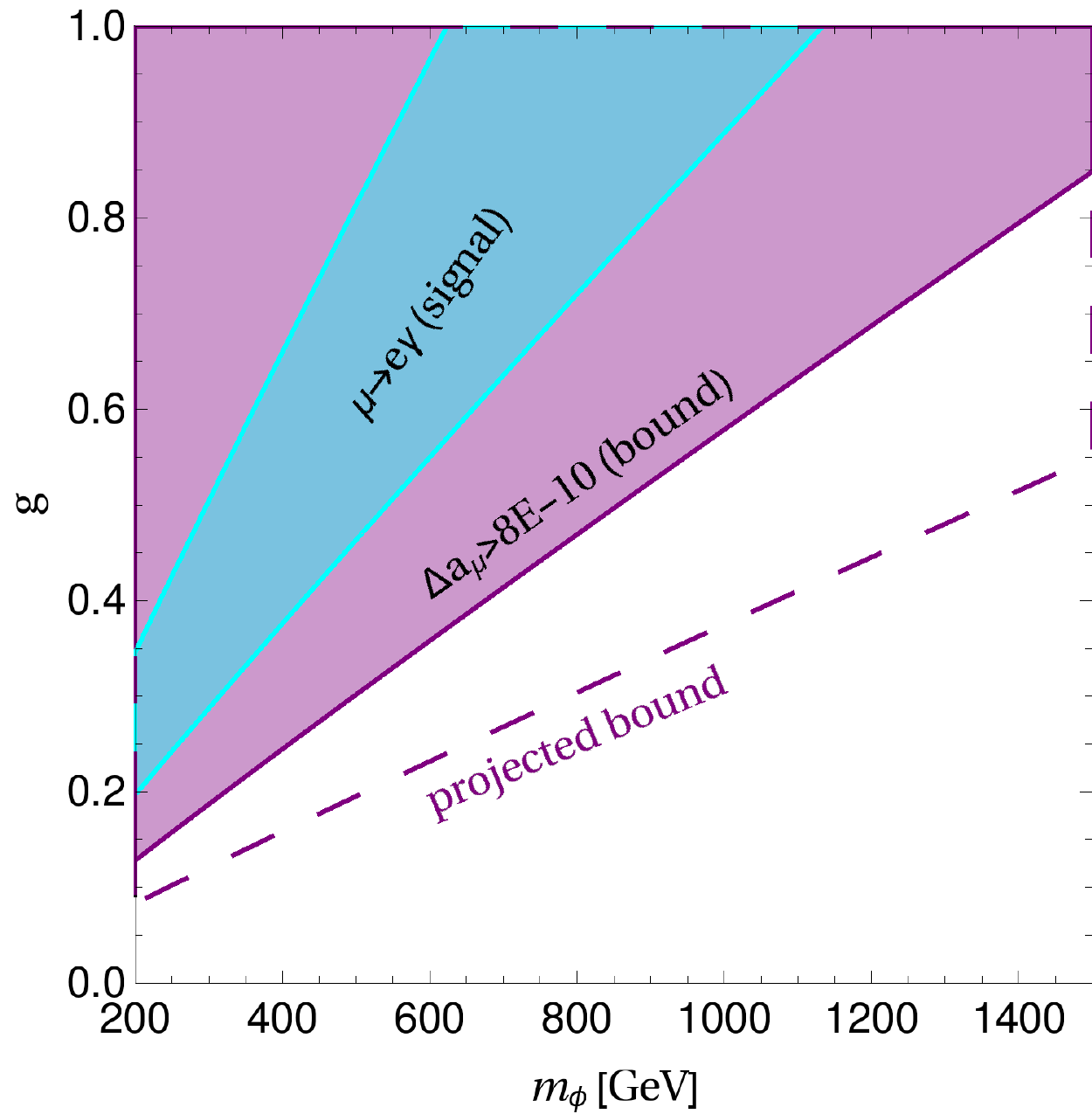}
      \subcaption{\label{fig:scalarDoublet1_Brstrong}strong hierarchy}
   \end{subfigure}
   \caption{\label{fig:results_scalarDoublet1Br}Reversing the argument. The light blue region represents the signal region for $\MEG$ between the current and the projected bound, while the violet region is excluded by $\Delta a_\mu$ assuming the anomaly is resolved. Again, the projected bound for $g-2$ is shown as a dashed line.}
\end{figure}

Since we have obtained these results already in Eqs.~\eqref{eq:Gamma1_result} and ~\eqref{eq:Gamma2_result}, there is no need to reproduce them. One can simply solve the integrals numerically to find the results shown in Fig.~\ref{fig:results_scalarDoublet1Da} and ~\ref{fig:results_scalarDoublet1Br}. 

In Fig.~\ref{fig:results_scalarDoublet1Da} we assume that the $g-2$ deviation persists and check whether this is consistent with the current and project limits on ${\rm BR}(\MEG)$ for the two aforementioned hierarchies. In both figures the green region represents the part of the parameter space (in the plane $g$ vs.~$m_{\phi}$), which could explain the $g-2$ deviation assuming the central value to be the same. Conversely, the shaded red region (dashed line) accounts for the current (projected) limit stemming from $\MEG$. 

In the left panel the result is exhibited assuming mild hierarchy and one can see that the region of parameter space which accommodates a signal in $g-2$ is excluded by the current limits on ${\rm BR}(\MEG)$. Interestingly, for a strong hierarchy (right panel), the region which explains the $g-2$ deviation falls within the projected sensitivity on the $\MEG$ decay. It should be noted that the charged scalar contribution must be sub-leading because its contribution to $\Delta a_\mu$ is negative.

Panels~\ref{fig:scalarDoublet1_Brmild} and~\ref{fig:scalarDoublet1_Brstrong} show the results with an orthogonal view. Assuming that $\Delta a_\mu$ is \emph{not} due to BSM physics, what are the constraints for $\MEG$? 

In both panels the blue area shows the region of parameter space in the plane $g$ vs  $m_{\phi}$ which could explain a signal in $\MEG$ with ${\rm BR (\MEG)} =4.2 \times 10^{-13} - 4 \times 10^{-14}$ which delimits the current and projected sensitivity. The shaded purple region (dashed line) delimits the current (projected) exclusion from the $g-2$ which is assumed to be otherwise resolved. 

In Fig.~\ref{fig:scalarDoublet1_Brmild} we adopt a mild hierarchy. As one might have anticipated, the limits from $g-2$ are very weak, even considering projected sensitivity. However, looking at Fig.~\ref{fig:scalarDoublet1_Brstrong} we may conclude that for a strong hierarchy a signal on $\MEG$ has already been ruled out by the $g-2$ constraint. The reason for this behavior is that $\Delta a_\mu$ grows with $g^2$, whereas $\mathrm{BR}(\MEG)$ grows with $g^4$. had we taken the signal in $\MEG$ to occur with a different ${\rm BR (\MEG)}$, the signal region for $\MEG$ would shift. In particular, if one takes a signal in $\MEG$ to happen at one order of magnitude below the projected sensitivity, i.e.~${\rm BR (\MEG)} =4 \times 10^{-15}$, it means that the coupling in Figs.~\ref{fig:scalarDoublet1_Brmild}-\ref{fig:scalarDoublet1_Brstrong} would have to be smaller roughly by a factor of two, since  ${\rm BR (\MEG)}$ goes with $g^4$, moving the signal region downwards.

Furthermore, we may observe from the results in Figs.~\ref{fig:results_scalarDoublet1Da} and~\ref{fig:results_scalarDoublet1Br} that the signal region/bounds of $\Delta a_\mu$ are rather insensitive to the chosen hierarchy, while for $\MEG$ the hierarchy is very decisive, illustrating that the $\Delta a_\mu$ is mostly sensitive to the flavor-diagonal couplings, while $\MEG$ probes the non-diagonal entries in $\Lambda$. Varying the hierarchies, we may naturally interpolate between both scenarios; however, a high degree of fine-tuning in the hierarchies would be needed to incorporate both viable signals $\MEG$ and $\Delta a_\mu$ in such a model. An example where the signal region for $g-2$ would have a significant change is when the $\tau-$lepton contribution becomes relevant for some reason.

In summary, the $g-2$ anomaly favors the large coupling and low mass regions, which are often disfavored by LFV searches cf.\ Fig.~\ref{figure3LFV}. Precisely for this reason, mild or strong hierarchies may yield signals in either $g-2$, or $\MEG$.

As a side note, keep in mind that a scalar doublet $\phi$ could also be of hypercharge $Y=-1/2$ such that it may be coupled to RH neutrinos in the following way
\begin{equation} \label{eq:up_type_Yukawa}
  \mathcal{L}_\mathrm{int} = {g_L}_{ij} \overline{N_R^i}\, \phi^\dag \cdot \ell_L^j + \mathrm{h.c.}
\end{equation}
However, glancing at Eq.~\eqref{eq:Delta_a_Singly_Scalar}, one can see that $\Delta a_\mu>0$ is possible only if we have a sizable RH neutrino mass $m_N$ and dominant pseudo-scalar coupling. However, in an $SU(2)_L$ invariant framework this will not be possible to obtain, and hence a hypercharge $-1/2$ scalar doublet cannot explain the $g-2$ anomaly.

\subsubsection{\label{Sec:scalartriplet} Scalar Triplet}

We conclude the discussion of scalar contributions to $\ell_i \to \ell_j \gamma$ with a model involving a $Y=1$ scalar triplet $\Delta$. Such a field contains a neutral, a singly and a doubly charged scalar component field with
\begin{equation}
  \Delta = \left( \begin{array}{cc}
		    \phi^+/\sqrt{2} 	& \phi^{++} \\
		    \phi^0 		& - \phi^+/\sqrt{2} 
                  \end{array}
	   \right).
\end{equation}
Such scalar triplets are arguably a signature of a type~II seesaw mechanism for generating neutrino masses. They can naturally appear in Left-Right symmetric models. However, they might also appear as the result of a broken scalar sextet in 331 models. The result is in principle a sum of \emph{four} diagrams, one for the neutral and charged component each, and the two diagrams shown in Fig.~\ref{fig:DoublychargedScalar} for the doubly charged field. Suppressing $SU(2)_L$ indices, the interaction term reads:
\begin{equation}
  \mathcal{L}_\mathrm{int} = g_L \overline{\ell_L^\mathcal{C}}^i i \sigma^2 \Delta \ell_L^j + \mathrm{h.c.}
\end{equation}
At this point, one should remark that due to electric charge conservation, the neutral component of $\Delta$ only couples to neutrinos, and hence has no effect on the charged leptons. On the other hand, we know from Sec.~\ref{sec:results} that both the singly and doubly charged scalars tend to yield a negative $\Delta a_\mu$, which cannot explain the observed excess. 

\begin{figure}[t]
   \centering
   \begin{subfigure}[b]{.45\textwidth}
      \centering
      \includegraphics[width=\textwidth]{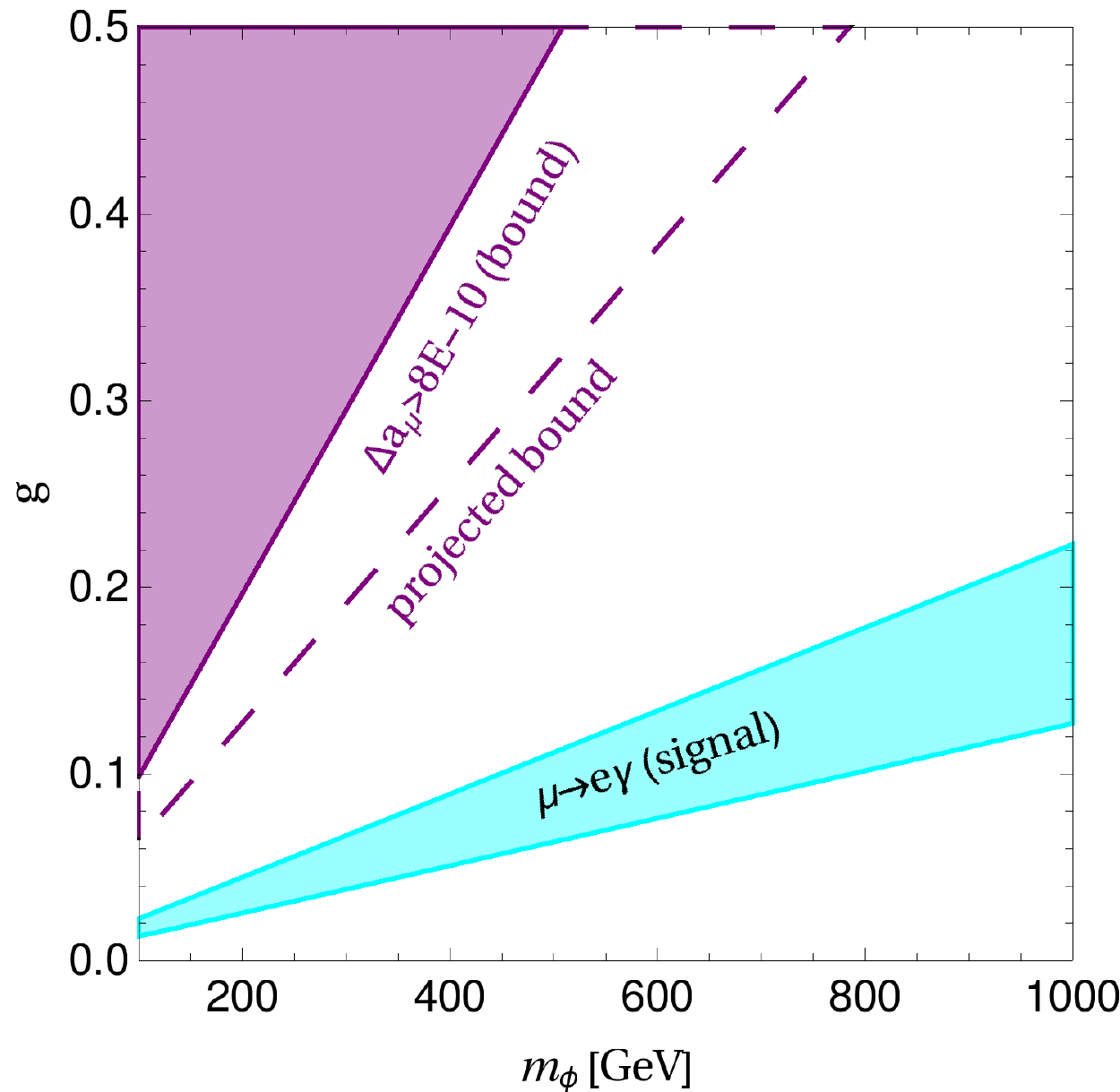}
      \subcaption{mild hierarchy}
   \end{subfigure}
   \hfill
   \begin{subfigure}[b]{.45\textwidth}
      \centering
      \includegraphics[width=\textwidth]{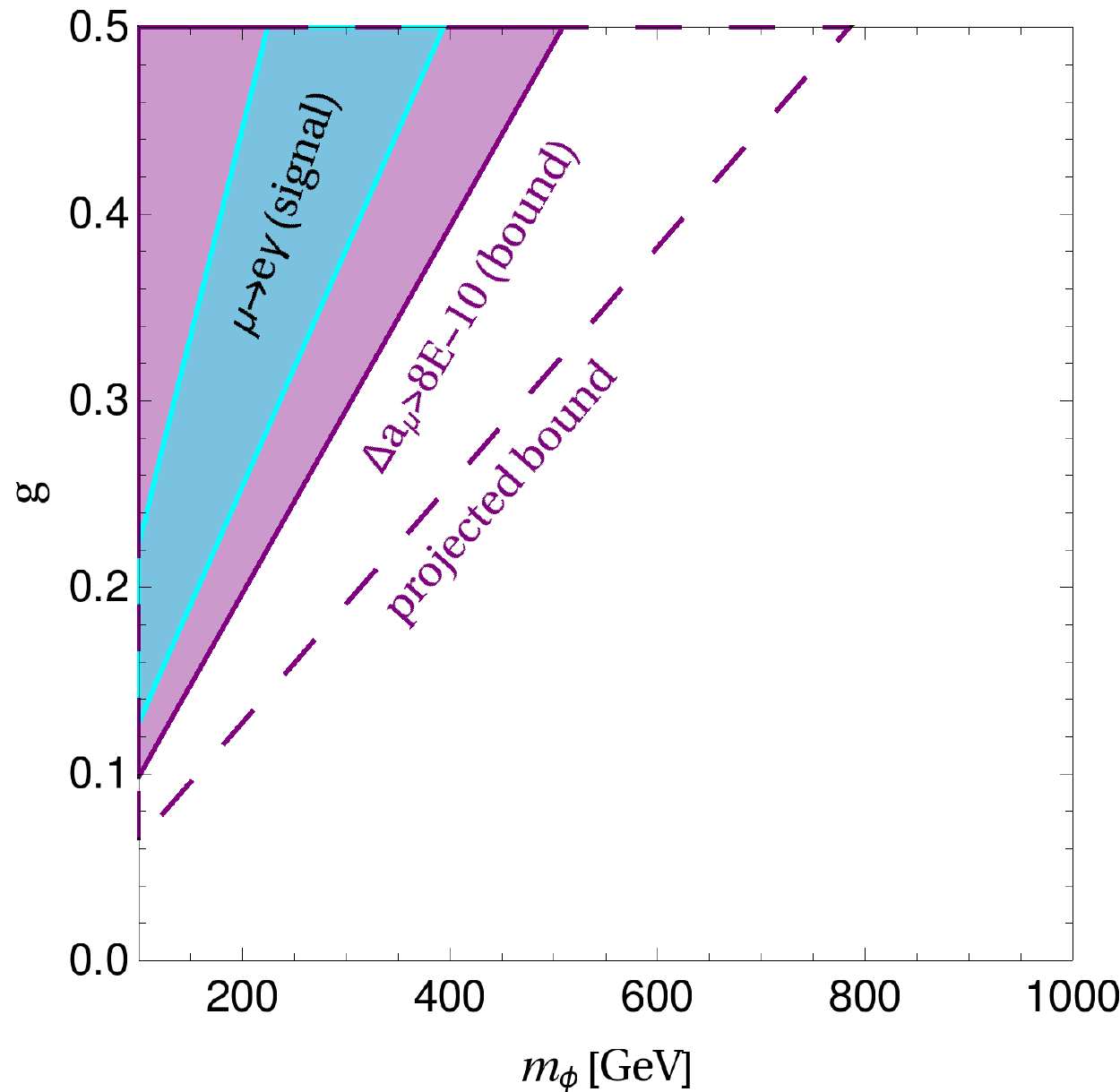}
      \subcaption{strong hierarchy}
   \end{subfigure}
   \caption{\label{fig:results_scalarTriplet}Results for possible LFV signals induced by a scalar triplet $\Delta$ coupling to SM leptons.}
\end{figure}

However, assuming there is a signal in $\MEG$ with ${\rm BR (\MEG)} =4.2\times 10^{-13}-4 \times 10^{-14}$, i.e.~within future sensitivity, we can draw the regions where such a signal is expected for a mild (strong) hierarchy as shown in the left (right) panel of Fig.~\ref{fig:results_scalarTriplet}. We observe from Fig.~\ref{fig:results_scalarTriplet} that there is a window for future observation of $\MEG$ if the Yukawa matrix has a mild hierarchy. In both panels the blue area shows the region of parameter space in the plane $g$ vs.~$m_{\phi}$ which could explain a signal in $\MEG$ and the shaded purple region (dashed line) represents the current (projected) exclusion from the $g-2$ which is assumed to be otherwise resolved. We emphasize that if we have adopted a signal in $\MEG$ to occur at ${\rm BR (\MEG)} =4 \times 10^{-15}$, for instance, the signal region would shift by a factor of two downwards, and in this case, even for the strong hierarchy case, a signal in $\MEG$ would be consistent with the current $g-2$ limit. However, one should keep in mind
that collider searches for such scenarios restrict the mass of a doubly charged scalar to be larger than $400\GeV$~\cite{Aad:2014hja}, already excluding a large region of the parameter space.

\subsection{Fermion Singlet Contributions}

Here, we will discuss the case in which fermionic $SU(2)_L$ singlet fields, with the electric charge equal to the field's hypercharge, contribute to $g-2$. Such fermions can be neutral, singly charged and even doubly charged as we explore below.

\subsubsection{Neutral Fermion Singlet}\label{sec:WprimeN}
\begin{figure}[p]
   \centering
   \noindent\makebox[\textwidth]{
   \begin{subfigure}[b]{.45\textwidth}
      \centering
      \includegraphics[width=\textwidth]{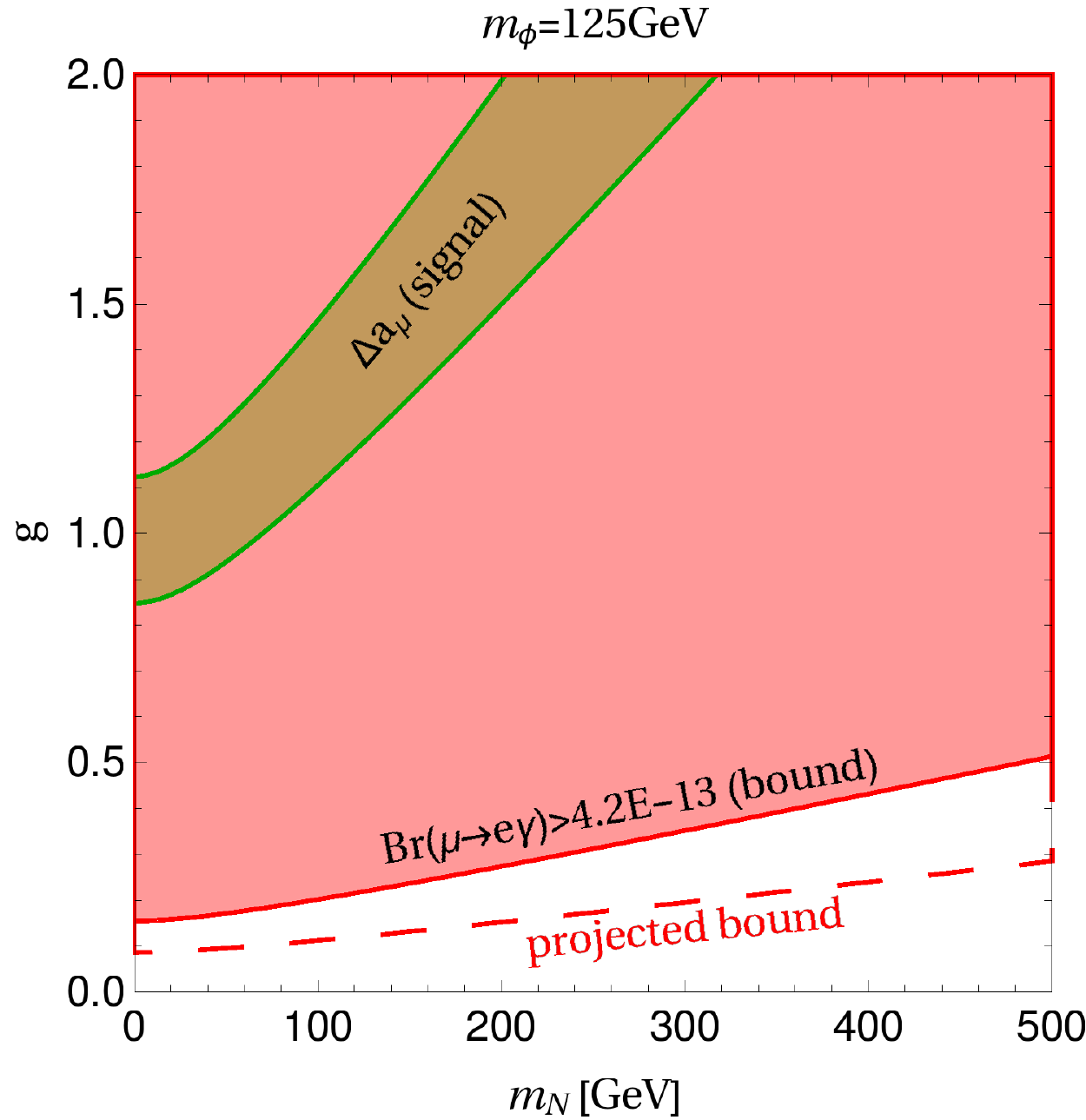}
      \subcaption{mild hierarchy}
   \end{subfigure}
   \hfill
   \begin{subfigure}[b]{.45\textwidth}
      \centering
      \includegraphics[width=\textwidth]{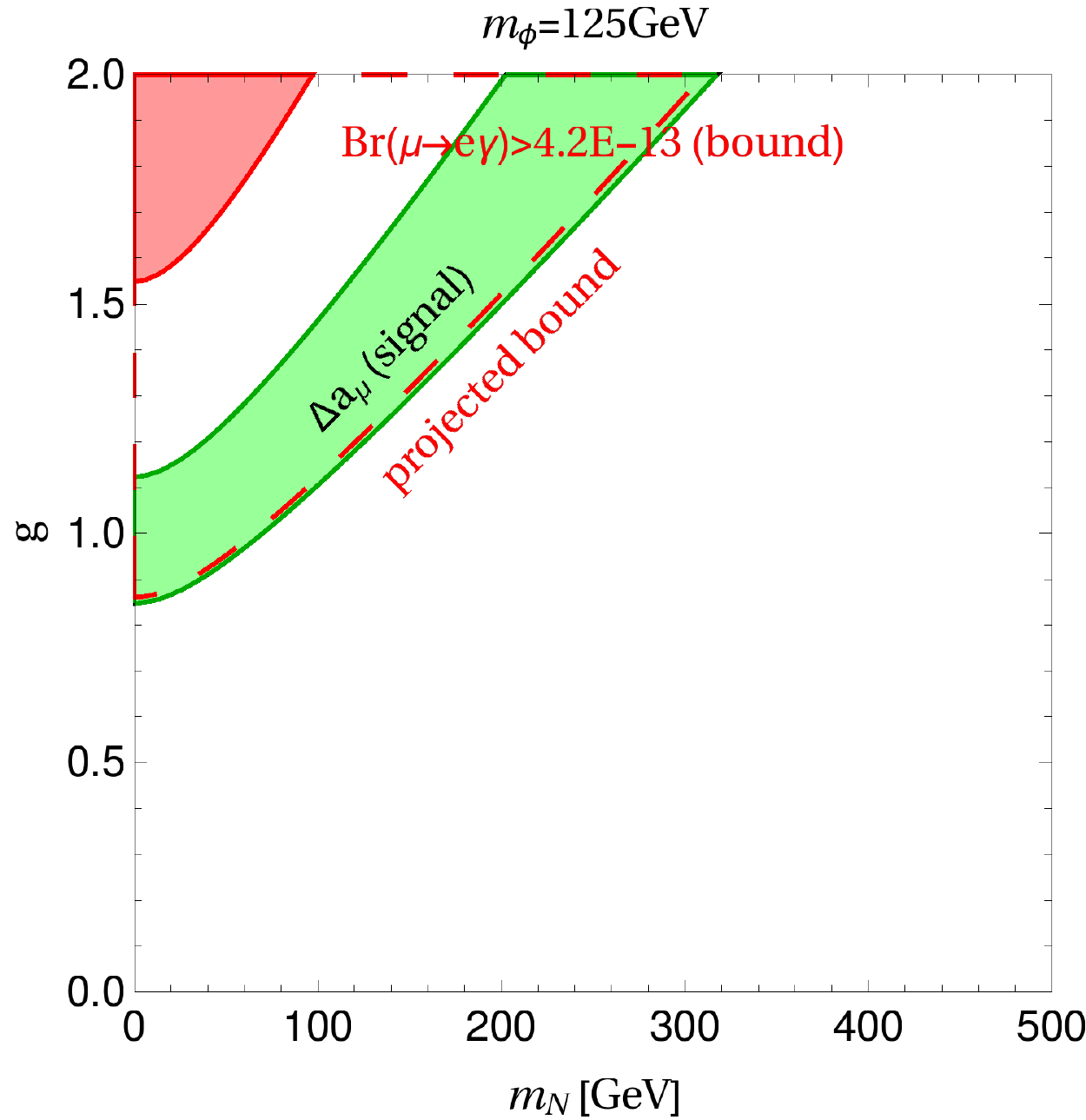}
      \subcaption{strong hierarchy}
   \end{subfigure}
   }\\
    \vspace{5mm}
   \noindent\makebox[\textwidth]{
   \begin{subfigure}[b]{.45\textwidth}
      \centering
      \includegraphics[width=\textwidth]{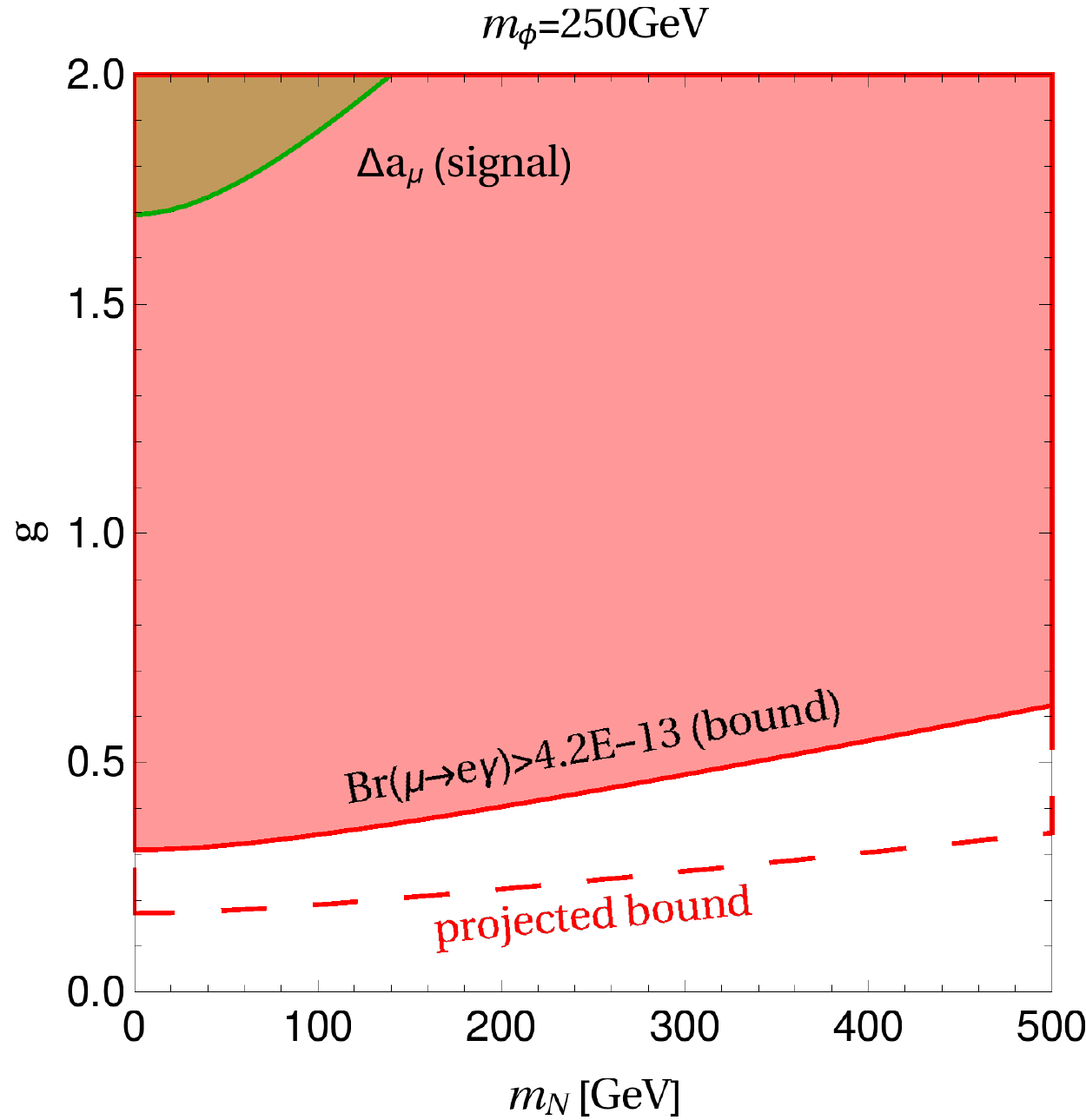}
      \subcaption{mild hierarchy}
   \end{subfigure}
   \hfill
   \begin{subfigure}[b]{.45\textwidth}
      \centering
      \includegraphics[width=\textwidth]{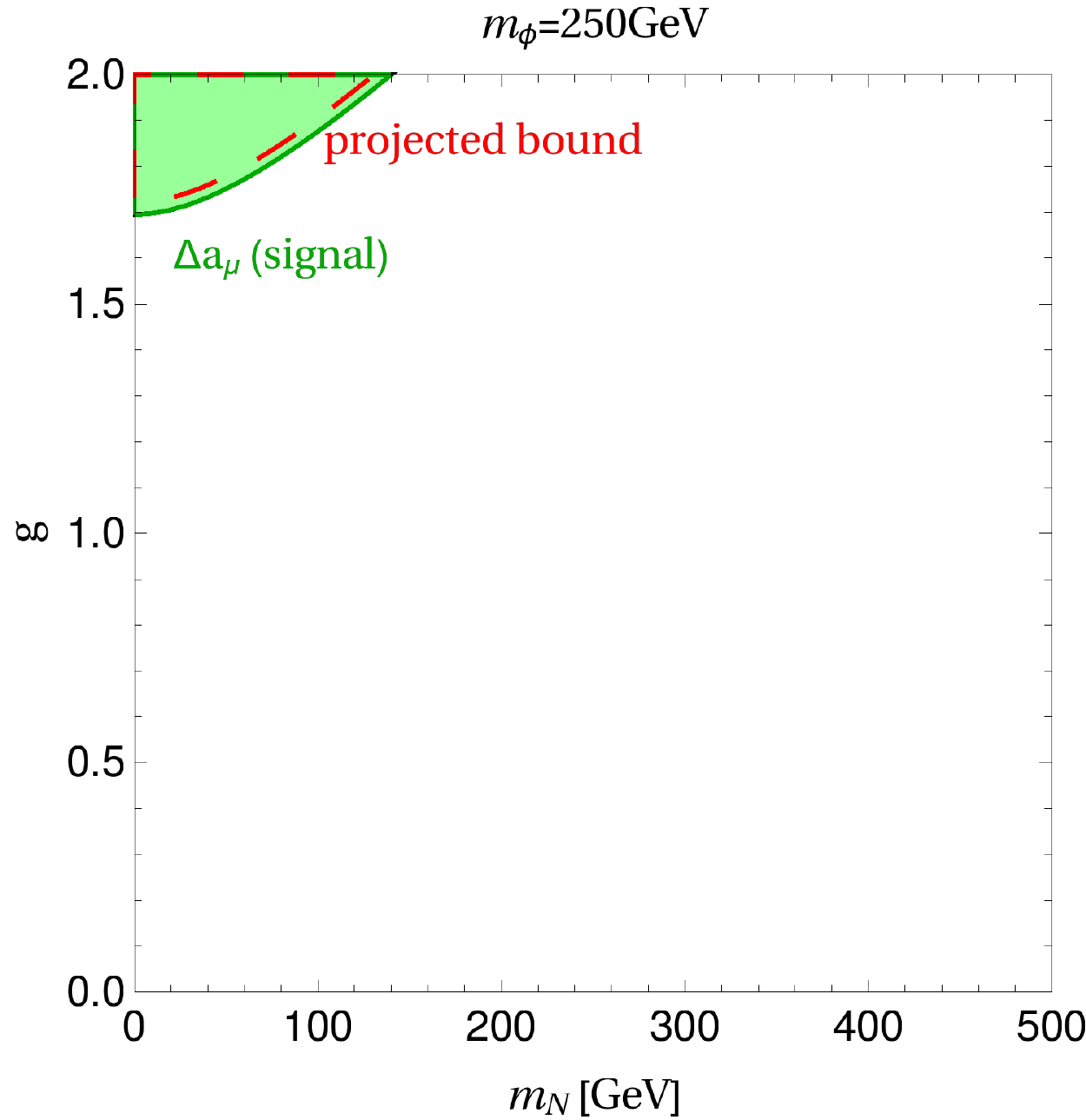}
      \subcaption{strong hierarchy}
   \end{subfigure}
   }
   \caption{\label{fig:results_fermionNeutral1Da}$g-2$ signal region for a neutral fermion $N$ coupling to the SM leptons via a scalar $\phi$ for different scalar masses $m_\phi = \left(125, 250\right) \GeV$.}
\end{figure}

\begin{figure}[p]
   \centering
   \noindent\makebox[\textwidth]{
   \begin{subfigure}[b]{.45\textwidth}
      \centering
      \includegraphics[width=\textwidth]{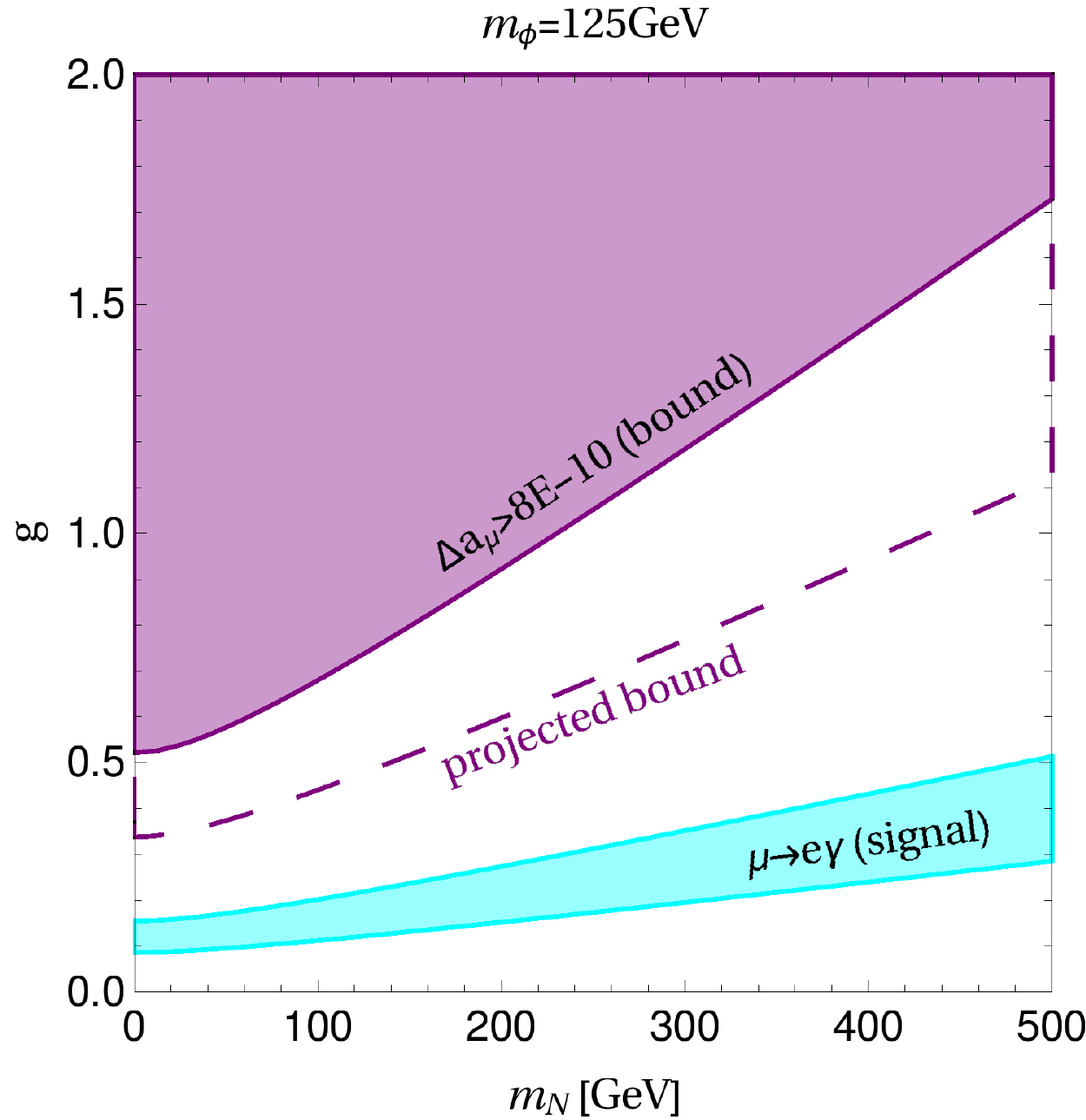}
      \subcaption{mild hierarchy}
   \end{subfigure}
   \hfill
   \begin{subfigure}[b]{.45\textwidth}
      \centering
      \includegraphics[width=\textwidth]{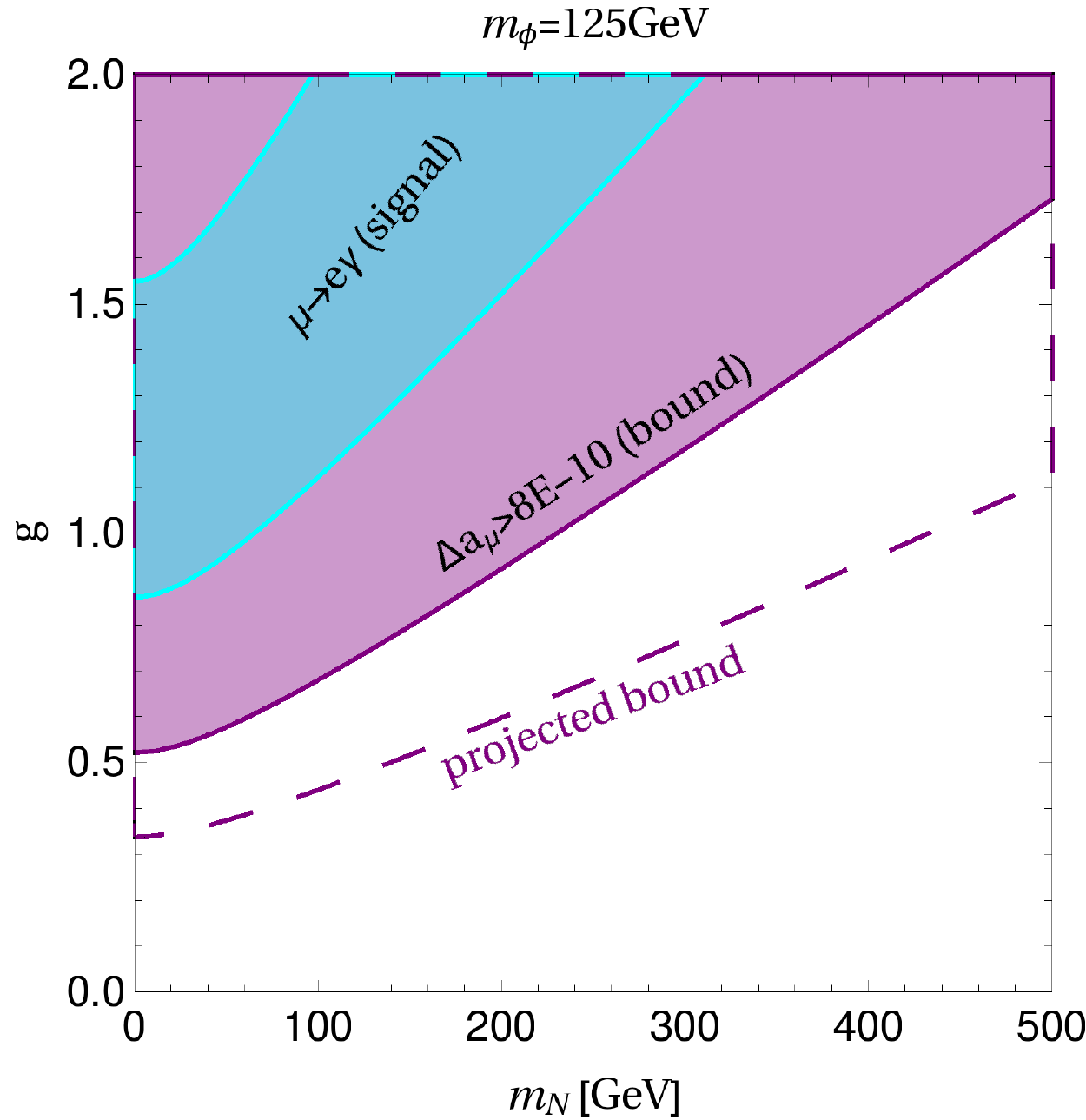}
      \subcaption{strong hierarchy}
   \end{subfigure}
   }\\
    \vspace{5mm}
   \noindent\makebox[\textwidth]{
   \begin{subfigure}[b]{.45\textwidth}
      \centering
      \includegraphics[width=\textwidth]{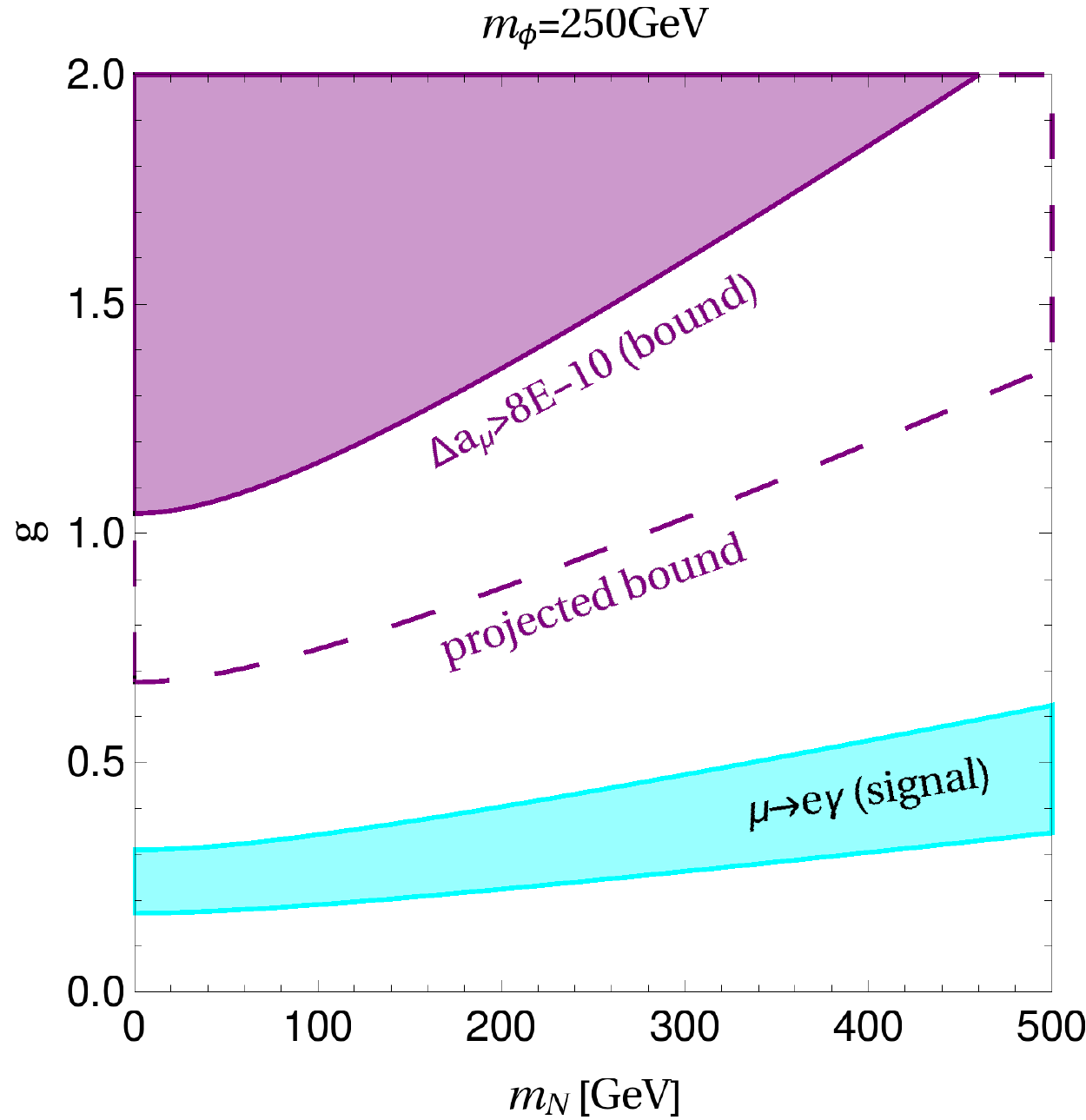}
      \subcaption{mild hierarchy}
   \end{subfigure}
   \hfill
   \begin{subfigure}[b]{.45\textwidth}
      \centering
      \includegraphics[width=\textwidth]{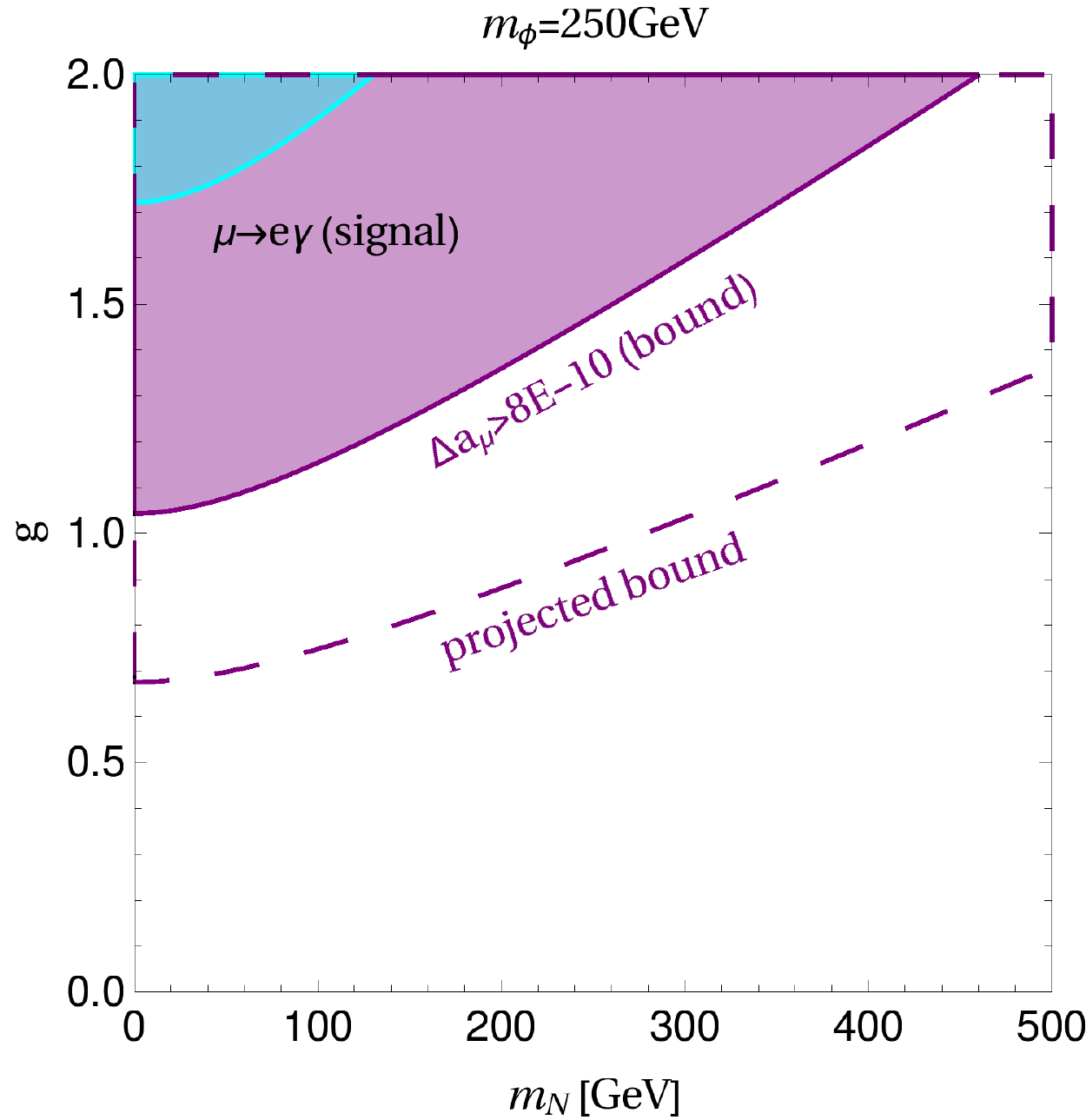}
      \subcaption{strong hierarchy}
   \end{subfigure}
   }
   \caption{\label{fig:results_fermionNeutral1Br}Signal region for $\MEG$ for a neutral fermion singlet.}
\end{figure}

Neutral fermions are present in many phenomenological studies of $g-2$ and in connection to dark matter~\cite{Agrawal:2014ufa}. For a new neutral fermion, we can have two potential interactions, one of which is the scalar field which is identical to the result already given in Eq.~\eqref{eq:up_type_Yukawa}. Repeating the previous analysis but now varying the fermion mass and fixing different scalar masses yields Figs.~\ref{fig:results_fermionNeutral1Da} and~\ref{fig:results_fermionNeutral1Br}. In the upper panels of Fig.~\ref{fig:results_fermionNeutral1Da} we set $m_{\phi}=125$~GeV  with the left graph for mild hierarchy and the right for strong hierarchy. In the bottom panels we fix $m_{\phi}=250$~GeV. In all these panels the green regions represent the parameter space in the plane $g$ vs.~$m_{N}$ which could explain the $g-2$ deviation assuming the same central value, whereas the shaded red region (dashed line) accounts for the current (projected) limit stemming from $\MEG$. 

Looking at Fig.~\ref{fig:results_fermionNeutral1Da}, one may conclude that the $g-2$ anomaly favors light mediators (both scalar and fermion need to be light in this case), or large couplings of $\mathcal{O}(1)$. For a mild hierarchy with $m_{\phi}=(125,250)$~GeV the deviation in $g-2$ is already ruled out by the $\MEG$ limit, whereas for a strong hierarchy the $g-2$ signal region lies within current and future sensitivity for the $\MEG$ decay. One can take an orthogonal view to these findings as shown in Fig.~\ref{fig:results_fermionNeutral1Br}.
Again assuming a signal in $\MEG$ with ${\rm BR (\MEG)} =4.2\times 10^{-13}-4 \times 10^{-14}$, we learn that for LFV decays the hierarchy in flavor space is more decisive than the masses of the particles themselves. In the upper panels of Fig.~\ref{fig:results_fermionNeutral1Br} again we set $m_{\phi}=125$~GeV, and in the bottom panels we keep  $m_{\phi}=250$~GeV. As expected, a signal in $\MEG$ has an opposite effect compared to $g-2$, where the mild hierarchy was excluded. This time for a mild hierarchy a signal in $\MEG$ is perfectly consistent with $g-2$ physics, whereas for the strong hierarchy case the scenario is widely excluded by the $g-2$ limit.


The more interesting scenario is obtained if the neutral fermion couples to a spin-1 field and the charged leptons via a RH charged current,
\begin{equation}
  \mathcal{L}_\mathrm{int} = {g_R}_i \overline{N_R} \gamma^\mu e_R^i W_\mu^\prime + \mathrm{h.c.}\,,
  \label{Eq:WprimeNgeneral}
\end{equation}
as it occurs in Left-Right models.

\begin{figure}[p]
   \centering
   \noindent\makebox[\textwidth]{
   \begin{subfigure}[b]{.45\textwidth}
      \centering
      \includegraphics[width=\textwidth]{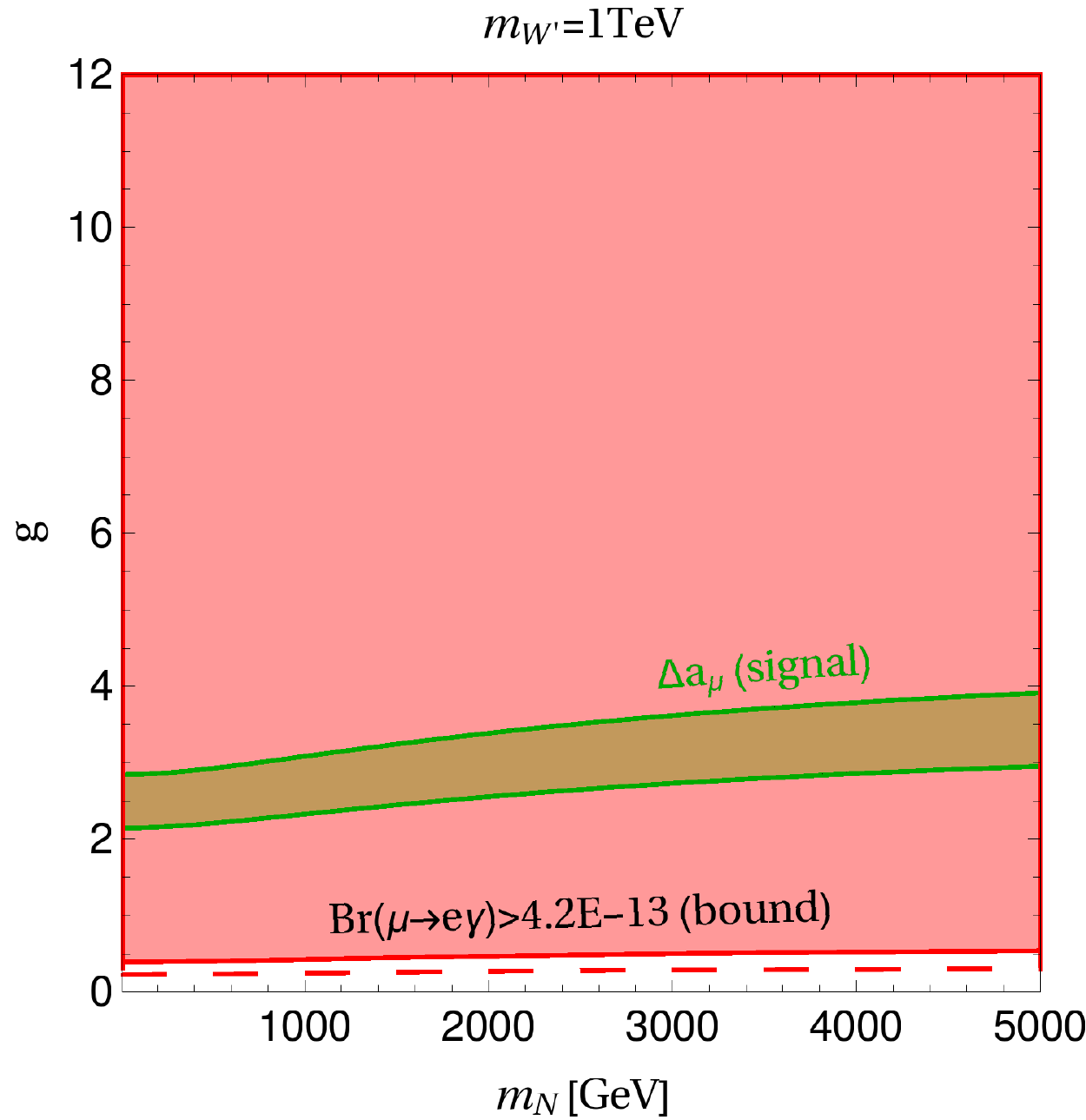}
   \end{subfigure}
   \hfill
   \begin{subfigure}[b]{.45\textwidth}
      \centering
      \includegraphics[width=\textwidth]{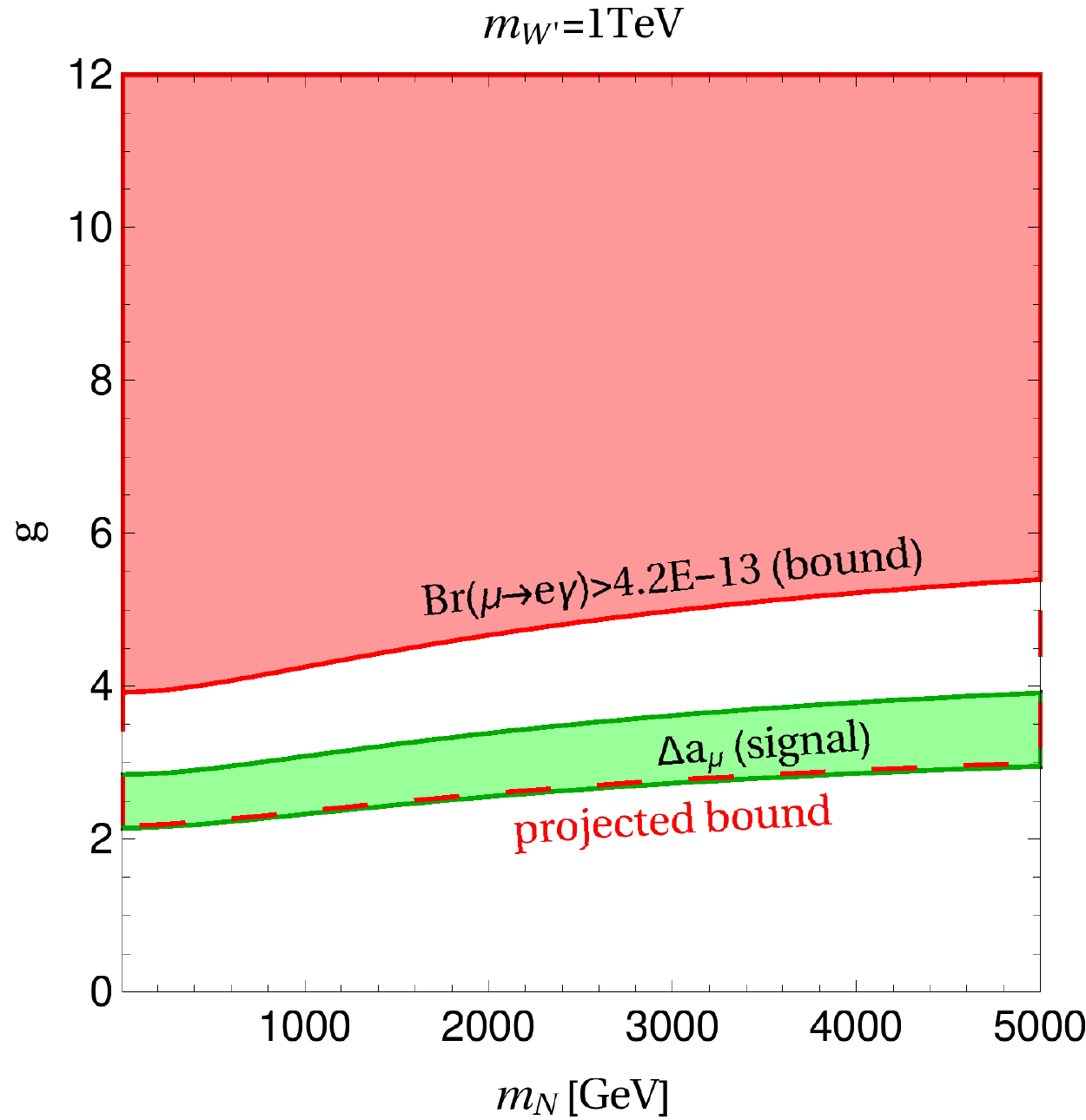}
   \end{subfigure}
   }\\
    \vspace{5mm}
   \noindent\makebox[\textwidth]{
   \begin{subfigure}[b]{.45\textwidth}
      \centering
      \includegraphics[width=\textwidth]{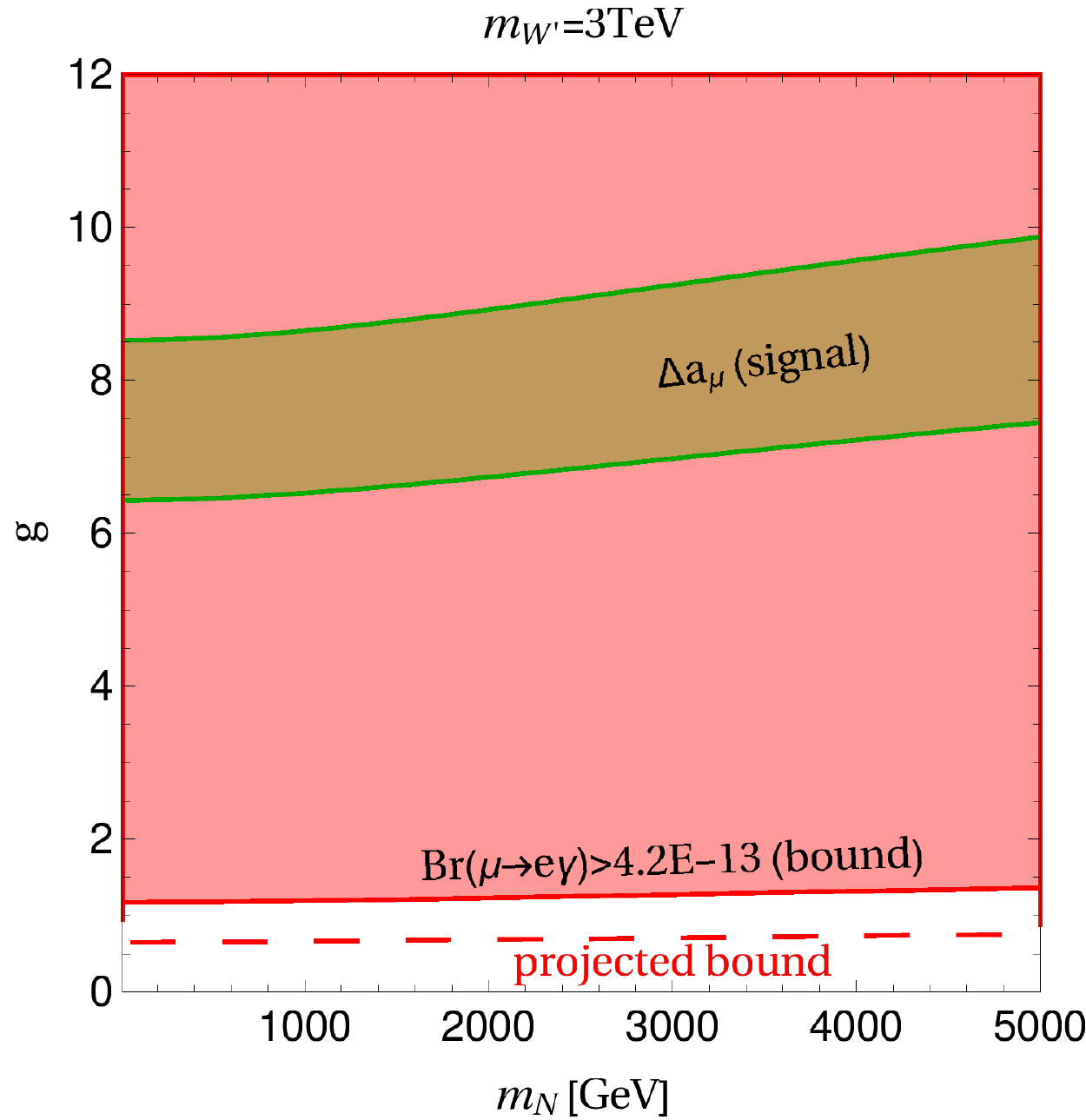}
   \end{subfigure}
   \hfill
   \begin{subfigure}[b]{.45\textwidth}
      \centering
      \includegraphics[width=\textwidth]{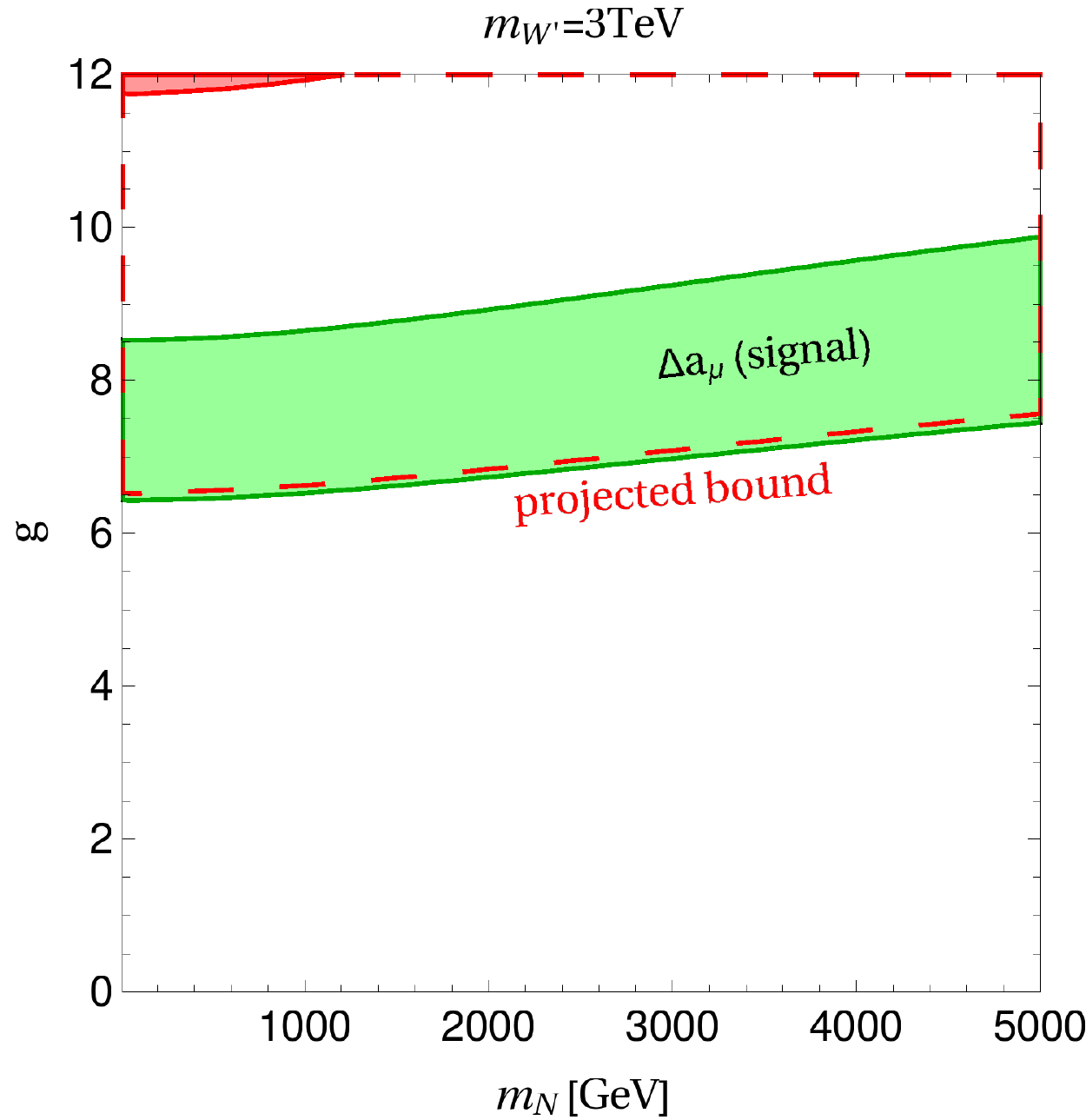}
   \end{subfigure}
   }\\
    \vspace{5mm}
   \noindent\makebox[\textwidth]{
   \begin{subfigure}[b]{.45\textwidth}
      \centering
      \includegraphics[width=\textwidth]{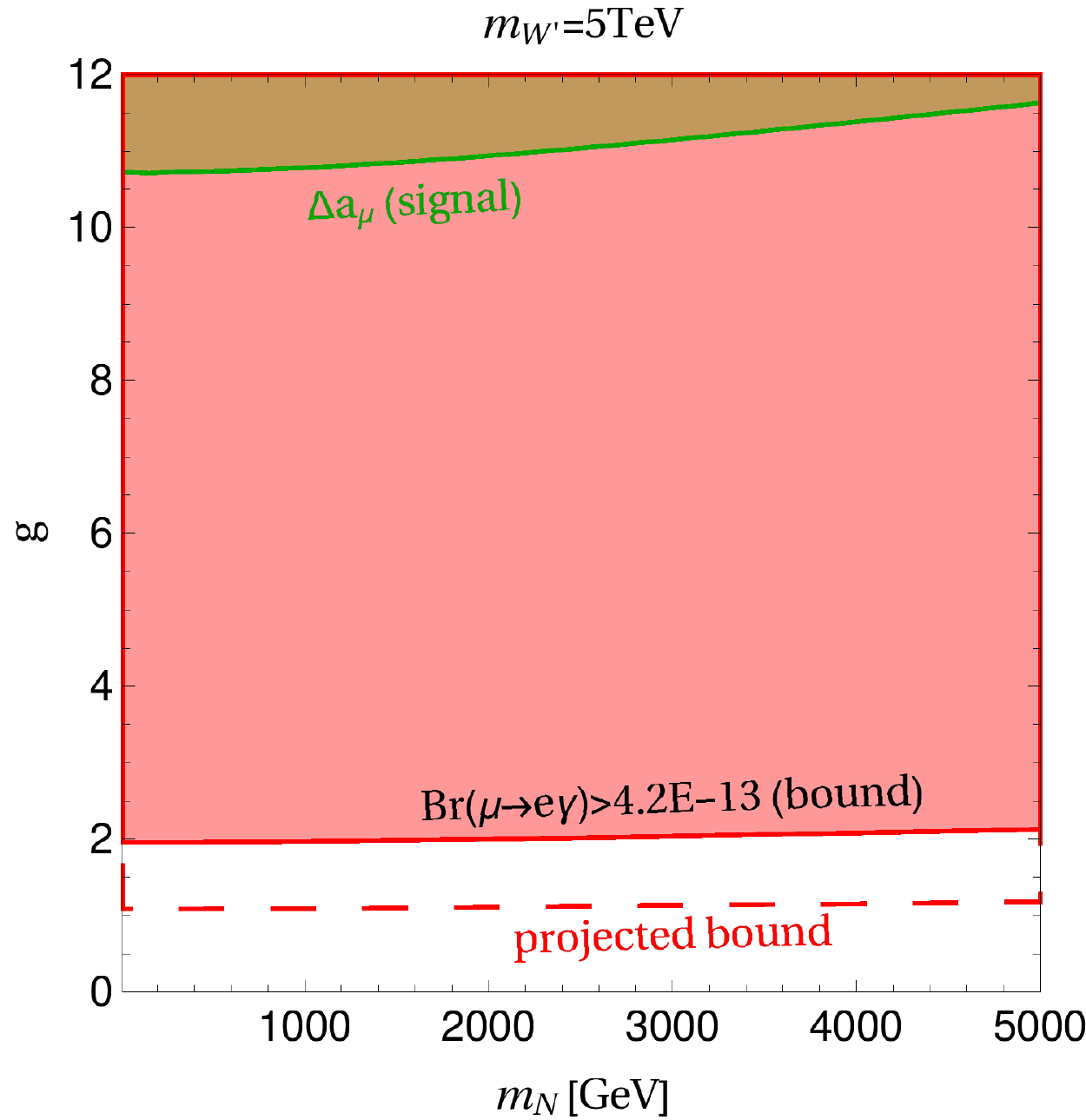}
      \subcaption{mild hierarchy}
   \end{subfigure}
   \hfill
   \begin{subfigure}[b]{.45\textwidth}
      \centering
      \includegraphics[width=\textwidth]{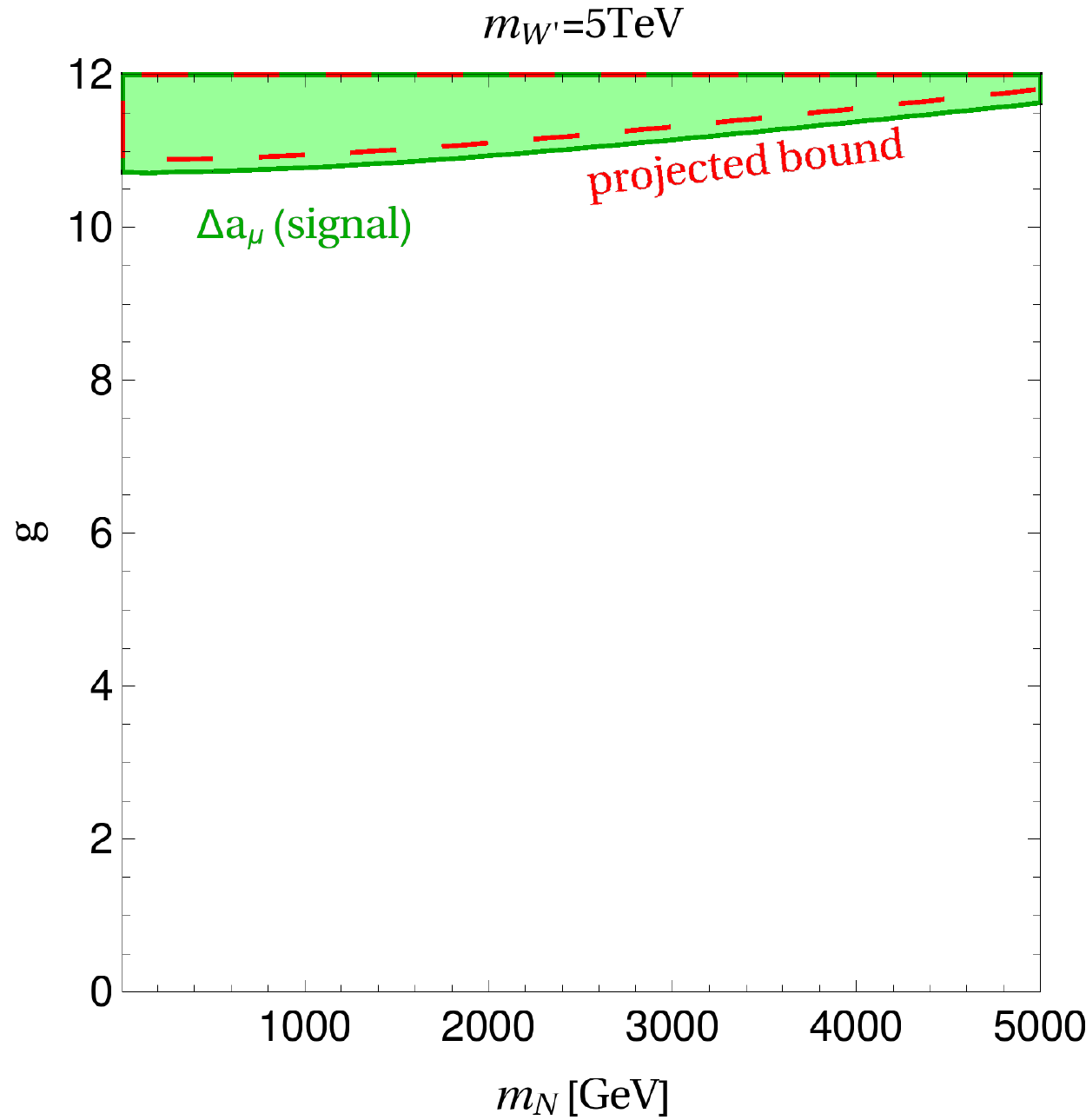}
      \subcaption{strong hierarchy}
   \end{subfigure}
   }
   \caption{\label{fig:results_fermionNeutral2Da}Contribution to the $g-2$ for a neutral fermion $N$ coupling to a $W^\prime$ boson and the SM leptons as described in Eq.~\eqref{Eq:WprimeNgeneral}.}
\end{figure}

The fully general result has been obtained in Eq.~\eqref{eq:P3def}, and limiting cases are given in Eq.~\eqref{eq:approxamu_W1} and Eq.~\eqref{eq:approxamu_W2}. In what follows we simply solve  Eq.~\eqref{eq:P3def} numerically and display the results in the plane $g$ vs $m_N$ in Fig.~\ref{fig:results_fermionNeutral2Da}, assuming the deviation in $g-2$ remains with the same central value, and overlay the limits from $\MEG$ using the same color scheme of the previous sections. In the left (right) graphs a mild (strong) hierarchy in the charged leptonic sector is adopted. In the upper panels of Fig.~\ref{fig:results_fermionNeutral2Da} we set $m_{W^{\prime}}=1$~TeV, in the middle ones $m_{W^{\prime}}=3$~TeV, and in the bottom $m_{W^{\prime}}=5$~TeV. For $m_{W^{\prime}}=1$~TeV we notice that the $g-2$ signal is consistent with $\MEG$ using strong hierarchies, and similarly for heavier masses but larger couplings are required. All cases with a mild hierarchy are excluded. Knowing that the coupling in Eq.~\eqref{Eq:WprimeNgeneral} is a gauge coupling, the plots exhibiting a large coupling (much larger than unity) in Fig.\ref{fig:results_fermionNeutral2Da} are rather unnatural.

We emphasize that we are not taking into account the existence of collider bounds, which are more restrictive for larger couplings. Currently they exclude $m_W^{\prime}$ masses up to $3-4$~TeV~\cite{Khachatryan:2014dka} with a projected sensitivity of up to $6$~TeV~\cite{Lindner:2016lxq}, assuming $g=e/s_W$, by sifting events with two leptons plus two jets or simply two jets. The former is rather sensitive to the $m_N$ though, and it weakens significantly when $m_W^{\prime} \simeq m_N$ or when $m_W^{\prime} \gg m_N$. So our findings have to be used with care. The latter, is not as sensitive but it does weaken when  $m_W^{\prime} > m_N$ since the branching ratio into jets is partially reduced. 

\begin{figure}[p]
   \centering
   \noindent\makebox[\textwidth]{
   \begin{subfigure}[b]{.45\textwidth}
      \centering
      \includegraphics[width=\textwidth]{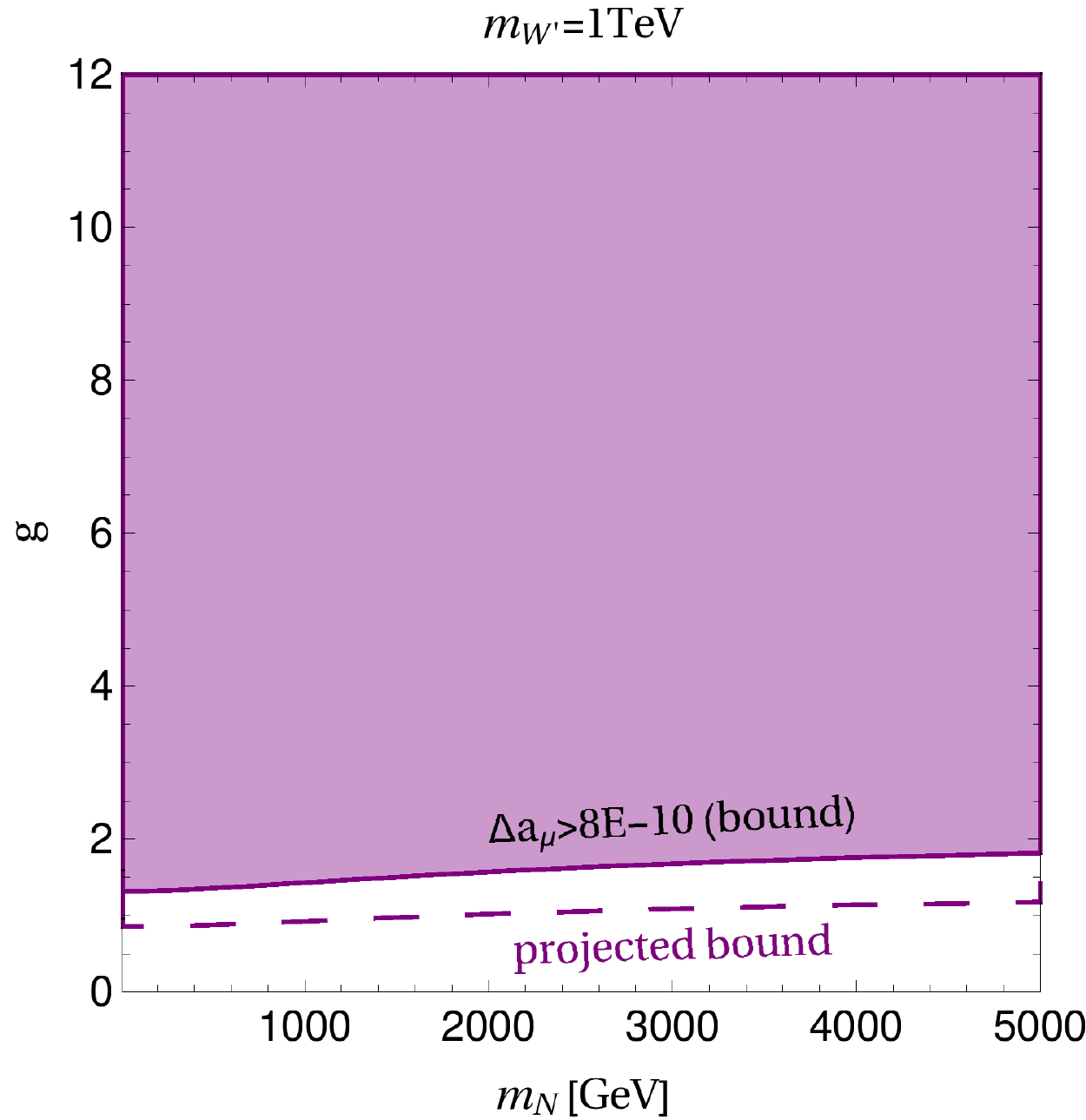}
   \end{subfigure}
   \hfill
   \begin{subfigure}[b]{.45\textwidth}
      \centering
      \includegraphics[width=\textwidth]{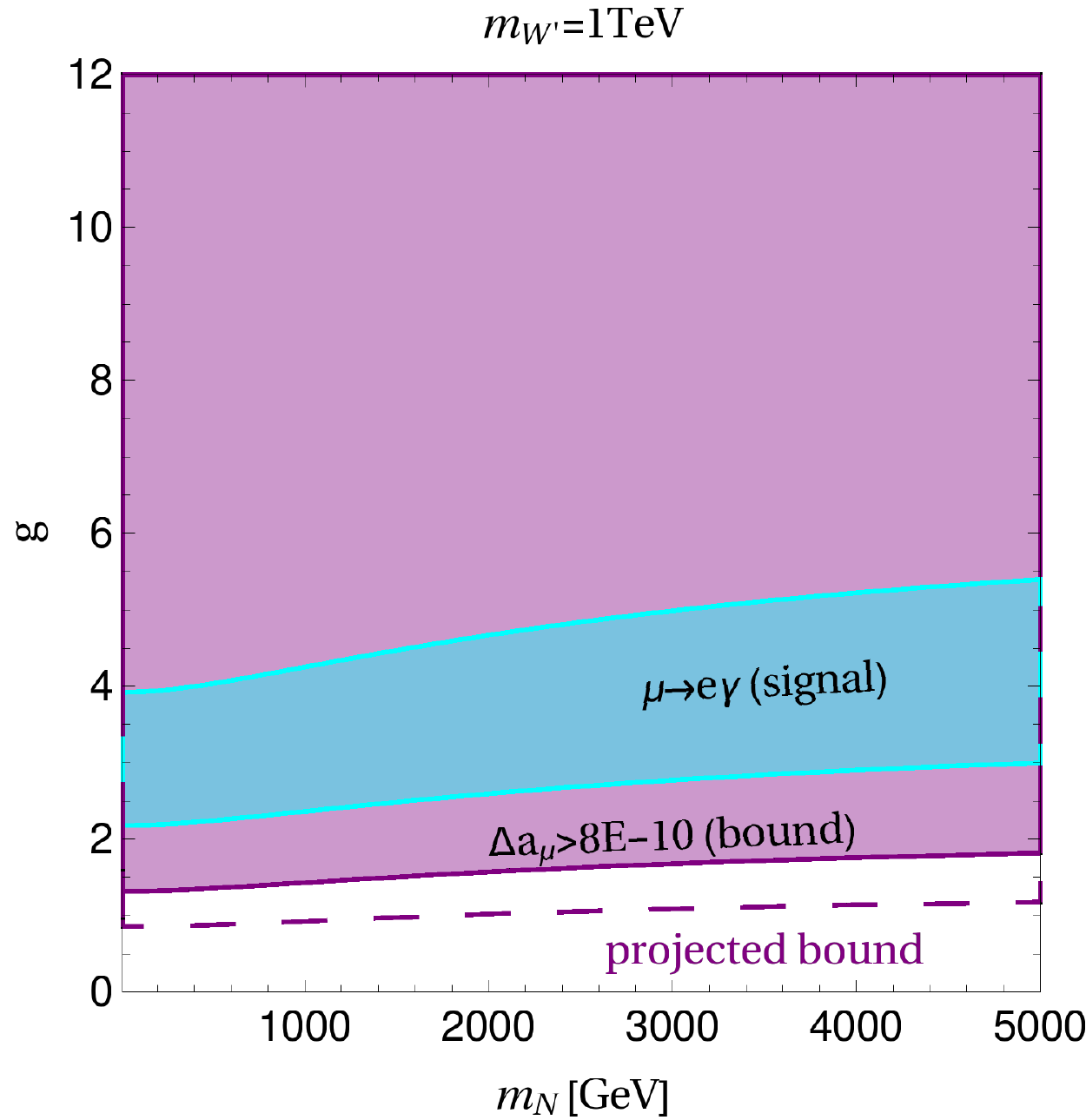}
   \end{subfigure}
   }\\
    \vspace{5mm}
   \noindent\makebox[\textwidth]{
   \begin{subfigure}[b]{.45\textwidth}
      \centering
      \includegraphics[width=\textwidth]{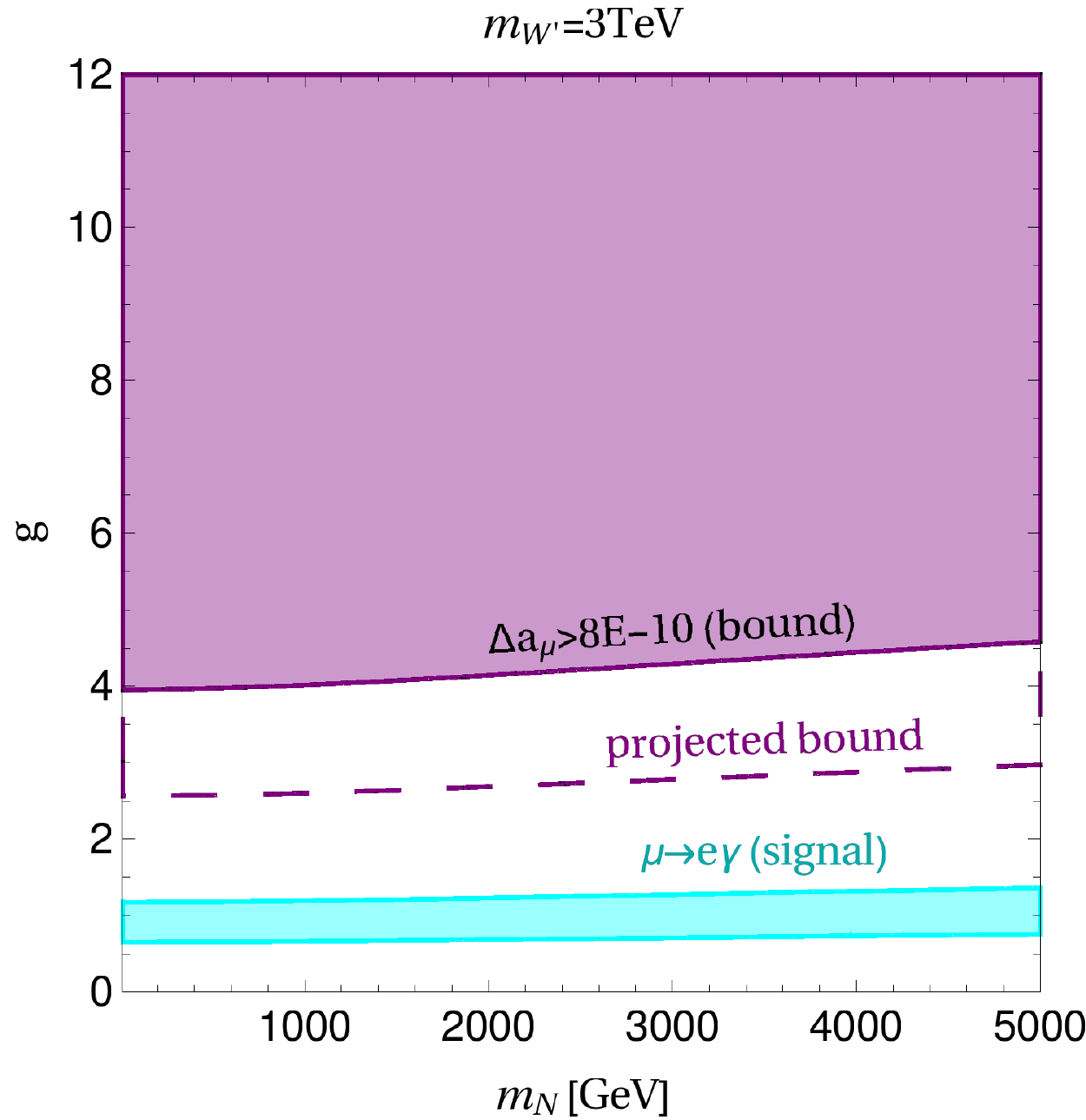}
   \end{subfigure}
   \hfill
   \begin{subfigure}[b]{.45\textwidth}
      \centering
      \includegraphics[width=\textwidth]{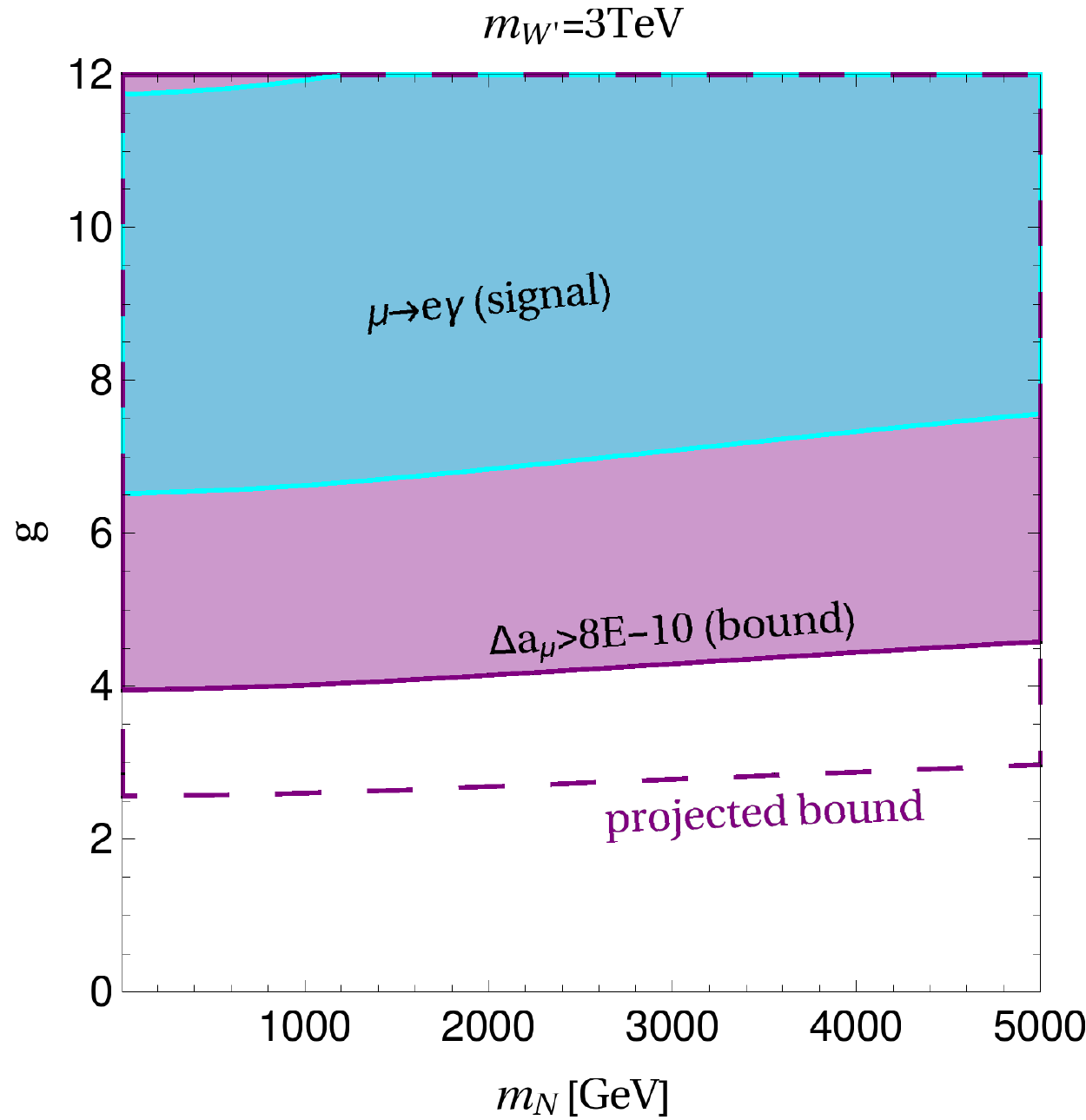}
   \end{subfigure}
   }\\
    \vspace{5mm}
   \noindent\makebox[\textwidth]{
   \begin{subfigure}[b]{.45\textwidth}
      \centering
      \includegraphics[width=\textwidth]{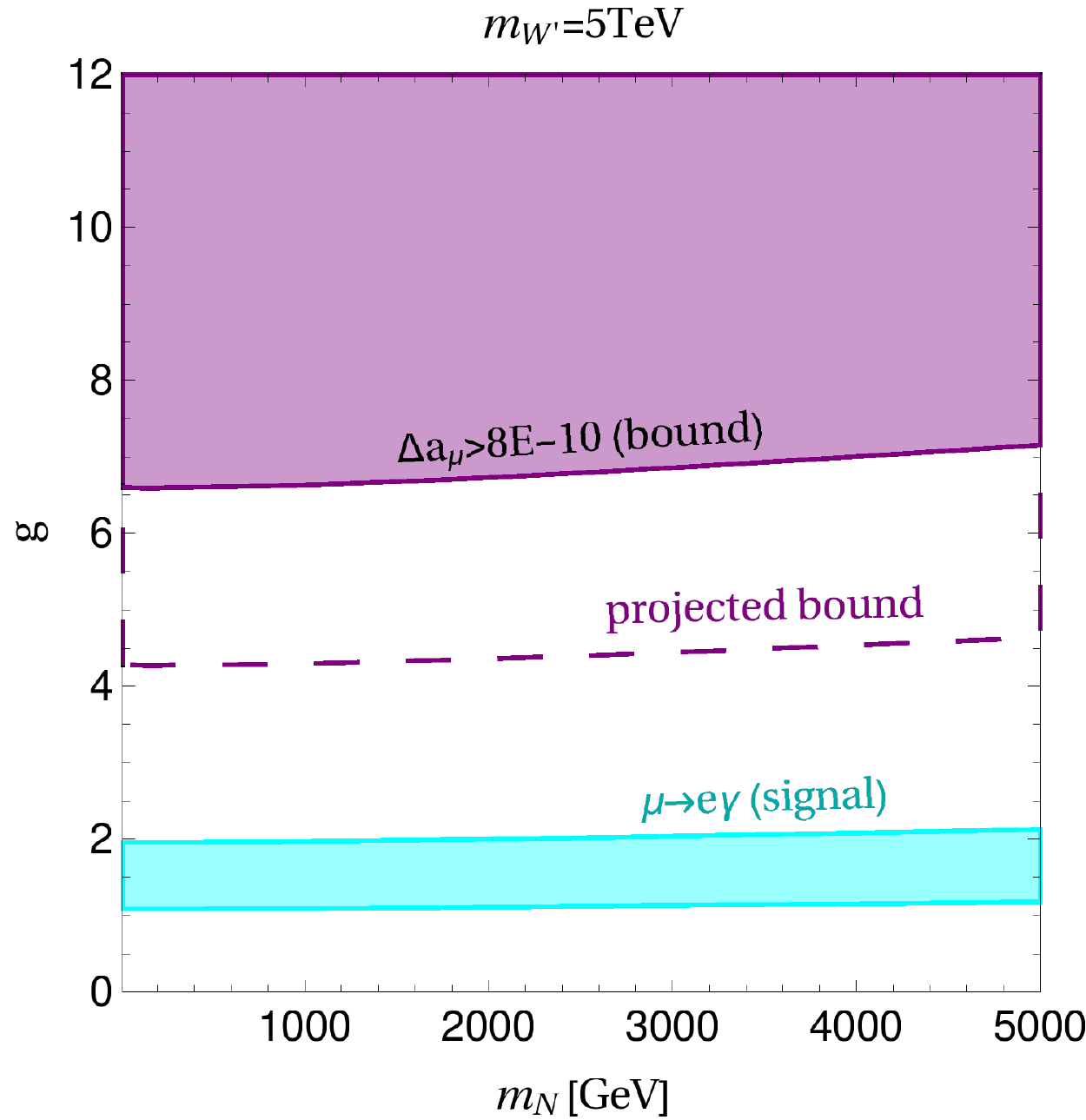}
      \subcaption{mild hierarchy}
   \end{subfigure}
   \hfill
   \begin{subfigure}[b]{.45\textwidth}
      \centering
      \includegraphics[width=\textwidth]{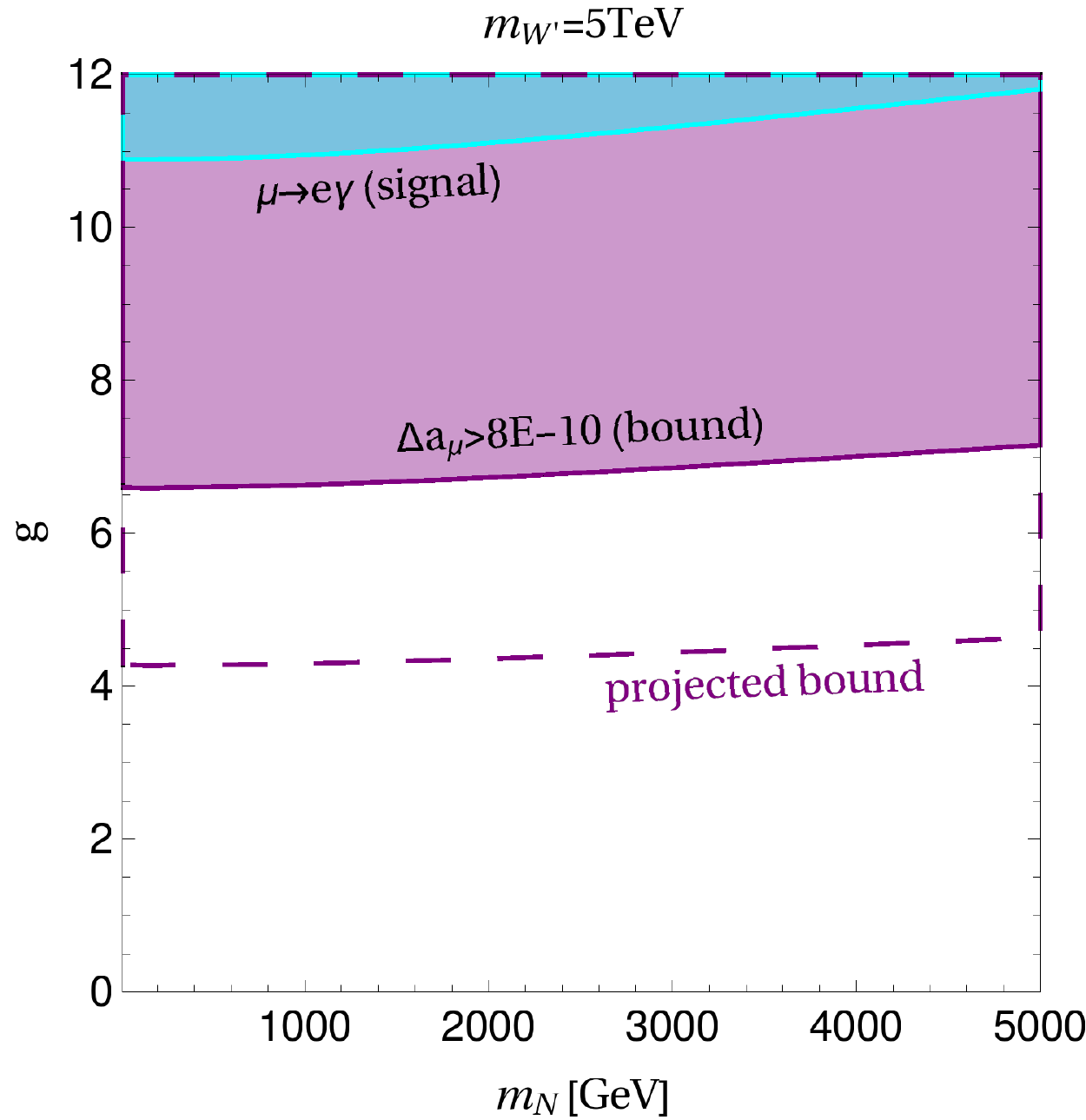}
      \subcaption{strong hierarchy}
   \end{subfigure}
   }
   \caption{\label{fig:results_fermionNeutral2Br}Potential $\MEG$ signal induced by a neutral fermion $N$ coupling to a $W^\prime$ boson and the SM leptons as dictated by Eq.~\eqref{Eq:WprimeNgeneral}.}
\end{figure}

In the converse approach, i.e.~assuming now a signal in $\MEG$ with ${\rm BR (\MEG)} =4.2\times 10^{-13}-4 \times 10^{-14}$ we obtain Fig.~\ref{fig:results_fermionNeutral2Br}, where the signal regions are delimited in blue, and the $g-2$ limits in violet as before. It is visible that a mild hierarchy and heavy $W^{\prime}$ are needed in order to reconcile a signal in $\MEG$ with the $g-2$ bound.  In particular, for  $m_W^{\prime}> 3$~TeV, one can easily evade current and projected limits on $g-2$. For $m_W^{\prime}=2\TeV$ (not shown in the figure) the signal region for $\MEG$ would partially fall within the current and projected sensitivity of $g-2$ measurements.

\subsubsection{Charged Fermion Singlet}

As we discussed previously [cf.~Eq.~\eqref{eq:LagrangianChargedFermion}], a charged fermion might interact with the muon via both scalar and vector mediators. In the former case, we consider
\begin{equation}
  \mathcal{L}_\mathrm{int} = {g_L}_i \overline{E_R}\, \phi^\dag \cdot \ell_L^i + \mathrm{h.c.} 
  \label{Eq:chargedleptonphi}
\end{equation}which resembles Eq.~\eqref{eq:down_type_Yukawa}.

\begin{figure}[p]
   \centering
   \noindent\makebox[\textwidth]{
   \begin{subfigure}[b]{.45\textwidth}
      \centering
      \makebox[\textwidth][l]{\includegraphics[width=\textwidth]{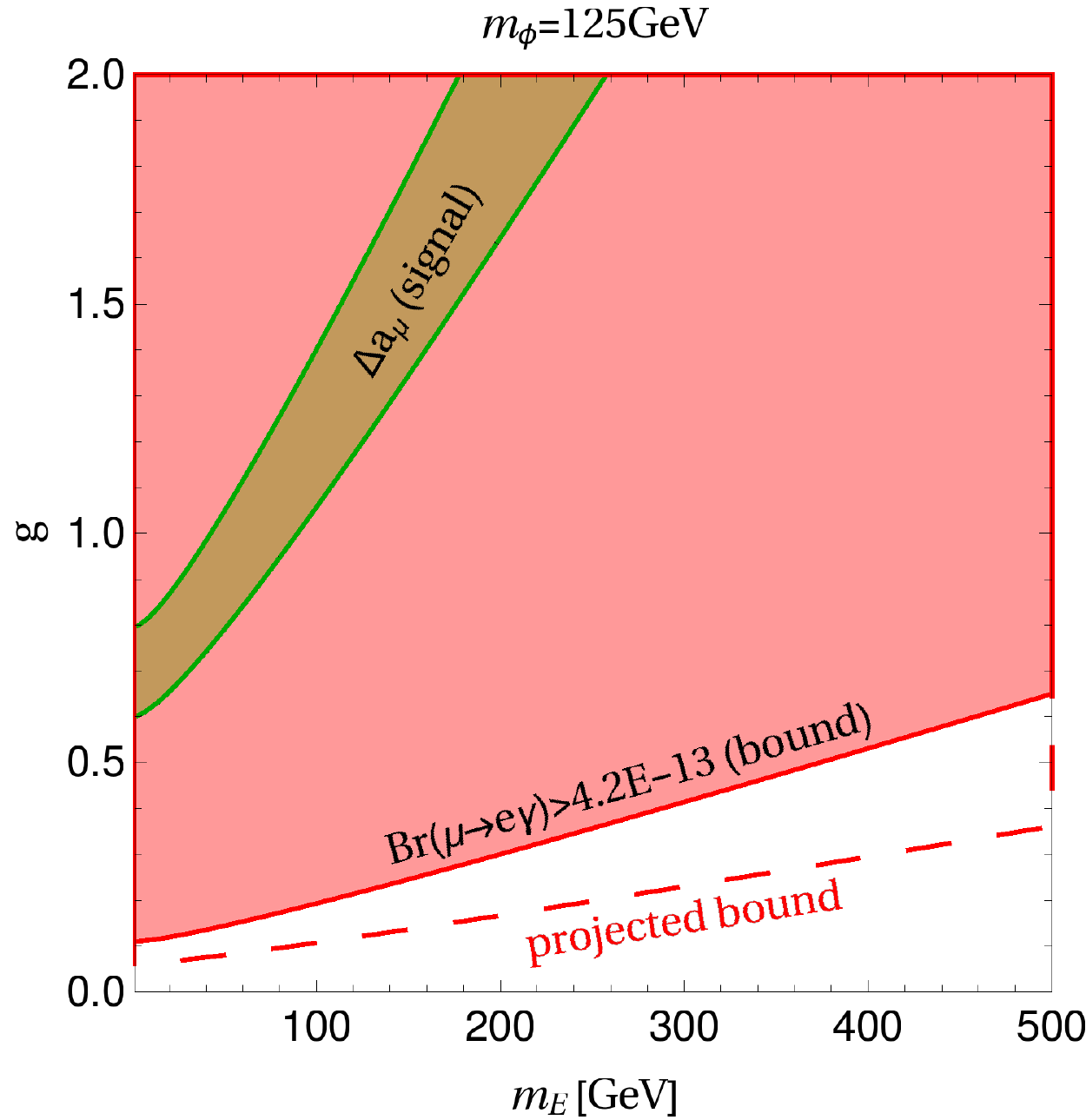}}
      \subcaption{\label{fig:fermionCharged1mildDa}mild hierarchy}
   \end{subfigure}
   \hfill
   \begin{subfigure}[b]{.45\textwidth}
      \centering
      \includegraphics[width=\textwidth]{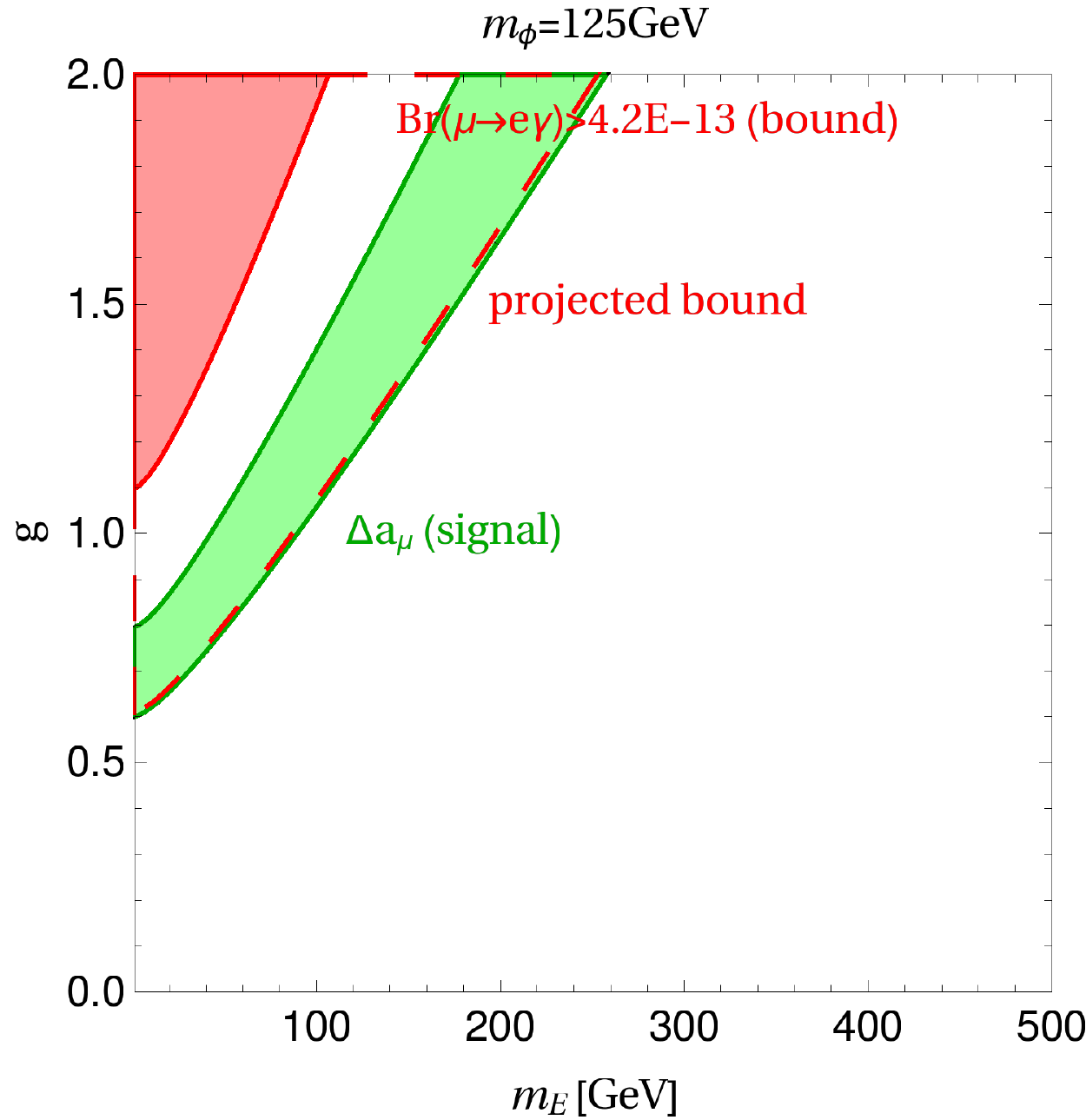}
      \subcaption{\label{fig:fermionCharged1strongDa}strong hierarchy}
   \end{subfigure}
   }\\
    \vspace{5mm}
   \noindent\makebox[\textwidth]{
   \begin{subfigure}[b]{.45\textwidth}
      \centering
      \makebox[\textwidth][l]{\includegraphics[width=\textwidth]{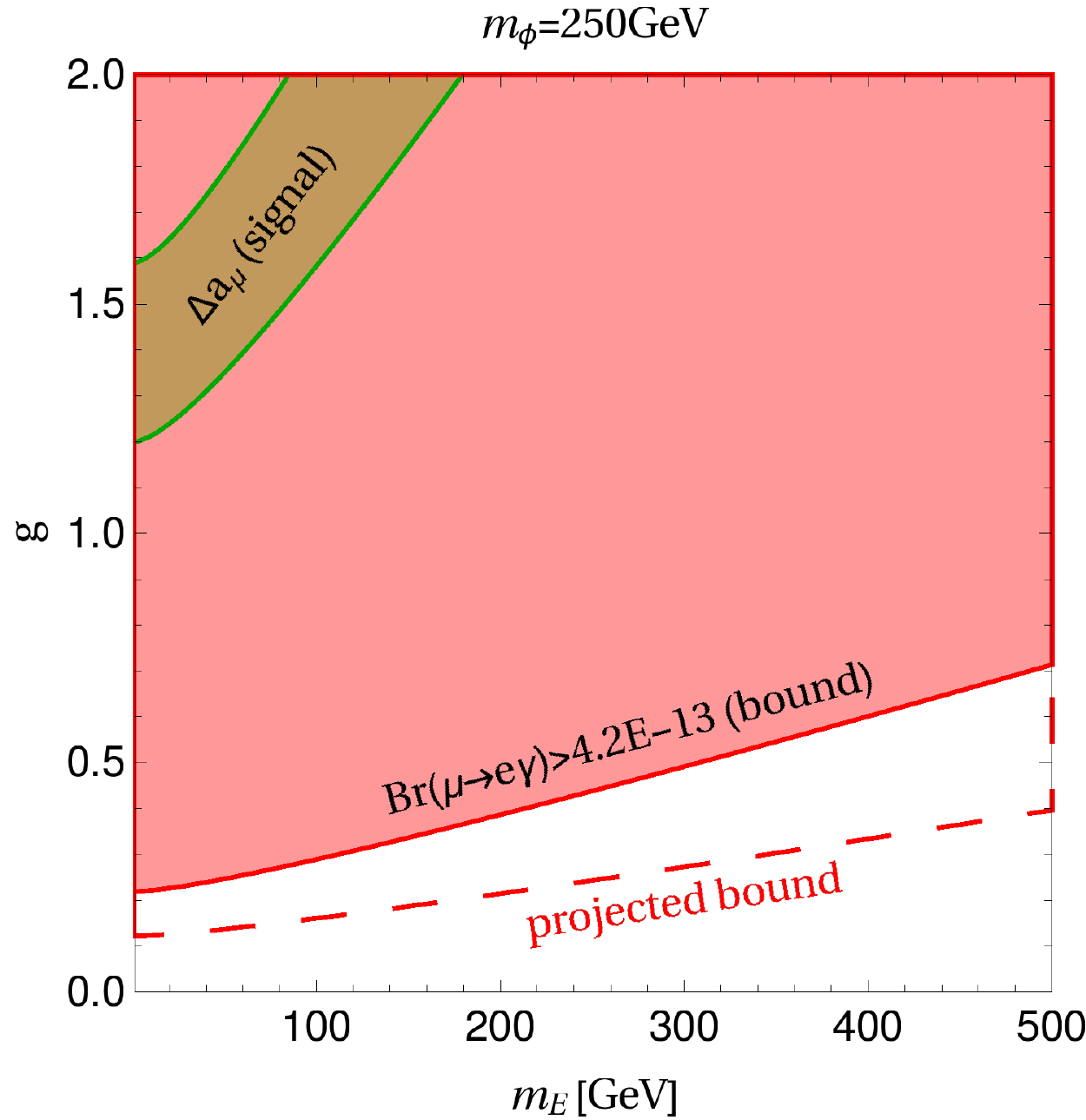}}
      \subcaption{mild hierarchy}
   \end{subfigure}
   \hfill
   \begin{subfigure}[b]{.45\textwidth}
      \centering
      \includegraphics[width=\textwidth]{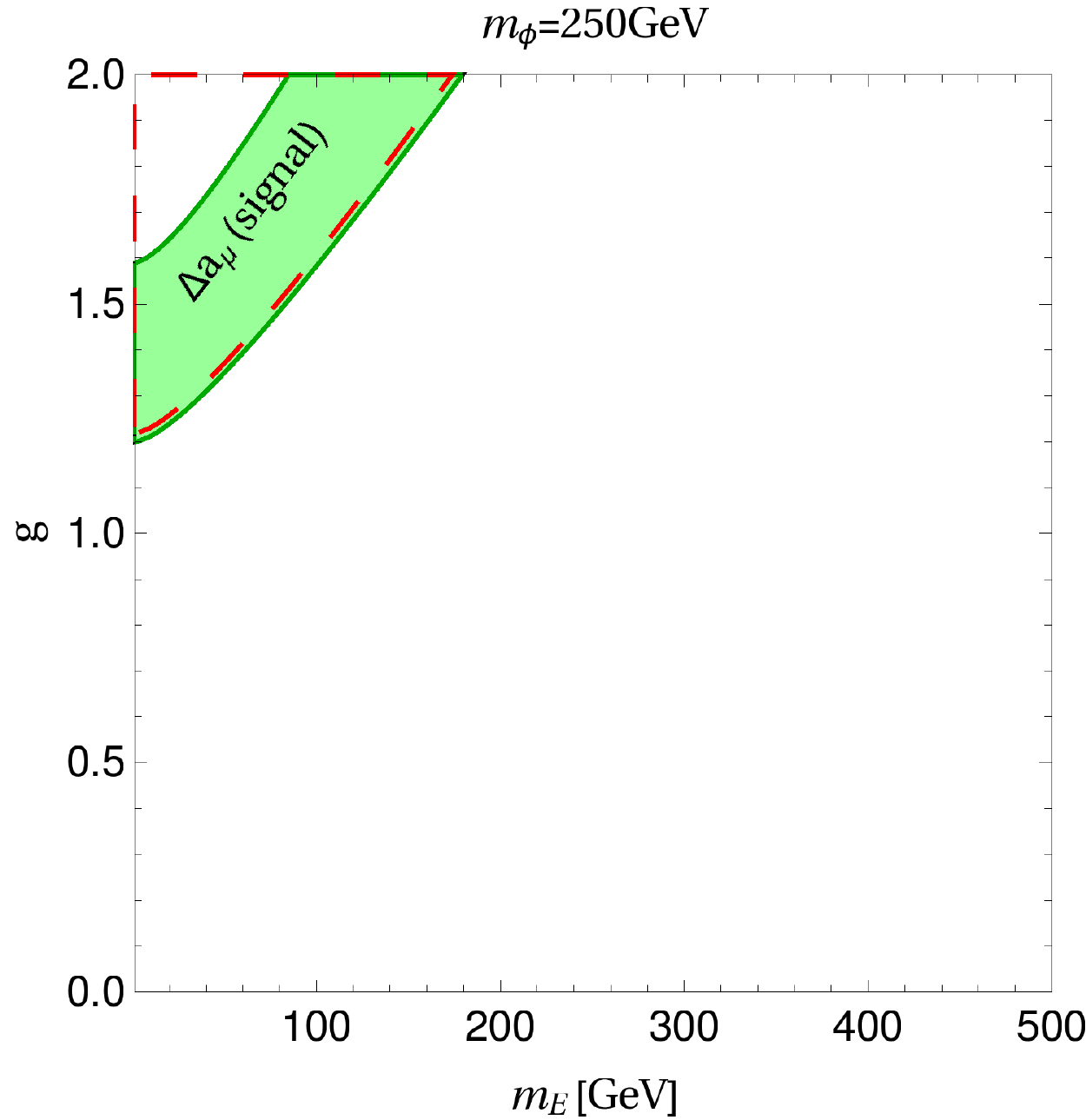}
      \subcaption{strong hierarchy}
   \end{subfigure}
   }
   \caption{\label{fig:results_fermionCharged1Da}$\Delta a_\mu$ for a fermion with unit electric charge coupling to the SM leptons via a scalar doublet $\phi$ according to Eq.~\eqref{Eq:chargedleptonphi}.}
\end{figure}

\begin{figure}[p]
   \centering
   \noindent\makebox[\textwidth]{
   \begin{subfigure}[b]{.45\textwidth}
      \centering
      \makebox[\textwidth][l]{\includegraphics[width=\textwidth]{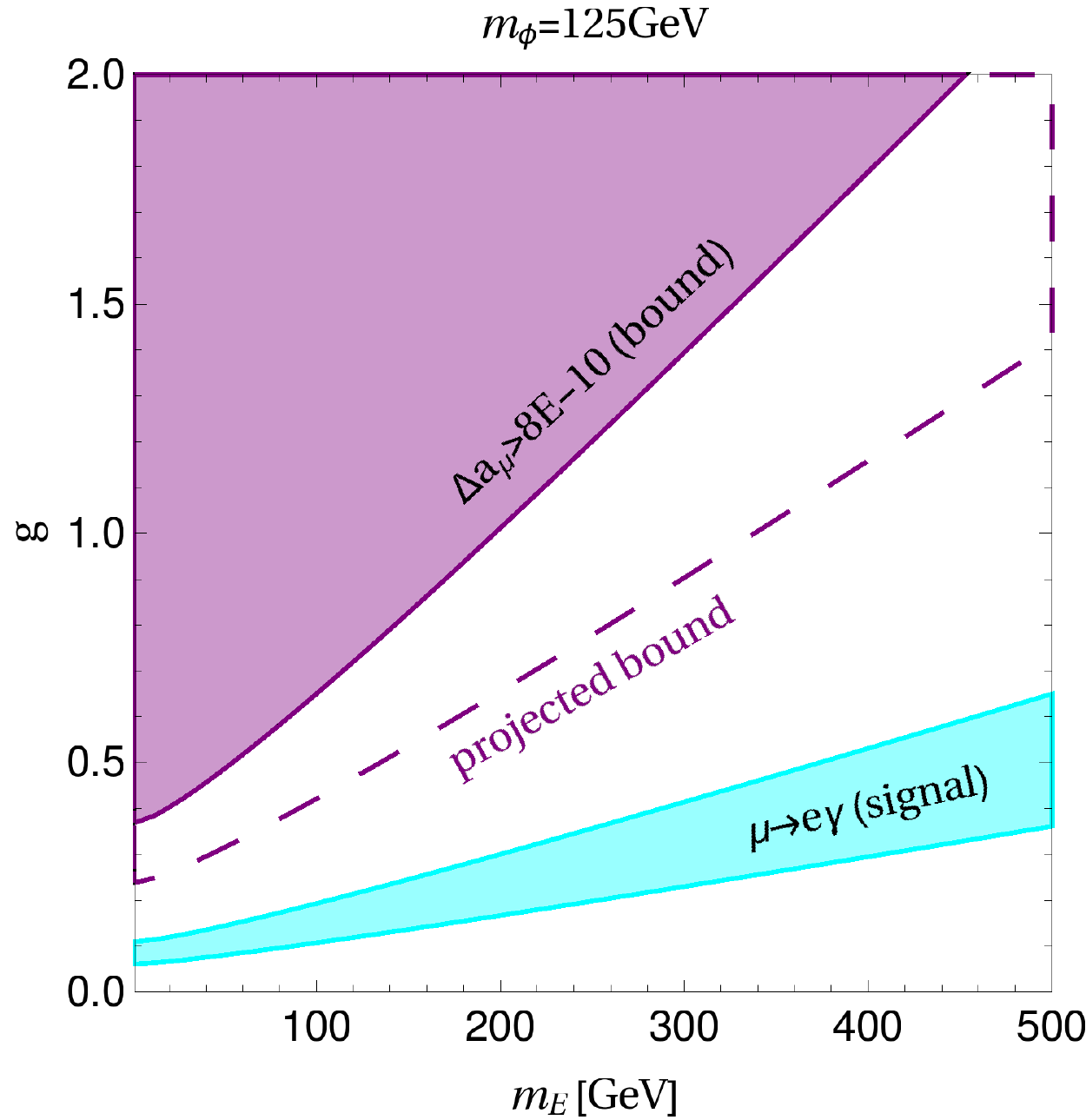}}
      \subcaption{\label{fig:fermionCharged1mildBr}mild hierarchy}
   \end{subfigure}
   \hfill
   \begin{subfigure}[b]{.45\textwidth}
      \centering
      \includegraphics[width=\textwidth]{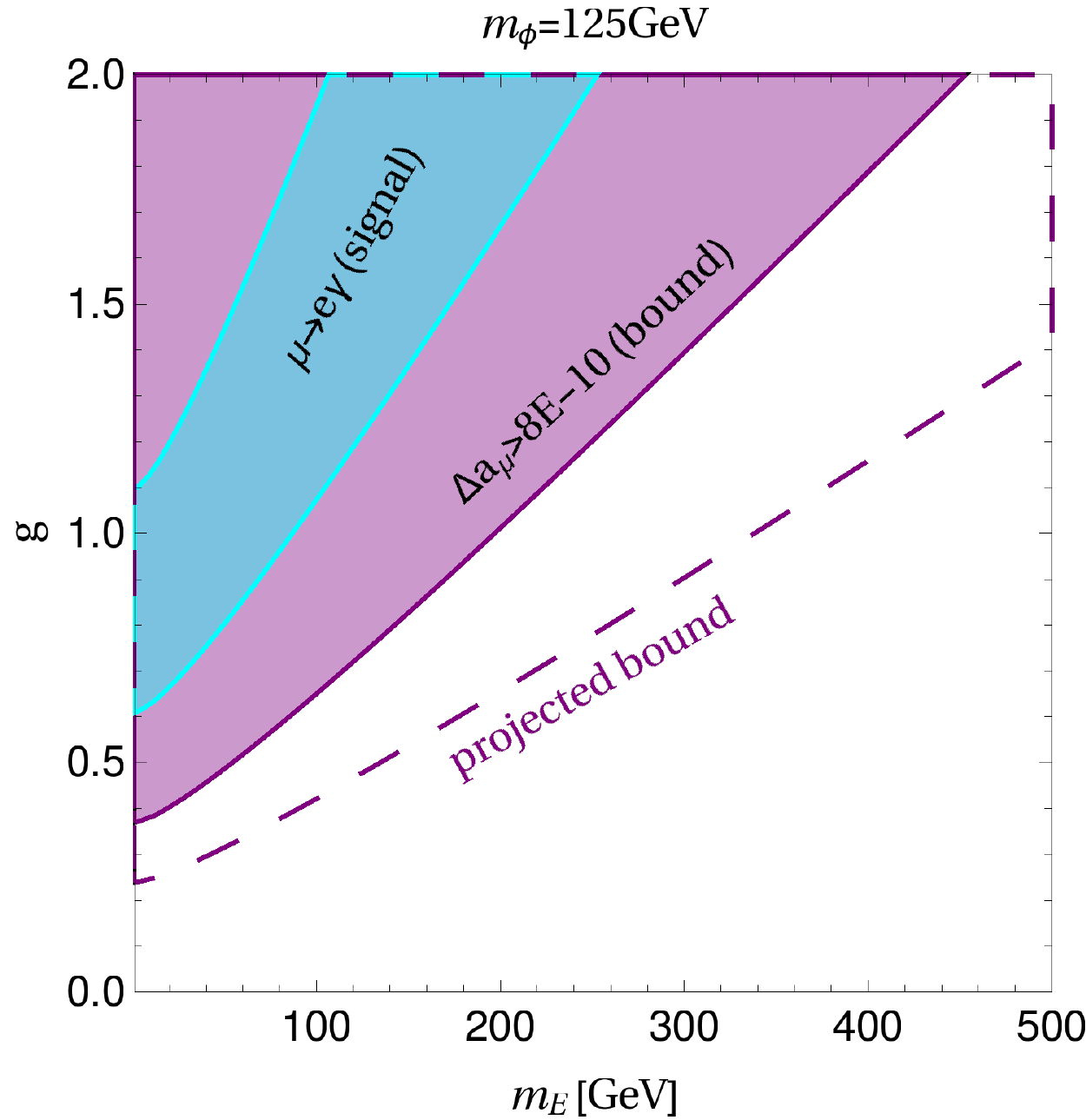}
      \subcaption{\label{fig:fermionCharged1strongBr}strong hierarchy}
   \end{subfigure}
   }\\
    \vspace{5mm}
   \noindent\makebox[\textwidth]{
   \begin{subfigure}[b]{.45\textwidth}
      \centering
      \makebox[\textwidth][l]{\includegraphics[width=\textwidth]{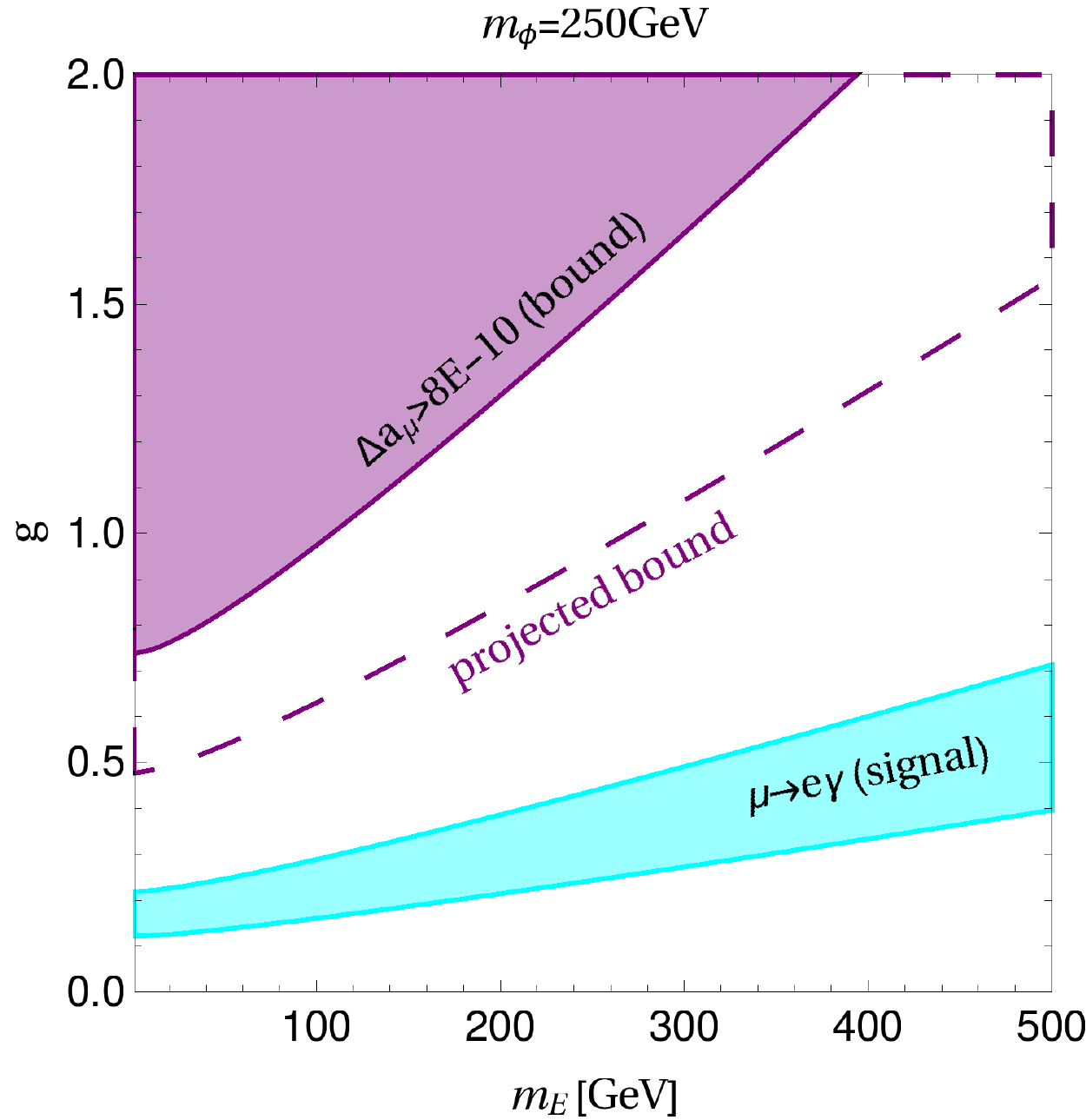}}
      \subcaption{mild hierarchy}
   \end{subfigure}
   \hfill
   \begin{subfigure}[b]{.45\textwidth}
      \centering
      \includegraphics[width= \textwidth]{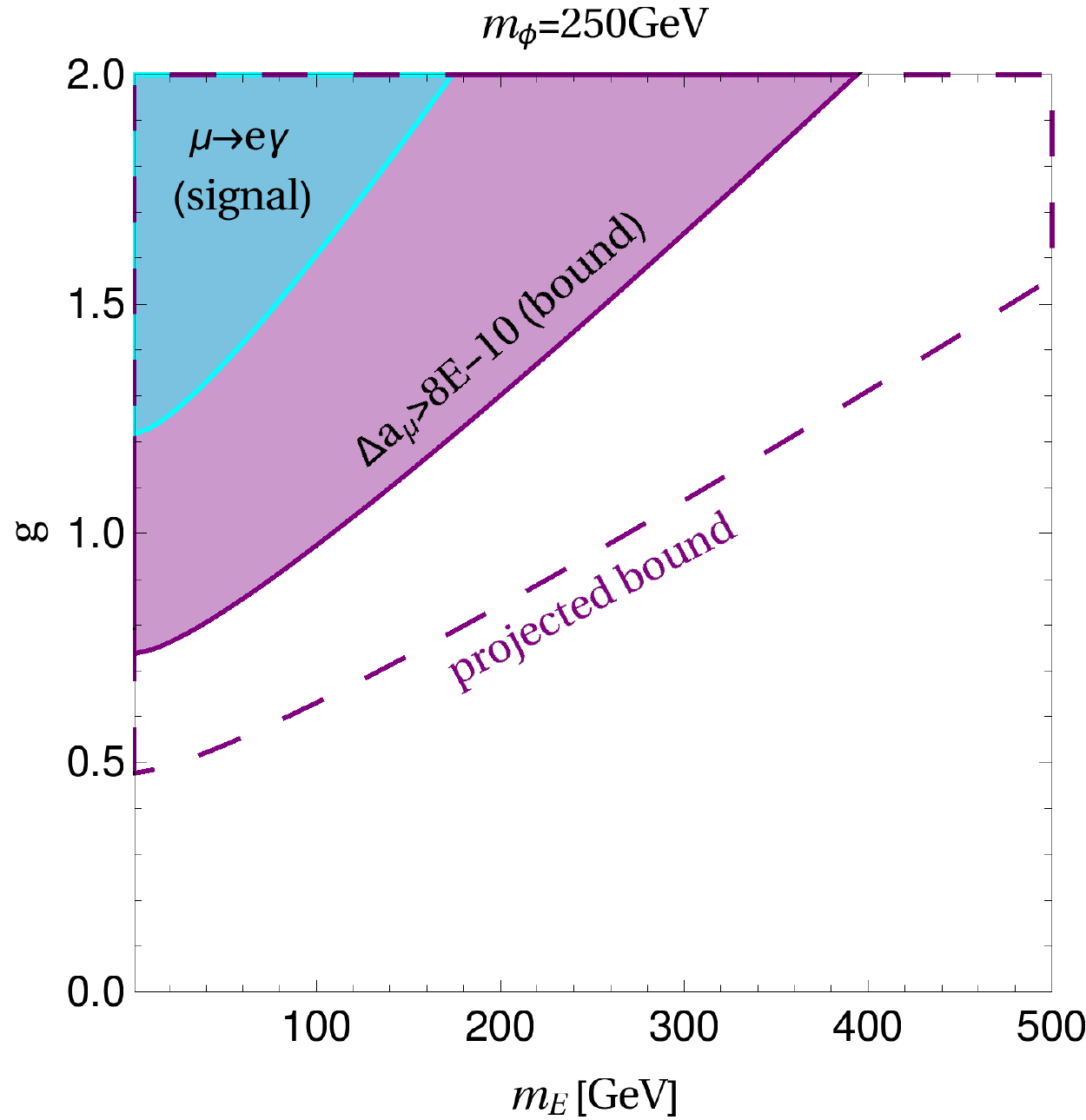}
      \subcaption{strong hierarchy}
   \end{subfigure}
   }
   \caption{\label{fig:results_fermionCharged1Br}Signal region for $\MEG$ for a fermion with unit electric charge coupling to the SM leptons via a scalar doublet $\phi$ governed by Eq.~\eqref{Eq:chargedleptonphi}.}
\end{figure}

Similar interactions appear in 331 models~\cite{Ponce:2006vw,Salazar:2007ym} and more exotic two Higgs doublet models~\cite{DePree:2008st}. We have treated this case before, so here we display only the results summarized in Figs.~\ref{fig:results_fermionCharged1Da} and~\ref{fig:results_fermionCharged1Br}.

In Fig.~\ref{fig:results_fermionCharged1Da} we assume the deviation in $g-2$ is confirmed with the same central value and display in the left (right) panels the findings for a mild (strong) hierarchy in the charged lepton sector. Setting $m_{\phi}=(125,250)$~GeV we see that the $g-2$ signal region has been excluded by $\MEG$, whereas for a strong hierarchy the $g-2$ signal region falls within current and projected sensitivity on the $\MEG$ decay. One may argue that we chose relatively light scalar masses, since for larger scalar masses the bounds from $\MEG$ weakens, but keep in mind that heavier scalars require larger couplings to still accommodate a signal in $g-2$. So generally heavier scalars are plausible, but much heavier scalars are problematic since very large couplings would be needed to accommodate a signal in $g-2$.

As usual, the picture is reversed if one is interested in addressing a potential signal in $\MEG$ as one can see in Fig.~\ref{fig:results_fermionCharged1Br}, since only for a mild hierarchy a signal in $\MEG$ is in agreement with $g-2$ measurements.

For a vector $Z^\prime$ mediator, one obtains results which are almost identical to the results for a neutral fermion coupling via a $W^\prime$ boson, and for this we skip this case.

\subsection{Fermion Multiplet Contributions}

\subsubsection{Fermion Doublet}
\begin{figure}[p]
   \centering
   \noindent\makebox[\textwidth]{
   \begin{subfigure}[b]{.45\textwidth}
      \centering
      \makebox[\textwidth][l]{\includegraphics[width=\textwidth]{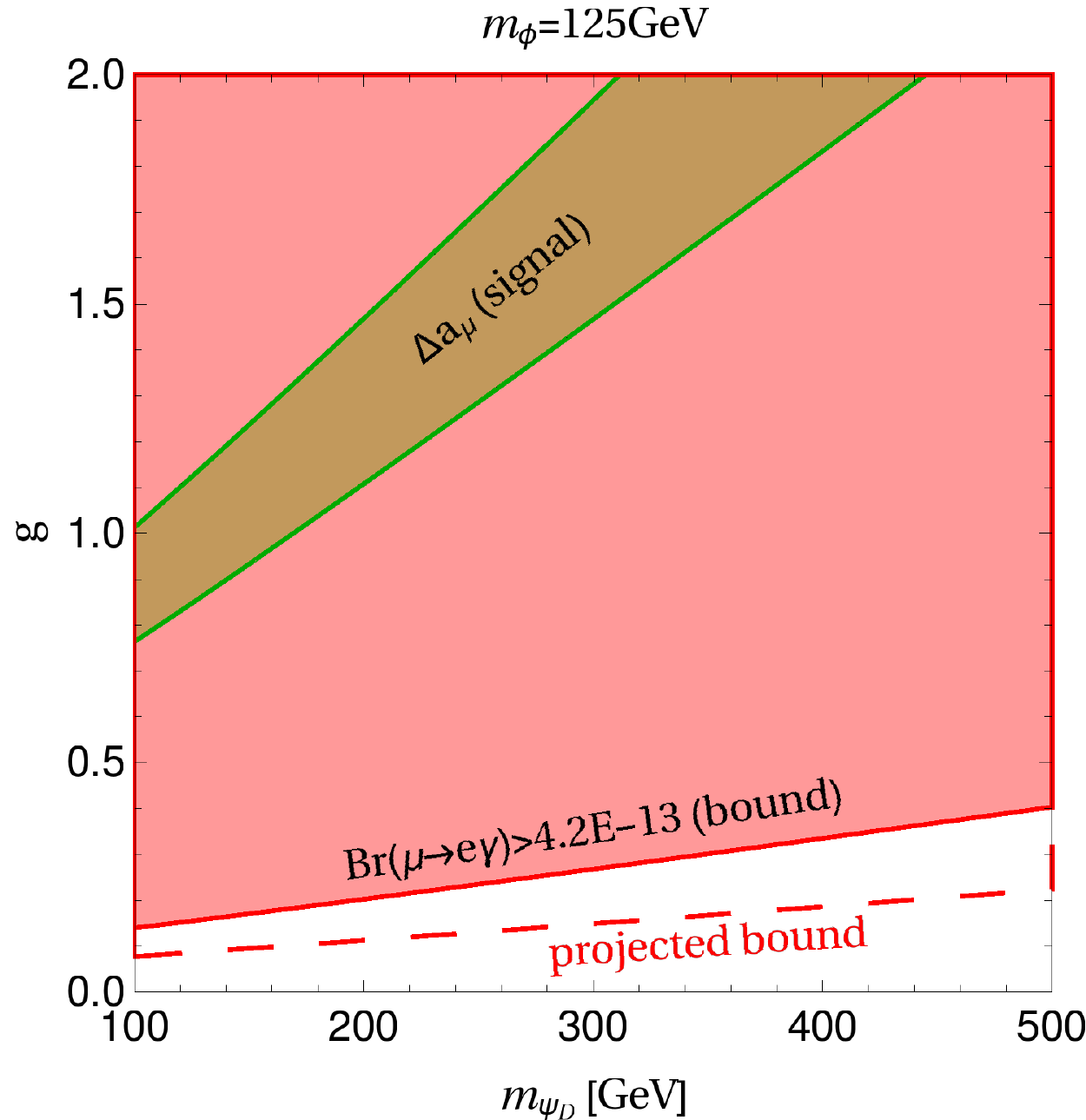}}
      \subcaption{mild hierarchy}
   \end{subfigure}
   \hfill
   \begin{subfigure}[b]{.45\textwidth}
      \centering
      \includegraphics[width=\textwidth]{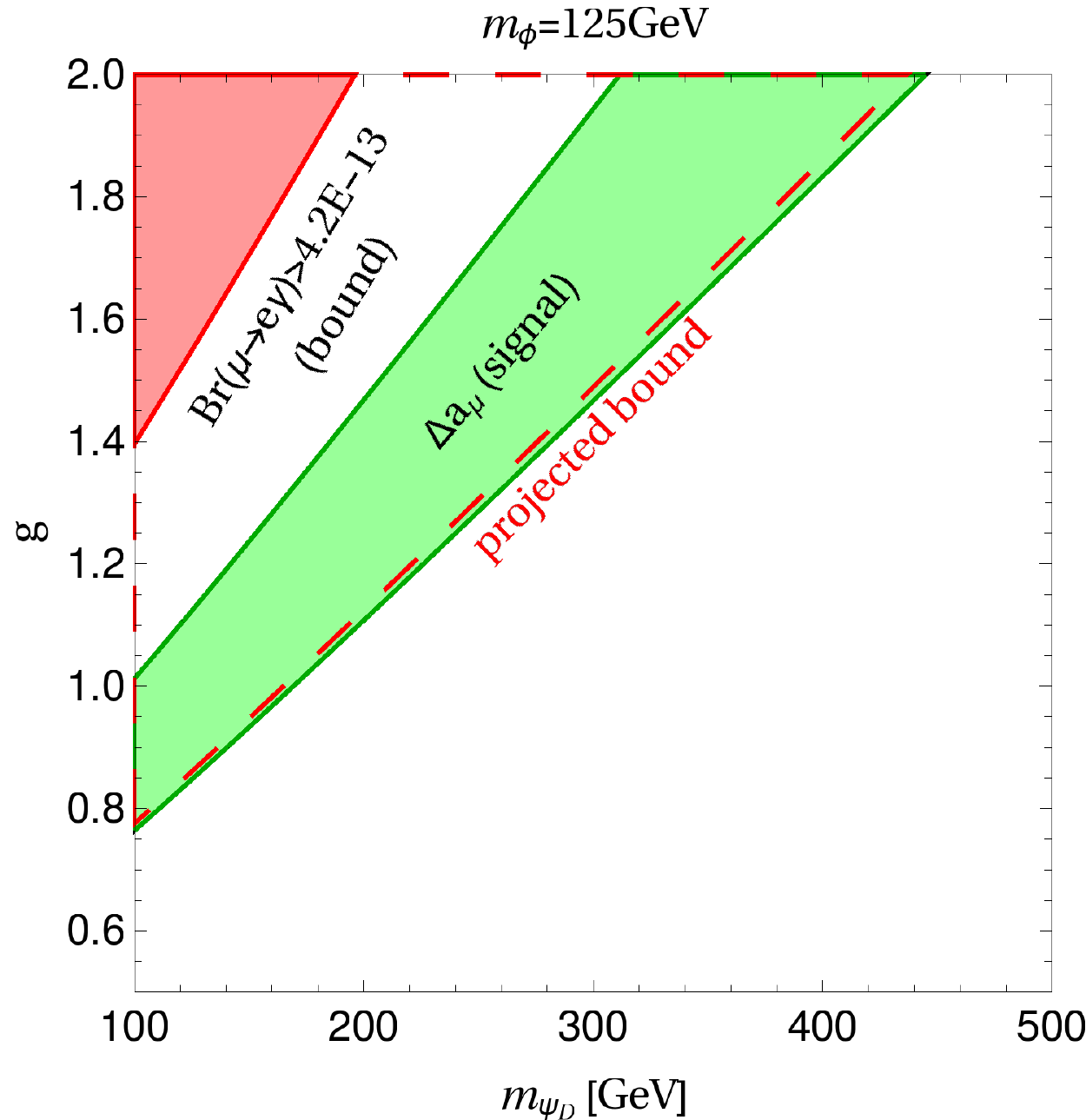}
      \subcaption{\label{fig:fermionDoubletBoth}strong hierarchy}
   \end{subfigure}
   }\\
    \vspace{5mm}
   \noindent\makebox[\textwidth]{
   \begin{subfigure}[b]{.45\textwidth}
      \centering
      \makebox[\textwidth][l]{\includegraphics[width=\textwidth]{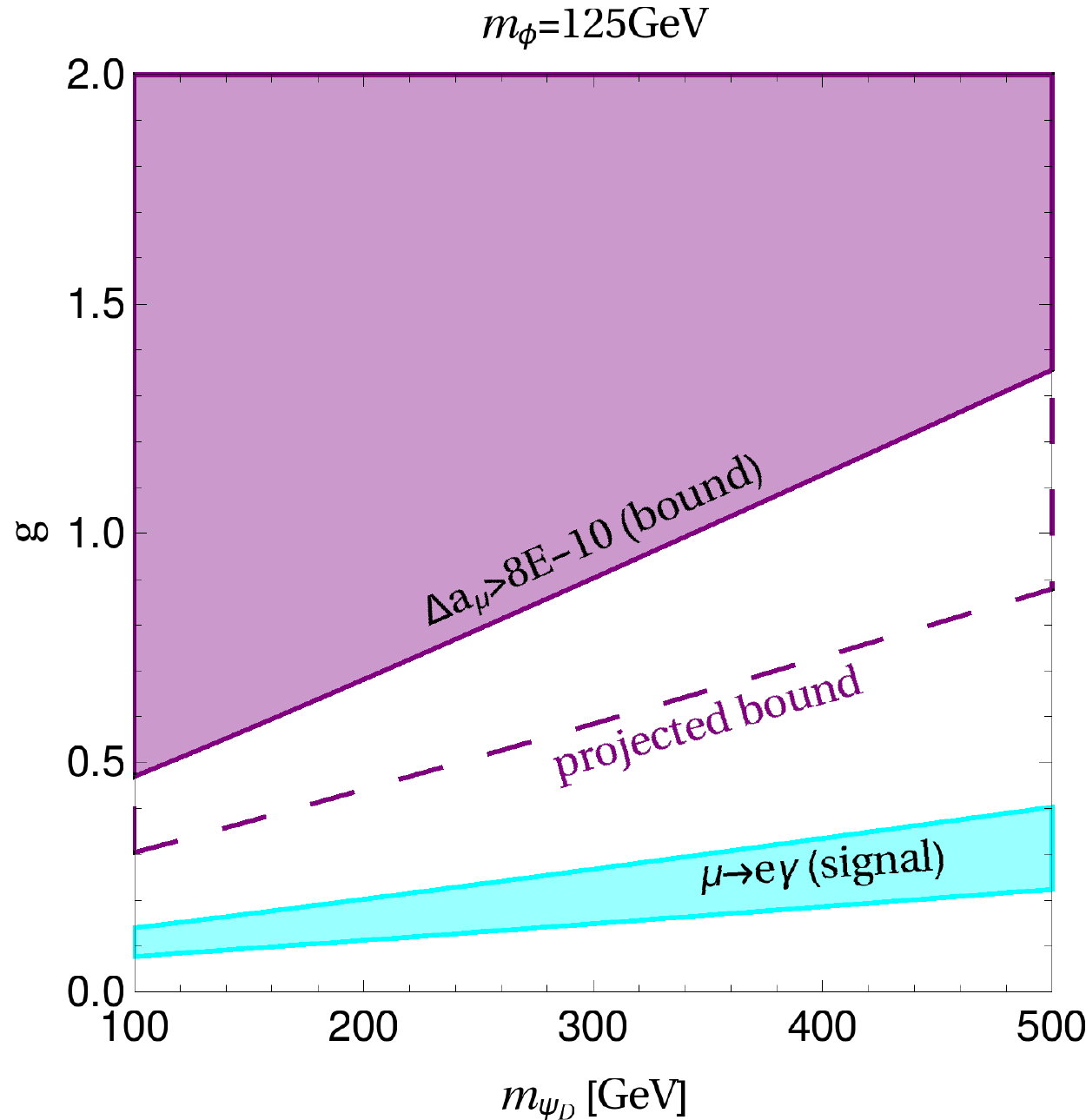}}
      \subcaption{mild hierarchy}
   \end{subfigure}
   \hfill
   \begin{subfigure}[b]{.45\textwidth}
      \centering
      \includegraphics[width= \textwidth]{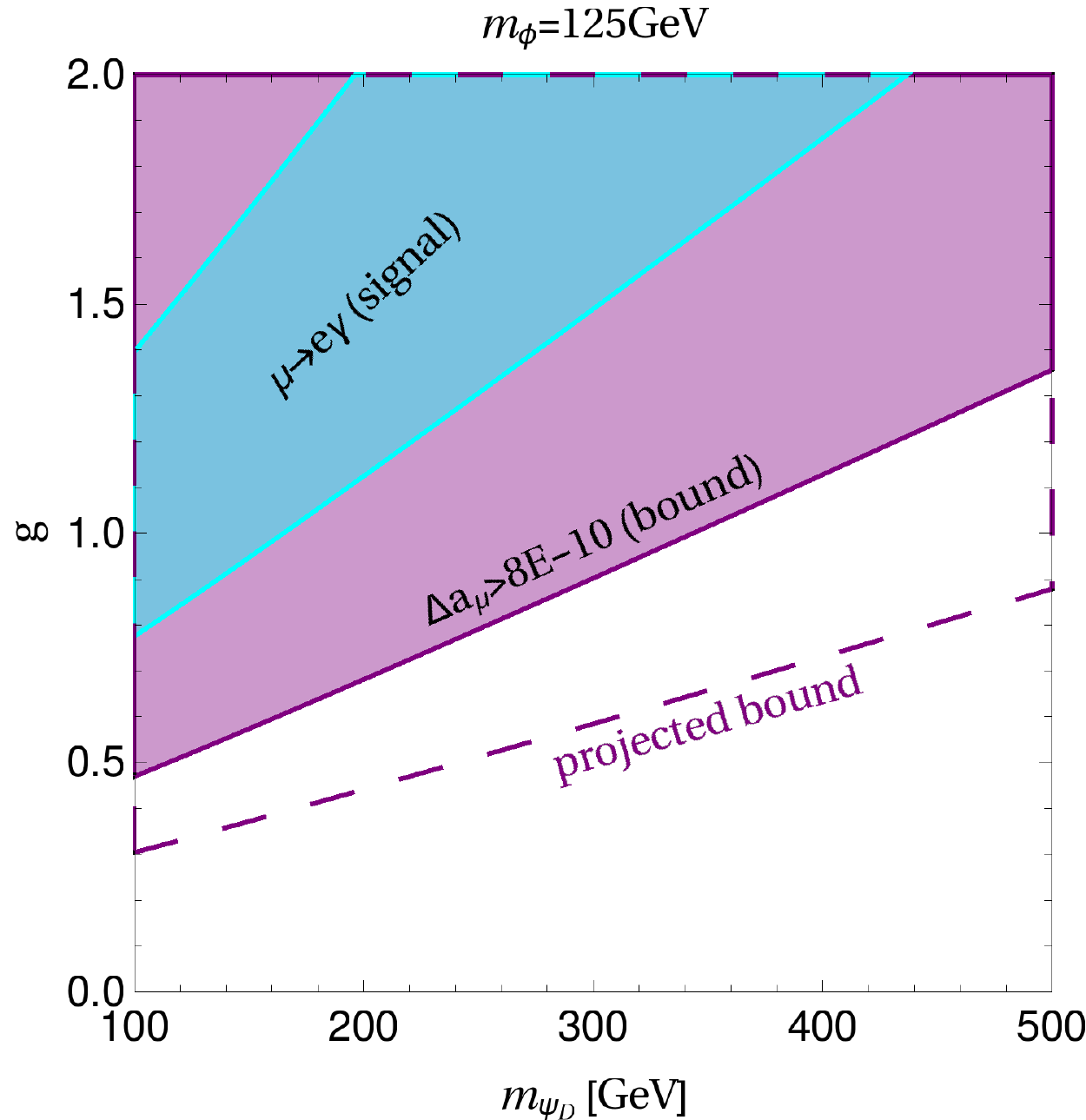}
      \subcaption{strong hierarchy}
   \end{subfigure}
   }
   \caption{\label{fig:results_fermionDoublet}Results for a fermion doublet $\psi_D$.}
\end{figure}

Let us consider the case of a new $SU(2)_L$ fermion doublet $\psi_D$ which interacts with the SM charged leptons as
\begin{equation}
  \mathcal{L}_\mathrm{int} = {g_L}_i\, \overline{e_R^i}\, \phi^\dag \cdot \psi_D.
\end{equation}
The parameter scan for this model is shown in Fig.~\ref{fig:results_fermionDoublet}. We exhibit the parameter space region in which a signal in $g-2$ is explained with a light scalar with mass of $125$~GeV. The result for a heavier Higgs can be easily scaled knowing that $\Delta a_{\mu}$ grows as $g^2/m_{\phi}^2$. It is noticeable that, in the right panel of Fig.~\ref{fig:results_fermionDoublet}, a signal in $g-2$ agrees with constraints stemming from $\MEG$. A signal in $\MEG$, on the other hand, is highly sensitive to the hierarchy as the bottom panels highlight, concluding that a mild hierarchy should be present to accommodate a signal in $\MEG$ while evading current and projected limits from the $g-2$. 

\subsubsection{Fermion Triplet}\label{Sec:fermiontriplet}

\begin{figure}[p]
   \centering
   \noindent\makebox[\textwidth]{
   \begin{subfigure}[b]{.45\textwidth}
      \centering
      \makebox[\textwidth][l]{\includegraphics[width=\textwidth]{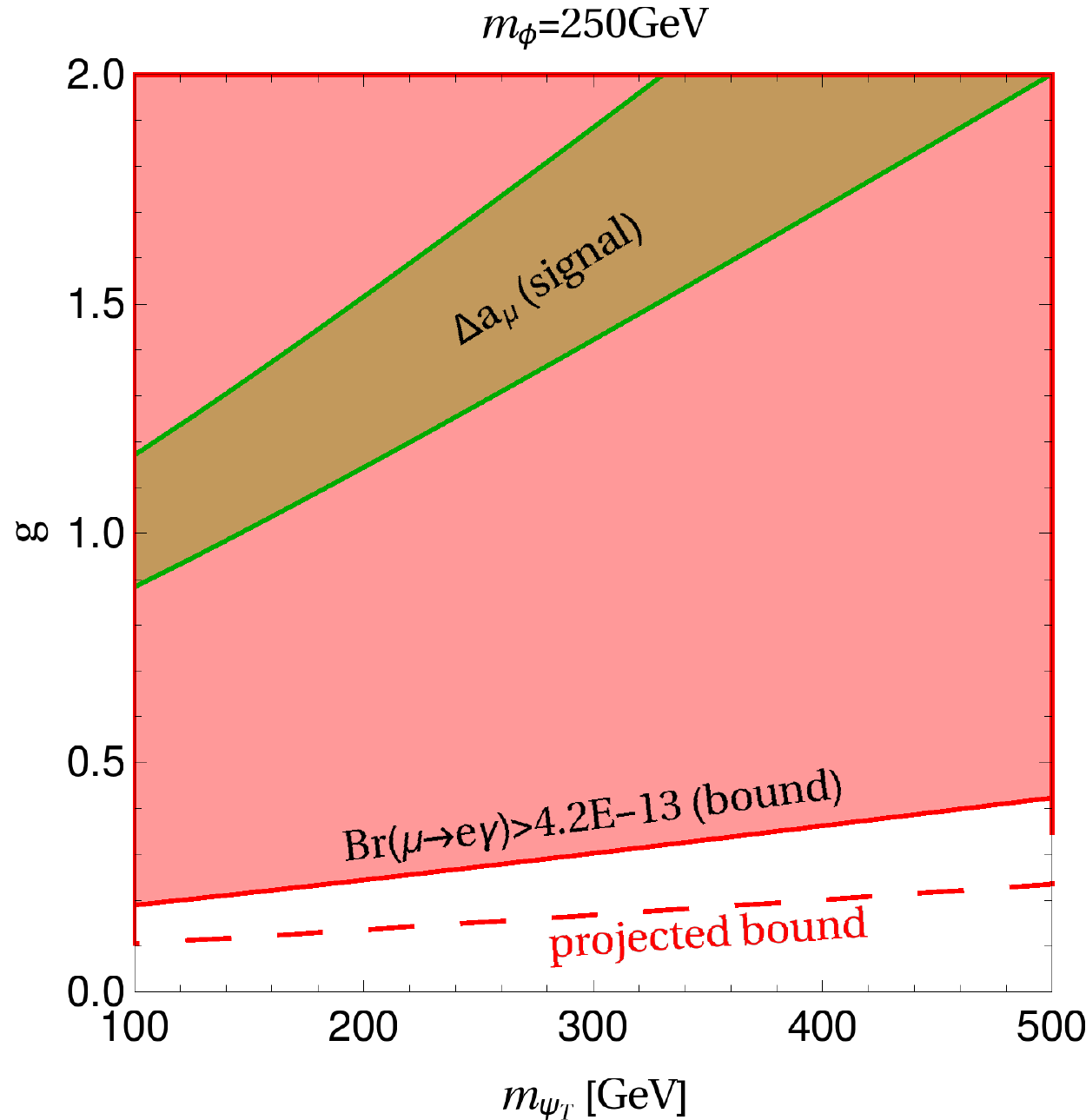}}
      \subcaption{mild hierarchy}
   \end{subfigure}
   \hfill
   \begin{subfigure}[b]{.45\textwidth}
      \centering
      \includegraphics[width=\textwidth]{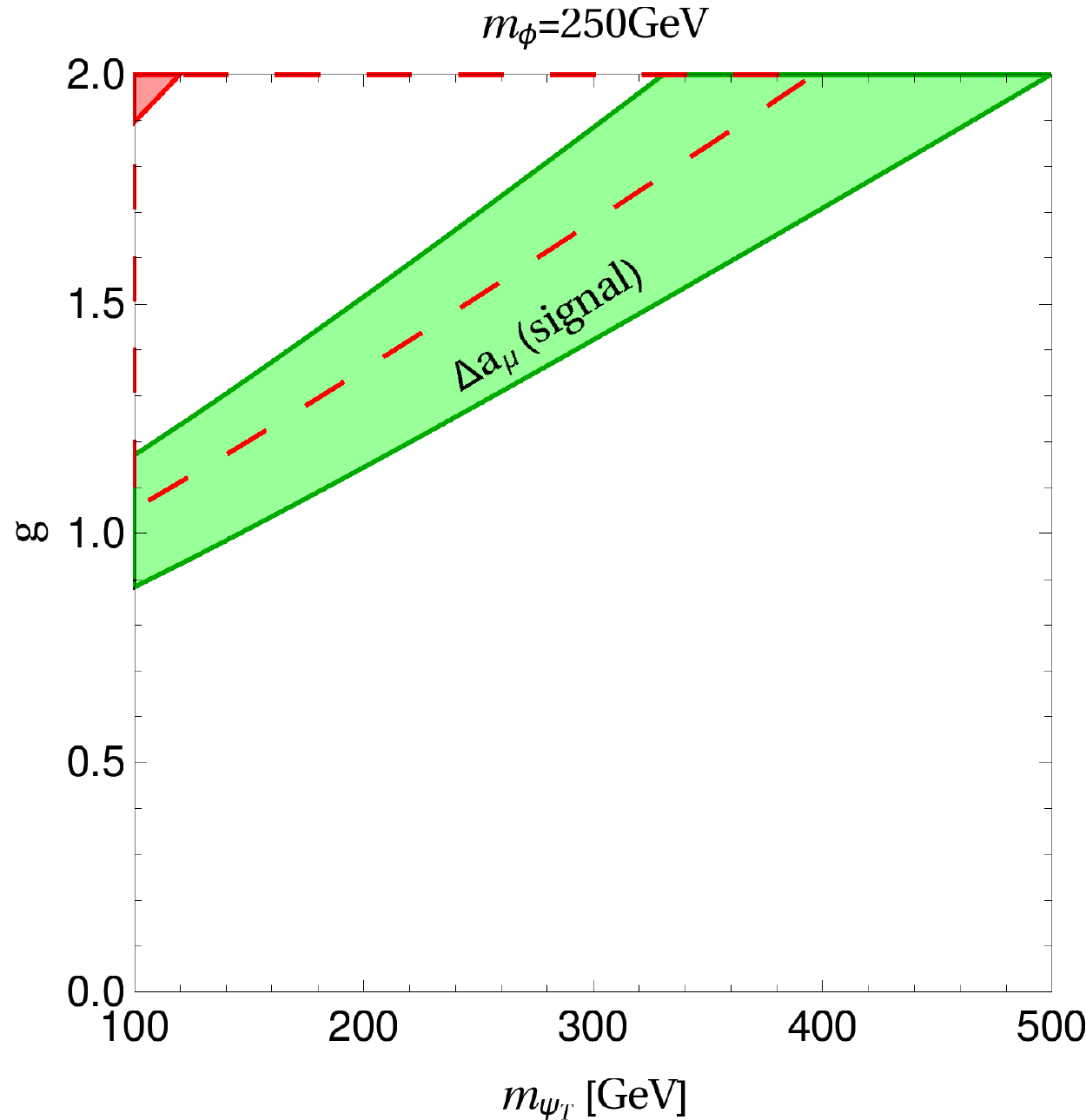}
      \subcaption{strong hierarchy}
   \end{subfigure}
   }\\
    \vspace{5mm}
   \noindent\makebox[\textwidth]{
   \begin{subfigure}[b]{.45\textwidth}
      \centering
      \makebox[\textwidth][l]{\includegraphics[width=\textwidth]{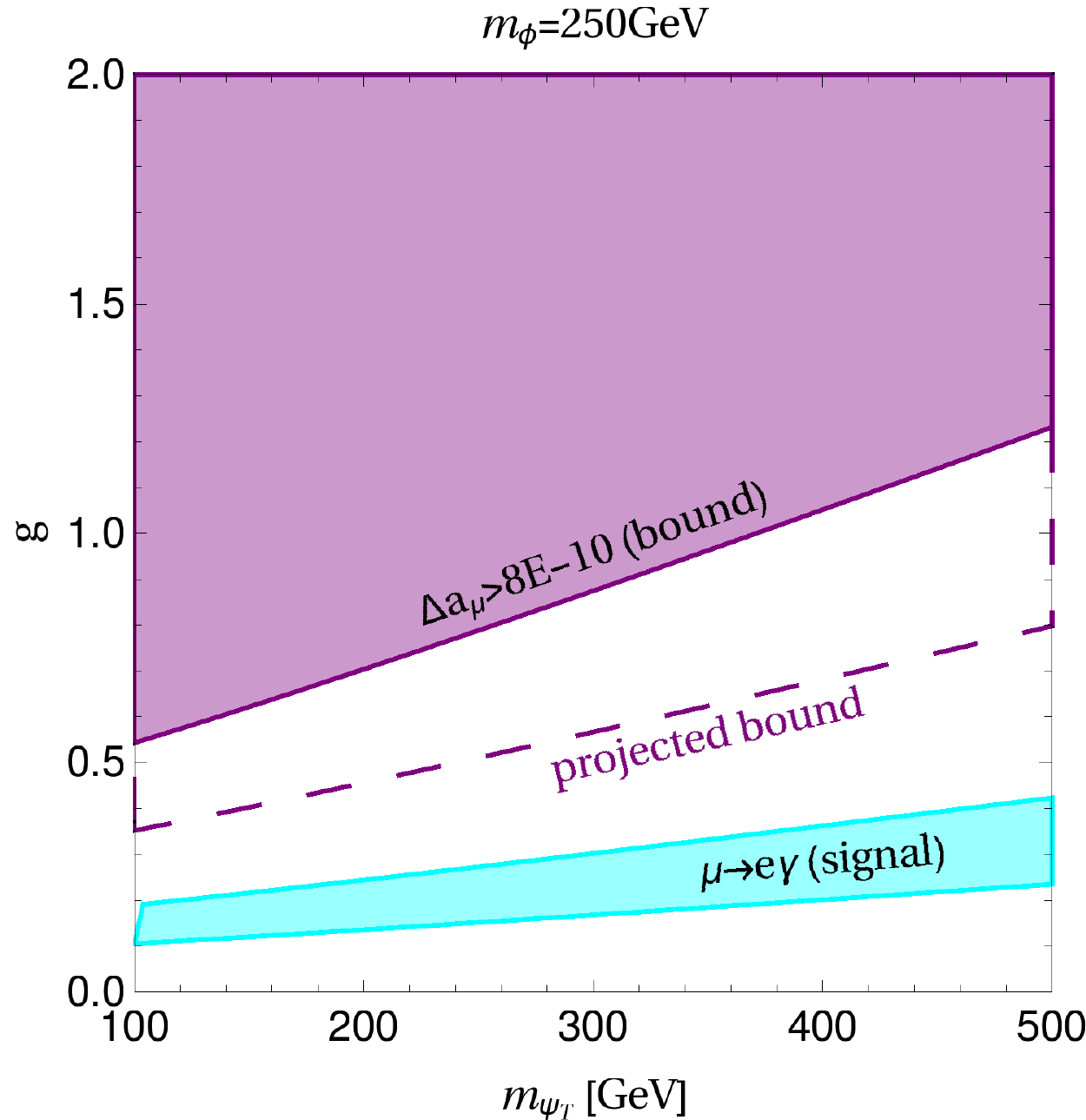}}
      \subcaption{mild hierarchy}
   \end{subfigure}
   \hfill
   \begin{subfigure}[b]{.45\textwidth}
      \centering
      \includegraphics[width= \textwidth]{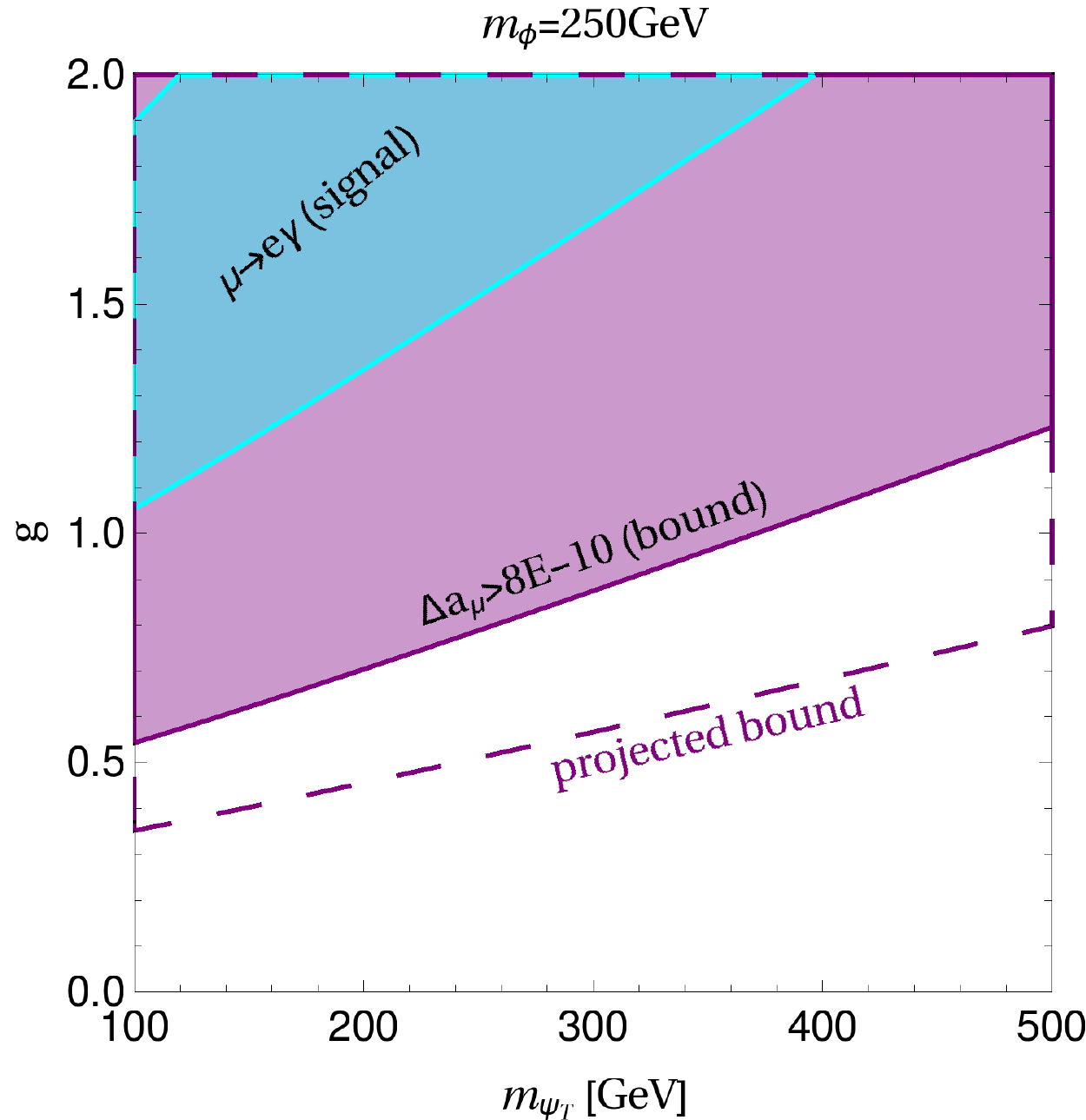}
      \subcaption{strong hierarchy}
   \end{subfigure}
   }
   \caption{\label{fig:results_fermionTriplet}Results for a fermion triplet $\psi_T$ of unit hypercharge.}
\end{figure}

Fermion triplets with zero hyperchage are the key feature of the type~III seesaw mechanism that we will discuss further. For now we will focus on a different type~of triplet fermion, one that has hypercharge $Y=-1$, and is therefore described by the following simplified Lagrangian,

\begin{equation}
  \mathcal{L}_\mathrm{int} = {g_L}_{i} \phi^\dag \overline{\psi_T} \ell_L^i + \mathrm{h.c.}\,,
\end{equation}
where the triplet $\psi_T$ may be written in component form as
\begin{equation}
  \psi_T = \left(
	    \begin{array}{cc}
	      \psi_T^-/\sqrt{2} & \psi_T^0 \\
	      \psi_T^{--} & -\psi_T^-/\sqrt{2}
	    \end{array}
	  \right).
\end{equation}

In the upper panels of Fig.~\ref{fig:results_fermionTriplet} we see that a mild hierarchy is excluded by $\MEG$. Interestingly, though, one can see that, for a strong hierarchy and $m_{\phi}=250$~GeV, the parameter space that addresses $g-2$ might be probed by future searches for $\MEG$. Furthermore, adopting a mild hierarchy and $m_{\phi}=250$~GeV a signal in $\MEG$ is consistent with current and projected limits on the $g-2$ measurement as shown in the bottom graphs. With a naive rescaling of the $\MEG$ signal region of the left-bottom panel, one can find lighter scalars producing a signal in $\MEG$ that can be probed with the upcoming $g-2$ experiments.

\subsection{Vector Contributions}
Finally, let us consider new vector fields with electrical charges $Q=0,1, \textrm{and } 2$  as mediators. 

\subsubsection{Neutral Vector Boson}\label{sec:Discussion_neutralVector}
\begin{figure}[t]
   \centering
   \noindent\makebox[\textwidth]{
   \begin{subfigure}[b]{.45\textwidth}
      \centering
      \includegraphics[width=\textwidth]{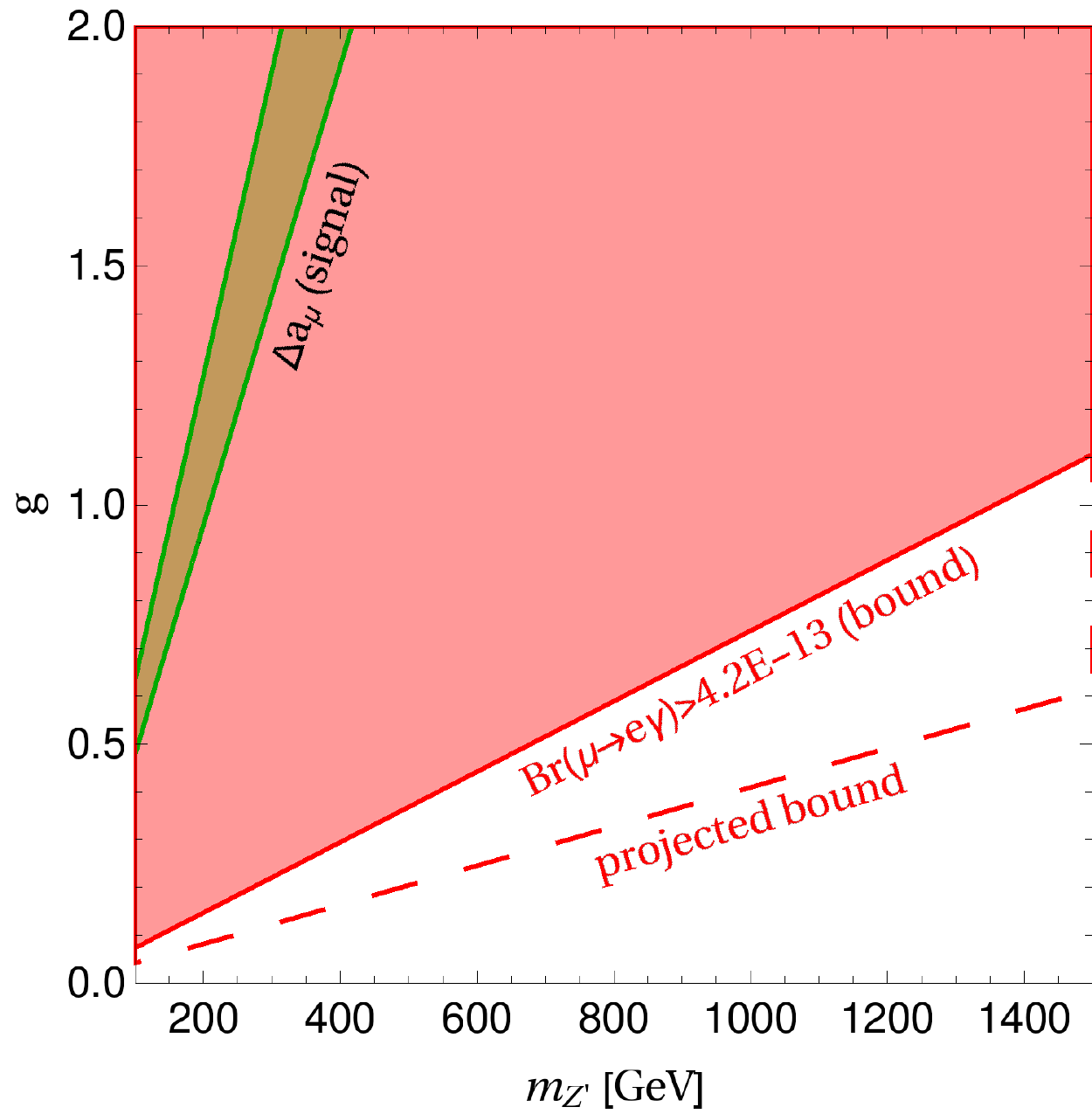}
      \subcaption{mild hierarchy}
   \end{subfigure}
   \hfill
   \begin{subfigure}[b]{.45\textwidth}
      \centering
      \includegraphics[width=\textwidth]{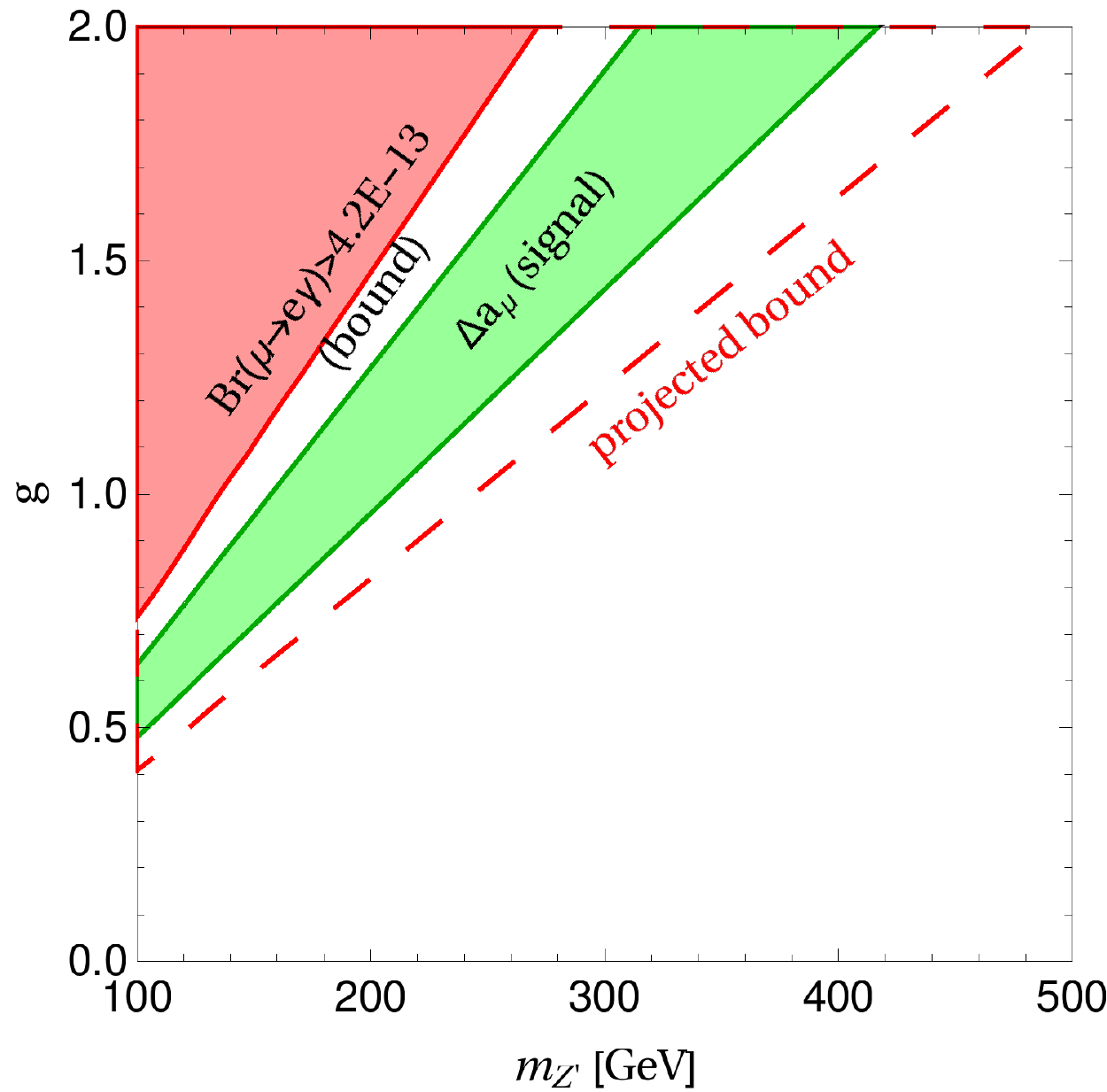}
      \subcaption{strong hierarchy}
   \end{subfigure}
   }
   \caption{\label{fig:results_vectorNeutral}A neutral $Z^\prime$ vector boson as mediator.}
\end{figure}

Heavy neutral gauge bosons ($Z^{\prime}$) arise in many popular models such as $B-L$, $L_{\mu}-L_{\tau}$, Left-Right models, and $SU(3)_c\times SU(3)_L \times U(1)_X$ models. Their masses can be generated via spontaneous symmetry breaking due to a scalar field charged under the new gauge symmetry, or via the Stueckelberg mechanism~\cite{Ruegg:2003ps,Feldman:2006wb,Feldman:2007wj,Feldman:2011ms}, where the existence of such a scalar field is not needed. In the former case, in addition to the Lagrangian we will describe below, interactions involving a scalar field would also show up, but we have already determined the $g-2$ and $\MEG$ contributions stemming from a scalar field. Thus, incorporating such interactions is straightforward by means of Eq.~\eqref{Eq:neutralscalar}. That said, we will restrict our attention to the Lagrangian which contains the $Z^{\prime}$ boson. Moreover, such vector boson may have either vector couplings such as in the case of the $B-L$ model,~\cite{Davidson:1978pm,Mohapatra:1980qe} axial-vector in some $U(1)_X$ models~\cite{Ismail:2016tod}, or more often both. Thus, we will keep our reasoning as general as possible by writing the $Z^{\prime}$ interactions as follows, 
\begin{equation}
  \mathcal{L}_\mathrm{int} = \left({g_L}_{ij} \overline{\ell_L}^i \gamma^\mu \ell_L^j + {g_R}_{ij} \overline{e_R}^i \gamma^\mu e_R^j\right) Z_\mu^\prime + \mathrm{h.c.}
\end{equation}
However, note that these interactions can always be made family-diagonal in case of universal couplings and consequently there cannot be any LFV involved. Therefore, only a signal in $g-2$ will be considered. Adopting the same procedure as before, in Fig.~\ref{fig:results_vectorNeutral} we exhibit the region of parameter space which can explain the $g-2$ anomaly (using $g_L=g_R$). We find that this would require very light boson masses $\ll 1\TeV$ which are ruled out by collider searches if they have sizable couplings to either electrons or quarks~\cite{Agashe:2014kda}. We point out that in scenarios of non-universal couplings, LFV can be present via a $Z^{\prime}$ as discussed in~\cite{Langacker:2000ju}. For such case one can easily solve our fully generic result in Eq.~\eqref{eq:P4def}.

\subsubsection{Charged Vector Boson}
The previous caveat is avoided for a charged vector boson, conventionally called $W^\prime$. In this case we cannot diagonalize the interaction in family space. We assume that this new vector boson couples only to left- or RH leptons, however, we remark that in the former case there exist other constraints from collider searches and electroweak precision tests~\cite{Aaboud:2016zkn,Khachatryan:2014tva} which we disregard here. The interaction under consideration reads
\begin{equation}	
  \mathcal{L}_\mathrm{int} = {g_R}_{i}\, \overline{\nu_R} \gamma^\mu e_R^i W_\mu^\prime + {g_L}_{i}\, \overline{\nu_L} \gamma^\mu e_L^i W_\mu^\prime + \mathrm{h.c.}
\end{equation}
The term proportional to $g_L$ is not relevant for our purposes because in this case it has to be accompanied by some small mixing, which would require very large couplings in order to produce a signal in $g-2$ or $\MEG$. The term proportional to $g_R$ has already been considered in Sec.~\ref{sec:WprimeN}, Eq.~\eqref{Eq:WprimeNgeneral}.

\subsubsection{Doubly Charged Vector Boson}
\begin{figure}[t]
   \centering
   \noindent\makebox[\textwidth]{
   \begin{subfigure}[b]{.45\textwidth}
      \centering
      \includegraphics[width=\textwidth]{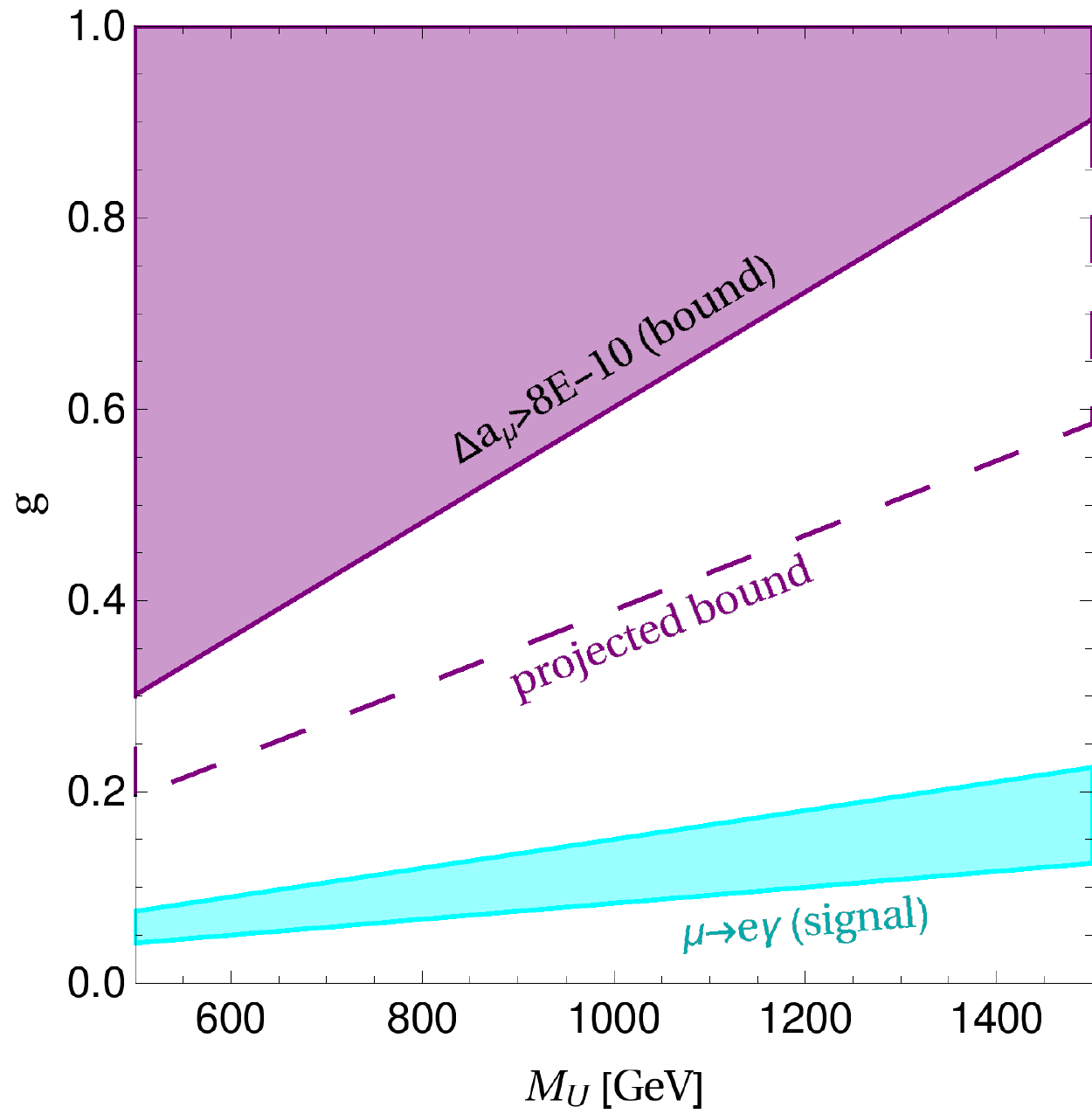}
      \subcaption{mild hierarchy}
   \end{subfigure}
   \hfill
   \begin{subfigure}[b]{.45\textwidth}
      \centering
      \includegraphics[width=\textwidth]{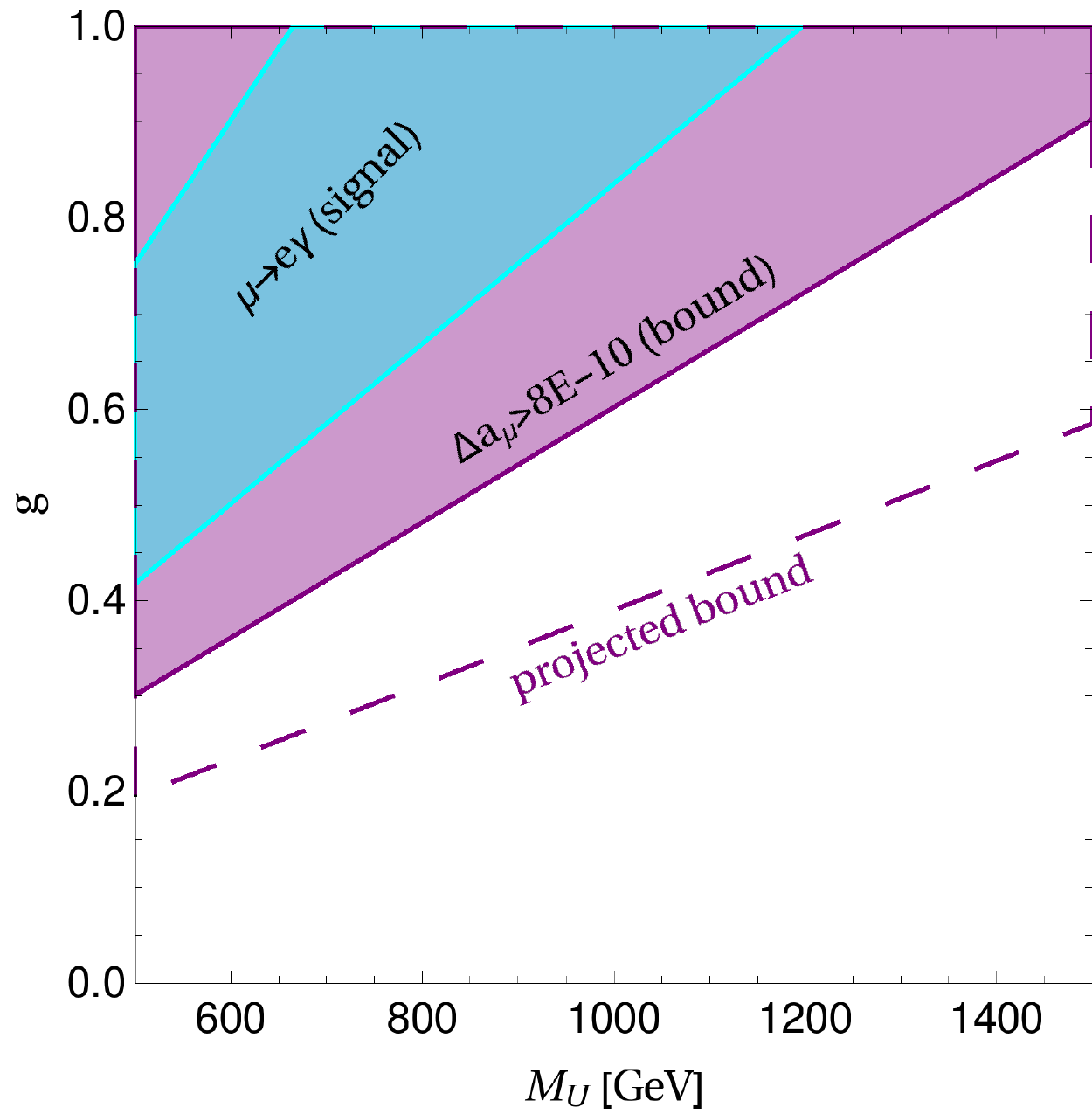}
      \subcaption{strong hierarchy}
   \end{subfigure}
   }
   \caption{\label{fig:results_vectorCharged2}Results for a doubly charged vector boson $U$ according to Eq.~\eqref{Eq:doublyVector}.}
\end{figure}

Doubly charged gauge bosons are a distinct signature of the minimal 331~model~\cite{Pisano:1991ee}. These particles are rather exotic, and rarely found in extensions of the SM. Such bosons appear in the minimal 331 model simply because there is a charged lepton with opposite electric charge in the fermion triplet of each generation. Concretely, a doubly charged vector boson $U$ induces interactions of the form
\begin{equation}
  \mathcal{L}_\mathrm{int} = \left({g_L}_{ij} \overline{\ell_L^\mathcal{C}}^i \gamma^\mu \ell_L^j + {g_R}_{ij} \overline{e_R^\mathcal{C}}^i \gamma^\mu e_R^j \right) U^{++}_\mu + \mathrm{h.c.}
  \label{Eq:doublyVector}
\end{equation}
Note however that, by virtue of the relation $\overline{\chi^\mathcal{C}} \gamma^\mu \psi = - \overline{\psi^\mathcal{C}}\gamma^\mu \psi$, the vector coupling matrix $g_v$ is anti-symmetric and therefore contains no diagonal entries. Since its contribution to $g-2$ is negative as shown in Eq.~\eqref{Eq:Delta_a_Doubly_Vectorlimit}, we show only the results for  $\mathrm{BR}(\MEG)$ in Fig.~\ref{fig:results_vectorCharged2}. From there we conclude that only with a mild hierarchy in the charged leptonic sector one can reconcile a signal in $\MEG$ with constraints from $g-2$. We emphasize that the constraint from $g-2$ comes from imposing that the $g-2$ correction is below the current and future $1\sigma$ error bar. It is also visible that even TeV scale masses offer a sizable contribution to $\MEG$ with gauge couplings of order of $\mathcal{O}(0.1)$.


\section{\label{sec:UVModels}UV Complete Models}

Our goal in this section is to revisit existing results in the literature in the context of UV complete models and show that one can apply our findings to well known extensions of the SM. We will discuss the Minimal Supersymmetric SM (MSSM), the Left-Right Symmetric model, as well as two classes of $B-L$ models. Furthermore, we discuss the scotogenic model, the two Higgs doublet model type~III, the minimal $L_{\mu}-L_{\tau}$ model, Zee-Babu model, and 331 model with RH neutrinos, minimal dark photon photon, and seesaw models type~I, II and III.  We start our discussion with the MSSM.

\subsection{Minimal Supersymmetric Standard Model}

Supersymmetry (SUSY) is one of the most compelling extensions of the SM since it uniquely relates fermions and bosons in a relativistic quantum field theory~\cite{Wess:1974tw}. Moreover, it cancels the quadratic divergences associated with the Higgs boson mass, stabilizing the weak scale against quantum corrections arising from high-energy scales~\cite{Veltman:1980mj}. Furthermore, it naturally leads to grand unification of the gauge couplings at high scales~\cite{Amaldi:1991cn,Langacker:1991an}. We have not observed any supersymmetric particle yet at high-energy colliders~\cite{Chatrchyan:2011zy,ATLAS:2016kts,Khachatryan:2016xdt}, however, we can probe SUSY models by using precision measurements in low energy experiments, especially those related to the muon anomalous magnetic moment and LFV~\cite{Cho:2000sf,Choi:2001pz,Feng:2001tr,Cheung:2001hz,Arnowitt:2001pm,Belanger:2001am,Byrne:2002cw,Baek:2002cc,
Stockinger:2006zn,RamseyMusolf:2006vr,Heinemeyer:2003dq,Cho:2011rk,Babu:2014lwa,Gomez:2014uha,Kobakhidze:2016mdx}. 

In this section we will discuss the correlation between $g-2$ and $\MEG$ using the MSSM. In order to follow our reasoning it is important to briefly describe the key features of the model. In Tab.~\ref{tab:MSSM} we show the particle content. One can see that the MSSM features two Higgs doublets ${\cal H}_{1,2}$ along with its supersymmetric partners called Higgsinos that are fermions, scalar SUSY partners
of each chiral SM fermion called sfermions $\tilde{f}_{L,R}$, and fermionic SUSY partners of the $U(1)_Y$, $SU(2)_L$ and $SU(3)_c$ SM gauge bosons, known respectively as bino ($\tilde{B}$), winos ($\tilde{W}^{\pm,3}$) and gluinos ($\tilde{g}$).

\begin{table}[t]
\begin{tabular}{ccccc}
\hline
& leptons & quarks & Higgs & gauge bosons\\
& sleptons & squarks & Higgsinos & gauginos\\
 & \null\ ${\displaystyle{{\nu_e}\choose{e}}_L}, e_R,\ldots$ \ \null & \null\ ${\displaystyle{u \choose d}_L}, u_R, d_R, \ldots$\ \null & ${\cal H}_1, {\cal H}_2$ 
& 
$B^\mu, W^{a\mu};\quad G^{a\mu}\hspace{-1ex}\null$
\\[-2.0ex]
&&&
\multicolumn{2}{c}{
\parbox{4.6cm}{
$\underbrace{\mbox{\parbox{4.1cm}{\ }}}$\\
\null\hfill$\gamma, Z, W^\pm, G^{0,\pm},$\ \ \hfill\null\\
\null\hfill
$h^0, H^0, A^0, H^\pm$\ \ \hfill\null
}
}
\\[1.5em]
&$ {\displaystyle{\tilde\nu_e\choose\tilde e}_L},
\tilde e_R,\ldots$
&
${\displaystyle{\tilde{u}_L\choose \tilde{d}_L}}, \tilde{u}_R, \tilde{d}_R, \ldots$
&
${\tilde H}_1, {\tilde H}_2 $
& 
$\tilde{B}, \tilde{W}^{a};\quad \tilde{g}^{a}$\\[-2.0ex]
&&&
\multicolumn{2}{c}{
\parbox{4.6cm}{
$\underbrace{\mbox{\parbox{4.1cm}{\ }}}$\\
\null\hfill$\chi^0_{1,2,3,4}$, $\chi^\pm_{1,2}$\ \ \hfill\null}
}
\\
\hline
\end{tabular}
\caption{Particle content of the MSSM. Only the first generation is explicitly shown.
}
\label{tab:MSSM}
\end{table}

As far $g-2$ and $\MEG$ are concerned, two parameters play a key role in the MSSM. The first is the ratio of the VEVs of the two Higgs doublets (${\cal H}_{1,2}$), defined as
\begin{eqnarray}
\tan\beta &=& \frac{v_2}{v_1}.
\end{eqnarray}
Since the muon Yukawa coupling $y_{\mu}$ is given by
\begin{equation}
y_{\mu}=\frac{m_{\mu}}{v_1}=\frac{m_{\mu} g_2}{\sqrt{2}M_W \cos_{\beta}},
\label{Eq:muonYukawa}
\end{equation}where $g_2=e/s_W$, and the chirality-flips relevant for $g-2$ and $\MEG$ are proportional to the muon mass, an enhancement to these observables might occur for large $\tan \beta \sim 1/\cos \beta$.

The second is the $\mu$ term which determines the Higgsino mass term and gives rise to the sfermions' interactions with ${\cal H}_{1,2}$ through the Lagrangian,
\begin{eqnarray}
\mu\tilde{H}_1\tilde{H}_2-\mu F_{H_1}{\cal H}_2-\mu F_{H_2}{\cal H}_1 +
\mathrm{h.c.} ,
\label{mudef}
\end{eqnarray}where $F_{H_{1,2}}$ are auxiliary fields. The auxiliary fields are eliminated to generate Yukawa interactions of the form ${\cal
 H}_2^0\tilde{\mu}_L\tilde{\mu}_R^\dagger$ for example. As we will see later on, the $\mu$ term and its sign are relevant to determine the neutralino and chargino contributions to $g-2$ and $\MEG$ since it appears in the mass matrices of both fields.

\begin{figure}[t]\centering
  \begin{subfigure}[b]{.45\textwidth}
    \centering
    \begin{tikzpicture}[x=1mm,y=1mm]
      \node[anchor=east] at (-20,0) (f1) {$\mu$};
      \node[anchor=west] at (20,0) (f2) {$\mu$};
      \node (V1) at (-10,0) {};
      \node (V2) at (10,0) {};
      \draw[fill=black] (V1) circle(.2);
      \draw[fill=black] (V2) circle(.2);
      \draw[decoration={markings, mark=at position 0.6 with {\arrow{latex}}}, postaction={decorate}] (f1) -- (V1.center);
      \draw (V2.center) -- (V1.center) node[midway, below] {$\chi^0$};	
      \draw[decoration={markings, mark=at position 0.6 with {\arrow{latex}}}, postaction={decorate}] (V2.center) -- (f2);
      \draw[mydash, decoration={markings, mark=at position 0.6 with {\arrow{latex}}}, postaction={decorate}] (V1.center) arc(180:90:10) 
      node (V3) {};
      \draw[fill] (V3.center) circle(.2);
      \draw[decorate, decoration={snake, segment length=3.3mm, amplitude=.5mm}] (V3.center) -- (20,20) node[anchor=west] {$\gamma$}; 
      \draw[mydash,decoration={markings, mark=at position 0.6 with {\arrow{latex}}}, postaction={decorate}] (V3.center) arc(90:0:10);
      \node at (-10,9) {$\widetilde{\mu}_j$};
      \node at (11,9) {$\widetilde{\mu}_j$};
    \end{tikzpicture}
  \end{subfigure}
  \hspace{1cm}
  \begin{subfigure}[b]{.45\textwidth}
    \centering
    \raisebox{1mm}{
    \begin{tikzpicture}[x=1mm,y=1mm]
      \node[anchor=east] at (-20,0) (f1) {$\mu$};
      \node[anchor=west] at (20,0) (f2) {$\mu$};
      \node (V1) at (-10,0) {};
      \node (V2) at (10,0) {};
      \draw[fill=black] (V1) circle(.2);
      \draw[fill=black] (V2) circle(.2);
      \draw[decoration={markings, mark=at position 0.6 with {\arrow{latex}}}, postaction={decorate}] (f1) -- (V1.center);
      \draw[mydash] (V2.center) -- (V1.center) node[midway, below] {$\widetilde{\nu}_\mu$};	
      \draw[decoration={markings, mark=at position 0.6 with {\arrow{latex}}}, postaction={decorate}] (V2.center) -- (f2);
      \draw[decoration={markings, mark=at position 0.6 with {\arrow{latex}}}, postaction={decorate}] (V1.center) arc(180:90:10) 
      node (V3) {};
      \draw[fill] (V3.center) circle(.2);
      \draw[decorate, decoration={snake, segment length=3.3mm, amplitude=.5mm}] (V3.center) -- (20,20) node[anchor=west] {$\gamma$}; 
      \draw[decoration={markings, mark=at position 0.6 with {\arrow{latex}}}, postaction={decorate}] (V3.center) arc(90:0:10);
      \node at (-9,9) {$\chi^-$};
      \node at (11,9) {$\chi^-$};
    \end{tikzpicture}
    }
  \end{subfigure}\\
\caption{\label{moritzfig}Diagrams contributing to the muon $g-2$ in the MSSM.}
\end{figure}
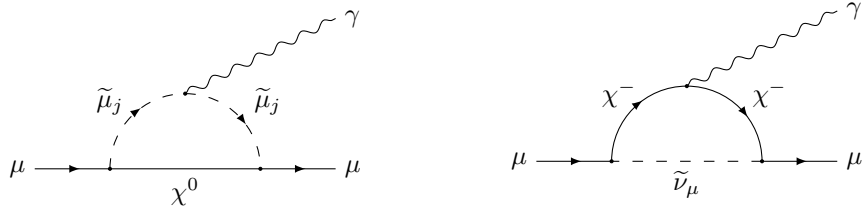

In order to provide an insight into the MSSM contributions to $g-2$ and $\mu \rightarrow e\gamma$, we will first work out the $g-2$ correction under some simplifying assumptions, revisiting well known results in certain regimes and then move to a very general approach and see whether signals in both observables can be accommodated. 

As far as $g-2$ is concerned there are basically  two sorts of diagrams (Fig.~\ref{moritzfig}) contributing to $g-2$: (i) muon-neutralino-smuon; (ii) muon-chargino-sneutrino, which are found to be,  respectively~\cite{Grifols:1982vx,Barbieri:1982aj,Ellis:1982by,Kosower:1983yw,Yuan:1984ww,
Romao:1984pn,Hisano:1995nq,Barbieri:1995tw,Ibrahim:1999aj,Deppisch:2002qw,Deppisch:2004xv,
Martin:2002eu,Girrbach:2009uy,Heinemeyer:2004py,Cirigliano:2004mv,Haestier:2006mg,Cheung:2009fc}
\begin{subequations}
\label{Eq:SUSYg2}         
\begin{eqnarray}
 \Delta a_\mu^{\chi^0} & = & \frac{m_\mu}{16\pi^2}
   \sum_{i,m}\left\{ -\frac{m_\mu}{ 12 m^2_{\tilde\mu_m}}
  (|n^L_{im}|^2+ |n^R_{im}|^2)F^N_1(x_{im}) + \right. \nonumber\\
&& \hspace{4.7cm} \left.  +\frac{m_{\chi^0_i}}{3 m^2_{\tilde \mu_m}}
    {\rm Re}[n^L_{im}n^R_{im}] F^N_2(x_{im})\right\},\\
\Delta a_{\mu}^{\chi^\pm} & = & \frac{m_\mu}{16\pi^2}\sum_k
  \left\{ \frac{m_\mu}{ 12 m^2_{\tilde\nu_\mu}}
   (|c^L_k|^2+ |c^R_k|^2)F^C_1(x_k)
 +\frac{2m_{\chi^\pm_k}}{3m^2_{\tilde\nu_\mu}}
         {\rm Re}[ c^L_kc^R_k] F^C_2(x_k)\right\},\nonumber\\
\end{eqnarray}
\end{subequations}
with $i=1,2,3,4$, and $k=1,2$ denoting the neutralino and chargino mass eigenstate indices, $m=1,2$ the smuon one, and the couplings given by
\begin{equation}\label{Eq:entries}
\begin{aligned}
  n^R_{im}  = &  \sqrt{2} g_1 N_{i1} X_{m2} + y_\mu N_{i3} X_{m1},\\
  n^L_{im}  = &  {1\over \sqrt{2}} \left (g_2 N_{i2} + g_1 N_{i1}
  \right ) X_{m1}^* - y_\mu N_{i3} X^*_{m2},\\
  c^R_k  = & y_\mu U_{k2},\\
  c^L_k = & -g_2V_{k1},
\end{aligned}
\end{equation}
where $g_1=e/c_W$, and $y_\mu$ is the muon Yukawa coupling defined in Eq.~\eqref{Eq:muonYukawa}. The kinematic loop functions, which are normalized to unity for $x=1$, depend on the variables $x_{im}=m^2_{\chi^0_i}/m^2_{\tilde\mu_m}$, 
$x_k=m^2_{\chi^\pm_k}/m^2_{\tilde\nu_\mu}$ and are found to be
\begin{subequations}
\label{Eq:FN}    
\begin{eqnarray}
F^N_1(x) & = &\frac{2}{(1-x)^4}\left[ 1-6x+3x^2+2x^3-6x^2\ln x\right], \\
F^N_2(x) & = &\frac{3}{(1-x)^3}\left[ 1-x^2+2x\ln x\right], \\
F^C_1(x) & = &\frac{2}{(1-x)^4}\left[ 2+ 3x - 6
x^2 + x^3
        +6x\ln x\right], \\
F^C_2(x) & = & -\frac{3}{2(1-x)^3}\left[ 3-4x+x^2
    +2\ln x\right].
\end{eqnarray}
\end{subequations}
With Eq.~\eqref{Eq:SUSYg2} one can compute the MSSM contribution to $g-2$ knowing the neutralino ($\chi^0$)and chargino ($\chi^\pm$) and smuon ($\tilde\mu$) mass matrices which are given by~\cite{Haber:1984rc,Martin:1997ns}
\begin{subequations}
\begin{equation}
M_{\chi^0} = \begin{pmatrix}
M_1 & 0 & - \cos \beta\, \sin W\, M_Z &
\sin \beta\, \sin W \, M_Z \\
0 & M_2 & \cos \beta\, \cos W\, M_Z & - \sin \beta\, \cos W\, M_Z \\
-\cos \beta \,\sin_W\, M_Z & \cos \beta\, \cos W\, M_Z & 0 & -\mu \\
\sin \beta\, \sin_W\, M_Z & - \sin \beta\, \cos_W \, M_Z& -\mu & 0 \\ \end{pmatrix},
\label{neutralinomassmatrix} 
\end{equation}
\begin{equation}
M_{\chi^\pm} = 
\begin{pmatrix}{cc}
M_2 & \sqrt{2} \sin \beta\, M_W\\
                              \sqrt{2} \cos \beta\, M_W & \mu \\
                              \end{pmatrix},
\label{charginomassmatrix}
\end{equation}and
\begin{eqnarray}
M^2_{\tilde\mu}=\left( \begin{array}{cc}
   m^2_L +(s_W^2 -\frac{1}{2})m_Z^2\cos\,  2\beta &
              m_\mu (A^*_{\tilde\mu}-\mu\tan\beta) \\
 m_\mu (A_{\tilde\mu}-\mu^*\tan\beta) &
    m^2_R -s_W^2 \, m_Z^2\cos\,  2\beta 
\end{array}\right),
\end{eqnarray}
\end{subequations}
where $A_{\tilde\mu}$ is the soft SUSY breaking parameter of the trilinear interaction $\tilde\mu_L-\tilde\mu_R$-Higgs, with the muon sneutrino mass being connected to the left-handed (LH) smuon mass parameter via
\begin{equation}
m_{\tilde{\nu}}^2 = m_L^2 + {1\over 2} M_Z^2 \cos 2\beta  .
\end{equation}
These matrices are diagonalized using four matrices $N$, $U$, $V$ and $X$ which define the entries in Eqs.~\eqref{Eq:entries}, and are determined as follows,
\begin{subequations}\label{Eq:neutralino}
\begin{eqnarray}
N^* M_{\chi^0} N^\dagger &=& {\rm diag}(
m_{\chi^0_1},
m_{\chi^0_2},
m_{\chi^0_3},
m_{\chi^0_4}),\\
U^* {M}_{\chi^\pm} V^\dagger &=& {\rm diag}(
m_{\chi^\pm_1},
m_{\chi^\pm_2}),\\
X M^2_{\tilde\mu}\, X^\dagger &=& 
{\rm diag}\, (m^2_{\tilde\mu_1}, m^2_{\tilde\mu_2}).
\end{eqnarray}
\end{subequations}
We have now gathered all relevant ingredients to compute the MSSM contribution to $g-2$ and have a grasp of the underlying physics. 

\subsubsection{Simplified Results}

\subsubsection*{Similar SUSY masses}
\begin{figure}[!t]
\centering
\includegraphics[scale=0.7]{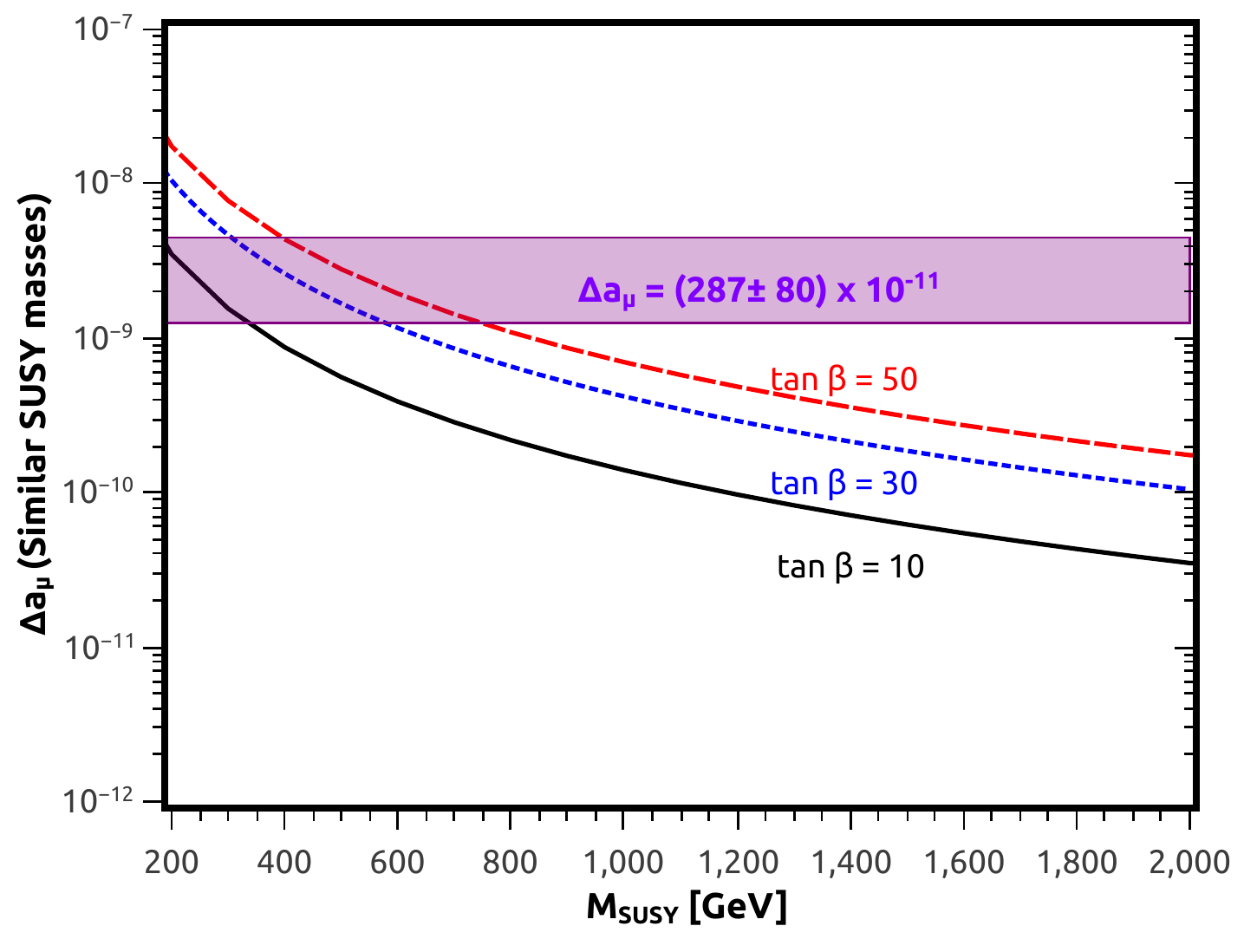}
\caption{MSSM contribution to $g-2$ for different values of $\tan \beta$ assuming all supersymmetric particles have masses equal to $M_{SUSY}$ according to Eq.~\eqref{msusy}.}
\label{plotMSUSY} 
\end{figure}

The simplest analytic result to obtain from Supersymmetry is to assume that
all superpartners have the same mass $M_{\rm SUSY}$, which leads to~\cite{Moroi:1995yh}
\begin{equation}
\begin{aligned}
\Delta a_{\mu}^{\chi^0} &=& \frac{\tan\beta}{192\pi^2}
    \frac{m_\mu^2}{M_{\rm SUSY}^2}\, (g_1^2-g_2^2),\\
\Delta a_{\mu}^{\chi^{\pm}} &=& \frac{\tan\beta}{32\pi^2}
    \frac{m_\mu^2}{M_{\rm SUSY}^2}\, g_2^2,
 \end{aligned}
\end{equation}
resulting in,
\begin{eqnarray}
\Delta a_\mu^{\rm SUSY} =
 \frac{\tan\beta}{192\pi^2}
    \frac{m_\mu^2}{M_{\rm SUSY}^2}\, (5g_2^2+g_1^2)
=
14 \tan\beta \left ( \frac{100\>{\rm GeV}}{M_{\rm SUSY}} \right )^2
10^{-10}.
\label{msusy}
\end{eqnarray}
In this regime the chargino contribution dominates~\cite{Moroi:1995yh}, and the effect of a large $\tan\beta$ is explicitly seen in Eq.~\eqref{msusy}. In Fig.~\ref{plotMSUSY} we display this dependence, where one can clearly see that relatively low masses are needed to account for the observed $g-2$ discrepancy. 

This result holds true for one-loop corrections only. Albeit, two-loop effects which are negative, lead to effects of the order of 10\%~\cite{Degrassi:1998es,Fargnoli:2013zia,Athron:2015rva,Bach:2015doa}:
\begin{equation}
\Delta a^{SUSY}_{\rm \mu\, 2-loop} = \Delta a^{SUSY}_{\rm \mu\, 1-loop} \left( 1 -\frac{4 \alpha_{em}}{\pi}\ln \frac{M_{SUSY}}{m_{\mu}}\right), 
\end{equation}thus not changing much the overall picture.

\subsubsection*{Large $\mu$}

Another important regime is obtained for large $\mu$, i.e.~$\mu,M_2 \gg M_1$, which implies that the diagrams with light bino and smuon are dominant. In this limit one finds~\cite{Martin:2001st}
\begin{eqnarray}
 a_\mu^{{\rm Large\, \mu}} &=&
{g_1^2 \over 48 \pi^2} {m^2_\mu  M_1 {\rm Re} [\mu \tan\beta - A_\mu^*]
\over m_{\tilde{\mu}_2}^2 - m_{\tilde{\mu}_1}^2 }
\left[ {F_2^N(x_{11}) \over m_{\tilde{\mu}_1}^2} 
- {F_2^N(x_{12})\over m^2_{\tilde{\mu}_2}} \right ],
\label{lightbino}
\end{eqnarray}which for $m_{\tilde{\mu}_1} \approx m_{\tilde{\mu}_2} =
2.0 M_1$ reduces to, 
\begin{eqnarray}
 a_\mu^{{\rm light}\> {\rm bino}} & = &
18 \tan\beta \left ( {100\>{\rm GeV} \over m_{\tilde{\mu} } } \right )^3
\left ( {\mu - A_\mu \cot\beta \over 1000 \>{\rm GeV}} \right ) 10^{-10},
\label{nouseforanumber}
\end{eqnarray}where $x_{1m} = M_1^2/m_{\tilde{\mu}_m}^2$. First, notice that both the value of the $\mu$ term and its sign are relevant to determine whether the MSSM gives rise to a negative or positive contribution to $g-2$, along with the value for $\tan \beta$. It is easy to see that there is plenty of room to accommodate the $g-2$ for several values of $\tan \beta, A_{\mu}$, and $m_{\tilde{\mu} }$. See Refs.~\cite{Martin:2001st,Stockinger:2006zn} for more details.

\subsubsection{Connecting $g-2$ and $\MEG$}

The correlation between $g-2$ and $\MEG$ has been investigated before in the context of the MSSM~\cite{Chacko:2001xd,Bi:2002ra}\footnote{In~\cite{Ibrahim:2015hva} the impact of complex phases were addressed.}. In order to connect $g-2$ and $\MEG$ we need to keep the discussion more general, leaving explicit the sneutrino mixings in the chargino contribution and the slepton mixings in the neutralino one~\cite{Kersten:2014xaa}.  We start with the chargino contribution.

{\bf Chargino contribution}

To do so, we first start by assuming the mixing of the third generation to be decoupled from the first two. One can find that the sneutrino mass eigenvalues $m_{\tilde\nu_i}$ and the mixing angle ($\theta_{\tilde\nu}$) are determined through the diagonalization of the sneutrino mass matrix
\begin{align}
\begin{pmatrix} 
\cos\theta_{\tilde{\nu}} & \sin\theta_{\tilde{\nu}} \\ 
-\sin\theta_{\tilde{\nu}} & \cos\theta_{\tilde{\nu}}
\end{pmatrix}
\begin{pmatrix}
m_{\tilde L_{11}}^2 + \mathcal{D}^\nu_L & m_{\tilde L_{12}}^2 \\ 
m_{\tilde L_{12}}^2 & m_{\tilde L_{22}}^2 + \mathcal{D}^\nu_L
\end{pmatrix}
\begin{pmatrix} 
\cos\theta_{\tilde{\nu}} & -\sin\theta_{\tilde{\nu}} \\
\sin\theta_{\tilde{\nu}} & \cos\theta_{\tilde{\nu}} 
\end{pmatrix}
=& \nonumber \\  = {\rm diag}(m_{\tilde\nu_1}^2,m_{\tilde\nu_2}^2)
\label{eq:SneutrinoMixing}
\end{align}which leads to
\begin{equation}
\tan 2 \theta_{\tilde\nu}=\frac{2 m_{\tilde L_{12}}^2 }{m_{\tilde L_{11}}^2 - m_{\tilde L_{22}}^2},
\end{equation}where $m^2_{\tilde L_{11}}$, $m^2_{\tilde L_{22}}$, $m_{\tilde L_{12}}^2$ are the flavor-diagonal and off-diagonal soft mass parameters with ${\cal D}_L^\nu$
being the $D$-term contributions to their masses. 

Now we can re-obtain the chargino contribution to $g-2$ leaving the dependence on $\theta_{\tilde\nu}$ explicit as follows:
\begin{align}
\label{eq:chargino1}
&a_\mu^{\tilde\chi^\pm_k} = \underbrace{\frac{m_\mu^2}{192\pi^2 m_{\tilde{\chi}^\pm_k}^2}
   \left( g_2^2 |V_{k1}|^2+    y_\mu^2 |U_{k2}|^2\right) \left[ \sin^2 \theta_{\tilde{\nu}} \, x_{k1} F_1^C(x_{k1}) 
    + \cos^2 \theta_{\tilde{\nu}} \, x_{k2} F_1^C(x_{k2}) \right]}_{A}+
\nonumber \\
&\underbrace{ -\frac{2m_\mu}{48\pi^2 m_{\tilde{\chi}^\pm_k}}
    \, g_2 y_\mu Re[V_{k1} U_{k2}] \left[ \sin^2 \theta_{\tilde{\nu}} \, x_{k1} F_2^C(x_{k1}) 
    + \sin^2 \theta_{\tilde{\nu}} \, x_{k2} F_2^C(x_{k2}) \right]}_{B}.
\end{align}
In the case of chargino dominance we find ${\rm BR}(\MEG)$ to be
\begin{eqnarray}
\label{eq:chargino2}
{\rm BR}(\MEG) &=&\frac{12\pi^3 \alpha_{em}}{G_F m_{\mu}^4} \bigg| \left(\frac{m_{\tilde L_{12}}^2}{m_{\tilde\nu_1}^2-m_{\tilde\nu_2}^2}\right)\left[ \,
\frac{A\, \Delta_1}{x_{k_2} F_I^C(x_{k2}) + \sin^2\theta_{\tilde\nu} \Delta_1}\right. \nonumber\\
&& + \left.  \frac{B\, \Delta_2}{x_{k_2} F_I^C(x_{k2}) + \sin^2\theta_{\tilde\nu} \Delta_2}\right] \bigg|^2 .\nonumber\\
\end{eqnarray}where $\Delta_i \equiv x_{k1} F_I^C(x_{k1}) - x_{k2} F_I^C(x_{k2})$, with $i=1,2$, and the $A$ and $B$ identified in Eq.~\eqref{eq:chargino1}. All coupling constants and functions have been defined previously.

\newpage
{\bf Neutralino contribution}

To compute the neutralino-slepton contributions to $g-2$ and $\MEG$ we need to work
on a more general foundation for the neutralino and slepton mixings. The diagonalization
procedure for the neutralino was described earlier in Eq.~\eqref{Eq:neutralino}. As for the sleptons, we follow the recipe of Ref.~\cite{Kersten:2014xaa}, where the full mixing structure was considered, and the
mass mixing matrix K with $K \mathcal{M}^2 K^{\dagger} =
{\rm diag}(m_{\tilde \ell_1}^2,\hdots,m_{\tilde\ell_6}^2)$ was derived. At the end one finds
\begin{eqnarray}
 a_\mu^{\chi^0} & = & \frac{m_\mu}{16\pi^2}
   \sum_{i,m}\left\{ -\frac{m_\mu}{ 12 m^2_{\tilde\mu_m}}
  (|n^{L}_{im}|^2+ |n^R_{im}|^2)F^N_1(x_{im}) \right. + \nonumber\\
&& \left.  +\frac{m_{\chi^0_i}}{3 m^2_{\tilde \mu_m}}
    {\rm Re}[n^L_{im}n^R_{im}] F^N_2(x_{im})\right\}
\label{Eq:neutralino1}         
\end{eqnarray}and
\begin{eqnarray}
\label{Eq:neutralino2}
{\rm BR}(\MEG) &=&\frac{12\pi^3 \alpha_{em}}{G_F m_{\mu}^4}\left( |A|^2 + |B|^2\right)
\end{eqnarray}for the case of neutralino-slepton dominance, where
\begin{eqnarray}
A & = & \frac{m_\mu}{16\pi^2}
   \sum_{m}\left\{ -\frac{m_\mu}{ 12 m^2_{\tilde\chi^0_i}}n^R_{\mu im} n^{R\ast}_{e im}x_{im} F_1^N(x_{im})+ \frac{1}{3 m_{\tilde\chi^0_i}}n^{L\ast}_{\mu im} n^{R\ast}_{e im}x_{im} F_2^N(x_{im})\right\},\nonumber\\
\label{Eq:neutralino3}         
\end{eqnarray}
\begin{eqnarray}
B & = & \frac{m_\mu}{16\pi^2}
   \sum_{m}\left\{ -\frac{m_\mu}{ 12 m^2_{\tilde\chi^0_i}}n^L_{\mu \ast im} n^{L}_{e im}x_{im} F_1^N(x_{im})+ \frac{1}{3 m_{\tilde\chi^0_i}}n^{R}_{\mu im} n^{L}_{e im}x_{im} F_2^N(x_{im})\right\},\nonumber\\
\label{Eq:neutralino4}         
\end{eqnarray}where $F_1^N(x)$ and $F_2^N(x)$ are defined in Eq.~\eqref{Eq:FN}, $x_{im}= m^2_{\tilde\chi^0_i}/m_{\tilde\ell_m}^2$ and
\begin{equation}
\begin{aligned}
n^L_{\ell im} &= \frac{1}{\sqrt{2}} \left( g_1 N_{i1} + g_2 N_{i2} \right) K^*_{m,\ell}
- y_\ell N_{i3} K_{m,\ell+3}^* ,\\
n^R_{\ell im} &=
\sqrt{2} g_1 N_{i1} K_{m,\ell+3} +
y_\ell N_{i3} K_{m,\ell}.
\end{aligned}
\end{equation}
Combining Eqs.~(\ref{eq:chargino1},\ref{eq:chargino2}) and Eqs.~(\ref{Eq:neutralino1},\ref{Eq:neutralino2}), we can explore the correlation between $g-2$ and $\MEG$ in the MSSM for different regimes, namely {\it similar SUSY masses} and {\it large $\mu$ term} as follows. We will assume that squarks and gluinos are much heavier than the sleptons, charginos and neutralinos, with masses sufficiently large (TeV scale) to avoid LHC bounds at 13 TeV~\cite{ATLAS:2016kts}.

\begin{figure}[t]
\centering
\includegraphics[scale=0.7]{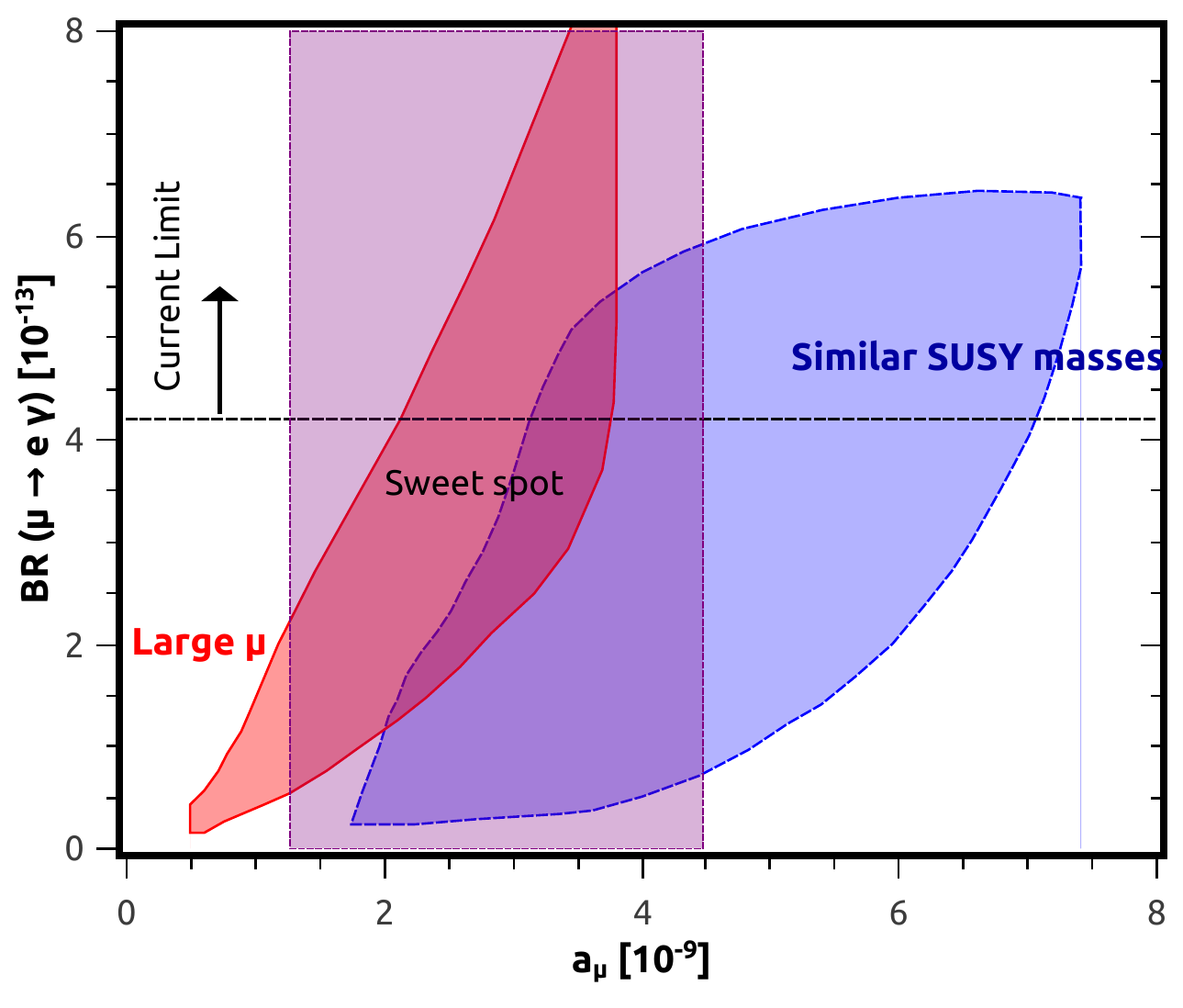}
\caption{Correlation between $g-2$ and $\MEG$ in the MSSM for two different regimes: (i) Similar masses; (ii) Large $\mu$ regime. We overlay the $2\sigma$ band for $g-2$ and current limits on $\MEG$. Notice that there is a sweet spot lying below the current limit on $\MEG$  and in the $2\sigma$ region of $g-2$, where a signal in both observables are compatible with each other.}
\label{plotMSSMcorrelation} 
\end{figure}
 
\subsubsection*{Similar SUSY masses}

We will assume that supersymmetric particles have the same mass, varying from $300$~GeV up to $800$~GeV, keeping $\tan \beta =50, A_{\mu}=0$. This is similar to~\cite{Kersten:2014xaa}, except that there the authors scanned up to masses of $600$~GeV only. Moreover, we adopted the hierarchies $m_{\tilde L_{12}}^2/\sqrt{m_{\tilde L_{11}}^2 m_{\tilde L_{22}}^2}=2\times 10^{-5}$ and $m_{\tilde R_{12}}^2/\sqrt{m_{\tilde R_{11}}^2 m_{\tilde R_{22}}^2 }=2\times 10^{-5}$. This is similar to~\cite{Kersten:2014xaa}, except that there the authors scanned up to masses of $600$~GeV. The result is shown in Fig.~\ref{plotMSSMcorrelation}.

\subsubsection*{Large $\mu$ Regime}

In this regime the most relevant parameters are $\left\lbrace\mu, \tan \beta, M_1, m_{{\tilde L}_{22}}^2, m_{{\tilde R}_{22}}^2\right\rbrace$. If $\tan \beta$ is relatively large, say $\tan \beta=50$, the neutralino contribution starts dominating and growing with $\mu$. Keeping the hierarchy $\mu > M_2 > M_1$, and varying $M_1, m_{{\tilde L}_{22}}^2$ and $m_{{\tilde R}_{22}}^2$ between $300-600$~GeV we find the result in Fig.~\ref{plotMSSMcorrelation}, which agrees well with~\cite{Kersten:2014xaa}. To obtain the correlation as shown in Fig.~\ref{plotMSSMcorrelation} we further assumed $|a_{\mu e L}/a_\mu| \simeq 2/3 |m_{{\tilde L}_{12}}^2|/ m_{{\tilde e}_{L}}^2$, and $|a_{\mu e R}/a_\mu| \simeq 2/3 |m_{{\tilde R}_{12}}^2|/ m_{{\tilde e}_{R}}^2$. For more details we refer to~\cite{Fargnoli:2013zda,Fargnoli:2013zia}.


Looking at Fig.~\ref{plotMSSMcorrelation} we can see that there is some degree of correlation between $g-2$ and $\MEG$ in the scan performed, which was converted into regions using an interpolation function. A large region of the parameter space in the {\it  similar SUSY masses} regime induces large contributions to $a_{\mu}$ incompatible with the data, whereas the {\it large $\mu$} typically yields corrections to $a_{\mu}$ in agreement with data. Interestingly, in both cases one can find a sweet spot within the $2\sigma$ band for $g-2$ shown in purple and below current limit on $\MEG$ where signals in observables can be made compatible with each other. The $2\sigma$ band for $g-2$ yields $\Delta a_{\mu}= (287 \pm 160)\times 10^{-11}$.

As for the other LFV observables such as $\mu \to eee$ and $\mu-e$ conversion, we will not address them here, because they are highly non-trivial to interplay with g-2, and leave this task for future works beyond the scope of this review.

{\bf Existing Limits}

There is a multitude of limits applicable to the MSSM, but instead of addressing each one in particular, we will simply assume that the squarks and gluinos are much heavier than the sleptons, charginos and neutralinos, with masses sufficiently large (TeV scale). This naive assumption is sufficient to avoid LHC bounds at 13 TeV~\cite{ATLAS:2016kts}.

\subsection{Left-Right Symmetry}
Left-Right symmetric models are based on the gauge group $SU(2)_L\times SU(2)_R\times
U(1)_{B-L}$ which under the addition of Left-Right parity means that the $SU(2)_L$ and $SU(2)_R$ couplings are identical, i.e.~$g_L=g_R=g$, where $g_L=e/s_W$.  These models may successfully be embedded in GUT theories, provide a natural environment for the seesaw mechanism~\cite{Minkowski:1977sc,Mohapatra:1979ia,Lazarides:1980nt,Mohapatra:1980yp,Schechter:1980gr}, and directly address parity violation at the weak scale~\cite{Senjanovic:1975rk,Senjanovic:1978ev}. The fermion and scalar content of the model reads
\begin{subequations}
\begin{gather}
Q_L =  \left (
\begin{array}{c}
u_L \\
d_L 
\end{array}
\right ), Q_R =  \left (
\begin{array}{c}
u_R \\
d_R 
\end{array}
\right ) ,\\
 l_L  =  \left (
\begin{array}{c}
\nu_L \\
e_L 
\end{array}
\right ),  l_R  =  \left (
\begin{array}{c}
N_R \\
e_R 
\end{array}
\right ),
\label{L}\\
\phi = \left (
\begin{array}{cc}
\phi_1^0 & \phi_1^+\\
\phi_2^- & \phi_2^0
\end{array}
\right ), \Delta_{L,R}  =  \left (
\begin{array}{cc}
\delta^+_{L,R}/\sqrt{2} & \delta^{++}_{L,R} \\
\delta^0_{L,R} & -\delta^+_{L,R}/\sqrt{2}
\end{array}
\right ),
\end{gather}
\end{subequations}
with the fields transforming under
parity and charge conjugation as follows, $P$: $Q_L \leftrightarrow Q_R, \phi \leftrightarrow  \phi^{\dagger}, \Delta_L \leftrightarrow  \Delta_R$; and $C$: $Q_L \leftrightarrow Q_R^c, \phi \leftrightarrow  \phi^{T}, \Delta_{L,R} \leftrightarrow  \Delta^{\ast}_{R,L}$. Here $\phi$ is a bi-doublet scalar not charged under $B-L$, whereas $\Delta_{L,R}$ are scalar triplets with $B-L=2$~\cite{Minkowski:1977sc,Mohapatra:1980yp,Chang:1984uy}. The scalar sector of the model can take different forms, but with little impact on our reasoning. 
The VEVs follow the pattern below
\begin{eqnarray}
\langle\phi\rangle&=&\left( \begin{array}{cc} \kappa_1/\sqrt{2} & 0\\
0 & \kappa_2/\sqrt{2}\end{array}\right),~
\langle\Delta_{L,R}\rangle=\left( \begin{array}{cc} 0 & 0
\\
v_{L,R} & 0 \end{array}\right).
\end{eqnarray}

There are two important scales relevant for our purposes, one is the electroweak scale, with $\kappa \sim \kappa_1 \sim
\kappa_2 \sim 246$ GeV, and the scale $v_R$ at which the symmetries $SU(2)_R$ and $U
(1)_{B-L}$ are spontaneously broken. After spontaneous symmetry breaking one finds
\begin{eqnarray}
\frac{M_{Z_{R}}}{M_{W_{R}}}=\frac{\sqrt{2} g_R/g_L}{\sqrt{(g_R/g_L)^2-\tan^2\theta_W}}\,,
\label{MZMW}
\end{eqnarray}with $M_{W_R} =g_R v_R$. Remember that we will be assuming $g_L= g_R$ throughout, which implies that $M_{Z_{R}}\simeq 1.7 M_{W_{R}}$, unless stated otherwise.
 
The existence of new gauge bosons is a consequence of the extended gauge symmetry. They lead to the neutral current involving the $Z^{\prime}$ gauge boson, 
\begin{eqnarray} \label{eq:coupl}
\frac{g_L}{\sqrt{1-\delta \tan^2\theta_W}}
\overline{f}\,\gamma_\mu \left(
     g_V^f - g_A^f \gamma^5 \right) \,f\,\, Z^{\prime \mu} \,,
\end{eqnarray}
with the couplings determined by 
\begin{eqnarray}
&&\hspace*{-0.5cm}g^f_V=\frac{1}{2}\left[\big\{\delta \tan^2\theta_W \left(T^{f}_{3L} 
     -\mbox{Q}^f \right)\big\} 
+ \big\{T^{f}_{3R}- \delta \tan^2\theta_W \mbox{Q}^f\big\}\right], \nonumber \\
&&\hspace*{-0.5cm}g^f_A=\frac{1}{2}\left[\big\{\delta \tan^2\theta_W \left(T^{f}_{3L} 
     -\mbox{Q}^f \right)\big\} 
- \big\{T^{f}_{3R}- \delta \tan^2\theta_W \mbox{Q}^f\big\}\right], \nonumber
\end{eqnarray} where $T^{f}_{3L,3R}= \pm 1/2$  
for $^{\rm up}_{\rm down}$-fermions,  $\delta=g^2_L/g^2_R$, and $Q^f$ being the corresponding electric charges. Moreover, the charged current is found to be,
\begin{equation}\label{eq:Lagrangian}
\begin{aligned}
\mathcal L  = & \frac{g_L}{\sqrt{2}} \left(
\bar l_L U_{L}^\dag  \slashed{W}\!_L l'_L +
\bar Q_L V_{L}^\dag  \slashed{W}\!_L Q'_L \right)
+\text{h.c.}\, +\\
&+\frac{g_R}{\sqrt{2}}  \left(
\bar l_R  U_{R}^\dag \slashed{W}\!_R l'_R+
\bar Q_R  V_{R}^\dag \slashed{W}\!_R Q'_R\right)
+\text{h.c.},
\end{aligned}
\end{equation}
where $U_{L/R}$ represent the PMNS mixing matrix for the LH and
RH leptons and $V_{L/R}$ is the Cabibbo-Kobayashi-Maskawa matrix for the LH and RH quarks.

Now that we have briefly reviewed the model, we compute the Left-Right contribution to the observables of interest. 

\subsubsection{Results in the Left-Right Model}

We will focus our discussion on the RH charged current, simply because the $Z^{\prime}$ contribution is dwindled, the scalar corrections are relatively small compared to the $W_R$ mediated ones~\cite{Cirigliano:2004mv}, and on top of that are sensitive to the scalar content of the model which can vary. Thus, in order to draw general conclusions we compute the one-loop processes that involve the $W_R$ gauge boson and heavy RH neutrinos ($N_R$). We compute their contribution for two different regimes.

(i) $M_{W_R} \gg M_{N_R}$

The charged current of the Left-Right model is identical to our simplified model with a gauge boson and a neutral fermion discussed in Sec.~\ref{sec:WprimeN}. Thus, we only need to adapt our findings knowing that $g_v$ and $g_a$ in Sec.~\ref{sec:WprimeN}, are now related as $g_v=g_a=g_R/\sqrt{2}\, U_R$, where $U_R$ is the PMNS matrix for the RH leptons. Thus, in this regime we find
\begin{subequations}
\begin{equation}
 \Delta a_\mu(N,W_R) = 2.2 \times 10^{-11} \left(\frac{g_R}{g_L}\right)^2 \left( \frac{1\TeV}{M_{W_R}}\right)^2 \sum_{N} |U_{R\mu N}|^2,
\end{equation}and
\begin{equation}
{\rm BR }(\MEG) \simeq 5 \times 10^{-8} \left(\frac{g_R}{g_L}\right)^4 \left(\frac{1\TeV}{M_{W_R}}\right)^4\times \sum_{N} |U_{ReN}^{\ast} U_{R\mu N}|^2,
\end{equation}
\end{subequations}
where the sum in $N$ runs over the three RH neutrino species. 

Notice that in order to explain $g-2$ the $W_R$ mass would have to lie below 1 TeV, but such low masses are excluded by the LHC.

Moreover, the current limit on $\mathrm{BR}(\MEG)$ enforces the product $|U_{ReN}^{\ast} U_{R\mu N}|$ to be less than $5\times 10^{-3}$ for $M_{W_R}$ masses at the TeV scale. Thus, one cannot reconcile possible signals in $g-2$ and $\MEG$.

(ii) $M_{W_R} \simeq M_{N_R}$

In this limit the results in a more general setting were derived in Eq.~\eqref{eq:approxamu_W2} and Eq.~\eqref{eq:BR4_1}. After computing the coupling constants and matching the vector and axial-vector couplings to the Left-Right charged current, as done above, we obtain,
\begin{subequations}
\begin{equation}
  \Delta a_\mu(N,W_R) \simeq 2.1 \times 10^{-11} \left(\frac{g_R}{g_L}\right)^2 \left( \frac{1\TeV}{M_{W_R}}\right)^2 \sum_{N} |U_{R\mu N}|^2,
\end{equation}
and
\begin{equation}
{\rm BR} (\MEG) \simeq 2 \times 10^{-7} \left(\frac{g_R}{g_L}\right)^4 \left(\frac{1\TeV}{M_{W_R}}\right)^4\times \sum_{N} |U_{ReN}^{\ast} U_{R\mu N}|^2,
\end{equation}
\end{subequations}which agrees well with the result in~\cite{Cirigliano:2004mv}.

The conclusion is similar to the previous regime, however ${\rm BR (\MEG)}$ is about one order of magnitude larger, yielding tighter constraints on the $W_R$ mass. 

As for the other LFV observables within the mass regime we get,

\begin{equation}
\mathrm{CR}(\mu-e)= 2\times 10^{-7} \left(\frac{g_R}{g_L}\right)^4 \left(\frac{1\TeV}{M_{\delta_R^{++}}}\right)^4\times \sum_{N} |U_{ReN}^{\ast} U_{R\mu N}|^2 \left[\log\left(\frac{M_{\delta_R^{++}}^2}{m_\mu^2}\right)\right]^2
\end{equation}with,

\begin{equation}
\mathrm{BR}(\mu \to 3e) \simeq 300 \times \mathrm{CR}(\mu-e)
\end{equation}for the Aluminum nucleus, where $M_{\delta_R^{++}}$ is the mass of the doubly charged scalar.

From the equations above, if $M_{\delta_R^{++}} \sim M_{W_R}$ it is clear the $\mu-e$ conversion and $\MEG$ feature similar rates. So what really dictates which observable offers the best probe is the experimental limit, which currently favors $\MEG$. For other interesting discussions of LFV in Left-Right models focused on collider features see~\cite{Das:2012ii,AguilarSaavedra:2012fu}.

{\bf Existing Limits}

There are several bounds applicable to the minimal Left-Right model. The most important ones arise for searches at the LHC for a RH current mediated by the $W_R$ gauge boson. Using the dijet plus dilepton data lower mass, bounds on the $W_R$ mass were obtained and those lie at the $3-4$~TeV scale depending on the RH neutrino mass. Moreover, there are others stemming from the exclusive use of either dilepton or dijet data which are not as restrictive as the former, but are rather insensitive to the RH neutrino mass, providing complementary limits. There are also competitive limits arising from meson oscillation studies. All these bounds can be found in~\cite{Nemevsek:2011hz,Helo:2013ika,Fowlie:2014mza,Aad:2014cka,Parida:2014dla,Khachatryan:2014dka,
Gao:2015irw,Deppisch:2015cua,Patra:2015bga,Helo:2015ffa,
Gluza:2015goa,Brehmer:2015cia,Lindner:2016lpp,Dev:2013oxa,Gluza:2016qqv}. Projected limits using the Large Hadron Electron Collider 
provide a complementary and very promising probe to the Left-Right symmetry for RH neutrino masses between $500-1$~TeV~\cite{Mondal:2016kof,Lindner:2016lxq,Queiroz:2016qmc}. These rule out the possibility of explaining $g-2$ but perfectly allow a signal in LFV observables in the near future.

\subsection{Two Higgs Doublet Model}

The addition of a scalar doublet is perfectly possible since it does not disturb the parameter $\rho$ determined in electroweak precision tests. In an  $SU(2)_L \times U(1)_Y$ gauge theory with $N$ scalar multiplets $\phi_i$, the $\rho$ parameter at tree-level is found to be~\cite{Langacker:1980js}
\begin{equation}
\rho = \frac{{\displaystyle \sum_{i=1}^n} \left[
I_i \left( I_i+1 \right) - \frac{1}{4}\, Y_i^2 \right] v_i}
{{\displaystyle \sum_{i=1}^n}\, \frac{1}{2}\, Y_i^2 v_i},
\label{jduei}
\end{equation}
where $I_i$ is the weak isospin, $Y_i$ the weak hypercharge, and $v_i$ are the VEVs of the neural fields. Since $\rho$ is measured to be nearly one~\cite{Olive:2016xmw}, we conclude that $SU(2)_L$ doublets  with $Y = \pm 1$ along with singlets with $Y = 0$ do not alter the value of $\rho$, knowing that $I \left( I+1 \right) = \frac{3}{4}\, Y^2$.

Thus, enlarging the SM with a scalar doublet is a natural framework, which is known as Two Higgs Doublet Model (2HDM)~\cite{Lee:1973iz}. The 2HDM has a rich phenomenology and possesses several nice features such as links to Supersymmetry~\cite{Haber:1984rc}, axion models~\cite{Peccei:1977hh,Kim:1986ax,Dasgupta:2013cwa,Alves:2016bib}, baryogenesis~\cite{Trodden:1998qg,Joyce:1994zt,Funakubo:1993jg,Davies:1994id,
Cline:1995dg,Laine:2000rm,Fromme:2006cm,Kozhushko:2011ea,
Cline:2011mm,Tranberg:2012qu,Ahmadvand:2013sna,Mou:2015aia} and even furnishes an environment for dark matter~
\cite{Bai:2012nv,Drozd:2014yla,Wang:2014elb,Adulpravitchai:2015mna,Liu:2015oaa}, among others~\cite{Guo:2016ixx,Chiang:2016vgf}. In what follows, we will be restricted to the non-supersymmetric 2HDM, see~\cite{Djouadi:2005gj} for an extensive discussion. Generally speaking, there are several types of 2HDMs (see~\cite{Branco:2011iw} for an excellent review). However, here we will focus on the type~III since there is a window for LFV~\cite{Davidson:2010xv,Sierra:2014nqa,Dorsner:2015mja}. In what follows, we will be assuming the Higgs potential to be CP invariant. That said, the two Higgs doublets are
\begin{equation}
\label{Eq:2HDMdoublets}
\Phi_1=\left(\begin{array}{c}
G^+ \\ {1\over\sqrt{2}}\left(v+H_1^0+iG^0\right)\end{array}
\right)\,,\qquad
\Phi_2=\left(\begin{array}{c}
H^+ \\ {1\over\sqrt{2}}\left(H_2^0+iA\right)\end{array}
\right)\,,
\end{equation}where the fields $G^{\pm},G^0$ are Goldstone bosons, $A$ is a CP-odd scalar, $H^\pm$ charged scalars, and finally we have $v= 246$ GeV. These doublets lead to a scalar potential which is found to be~\cite{Davidson:2005cw,Haber:2015pua}
\begin{eqnarray}
V&=& M_{11}^2\Phi_1^\dagger \Phi_1+M^2 \Phi_2^\dagger \Phi_2
-[M_{12}^2 \Phi_1^\dagger \Phi_2+{\rm h.c.}]
\nonumber\\[6pt]
&&\quad +1/2\Lambda_1(\Phi_1^\dagger \Phi_1)^2
+1/2\Lambda_2(\Phi_2^\dagger \Phi_2)^2
+\Lambda_3(\Phi_1^\dagger \Phi_1)(\Phi_2^\dagger \Phi_2)
+\Lambda_4(\Phi_1^\dagger \Phi_2)(\Phi_2^\dagger \Phi_1)
\nonumber\\[6pt]
&&\quad +\left\{1/2\Lambda_5(\Phi_1^\dagger \Phi_2)^2
+\big[\Lambda_6(\Phi_1^\dagger \Phi_1)
+\Lambda_7(\Phi_2^\dagger \Phi_2)\big]
\Phi_1^\dagger \Phi_2+{\rm h.c.}\right\}.
\label{Eq:2HDMpotential}
\end{eqnarray}
This scalar potential, with the symmetry breaking pattern of the two scalar doublets in Eq.~\eqref{Eq:2HDMdoublets}, leads to a mixing between the neutral components which reads
\begin{eqnarray}
h &=& H_1^0~s_{\beta-\alpha}+ H_2^0\,c_{\beta-\alpha}\,,\nonumber\\
H &=& H_1^0~c_{\beta-\alpha}-
H_2^0\,s_{\beta-\alpha}\,,
\label{Eq:2HDMneutralHiggses}
\end{eqnarray}with $\cos_{\beta-\alpha}\equiv \cos [(\beta-\alpha)]$ and $\sin_{\beta-\alpha}\equiv \sin [(\beta-\alpha)]$, where~\cite{Davidson:2005cw}
\begin{eqnarray} \label{exactbma}
\sin_{2(\beta-\alpha)}&=&{-2\Lambda_6 v^2\over m_H^2-m_h^2}.
\end{eqnarray}
Moreover, the scalar potential gives rise to the scalar masses
\begin{subequations}\label{Eq:2HDMmasses}
\begin{align}
m_{H^\pm}^2 & =  M^2 + \frac{v^2}{2} \Lambda_3, \\
m_A^2 - m_{H^\pm}^2&=  -\frac{v^2}{2} (\Lambda_5 - \Lambda_4), \\
m_H^2 + m_h^2-  m_A^2 & =  + v^2 (\Lambda_1 + \Lambda_5), \\ 
(m_H^2 - m_h^2)^2 & =  
[m_A^2 + (\Lambda_5 - \Lambda_1)v^2]^2 + 4 \Lambda_6^2 v^4.
\end{align}
\end{subequations}
The Yuwaka Lagrangian of the 2HDM hosts the key information for the $g-2$ and $\MEG$ observables and it is found to be
\begin{eqnarray}
 \label{Eq:2HDMYukawa}
\mathcal{L}_Y&=&  
\overline{Q}_{j}  \widetilde{\Phi}_1 K^*_{ij} y^{u}_i u_{Ri}   +
\overline{Q}_i \Phi_1  y^{d}_{i}  d_{Ri}
+\overline{L}_i \Phi_1  y^{e}_{i}  e_{Ri}
\nonumber \\
&&
+\overline{Q}_{i}  \widetilde{\Phi}_2 [K^\dagger w^{u}]_{ij} u_{Rj}   +
\overline{Q}_i \Phi_2 [w^{d}]_{ij} d_{Rj}
+\overline{L}_i \Phi_2 [w^{e}]_{ij} e_{Rj}
+{\rm h.c.}\,,\nonumber\\
\end{eqnarray}where $\widetilde{\Phi}_i = i \sigma_2  \Phi_i^*$, $Q,L$ are the quark and lepton $SU(2)_L$ doublets, $K$ is the CKM matrix, $y$ and $w$ are the Yukawa couplings, and $i,j=1,2,3$ run through the fermion generations. As in the SM, $y$ is flavor conserving with
\begin{equation}
y_{ij}=\sqrt{2} m_f/v \delta_{ij},
\end{equation}where $m_f$ is the fermion mass, but $w$ can have non-zero flavor changing entries relevant for $\MEG$. From Eq.~\eqref{Eq:2HDMYukawa} we get,
\begin{eqnarray}
\label{Eq:2HDMYukawasimple}
\mathcal{L}_Y& \supset & \bar{e} \frac{ 1  }{\sqrt{2}} 
\left[ y^e(P_R + P_L) s_{\beta-\alpha}+
(w^e P_R+{w^e}^\dagger P_L)c_{\beta-\alpha}\right]e\, h + \nonumber \\
&&+ \bar{e} 
\frac{ 1  }{\sqrt{2}}
\left[ y^e(P_R + P_L)c_{\beta-\alpha}-
(w^e P_R+{w^e}^\dagger P_L)s_{\beta-\alpha}\right]e\, H+ \nonumber\\
&& +\frac{i}{\sqrt{2}}\bar{e}\left( w^e P_R-{w^{e^\dagger}} P_L\right)e\,A+ \nonumber\\
&&+ \bar{ \nu} \left( w^e P_R \right) e\, H^+ + {\rm h.c.}
\end{eqnarray}
Eq.~\eqref{Eq:2HDMYukawasimple} gathers all the information needed to compute $\Delta a_\mu$ and $\MEG$ in the 2HDM type~III at the one-loop level.

\begin{figure}[!t]
\centering
\includegraphics[scale=0.7]{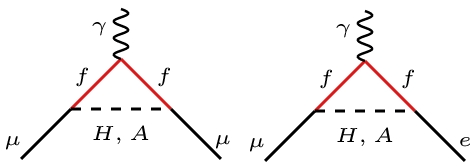}
\caption{Feynman diagrams contributing to $g-2$ and $\MEG$ at one-loop level.}
\label{plot2HDMdiagram} 
\end{figure}

\subsubsection{Results}
The Feynman diagrams that lead to corrections to $g-2$ and ${\rm BR}(\MEG)$ are displayed in Fig.~\ref{plot2HDMdiagram}. The $g-2$ contribution has already been obtained in Eq.~\eqref{eq:Delta_a_Neutral_Scalar_approx}, and with a straightforward replacement of the (real) scalar couplings one finds
\begin{equation} \label{eq:2HDMamu}
  \begin{split}
	\Delta a_\mu {\rm (H+A)} \simeq \frac{1}{4\pi^2}\frac{m_\mu^2}{m_\phi^2} \sum_i \left[\left(y^H_{2i}\right)^2 \left( \frac{1}{6} - \epsilon_f\left(\frac{3}{4} + \log(\epsilon_f\lambda_H)\right)\right)\right. + \\
	 + \left. |y_{2i}^A|^2 \left( \frac{1}{6} + \epsilon_f\left(\frac{3}{4} + \log(\epsilon_i\lambda_A)\right) \right)  \right],
  \end{split}
\end{equation}where $i=1,2,3$ generally runs through all generations of charged leptons of mass $m_f$ if one consider non-vanishing flavor violating mixings. Here, $\epsilon_f= m_f/m_{\mu}$, $\lambda_{\phi}=m_{\mu}/m_{\phi}\, (\phi=H,A)$ and,
\begin{equation}
\begin{aligned}
 y_{ij}^{H}& \equiv  \frac{1}{\sqrt{2}}\left( y^e_{ij} \cos_{\beta-\alpha} -w^e_{ij} \sin_{\beta-\alpha}\right),\\
 y_{ij}^{A}& \equiv  i \frac{w^e_{ij}}{\sqrt{2}},
\end{aligned}
\end{equation}which are easily identified from Eq.~\eqref{Eq:2HDMYukawasimple}. Keeping in mind that in our notation $y^e_{ij}$ is the Yukawa coupling appearing in front of the $\bar{e_i}e_j\, H$  interaction, whereas $w^e_{ij}$ refers to the Yukawas of the interactions $\bar{e_i}e_j\, A$. 

Eq.~\eqref{eq:2HDMamu} encompasses the one-loop contributions of neutral scalars in the 2HDM type~III. Since the singly charged scalar correction is negative and rather suppressed, Eq.~\eqref{eq:2HDMamu} represents basically the overall prediction of the model. Moreover, the Higgs may also correct $g-2$ differently than in the SM. Such a correction can easily be extracted from the first term in Eq.~\eqref{Eq:2HDMYukawasimple}, with $y_{ij}^{h} \equiv  1/\sqrt{2} ( y^e_{ij} \sin_{\beta-\alpha} +w^e_{ij} \cos_{\beta-\alpha} )$, when plugged in Eq.~\eqref{eq:2HDMamu} along with the $y_{ij}^{H}$ term. However, notice that in the decoupling limit, i.e.~$\sin_{\beta-\alpha} \sim 1$ and $\cos_{\beta-\alpha} \sim -\Lambda_6 v^2/M^2 + O(v^4/M^4)$, the Higgs contribution is negligible since it will scale with the Yukawa coupling $y^e_{22}$, which is proportional to $m_{\mu}/v$. Thus, the leading corrections stem from $H$ and $A$. That said one gets
\begin{equation}
\Delta a_{\mu} (H+A) = 4.7 \times 10^{-11} \left(\frac{w^e_{2j}}{0.1}\right)^2\left(\frac{100 GeV}{m_{\phi}^2}\right)^2.
\label{eq:2HDMamusimplified}
\end{equation}
As for the ${\rm BR}(\MEG)$ we find,
\begin{equation}
\label{eq:2HDMmutoe}
    \mathrm{Br}(\MEG) \approx \frac{3(4\pi)^3 \alpha_\mathrm{em}}{4 G_F^2}\left( |A_{e\mu}^M|^2 + |A_{e\mu}^E|^2 \right),
\end{equation}where,
  \begin{equation}
  \begin{split}
    A_{e\mu}^M= \frac{1}{16\pi^2 m_\phi^2 }\sum_i \left\lbrace y_{2i}^{H} y_{1i}^{H} \left[ \frac{1}{6}  - \epsilon_i \left(\frac{3}{2} +\log(\epsilon_f^2 \lambda^2) \right)\right]\right. \\
    \left. \hfill +y_{2i}^{A}y_{1i}^{A}  \left[ \frac{1}{6}  + \epsilon_f \left(\frac{3}{2} +\log(\epsilon_i^2 \lambda^2) \right)\right] \right\rbrace,
  \end{split}
  \label{eq:2HDMmutoe1}
  \end{equation}and 
  \begin{equation}
  \begin{split}
    A_{e\mu}^E= \frac{1}{16\pi^2 m_\phi^2 }\sum_f \left\lbrace y_{2i}^{H}y_{1i}^{A} \left[ \frac{1}{6}  - \epsilon_i \left(\frac{3}{2} +\log(\epsilon_f^2 \lambda^2) \right)\right]\right.  \\
    \left. \hfill - y_{1i}^{H}y_{2i}^{A}  \left[ \frac{1}{6}  + \epsilon_f \left(\frac{3}{2} +\log(\epsilon_i^2 \lambda^2) \right)\right] \right\rbrace.
  \end{split}
  \label{eq:2HDMmutoe2}
  \end{equation}
In summary, Eqs.~(\ref{eq:2HDMmutoe}-\ref{eq:2HDMmutoe2}) represent the exact results for the $\MEG$ contribution in the 2HDM type~III at the one-loop level. Our results are more general than those presented in~\cite{Davidson:2016utf} which focused on the decoupling limit, i.e.~when $\sin_{\beta-\alpha} \rightarrow 1$. Discussions of $\MEG$ and $g-2$ were also presented in~\cite{Diaz:2000cm,Diaz:2002uk}. 

\begin{figure}[t]
\centering
\includegraphics[width=0.45\textwidth]{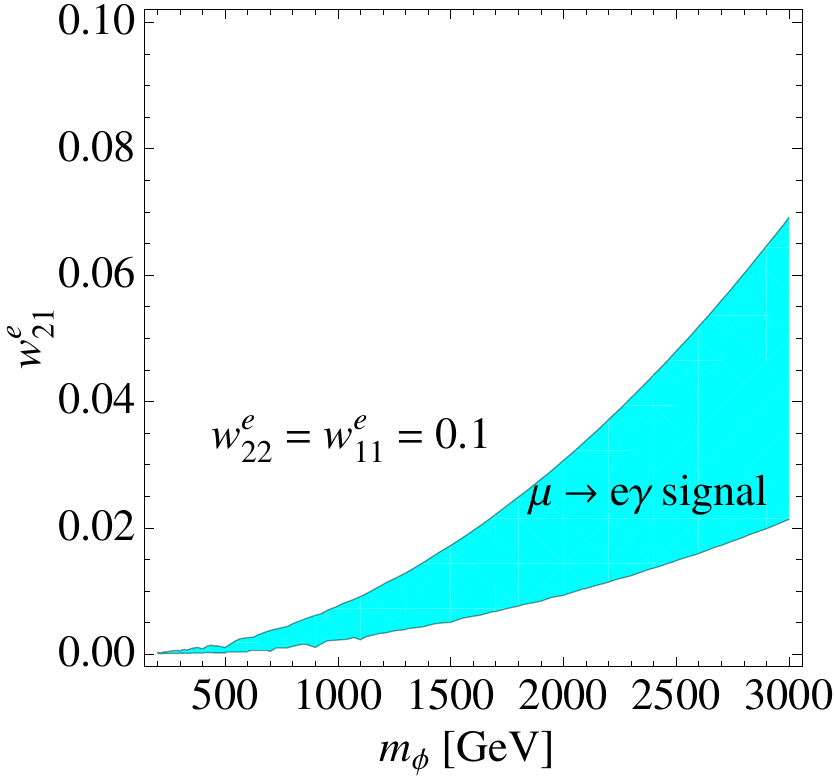}
\includegraphics[width=0.45\textwidth]{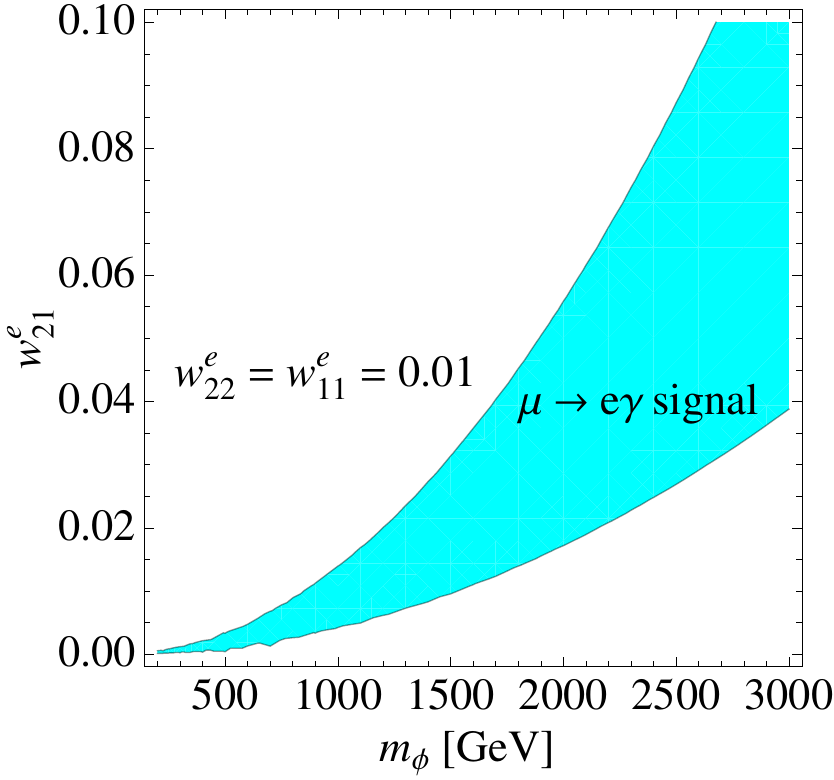}
\caption{Region of parameter space in which the 2HDM type~III could address a signal in $\MEG$ with ${\rm BR(\MEG) = 4.2\times 10^{-13}-4\times 10^{-14}}$ for a specific set of couplings with $w^e_{22}=w^e_{11}=0.1$ in the left panel, and $w^e_{22}=w^e_{11}=0.01$.}
\label{plot2HDM} 
\end{figure}

It is clear that the degree of complementarity between $g-2$ and $\MEG$ is rather arbitrary since they depend on the values used for the Yukawa couplings. In Fig.~\ref{plot2HDM} we show that the 2HDM type~III can accommodate a signal in the $\MEG$ decay, delimited by the blue region, while avoiding $g-2$ constraints for two different choices of couplings. One can easily see that for these choices $\Delta a_{\mu}$ is very small using Eq.~\eqref{eq:2HDMamusimplified}. 

As for the other LFV observables such as $\mu \to 3e$ and $\mu-e$ conversion, the rule of thumb relations are expected to be valid in this case, with
${\rm CR}(\mu\, \mbox{Al}-e\, \mbox{Al}) \sim 1/350\, {\rm BR}(\MEG)$, and $ {\rm BR}(\mu \to 3e) \sim 1/160\,  {\rm BR}(\MEG)$. Therefore, $\MEG$ decay is indeed the most effective way to probe the model, with nuclear $\mu-e$ conversion taking lead far in the future when a sensitivity of $10^{-17}-10^{-18}$ is reached for the conversion rate.

There are caveats in our reasoning which are worth pointing out: There are two-loop diagrams which also yield sizable contributions for some regions of the parameter space involving quarks and gauge bosons, as already pointed out in~\cite{Bjorken:1977vt,Langacker:1988cm,Leigh:1990kf} and discussed further in~\cite{Broggio:2014mna,Han:2015yys,Davidson:2016utf,Cherchiglia:2016eui} which may affect our results; There are arguments concerning vacuum stability which might be relevant to constrain the set up we just described, see~\cite{Ferreira:2009wh,Haber:2010bw,Belanger:2012tt,Dumont:2014wha,Ferreira:2015rha,
Bernon:2015wef}, 

{\bf Existing Limits}

LEP limits the charged scalar to be heavier than $78$~GeV~\cite{Beringer:1900zz}. There are bounds stemming from the kaon mass difference system which can rule out the charged scalar mass up to $100-500$~GeV~\cite{Cho:2017jym}. There are also limits based on searches for a heavy Higgs in this model which, depending on the Yukawa couplings, may excluded heavy Higgs masses up to $200$~GeV~\cite{Primulando:2016eod}. All these limits are not very restrictive and do not have an impact on the possibility of observing a signal in $\MEG$ in the near future.

\subsection{Scotogenic Model}

The scotogenic model is a scenario proposed in~\cite{Ma:2006km,Ma:2012if}, in which neutrinos acquire masses via their interactions with dark matter at the one-loop level. One simply extends the SM by a number of singlet fermions, conventionally dubbed RH neutrinos $N_R^i$, and a second scalar $SU(2)_L$ doublet $\eta$. In addition, a discrete $\mathbb{Z}_2$ symmetry is imposed under which the SM fields are even and both types of new fields are odd. This symmetry guarantees a number of important facts to hold true in the model: First, the new doublet does not acquire a VEV and consequently no tree-level neutrino masses arise. Second, there is no mixing between the new scalar particles and the SM Higgs. Lastly, the lightest particles charged under $\mathbb{Z}_2$ is either a fermion or a neutral scalar, making it stable.\footnote{See~\cite{Kashiwase:2013uy,Klasen:2013jpa,Toma:2013zsa,Vicente:2014wga,Lindner:2016kqk} for some recent studies.} The Lagrangian of the model is given by,
\begin{equation}\label{eq:ScotogenicLagrangian}
  \mathcal{L} \supset - \frac{1}{2} M_i \overline{N_R^c}^i N_R^i - y_{i\alpha} \overline{N_R^i} \widetilde{\eta}^\dag L_L^\alpha + \mathrm{h.c.} - V(\phi, \eta),
\end{equation}
where the scalar potential is
\begin{equation}
\begin{aligned}
  V(\phi,\eta) =&
  m_{\phi}^2\phi^\dag\phi+m_\eta^2\eta^\dag\eta+
  \frac{\lambda_1}{2}\left(\phi^\dag\phi\right)^2+
  \frac{\lambda_2}{2}\left(\eta^\dag\eta\right)^2+
  \lambda_3\left(\phi^\dag\phi\right)\left(\eta^\dag\eta\right)\\
  &+
  \lambda_4\left(\phi^\dag\eta\right)\left(\eta^\dag\phi\right)+
  \frac{\lambda_5}{2}\left[\left(\phi^\dag\eta\right)^2+
  \left(\eta^\dag\phi\right)^2\right].
\end{aligned}
\end{equation}
Indeed, we observe that unless $\eta$ develops a VEV, neutrinos are massless at tree-level. Upon electroweak symmetry breaking, the scalar sector contains, besides a Higgs boson shifted by its VEV $v$, four scalar degrees of freedom $\eta_R, \eta_I,$ and $\eta^\pm$ with masses
\begin{subequations}
  \begin{align}
    m_{\pm}^2 &= m_\eta^2 + v^2 \lambda_3,\\
    m_{R}^2 &= m_\eta^2 + v^2 (\lambda_3 + \lambda_4 + \lambda_5),\\
    m_{I}^2 &= m_\eta^2 + v^2 (\lambda_3 + \lambda_4 - \lambda_5).
  \end{align}
\end{subequations}
One may calculate the one-loop correction to the neutrino masses, which amounts to~\cite{Ma:2006km,Merle:2015ica}
\begin{equation}
  \mathcal{M}^{(\nu)}_{\alpha\beta} = \sum_{k=1}^3 \frac{y_{k\alpha} y_{k\beta} M_k }{32\pi^2} \left[ \frac{m_R^2}{m_R^2-M_k^2} \log\left(\frac{m_R^2}{M_k^2} \right) -  \frac{m_R^2}{m_I^2-M_k^2} \log\left(\frac{m_I^2}{M_k^2} \right) \right].
\end{equation}
Note that, in order to have at least two massive light neutrinos, we need at least two neutral fermions $N_R^i$.

The Yukawa interaction in Eq.~\eqref{eq:ScotogenicLagrangian} comprises an interaction of the form $y_{i\alpha}^* \overline{\ell_L^\alpha} \eta^- N_R^i$, very similar to our Eq.~\eqref{eq:singlyscalar}. Thus, the model gives rise to both LFV decays and a correction to $g-2$, the latter is however negative, as discussed below. The decay $\MEG$ is conventionally parametrized as~\cite{Vicente:2014wga}
\begin{equation}
\mathrm{BR}\left(\ell_\alpha\to\ell_\beta\gamma\right)=
\frac{3(4\pi)^3 \alpha_{\mathrm{em}}}{4G_F^2} 
|A_{D,\, \beta\alpha}|^2
\mathrm{Br}\left(\ell_\alpha\to\ell_\beta\nu_\alpha\overline{\nu_\beta}\right) ,
\end{equation}
with the amplitude
\begin{equation}
A_{D,\, \beta\alpha} = \sum_{k=1}^3\frac{y_{k\beta}^*y_{k\alpha}} {2(4\pi)^2}\frac{1}{m_{\pm}^2} F_2(x_i) \, ,
\label{eq:AD}
\end{equation}
and $x_i= M_{N_i}^2/m_{\eta^\pm}^2$. 

Starting from our exact expression~\eqref{eq:BR2} for the charged scalar, using $g_p = - g_s$, and approximating $m_e \ll m_\mu \ll m_{\eta^+/N}$, one finds,
\begin{equation}
F_2(x) =\frac{1-6x+3x^2+2x^3-6x^2 \log x}{6(1-x)^4},
\end{equation}
in agreement with Refs.~\cite{Ma:2001mr,Toma:2013zsa}. For the contribution to $g-2$, we may consult Eq.~\eqref{eq:Delta_a_Singly_Scalar_approx} to observe that the contribution is negative. Let us therefore focus on LFV in the scotogenic model.

{\bf Existing Limits}

\begin{figure}[t]
   \centering
   \noindent\makebox[\textwidth]{
   \begin{subfigure}[b]{.45\textwidth}
      \centering
      \includegraphics[width=\textwidth]{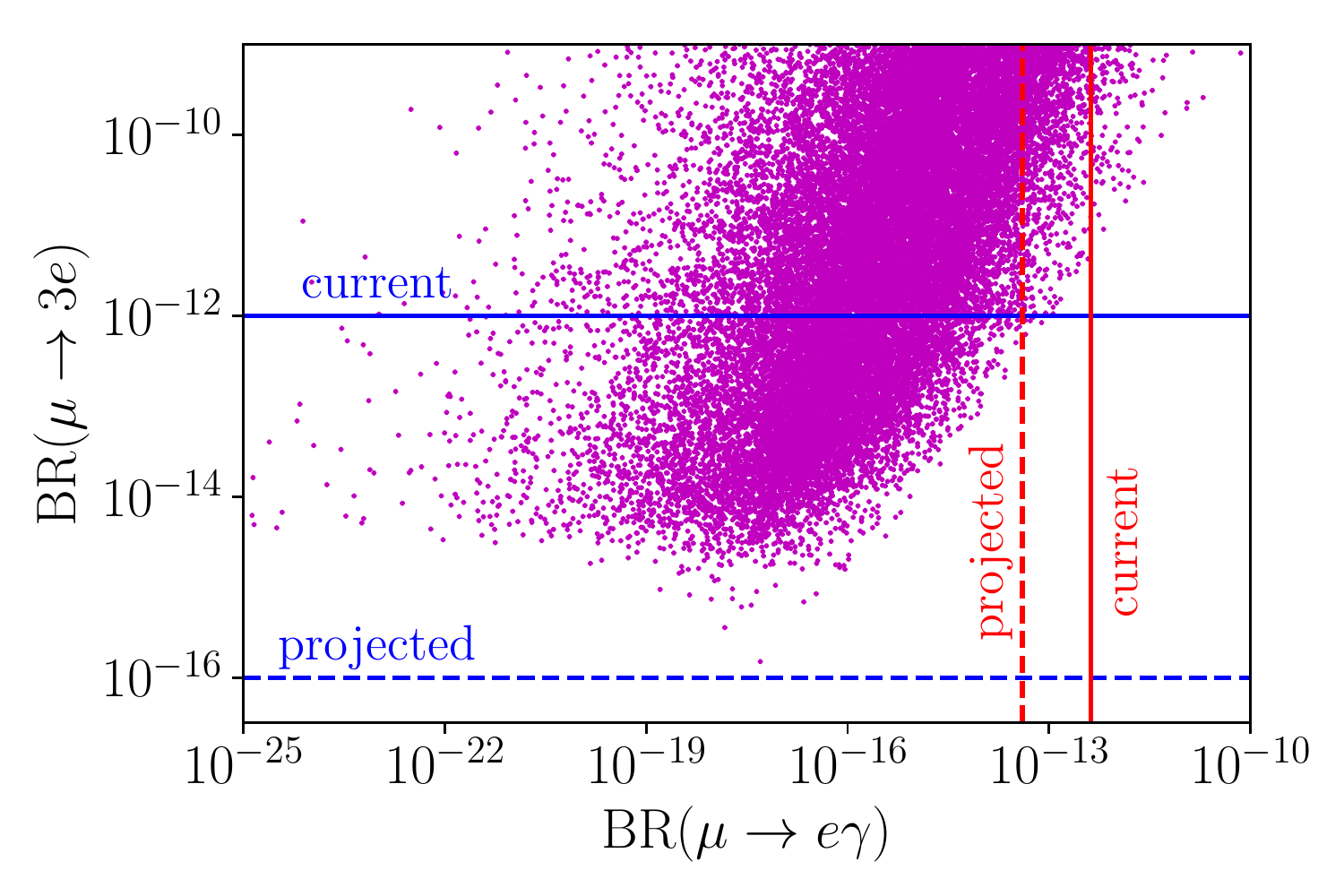}
	  \subcaption{\label{fig:scotogenic_mu3eNO}normal ordering}
   \end{subfigure}
   \hfill
   \begin{subfigure}[b]{.45\textwidth}
     \includegraphics[width=\textwidth]{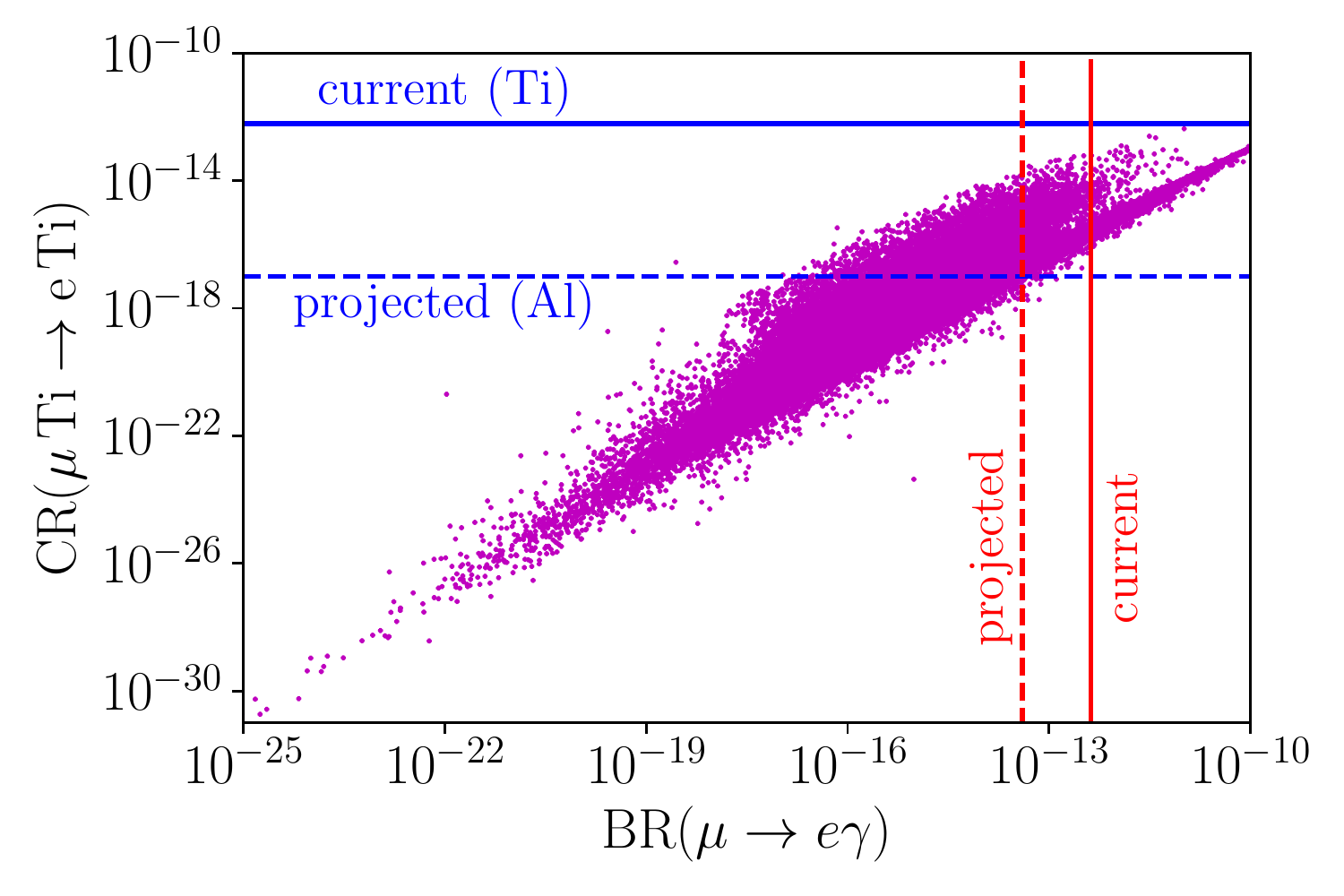}
     \subcaption{normal ordering}
   \end{subfigure}
   }\\
    \vspace{5mm}
   \noindent\makebox[\textwidth]{
   \begin{subfigure}[b]{.45\textwidth}
      \centering
      \includegraphics[width=\textwidth]{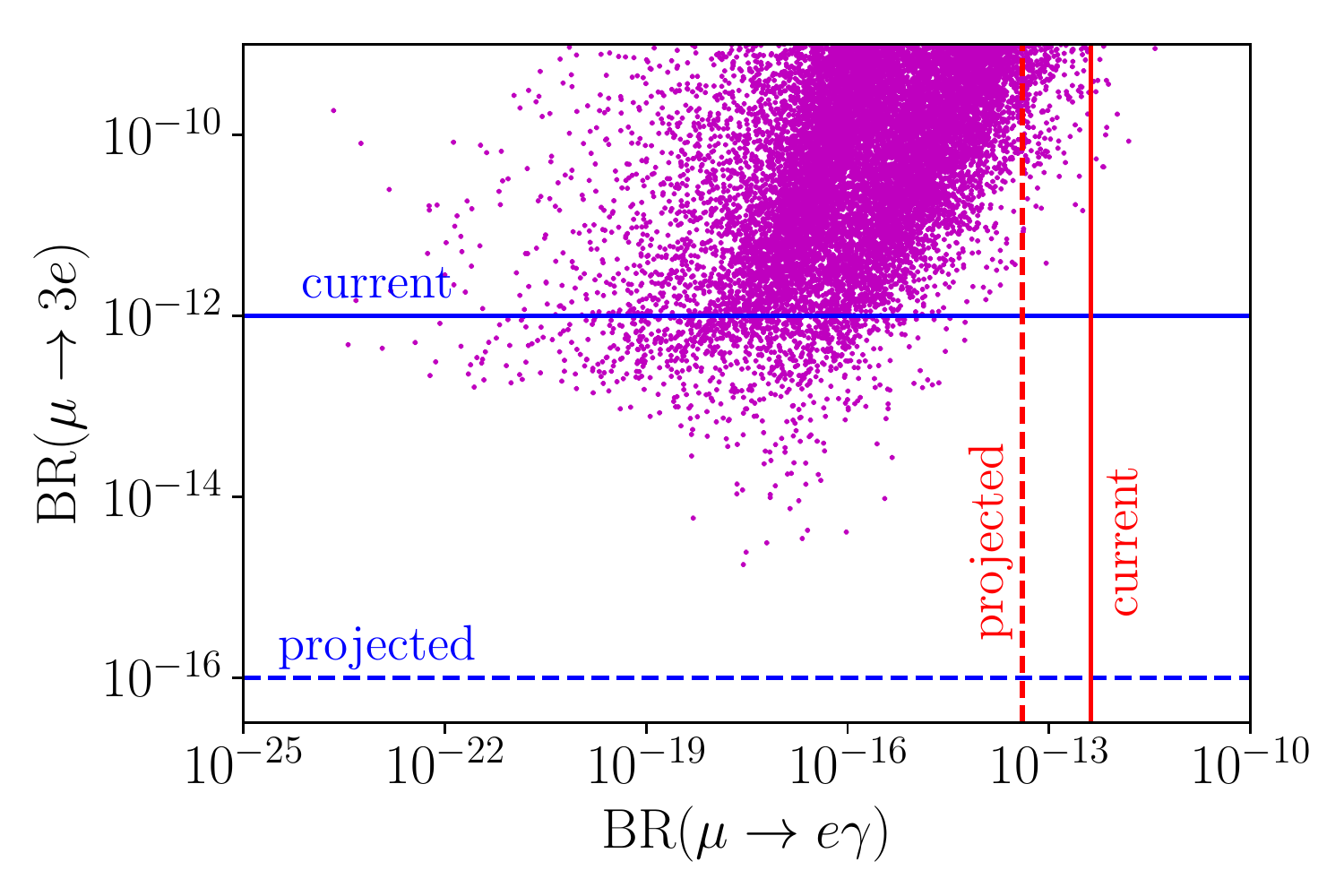}
      \subcaption{\label{fig:scotogenic_mu3eIO}inverted ordering}
   \end{subfigure}
   \hfill
   \begin{subfigure}[b]{.45\textwidth}
      \centering
      \includegraphics[width=\textwidth]{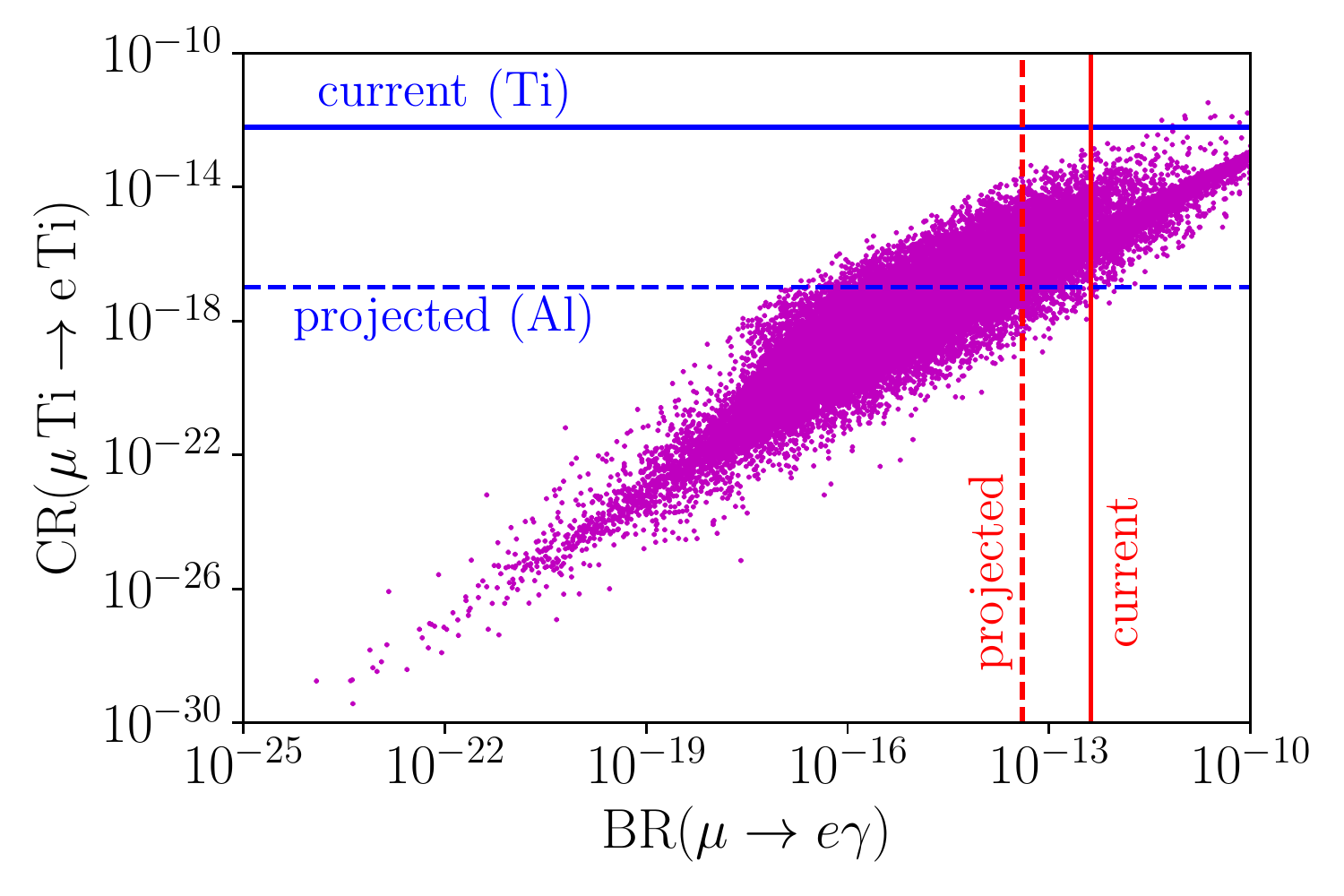}
      \subcaption{inverted ordering}
   \end{subfigure}
   }
   \caption{\label{fig:results_scotogenic}Results and bounds for the scotogenic model for different LFV violation processes in normal as well as inverted neutrino mass ordering. Note that the projected limit employs a different nucleus (Al) than the current limit (Ti) on $\mu\, N \to e\, N$; however, we only show the results for one nucleus (Ti) since they deviate only marginally for the different nuclei.}
\end{figure}

Owing to the $\mathbb{Z}_2$ symmetry and the resulting absence of mixing between the scalar doublet and the SM Higgs, it has been argued in the literature that the strongest constraints on the model are indeed due to LFV. Following Ref.~\cite{Vicente:2014wga}, we find that it is the decay $\mu \to 3 e$ and not $\MEG$ that will probe most of the available parameter space in the near future. For completeness, we also consider the conversion of a muon to an electron in a nucleus. The relevant expressions for this decay read:
\begin{subequations}\allowdisplaybreaks
\begin{align}
	\mathrm{BR}\left(\ell_\alpha\to \ell_\beta \overline{\ell_\beta}\ell_\beta\right) =& \frac{3(4\pi)^3 \alpha_{\mathrm{em}}}{8G_F^2} \left[ |A_{ND,\, \beta\alpha}|^2  +  |A_{D,\, \beta\alpha}|^2 \left( \frac{16}{3} \log\left(\frac{m_\alpha}{m_\beta}\right) - \frac{22}{3}\right) + \right. \nonumber\\
	& + \left. \frac{1}{6} |B|^2 + \left(-2 A_{ND,\, \beta\alpha} A_{D,\, \beta\alpha}^* + \frac{1}{3} A_{ND,\, \beta\alpha} B^* - \right.\right.\nonumber \\
	& - \frac{2}{3} A_{D,\, \beta\alpha} B^* +  \mathrm{c.c.} \bigg) \bigg]\times \mathrm{Br}\left(\ell_i\to\ell_\beta\nu_i\overline{\nu_j}\right) ,\label{eq:BRmu3e_scoto}\\
	\mathrm{CR}\left(\ell_\alpha\, N\to \ell_\beta \, N\right) =& \frac{p_\beta E_\beta m_\alpha^3 G_F^2 \alpha_\text{em}^3 Z_\text{eff}^4 F_p^2}{8 \pi^2 Z 
	\Gamma_\text{capt}} \left\lbrace \left| (Z+N) (g_{LV}^{(0)} + g_{LS}^{(0)}) + \right. \right. \nonumber\\
	& \hspace{17mm}+ \left.\left. (Z-N) (g_{LV}^{(1)} + g_{LS}^{(1)})\right|^2 + (L \leftrightarrow R)\right\rbrace.
\end{align}
\end{subequations}
with
\begin{subequations}
\begin{gather} \label{eq:AND}
	A_{ND,\, \beta\alpha} = \sum_{k=1}^3\frac{y_{k\beta}^*y_{k\alpha}} {6(4\pi)^2}\frac{1}{m_{\pm}^2} G_2(x_i) ,\\
	e^2 B = \frac{1}{(4\pi)^2m_{\pm}} \sum_{i,j=1}^3 \left( \frac{1}{2} D_1(x_i, x_j) y_{j_\beta}y_{j_\beta}^* y_{i\beta}^* y_{i\alpha} + \sqrt{x_i x_j} D_2(x_i,x_j) y_{j_\beta}^* y_{j_\beta}^* y_{i\beta}^* y_{i\alpha}  \right).
\end{gather}
\end{subequations}
We do not reproduce the numerical input values required for the nuclear matrix element involved in the $\mu - e$ conversion rate. The values of and details on the quantities $Z_\text{eff}$, $F_p$, $\Gamma_\text{capt}$, and the $g^{(i)}_{L/R \, S/V}$ can be found in Refs.~\cite{Vicente:2014wga,Arganda:2007jw}. The additional loop functions are given by
\begin{subequations}
\begin{align}
	G_2(x) =& \frac{1-9x+18x^2-11x^3+6x^3 \log x}{6(1-x)^4},\\
	D_1(x,y) =& \frac{-1}{(1-x)(1-y)} - \frac{x^2\log(x)}{(1-x)^2 (x-y)} - \frac{y^2\log(y)}{(1-y)^2 (y-x)},\\
	D_2(x,y) =& \frac{-1}{(1-x)(1-y)} - \frac{x\log(x)}{(1-x)^2 (x-y)} - \frac{y\log(y)}{(1-y)^2 (y-x)}.
\end{align}
\end{subequations}
We have combined all these results and, by means of an adapted Casas-Ibarra parametrization for the Yukawa couplings~\cite{Casas:2001sr}, we can scan over the remaining free parameters. Confining the RH neutrino and inert scalar masses to a range $[100\GeV,100\TeV]$ and the coupling $\lambda_5 < 10^{-8}$, we obtain Fig.~\ref{fig:results_scotogenic}.

We observe that first of all the current limits on $\MEG$ and $\mu\to 3e$ give very tight constraints on the available parameter space since most points lie above the current limits. Note in particular that generally $\mathrm{BR}(\mu\to 3e) > \mathrm{BR}(\MEG)$, and the approximation~\eqref{eq:Mu3eEstimate} fails. This fact becomes obvious when observing that $A_D$ and $A_{ND}$, as defined in Eq.~\eqref{eq:AD} and~\eqref{eq:AND},  are of the same order of magnitude and enter Eq.~\eqref{eq:BRmu3e_scoto} with the same order prefactors. Thus, we cannot expect the photonic dipole contribution to dominate, which is necessary for the approximation~\eqref{eq:Mu3eEstimate} to hold.
 
Simultaneously, the conversion $\mu \to e$ in a (Ti-) nucleus gives hardly any constraints at all with the current sensitivity.\footnote{We do not display the process with a different nucleus since the conversion rates turn out almost identical.} As a rule of thumb, we can estimate the conversion rate from the $\mathrm{BR}(\MEG)$ as 
\begin{equation}
	\mathrm{CR}(\mu \, \mathrm{Ti} \to e\, \rm Ti) \gtrsim 10^{-3}\, \mathrm{BR}(\MEG),
\end{equation}
in very rough agreement with Eq.~\eqref{eq:MuEconvEstimate}. 

In spite of the weak constrains at present, the right panels of Fig.~\ref{fig:results_scotogenic} tell us that this will change significantly with the projected sensitivities that allow one to probe a large fraction of the scotogenic model's viable parameter space. Returning to the left panels, Figs.~\ref{fig:scotogenic_mu3eNO} and~\ref{fig:scotogenic_mu3eIO}, we see that the next generation experiments searching for the decay $\mu \to 3 e$ will probe the remaining viable parameter space entirely, while $\MEG$ will give only a slight improvement. Similarly, the next generation experiments searching the nuclear reaction $\mu^-\, \text{Al} \to e^-\, \text{Al}$ are not as restrictive as $\mu \to 3e$.

Thus, we conclude that the next generation LFV experiments will either detect LFV or, in case of a null result, rule out most, if not all of the scotogenic model's parameter space.

\subsection{Zee-Babu Model}

The Zee-Babu model~\cite{Zee:1985rj,Zee:1985id,Babu:1988ki} is yet another scenario which realizes neutrino masses at the loop-level. However, in this particular scenario, neutrinos remain massless up to two-loop order, where the singly and doubly charged scalars added to the SM field content induce a small Majorana mass. The Lagrangian of the model reads
\begin{equation}
  \mathcal{L} \supset f_{\alpha\beta} \overline{\ell_L^c}^\alpha_a \epsilon^{ab} {\ell_L^c}^\beta_b h^+ + g_{\alpha\beta} \overline{e_R^c}^\alpha {e_R}^\beta k^{++} + \mathrm{h.c.}\, ,
\end{equation}
where $g$ is symmetric and $f$ is anti-symmetric under the exchange of $i \leftrightarrow j$, and $\psi^c$ denotes the charge conjugate spinor. While the above interaction can be made invariant under lepton number transformations if one assigns lepton numbers $L(h^+,k^{++})=2$, the scalar potential contains a coupling $\mu\, h^+ h^+ k^{--}$, which explicitly violates this symmetry by two units and induces Majorana neutrino masses. The expression for the neutrino mass matrix is
\begin{equation}\label{eq:massZeeBabu}
  \mathcal{M}_{\alpha\beta}^{(\nu)} = 16\, \mu\, f_{\alpha k}\, m_k\, g^*_{kl}\, I_{ln}\, m_n\, f_{n \beta}, \textrm{ where } I_{ln} \simeq \frac{\delta_{ln}}{(16\pi)^2 M^2} \frac{\pi^2}{3} \times \mathcal{O}(1) 
\end{equation}
and the factor of $\mathcal{O}(1)$ is due to a two-loop integral which must be evaluated numerically~\cite{McDonald:2003zj}. Furthermore, $m_n$ are the charged lepton masses and $M = \max (m_h, m_k)$. Using the fact that $f$ is anti-symmetric, one can derive the following relations:~\cite{Babu:2002uu,AristizabalSierra:2006gb,Nebot:2007bc}
\begin{subequations}\label{eq:ZeeBabuRelations}
\begin{gather}
	\frac{f_{e \tau}}{f_{\mu\tau}} = \frac{s_{12}c_{23}}{c_{12}c_{13}} + \frac{s_{13}s_{23}}{c_{13}}e^{-i \delta}, \quad  
	\frac{f_{e \mu}}{f_{\mu\tau}} = \frac{s_{12}s_{23}}{c_{12}c_{13}} + \frac{s_{13}c_{23}}{c_{13}}e^{-i \delta} \quad \text{ for normal ordering,}\\
	\frac{f_{e \tau}}{f_{\mu\tau}} = -\frac{s_{23}c_{13}}{s_{13}}e^{-i \delta}, \quad  
	\frac{f_{e \mu}}{f_{\mu\tau}} = \frac{c_{23}c_{23}}{s_{13}}e^{-i \delta} \quad \text{for inverted ordering},
\end{gather}
\end{subequations}
which imply that $f_{e\mu}\approx f_{e\tau} \approx f_{\mu\tau}/2$ (NO) or $|f_{e\mu}|\approx |f_{e\tau}|$ and $|f_{\mu\tau}| \approx |f_{e\tau}| \frac{s_{13}}{s_{23}}$ (IO).

Combining our results in Eqs.~\eqref{eq:Delta_a_Neutral_Scalar_approx}, \eqref{eq:Delta_a_Singly_Scalar_approx}, and \eqref{eq:Delta_a_Doubly_Scalar}, we may obtain an expression for the Zee-Babu fields' contribution to the $g-2$, as well as $\MEG$:\footnote{Note that there is an extra factor of $2$ coming with each insertion of $f$ and a factor of $1/2$ coming with each projector $P_{L/R}$.}
\begin{align}
  \Delta a_\mu (h^+,k^{++}) =& -\frac{m_\mu^2}{24\pi^2} \left( \frac{(f^\dag f)_{\mu\mu}}{m_h^2} + 4 \frac{(g^\dag g)_{\mu\mu}}{m_k^2}\right),\\
  \mathrm{BR} (\MEG) =& \frac{\alpha_\mathrm{em}}{48 \pi G_F^2} \left( \left|\frac{(f^\dag f)_{e\mu}}{m_h^2}\right|^2 + 16 \left|\frac{(g^\dag g)_{e\mu}}{m_k^2} \right|^2 \right),
\end{align}
valid in the limit $m_{h,k} \gg m_{\mu,e}$. This is in agreement with the results found in the literature~\cite{Babu:2002uu,AristizabalSierra:2006gb,Nebot:2007bc,Ohlsson:2009vk,Araki:2010kq,Schmidt:2014zoa,Herrero-Garcia:2014hfa,Okada:2014qsa,Herrero-Garcia:2014usa,Nomura:2016rjf,Nomura:2016ask,Chang:2016zll}. Note that the Zee-Babu setting provides no explanation for the $g-2$ anomaly since the contribution is negative, nevertheless we may use it to derive constraints on the the parameter space by enforcing its contribution to be below the error bars.

\begin{table}[t]
\centering
\makebox[10cm][c]{
	\begin{tabular}{|l|r|}
		\hline
		current limit 	&	 resulting bound\\
		\hline \hline 
		Lepton Flavor Violation & \\
		\hline
		$\mathrm{BR}(\mu \to e^+ e^- e^-) < 1 \cdot 10^{-12}$ 					& $|g_{e\mu} g_{ee}^*| < 2.3 \cdot 10^{-5}(m_k / \TeV)^2$\\
		$\mathrm{BR}(\tau \to e^+ e^- e^-) < 2.7 \cdot 10^{-8}$ 					& $|g_{e\tau} g_{ee}^*| < 9 \cdot 10^{-3} (m_k / \TeV)^2$\\
		$\mathrm{BR}(\tau \to e^+ e^- \mu^-) < 1.8 \cdot 10^{-8} $ 			& $|g_{e\tau} g_{e\mu}^*| < 5 \cdot 10^{3} (m_k / \TeV)^2$\\
		$\mathrm{BR}(\tau \to e^+ \mu^- \mu^-) < 1.7 \cdot 10^{-8} $ 		& $|g_{e\tau} g_{\mu\mu}^*| < 7 \cdot 10^{-3} (m_k / \TeV)^2$\\
		$\mathrm{BR}(\tau \to \mu^+ e^- e^-) < 1.5 \cdot 10^{-8} $ 			& $|g_{\mu\tau} g_{ee}^*| < 7 \cdot 10^{-3} (m_k / \TeV)^2$\\
		$\mathrm{BR}(\tau \to \mu^+ e^- \mu^-) < 2.7 \cdot 10^{-8} $ 		& $|g_{\mu\tau} g_{e\mu}^*| < 6 \cdot 10^{-3} (m_k / \TeV)^2$\\	
		$\mathrm{BR}(\tau \to \mu^+ \mu^- \mu^-) < 2.1 \cdot 10^{-8} $ 	& $|g_{\mu\tau} g_{\mu\mu}^*| < 8 \cdot 10^{-3} (m_k / \TeV)^2$\\
		\hline
		$\mathrm{BR}(\MEG) < 4.2 \cdot 10^{-13}$ 										& $\left|\frac{m_k^2}{m_h^2} f_{e\tau}^* f_{\mu\tau}\right|^2 + 
		16 \left|(g^\dag g)_{e\mu}\right|^2 < 1.1 \cdot 10^{-6 } (m_k / \TeV)^4$\\
		$\mathrm{BR}(\tau \to e \gamma) < 3.3 \cdot 10^{-8}$ 					& $\left|\frac{m_k^2}{m_h^2} f_{e\mu}^* f_{\mu\tau}\right|^2 + 
		16 \left|(g^\dag g)_{e\tau}\right|^2 < 0.52\, (m_k / \TeV)^4$\\
		$\mathrm{BR}(\tau \to \mu \gamma) < 4.5\cdot10^{-8}$ 					& $\left|\frac{m_k^2}{m_h^2} f_{e\mu}^* f_{e\tau}\right|^2 + 
		16 \left|(g^\dag g)_{\mu\tau}\right|^2 < 0.71\, (m_k / \TeV)^4$\\
		\hline
		$\mathrm{CR}(\mu\, \mathrm{Ti} \to e\, \rm Ti) < 6.1 \cdot 10^{-13}$ 		& See Eqs.~\eqref{eq:CRZeeBabu}\\
		\hline \hline
		Universality bounds & \\
		\hline
		$\sum_{q=u,d,s} |V_{uq}|^2 = 0.9999 \pm 0.0006$ 				& $|f_{e\mu}|^2 < 0.014 (m_h / \TeV)^2$ \\
		$\frac{G_{\tau\mu}}{G_{\tau e}} = 1.0001 \pm 0.0020$		& $\left| |f_{\mu \tau}|^2 - |f_{e \tau}|^2 \right| < 0.05 (m_h / \TeV)^2$\\	
		$\frac{G_{\tau e}}{G_{\mu e}} = 1.0004 \pm 0.0022$			& $\left| |f_{e \tau}|^2 - |f_{e \mu}|^2 \right| < 0.06 (m_h / \TeV)^2$\\
		$\frac{G_{\tau\mu}}{G_{\tau e}} = 1.0004 \pm 0.0023$ 		& $\left| |f_{\mu \tau}|^2 - |f_{e \mu}|^2 \right| < 0.06 (m_h / \TeV)^2$\\
		\hline
	\end{tabular}
}
\caption{\label{tab:ZeeBabu}Existing limits on the parameter space of the Zee-Babu model. See Refs.~\cite{Schmidt:2014zoa,Herrero-Garcia:2014hfa} for further details.}
\end{table}

\begin{figure}[t]
   \centering
   \noindent\makebox[\textwidth]{
   \begin{subfigure}[b]{.45\textwidth}
      \centering
      \includegraphics[width=\textwidth]{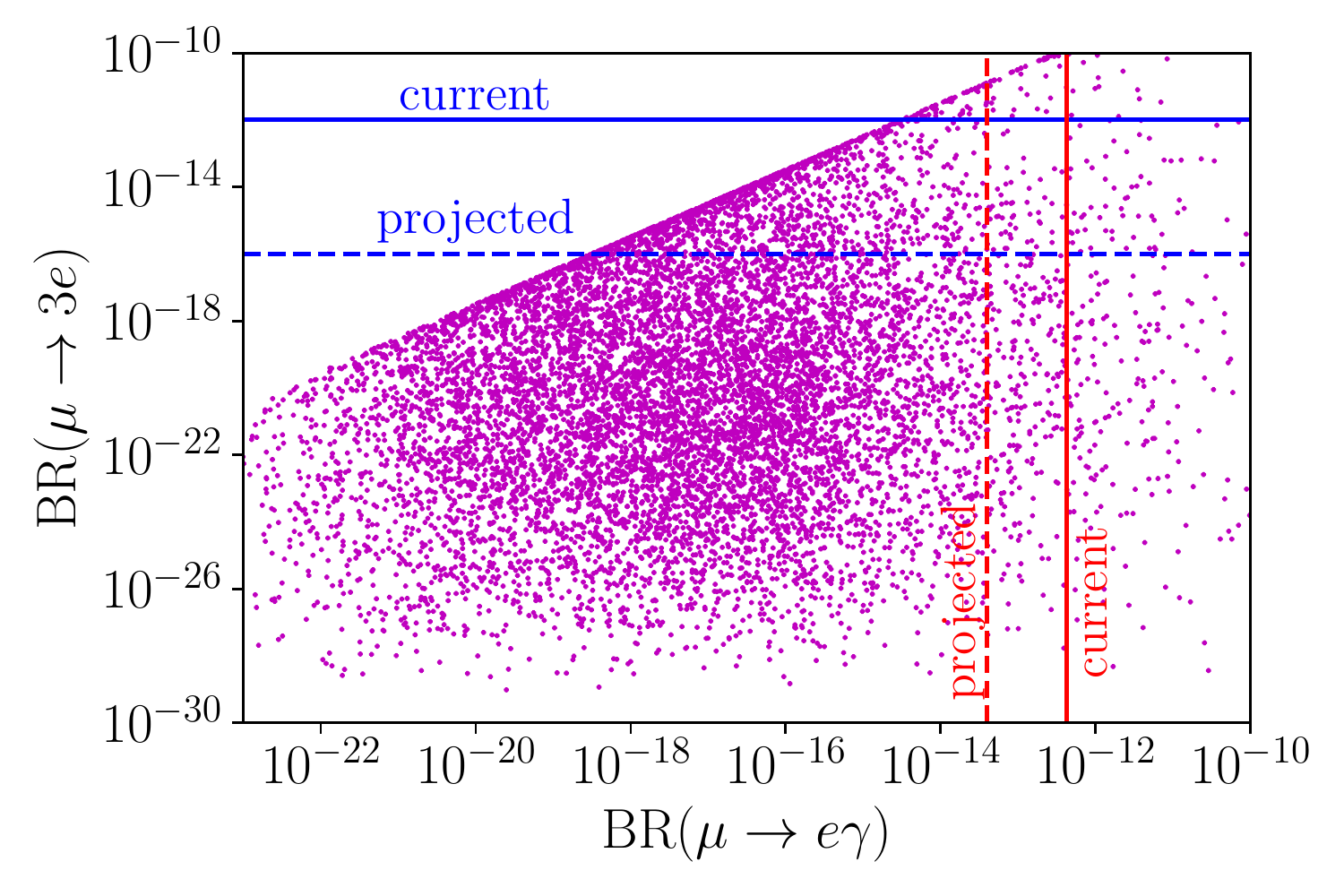}
	  \subcaption{\label{fig:ZeeBabu_mu3eNO}normal ordering}
   \end{subfigure}
   \hfill
   \begin{subfigure}[b]{.45\textwidth}
     \includegraphics[width=\textwidth]{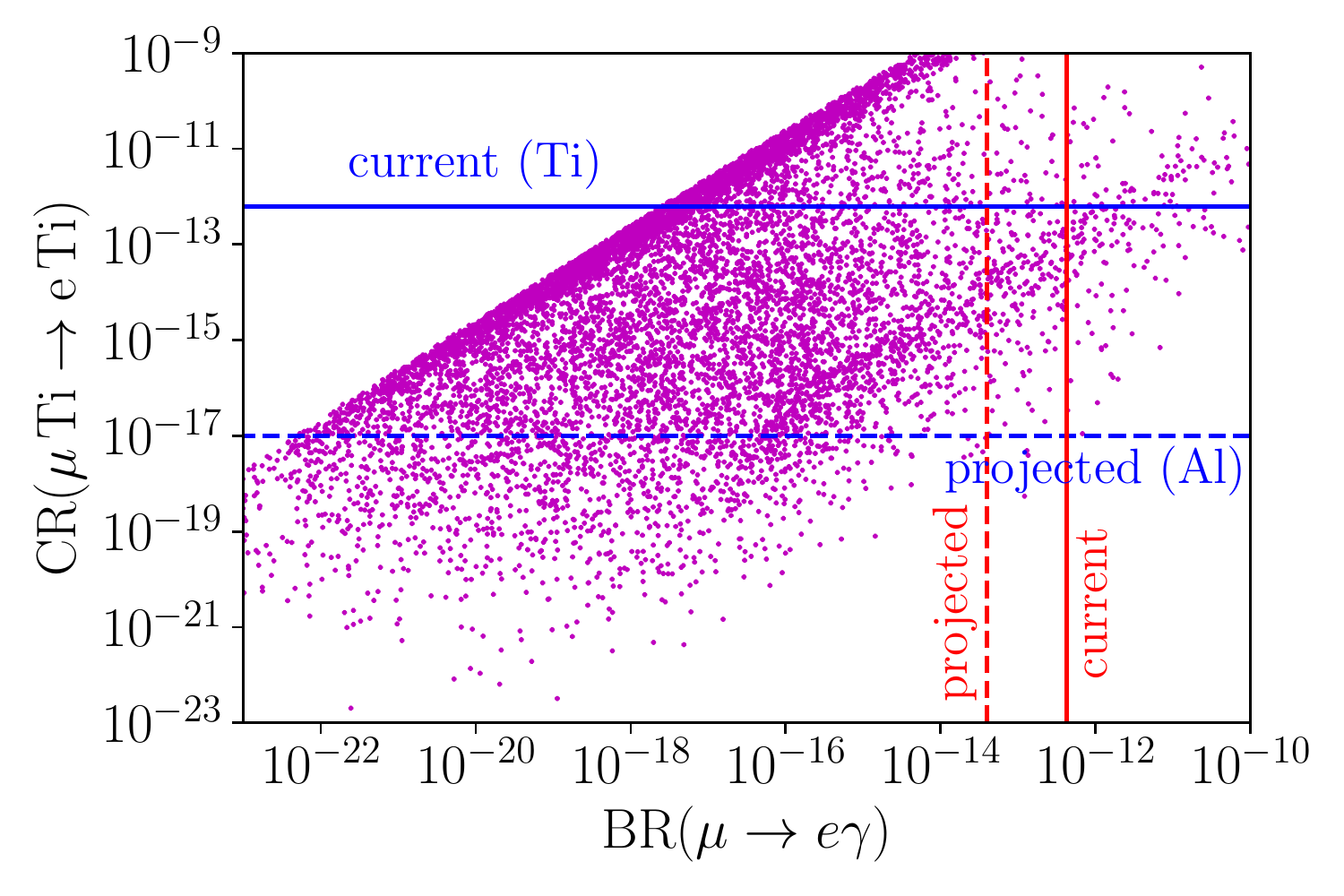}
     \subcaption{normal ordering}
   \end{subfigure}
   }\\
    \vspace{5mm}
   \noindent\makebox[\textwidth]{
   \begin{subfigure}[b]{.45\textwidth}
      \centering
      \includegraphics[width=\textwidth]{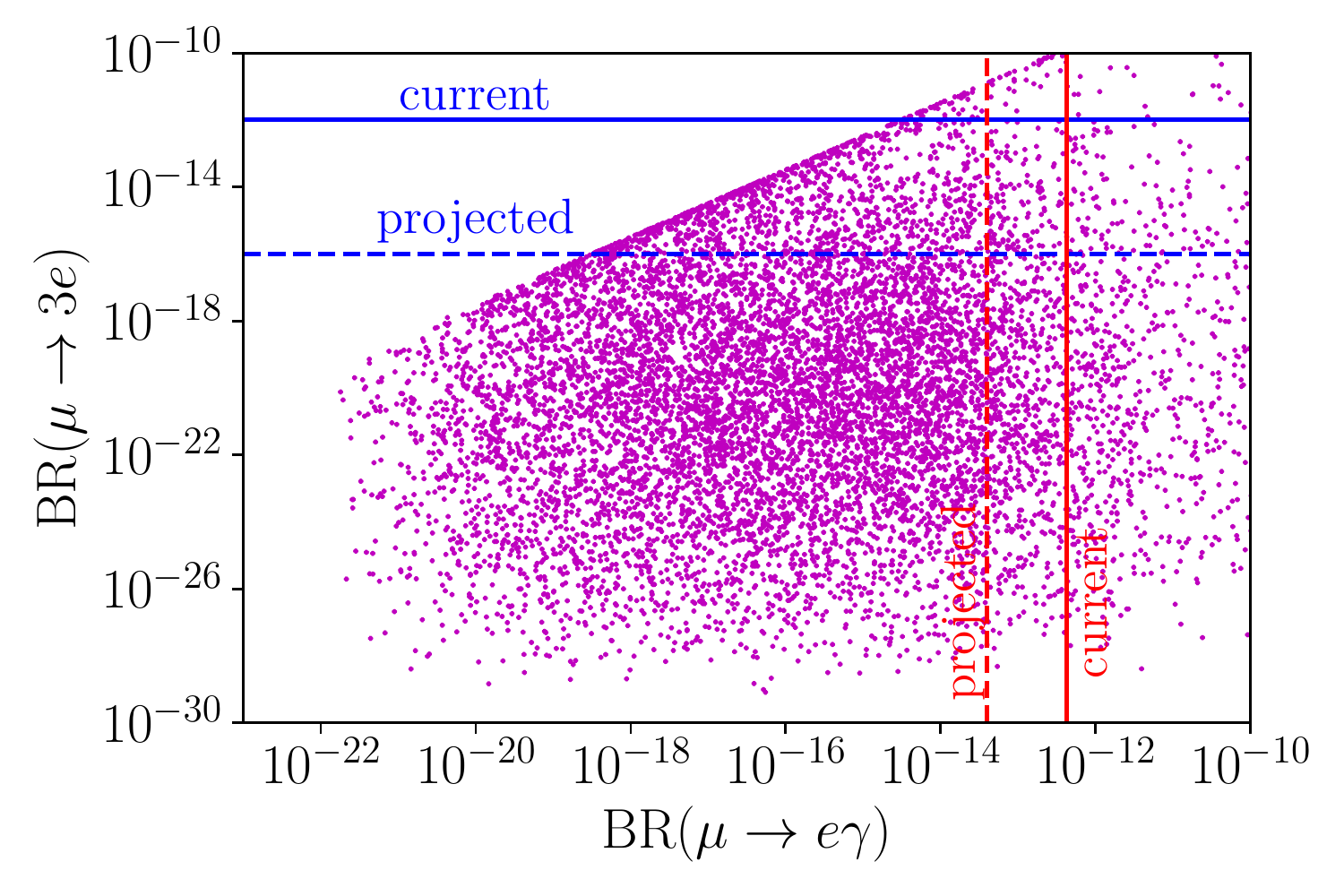}
      \subcaption{\label{fig:ZeeBabu_mu3eIO}inverted ordering}
   \end{subfigure}
   \hfill
   \begin{subfigure}[b]{.45\textwidth}
      \centering
      \includegraphics[width=\textwidth]{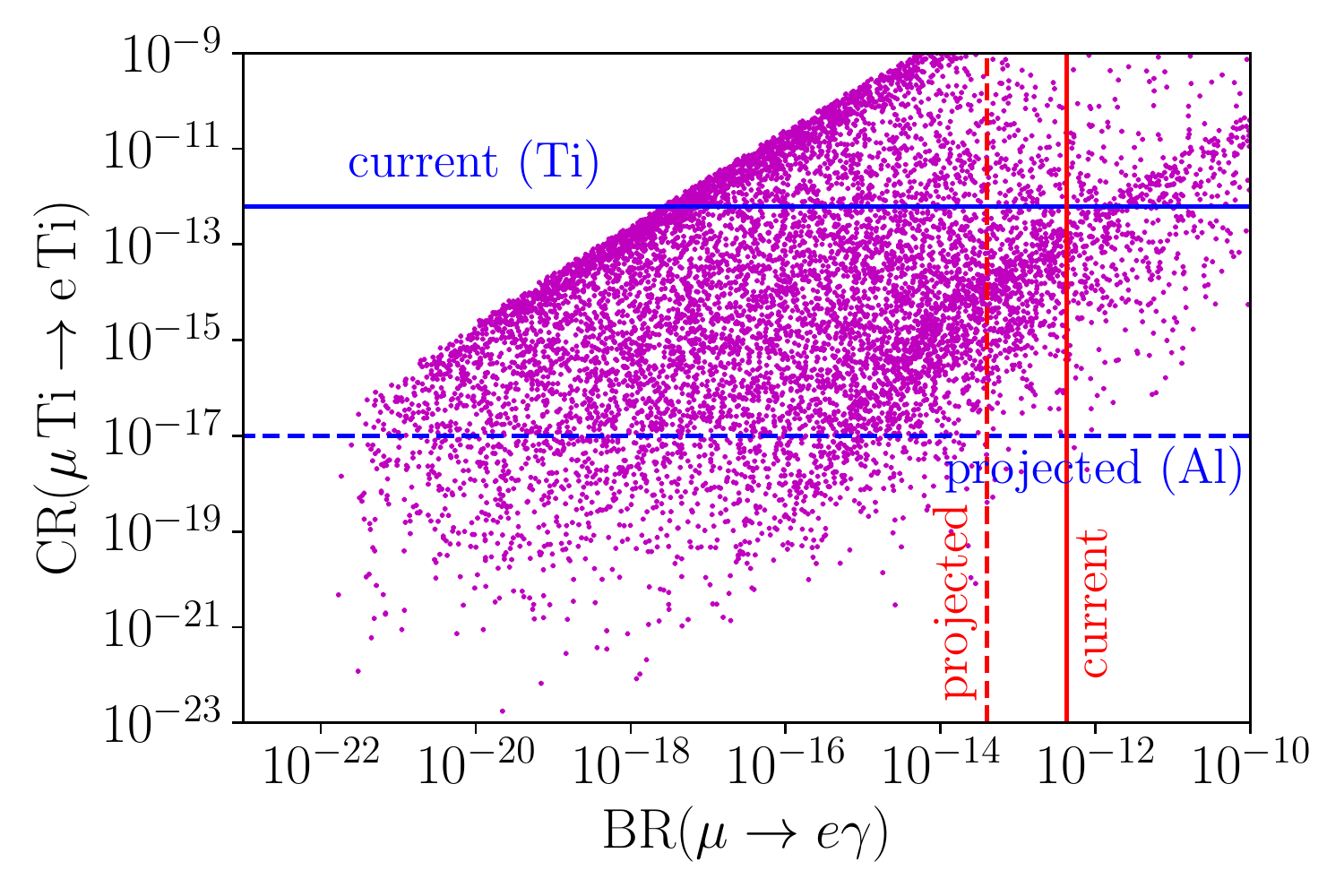}
      \subcaption{inverted ordering}
   \end{subfigure}
   }
   \caption{\label{fig:results_ZeeBabu}Results and bounds for the Zee-Babu model for different LFV violation processes in normal as well as inverted neutrino mass ordering. Note that the projected limit employs a different nucleus (Al) than the current limit (Ti) on $\mu\, N \to e\, N$; however, we only show the results for one nucleus (Ti) since they deviate only marginally for the different nuclei.}
\end{figure}

{\bf Existing Limits}

As far as other limits are concerned, the most important ones are again due to LFV, since the models extends only the  leptonic sector of the SM. Therefore, the collider constraints are rather mild, $m_{k,h} \gtrsim \mathcal{O}(100\GeV)$~\cite{Herrero-Garcia:2014hfa}. Note that the doubly charged scalar can mediate tree-level decays of leptons into 3-body final states, i.e.~$\ell_\alpha^- \to \ell_\beta^+ \ell_\gamma^- \ell_\epsilon^-$. The corresponding branching ratio reads~\cite{Schmidt:2014zoa}
\begin{equation}
	\mathrm{BR}\left(\ell_\alpha^- \to \ell_\beta^+ \ell_\gamma^- \ell_\epsilon^-\right) = \frac{1}{2 (1+ \delta_{\gamma,\epsilon})} \left| \frac{g_{\alpha\beta} g_{\gamma\epsilon}^*}{G_F m_k^2}\right|^2  \times \mathrm{BR}( \ell \to \ell \overline{\nu} \nu ),
\end{equation}
which results in a limit $|g_{e\mu} g_{ee}^*| < 2.3 \cdot 10^{-5} (m_k / \TeV)^2$ with current and $|g_{e\mu} g_{ee}^*| < 2.3 \cdot 10^{-7} (m_k / \TeV)^2$ with future sensitivity on $\mu \to 3e$. One can always choose the relevant couplings $g_{\alpha\beta}$ small enough to evade these bounds. However, by virtue of the neutrino mass relation~\eqref{eq:massZeeBabu}, the entries of $f$ would have to grow correspondingly. These will then be constrained via the $\MEG$ process, as $\left|(f^\dag f)_{e\mu}\right| < 1.1 \cdot 10^{-3} (3.3 \cdot 10^{-4}) (m_h / \TeV)^2$. Another important constraint arises due to $\mu - e$ conversion. The relevant expressions read (in the conventions of Refs.~\cite{Schmidt:2014zoa,Kitano:2002mt,Dinh:2012bp})
\begin{subequations}\label{eq:CRZeeBabu}\allowdisplaybreaks
\begin{equation}
	\mathrm{CR} (\mu\, N \to e\, N) = \frac{2 e^2 G_F^2}{\Gamma_\text{capt}} \left(\left|A^h_R D + e A^h_L V^{(p)} \right|^2 + \left|A^k_R D + e A^k_L V^{(p)} \right|^2\right) 
\end{equation}
with
\begin{align}
	A^h_R =& - \frac{\left(f^\dag f\right)_{e\mu}}{768 \sqrt{2} \pi^2 G_F m_h^2}\,, 
	&&A^k_R = - \frac{\left(g^\dag g\right)_{e\mu}}{48 \sqrt{2} \pi^2 G_F m_k^2}\,,\\
	A^h_L =& - \frac{\left(f^\dag f\right)_{e\mu}}{144 \sqrt{2} \pi^2 G_F m_h^2}\,, 
	&&A^k_L = -\sum_{a=e\mu\tau} \frac{\left(g^*_{a e} g_{a\mu}\right)}{6 \sqrt{2} \pi^2 G_F m_k^2} F\left(\frac{q^2}{m_k^2},\frac{m_a^2}{m_k^2}\right) \,,
\end{align}
where
\begin{equation}
F(x,y) \equiv  \frac{4 y}{x} + \log (y) + \left(1 - \frac{2y}{x}\right) \sqrt{1 + \frac{4y}{x}} \log\left(\frac{\sqrt{x+4y} + \sqrt{x}}{\sqrt{x+4y}-\sqrt{x}}\right).
\end{equation}
\end{subequations}
The nuclear form factors can be found in Tab.~\ref{nuclfactorsmueconv}. We have summarized the relevant constraints on the model parameter space in Table~\ref{tab:ZeeBabu}. Finally, in Fig.~\ref{fig:results_ZeeBabu}, we show the results of a parameter scan for values of the parameters $g_{\alpha\beta}, f_{\alpha\beta} \in [10^{-5}, 10^{-1}]$ and masses $300\GeV < m_{h,k} <50\TeV$, in agreement with collider bounds.  Notice that neither $\MEG$, nor $\mu\to3e$ can probe a significant part of the parameter space, while $\mu^-\, N \to e^-\, N$ has the potential to rule out much of the currently viable parameter space in the near future, where the nucleus in use will be Aluminum. This result is in a sense complementary to that of the scotogenic model in the previous section, where it was $\mu\to3e$ that will probe the available parameter space with the next generation experiments. Another important difference is that in the previous section \emph{all} of the parameter space can be probed, while in the Zee-Babu model, one can always go to smaller couplings/larger masses to evade the bounds. 

\subsection{B-L Model}
Since both baryon and lepton numbers are global symmetries in the SM, a natural and well motivated extension of the SM is the gauge group $SU(3)_C\times SU(2)_L\times U(1)_Y\times U(1)_{B-L}$, which requires the addition of three RH neutrinos to cancel the triangle anomalies.\footnote{In fact, only the linear combination $B-L$ is anomaly-free, while $B+L$ is broken non-perturbatively.~\cite{'tHooft:1976up,Klinkhamer:1984di,Kuzmin:1985mm,Fukugita:1986hr,Arnold:1987mh}} 

\begin{figure}[t]
\centering
\includegraphics[scale=0.5]{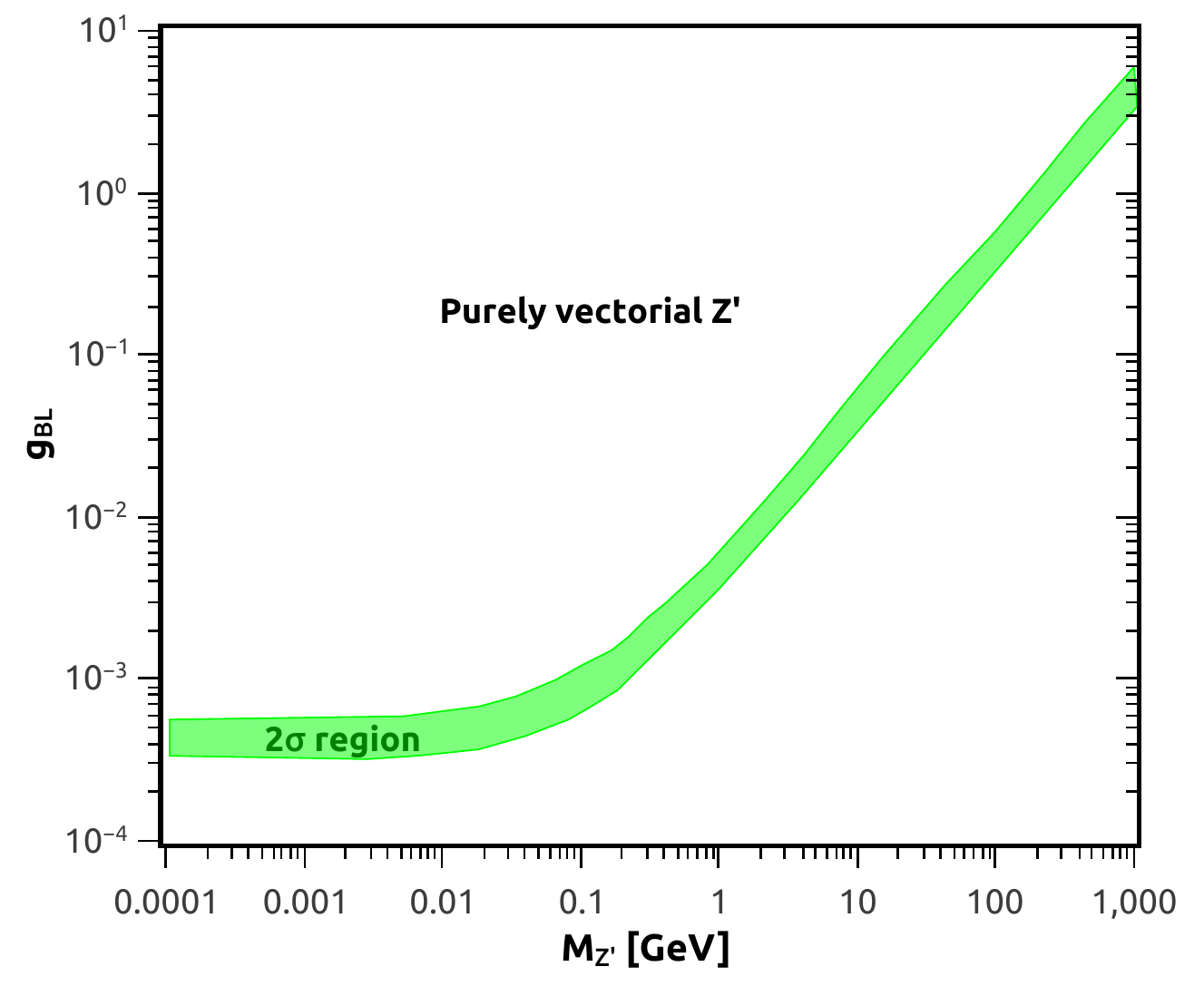}
\caption{$2\sigma$ favored region for $g-2$ for a neutral vector boson with purely vectorial couplings to leptons.} 
\label{fig:g2neutralvector}
\end{figure}

In the $B-L$ model, the $Z^{\prime}$ possesses purely vectorial couplings to the fermions with
\begin{equation}
\mathcal{L} \supset g_{BL} \sum_{i=1}^3  \left( \bar{\ell}_i\gamma^{\mu}\ell_i + \bar{\nu}_i \gamma^{\mu}\nu_i\right) Z^{\prime}_{\mu}.
\label{Eq:BLmodel}
\end{equation}
The $Z^{\prime}$ gauge bosons gain mass either through a Stueckelberg mechanism ~\cite{Feldman:2006wb} or a spontaneous symmetry breaking governed by a singlet scalar charged under $B-L$~\cite{Alves:2015mua,Rodejohann:2015lca,Okada:2016gsh,Klasen:2016qux}. In the former case the $B-L$ symmetry remains unbroken. Either way, there is no $Z-Z^{\prime}$ mass mixing at tree-level, and one can set the kinetic mixing to zero. That said, such vectorial interactions with charged leptons yield a contribution to $g-2$ but none to $\MEG$. The $g-2$ correction has been generally determined in Eq.~\eqref{Eq:doublyEamu}. Since there is no mixing among lepton flavors, one can straightforwardly solve Eq.~\eqref{Eq:doublyEamu} to find the $2\sigma$ region for $g-2$ as drawn in Fig.~\ref{fig:g2neutralvector}. Notice that we have scanned over several orders of magnitude in the $Z^{\prime}$ mass and $g_{BL}$ coupling, reaching $10$~MeV mass. We point out that this result is applicable to any purely-vectorial $Z^{\prime}$ boson since the term in the Lagrangian Eq.~\eqref{Eq:BLmodel} is rather general. Although, for sub-GeV masses one-loop corrections may induce a $Z-Z^{\prime}$ kinetic mixing which could shift the favored $g-2$ region upwards. 

{\bf Existing Limits}

The model described above cannot explain $g-2$ for heavy $Z^\prime$ masses, say above 100 GeV, due to existing collider limits on the $Z^{\prime}$ based on dilepton resonance searches which impose the $Z^\prime$ to lie above several TeV already~\cite{Klasen:2016qux}. As for light $Z^\prime$, say below the $Z$ mass, contraints from neutrino-electron scattering also exclude the $g-2$ favored region~\cite{Batell:2014mga}. Therefore, one cannot address the $g-2$ anomaly in this model, which features non LFV observable. Nevertheless, one can extend this minimal $B-L$ scenario to accommodate a signal in $\MEG$ as we describe now.

\subsection{B-L Model with Inverse seesaw}

In the previous $B-L$ model a signal in $g-2$ could not be addressed in the light of existing constraints on the $Z^{\prime}$ mass. However, the canonical seesaw type~I with heavy RH neutrinos, which is naturally incorporated in the minimal $B-L$ model, gives rise to marginal contributions to $g-2$, and none to $\MEG$. Albeit, there is an alternative solution to accommodate a possible signal in $\mu \rightarrow e\gamma$ via a different type~of seesaw mechanism known as inverse or low-scale seesaw where the mixing between the RH and active neutrinos is not so small. Thus, larger corrections to $\MEG$ are possible ~\cite{Abdallah:2011ew}. For some recent studies of a supersymmetric version of this model see also~\cite{Khalil:2015wua}.

\begin{table}[t]
\centering
\begin{tabular}{|c|c|c|c|c|c|c|c|c|c|c|c|c|}
\hline particle & ~$Q$~ & ~$u_R$~& ~$d_R$~& ~$L$~ & ~$e_{R}$~ & ~$\nu _{R}$ ~& ~$\chi_1$~ & ~$\chi_2$ ~ & ~$\phi$~ & ~$s$~ & $Z^\prime$\\
\hline\hline spin & $\frac{1}{2}$ & $\frac{1}{2}$ & $\frac{1}{2}$ & $\frac{1}{2}$ & $\frac{1}{2}$ & $\frac{1}{2}$ & $\frac{1}{2}$ & $\frac{1}{2}$ & $0$ & $0$ & $1$ \\ 
\hline $Y_{B-L}$& $1/3$  & $1/3$ & $1/3$ & $-1$ & $-1$ & $-1$ & $-2$ & $+2$ & $0$ & $-1$ & $0$ \\
\hline
\end{tabular}%
 \caption{Particle content and quantum numbers under $B-L$ symmetry.}
\label{tab:b-l}
\end{table}

A possible realization of the inverse seesaw within the $B-L$ symmetry occurs by adding two singlet fermions $\chi_{1,2}$ per generation. The $B-L$ symmetry is broken spontaneously by introducing an $SU(2)_L$ singlet scalar with hypercharge $Y_{B-L}= -1$, which generates a mass for the new gauge boson, $Z^{\prime}$, associated with the $U(1)_{B-L}$ gauge group. The particle content is summarized in Tab.~\ref{tab:b-l} for clarity. In addition to the $B-L$ symmetry, ones needs to impose upon the three singlet fermions represented by $\chi_1$ a discrete $\mathbb{Z}_2$ symmetry to avoid mass terms such as $m \overline{\chi}_1 \chi_2$.

The relevant part of the Lagrangian in this model is given by
\begin{equation}\label{eq:Lagrangian_BL}
  \begin{split}  
    \mathcal{L}_{B-L} \supset - \left(\lambda_e \overline{L_L} \phi e_R + \lambda_{\nu} \overline{L_L} \tilde{\phi} \nu_R +
    \lambda_{\chi} \overline{\nu}^c_R s \chi_2 + \mathrm{h.c.}\right) -\\
     - \frac{1}{\Lambda^3}\overline{\chi^c_1} {s^\dag}^{4} \chi_{1}-\frac{1}{\Lambda^3}\overline{\chi_2^c} {s}^{4} \chi_2 - V(\phi,s),
  \end{split}
\end{equation}
with 
\begin{equation}%
V(\phi,s)=m_1^2 \phi^\dagger \phi+m_2^2 s^\dagger
s+\lambda_1 (\phi^\dagger\phi)^2+\lambda_2
(s^\dagger s)^2
+\lambda_3
(s^\dagger s)(\phi^\dagger\phi),
\end{equation}
and $F'_{\mu\nu} = \partial_{\mu} Z'_{\nu} - \partial_{\nu} Z'_{\mu}$ being the field strength of the $U(1)_{B-L}$ gauge boson, which is minimally coupled via the covariant derivative
\begin{eqnarray}%
D_{\mu} = \partial_\mu - i g_s  T^a G_\mu^a- i g \frac{\tau^i}{2} W_\mu^i -
i g' Y B_\mu -  i g_{BL} Y_{BL} Z'_{\mu}.
\end{eqnarray}
The last
two terms in ${\cal L}_{B-L}$ are non-renormalizable terms, which
are allowed by the symmetries and relevant for generating a TeV-scale
mass for $\chi_1$ and $\chi_2$, as well as being required for the inverse seesaw
mechanism. The scale $\Lambda$ in these terms is a cut-off for the validity of the $B-L$ model.



As for neutrino masses, they are generated after spontaneous symmetry breaking via the Lagrangian
\begin{equation}
{\cal L}_m^{\nu} =\mu_s \overline{\chi_2^c} \chi_2 +(m_D \bar{\nu}_L \nu_R + M_N \bar{\nu}^c_R \chi_2 +\mathrm{h.c.})\, ,%
\end{equation}where $m_D=\frac{1}{\sqrt{2}}\lambda_\nu v$ and $ M_N =
\frac{1}{\sqrt 2}\lambda_{S} v_s $, with $\mu_s=\frac{v_s^4}{4 \Lambda^3}$.

Writing $\psi=(\nu_L^c ,\nu_R, \chi_2)$, we can recast the Lagrangian above as $\overline{\psi^c} {\cal M}_{\nu} \psi$ with
\begin{equation}\label{eq:neutrinoMass_BL}
 {\cal M}^{(\nu)}=
\left(%
\begin{array}{ccc}
  0 & m_D & 0\\
  m^T_D & 0 & M_N \\
  0 & M^T_N & \mu_s\\
\end{array}%
\right). %
\end{equation}
Diagonalizing, this gives approximate mass matrices
\begin{equation}\label{eq:neutrinoMassLight_BL}
  m_{\nu,\,\mathrm{light}} \approx m_D\, M_N^{-1} \mu_s\, {M_N^{-1}}^T {m_D}^T \quad \textrm{and}\quad m_{\nu,\,\mathrm{heavy}} \approx M_N \pm \mu_s.
\end{equation}
For a cut-off $\Lambda$ around $10^7\GeV$, neutrino masses of the order eV can be achieved with Yukawa couplings of order one, $\mu_s \sim 10^{-9}\GeV$, and TeV scale $M_N$.\footnote{Even without such a large cut-off, one may argue that $\mu_s$ is naturally small in 't Hooft's sense~\cite{'tHooft:1979bh}, since in the limit $\mu_s \to 0$ a global $U(1)$ lepton number symmetry can be defined.} Therefore, the Yukawa coupling $\lambda_\nu$ is no longer required to be tiny, which would preclude any experimental test of the scenario.

Note that, while the full $9\times 9$ mass matrix in Eq.~\eqref{eq:neutrinoMass_BL} is diagonalized by a unitary matrix $U$ according to $\mathcal{M}^{(\nu)}_\mathrm{diag} = U^\dag \mathcal{M}^{(\nu)} U^*$, the light neutrino mass matrix in Eq.~\eqref{eq:neutrinoMassLight_BL} is only a $3\times 3$ sub-matrix that is not necessarily diagonalized in that way~\cite{Dev:2009aw}. Thus, the model predicts that in general the leptonic mixing matrix $U_\mathrm{PMNS}$ is non-unitary. Most generally, we have
\begin{equation}
  U = \left(
  \begin{array}{cc}
   V_{3\times 3} & V_{3 \times 6}\\
   V_{6 \times 3} & V_{6 \times 6}
  \end{array}\right),
\end{equation}
where one conventionally parametrizes $V_{3\times 3} = \left( \mathbf{1} - \frac{1}{2} F F^\dag\right)U_\mathrm{PMNS}$ such that the non-unitarity is measured exclusively by $F \simeq m_D M_N^{-1}$ instead of $U_\mathrm{PMNS}$ itself~\cite{Kanaya:1980cw,Altarelli:2008yr}. Furthermore, the mixing of light and heavy states is given by $V_{3\times 6} \simeq ( I_{3\times3},F) V_{6\times 6}$. Finally, $V_{6\times 6}$ is the matrix that diagonalizes the $(\nu_R, \chi_2)$ subspace.  As for the scalar sector, the Higgs boson $h$ becomes a linear combination of $\phi$ and $s$~\cite{Khalil:2006yi}.

We have assembled the basic building blocks of the model to compute $g-2$ and $\MEG$ in the model. The $g-2$ contribution stems from: (i) the $W$ exchange via $\nu-\nu_R$ mixing; (ii) the $B-L$ gauge boson $Z^{\prime}$; (iii) the heavy Higgs, the latter of which is suppressed. The $Z^\prime$ contribution has been computed before, and the $W$ exchange has been generally given in Eq.~\eqref{eq:approxamu_W1}. 

We have seen before that one cannot explain $g-2$ via a $Z^\prime$ in the context of $B-L$ due the existing collider bounds at high masses and neutrino-electron scattering at low masses.

As for $\MEG$ the contribution to this decay is via $W$ exchange, induced by the $\nu-\nu_R$ mixing. Again this calculation has been performed in Eq.~\eqref{eq:BR4}, and after adapting the couplings to this specific model, one gets,
\begin{equation}
\big| \sum_N V_{\mu N} V_{e N} \big| \lesssim 10^{-5},
\end{equation}where $N=\nu_{R}^1,\nu_{R}^2,\nu_{R}^3$. 

This is the strongest bound on the model. The other LFV observables such as $\mu \to 3e$ and $\mu-e$ conversion are, as it turns out, not very constraining.

\subsection{$SU(3)_c \times SU(3)_L \times U(1)_X$ Model}

This class of models represents electroweak extensions of the SM based on the gauge group $SU(3)_c \times SU(3)_L \times U(1)_X$, shortly referred as 331 models. Due to the enlarged gauge symmetry, the fermionic generations are accommodated in the fundamental representation of $SU(3)_L$, i.e triplets. Since the SM spectrum should be reproduced, the triplet must contain the SM doublet, but the arbitrariness of   the third component leads to a multitude of models based on this gauge symmetry~\cite{Pisano:1991ee,Foot:1992rh,Pleitez:1992xh,Foot:1994ym,Hoang:1995vq,
Montero:2000ng,Dias:2003iq,VanDong:2005pi,
Dong:2006mg,Cogollo:2009yi,Mizukoshi:2010ky,
Queiroz:2010rj,Ferreira:2011hm,Alves:2011kc,Dong:2011vb,
Huyen:2012uk,Huong:2012pg,Alvares:2012qv,
Kelso:2013nwa,Kelso:2014qka,Pires:2016vek,Pires:2016dqq}.

Generally speaking, 331 models send the appealing message of solving the puzzle of why there are three generations of fermions in Nature. The models are only self-consistent if there exist exactly three generations of fermions as a result of the triangle gauge anomalies and QCD asymptotic freedom~\cite{Singer:1980sw,Pisano:1991ee,Foot:1992rh}. Moreover, they can host a dark matter candidate~\cite{deS.Pires:2007gi,Mizukoshi:2010ky,Alvares:2012qv,Queiroz:2013lca,
Kelso:2013nwa,Profumo:2013sca,Kelso:2014qka,Dong:2014esa,Dong:2014wsa,
Cogollo:2014jia,Martinez:2014ova,Martinez:2014rea,daSilva:2014qba,
Dong:2015rka,Martinez:2015wrp,Huong:2016ybt,Pires:2016vek}, generate neutrino masses~\cite{Minkowski:1977sc,Mohapatra:1979ia,Schechter:1980gr,Mohapatra:1980yp,
Lazarides:1980nt,Schechter:1981cv,Keung:1983uu}, among other things ~\cite{Dias:2002gg,Dias:2003zt,Cogollo:2007qx,Phong:2014yca,DeConto:2016osh,
Hernandez:2016eod,Machado:2016jzb,Fonseca:2016tbn,Buras:2016dxz}. 

Since we are focussing on $g-2$ and LFV we will adopt a model known as 331 model with right handed neutrinos, $331r.h.n.$ for short~\cite{Hoang:1995vq}, where the third component of the fermion triplet is a RH neutrino as follows,
\begin{equation}
f^a_L=\left(
\begin{array}{c}
\nu^a_l\\
e^a_l\\
(\nu^c_R)^a
\end{array}\right) \sim (1,3,-1/3),\ e_R^a \sim (1,1,-1), 
\end{equation} where $a=1,\, 2,\, 3$.

We will set the hadronic sector aside since it is irrelevant for our purposes (see~\cite{Hoang:1995vq} for a more detailed discussion). In order to successfully generate masses for the fermions, one needs to invoke the presence of three scalar triplets and a sextet as follows
\begin{equation}
\chi=\left(
\begin{array}{c}
\chi_1^0\\
\chi_2^-\\
\chi_3^0
\end{array}\right),\,
\eta=\left(
\begin{array}{c}
\eta_1^0\\
\eta_2^-\\
\eta_3^0
\end{array}\right),\,
\rho=\left(
\begin{array}{c}
\rho_1^+\\
\rho_2^0\\
\rho_3^+
\end{array}\right), \,
S=\left(
\begin{array}{ccc}
S^0_{11} & S^-_{12} & S^0_{13} \\
S^-_{12} & S^{--}_{22} & S^-_{23} \\
S^0_{13} & S^-_{23} & S^0_{33} \\
\end{array}\right).
\label{Eq:scalarfields}
\end{equation} 
They have the following quantum numbers under the gauge group: $\chi \sim (1,3,-1/3)$, $\eta \sim (1,3,-1/3)$, $\rho \sim (1,3,2/3)$, $S \sim (1,6,-2/3)$. The spontaneous symmetry breaking pattern of the model proceeds via the triplet $\chi$ developing a non-trivial VEV, breaking $\mathrm{SU}(3)_L\times \mathrm{U}(1)_X \rightarrow
\mathrm{SU}(2)_L\times\mathrm{U}(1)_Y$. This is followed by the breaking 
$\mathrm{SU}(2)_L\times\mathrm{U}(1)_Y \rightarrow U(1)_{QED}$ via the VEVs of $\rho$ and $\eta$, as indicated below:
\begin{equation}
\left\langle\chi\right\rangle=\left(
\begin{array}{c}
u^{\prime}/\sqrt{2}\\
0\\
w/\sqrt{2}
\end{array}\right),\,
\left\langle\eta\right\rangle=\left(
\begin{array}{c}
u/\sqrt{2}\\
0\\
w^{\prime}/\sqrt{2}
\end{array}\right),\,
\left\langle\rho\right\rangle=\left(
\begin{array}{c}
0\\
v\sqrt{2}\\
0
\end{array}\right),
\end{equation} and
\begin{equation}
\left\langle S\right\rangle=\left(
\begin{array}{ccc}
v_{s1} & 0 & v_{s3} \\
0 & 0 & 0 \\
v_{s3} & 0 & \Lambda \\
\end{array}\right).
\end{equation} 
The role of the sextet is to give masses to the neutrinos, and in order to keep the symmetry breaking consistent, some conditions have to be satisfied, namely $\Lambda,w \gg v_{s3},v,u \gg v_{s1} u^{\prime}$, along with $w \gg w^{\prime}$. In this way the SM gauge boson masses are correctly obtained, the $\rho$ parameter remains close to unity, and the fermions acquire masses through the Yukawa Lagrangian which is divided into pieces, one where lepton flavor is conserved (LFC) and other where it is violated (LFV):
\begin{subequations}
 \bea
 \label{Eq:LNV331}
{\mathcal L}_{\mathrm{LFC}}& \supset & h^l_{ab}\bar{\psi}_{aL}\rho
l_{bR}+h^\nu_{ab}\bar{\psi}^c_{aL}\psi_{bL}\rho+
\mathrm{h.c.}\,,\\ {\mathcal
L}_{\mathrm{LFV}}& \supset & f^\nu_{ab}(\bar{\psi}^c_{aL})_m(\psi_{bL})_n(S^*)_{mn}+ \mathrm{h.c.}\,,
\eea
\end{subequations} 
 where $a,b=1,2,3$ account for the three generations, $m,n=1,2,3$ indicate the entries of the sextet, and $f_{ab}$ is symmetric. The last term in Eq.~\eqref{Eq:LNV331} gives rise to LFV interactions:
\begin{equation}
f_{12} \overline{(e_L)^c}\, \mu_L\, S^{++}=\frac{f_{12}}{2}\overline{e^c}\, \mu\, S^{++} -
\frac{f_{12}}{2}\overline{e^c}\,\gamma_5 \mu\, S^{++},
\end{equation}which contributes to $\MEG$ via the presence of the doubly charged scalar in the sextet in Eq.~\eqref{Eq:scalarfields} with $S^{++} \equiv S^{++}_{22}$. One can construct a similar term but proportional to $f_{22}$ correcting $g-2$. Keep in mind that other charged leptons might run in the loop for $g-2$ and $\MEG$, and therefore  the overall corrections in the $331r.h.n$ model have to be summed over all charged lepton flavors in general. We have computed both observables already in Eq.~\eqref{eq:Delta_a_Doubly_Scalar}. Applying the results to the $331r.h.n$ under study we get
\begin{equation}
  \Delta a_\mu\left(S^{++}\right) =-\frac{1}{4\pi^2} \frac{m_\mu^2}{m_{S^{++}}^2} \sum_b \left[ \left|f_{2b}/2\right|^2 \left(\frac{4}{3}-\epsilon_b\right) + \left|f_{2b}/2\right|^2 \left(\frac{4}{3}+\epsilon_b\right) \right].
  \label{Eq:331doubly}
\end{equation}where $\epsilon_b\equiv\frac{m_b}{m_\mu}$, and the sum runs over all charged leptons of mass $m_b$  in the loop, remembering that $b$ is a fermion generation index. If one considers no mixing between the charged leptons, then $m_b \equiv m_{\mu}$ and $\epsilon=1$ (without the sum). In general, however, there might be a mixing with other charged leptons, and in that case one needs to sum over all fermion masses. This sum is only relevant in case there is mixing with the $\tau$ lepton. 

We highlight that there are other diagrams that contribute to $g-2$, one mediated by a $W^\prime$ gauge boson and another through a $Z^\prime$ vector. Their contribution to $g-2$ are of the same order but with opposite signs. The $W^\prime$ contribution arises from the charged current whereas the $Z^\prime$ one from the neutral current as follows
\begin{eqnarray}
L \supset \frac{g}{2\sqrt{2}} \overline{\nu_{\mu R}^c}\gamma^\mu(1-\gamma_5)\mu\, W^\prime_\mu ,\quad
L \supset \bar{\mu} \gamma^\mu (g_V + g_A \gamma_5 ) \mu\, Z^\prime,
\end{eqnarray}
where
\begin{eqnarray}
g_V=\frac{g}{4c_W} \frac{1-4 s_W^2}{\sqrt{3-4s_W^2}},\quad
g_A=\frac{-g}{4c_W} \frac{1}{\sqrt{3-4s_W^2}}.
\end{eqnarray}
Their contribution to $g-2$ can be straightforwardly computed using Eq.~\eqref{eq:approxamu_W1} and Eq.~\eqref{eq:neutralGBapprox}. In this model, the masses of these gauge bosons are directly tied to the scale of symmetry breaking of the model, $w$. Indeed, they read $m_{Z^\prime} =0.4 w$ and $m_{W^\prime} =0.3 w$. Hence, their contributions to $g-2$ are directly governed by the scale of symmetry breaking of the model. It is easy to show that in order to get a contribution to $g-2$ of order of $10^{-9}$ one needs $w \sim 2$~TeV. That, in turn,  implies that $m_{Z^\prime} \sim 800$~GeV and $m_{W^\prime} \sim 600$~GeV~\cite{Kelso:2014qka}. We will see further below that such low masses for the gauge bosons is prohibited due to the existing flavor and collider limits. 

Therefore, in what follows, we concentrate on the doubly charged scalar and its interplay with LFV. The doubly charged scalar gives a negative contribution to $g-2$, making it impossible to accommodate the $g-2$ anomaly since it constitutes a positive excess over the SM expectation. 

That said, looking at Eq.~\eqref{Eq:331doubly} one can see that the doubly charged contribution to $g-2$ can be in principle large enough to explain the $g-2$ anomaly, bu, as we emphasized above, its contribution to $g-2$ has the opposite sign, and is thus irrelevant. Setting $f_{22}\sim 2$ and the other off-diagonal couplings to be very small, $g-2$ imposes a lower mass bound on the charged scalar of $m_S^{++} \sim 500$~GeV. Lets check if this scenario is consistent with LFV probes.

Regarding lepton flavor violation observables we start our discussion with the $\MEG$ decay. In this model, we get
\begin{equation}
  \mathrm{BR}(\MEG) \simeq \frac{\alpha_\mathrm{em} \left|(f_{1a}^* f_{2a})_{e\mu}\right|^2}{3 \pi G_F^2 m_{S^{++}}^4}.\
  \label{Eq:331doublyBRmutoe}
\end{equation}
Ignoring the tau-flavor, notice that the decay $\MEG$ is proportional to $(f_{11}f_{21}+f_{12}f_{22})^2$ whereas $g-2$ goes with $f_{21}^2+f_{22}^2$. Knowing that the bounds on $\mathrm{BR}(\MEG)$ forces a hierarchy between the diagonal and off-diagonal couplings; lets suppress $f_{12}$ while keeping $f_{22}$ sufficiently large for now. It so happens that keeping $f_{22} \sim 2$ and $m_S^{++} \sim 500$~GeV, as adopted above, and $f_{12}=f_{21}=10^{-6},f_{11}=10^{-3}$ we obtain  $\mathrm{BR}(\MEG)\sim 3\times 10^{-16}$, which is below current limits. See~\cite{Cabarcas:2013jba,Boucenna:2015zwa,Hue:2015fbb} for further discussions. Therefore, the presence of the scalar sextet, which is crucial to generate neutrino masses in the 331 model in a successful way, also gives rise to an interesting possibility to have a doubly charged scalar below the TeV scale while being consistent with bounds from $g-2$ and $\MEG$.

Regarding other LFV signals such as $\mu \to 3e$ and $\mu-e$ conversion, a rule of thumb relation that gives a rough estimate on the size and correlation between $\mu \to 3e$, $\mu-e$ conversion and $\MEG$ reads
\begin{subequations}\label{Eq331mue}
\begin{align}
{\rm CR}(\mu \, \mbox{Al} \to e\, \mbox{Al}) \simeq& \frac{1}{350} {\rm BR}(\MEG),\\
{\rm BR} (\mu \to 3e) \simeq& \frac{1}{160} {\rm BR}(\MEG).
\end{align}
\end{subequations}
With Eq.~\eqref{Eq331mue} at hand, we conclude that for $\mu-e$ conversion to be more sensitive than the $\MEG$ decay as far as $331r.h.n$ model is concerned, the bound on $\mu-e$ conversion has to be rouhgly two orders of magnitude more restrictive. This will not be conceivable in the near future. However, in the long run, the bound on ${\rm CR}(\mu \, \mbox{Al} -e\, \mbox{Al})$ will be significantly improved reaching an expected bound on the order of $10^{-17}$ using Aluminum (see Table \ref{tab:mu-eOverview}), while no further improvement is expected for $\MEG$ beyond ${\rm BR}(\MEG)< 4 \times 10^{-14}$. 

Therefore, $\mu-e$ conversion will be more sensitive than $\MEG$. In the benchmark model we chose $\mathrm{BR}(\MEG)\sim 3\times 10^{-16}$  which yields  $\mathrm{CR} (\mu \, \mbox{Al}-e \, \mbox{Al}) \sim 3 \times 10^{-18}$, falling below projected sensitivity. Therefore, this model can accommodate a signal in $\MEG$ while predicting no signal in the next generation of experiments in the search for $\mu-e$ conversion. Albeit, if Mu2e and COMET experiments are capable of further reducing their sensitivity to $10^{-18}$ \cite{Bernstein:2013hba,deGouvea:2013zba} then the $\mu-e$ probe nicely this scenario. This UV complete model constitutes a compelling case for these experiments since it shows that one can have new physics effects taking place below the TeV scale while being consistent with flavor bounds. 

Another way to look at these observables is to assume that no new physics effects take place. In this case one can use LFV to place strong limits on the mass of the doubly charged scalar. The bounds we derive strongly depend on the coupling choices. For example, assuming $f_{11}=f_{22}=0.1$ and $f_{21}=f_{12}=10^{-3}$ the non observation of $\mu-e$ conversion with a sensitivity of $10^{-1}$ on the branching ratio yields a lower mass bound of $m_S^{++} > 5$~TeV, whereas if $f_{11}=f_{22}=1$ and $f_{21}=f_{12}=10^{-3}$ we get $m_S^{++} > 15$~TeV.  The latter lower mass bound is stronger than any constraint arising from current and planned particle colliders. Hence, $\mu-e$ conversion provides a unique opportunity to probe new physics models beyond accelerator capabilities.

Obviously, the bound is rather dependent on the couplings we assume, but the same would happen for collider  searches for doubly charged scalars since the production cross section also depends on how strongly the doudbly charged scalar couples to fermions. The couplings that enter in these two probes are different though. The key difference between collider and LFV probes is that for the latter to exist we need non-zero off-diagonal couplings between lepton flavors. In principle, one can simply set these off-diagonal couplings to zero and be subject to only collider bounds, rendering the interplay between $g-2$ and LFV irrelevant.

{\bf Existing Limits}

There are several bounds applicable to the 331r.h.n.~model. The most relevant ones arise from the non-observation of signal related to the presence of exotic gauge bosons that are necessarily present in these models due to the enlarged gauge sector. Since the masses of these gauge bosons are directly connected to the scale of symmetry breaking of the model, a lower mass bound on the mass of a given gauge boson translates into a lower mass bound on the entire mass spectrum of the model. All 331 models feature the presence of a $W^\prime$ and $Z^\prime$ gauge boson. The former cannot be singly produced at the LHC because its interactions with SM quarks are accompanied by the exotic quarks. Therefore they need to be pair produced. Hence, the most efficient way to test the 331 symmetry is by looking at observables connected to the $Z^\prime$ gauge boson. That said, the most promising observables come from flavor changing neutral current effects and from collider probes. The bounds from electroweak precision tests are weak~\cite{Dong:2014wsa}. The most relevant flavor bounds in this model arise due to the LFV
character of the $Z^\prime$ gauge boson.~\cite{Benavides:2009cn} Because the three generations of quarks do not transform in the same way under $SU(3)_L$, the $Z^\prime$ interactions with SM fermions are not flavor diagonal. In particular, the meson system $B^0_d- \bar{B}^0_d$ gives to tight constraints on this model since the $Z^\prime$ gauge boson leads to the $d-b$ transition. That said, it has been found that~\cite{Queiroz:2016gif}
\begin{equation}
\Delta m_{B_d}-\frac{4\sqrt{2} G_F C_W^4}{3-4S_W^2}\frac{m_Z^2}{m_{Z^\prime 2}} |(V_L)^{ast}_{31} (V_L)_{33}|^2 f_{B_d}^2 B_{B_d} \eta_{B_d} m_{B_d}
\label{eqBd}
\end{equation}where $m_Z$ is the SM Z mass,$f_B, B_{B_d}$ and $\eta_{B_d}$ are the called bag parameters with, $f_{B_d}^2\, B_{B_d}=43264\,\mathrm{MeV}^2$, and $\eta_B=0.55$ and $m_B=5279.5$~MeV being the $B_d$ mass. The matrix $V_L$ is the mixing matrix that related the flavor and mass-eigenstate basis of the $d, s, b$ quarks. This matrix is not well constrained, since the only requirement it has to obey is that $U_L^\dagger$ $V_L=V_{CKM}$, where $U_L$ is the mixing matrix for the $u,c,t$ quarks. As long as the product of these matrices reproduce the measured CKM matrix the individual entries are in principle arbitrary. Different parametrization for these matrices might lead to very different outcomes. Knowing that the current measurement implies $\Delta m_{B_d}=3.33\times 10^{-10}$~ MeV, one can find either a conservative or optimistic bound on the $Z^\prime$ mass using Eq.~\eqref{eqBd} that reads $m_{Z^\prime}> 3$~TeV and  $m_{Z^\prime}> 4$~TeV, respectively.  No further improvement is expected on these bounds.

As for the collider bounds, searches for bumps in the invariant mass of dilepton final states are the golden channel for spotting $Z^\prime$ gauge bosons. In the context of the $331r.h.n$, it has been shown that LHC with $13$~TeV of center-of-mass energy and $3.2fb^{-1}$ of integrated luminosity impose $m_{Z^\prime} > 3$~TeV. This result is now outdated since LHC has collected already about $36 fb^{-1}$ of data. To keep our conclusions up-to-date and give a prospect for the future, we implemented the $Z^\prime$ interactions in MadGraph5~\cite{Alwall:2014hca}. Using the CTEQ6L parton distribution function~\cite{Kretzer:2003it} and DELPHES~\cite{deFavereau:2013fsa} to account for detector effects we performed the same analysis using the signal selection cuts
\begin{equation}
\begin{aligned}
&& E_T(e_1) > 30 GeV, E_T(e2) > 30 GeV, |\eta_e|<2.5\,,\\
&& p_T(\mu_1) > 30 GeV, p_T(\mu_2) > 30 GeV, |\eta_\mu|<2.5\,,
\end{aligned}
\end{equation}
with the dilepton invariant mass in the $500$~GeV-$6000$~GeV mass range as recommended by ATLAS~\cite{Aaboud:2016ejt,Aaboud:2017buh,ATLAS:2017wce}. We were able to reproduce the same lower mass bound on the $Z^\prime$ mass of 3~TeV with $3.2fb^{-1}$. Using $36.1fb^{-1}$ of data, we improve this limit to $m_{Z^\prime} > 4$~TeV. We emphasize that we have combined the dielectron and dimuon channels. Moreover, with a center-of-mass energy upgrade to 14~TeV we expect $m_{Z^\prime} > 4.8$~TeV (for $100fb^{-1}$) and $5.9$~TeV (for $1000fb^{-1}$) and $m_{Z^\prime} > 6.4$~TeV (for $3000fb^{-1}$) under the assumption of null results.

Having these bounds in mind, we decisively rule out the possibility of explaining $g-2$ in the 331r.h.n.~model via gauge bosons, their masses would have to be lighter than $1$~TeV in direct contradiction with collider and flavor probes.
 
Perhaps with the addition of more scalars in the model one could possibly change this picture.

\subsection{$L_{\mu}-L_{\tau}$}

Lepton number is an accidental global symmetry of the SM, which is however broken by quantum corrections. It has been noted, however, that gauging any \emph{difference} between two lepton family numbers with an Abelian group leads to an anomaly free theory~\cite{Foot:1990mn,He:1991qd,He:1991qd,Foot:1994vd}. $L_{\mu}-L_{\tau}$ is an explicit example which has been investigated in detail in~\cite{Bi:2009uj,Heeck:2010pg,Heeck:2011wj,Altmannshofer:2014cfa,Altmannshofer:2016brv,
Crivellin:2015mga,Crivellin:2015hha,Kim:2015fpa,Baek:2015fea,
Altmannshofer:2016jzy,Biswas:2016yan}. The gauge group $SU(3)_c \times SU(2)_L \times U(1)_Y\times U(1)_{L_{\mu}-L_{\tau}}$ implies that only the second and third lepton generations are charged under the new Abelian gauge symmetry, under which they carry opposite charges. As usual, the new Abelian gauge group leads to the existence of a new massive gauge bosons, $Z^{\prime}$, which can acquire a mass either via spontaneous symmetry breaking governed by a new scalar field, or through the Stueckelberg mechanism~\cite{Ruegg:2003ps,Feldman:2006wb}. Either way, the new boson couples to the SM lepton doublets via the term $\bar{L}\gamma^{\alpha}D_{\alpha}L$, where the covariant derivative is $D_{\alpha} =\partial_{\alpha} + ig^{\prime}\, q\, Z^{\prime}_{\alpha}$, with $g^{\prime}$ being the new gauge coupling of the $U(1)_{L_\mu-L_\tau}$ symmetry and $q$ the corresponding charge ($q_{\mu,\nu_{\mu}}=1$,$q_{\tau,\nu_{\tau}}=-1$). Writing down explicitly this term we get
\begin{equation}
{\cal L}_{\rm fermions} \supset g^{\prime} \left( \bar{\mu} \gamma_{\alpha} \mu - \bar{\tau} \gamma_{\alpha} \tau + \bar{\nu_{\mu}} \gamma_{\alpha} P_L \nu_{\mu} - \bar{\nu_{\tau}} \gamma_{\alpha} P_L \nu_{\tau} \right)Z^{\prime\alpha} \,.
\label{Eq:mu-tauLagran}
\end{equation}
The first term in Eq.~\eqref{Eq:mu-tauLagran} gives rise to a contribution to $g-2$, which we find to be [cf.\ Eq.~\eqref{Eq:doublyEamu}]
\begin{equation}\label{eq:mu-tauamu}
  \begin{aligned}
    \Delta a_\mu\left(Z^\prime\right) &= \frac{g^{\prime\, 2}}{8\pi^2} \frac{m_\mu^2}{m_{Z^\prime}^2} \int_0^1 \mathrm{d}x \frac{P_4^+(x)}{(1-x)\left(1-\lambda^2 x\right)+\epsilon_f^2\lambda^2 x},\\
    \textrm{where } P_4^+ &= 2x^2(1-x),\textrm{ with } \lambda \equiv \frac{m_\mu}{m_{Z^\prime}},\ \epsilon_f \equiv \frac{m_f}{m_{\mu}},
  \end{aligned}
\end{equation}
in agreement with~\cite{Leveille:1977rc, Jegerlehner:2009ry}.
 
Approximating to leading order for a $Z^{\prime}$ much heavier than the muon, one finds
\begin{equation} 
\label{eq:mu-tauamu2}
	\Delta a_\mu\left(Z^\prime\right) = 
	\frac{g^{\prime\,2}}{12\pi^2}\frac{m_\mu^2}{m_{Z^\prime}^2} \simeq 4 \times 10^{-9} \left(\frac{{\rm 304\, GeV}}{m_{Z^\prime}^2}\right)^2\left(\frac{g^{\prime}}{0.5}\right)^2,
\end{equation}which can potentially address the $g-2$ measurement within $2\sigma$. We emphasize that the result in Eq.~\eqref{eq:mu-tauamu} is completely general and applicable to any model with a gauge boson with purely vectorial couplings to muons. A more general result, including both vector and axial-vector couplings, as well as possible charged lepton mixings, was obtained in Eq.~\eqref{eq:P4def}.

{\bf Existing Limits}

From the result found in Eq.~\eqref{eq:mu-tauamu2} in the context of heavy vector mediators, one is led to the conclusion that $Z^\prime$ gauge bosons can explain the $g-2$ anomaly; however, it has been noted that in the context of the $L_\mu-L_\tau$ model this is not the case due to the existence of collider bounds based on the neutrino-trident production and the measurement of the $Z$ decay width into four leptons, as outlined below.

\begin{figure}
\includegraphics[scale=0.45]{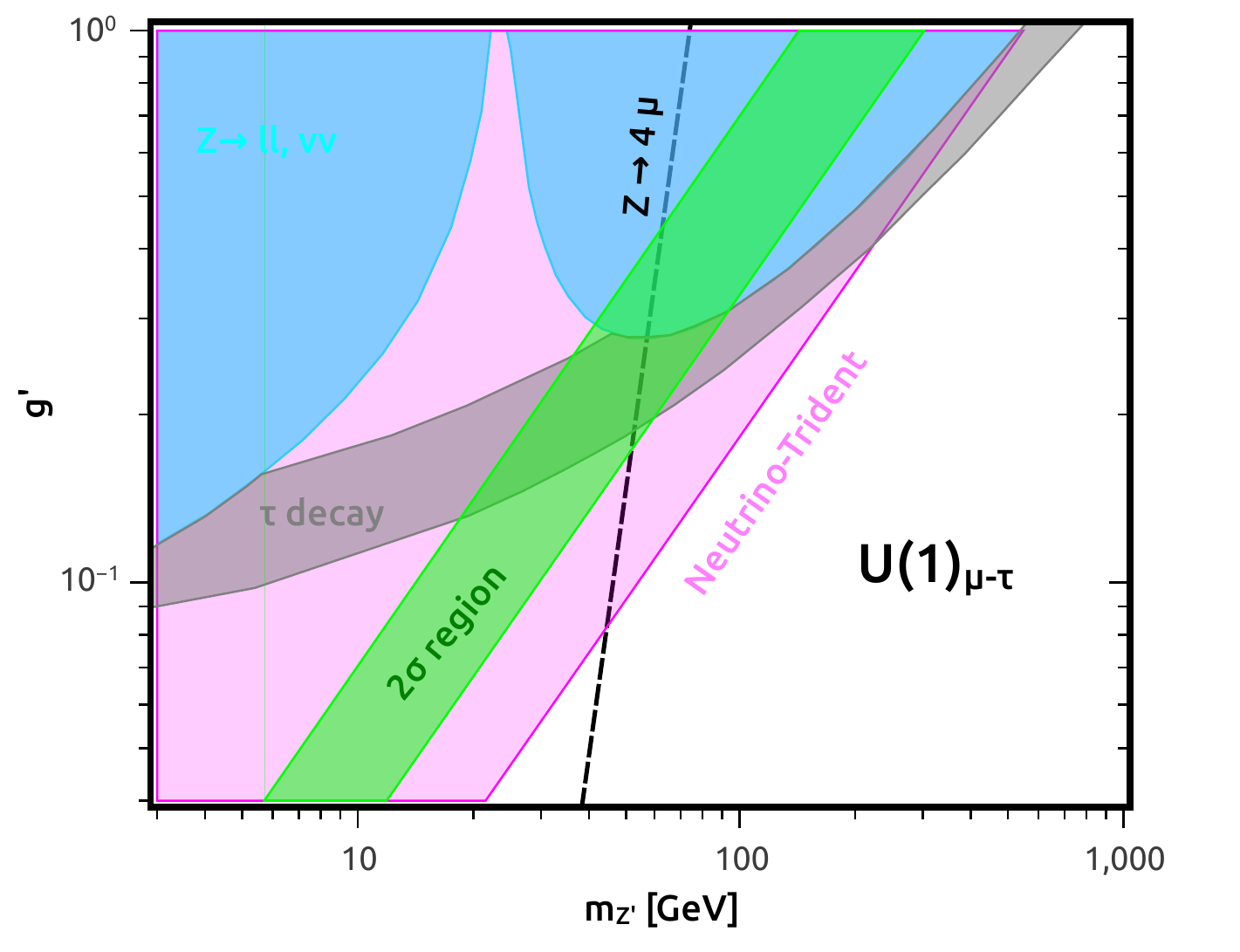}
\includegraphics[scale=0.45]{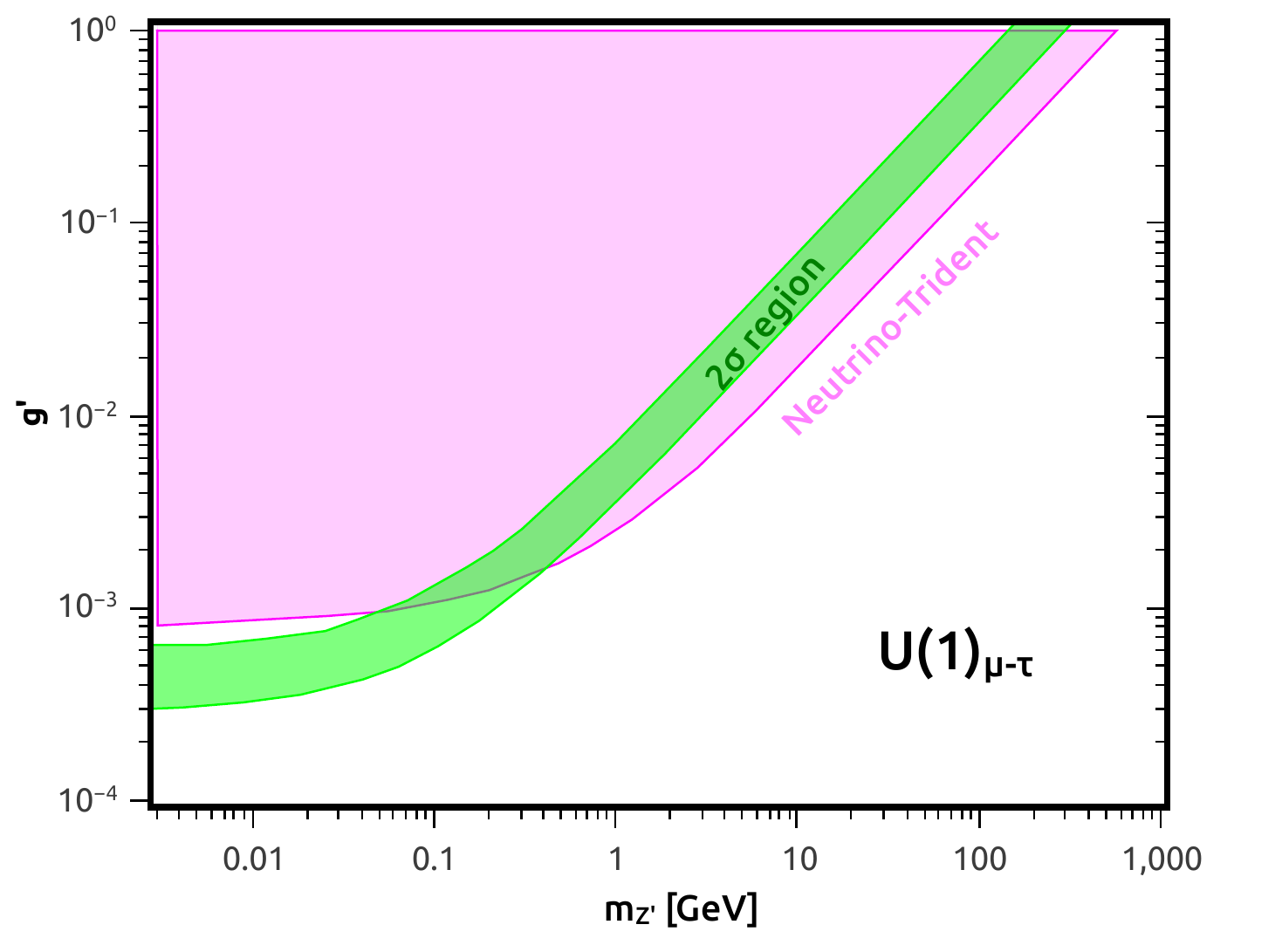}
\caption{Summary plot for the $L_\mu-L_\tau$ model. In the left plot we are concentrated on the large $Z^\prime$ mass regime showing that the existing constraints rule out the possibility to explain the $g-2$ anomaly in this model. In the right-panel we extend our findings to lower $Z^\prime$ masses showing that a small region of the parameter space is still alive for $m_{Z^\prime} < 300$~MeV and $g^\prime \sim 4\times 10^{-3}$. }
\label{mu-taufig}
\end{figure}

Assuming the $Z^\prime$ mass is generated through the VEV of a singlet scalar $v_{\phi}$, its mass is found to be $m_{Z^\prime} g^\prime v_{\phi}$. The bound arising from the neutrino-trident production imposes $m_{Z^\prime}/g^\prime > 750$~GeV, whereas the  limit from the measurement of the $Z$ decay width into four leptons required $m_{Z^\prime} \gtrsim 40$~GeV. These constraints decisively rule out the $L_\mu-L_\tau$ model as a possible explanation for the $g-2$ anomaly in this minimal setup for $m_{Z^\prime}$ masses larger than $300$~MeV. However, when the $Z^\prime$ becomes sufficiently light the neutrino-trident production bound loosens leaving a window for $L_\mu-L_\tau$ model to explain the $g-2$ anomaly. These findings are clearly exhibited in Fig.~\ref{mu-taufig}. A more detail discussion of these limits can be found in~\cite{Altmannshofer:2014cfa,Altmannshofer:2014pba,Allanach:2015gkd}.

\subsection{Dark Photon}

Dark photon models refer to an Abelian extension ($U(1)_X$) of the SM where the kinetic mixing dictates the observables~\cite{Holdom:1985ag,Galison:1983pa,Boehm:2003ha,Pospelov:2007mp,ArkaniHamed:2008qn,Pospelov:2008zw}. In the minimal setup the model only contains a new vector boson which interacts with the SM particles via kinetic mixing as follows
\begin{equation}
\mathcal{ L} =-\frac{1}{4} B_{\mu\nu} B^{\mu\nu} -\frac{1}{4}F'_{\mu\nu} F'^{\mu\nu} + \frac{\epsilon}{2} F'_{\mu\nu} B^{\mu\nu}+
\frac{1}{2} m_{A^\prime}^2 A^\prime_\mu A^{\prime \mu},
\end{equation}
where $F'_{{\mu\nu}}$ and $B_{{\mu\nu}}$ are the field strengths of the $U(1)_X$ and $U(1)_Y$ gauge groups, respectively; $\epsilon$ is the kinetic mixing parameter. Note that the mass term breaks $U(1)_X$ explicitly.

The kinetic mixing is the key input of the model and it governs the strength of the dark photon coupling with the SM particles. Such kinetic mixing is often taken to be zero, but it is generated via loops if there are new particles charged under $U(1)_X$. Changing to the basis of mass eigenstates, the $U(1)_X$ gauge boson, $A^{\prime}$, is much lighter than the SM $Z$ boson. For $\epsilon \ll 1$ we get,
\begin{equation}
\mathcal{L} = \epsilon^{\prime}\, e\, Q \bar{f} \gamma^{\mu} f A^{\prime},
\end{equation}where $\epsilon^{\prime}=\epsilon/c_W$ and $Q$ is electric charge of the fermion $f$.

Notice this interaction is identical to the QED Lagrangian except for the presence of the kinetic mixing parameter $\epsilon^\prime$. In this model, all interactions are flavor diagonal, and for this reason we have no LFV observables. If one embeds the dark photon model into two-Higgs doublet models, LFV processes might arise~\cite{Campos:2017dgc}.

\begin{figure}
\centering
\includegraphics[scale=0.4]{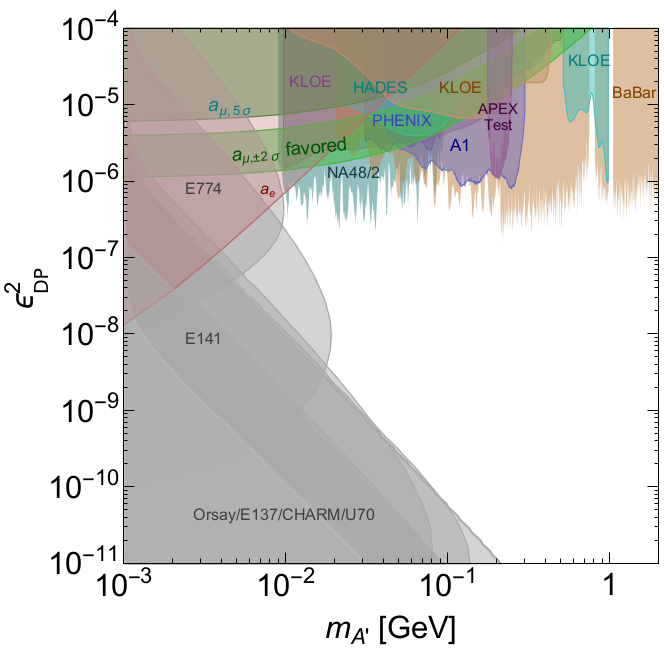}
\caption{Parameter space of the dark photon model in the kinetic mixing vs dark photon mass plane. The $\epsilon^2$ $2\sigma$ favored region to explain the muon anomalous magnetic moment is also shown.}
\label{DPfig}
\end{figure}

As far as the muon magnetic moment is concerned, the purely vectorial couplings to charged leptons have already been investigated before in the context of the $B-L$ and $L_{\mu}-L_{\tau}$ models. Therefore, the same results apply here. Altough, keep in mind that dark photon models typically are studied in the regime in which the new gauge boson is much lighter than the $Z$ boson, with masses at the MeV scale. For this reason the integration of Eq.~\eqref{Eq:doublyEamu} ought to be handled numerically. In Fig.~\ref{DPfig} we show the result for the dark photon which agrees well with the results presented in~\cite{Alexander:2016aln} and we also overlay with current collider and accelerator probes described below:

{\bf Existing Limits}

{\it Electron-Beam-Dump Experiments}

In 1987, an electron-beam-dump search reported null results in the search for neutral particles with masses in the range 1 to 15 MeV~\cite{Riordan:1987aw}. As any other electron-beam-dump experiment, the electron beam energy, the target, and the decay length are some of the key parameters. The particle, in this case the dark photon, needs to decay inside the detector. The decay width in the dark photon model, for instance, is found to be~\cite{Pospelov:2008zw}
\begin{equation}
\Gamma_{DP}=\frac{\alpha_{em}\epsilon^2}{3} m_A^\prime \left( 1+2\frac{m_l}{m_A^{\prime 2}}\right)\sqrt{1-4\frac{m_l^2}{m_A^{\prime 2}}}.
\end{equation}
Taking into account the geometry of the detector, the decay length, the dark photon mass, and differential production cross section of the dark photon that features a dependence on the target atomic mass and number and energy of the electron beam, one finds the exclusion limit in Fig.~\ref{DPfig}. Other electron-beam-dump experiments are dictated by the same physics and for this reason the exclusion limits look alike in shape. The dark photon mass reach is dictated by the beam energy, but falls rapidly with mass.
 
{\it Electron-Positron Colliders}

Low energy electron-positron colliders may produce dark photons via the mode $e^-e^+ \rightarrow A^\prime \gamma$. This search provides a very efficient probe for both visible and potential invisible decay modes of the dark photon. The mass reach is determined by the center of mass energy. Experiments such as KLOE and BaBar fall into this category.
 
 {\it Meson Decays}

 Rare meson decays such $K \rightarrow \pi A^\prime$ and $\pi^0 \gamma A^\prime$ are great laboratories to probe dark photon models. These mesons are produced in an enormous amount in accelerators and beam-dump experiments. A clear example is the Super Proton Synchrotron accelerator at the LHC which produces about $10^{11}$ charged kaons in the fiducial volume~\cite{Batley:2015lha}. In this case the dark photon mass is set by the parent meson particle.\\

In summary, from Fig.~\ref{DPfig} one can conclude that the original dark photon model can no longer address the $g-2$ anomaly in light of recent bounds. If one departs from the canonical dark photon, and adds invisible decays, one might alleviate the experimental bounds and find a small room to explain the $g-2$ anomaly~\cite{Davoudiasl:2014kua}. In this model, no LFV observables arise.

\subsection{Type~I Seesaw}

The type~I seesaw offers a plausible explanation for the lightness of the neutrino masses by making their masses inversely proportional to the masses of RH neutrinos supposibly heavy~\cite{Minkowski:1977sc,Mohapatra:1979ia,Schechter:1980gr,Keung:1983uu}.

We will briefly revise this mechanism and then provide the general expressions for the charged and neutral currents as well as the Yukawa terms involving the Higgs boson, since these are the ingredients for our phenomenological agenda. 

In the type~I seesaw mechanism, one adds three RH neutrinos ($N_R$) to the SM spectrum~\cite{Minkowski:1977sc,Mohapatra:1979ia,Schechter:1980gr,Keung:1983uu}. In this case the relevant Lagrangian is
\begin{equation}
\mathcal{L}=\mathcal{L}_{SM}+i\overline{N_R}\slashed \partial N_R-\left[ y_N  \overline{N_R}\tilde{H}^\dagger \ell_L +\frac{1}{2}M_R \overline{N_R} {N_R}^c+{\rm h.c.}\right] \,,
\label{eqseetypeI}
\end{equation}where $H$ is the SM Higgs doublet and $\tilde{H}$ the isospin transformation of the Higgs doublet,
\begin{equation}
H=\left(\begin{array}{c}
\phi^-\\
\frac{1}{\sqrt{2}}\left(v+h+i\phi^3\right)
\end{array}\right)\,,
\end{equation}where $\tilde{H}=i\tau_2H^*$ with $v= 246$~GeV. 

The presence of RH neutrinos will induce a mass mixing with the active neutrinos. Keeping the charged lepton mass matrix diagonal, without loss of generality, we can transform the neutrino fields into the basis where the mass matrix is diagonal via a unitary mixing matrix of size $(3+k)\times (3+k)$, where $k$ is number of RH neutrinos added in Eq.~\eqref{eqseetypeI},
\begin{equation}
\left(
\begin{array}{c}
\nu_L\\
{N_R}^{c}\\
\end{array}
\right)= U \, P_L\, n \equiv U \, P_L\,
 \left(\begin{array}{c} \nu_{1}\\ \nu_{2}\\ \nu_3\\N_1\\ \vdots \\ N_k\\ \end{array}\right)\,.
\label{Udefseesaw}
\end{equation}where $P_L=(1-\gamma_5)/2$ is the LH projection operator.

Notice that $n$ in Eq.~\eqref{Udefseesaw} is a vector containing all neutrino mass-eigenstates.  Since these neutrinos mix with each other, they are automatically Majorana fermions, i.e.~$n=n^c$. This mixing changes the charged and neutral currents, and this change is proportional to the mixing matrix.

\begin{figure}
\centering
\includegraphics[scale=0.3]{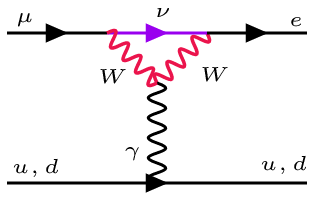}
\includegraphics[scale=0.4]{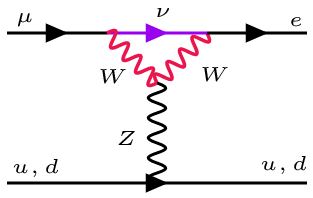}
\includegraphics[scale=0.3]{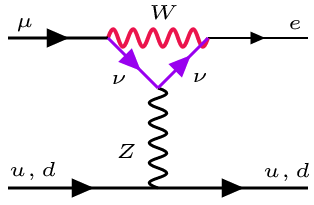}
\includegraphics[scale=0.3]{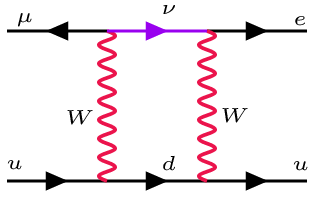}
\includegraphics[scale=0.3]{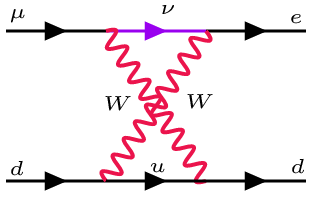}
\caption{Feynmann diagrams contributing to $\mu-e$ conversion in the type~I seesaw model. These five diagrams are known respectively as (i) photon penguin, (ii) $Z$ penguin, (iii) $Z$ penguin, (iv) $W$ box diagram, (v) $W$ box diagram. }
\label{typeIseesawfeyn}
\end{figure}

To make our expressions more compact in what follows, we will define the product of the mixing matrix as
\begin{equation}
C_{ij}\equiv\sum_{\alpha=1}^3 U^\dagger_{i \alpha}U_{\alpha j }.
\end{equation}
Under the assumption that $y_N v/m_N$ is much smaller than unit, the mixing matrix U can written as
\begin{equation}
U=
\begin{pmatrix}
		U_{\nu\nu}& U_{\nu N} \\
		U_{N\nu}  & U_{NN}
\end{pmatrix}\,,
\label{Umatrix}
\end{equation}with
\begin{subequations}\label{approxUs}
\begin{align}
U_{\nu\nu} =& (1-\frac{\epsilon}{2})U_{{\rm PMNS}}\, , \quad
&&U_{\nu N} = y_N^\dagger (M)^{-1}\frac{v}{\sqrt{2}}\,, \\
U_{N\nu} =& -M^{-1}y_N \frac{v}{\sqrt{2}} U_{\nu\nu}\,, \quad
&&U_{NN}=(1-\epsilon'/2),
\end{align}
\end{subequations}
where $U_{{\rm PMNS}}$ is the usual PMNS $3\times 3$ mixing matrix appearing in the charged current, and 
$\epsilon$ and $\epsilon^\prime$ account for second order corrections of the order of $(y_N v/m_N)^2$, and are found to be respectively,
\begin{equation}
\epsilon\equiv\frac{v^2}{2}y_N^\dagger M^{-2}y_N,\quad
\epsilon'\equiv\frac{v^2}{2} M^{-1}Y_N y_N^\dagger M^{-1}.
\end{equation}
Now that we have defined the mixing matrices, there is one thing left for us to do before presenting the Lagrangians in the flavor basis relevant for the observables we will investigate, that is, the mass matrix defined as
\begin{equation}
m_n=\mbox{Diag}\left(m_{n_i}\right)=\mbox{Diag}(m_{\nu_1},m_{\nu_2},m_{\nu_3},m_{N_1},...,m_{N_k},)
\end{equation}where the first three entries of the mass matrix, $m_n$, represent the light neutrino masses, while the others that carry index $k$, the heavy masses. In particular, the active neutrino mass matrix are given by
\begin{equation}
 m_\nu=-\frac{v^2}{2}y_N^T\frac{1}{M}y_N\,.
\end{equation} 
With these at hand, one can write the Yukawa Lagrangian and currents in the unitary gauge
\begin{subequations}
\begin{align}
\mathcal{L}^{W^{\pm}} =&   \frac{g}{\sqrt{2}} W^-_\mu \ \overline{l}_\alpha \gamma^\mu U_{\alpha i} P_L n_i + h.c. \,,\\
\mathcal{L}^{Z} =&   \frac{g}{2 c_W} Z_\mu \ \overline{n}_i \gamma^\mu C_{ ij} P_L n_j = \frac{g}{4 c_W} Z_\mu \ \overline{n}_i \gamma^\mu \left[C_{ ij} P_L - C^*_{ ij} P_R\right] n_j \,,\\
\mathcal{L}^{h} =&      -\frac{  g}{2 m_W} h \ \overline{n}_i    C_{ ij} \left(  m_{n_i} P_L +m_{n_j}   P_R \right)  n_j \,,\\
=&-\frac{  g}{4 m_W} h \ \overline{n}_i   \left[ C_{ ij} \left(  m_{n_i} P_L +m_{n_j}   P_R \right)+C^*_{ ij} \left(  m_{n_i} P_R +m_{n_j}   P_L \right)\right] n_j\,, \nonumber
\end{align}
\end{subequations}
where $c_W$ is the cosine of the weak mixing angle, $\alpha,\beta$ runs through the three flavors, and $i,j$ through the $3+k$ mass eigenstates.

The Lagrangian above captures all relevant information needed to discuss the $g-2$, the muon decays $\mu \to e\gamma$, $\mu \to 3e$ and $\mu-e$ conversion.

{\bf Lepton Flavor Violating Muon decays}

The lepton flavor violating muon decay processes arise at one loop level as shown in the first three Feynman diagram in Fig.~\ref{typeIseesawfeyn} (please ignore the bottom Feynman diagrams which are only relevant for $\mu-e$ conversion). These five diagrams are known respectively as (i) photon penguin, (ii) $Z$ penguin, (iii) $Z$ penguin, (iv) $W$ box diagram, (v) $W$ box-crossed diagram. 

We shall start  with the $\mu \to e\gamma$ decay whose amplitude is
\begin{equation}
  i  \mathcal{M}=  \frac{ ieg_W^2}{2 (4\pi)^2 M^2_W} \epsilon^\mu_\lambda(q) \overline{u}_e(p')\Big[ F^{\mu e }_\gamma (q^2\gamma_\mu - {\not}q q_\mu) P_L   - i \sigma_{\mu\nu} q^\nu G^{\mu e}_\gamma (m_eP_L+m_\mu P_R) \Big] u_\mu(p)\,,
    \label{muegammageneral}
\end{equation}where $q=p-p'$ is the photon momentum.

The amplitude has the usual form where one term is proportional to the monopole $F^{\mu e}$ and other proportional to the electric dipole moment $G^{\mu e}$, with the former accounting for the off-shell photon emission. For the $\mu \to e\gamma$ decay, the on-shell contribution is the relevant one and it leads to~\cite{Minkowski:1977sc,Cheng:1980tp},
\begin{equation}
 {\rm BR}(\mu \to e \gamma)=\frac{\alpha^3_W s^2_W}{256 \pi^2}\frac{m^4_\mu}{m^4_W}\frac{m_\mu}{\Gamma_\mu}\big| G^{\mu e}_\gamma \big|^2 \,,  
    \label{muegexactrate}
\end{equation}where $\alpha_W= g^2/4\pi$, and 
\begin{equation}
G^{\mu e }_\gamma = \sum_{i=1}^{3+k} U_{ei}U^*_{\mu i} G_\gamma(x_{i}) = \sum_{i=1}^{k} U_{e N_i}U^*_{\mu N_i} G_\gamma(x_{N_i}),
\end{equation}with $x_{N_i}=m_{N_i}^2/m_W^2$ and
\begin{equation}
G_\gamma (x) = -\frac{x (2x^2+5x-1)}{4(1-x)^3}-\frac{3x^3}{2(1-x)^4} \ln(x).
\end{equation}
Hence, in the limiting cases $x\gg 1$ and $x \ll 1$, we get
\begin{eqnarray}
G^{\mu e }_\gamma (x_{N_i})=\sum_{i=1}^{k} U_{e N_i}U^*_{\mu N_i} G_\gamma(x_{N_i}) \frac{x_{N_i}}{4}\,\,\mbox{for} \, \,, x \ll1, \nonumber\\
G^{\mu e }_\gamma (x_{N_i})=\sum_{i=1}^{k} U_{e N_i}U^*_{\mu N_i} G_\gamma(x_{N_i}) \frac{1}{2}\,\, \mbox{for} \, \,, x \gg .1 \nonumber\\
\end{eqnarray}
\begin{table}[t]
\centering
\begin{tabular}{c | c | c | c | c |c |c }
\hline\hline
Nucleus $^A_Z \mbox{N}$          &$V^{(p)}$& $V^{(n)}$& $D$      & $Z_\mathrm{eff}$   & $|F_p(-m^2_\mu)|$ &$ \Gamma_{capt} $ ($10^6s^{-1}$)\\
\hline
$_{13}^{27}\mbox{Al}$ & 0.0161   & 0.0173    & 0.0362 & $11.5$        &$0.64$                       &  0.7054\\
$_{22}^{48}\mbox{Ti}$ & 0.0396   & 0.0468    & 0.0864 & $17.6$         &$0.54$                      & 2.59\\
$^{197}_{79} \mbox{Au}        $  & 0.0974   & 0.146      & 0.189    & $33.5$         &$0.16$                     & 13.07\\
$^{208}_{82} \mbox{Pb }       $  & 0.0834   & 0.128      &  0.161   & $34.0$         &$0.15$                     & 13.45\\
\hline
\end{tabular} 
\caption{Nuclear form factors and capture rates as presented in~\cite{Kitano:2002mt,Suzuki:1987jf}.}
\label{nuclfactorsmueconv}
\end{table}
As for the $\mu-e$ conversion, we have both the monopole and the dipole moment terms. The relevant quantity is the branching ratio defined in Eq.~\eqref{defmueconversion} and it is found to be
\begin{equation}
{\rm CR} (\mu - e)=\frac{2G_F^2\alpha_{W}^2m_\mu^5}{(4\pi)^2 \Gamma_{capt}
}\left|4V^{(p)}\left(2 \tilde{F}_{u}^{\mu e}+\tilde{F}_{d}^{\mu e}\right)+4V^{(n)}\left(\tilde{F}_u^{\mu e}+2\tilde{F}_{d}^{\mu e}\right)+ s^2_w G^{\mu e}_{\gamma} D/(2e)  \right|^2
\label{BRmueexacteq}
\end{equation}
with  
\begin{equation}
\tilde F_q^{\mu e}=Q_q s_W^2 F^{\mu e}_\gamma+F^{\mu e}_Z\left(\frac{{\mathcal I}^3_q}{2}-Q_qs_W^2\right)+\frac{1}{4}F^{\mu eqq}_{box}\,,
\label{formfact}
\end{equation}where $q=u,d$, $Q_q$ is the quark electric charge ($Q_u=2/3, Q_d=-1/3$), $\mathcal{I}^3_q$ is the weak isospin ($\mathcal{I}^3_u=1/2\,,\, \mathcal{I}^3_d=-1/2$). The nuclear form factors are given in Table \ref{nuclfactorsmueconv}, and other form factors appearing in Eq.~\eqref{formfact} arise from loop calculations and are defined as
\begin{subequations}\allowdisplaybreaks\label{FormfactmueZ}
\begin{equation}
	 F^{\mu e }_\gamma = \sum_{i=1}^{3+k} U_{ei}U^*_{\mu i} F_\gamma(x_i)   = \sum_{i=1}^{k} U_{e N_i}U^*_{\mu N_i} F_\gamma(x_{N_i})\,,
	 \label{Fmuegamma}\\
\end{equation}
\begin{eqnarray}
	 F^{\mu e }_Z			 &=& \sum_{i,j=1}^{3+k} U_{ei}U^*_{\mu j} \left[\delta_{ij} F_Z(x_i) + C_{ij} G_Z(x_i,x_j) + C^*_{ij} H_Z(x_i,x_j)   \right]\,, \\
F^{\mu e }_Z	 &=& \sum_{i,j=1}^{k} U_{e N_i}U^*_{\mu N_j} \Big[\delta_{N_iN_j} \left(F_Z(x_{N_i}) +2 G_Z(0,x_{N_i})	\right) + \\
&& +C_{N_iN_j} \left(G_Z(x_{N_i},x_{N_j})- G_Z(0,x_{N_i})-G_Z(0,x_{N_j})	\right) + C^*_{N_iN_j}H_Z  (x_{N_i},x_{N_j}) \Big]\,,\nonumber
	 \end{eqnarray}
\begin{eqnarray}	 
 F^{\mu e uu}_{Box}&=&\sum_{i=1}^{3+k}\sum_{d_i=d,s,b} U_{ei}U^*_{\mu i} V_{u d_i} V^*_{u d_i} F_{Box}(x_i,x_{d_i}) \simeq \sum_{i=1}^{3+k} U_{ei}U^*_{\mu i} F_{Box}(x_i,0)  \nonumber\\
	 &=&\sum_{i=1}^{k} U_{eN_i}U^*_{\mu N_i} \left[F_{Box}(x_{N_i},0)-F_{Box}(0,0) \right]\,,
	  \label{Formfactmueuu}
\end{eqnarray}
\begin{eqnarray}	 
	 F^{\mu e dd}_{Box}&=&  \sum_{i=1}^{3+k}\sum_{u_i=u,c,t} U_{ei}U^*_{\mu i} V_{d  u_i } V^*_{d u_i} F_{XBox}(x_i,x_{u_i}) \simeq         \sum_{i=1}^{3+k} U_{ei}U^*_{\mu i}   F_{XBox}(x_i,0) \nonumber\\
	 &=&\sum_{i=1}^{k} U_{eN_i}U^*_{\mu N_i} \left[F_{XBox}(x_{N_i},0)-F_{XBox}(0,0) \right]\, ,
	 \label{Formfactmuedd}
\end{eqnarray}
\begin{eqnarray}	 
	 	 F^{\mu eee}_{Box}&=&  \sum_{i,j=1}^{3+k} U_{ei}U^*_{\mu j}\left[U_{ei}U^*_{ej}G_{Box}(x_i,x_j)-2\,U^*_{ei}U_{ej}F_{XBox}(x_i,x_j)\right]  \\ \nonumber
	 &=&-2\sum_{i=1}^{k} U_{eN_i}U^*_{\mu N_i} \left[F_{XBox}(x_{N_i},0)-F_{XBox}(0,0) \right]\\ \nonumber
	 &&
	+\sum_{i,j=1}^{k}U_{eN_i}U^*_{\mu N_j}\Big\{ U_{eN_i}U^*_{eN_{j}}G_{Box}(x_{N_{i}},x_{N_j})-2\, U^*_{eN_i} U_{e N_j}\big[F_{XBox}(x_{N_i},x_{N_j})\\ &&- F_{XBox}(0,x_{N_j})
- F_{XBox}(x_{N_i},0)+F_{XBox}(0,0) \big]\Big\},
	 \label{Formfactmueee}
\end{eqnarray}
\end{subequations}
where $x_i$ is a vector that encompasses three active neutrinos and $k$ heavy neutrinos. That said, $x_{1,2,3}\equiv x_{\nu_{1,2,3}}\equiv m^2_{\nu_{1,2,3}}/m^2_W$,  $x_{4,...,3+k}\equiv x_{N_{1,...,k}}= m^2_{N_{1,...,k}}/m^2_W$, $x_q\equiv m^2_{q}/m^2_W$, with $U$ being the neutrino mass mixing matrix defined in Eq.~\eqref{Umatrix} and $V$ the CKM matrix. The remaining functions are identified as
\begin{subequations}\allowdisplaybreaks
\begin{eqnarray}
F_\gamma(x)&=& 	\frac{x(7x^2-x-12)}{12(1-x)^3} - \frac{x^2(x^2-10x+12)}{6(1-x)^4} \ln x	\, ,\\
F_Z(x)&=& -\frac{5x}{2(1-x)}-\frac{5x^2}{2(1-x)^2}\ln x \, , \\
G_Z(x,y)&=& -\frac{1}{2(x-y)}\left[	\frac{x^2(1-y)}{1-x}\ln x - \frac{y^2(1-x)}{1-y}\ln y	\right]\, , \\
H_Z(x,y)&=&  \frac{\sqrt{xy}}{4(x-y)}\left[	\frac{x^2-4x}{1-x}\ln x - \frac{y^2-4y}{1-y}\ln y	\right] \, ,\\
F_{Box}(x,y)&=&\frac{1}{x-y}\Big\{\left(	4+\frac{xy}{4} 	\right) \left[\frac{1}{1-x}+\frac{x^2}{(1-x)^2} \ln x - \frac{1}{1-y}-\frac{y^2}{(1-y)^2} \ln y\right] \nonumber \\
 	 	&&-2 xy \left[\frac{1}{1-x}+\frac{x }{(1-x)^2} \ln x - \frac{1}{1-y}-\frac{y }{(1-y)^2} \ln y\right]\Big\} \, , \\
F_{XBox}(x,y)&=&\frac{-1}{x-y}\Big\{\left(	1+\frac{xy}{4} 	\right) \left[\frac{1}{1-x}+\frac{x^2}{(1-x)^2} \ln x - \frac{1}{1-y}-\frac{y^2}{(1-y)^2} \ln y\right] \nonumber \\
 	 	&&-2 xy \left[\frac{1}{1-x}+\frac{x }{(1-x)^2} \ln x - \frac{1}{1-y}-\frac{y }{(1-y)^2} \ln y\right]\Big\} \, ,
		\\ G_{Box}(x,y)&=&\frac{-\sqrt{xy}}{x-y}\Big\{\left(	4+xy 	\right) \left[\frac{1}{1-x}+\frac{x}{(1-x)^2} \ln x - \frac{1}{1-y}-\frac{y}{(1-y)^2} \ln y\right] \nonumber \\
 	 	&&-2 \left[\frac{1}{1-x}+\frac{x^2 }{(1-x)^2} \ln x - \frac{1}{1-y}-\frac{y^2 }{(1-y)^2} \ln y\right]\Big\} \, .
 \end{eqnarray}
\end{subequations}
One can already see how much more complicated the $\mu-e$ conversion is compared to the $\MEG$ decay. There are several form factors related to nuclear dependence and more Feynmann diagrams that are relevant for a rigorous computation of this LFV observable. If one is interested in asymptotic behavior  to simplify quite a bit the calculation, the limiting cases $x\gg 1$ and $x\ll 1$ are of interest. In these cases the functions read
\begin{subequations}\allowdisplaybreaks
\begin{align}
G_Z(0,x) & =  -\frac{x}{2(1-x)} \ln x  \, , & H_Z(0,x)  &= G_{Box}(0,x)=0 \,\\
 F_{Box}(0,x)	 &  =    \frac{4}{ 1-x } + \frac{4x}{(1-x)^2}\ln x\, ,&
F_{XBox}(0,x)  &=   - \frac{1}{ 1-x } - \frac{x}{(1-x)^2}\ln x \, ,\\
F_\gamma(x ) & \xrightarrow[x\ll 1]{}  -x \,, &  F_\gamma(x )  &  \xrightarrow[x\gg 1]{}    - \frac{7}{12}-\frac{1}{6} \ln x\, , \\
G_\gamma(x )  & \xrightarrow[x\ll 1]{}  \frac{x}{4}\, ,&  G_\gamma(x )  &  \xrightarrow[x\gg 1]{}    \frac{1}{2} \,\\
F_Z(x) &  \xrightarrow[x\ll 1]{}    -\frac{5x}{2} \, ,  & F_Z(x )		  &  \xrightarrow[x\gg 1]{}   \frac{5}{2}-\frac{5}{2} \ln x\, , \\
G_Z(0,x ) & \xrightarrow[x\ll 1]{}  -\frac{1}{2} x \ln x
 \, , & G_Z(0,x ) & \xrightarrow[x\gg 1]{}  \frac{1}{2} \ln x  \, ,\\
 F_{Box}(0,x )& \xrightarrow[x\ll 1]{}  4\left(1+x\left(1+\ln x\right)\right)
  \, , & F_{Box}(0,x ) & \xrightarrow[x\gg 1]{}  0 \, , \\
 F_{XBox}(0,x )& \xrightarrow[x\ll 1]{}  -1-x\left(1+\ln x\right)
 \, , & F_{XBox}(0,x ) & \xrightarrow[x\gg 1]{}  0 \, .\label{limitval2}
\end{align}
\end{subequations}
Indeed, it is more intuitive to consider a limiting case so we can understand the key features of the $\mu-e$ conversion process. For a light nucleus of atomic mass and number equal to $A$ and $Z$ respectively, and assuming that the RH neutrinos have the same mass, $m_N$, we find
\begin{equation}
{\rm CR} (\mu -e)=\frac{G_F^2 \alpha_W^2\alpha^3m_\mu^5}{8\pi^4\Gamma_{ {\rm capt}}}\frac{Z_\mathrm{eff}^4}{Z}F_p^2  J(A,Z) \left|\sum_{i=1}^{k} U_{e N_i}U^*_{\mu N_i} \right|^2\, ,
 \label{BRmueconlimit}
\end{equation}
where
\begin{equation}
J(A,Z)= \left[ (A+Z) F_u(x_N) + (2A-Z)F_d(x_N)\right]^2,
\label{EqJAZ}
\end{equation}
with
\begin{equation}
F_u (x)= \frac{(32 \log(x)-62)s_W^2}{36} -\frac{(3+3\log(x))}{8},
\end{equation}
\begin{equation}
F_d (x) = \frac{(16 \log(x)-31)s_W^2}{36} -\frac{(3-3\log(x))}{8},
\end{equation}
which should be compared with the $\MEG$ decay under the same assumptions:
\begin{equation}
 {\rm BR} (\mu \to e \gamma)=\frac{\alpha^3_W s^2_W}{256 \pi^2}\frac{m^4_\mu}{m^4_W}\frac{m_\mu}{\Gamma_\mu} G^{2}_\gamma (x_N) \left|\sum_{i=1}^{k} U_{e N_i}U^*_{\mu N_i} \right|^2\,.
    \label{muedecaylimit}
\end{equation}

As expected $\mu-e$ conversion and the $\MEG$ decay depend on the same mixing matrices, i.e.~the same neutrino physics. The effective atomic number, $Z_\mathrm{eff}$, the nuclear form factor $F_p$, and the capture rate $\Gamma_{{\rm capt}}$ appearing
in Eq.~\eqref{EqJAZ} are given in Tab.~ \ref{nuclfactorsmueconv}. The first two are related since
$V^{(p)}/\sqrt{Z}= (4\pi)^{-1} Z_\mathrm{eff}^2 F_p \alpha^{3/2}$~\cite{Kitano:2002mt,Suzuki:1987jf}.

Having Eq.~\eqref{BRmueconlimit} and Eq.~\eqref{muedecaylimit} at hand, we conclude that the ratio ${\rm CR} (\mu \to e)/{\rm BR}(\MEG)$ is determined up to the RH neutrino mass for a given nucleus. For heavy nuclei we cannot use these equations though. We actually need to go back to the general and more complex expressions. Once one computes the ratio between these two observables, the dependence on the mixing matrices disappears. Therefore, one can draw this ratio as a function of the RH neutrino mass as shown Fig.~\ref{mutoeratio1}, where one can notice several features:

\begin{itemize}

\item {\it Light RH neutrinos}: One can see that for RH neutrino masses below $100$~GeV the $\mu-e$ conversion rate is larger than the $\MEG$ branching ratio;

\item {\it Heavy RH Neutrinos:} For RH neutrinos heavier than $100$~TeV again the $\mu-e$ conversion rate is larger than the $\MEG$ branching ratio;

\item {\it Cancellation Mechanism:} For every single nucleus there is a RH neutrino mass where the $\mu-e$ conversion rate vanishes explaining the behavior seen in the figure. This occurs because $\tilde{F}_u^{\mu e}$ and $\tilde{F}_u^{\mu e}$ in Eq.~\eqref{BRmueexacteq} have opposite signs, and for a given RH neutrino mass they cancel each other. From Eq.~\eqref{BRmueexacteq} it is hard to see this effect. However looking at the Eq.~\eqref{EqJAZ} which is valid for light nuclei, one can easily see that this cancellation happens for
\begin{equation}
\frac{F_u}{F_d}= -\frac{2A-Z}{A+Z}.
\end{equation}
For heavy nuclei, the relation is more complicated and a more general treatment is needed. Generally for light or heavy nuclei the $\mu-e$ conversion the rate vanishes if
 \begin{equation}
m_N^2\Big|_0=M_W^2\,\mbox{exp}\left(\frac{\frac{9}{8} V^{(n)}+\left(\frac{9}{8}+\frac{37 s_W^2}{12}\right) V^{(p)}-\frac{ s_W^2}{16 e}D}{\frac{3 }{8}V^{(n)}+\left(\frac{4 s_W^2}{3}-\frac{3}{8}\right) V^{(p)}}\right).
\label{vanishmassfull}
\end{equation}
Eq.~\eqref{vanishmassfull} explicitly shows that different nuclei lead to a cancellation at different RH neutrino masses in agreement with~\cite{Alonso:2012ji}. Nevertheless, Eq.~\eqref{vanishmassfull} is not so robust because there are uncertainties of the order of 10\% surrounding the form factors and these have a direct impact on which RH neutrino mass leads to a vanishing $\mu-e$ conversion rate.

\item {\it Overall Sensitivity}: In order to assess the overall sensitivity of $\mu-e$ conversion or $\MEG$ in the type~I seesaw mechanism, one needs to have in mind the experimental bounds on the individual branching ratios. We have seen above that depending on the RH neutrino mass the ratio ${\rm CR}(\mu-e)/{\rm BR}(\MEG)$ varies a lot, but since the current experimental limit on the $\MEG$ is much more restrictive than the one on ${\rm CR}(\mu-e),  \MEG$ offers a much more effective probe to the type~I seesaw mechanism. Be that as it may, further into the future, we expect a tremendous improvement on the bounds on ${\rm CR}(\mu-e)$ which will eventually invert the picture making ${\rm CR}(\mu-e)$ more constraining than $\MEG$.

\end{itemize}

\begin{figure}
\centering
\includegraphics[scale=0.6]{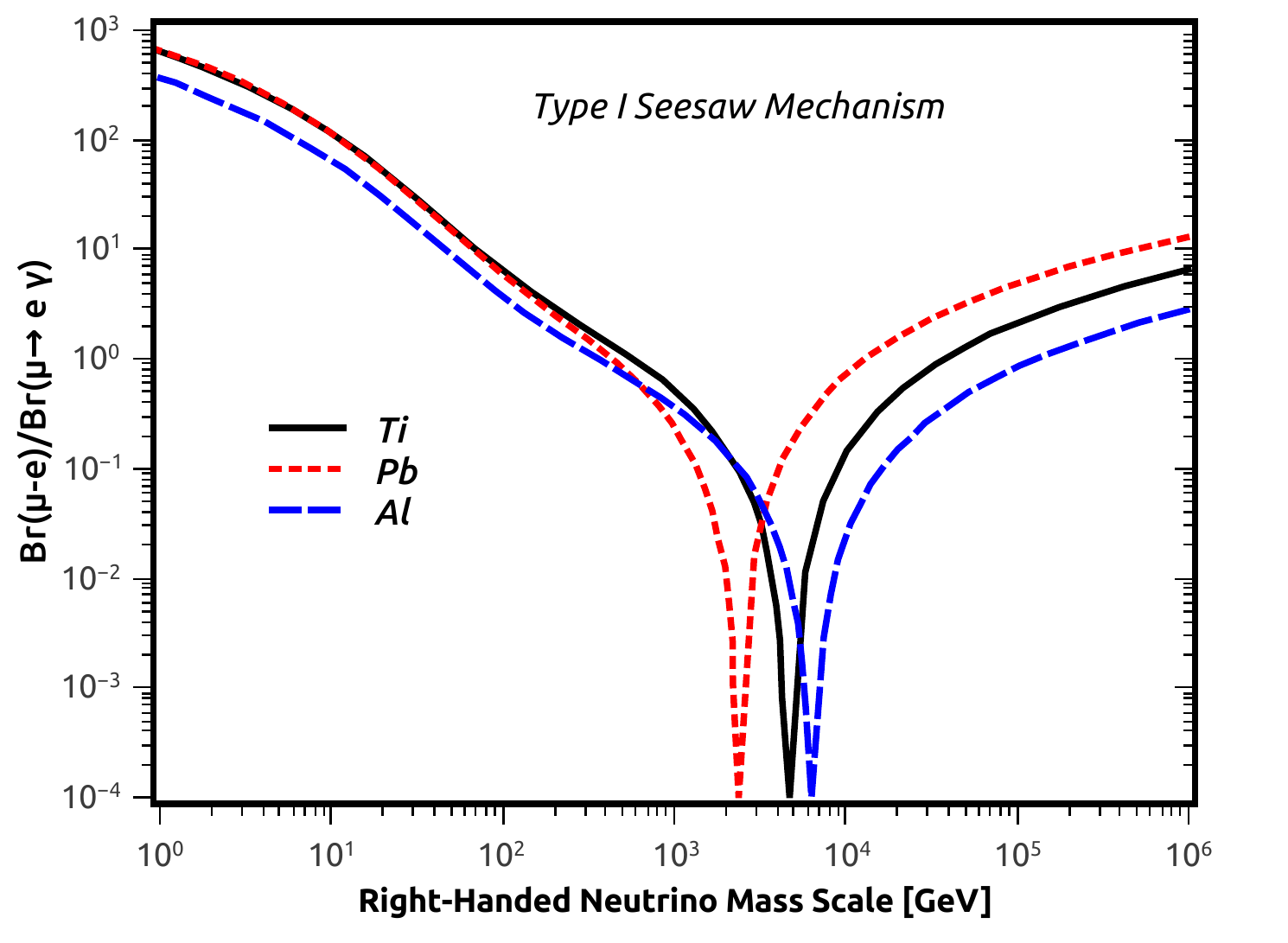}
\caption{Ratio ${\rm CR}(\mu-e)/{\rm BR}(\MEG)$ as a function of the RH neutrino mass scale under the assumption that all RH neutrino have the same mass  for three different nuclei: Al (blue), Pb (red) and Ti (black).}
\label{mutoeratio1}
\end{figure}

Now going back to Eq.~\eqref{muegammageneral}, one can use it to obtain the branching ratio for the $\mu \to eee$ decay after taking into account that more diagrams arise due to the off-shell emission of a photon and $Z$ boson to find~\cite{Ilakovac:1994kj}
\begin{align}
{\rm BR}(\mu   \rightarrow eee)=& \frac{\alpha^4_w }{24576\pi^3}\frac{m^4_\mu}{M^4_W}\frac{m_\mu}{\Gamma_\mu} \nonumber \\
		&\times \Bigg\{ 2 \left|\frac{1}{2}F^{\mu eee}_{Box}+F^{\mu e}_Z-2s^2_w(F^{\mu e}_Z-F^{\mu e}_\gamma)\right|^2+4 s^4_w \left|F^{\mu e}_Z-F^{\mu e}_\gamma\right|^2 \nonumber \\
		&+ 16 s^2_w Re\left[	(F^{\mu e}_Z +\frac{1}{2}F^{\mu eee}_{Box})	G^{\mu e*}_\gamma 			\right]		- 48 s^4_w Re\left[	(F^{\mu e}_Z-F^{\mu e}_\gamma)	G^{\mu e*}_\gamma 			\right]	\nonumber 		\\
		&+32 s^4_w |G^{\mu e}_\gamma|^2\left[		\ln \frac{m^2_\mu}{m^2_{e}} -\frac{11}{4}		\right]		\Bigg\}\,,  \label{mueee}
\end{align}where all these functions have already been defined above. 

The comparison between $\mu-e$ conversion and $\mu \to eee$ is no longer  trivial, because ${\rm CR}(\mu-e)$  depends on $G^{\mu e}_\gamma$ only, whereas ${\rm BR}(\mu \to eee)$ relies on several functions. Although, if one looks carefully the functions $F_{Box}^{\mu eee}$, $F_Z^{\mu e}$, and $F_\gamma^{\mu e}$ defined in Eqs.~\eqref{FormfactmueZ}, it will be visible that they all depend on the mixing matrices $U_{eN}U_{\mu N}$ after all, which is identified with $G^{\mu e}_\gamma$. Therefore, the ratio ${\rm CR}(\mu-e)/{\rm BR}( \mu \to eee)$ is neither dependent on $G^{\mu e}_\gamma$, nor on the mixing matrices, and thus the ratio varies only with the RH neutrino mass scale.  

In Fig.~\ref{mutoeratio1} we plot this ratio as a function of the RH neutrino mass. The same features presented before operate here to explain the behavior of the ratio ${\rm CR}(\mu-e)/{\rm BR}(\MEG)$. 

In summary, in the type~I seesaw mechanism, the $\MEG$ decay offers the most restrictive bounds, however in the very long run when a huge improvement is expected on the experimental sensitivity over $\mu-e$ conversion and $\mu \to eee$ observables, the picture changes. 

An important aspect that should be highlighted is that all this information could be used to discover the type~I seesaw mechanism. This could in in principle be done because the ratios ${\rm CR}(\mu-e)/{\rm BR}(\MEG)$ and ${\rm CR}(\mu-e)/{\rm BR}(\mu \to eee)$ are fixed for a given RH neutrino mass scale. If RH neutrinos are discovered at collider experiments, one can pin point what are these ratios for the type~I seesaw mechanism and discriminate it from other neutrino mass generation mechanisms.

Now we have addressed LFV observables in the type~I seesaw mechanism it is a good timing to discuss the existence of several other bounds on this mechanism to have an overall picture of this model.

\begin{figure}
\centering
\includegraphics[scale=0.6]{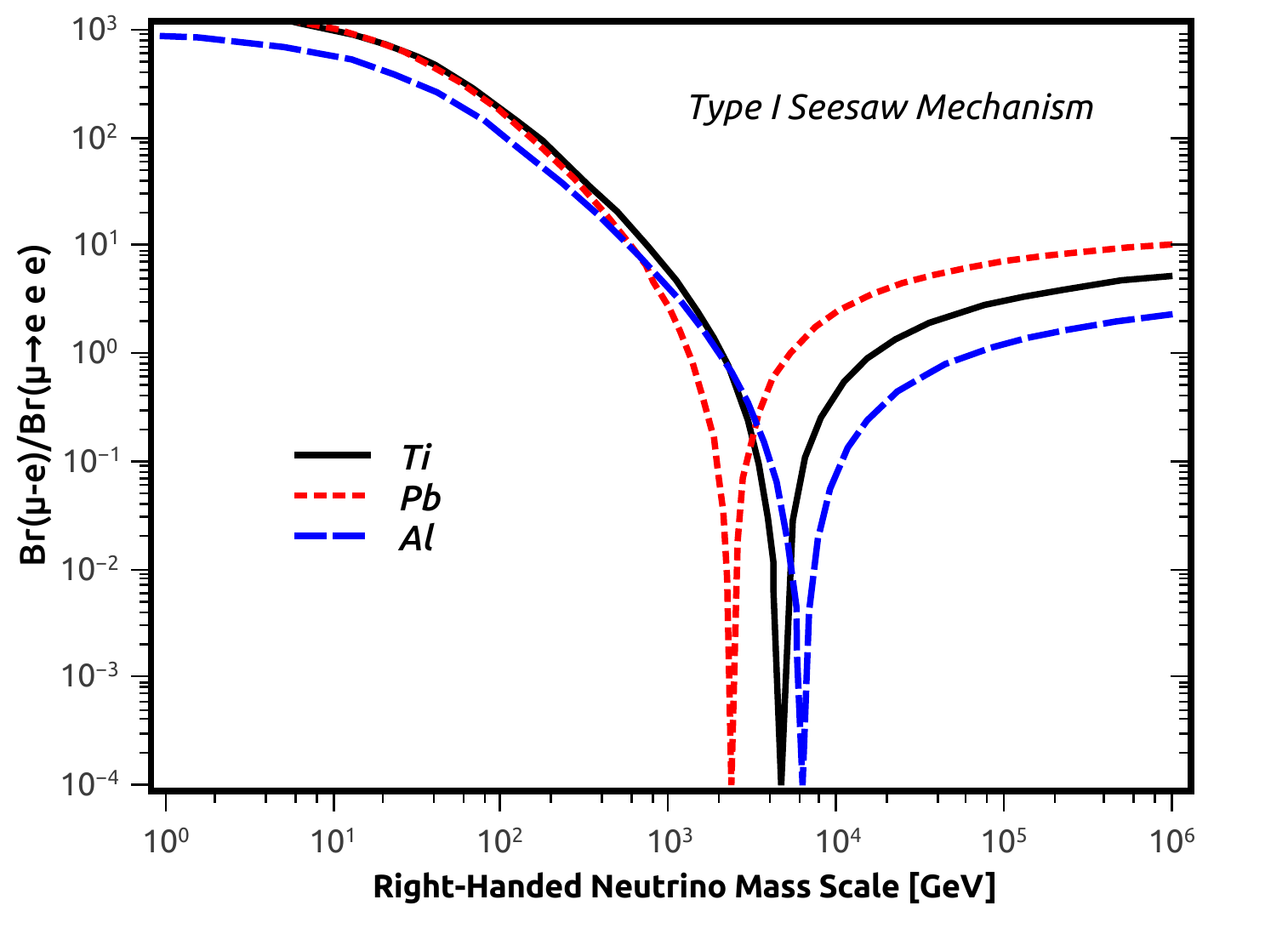}
\caption{Ratio ${\rm CR}(\mu-e)/{\rm BR}(\mu \to eee)$ as a function of the RH neutrino mass scale under the assumption that all RH neutrino have the same mass  for three different nuclei: Al (blue), Pb (red) and Ti (black).}
\label{mutoeratio1}
\end{figure}

{\bf Existing Limits}

Naively, one expects that the mixing matrices should follow~\cite{Deppisch:2015qwa},
\begin{equation}
U_{lN} \sim 10^{-6} \sqrt{\frac{100 GeV}{m_N}}.
\label{eqseesawvanilla}
\end{equation}
Therefore, one could test the type I seesaw mechanism by searching for RH neutrinos that feature mixing angles as governed by Eq.~\eqref{eqseesawvanilla}. However, we will see current experiments do not have the sensitivity to test the natural scale of the type~I seesaw mechanism, but they have already ruled out a significant region of the parameter space with large mixing angles. Future searches, on the other hand, feature a sensitivity somewhat near the vanilla seesaw scale. In other words, the type~I seesaw mechanism is still an interesting and viable framework to generate neutrino masses.

\begin{figure}
\centering
\includegraphics[scale=0.7]{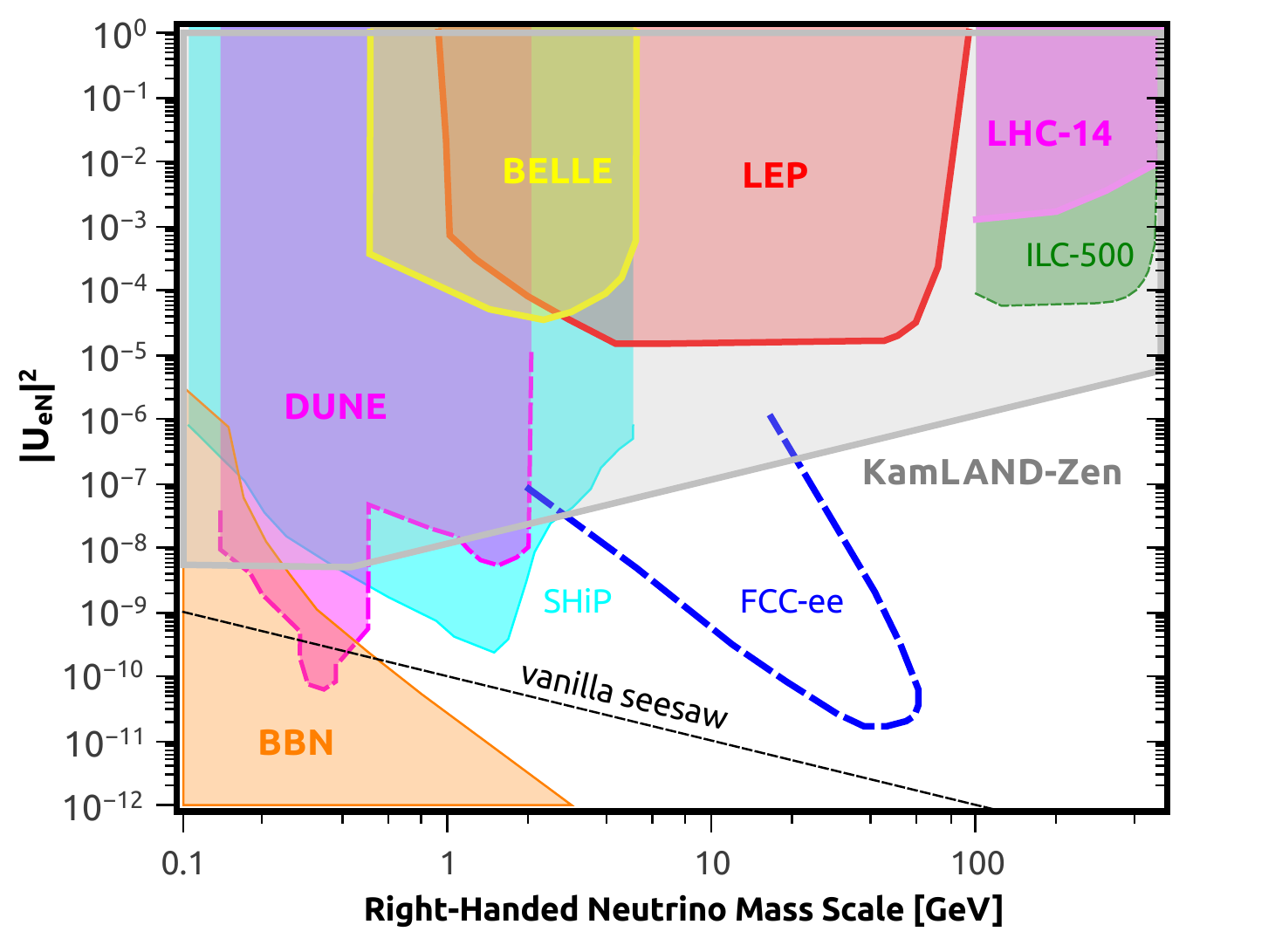}
\caption{Summary of existing and projected bounds on the mixing matrix $U_{e N}$ as function of the RH neutrino mass scale. See text for details.}
\label{seecollider1}
\end{figure}

\begin{figure}
\centering
\includegraphics[scale=0.7]{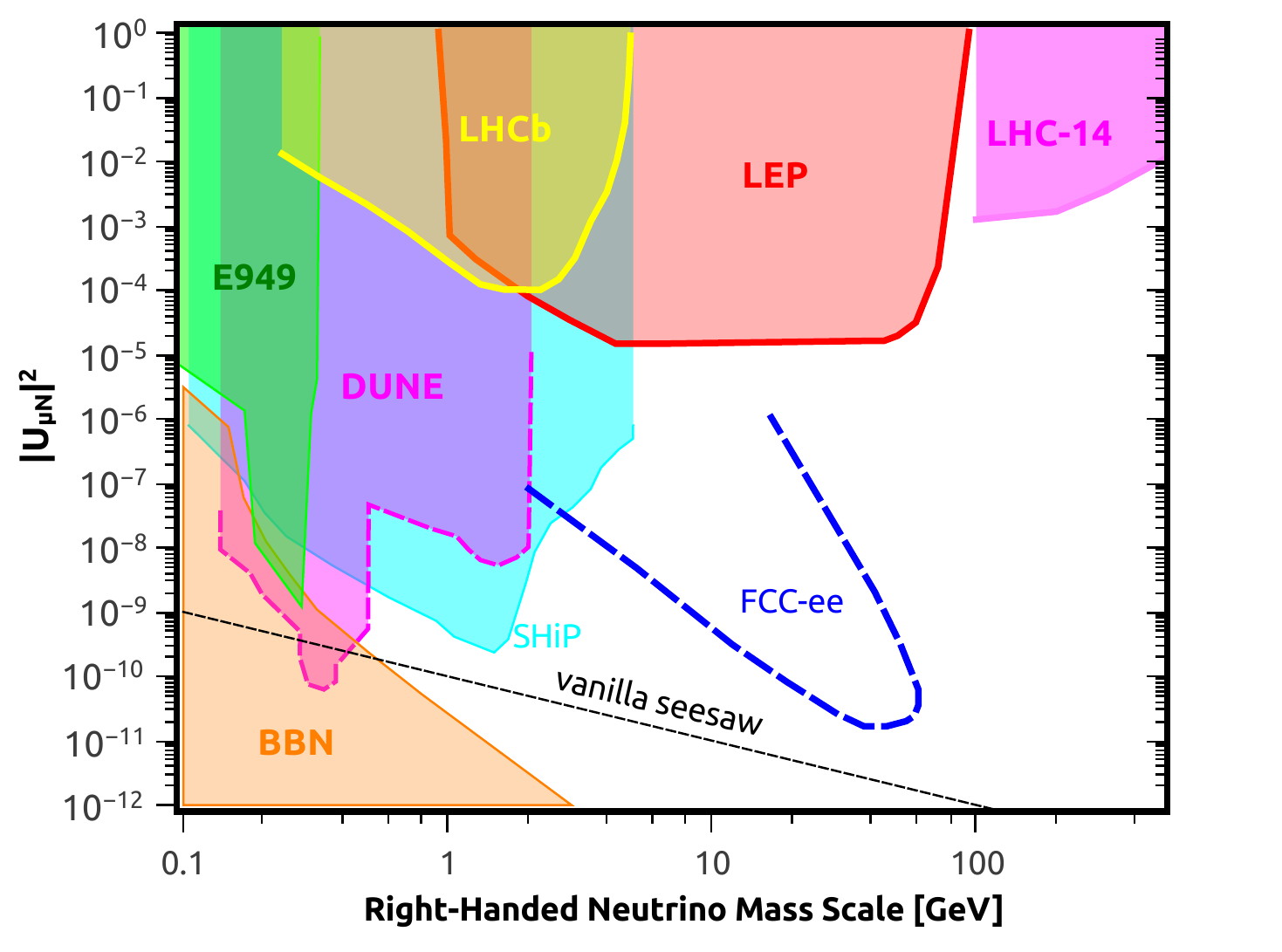}
\caption{Summary of existing and projected bounds on the mixing matrix $U_{\mu N}$ as function of the RH neutrino mass scale. See text for details.}
\label{seecollider2}
\end{figure}

\begin{figure}
\centering
\includegraphics[scale=0.7]{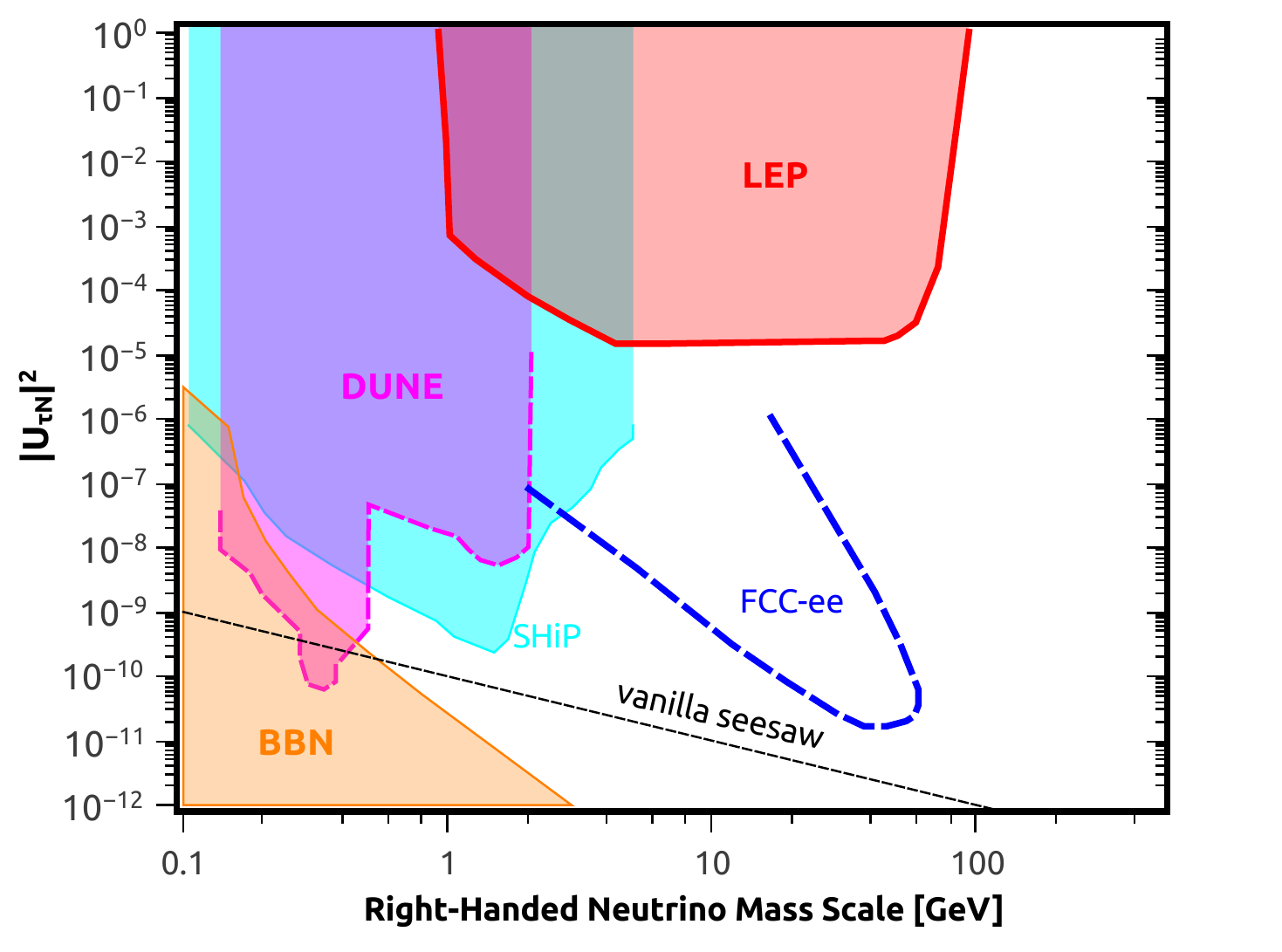}
\caption{Summary of existing and projected bounds on the mixing matrix $U_{\tau N}$ as function of the RH neutrino mass scale. See text for details.}
\label{seecollider3}
\end{figure}

Existing and projected bounds on RH neutrinos in the context of low scale seesaw mechanisms can probe RH neutrino masses up to $\sim 500$~ GeV. Beyond that, the sensitivity drastically worsens. Some peculiar setups where these RH neutrinos are not mixed with the active neutrinos, as predicted by the canonical type~I seesaw mechanism, are subject to different bounds as, e.g., in the left-right model~\cite{Mondal:2015zba,Lindner:2016lxq,Lindner:2016lpp,Queiroz:2016qmc,Mondal:2016kof,Biswal:2017nfl}. In Fig.~\ref{seecollider1}-~\ref{seecollider3} we display the existing limits on the left-right neutrino mixing matrices $U_{e N}$, $U_{\mu N}$, and $U_{\tau N}$ as a function of the RH neutrino mass scale. For a more detailed discussion we recommend~\cite{Banerjee:2015gca,Deppisch:2015qwa}. The bounds shown are described as follows:

{\it BBN:} In order not to spoil the light element abundances one should naively enforce the RH neutrino lifetime to be smaller than $1$~second resulting into the bound shown~\cite{Gorbunov:2007ak}.  

{\it KamLAND-Zen:} It is based on the non-observation of neutrinoless double beta decay in Xenon which implies a bound on the half lifetime of $T_{1/2}^{0\nu} (^{136} Xe) > 2.6 \times 10^{25}$~yr~\cite{Asakura:2014lma}. A further improvement down to $10^{27}$~yr is expected in the long run~\cite{DellOro:2016tmg}. It is already clear that neutrinoless double beta decay is the most powerful probe for RH neutrinos that mix with  electron-neutrinos. There are cancellation mechanisms that can be used to severely weaken KamLAND-Zen sensitivity motivating the complementary search for RH neutrinos though.

{\it DUNE:} The DUNE experiment could probe heavy-right handed neutrinos produced from charm meson decays down to much smaller mixings using a near detector~\cite{Adams:2013qkq}.

{\it BELLE/SHiP/LHCb:} By probing rare meson decays such as the $K^+ \to l^+ l^+ \pi^-$ decay. In this case one of the charged leptons could have originated from a RH neutrino decay~\cite{Adams:2013qkq,Anelli:2015pba}.

{\it LEP/FCC-ee:} The bound from LEP is based on the $Z$ decay into $\nu N$ via neutral current. Therefore it is applicable only to RH neutrino masses below the $Z$ mass~\cite{Dittmar:1989yg}. A huge improvement is expected in this direction with the planned FCC-ee experiment as shown in the figures~\cite{Blondel:2014bra,Abada:2014cca}.

\subsection{Type~II Seesaw}

Type~II seesaw mechanism refers to a framework where the active neutrino masses are generated via the addition of an $SU(2)_L$ scalar triplet $\Delta$ with weak hypercharge $Y_\Delta=2$ as follows~\cite{Magg:1980ut,Lazarides:1980nt,Mohapatra:1980yp}
\begin{equation}
	\Delta\;=\;\left(
			\begin{array}{cc}
				\Delta^{+}/\sqrt{2} & \Delta^{++} \\
				\Delta^{0}	& -\Delta^{+}/\sqrt{2}
			\end{array}\right)\,,
\end{equation}
which leads to the Lagrangian~\cite{Han:2006ip,delAguila:2007qnc,Akeroyd:2009nu}
\begin{equation}
	\mathcal{L}^{\rm II}_{\rm seesaw} = -M_{\Delta}^{2}\,{\rm Tr}\left(\Delta^{\dagger}\Delta\right)
		-\left(h_{\ell\ell^{\prime}}\,\overline{\psi^{C}}_{\ell L}\,i\tau_{2}\,\Delta\,\psi_{\ell^{\prime}L}\,+\,\mu_{\Delta}\, H^{T}\,i\tau_{2}\,\Delta^{\dagger}\,H\,+\,{\rm h.c.}\right)\,,
\label{eq.typeII}
\end{equation}
where $\psi_L$ is the SM $SU(2)_L$ doublet and $H$ the Higgs doublet. Note that $(\psi_{\ell L})^T \equiv (\nu^T_{\ell L}~~\ell^T_{L})$,
$\overline{\psi^{C}}_{\ell L}
\equiv ( -\,\nu^T_{\ell L}C^{-1}~~-\,\ell^T_{L}C^{-1})$, with $C$ being the charge conjugation matrix, and $\mu_{\Delta}$ is a parameter that accounts for the  lepton charge violation.

Neutrino masses are generated via the second term in Eq.~\eqref{eq.typeII} when $\Delta^0$ develops a non zero VEV. In order to generate tiny neutrino masses naturally, at the eV scale, $v_\Delta$ should be small, possibly proportional to $\mu_\Delta$, which also should be small since it accounts for the violation of lepton number. In this context we get
\begin{equation}
\left(m_{\nu}\right)_{\ell\ell^{\prime}}\, \equiv m_{\ell\ell^{\prime}}\,
\simeq\;2\,h_{\ell\ell^{\prime}}\,
v_{\Delta},
\label{mneutrinotypeII}
\end{equation}where,
\begin{equation}
h_{\ell\ell^\prime}\;\equiv\; \frac{1}{2v_\Delta}\left(U^*_{PMNS}\,
(m_1,m_2,m_3)\,U^\dagger_{{\rm PMNS}} \right)_{\ell\ell^\prime}\,.
\label{hll}
\end{equation}
The matrix $U_{{\rm PMNS}}$ can be identified with the $U_{\nu\nu}$ matrix in the type~I seesaw mechanism defined in Eq.~\eqref{Umatrix}, in the regime of heavy RH neutrinos. The explicit form of the PMNS matrix can also be found in the PDG review of particle physics~\cite{Amsler:2008zzb}. Thus, tuning $v_\Delta$, one can nicely explain the neutrino masses. However, the addition of a $SU(2)_L$ triplet is known to alter the $\rho$ parameter, with $\rho=m_W^2/(m_Z^2 \cos^2_W)$. Since the $\rho$ parameter is measured to be one with error bar at the percent level, one can use this precision to constraint $v_\Delta$, concluding that $v_\Delta < 5$~GeV~\cite{Arhrib:2011uy,Akeroyd:2012nd}.

The key feature of the type~II seesaw mechanism as far as LFV and $g-2$ observables are concerned is the presence of a singly and doubly charged scalar. These arise from the terms $\nu^c l \Delta^+$  and $\overline{l^c} l \Delta^{++}$. These fields cannot explain $g-2$ because they give a negative contribution.

The corrections to these observables stemming from a singlet and doubly charged scalar have been computed in Eq.~\eqref{eq:Delta_a_Singly_Scalar_approx} and Eq.~\eqref{Eq:BRsinglycharged} for the singly charged scalar, and in Eq.~\eqref{Eq:doublychargedamu} and Eq.~\eqref{Eq:doublyBRmutoe} for the doubly charged scalar. Using these equations one can perform a more detailed study of the $g-2$ and $\MEG$ quantities in the type~II seesaw model. Results for this model were already derived in Sec.\ref{Sec:scalartriplet}. We now simply put these results into context with $\mu \to 3e$ and $\mu-e$ conversion.

As for LFV observables we find~\cite{Akeroyd:2009nu}
\begin{eqnarray}
{\rm BR}(\mu\to e \gamma) & \cong &
384\,\pi^2\,(4\pi\,\alpha_{\rm em}) \left|A_R\right|^2\;
=\; \frac{\alpha_{\rm em}}{192\,\pi}\,
\frac{\left|\left(h^{\dagger}h\right)_{e \mu}\right|^{2}}
{G_{F}^{2}}\,\left(
\frac{1}{m^2_{\Delta^{+}}} +
\frac{8}{m^2_{\Delta^{++}}}
\right)^{2},\nonumber\\
\label{mutoetypeII}
\end{eqnarray}where
\begin{equation}
|h^\dagger h_{e\mu}| < 1 \times 10^{-6} \left(\frac{M_\Delta}{100 GeV}\right)^2,
\end{equation} as extracted from Eq.~\eqref{hll}.

Moreover, we get
\begin{equation}
{\rm BR}(\mu\rightarrow 3e) =
\frac{1}{G_F^2}\frac{|(h^\dagger)_{ee}(h)_{\mu e}|^2}{m^4_{\Delta^{++}}}\,,
\label{muto3etypeII}
\end{equation}
which implies,
\begin{equation}
|h^\dagger_{ee}h_{\mu e}| < 1.2 \times 10^{-7} \left(\frac{m_{\Delta^{++}}}{100 GeV}\right)^2.
\end{equation}
Furthermore, for the $\mu-e$ conversion we find
\begin{eqnarray}
	 {\rm CR}(\mu - e) &\cong&
\frac{\alpha^5_{\rm em}}{36\,\pi^4}\,
\frac{m_\mu^5}{\Gamma_{\rm capt}}\,
Z_\mathrm{eff}^4\,Z\,F^2(-m_{\mu}^{2})\,
\left| \left(h^{\dagger}h\right)_{e\mu }\,
\left [ \frac{5}{24\,m^2_{\Delta^{+}}}
+ \frac{1}{m^2_{\Delta^{++}}}\right ] \right.\nonumber \\
  & + & \left. \frac{1}{m^2_{\Delta^{++}}}\,
\sum_{l=e,\mu,\tau} h^{\dagger}_{e l}\, h_{l\mu} F(m_l,m_{\Delta^{++}})
\right |^2\,,
\label{CRtypeII2}
\end{eqnarray}
where,
\begin{eqnarray}
&& F(m_l,m_{\Delta^{++}})= \log\left(\frac{m_l^2}{m_{\Delta^{++}}^2}\right)\nonumber\\
&&
 +\left(1- 2 \frac{m_l^2}{m_\mu^2} \right)\sqrt{1+4\frac{m_l^2}{m_\mu^2}}\log\left(\frac{ \sqrt{m_\mu^2 m_{\Delta^{++}}^{-2} }+\sqrt{ m_\mu^2 m_{\Delta^{++}}^{-2} + 4 m_l^2 m_{\Delta^{++}}^{-2} }}{\sqrt{m_\mu^2 m_{\Delta^{++}}^{-2} }-\sqrt{ m_\mu^2 m_{\Delta^{++}}^{-2} + 4 m_l^2 m_{\Delta^{++}}^{-2} }} \right).  \nonumber\\
\end{eqnarray}
\begin{table}[!h]
\centering
\small
\begin{tabular}{|c | c | c|}
\hline
Branching ratio &  $13$~TeV - Bound &  $14$~TeV - $1000 fb^{-1}$\\
\hline
${\rm BR}(\Delta^{++} \to e^{\pm} e^{\pm})=1$ & $m_\Delta^{\pm\pm} > 800 $~GeV & $m_\Delta^{\pm\pm} > 2.3$~TeV \\
\hline
${\rm BR}(\Delta^{++}  \to \mu^{\pm} \mu^{\pm})=1$ & $m_\Delta^{\pm\pm} > 850 $~GeV &  $m_\Delta^{\pm\pm} > 1.94$~TeV \\
\hline
${\rm BR}(\Delta^{++}  \to e^{\pm} \mu^{\pm})=1$ & $m_\Delta^{\pm\pm} > 860 $~GeV & $m_\Delta^{\pm\pm} > 1.96$~TeV \\
\hline
${\rm BR}(\Delta^{++}  \to e^{\pm} e^{\pm})=0.3$ & & \\
${\rm BR}(\Delta^{++}  \to e^{\pm} \mu^{\pm})=0.4$ & $m_\Delta^{\pm\pm} > 750 $~GeV & $m_\Delta^{\pm\pm} > 1.75$~TeV\\
${\rm BR}(\Delta^{++}  \to \mu^{\pm} \mu^{\pm})=0.3$ &  &  $m_\Delta^{\pm\pm} > 2.13$~TeV\\
\hline
${\rm BR}(\Delta^{++}  \to e^{\pm} \tau^{\pm})=1$ & $m_\Delta^{\pm\pm} > 714 $~GeV & $m_\Delta^{\pm\pm} > 2.13$~TeV \\
\hline
${\rm BR}(\Delta^{++}  \to \mu^{\pm} \tau^{\pm})=1$ & $m_\Delta^{\pm\pm} > 643 $~GeV &  $m_\Delta^{\pm\pm} > 1.96$~TeV\\
\hline
${\rm BR}(\Delta^{++}  \to \tau^{\pm} \tau^{\pm})=1$ & $m_\Delta^{\pm\pm} > 535 $~GeV & $m_\Delta^{\pm\pm} > 1.69$~TeV \\
\hline
${\rm BR}(\Delta^{++}  \to e^{\pm} \mu^{\pm})=0.01$ & & \\
${\rm BR}(\Delta^{++}  \to e^{\pm} \tau^{\pm})=0.01$ & $m_\Delta^{\pm\pm} > 723 $~GeV & $m_\Delta^{\pm\pm} > 2.16$~TeV\\
${\rm BR}(\Delta^{++}  \to \mu^{\pm} \mu^{\pm})=0.3$ &  &  \\
${\rm BR}(\Delta^{++}  \to \mu^{\pm} \tau^{\pm})=0.38$ &  &  \\
${\rm BR}(\Delta^{++}  \to \tau^{\pm} \tau^{\pm})=0.3$ &  &  \\
\hline
${\rm BR}(\Delta^{++}  \to e^{\pm} e^{\pm})=0.5$ & & \\
${\rm BR}(\Delta^{++}  \to \mu^{\pm} \mu^{\pm})=0.125$ & $m_\Delta^{\pm\pm} > 716 $~GeV & $m_\Delta^{\pm\pm} > 2.14$~TeV  \\
${\rm BR}(\Delta^{++}  \to \mu^{\pm} \tau^{\pm})=0.25$ &  &  \\
${\rm BR}(\Delta^{++}  \to \tau^{\pm} \tau^{\pm})=0.125$ &  &  \\
\hline
${\rm BR}(\Delta^{++}  \to e^{\pm} e^{\pm})=0.33$ & & \\
${\rm BR}(\Delta^{++}  \to \mu^{\pm} \mu^{\pm})=0.33$ &  $m_\Delta^{\pm\pm} > 761 $~GeV  &  $m_\Delta^{\pm\pm} > 2.24$~TeV \\
${\rm BR}(\Delta^{++}  \to \tau^{\pm} \tau^{\pm})=0.33$ &  &  \\
\hline
${\rm BR}(\Delta^{++}  \to e^{\pm} e^{\pm})=0.166$ & & \\
${\rm BR}(\Delta^{++}  \to e^{\pm} \mu^{\pm})=0.166$ & & \\
${\rm BR}(\Delta^{++}  \to e^{\pm} \tau^{\pm})=0.166$ & $m_\Delta^{\pm\pm} > 722 $~GeV & $m_\Delta^{\pm\pm} > 2.15$~TeV\\
${\rm BR}(\Delta^{++}  \to \mu^{\pm} \mu^{\pm})=0.166$ &  &  \\
${\rm BR}(\Delta^{++}  \to \mu^{\pm} \tau^{\pm})=0.166$ &  &  \\
${\rm BR}(\Delta^{++}  \to \tau^{\pm} \tau^{\pm})=0.166$ &  &  \\
\hline
\end{tabular}
\caption{Bounds on the doubly charged scalar.}
\label{tabletypeIIseesaw}
\end{table}
All the other parameters such the capture rate $\Gamma_{{\rm capt}}$, $Z_\mathrm{eff}$, $F^2(-m_\mu^2)$ are given in Tab.~\ref{nuclfactorsmueconv}.

The bounds arising from $\mu-e$ conversion cannot be easily translated into a limit on the couplings as we have done for $\MEG$ and $\mu \to 3e$ due to the presence of several form factors. The projected sensitivity using the Aluminum of $10^{-17}$ does not look promissing in this model. Even considering a very optimist projected sensitivity of $10^{-18}$ on the $\mu-e$ conversion branching ratio, $\mu-e$ will not be more sensitive than $\mu \to 3e$ for the type~II seesaw model.\\

{\bf Existing Limits}

Existing and projected limits on the type~II seesaw mechanism do not rule out the possibility of successfully explaining neutrino masses because one can in principle push the mass of the doubly charged scalar above the TeV scale without prejudice. Conversely, the observation of such a doubly charged scalar would constitute a strong case for the type~II seesaw mechanism. In what follows we introduce the main limits. 

A key signature of the type~II seesaw mechanism at the LHC would be the observation of the associated production channel $\Delta^{\pm \pm} \Delta^{\mp}$ at the LHC~\cite{Perez:2008ha}.  The ATLAS collaboration has carried independent searches for doubly charged and charged scalars that decays entirely into charged leptons~\cite{ATLAS:2017iqw}, while the CMS collaboration, besides these independent searches, has performed searches for the associated production channel above~\cite{CMS:2017pet}. We have implemented the model into MadGraph and reproduced with good agreement the results reported by the ATLAS and CMS collaborations using the same selection cuts used in~\cite{ATLAS:2017iqw,CMS:2017pet}. Moreover, under the null result hypothesis we derived projected limits for the high-luminosity LHC. In summary these limits are presented in Tab.~\ref{tabletypeIIseesaw}.

\subsection{Type~III seesaw}

The type~III seesaw model refers to a mechanism in which neutrino masses are generated via the addition of $SU(2)_L$ fermion triplets, i.e.~in the adjoint representation of $SU(2)_L$, with zero hypercharge~\cite{Foot:1988aq,Bonilla:2015eha}. However, only two fermion triplets are required to successfully generate two non-vanishing neutrino masses. That said, the Lagrangian is found to be
\begin{equation}
{\cal{L}} \supset -y_\Sigma \, \tilde{H}^{\dagger} \, \overline{\Sigma}\, L \ + {\rm h.c.}\,, 
\label{Eq:seesawtypeIII} 
\end{equation}
where
\begin{equation}
\Sigma = \left(\begin{array}{cc} 
\Sigma^0/\sqrt{2} & \Sigma^+ \\ 
\Sigma^- & -\Sigma^0/\sqrt{2} 
\end{array} \right), \
\Sigma^c=
\left(
\begin{array}{ cc}
   \Sigma^{0c}/\sqrt{2}  &   \Sigma^{-c} \\
     \Sigma^{+c} &  -\Sigma^{0c}/\sqrt{2} 
\end{array}
\right), \label{Sigma}
\end{equation}
and $\Sigma^c \equiv C \overline{\Sigma}^T$. Taking $\Psi \equiv {(\Sigma^{+c} + i \Sigma^-)}{\sqrt{2}} $ and expanding the Lagrangian, we get in the unitary gauge,
\begin{equation}
{\cal{L}} \supset -  y_\Sigma\, \overline{\Sigma^0}\, \nu_{L} \phi^0+  y_\Sigma \, \overline{\Psi}\, L \, \phi^0 
+ \text{h.c.}\,,
\end{equation}
where $\phi^0$ is simply the neutral component of the Higgs doublet $H$. In this way, the new charged fermions ($\Psi$) will mix with the charged leptons, and the neutral fermions ($\Sigma^0$) will mix with neutrinos. Therefore, new contribution to $g-2$ and $\MEG$ will arise: (i) via $W$ exchange due to the $\nu-\Sigma^0$ mixing; (ii) via $Z$ exchange as result of the charged lepton-$\Psi$ mixing; (iii) through Higgs exchange again as result of charged lepton-$\Psi$ mixing. Fully general results for $W$-neutral fermion contribution, neutral vector and neutral scalar exchanges were obtained in Sec.~\ref{sec:numerics}, and can be straightforwardly applied to this seesaw model. Furthermore, in Ref.~\cite{Abada:2008ea} the authors have discussed these observables in more detail, while Ref.~\cite{Franceschini:2008pz} derived collider bounds.

Let us discuss the results more quantitatively now. Since the fermion triplet contains a charged component, collider bounds are generally expected to be strong.  Therefore, we assume the limit where $M_\Sigma \gg m_{H,W,Z}$ in order to evade these bounds. The general expression for the branching fraction for $\MEG$ in this limit is given by~\cite{Abada:2008ea}
\begin{equation}
	\mathrm{BR}(\MEG) = \frac{3 \alpha_\text{em}}{32\pi} \left| \left(\frac{13}{3} - 6.56 \right) \epsilon_{e \mu} - \sum_i \frac{m_{\nu_i}^2}{m_W^2} (U_\text{PMNS})_{ei} (U^\dag_\text{PMNS})_{i\mu}\right|^2,
\end{equation}
with $\epsilon \equiv \frac{v^2}{2} y_\Sigma^\dag M_\Sigma^{-2} y_\Sigma$ being the effective low-energy coupling mixing the SM lepton flavors once the heavy triplets are integrated out. Also, note that the second contribution is nothing but the usual massive neutrino contribution which is usually negligible. In the same limit one can derive relations between $\MEG$ and the other LFV processes:
\begin{subequations}
\begin{align}
	\mathrm{BR} (\MEG) &= 1.3 \cdot 10^{-3} \times \mathrm{BR} (\mu \to eee),\\
	\mathrm{BR} (\tau \to \mu \gamma) &= 1.3 \cdot 10^{-3} \times \mathrm{BR} (\tau \to \mu\mu\mu),\\
	\mathrm{BR} (\tau \to e \gamma) &= 1.3 \cdot 10^{-3} \times \mathrm{BR} (\tau \to eee).
\end{align}
\end{subequations}

Moreover, for the relation with $\mu-e$ conversion reads, 
\begin{equation}
	\mathrm{BR} (\MEG) = 3.1 \cdot 10^{-4} \times \mathrm{CR} (\mu\, \text{Ti} \to e\, \text{Ti}),
\end{equation}

\begin{equation}
	\mathrm{BR} (\MEG) = 2 \cdot 10^{-4} \times \mathrm{CR} (\mu\, \text{Al} \to e\, \text{Al}).
\end{equation}

Going back to Tab.~\ref{tab:mu-eOverview}, we see that currently and in the future the strongest constraints come from the nuclear $\mu - e$ transition. Finally, we note that the contribution to the muon $g-2$ is obtained from Sec.~\ref{sec:results}.

{\bf Existing Limits}

Compared to the previous seesaw mechanisms there are fewer existing limits applicable to this model. Arguably the most robust bound stems from LHC searches for the pair production of this heavy fermions. The lower mass bound on the heavy fermions vary from $400$~GeV for exclusive decays into $\tau$ leptons to $930$~GeV to decays entirely into electrons. These bounds, regardless of the assumptions used for the decay modes, limit the parameter space of the type~III seesaw model. Be that as it may, it remains a plausible mechanism for generating neutrino masses. Now lets discuss in a bit more detail these limits.

A recent analysis has been conducted by the CMS collaboration~\cite{Sirunyan:2017qkz} with $13$~TeV of center-of-mass energy and $35.9fb^{-1}$ of integrated luminosity. We have implemented the model in Madgraph and assumed that all exotic fermions are degenerate in mass, $m_\Sigma$, and feature 100\% branching ratio into charged leptons, similarly to what has been done in~\cite{Sirunyan:2017qkz}. We simulated the events using similar selection criteria as in~\cite{Sirunyan:2017qkz}. Our lower mass bound on $m_\Sigma$ is slightly overoptimistic compared to the one found by CMS collaboration: $m_\Sigma > 900$~GeV (ours), $m_\Sigma > 840$~GeV (CMS) for the benchmark model where the charged fermions decay democratically into charged leptons. We have taken into account detector effects using Delphes~\cite{deFavereau:2013fsa}, whereas the CMS collaboration used GEANT~\cite{Agostinelli:2002hh}, perhaps explaining the difference in the lower mass bounds.

Besides this scenario we also explored different benchmark models and also derived the projection for the high-luminosity LHC as summarized in Tab.~\ref{tabletypeIIIseesaw}. We analysed only the $\Sigma^+ \Sigma^-$ production channel with the possible decay chains containing at least three charged leptons as enforced in~\cite{Sirunyan:2017qkz}:
\begin{eqnarray}
\Sigma^\pm\Sigma^\mp \to (Z l^\pm) Z l^\mp\,; \qquad
\Sigma^\pm\Sigma^\mp \to (h l^\pm) h l^\mp
\label{brtypeIII} 
\end{eqnarray}
with at least one of the bosons, $(Z,h)$, decaying into charged leptons.  

In Tab.~\ref{tabletypeIIIseesaw} ${\rm BR}_e$ (electron), ${\rm BR}_\mu$ (muon),${\rm BR}_\tau$ (tau) stand for the branching ratios of the exotic fermions into charged leptons of a given flavor. For instance, if ${\rm BR}_e=1$, it means that all charged leptons in Eq.~\eqref{brtypeIII} are electrons.

\begin{table}[!h]
\centering
\small
\begin{tabular}{|c | c | c|}
\hline
Branching ratio &  $13$~TeV - Bound &  $14$~TeV - $1000 fb^{-1}$\\
\hline
${\rm BR}_e=1,{\rm BR}_\mu=0,{\rm BR}_\tau=0$ & $m_\Sigma > 960 $~GeV & $m_\Sigma > 2.16 $~TeV \\
\hline
${\rm BR}_e=0,{\rm BR}_\mu=1,{\rm BR}_\tau=0$ & $m_\Sigma > 1 $~TeV & $m_\Sigma > 2.23 $~TeV \\
\hline
${\rm BR}_e=0,{\rm BR}_\mu=0,{\rm BR}_\tau=1$ & $m_\Sigma > 450 $~GeV & $m_\Sigma > 1.12 $~TeV \\
\hline
${\rm BR}_e=0,{\rm BR}_\mu=1,{\rm BR}_\tau=0$ & $m_\Sigma > 1 $~TeV & $m_\Sigma > 2.23 $~TeV \\
\hline
${\rm BR}_e=0,{\rm BR}_\mu=0.5,{\rm BR}_\tau=0.5$ & $m_\Sigma > 850 $~GeV & $m_\Sigma > 1.95 $~TeV \\
\hline
${\rm BR}_e=0,{\rm BR}_\mu=0.8,{\rm BR}_\tau=0.2$ & $m_\Sigma > 930 $~GeV & $m_\Sigma > 2.1 $~TeV \\
\hline
${\rm BR}_e=0,{\rm BR}_\mu=0.2,{\rm BR}_\tau=0.8$ & $m_\Sigma > 700 $~GeV & $m_\Sigma > 1.65 $~TeV \\
\hline
${\rm BR}_e=0,{\rm BR}_\mu=0.6,{\rm BR}_\tau=0.4$ & $m_\Sigma > 870 $~GeV & $m_\Sigma > 1.98 $~TeV \\
\hline
${\rm BR}_e=0,{\rm BR}_\mu=0.4,{\rm BR}_\tau=0.6$ & $m_\Sigma > 800 $~GeV & $m_\Sigma > 1.85 $~TeV \\
\hline
${\rm BR}_e=0.33,{\rm BR}_\mu=0.33,{\rm BR}_\tau=0.33$ & $m_\Sigma > 900 $~GeV & $m_\Sigma > 2.04 $~TeV \\
\hline
${\rm BR}_e=0.5,{\rm BR}_\mu=0.5,{\rm BR}_\tau=0$ & $m_\Sigma > 990 $~GeV & $m_\Sigma > 2.22 $~TeV \\
\hline
${\rm BR}_e=0.5,{\rm BR}_\mu=0,{\rm BR}_\tau=0.5$ & $m_\Sigma > 820 $~GeV & $m_\Sigma > 1.89 $~TeV \\
\hline
${\rm BR}_e=0.5,{\rm BR}_\mu=0.25,{\rm BR}_\tau=0.25$ & $m_\Sigma > 950 $~GeV & $m_\Sigma > 2.14 $~TeV \\
\hline
${\rm BR}_e=0.5,{\rm BR}_\mu=0.1,{\rm BR}_\tau=0.4$ & $m_\Sigma > 870 $~GeV & $m_\Sigma > 1.98 $~TeV \\
\hline
${\rm BR}_e=0.5,{\rm BR}_\mu=0.4,{\rm BR}_\tau=0.1$ & $m_\Sigma > 970 $~GeV & $m_\Sigma > 2.18 $~TeV \\
\hline
\end{tabular}
\caption{95\% C.L. bounds on the mass of the exotic fermions in the type~III seesaw model assuming they are mass degenerate.}
\label{tabletypeIIIseesaw}
\end{table}

From Tab.\ref{tabletypeIIIseesaw} we can see that LHC is expected to exclude the type III seesaw mechanism featuring masses below the TeV scale, being able to probe masses up to $\sim 2.2$~TeV in certain cases. The potential discovery of the associated production or pair production of these exotic fermions in the future would constitute a smoking gun signature of this model compared to the other neutrino mass generation mechanisms.

\subsection{Inverse and Linear seesaw}

There are other ways to generate neutrinos masses known as inverse~\cite{Mohapatra:1986bd} and linear seesaw mechanisms~\cite{Malinsky:2005bi}.  In both cases the smallness of the light neutrino masses is addressed via the combination of new heavy states and a mass parameter which is arguably naturally small, inspired by grand unified theories~\cite{Wyler:1982dd, 
Witten:1985xc,Mohapatra:1986bd}. They typically include two singlet neutral leptons per generation, namely $\nu_{i R}$ and $S_i$. In the basis $(\nu_L^{\ c}, \nu_R, S)$, the mass matrix takes the form,
\begin{equation}
{\cal L}_m =  - \frac{1}{2}  (\overline{\nu_L}, \overline{\nu_R^{\ c}}, \overline{S^c})    \left( \begin{array}{ccc}
0 & m_D & {\color{blue}\epsilon }\\
m_D^T &0 & M_R\\
{\color{blue} \epsilon^T} & M_R^T & {\color{red} \mu} \end{array} \right)  
\left( \begin{array}{c}
\nu_L^{\ c} \\
\nu_R
\\
S
 \end{array} \right) + h.c. 
\label{lowscale}
\end{equation}
Taking $\mu \equiv 0$, we obtain to the so called linear seesaw with $m_{\nu} \sim \mu m_D^2/M_R^2$, whereas if $\epsilon\equiv 0$, we obtain the inverse seesaw mechanism with $m_{\nu} \sim \epsilon m_D/M_R$.  That said, these mechanisms can be naturally embedded in a gauged $B-L$ symmetry, which gives rise to the following Yukawa terms
\begin{equation}
{\cal L} \supset -  y_{\epsilon} \overline{L} \Phi^c S +  y_{D} \overline{L} H^c \nu_R,
\end{equation}with both contributing to $g-2$ and $\MEG$. We have faced terms of this type several times in this work and their corrections to $g-2$ and $\MEG$ were summarized in Eq.~\eqref{eq:singlyscalar}. One can simply use the results of that section to compute, in fully generality, the contributions to these observables. The key aspect of these models, as far as $g-2$ and $\MEG$ are concerned, is presence of non-suppressed mixing, different from the type~I seesaw mechanism. The interplay with $\mu-e$ conversion resembles the type I seesaw mechanim if $y_{\epsilon} \ll y_D$. However, the presence of the exotic field S may induces significant changes. We leave this discusssion for future works. 

{\bf Existing Limits}

Since one one can generate sizable corrections to $g-2$ and LFV observables in this model, equally restrictive constraints arise from collider searches. One way to test this mechanism is to search for RH neutrinos as presented in the type~I seesaw mechanism. Moreover, we point out that bounds stemming from lepton universality in kaon decays most strongly limit interpretations of the $g-2$ deviation~\cite{Abada:2014nwa}.

\section{\label{sec:conclusion}Summary and Outlook}
In this article we have reviewed key observables of modern particle physics, namely the muon anomalous magnetic moment and lepton flavor violation in muon decays. While recent measurements of the former observable may point towards new physics being around the corner, the latter gives rise to strong constraints on models beyond the Standard Model. We have reviewed the current experimental status of these observables in the light of the upcoming flagship experiments which will hopefully set a new direction in particle physics.

In the subsequent discussion, we have derived fully general expressions that allow the reader to compute the contribution of new physics to the rate of the process $\ell_i \to \ell_j \gamma$ as well as the anomalous magnetic dipole moment of a given lepton. We have studied these expressions extensively in the context of simplified and $SU(2)_L$ invariant extensions of the Standard Model. For definiteness, we have focused on the decay $\MEG$ and the anomalous magnetic moment of the muon. We have discovered that one may accommodate a signal in either observable while circumventing constrains from the other. In certain scenarios it is even feasible to address signals in both phenomena in the next generation of experiments. 

In a final step, we have applied our findings to well-known UV-completions of the Standard Model to illustrate the broad applicability of our results to inspire any inclined reader to use those results for phenomenological studies in their favorite model. In the scope of UV complete models we also discussed existing limits and the interplay with $\mu \to eee$ and $\mu-e$ conversion. Moreover, in some cases we derived the collider limits by implementing the models in collider tools such as Madgraph while accounting for detector effects. Therefore, we could draw conclusions having a global picture of the UV complete models.

We hope that, with this review, we have paved the road to new model building endeavors and motivated the interest in the complementarity between $g-2$ and lepton flavor violation.

\section*{Acknowledgements}

MP is supported by the IMPRS-PTFS. The authors are indebted to Hiren~Patel for clarifications and comments on the use of \textsl{Package-X}. We are grateful to Carlos~Yaguna, Werner~Rodejohann, Giorgio Arcadi, Evgeny Akhmedov, Alex Dias, Tanja Geib and Alexander Merle for enlightening discussions. We very much thank comments from Pran Nath, Avelino Vicente, Jos\'e Valle, Talal Ahmed, Frank Deppisch, Albert Petrov, Mariano Quiros, Dominik Stoeckinger, Roberto Martinez, Hiroshi Okada, Choong Sun Kim. We are also greatly thankful to Michael Ramsey-Musolf for the invitation.



\appendix
\renewcommand*{\appendixname}{}

\section{Master integrals}

\renewcommand{\theequation}{A-\arabic{equation}}
\setcounter{equation}{0}  

In this appendix, we list the fully analytical loop integrals for the amplitude $\ell_i \to \ell_j \gamma$ for the different new particle contributions. All results have been numerically cross-checked with the Mathematica Package-X~\cite{Patel:2015tea}. We begin with the neutral scalar integral represented  by the graph in Fig.~\ref{fig:neutralScalar}. It is given by
\begin{align}\label{eq:I1_def}
I^{(\pm)_1\,(\pm)_2}_{f,\, 1} &\equiv I_{f,\,1}\left[m_i,(\pm)_1 m_j, (\pm)_2 m_f,m_\phi\right] \nonumber\\
  &= \int \underbrace{\mathrm{d}x\mathrm{d}y\mathrm{d}z\, \delta\!\left({1\!-\!x\!-\!y\!-\!z}\right)}_{\equiv\mathrm{d}^3\mathbf{X}} \frac{x\left(y +(\pm)_1 z\, \frac{m_j}{m_i}\right) + (\pm)_2 (1-x)\frac{m_f}{m_i}}{-xy\, m_i^2 - xz\, m_j^2 +x\, m_\phi^2 + (1-x) m_f^2}. 
\end{align}
The equivalent diagram involving a charged internal scalar yields
\begin{equation}\label{eq:I2_def}
  I^{(\pm)_1\,(\pm)_2}_{f,\, 2} = \int \mathrm{d}^3\mathbf{X} \frac{x\left(y +(\pm)_1 z\, \frac{m_j}{m_i} +(\pm)_1 \frac{m_{\nu_f}}{m_i}\right)}{-xy\, m_i^2 - xz\, m_j^2 +(1-x) m_{\phi^+}^2 + x\, m_{\nu_f}^2}.
\end{equation}
This auxiliary function may be approximated for $m_j \ll m_i$ as
\begin{equation}
  I^{(\pm)_1\,(\pm)_2}_{f,\, 2} \simeq \frac{1}{m_{\phi^+}^2}\int_0^1 \mathrm{d}x \int_0^1 \mathrm{d}y \, x(1-x) \frac{xy +(\pm)_2 \epsilon_f}
  {\epsilon_f^2\lambda^2(1-x)\left(1-\epsilon_f^{-2}xy\right)+x}.
\end{equation}
Assuming in addition a very heavy mediator, i.e.\ $\lambda \to 0$, we find that
\begin{equation}
  I^{(\pm)_1\,(\pm)_2}_{f,\, 2} \simeq \frac{1}{m_{\phi^+}^2} \left[ \frac{1}{12} + (\pm)_2 \frac{\epsilon_f}{2} \right].
\end{equation}
When there is a charged vector boson propagating in the loop, the occurrence of two gauge boson propagators complicates the calculation significantly. The result is rather lengthy and reads:
\begingroup
\allowdisplaybreaks
\begin{align} \label{eq:I3_def}
  I_{f,\,3}^{(\pm)_1\,(\pm)_2} &\equiv I_{f,\,3}\left[m_i,(\pm)_1 m_j,(\pm)_2 m_{N_f},m_W\right] \nonumber\\
  =& \int \mathrm{d}^3\mathbf{X} \Bigg[\left[-xz\, m_i^2 - xy\, m_j^2 +(1-x) m_W^2 + x\, m_{N_f}^2\right]^{-1} \nonumber\times\\
  & \times\bigg\lbrace -(\pm)_2 3 (1-x) \frac{m_{N_f}}{m_i} + (y+2z(1-x)) +(\pm)_1 \frac{m_j}{m_i}(z+2y(1-x)) \nonumber \\
  & +  \frac{m_i^2}{m_W^2}\bigg[ x\left( (\pm)_1 (1-y)\frac{m_j}{m_i}-z\right)\left(z +(\pm)_1 y\,\frac{m_j}{m_i} +(\pm)_2 \frac{m_{N_f}}{m_i}\right)\left( (\pm)_1 y\,\frac{m_j}{m_i}-(1-z)\right) \nonumber \\
  & + xy\left(1 -(\pm)_2  \frac{m_{N_f}}{m_i}\right)\left(\frac{m_j^2}{m_i^2}\,(1-y) -z\right) \nonumber\\
  &  + xz\left(+(\pm)_1 \frac{m_j}{m_i} -(\pm)_2  \frac{m_{N_f}}{m_i}\right)\left((1-z)-y\frac{m_j^2}{m_i^2}\right) \bigg] \bigg\rbrace \nonumber\\
  & + m_W^{-2} \left[ x (1-z) (\pm)_1 x(1-y) \frac{m_j}{m_i} -(\pm)_2  x \frac{m_{N_f}}{m_i} + y\left(1 -(\pm)_2  \frac{m_{N_f}}{m_i}\right)\right. + \nonumber\\
  & + \left.z \left((\pm)_1  \frac{m_j}{m_i} -(\pm)_2  \frac{m_{N_f}}{m_i}\right)\right] \nonumber\\
  & - m_W^{-2} \bigg[ (1 - 3 x) \left((\pm)_2\frac{m_{N_f}}{m_i} - (1-z) -(\pm)_1 (1 - y) \frac{m_j}{m_i}\right) - x z -(\pm)_1 x y \frac{m_j}{m_i}  \nonumber\\ 	
   &+ \left((\pm)_2\frac{m_{N_f}}{m_i} - 1\right)(1 - 3 y) \left((\pm)_2\frac{m_{N_f}}{m_i} -(\pm)_1 \frac{m_j}{m_i}\right) (1 + 3 z) \bigg] \times \nonumber\\
   & \times\log \left(\frac{m_W^2}{-xz\, m_i^2 - xy\, m_j^2 +(1-x) m_W^2 + x\, m_{N_f}^2}\right) \Bigg] 
%
\end{align}
\endgroup
The expression involving a neutral vector boson in the loop is similarly complicated:
\begingroup
\allowdisplaybreaks
\begin{align} \label{eq:I4_def}
  I_{f,\,4}^{(\pm)_1\,(\pm)_2} &\equiv I_{f,\,4}[m_i,(\pm)_1 m_j,(\pm)_2 m_{E_f},m_Z] \nonumber\\
  &=  \int \mathrm{d}^3\mathbf{X} \Bigg(\left[-xz\, m_i^2 - xy\, m_j^2 +(1-x) m_{E_f}^2 + x\, m_Z^2\right]^{-1} \times \nonumber\\
  &\quad \times\Bigg\lbrace 2x\left( (1-z) + (\pm)_1 (1-y)\frac{m_j}{m_i} -(\pm)_2 2 \frac{m_{E_f}}{m_i}\right) \nonumber\\
  &\quad + \frac{m_i^2}{m_Z^2}\bigg[ (x-1) \left( (\pm)_1 \frac{m_j}{m_i} -(\pm)_2 \frac{m_{E_f}}{m_i}\right)\left(z + (\pm)_1 y\,\frac{m_j}{m_i}\right) \left(1 -(\pm)_2 \frac{m_{E_f}}{m_i}\right) \nonumber\\
  &\quad - z\left( (\pm)_1 \frac{m_j}{m_i} -(\pm)_2 \frac{m_{E_f}}{m_i}\right)\left(xy\,\frac{m_j^2}{m_i^2}+(1-x+xz)\right) \nonumber\\
  &\quad - y(1 -(\pm)_2 \frac{m_{E_f}}{m_i})\left(xz+(1-x+xy)\frac{m_j^2}{m_i^2}\right) \bigg]\Bigg\rbrace \nonumber\\
  &\quad + m_Z^{-2} \left((\pm)_1\frac{m_j}{m_i} -(\pm)_2\frac{m_{E_f}}{m_i}\right)  \times\nonumber\\
  &\qquad \times \left[ z + (1-3z)\log \left(\frac{m_Z^2}{-xz\, m_i^2 - xy\, m_j^2 +(1-x) m_{E_f}^2 + x\, m_Z^2}\right)\right] \nonumber\\
  &\quad + m_Z^{-2} \left(1 -(\pm)_2\frac{m_{E_f}}{m_i}\right) \times\nonumber\\
  &\qquad \times \left[ y + (1-3y)\log \left(\frac{m_Z^2}{-xz\, m_i^2 - xy\, m_j^2 +(1-x) m_{E_f}^2 + x\, m_Z^2}\right)\right]
  \Bigg).
%
\end{align}
\endgroup

\bibliographystyle{elsarticle-num}
\bibliography{literature}

\begin{thebibliography}{100}
\expandafter\ifx\csname url\endcsname\relax
  \def\url#1{\texttt{#1}}\fi
\expandafter\ifx\csname urlprefix\endcsname\relax\def\urlprefix{URL }\fi
\expandafter\ifx\csname href\endcsname\relax
  \def\href#1#2{#2} \def\path#1{#1}\fi

\bibitem{GellMann:1954kc}
M.~Gell-Mann, M.~L. Goldberger, {Scattering of low-energy photons by particles
  of spin 1/2}, Phys. Rev. 96 (1954) 1433--1438.
\newblock \href {http://dx.doi.org/10.1103/PhysRev.96.1433}
  {\path{doi:10.1103/PhysRev.96.1433}}.

\bibitem{Blum:2013xva}
T.~Blum, A.~Denig, I.~Logashenko, E.~de~Rafael, B.~Lee~Roberts, T.~Teubner,
  G.~Venanzoni, {The Muon (g-2) Theory Value: Present and Future}\href
  {http://arxiv.org/abs/1311.2198} {\path{arXiv:1311.2198}}.

\bibitem{Fukuda:1998mi}
Y.~Fukuda, et~al., {Evidence for oscillation of atmospheric neutrinos}, Phys.
  Rev. Lett. 81 (1998) 1562--1567.
\newblock \href {http://arxiv.org/abs/hep-ex/9807003}
  {\path{arXiv:hep-ex/9807003}}, \href
  {http://dx.doi.org/10.1103/PhysRevLett.81.1562}
  {\path{doi:10.1103/PhysRevLett.81.1562}}.

\bibitem{Ahmad:2002jz}
Q.~R. Ahmad, et~al., {Direct evidence for neutrino flavor transformation from
  neutral current interactions in the Sudbury Neutrino Observatory}, Phys. Rev.
  Lett. 89 (2002) 011301.
\newblock \href {http://arxiv.org/abs/nucl-ex/0204008}
  {\path{arXiv:nucl-ex/0204008}}, \href
  {http://dx.doi.org/10.1103/PhysRevLett.89.011301}
  {\path{doi:10.1103/PhysRevLett.89.011301}}.

\bibitem{Garwin:1960zz}
R.~L. Garwin, D.~P. Hutchinson, S.~Penman, G.~Shapiro, {Accurate Determination
  of the mu+ Magnetic Moment}, Phys. Rev. 118 (1960) 271--283.
\newblock \href {http://dx.doi.org/10.1103/PhysRev.118.271}
  {\path{doi:10.1103/PhysRev.118.271}}.

\bibitem{Burnett:1967zfb}
T.~Burnett, M.~J. Levine, {Intermediate vector boson contribution to the muon's
  anomalous magnetic moment}, Phys. Lett. B24 (1967) 467--468.
\newblock \href {http://dx.doi.org/10.1016/0370-2693(67)90274-2}
  {\path{doi:10.1016/0370-2693(67)90274-2}}.

\bibitem{Kinoshita:1967txv}
T.~Kinoshita, R.~J. Oakes, {Hadronic contributions to the muon magnetic
  moment}, Phys. Lett. B25 (1967) 143--145.
\newblock \href {http://dx.doi.org/10.1016/0370-2693(67)90209-2}
  {\path{doi:10.1016/0370-2693(67)90209-2}}.

\bibitem{Terazawa:1968jh}
H.~Terazawa, {All the hadronic contributions to the anomalous magnetic moment
  of the muon and the lamb shift in the hydrogen atom}, Prog. Theor. Phys. 39
  (1968) 1326--1332.
\newblock \href {http://dx.doi.org/10.1143/PTP.39.1326}
  {\path{doi:10.1143/PTP.39.1326}}.

\bibitem{Lautrup:1969fr}
B.~E. Lautrup, E.~De~Rafael, {Calculation of the sixth-order contribution from
  the fourth-order vacuum polarization to the difference of the anomalous
  magnetic moments of muon and electron}, Phys. Rev. 174 (1968) 1835--1842.
\newblock \href {http://dx.doi.org/10.1103/PhysRev.174.1835}
  {\path{doi:10.1103/PhysRev.174.1835}}.

\bibitem{Aldins:1970id}
J.~Aldins, T.~Kinoshita, S.~J. Brodsky, A.~J. Dufner, {Photon-Photon Scattering
  Contriution to the sixth order magnetic moments of the muon and electron},
  Phys. Rev. D1 (1970) 2378.
\newblock \href {http://dx.doi.org/10.1103/PhysRevD.1.2378}
  {\path{doi:10.1103/PhysRevD.1.2378}}.

\bibitem{Lautrup:1972iw}
B.~E. Lautrup, {On the order of magnitude of 8th order corrections to the
  anomalous magnetic moment of the muon}, Phys. Lett. B38 (1972) 408--410.
\newblock \href {http://dx.doi.org/10.1016/0370-2693(72)90168-2}
  {\path{doi:10.1016/0370-2693(72)90168-2}}.

\bibitem{Bramon:1972bc}
A.~Bramon, E.~Etim, M.~Greco, {Hadronic contributions to the muon anomalous
  magnetic moment}, Phys. Lett. B39 (1972) 514--516.
\newblock \href {http://dx.doi.org/10.1016/0370-2693(72)90334-6}
  {\path{doi:10.1016/0370-2693(72)90334-6}}.

\bibitem{Auberson:1972xd}
G.~Auberson, L.~Ling-Fong, {Lower bound for the hadronic contribution to the
  muon magnetic moment}, Phys. Rev. D5 (1972) 2269--2276.
\newblock \href {http://dx.doi.org/10.1103/PhysRevD.5.2269}
  {\path{doi:10.1103/PhysRevD.5.2269}}.

\bibitem{Hagiwara:2011af}
K.~Hagiwara, R.~Liao, A.~D. Martin, D.~Nomura, T.~Teubner, {$(g-2)_{\mu}$ and
  $\alpha(M_Z^2)$ re-evaluated using new precise data}, J. Phys. G38 (2011)
  085003.
\newblock \href {http://arxiv.org/abs/1105.3149} {\path{arXiv:1105.3149}},
  \href {http://dx.doi.org/10.1088/0954-3899/38/8/085003}
  {\path{doi:10.1088/0954-3899/38/8/085003}}.

\bibitem{Amsler:2008zzb}
C.~Amsler, et~al., {Review of Particle Physics}, Phys. Lett. B667 (2008)
  1--1340.
\newblock \href {http://dx.doi.org/10.1016/j.physletb.2008.07.018}
  {\path{doi:10.1016/j.physletb.2008.07.018}}.

\bibitem{Ellis:1994qf}
J.~R. Ellis, M.~Karliner, M.~A. Samuel, E.~Steinfelds, {The Anomalous magnetic
  moments of the electron and the muon: Improved QED predictions using Pade
  approximants}\href {http://arxiv.org/abs/hep-ph/9409376}
  {\path{arXiv:hep-ph/9409376}}.

\bibitem{Bijnens:1995xf}
J.~Bijnens, E.~Pallante, J.~Prades, {Analysis of the hadronic light by light
  contributions to the muon g-2}, Nucl. Phys. B474 (1996) 379--420.
\newblock \href {http://arxiv.org/abs/hep-ph/9511388}
  {\path{arXiv:hep-ph/9511388}}, \href
  {http://dx.doi.org/10.1016/0550-3213(96)00288-X}
  {\path{doi:10.1016/0550-3213(96)00288-X}}.

\bibitem{Alemany:1997tn}
R.~Alemany, M.~Davier, A.~Hocker, {Improved determination of the hadronic
  contribution to the muon (g-2) and to alpha (M(z)) using new data from
  hadronic tau decays}, Eur. Phys. J. C2 (1998) 123--135.
\newblock \href {http://arxiv.org/abs/hep-ph/9703220}
  {\path{arXiv:hep-ph/9703220}}, \href
  {http://dx.doi.org/10.1007/s100520050127} {\path{doi:10.1007/s100520050127}}.

\bibitem{Hayakawa:1997rq}
M.~Hayakawa, T.~Kinoshita, {Pseudoscalar pole terms in the hadronic light by
  light scattering contribution to muon g - 2}, Phys. Rev. D57 (1998) 465--477,
  [Erratum: Phys. Rev.D66,019902(2002)].
\newblock \href {http://arxiv.org/abs/hep-ph/9708227}
  {\path{arXiv:hep-ph/9708227}}, \href
  {http://dx.doi.org/10.1103/PhysRevD.57.465, 10.1103/PhysRevD.66.019902}
  {\path{doi:10.1103/PhysRevD.57.465, 10.1103/PhysRevD.66.019902}}.

\bibitem{Knecht:2001qf}
M.~Knecht, A.~Nyffeler, {Hadronic light by light corrections to the muon g-2:
  The Pion pole contribution}, Phys. Rev. D65 (2002) 073034.
\newblock \href {http://arxiv.org/abs/hep-ph/0111058}
  {\path{arXiv:hep-ph/0111058}}, \href
  {http://dx.doi.org/10.1103/PhysRevD.65.073034}
  {\path{doi:10.1103/PhysRevD.65.073034}}.

\bibitem{Blokland:2001pb}
I.~R. Blokland, A.~Czarnecki, K.~Melnikov, {Pion pole contribution to hadronic
  light by light scattering and muon anomalous magnetic moment}, Phys. Rev.
  Lett. 88 (2002) 071803.
\newblock \href {http://arxiv.org/abs/hep-ph/0112117}
  {\path{arXiv:hep-ph/0112117}}, \href
  {http://dx.doi.org/10.1103/PhysRevLett.88.071803}
  {\path{doi:10.1103/PhysRevLett.88.071803}}.

\bibitem{Melnikov:2003xd}
K.~Melnikov, A.~Vainshtein, {Hadronic light-by-light scattering contribution to
  the muon anomalous magnetic moment revisited}, Phys. Rev. D70 (2004) 113006.
\newblock \href {http://arxiv.org/abs/hep-ph/0312226}
  {\path{arXiv:hep-ph/0312226}}, \href
  {http://dx.doi.org/10.1103/PhysRevD.70.113006}
  {\path{doi:10.1103/PhysRevD.70.113006}}.

\bibitem{deTroconiz:2004yzs}
J.~F. de~Troconiz, F.~J. Yndurain, {The Hadronic contributions to the anomalous
  magnetic moment of the muon}, Phys. Rev. D71 (2005) 073008.
\newblock \href {http://arxiv.org/abs/hep-ph/0402285}
  {\path{arXiv:hep-ph/0402285}}, \href
  {http://dx.doi.org/10.1103/PhysRevD.71.073008}
  {\path{doi:10.1103/PhysRevD.71.073008}}.

\bibitem{Kinoshita:2005sm}
T.~Kinoshita, M.~Nio, {The Tenth-order QED contribution to the lepton g-2:
  Evaluation of dominant $\alpha^5$ terms of muon g-2}, Phys. Rev. D73 (2006)
  053007.
\newblock \href {http://arxiv.org/abs/hep-ph/0512330}
  {\path{arXiv:hep-ph/0512330}}, \href
  {http://dx.doi.org/10.1103/PhysRevD.73.053007}
  {\path{doi:10.1103/PhysRevD.73.053007}}.

\bibitem{Kataev:2006yh}
A.~L. Kataev, {Reconsidered estimates of the 10th order QED contributions to
  the muon anomaly}, Phys. Rev. D74 (2006) 073011.
\newblock \href {http://arxiv.org/abs/hep-ph/0608120}
  {\path{arXiv:hep-ph/0608120}}, \href
  {http://dx.doi.org/10.1103/PhysRevD.74.073011}
  {\path{doi:10.1103/PhysRevD.74.073011}}.

\bibitem{Aoyama:2008gy}
T.~Aoyama, M.~Hayakawa, T.~Kinoshita, M.~Nio, N.~Watanabe, {Eighth-Order
  Vacuum-Polarization Function Formed by Two Light-by-Light-Scattering Diagrams
  and its Contribution to the Tenth-Order Electron g-2}, Phys. Rev. D78 (2008)
  053005.
\newblock \href {http://arxiv.org/abs/0806.3390} {\path{arXiv:0806.3390}},
  \href {http://dx.doi.org/10.1103/PhysRevD.78.053005}
  {\path{doi:10.1103/PhysRevD.78.053005}}.

\bibitem{Prades:2009tw}
J.~Prades, E.~de~Rafael, A.~Vainshtein, {The Hadronic Light-by-Light Scattering
  Contribution to the Muon and Electron Anomalous Magnetic Moments}, Adv. Ser.
  Direct. High Energy Phys. 20 (2009) 303--317.
\newblock \href {http://arxiv.org/abs/0901.0306} {\path{arXiv:0901.0306}},
  \href {http://dx.doi.org/10.1142/9789814271844_0009}
  {\path{doi:10.1142/9789814271844_0009}}.

\bibitem{Goecke:2010if}
T.~Goecke, C.~S. Fischer, R.~Williams, {Hadronic light-by-light scattering in
  the muon g-2: a Dyson-Schwinger equation approach}, Phys. Rev. D83 (2011)
  094006, [Erratum: Phys. Rev.D86,099901(2012)].
\newblock \href {http://arxiv.org/abs/1012.3886} {\path{arXiv:1012.3886}},
  \href {http://dx.doi.org/10.1103/PhysRevD.83.094006,
  10.1103/PhysRevD.86.099901} {\path{doi:10.1103/PhysRevD.83.094006,
  10.1103/PhysRevD.86.099901}}.

\bibitem{Davier:2010nc}
M.~Davier, A.~Hoecker, B.~Malaescu, Z.~Zhang, {Reevaluation of the Hadronic
  Contributions to the Muon g-2 and to alpha(MZ)}, Eur. Phys. J. C71 (2011)
  1515, [Erratum: Eur. Phys. J.C72,1874(2012)].
\newblock \href {http://arxiv.org/abs/1010.4180} {\path{arXiv:1010.4180}},
  \href {http://dx.doi.org/10.1140/epjc/s10052-012-1874-8,
  10.1140/epjc/s10052-010-1515-z} {\path{doi:10.1140/epjc/s10052-012-1874-8,
  10.1140/epjc/s10052-010-1515-z}}.

\bibitem{Boughezal:2011vw}
R.~Boughezal, K.~Melnikov, {Hadronic light-by-light scattering contribution to
  the muon magnetic anomaly: constituent quark loops and QCD effects}, Phys.
  Lett. B704 (2011) 193--196.
\newblock \href {http://arxiv.org/abs/1104.4510} {\path{arXiv:1104.4510}},
  \href {http://dx.doi.org/10.1016/j.physletb.2011.09.001}
  {\path{doi:10.1016/j.physletb.2011.09.001}}.

\bibitem{Aoyama:2011dy}
T.~Aoyama, M.~Hayakawa, T.~Kinoshita, M.~Nio, {Tenth-Order QED Lepton Anomalous
  Magnetic Moment --- Eighth-Order Vertices Containing a Second-Order Vacuum
  Polarization}, Phys. Rev. D85 (2012) 033007.
\newblock \href {http://arxiv.org/abs/1110.2826} {\path{arXiv:1110.2826}},
  \href {http://dx.doi.org/10.1103/PhysRevD.85.033007}
  {\path{doi:10.1103/PhysRevD.85.033007}}.

\bibitem{Goecke:2012qm}
T.~Goecke, C.~S. Fischer, R.~Williams, {Role of momentum dependent dressing
  functions and vector meson dominance in hadronic light-by-light contributions
  to the muon $g-2$}, Phys. Rev. D87~(3) (2013) 034013.
\newblock \href {http://arxiv.org/abs/1210.1759} {\path{arXiv:1210.1759}},
  \href {http://dx.doi.org/10.1103/PhysRevD.87.034013}
  {\path{doi:10.1103/PhysRevD.87.034013}}.

\bibitem{Bijnens:2012an}
J.~Bijnens, M.~Z. Abyaneh, {The hadronic light-by-light contribution to the
  muon anomalous magnetic moment and renormalization group for EFT}, EPJ Web
  Conf. 37 (2012) 01007.
\newblock \href {http://arxiv.org/abs/1208.3548} {\path{arXiv:1208.3548}},
  \href {http://dx.doi.org/10.1051/epjconf/20123701007}
  {\path{doi:10.1051/epjconf/20123701007}}.

\bibitem{Dorokhov:2012qa}
A.~E. Dorokhov, A.~E. Radzhabov, A.~S. Zhevlakov, {The Light-by-Light
  Contribution to the Muon (g-2) from Lightest Pseudoscalar and Scalar Mesons
  within Nonlocal Chiral Quark Model}, Eur. Phys. J. C72 (2012) 2227.
\newblock \href {http://arxiv.org/abs/1204.3729} {\path{arXiv:1204.3729}},
  \href {http://dx.doi.org/10.1140/epjc/s10052-012-2227-3}
  {\path{doi:10.1140/epjc/s10052-012-2227-3}}.

\bibitem{Aoyama:2012wk}
T.~Aoyama, M.~Hayakawa, T.~Kinoshita, M.~Nio, {Complete Tenth-Order QED
  Contribution to the Muon g-2}, Phys. Rev. Lett. 109 (2012) 111808.
\newblock \href {http://arxiv.org/abs/1205.5370} {\path{arXiv:1205.5370}},
  \href {http://dx.doi.org/10.1103/PhysRevLett.109.111808}
  {\path{doi:10.1103/PhysRevLett.109.111808}}.

\bibitem{Williams:2013tia}
R.~Williams, C.~S. Fischer, T.~Goecke, {Non-contact Interactions and the
  Hadronic Light-by-light Contribution to the Muon $g-2$}, Acta Phys. Polon.
  Supp. 6~(3) (2013) 785--790.
\newblock \href {http://arxiv.org/abs/1304.4347} {\path{arXiv:1304.4347}},
  \href {http://dx.doi.org/10.5506/APhysPolBSupp.6.785}
  {\path{doi:10.5506/APhysPolBSupp.6.785}}.

\bibitem{Kurz:2014wya}
A.~Kurz, T.~Liu, P.~Marquard, M.~Steinhauser, {Hadronic contribution to the
  muon anomalous magnetic moment to next-to-next-to-leading order}, Phys. Lett.
  B734 (2014) 144--147.
\newblock \href {http://arxiv.org/abs/1403.6400} {\path{arXiv:1403.6400}},
  \href {http://dx.doi.org/10.1016/j.physletb.2014.05.043}
  {\path{doi:10.1016/j.physletb.2014.05.043}}.

\bibitem{Colangelo:2014qya}
G.~Colangelo, M.~Hoferichter, A.~Nyffeler, M.~Passera, P.~Stoffer, {Remarks on
  higher-order hadronic corrections to the muon g−2}, Phys. Lett. B735 (2014)
  90--91.
\newblock \href {http://arxiv.org/abs/1403.7512} {\path{arXiv:1403.7512}},
  \href {http://dx.doi.org/10.1016/j.physletb.2014.06.012}
  {\path{doi:10.1016/j.physletb.2014.06.012}}.

\bibitem{Masjuan:2014rea}
P.~Masjuan, {Overview of the hadronic light-by-light contribution to the muon
  (g−2)}, Nucl. Part. Phys. Proc. 260 (2015) 111--115.
\newblock \href {http://arxiv.org/abs/1411.6397} {\path{arXiv:1411.6397}},
  \href {http://dx.doi.org/10.1016/j.nuclphysbps.2015.02.023}
  {\path{doi:10.1016/j.nuclphysbps.2015.02.023}}.

\bibitem{Dorokhov:2014iva}
A.~E. Dorokhov, A.~E. Radzhabov, A.~S. Zhevlakov, {Status of the lepton $g-2$
  and effects of hadronic corrections}, JETP Lett. 100~(2) (2014) 133--143,
  [Pisma Zh. Eksp. Teor. Fiz.100,no.2,141(2014)].
\newblock \href {http://arxiv.org/abs/1406.1019} {\path{arXiv:1406.1019}},
  \href {http://dx.doi.org/10.1134/S0021364014140045}
  {\path{doi:10.1134/S0021364014140045}}.

\bibitem{Roig:2014dya}
P.~Roig, A.~Guevara, G.~Lopez~Castro, {Lightest pseudoscalar exchange
  contribution to light-by-light scattering piece of the muon g − 2}, Nucl.
  Part. Phys. Proc. 273-275 (2016) 2702--2704.
\newblock \href {http://arxiv.org/abs/1409.2174} {\path{arXiv:1409.2174}},
  \href {http://dx.doi.org/10.1016/j.nuclphysbps.2015.10.036}
  {\path{doi:10.1016/j.nuclphysbps.2015.10.036}}.

\bibitem{Blum:2014oka}
T.~Blum, S.~Chowdhury, M.~Hayakawa, T.~Izubuchi, {Hadronic light-by-light
  scattering contribution to the muon anomalous magnetic moment from lattice
  QCD}, Phys. Rev. Lett. 114~(1) (2015) 012001.
\newblock \href {http://arxiv.org/abs/1407.2923} {\path{arXiv:1407.2923}},
  \href {http://dx.doi.org/10.1103/PhysRevLett.114.012001}
  {\path{doi:10.1103/PhysRevLett.114.012001}}.

\bibitem{Blum:2015gfa}
T.~Blum, N.~Christ, M.~Hayakawa, T.~Izubuchi, L.~Jin, C.~Lehner, {Lattice
  Calculation of Hadronic Light-by-Light Contribution to the Muon Anomalous
  Magnetic Moment}, Phys. Rev. D93~(1) (2016) 014503.
\newblock \href {http://arxiv.org/abs/1510.07100} {\path{arXiv:1510.07100}},
  \href {http://dx.doi.org/10.1103/PhysRevD.93.014503}
  {\path{doi:10.1103/PhysRevD.93.014503}}.

\bibitem{Eidelman:2015isu}
S.~Eidelman, {Muon g-2 and hadronic vacuum polarization: Recent developments},
  PoS CD15 (2015) 013.

\bibitem{Bijnens:2016hgx}
J.~Bijnens, J.~Relefors, {Pion light-by-light contributions to the muon $g-2$},
  JHEP 09 (2016) 113.
\newblock \href {http://arxiv.org/abs/1608.01454} {\path{arXiv:1608.01454}},
  \href {http://dx.doi.org/10.1007/JHEP09(2016)113}
  {\path{doi:10.1007/JHEP09(2016)113}}.

\bibitem{Sanchez-Puertas:2016mmz}
P.~Sanchez-Puertas, P.~Masjuan, {Updated pseudoscalar contributions to the
  hadronic light-by-light of the muon (g-2)}, Mod. Phys. Lett. A31 (2016)
  1630034.
\newblock \href {http://arxiv.org/abs/1606.02704} {\path{arXiv:1606.02704}},
  \href {http://dx.doi.org/10.1142/S0217732316300342}
  {\path{doi:10.1142/S0217732316300342}}.

\bibitem{Nyffeler:2016xul}
A.~Nyffeler, {On the precision of a data-driven estimate of the
  pseudoscalar-pole contribution to hadronic light-by-light scattering in the
  muon $g−2$}, EPJ Web Conf. 118 (2016) 01024.
\newblock \href {http://arxiv.org/abs/1602.03737} {\path{arXiv:1602.03737}},
  \href {http://dx.doi.org/10.1051/epjconf/201611801024}
  {\path{doi:10.1051/epjconf/201611801024}}.

\bibitem{Bernard:2016tlj}
D.~Bernard, {Low-energy hadronic cross sections measurements at BaBar and $g
  − 2$ of the muon}, Nucl. Part. Phys. Proc. 282-284 (2017) 132--138.
\newblock \href {http://arxiv.org/abs/1607.07181} {\path{arXiv:1607.07181}},
  \href {http://dx.doi.org/10.1016/j.nuclphysbps.2016.12.025}
  {\path{doi:10.1016/j.nuclphysbps.2016.12.025}}.

\bibitem{Dominguez:2016eol}
C.~A. Dominguez, K.~Schilcher, H.~Spiesberger, {Theoretical determination of
  the hadronic g − 2 of the muon}, Mod. Phys. Lett. A31~(32) (2016) 1630035.
\newblock \href {http://arxiv.org/abs/1605.07903} {\path{arXiv:1605.07903}},
  \href {http://dx.doi.org/10.1142/S0217732316300354}
  {\path{doi:10.1142/S0217732316300354}}.

\bibitem{Ananthanarayan:2016mns}
B.~Ananthanarayan, I.~Caprini, D.~Das, I.~Sentitemsu~Imsong, {Precise
  determination of the low-energy hadronic contribution to the muon $g-2$ from
  analyticity and unitarity: An improved analysis}, Phys. Rev. D93~(11) (2016)
  116007.
\newblock \href {http://arxiv.org/abs/1605.00202} {\path{arXiv:1605.00202}},
  \href {http://dx.doi.org/10.1103/PhysRevD.93.116007}
  {\path{doi:10.1103/PhysRevD.93.116007}}.

\bibitem{Knecht:2014sea}
M.~Knecht, {The Muon Anomalous Magnetic Moment}, Nucl. Part. Phys. Proc.
  258-259 (2015) 235--240.
\newblock \href {http://arxiv.org/abs/1412.1228} {\path{arXiv:1412.1228}},
  \href {http://dx.doi.org/10.1016/j.nuclphysbps.2015.01.050}
  {\path{doi:10.1016/j.nuclphysbps.2015.01.050}}.

\bibitem{KNECHT:2014bsa}
M.~Knecht, {Status of Standard Model Calculation of Lepton g - 2}, Int. J. Mod.
  Phys. Conf. Ser. 35 (2014) 1460405.
\newblock \href {http://dx.doi.org/10.1142/S2010194514604050}
  {\path{doi:10.1142/S2010194514604050}}.

\bibitem{Colangelo:2014pva}
G.~Colangelo, M.~Hoferichter, B.~Kubis, M.~Procura, P.~Stoffer, {Towards a
  data-driven analysis of hadronic light-by-light scattering}, Phys. Lett. B738
  (2014) 6--12.
\newblock \href {http://arxiv.org/abs/1408.2517} {\path{arXiv:1408.2517}},
  \href {http://dx.doi.org/10.1016/j.physletb.2014.09.021}
  {\path{doi:10.1016/j.physletb.2014.09.021}}.

\bibitem{Kurz:2015bia}
A.~Kurz, T.~Liu, P.~Marquard, A.~V. Smirnov, V.~A. Smirnov, M.~Steinhauser,
  {Light-by-light-type corrections to the muon anomalous magnetic moment at
  four-loop order}, Phys. Rev. D92~(7) (2015) 073019.
\newblock \href {http://arxiv.org/abs/1508.00901} {\path{arXiv:1508.00901}},
  \href {http://dx.doi.org/10.1103/PhysRevD.92.073019}
  {\path{doi:10.1103/PhysRevD.92.073019}}.

\bibitem{Stoffer:2015fvt}
P.~Stoffer, G.~Colangelo, M.~Hoferichter, M.~Procura, {Hadronic light-by-light
  scattering and the muon g−2}, Nuovo Cim. C38~(4) (2016) 135.
\newblock \href {http://dx.doi.org/10.1393/ncc/i2015-15135-9}
  {\path{doi:10.1393/ncc/i2015-15135-9}}.

\bibitem{Green:2015mva}
J.~Green, N.~Asmussen, O.~Gryniuk, G.~von Hippel, H.~B. Meyer, A.~Nyffeler,
  V.~Pascalutsa, {Direct calculation of hadronic light-by-light scattering},
  PoS LATTICE2015 (2016) 109.
\newblock \href {http://arxiv.org/abs/1510.08384} {\path{arXiv:1510.08384}}.

\bibitem{Jin:2015bty}
L.~Jin, T.~Blum, N.~Christ, M.~Hayakawa, T.~Izubuchi, C.~Lehner, {Hadronic
  Light by Light Contributions to the Muon Anomalous Magnetic Moment With
  Physical Pions}, PoS LATTICE2015 (2016) 103.
\newblock \href {http://arxiv.org/abs/1511.05198} {\path{arXiv:1511.05198}}.

\bibitem{Nyffeler:2016gnb}
A.~Nyffeler, {Precision of a data-driven estimate of hadronic light-by-light
  scattering in the muon $g-2$: Pseudoscalar-pole contribution}, Phys. Rev.
  D94~(5) (2016) 053006.
\newblock \href {http://arxiv.org/abs/1602.03398} {\path{arXiv:1602.03398}},
  \href {http://dx.doi.org/10.1103/PhysRevD.94.053006}
  {\path{doi:10.1103/PhysRevD.94.053006}}.

\bibitem{Asmussen:2016lse}
N.~Asmussen, J.~Green, H.~B. Meyer, A.~Nyffeler, {Position-space approach to
  hadronic light-by-light scattering in the muon $g-2$ on the lattice}, PoS
  LATTICE2016 (2016) 164.
\newblock \href {http://arxiv.org/abs/1609.08454} {\path{arXiv:1609.08454}}.

\bibitem{Kinoshita:1981vs}
T.~Kinoshita, W.~B. Lindquist, {Eighth Order Anomalous Magnetic Moment of the
  electron}, Phys. Rev. Lett. 47 (1981) 1573.
\newblock \href {http://dx.doi.org/10.1103/PhysRevLett.47.1573}
  {\path{doi:10.1103/PhysRevLett.47.1573}}.

\bibitem{Kinoshita:1981ww}
T.~Kinoshita, W.~B. Lindquist, {Eighth Order Magnetic Moment of the Electron.
  4. Vertex Diagrams Containing Photon - Photon Scattering Subdiagrams}, Phys.
  Rev. D39 (1989) 2407.
\newblock \href {http://dx.doi.org/10.1103/PhysRevD.39.2407}
  {\path{doi:10.1103/PhysRevD.39.2407}}.

\bibitem{Kinoshita:1981wm}
T.~Kinoshita, W.~B. Lindquist, {Eighth Order Magnetic Moment of the Electron.
  5. Diagrams Containing No Vacuum Polarization Loop}, Phys. Rev. D42 (1990)
  636--655.
\newblock \href {http://dx.doi.org/10.1103/PhysRevD.42.636}
  {\path{doi:10.1103/PhysRevD.42.636}}.

\bibitem{Odom:2006zz}
B.~C. Odom, D.~Hanneke, B.~D'Urso, G.~Gabrielse, {New Measurement of the
  Electron Magnetic Moment Using a One-Electron Quantum Cyclotron}, Phys. Rev.
  Lett. 97 (2006) 030801, [Erratum: Phys. Rev. Lett.99,039902(2007)].
\newblock \href {http://dx.doi.org/10.1103/PhysRevLett.97.030801}
  {\path{doi:10.1103/PhysRevLett.97.030801}}.

\bibitem{Hanneke:2008tm}
D.~Hanneke, S.~Fogwell, G.~Gabrielse, {New Measurement of the Electron Magnetic
  Moment and the Fine Structure Constant}, Phys. Rev. Lett. 100 (2008) 120801.
\newblock \href {http://arxiv.org/abs/0801.1134} {\path{arXiv:0801.1134}},
  \href {http://dx.doi.org/10.1103/PhysRevLett.100.120801}
  {\path{doi:10.1103/PhysRevLett.100.120801}}.

\bibitem{Hanneke:2010au}
D.~Hanneke, S.~F. Hoogerheide, G.~Gabrielse, {Cavity Control of a
  Single-Electron Quantum Cyclotron: Measuring the Electron Magnetic Moment},
  Phys. Rev. A83 (2011) 052122.
\newblock \href {http://arxiv.org/abs/1009.4831} {\path{arXiv:1009.4831}},
  \href {http://dx.doi.org/10.1103/PhysRevA.83.052122}
  {\path{doi:10.1103/PhysRevA.83.052122}}.

\bibitem{Haffner:2000zzi}
H.~Häffner, T.~Beier, N.~Hermanspahn, H.~J. Kluge, W.~Quint, S.~Stahl,
  J.~Verdú, G.~Werth, {High-Accuracy Measurement of the Magnetic Moment
  Anomaly of the Electron Bound in Hydrogenlike Carbon}, Phys. Rev. Lett.
  85~(25) (2000) 5308.
\newblock \href {http://dx.doi.org/10.1103/PhysRevLett.85.5308}
  {\path{doi:10.1103/PhysRevLett.85.5308}}.

\bibitem{Bouchendira:2010es}
R.~Bouchendira, P.~Clade, S.~Guellati-Khelifa, F.~Nez, F.~Biraben, {New
  determination of the fine structure constant and test of the quantum
  electrodynamics}, Phys. Rev. Lett. 106 (2011) 080801.
\newblock \href {http://arxiv.org/abs/1012.3627} {\path{arXiv:1012.3627}},
  \href {http://dx.doi.org/10.1103/PhysRevLett.106.080801}
  {\path{doi:10.1103/PhysRevLett.106.080801}}.

\bibitem{Aoyama:2014sxa}
T.~Aoyama, M.~Hayakawa, T.~Kinoshita, M.~Nio, {Tenth-Order Electron Anomalous
  Magnetic Moment Contribution of Diagrams without Closed Lepton Loops}, Phys.
  Rev. D91~(3) (2015) 033006.
\newblock \href {http://arxiv.org/abs/1412.8284} {\path{arXiv:1412.8284}},
  \href {http://dx.doi.org/10.1103/PhysRevD.91.033006}
  {\path{doi:10.1103/PhysRevD.91.033006}}.

\bibitem{Giudice:2012ms}
G.~F. Giudice, P.~Paradisi, M.~Passera, {Testing new physics with the electron
  g-2}, JHEP 11 (2012) 113.
\newblock \href {http://arxiv.org/abs/1208.6583} {\path{arXiv:1208.6583}},
  \href {http://dx.doi.org/10.1007/JHEP11(2012)113}
  {\path{doi:10.1007/JHEP11(2012)113}}.

\bibitem{Aboubrahim:2014hya}
A.~Aboubrahim, T.~Ibrahim, P.~Nath, {Probe of New Physics using Precision
  Measurement of the Electron Magnetic Moment}, Phys. Rev. D89~(9) (2014)
  093016.
\newblock \href {http://arxiv.org/abs/1403.6448} {\path{arXiv:1403.6448}},
  \href {http://dx.doi.org/10.1103/PhysRevD.89.093016}
  {\path{doi:10.1103/PhysRevD.89.093016}}.

\bibitem{Eidelman:2007sb}
S.~Eidelman, M.~Passera, {Theory of the tau lepton anomalous magnetic moment},
  Mod. Phys. Lett. A22 (2007) 159--179.
\newblock \href {http://arxiv.org/abs/hep-ph/0701260}
  {\path{arXiv:hep-ph/0701260}}, \href
  {http://dx.doi.org/10.1142/S0217732307022694}
  {\path{doi:10.1142/S0217732307022694}}.

\bibitem{Eidelman:2007fj}
S.~Eidelman, M.~Giacomini, F.~V. Ignatov, M.~Passera, {The tau lepton anomalous
  magnetic moment}, Nucl. Phys. Proc. Suppl. 169 (2007) 226--231, [,226(2007)].
\newblock \href {http://arxiv.org/abs/hep-ph/0702026}
  {\path{arXiv:hep-ph/0702026}}, \href
  {http://dx.doi.org/10.1016/j.nuclphysbps.2007.03.002}
  {\path{doi:10.1016/j.nuclphysbps.2007.03.002}}.

\bibitem{Abdallah:2003xd}
J.~Abdallah, et~al., {Study of tau-pair production in photon-photon collisions
  at LEP and limits on the anomalous electromagnetic moments of the tau
  lepton}, Eur. Phys. J. C35 (2004) 159--170.
\newblock \href {http://arxiv.org/abs/hep-ex/0406010}
  {\path{arXiv:hep-ex/0406010}}, \href
  {http://dx.doi.org/10.1140/epjc/s2004-01852-y}
  {\path{doi:10.1140/epjc/s2004-01852-y}}.

\bibitem{Silverman:1982ft}
D.~J. Silverman, G.~L. Shaw, {Limits on the Composite Structure of the Tau
  Lepton and Quarks From Anomalous Magnetic Moment Measurements in $e^+ e^-$
  Annihilation}, Phys. Rev. D27 (1983) 1196.
\newblock \href {http://dx.doi.org/10.1103/PhysRevD.27.1196}
  {\path{doi:10.1103/PhysRevD.27.1196}}.

\bibitem{Domokos:1985rp}
G.~Domokos, S.~Kovesi-Domokos, C.~Vaz, D.~Wurmser, {Magnetic Moments of Heavy
  Quarks and Leptons}, Phys. Rev. D32 (1985) 247.
\newblock \href {http://dx.doi.org/10.1103/PhysRevD.32.247}
  {\path{doi:10.1103/PhysRevD.32.247}}.

\bibitem{Grifols:1990ha}
J.~A. Grifols, A.~Mendez, {Electromagnetic properties of the tau lepton from Z0
  decay}, Phys. Lett. B255 (1991) 611--612, [Erratum: Phys.
  Lett.B259,512(1991)].
\newblock \href {http://dx.doi.org/10.1016/0370-2693(91)90276-V}
  {\path{doi:10.1016/0370-2693(91)90276-V}}.

\bibitem{delAguila:1990jg}
F.~del Aguila, M.~Sher, {The Electric dipole moment of the tau}, Phys. Lett.
  B252 (1990) 116--118.
\newblock \href {http://dx.doi.org/10.1016/0370-2693(90)91091-O}
  {\path{doi:10.1016/0370-2693(90)91091-O}}.

\bibitem{Escribano:1996wp}
R.~Escribano, E.~Masso, {Improved bounds on the electromagnetic dipole moments
  of the $\tau$ lepton}, Phys. Lett. B395 (1997) 369--372.
\newblock \href {http://arxiv.org/abs/hep-ph/9609423}
  {\path{arXiv:hep-ph/9609423}}, \href
  {http://dx.doi.org/10.1016/S0370-2693(97)00059-2}
  {\path{doi:10.1016/S0370-2693(97)00059-2}}.

\bibitem{Ackerstaff:1998mt}
K.~Ackerstaff, et~al., {An Upper limit on the anomalous magnetic moment of the
  tau lepton}, Phys. Lett. B431 (1998) 188--198.
\newblock \href {http://arxiv.org/abs/hep-ex/9803020}
  {\path{arXiv:hep-ex/9803020}}, \href
  {http://dx.doi.org/10.1016/S0370-2693(98)00520-6}
  {\path{doi:10.1016/S0370-2693(98)00520-6}}.

\bibitem{Acciarri:1998iv}
M.~Acciarri, et~al., {Measurement of the anomalous magnetic and electric dipole
  moments of the tau lepton}, Phys. Lett. B434 (1998) 169--179.
\newblock \href {http://dx.doi.org/10.1016/S0370-2693(98)00736-9}
  {\path{doi:10.1016/S0370-2693(98)00736-9}}.

\bibitem{GonzalezSprinberg:2000mk}
G.~A. Gonzalez-Sprinberg, A.~Santamaria, J.~Vidal, {Model independent bounds on
  the tau lepton electromagnetic and weak magnetic moments}, Nucl. Phys. B582
  (2000) 3--18.
\newblock \href {http://arxiv.org/abs/hep-ph/0002203}
  {\path{arXiv:hep-ph/0002203}}, \href
  {http://dx.doi.org/10.1016/S0550-3213(00)00275-3}
  {\path{doi:10.1016/S0550-3213(00)00275-3}}.

\bibitem{Passera:2007fk}
M.~Passera, {Electron, muon and tau magnetic moments: A Theoretical update},
  Nucl. Phys. Proc. Suppl. 169 (2007) 213--225, [,213(2007)].
\newblock \href {http://arxiv.org/abs/hep-ph/0702027}
  {\path{arXiv:hep-ph/0702027}}, \href
  {http://dx.doi.org/10.1016/j.nuclphysbps.2007.03.001}
  {\path{doi:10.1016/j.nuclphysbps.2007.03.001}}.

\bibitem{nevis57}
R.~L. Garwin, L.~M. Lederman, M.~Weinrich, {Observations of the Failure of
  Conservation of Parity and Charge Conjugation in Meson Decays: The Magnetic
  Moment of the Free Muon}, Phys. Rev. 105 (1957) 1415--1417.
\newblock \href {http://dx.doi.org/10.1103/PhysRev.105.1415}
  {\path{doi:10.1103/PhysRev.105.1415}}.

\bibitem{nevis59}
R.~L. Garwin, D.~P. Hutchinson, S.~Penman, G.~Shapiro, {Accurate Determination
  of the mu+ Magnetic Moment}, Phys. Rev. 118 (1960) 271--283.
\newblock \href {http://dx.doi.org/10.1103/PhysRev.118.271}
  {\path{doi:10.1103/PhysRev.118.271}}.

\bibitem{cern1-61}
G.~Charpak, F.~J.~M. Farley, R.~L. Garwin, T.~Muller, J.~C. Sens, V.~L.
  Telegdi, A.~Zichichi, {Measurement of the anomalous magnetic moment of the
  muon}, Phys. Rev. Lett. 6 (1961) 128--132.
\newblock \href {http://dx.doi.org/10.1103/PhysRevLett.6.128}
  {\path{doi:10.1103/PhysRevLett.6.128}}.

\bibitem{cern2-61}
G.~Charpak, P.~J.~M. Farley, E.~L. Garwin, T.~Muller, J.~C. Sens, A.~Zichichi,
  \href{http://dx.doi.org/10.1007/BF02783344}{The anomalous magnetic moment of
  the muon}, Il Nuovo Cimento (1955-1965) 37~(4) (1965) 1241--1363.
\newblock \href {http://dx.doi.org/10.1007/BF02783344}
  {\path{doi:10.1007/BF02783344}}.
\newline\urlprefix\url{http://dx.doi.org/10.1007/BF02783344}

\bibitem{cern1-62}
G.~Charpak, F.~J.~M. Farley, R.~L. Garwin, {A New Measurement of the Anomalous
  Magnetic Moment of the Muon}, Phys. Lett. 1 (1962) 16.
\newblock \href {http://dx.doi.org/10.1016/0031-9163(62)90263-9}
  {\path{doi:10.1016/0031-9163(62)90263-9}}.

\bibitem{cern2-68}
J.~Bailey, W.~Bartl, G.~Von~Bochmann, R.~C.~A. Brown, F.~J.~M. Farley,
  H.~Joestlein, E.~Picasso, R.~W. Williams, {Precision measurement of the
  anomalous magnetic moment of the muon}, Phys. Lett. B28 (1968) 287--290.
\newblock \href {http://dx.doi.org/10.1016/0370-2693(68)90261-X}
  {\path{doi:10.1016/0370-2693(68)90261-X}}.

\bibitem{cern3-75}
J.~Bailey, et~al., {New Measurement of (G-2) of the Muon}, Phys. Lett. B55
  (1975) 420--424.
\newblock \href {http://dx.doi.org/10.1016/0370-2693(75)90374-3}
  {\path{doi:10.1016/0370-2693(75)90374-3}}.

\bibitem{cern3-79}
J.~Bailey, et~al., {Final Report on the CERN Muon Storage Ring Including the
  Anomalous Magnetic Moment and the Electric Dipole Moment of the Muon, and a
  Direct Test of Relativistic Time Dilation}, Nucl. Phys. B150 (1979) 1--75.
\newblock \href {http://dx.doi.org/10.1016/0550-3213(79)90292-X}
  {\path{doi:10.1016/0550-3213(79)90292-X}}.

\bibitem{bnl00}
H.~N. Brown, et~al., {Improved measurement of the positive muon anomalous
  magnetic moment}, Phys. Rev. D62 (2000) 091101.
\newblock \href {http://arxiv.org/abs/hep-ex/0009029}
  {\path{arXiv:hep-ex/0009029}}, \href
  {http://dx.doi.org/10.1103/PhysRevD.62.091101}
  {\path{doi:10.1103/PhysRevD.62.091101}}.

\bibitem{bnl01}
H.~N. Brown, et~al., {Precise measurement of the positive muon anomalous
  magnetic moment}, Phys. Rev. Lett. 86 (2001) 2227--2231.
\newblock \href {http://arxiv.org/abs/hep-ex/0102017}
  {\path{arXiv:hep-ex/0102017}}, \href
  {http://dx.doi.org/10.1103/PhysRevLett.86.2227}
  {\path{doi:10.1103/PhysRevLett.86.2227}}.

\bibitem{bnl02}
G.~W. Bennett, et~al., {Measurement of the positive muon anomalous magnetic
  moment to 0.7 ppm}, Phys. Rev. Lett. 89 (2002) 101804, [Erratum: Phys. Rev.
  Lett.89,129903(2002)].
\newblock \href {http://arxiv.org/abs/hep-ex/0208001}
  {\path{arXiv:hep-ex/0208001}}, \href
  {http://dx.doi.org/10.1103/PhysRevLett.89.101804}
  {\path{doi:10.1103/PhysRevLett.89.101804}}.

\bibitem{bnl04}
G.~W. Bennett, et~al., {Measurement of the negative muon anomalous magnetic
  moment to 0.7 ppm}, Phys. Rev. Lett. 92 (2004) 161802.
\newblock \href {http://arxiv.org/abs/hep-ex/0401008}
  {\path{arXiv:hep-ex/0401008}}, \href
  {http://dx.doi.org/10.1103/PhysRevLett.92.161802}
  {\path{doi:10.1103/PhysRevLett.92.161802}}.

\bibitem{bnl06}
G.~W. Bennett, et~al., {Final Report of the Muon E821 Anomalous Magnetic Moment
  Measurement at BNL}, Phys. Rev. D73 (2006) 072003.
\newblock \href {http://arxiv.org/abs/hep-ex/0602035}
  {\path{arXiv:hep-ex/0602035}}, \href
  {http://dx.doi.org/10.1103/PhysRevD.73.072003}
  {\path{doi:10.1103/PhysRevD.73.072003}}.

\bibitem{Agashe:2014kda}
K.~A. Olive, et~al., {Review of Particle Physics}, Chin. Phys. C38 (2014)
  090001.
\newblock \href {http://dx.doi.org/10.1088/1674-1137/38/9/090001}
  {\path{doi:10.1088/1674-1137/38/9/090001}}.

\bibitem{Carey:2009zzb}
R.~M. Carey, et~al., {The New (g-2) Experiment: A proposal to measure the muon
  anomalous magnetic moment to +-0.14 ppm precision}.

\bibitem{Beringer:1900zz}
J.~Beringer, et~al., {Review of Particle Physics (RPP)}, Phys. Rev. D86 (2012)
  010001.
\newblock \href {http://dx.doi.org/10.1103/PhysRevD.86.010001}
  {\path{doi:10.1103/PhysRevD.86.010001}}.

\bibitem{Zhang:2008pka}
Z.~Zhang, {Muon g-2: A Mini review}, in: {Electroweak Interactions and Unifield
  Theories: Proceedings, 42nd Rencontres de Moriond, La Thuile, Italy, March
  10-17, 2007}, 2008, pp. 457--466.
\newblock \href {http://arxiv.org/abs/0801.4905} {\path{arXiv:0801.4905}}.

\bibitem{Bargmann:1959gz}
V.~Bargmann, L.~Michel, V.~L. Telegdi, {Precession of the polarization of
  particles moving in a homogeneous electromagnetic field}, Phys. Rev. Lett. 2
  (1959) 435--436, [92(1959)].
\newblock \href {http://dx.doi.org/10.1103/PhysRevLett.2.435}
  {\path{doi:10.1103/PhysRevLett.2.435}}.

\bibitem{Venanzoni:2014ixa}
G.~Venanzoni, {The New Muon g−2 experiment at Fermilab}, Nucl. Part. Phys.
  Proc. 273-275 (2016) 584--588.
\newblock \href {http://arxiv.org/abs/1411.2555} {\path{arXiv:1411.2555}},
  \href {http://dx.doi.org/10.1016/j.nuclphysbps.2015.09.087}
  {\path{doi:10.1016/j.nuclphysbps.2015.09.087}}.

\bibitem{Anastasi:2015oea}
A.~Anastasi, {The Muon g-2 experiment at Fermilab}, EPJ Web Conf. 96 (2015)
  01002.
\newblock \href {http://dx.doi.org/10.1051/epjconf/20159601002}
  {\path{doi:10.1051/epjconf/20159601002}}.

\bibitem{Korostelev:2016qaf}
M.~Korostelev, I.~Bailey, A.~Herrod, J.~Morgan, W.~Morse, D.~Stratakis,
  V.~Tishchenko, A.~Wolski, {End-to-End Beam Simulations for the New Muon G-2
  Experiment at Fermilab}, in: {Proceedings, 7th International Particle
  Accelerator Conference (IPAC 2016): Busan, Korea, May 8-13, 2016}, 2016, p.
  WEPMW001.
\newblock \href {http://dx.doi.org/10.18429/JACoW-IPAC2016-WEPMW001}
  {\path{doi:10.18429/JACoW-IPAC2016-WEPMW001}}.

\bibitem{Gray:2015qna}
F.~Gray,
  \href{http://inspirehep.net/record/1395618/files/arXiv:1510.00346.pdf}{{Muon
  g-2 Experiment at Fermilab}}, in: {Proceedings, 12th Conference on the
  Intersections of Particle and Nuclear Physics (CIPANP 2015): Vail, Colorado,
  USA, May 19-24, 2015}, 2015.
\newblock \href {http://arxiv.org/abs/1510.00346} {\path{arXiv:1510.00346}}.
\newline\urlprefix\url{http://inspirehep.net/record/1395618/files/arXiv:1510.00346.pdf}

\bibitem{FNALtalk}
Experimental prospects on muon g-2,
  \url{https://indico.cern.ch/event/492464/contributions/2201359/attachments/1345286/2028020/ML-g-2.pdf},
  accessed: 2016-09-30.

\bibitem{Ishida:2009zz}
K.~Ishida, {Ultra slow muon source for new muon g-2 experiment}, AIP Conf.
  Proc. 1222 (2010) 396--399.
\newblock \href {http://dx.doi.org/10.1063/1.3399351}
  {\path{doi:10.1063/1.3399351}}.

\bibitem{Mibe:2010zz}
T.~Mibe, {New g-2 experiment at J-PARC}, Chin. Phys. C34 (2010) 745--748.
\newblock \href {http://dx.doi.org/10.1088/1674-1137/34/6/022}
  {\path{doi:10.1088/1674-1137/34/6/022}}.

\bibitem{Iinuma:2011zz}
H.~Iinuma, {New approach to the muon g-2 and EDM experiment at J-PARC}, J.
  Phys. Conf. Ser. 295 (2011) 012032.
\newblock \href {http://dx.doi.org/10.1088/1742-6596/295/1/012032}
  {\path{doi:10.1088/1742-6596/295/1/012032}}.

\bibitem{Saito:2012zz}
N.~Saito, {A novel precision measurement of muon g-2 and EDM at J-PARC}, AIP
  Conf. Proc. 1467 (2012) 45--56.
\newblock \href {http://dx.doi.org/10.1063/1.4742078}
  {\path{doi:10.1063/1.4742078}}.

\bibitem{Eads:2015arb}
M.~Eads, {New Experiments to Measure the Muon Anomalous Gyromagnetic MomentNew
  Experiments to Measure the Muon Anomalous Gyromagnetic Ratio}, PoS FPCP2015
  (2015) 046.
\newblock \href {http://arxiv.org/abs/1512.07214} {\path{arXiv:1512.07214}}.

\bibitem{Mibe:2011zz}
T.~Mibe, {Measurement of muon g-2 and EDM with an ultra-cold muon beam at
  J-PARC}, Nucl. Phys. Proc. Suppl. 218 (2011) 242--246.
\newblock \href {http://dx.doi.org/10.1016/j.nuclphysbps.2011.06.039}
  {\path{doi:10.1016/j.nuclphysbps.2011.06.039}}.

\bibitem{JPARCproposal}
New measurement of muon anomalous magnetic moment g-2 and electric dipole
  moment at j-parc,
  \url{http://j-parc.jp/researcher/Hadron/en/pac_0907/pdf/LOI_Saito.pdf},
  accessed: 2016-09-30.

\bibitem{Otani:2015jra}
M.~Otani, {Status of the Muon g-2/EDM Experiment at J-PARC (E34)}, JPS Conf.
  Proc. 8 (2015) 025008.
\newblock \href {http://dx.doi.org/10.7566/JPSCP.8.025008}
  {\path{doi:10.7566/JPSCP.8.025008}}.

\bibitem{Tanida:2016ryv}
K.~Tanida, {Overview of Hadron Physics at J-PARC}, JPS Conf. Proc. 10 (2016)
  010019.
\newblock \href {http://dx.doi.org/10.7566/JPSCP.10.010019}
  {\path{doi:10.7566/JPSCP.10.010019}}.

\bibitem{Benayoun:2012wc}
M.~Benayoun, P.~David, L.~DelBuono, F.~Jegerlehner, {An Update of the HLS
  Estimate of the Muon g-2}, Eur. Phys. J. C73 (2013) 2453.
\newblock \href {http://arxiv.org/abs/1210.7184} {\path{arXiv:1210.7184}},
  \href {http://dx.doi.org/10.1140/epjc/s10052-013-2453-3}
  {\path{doi:10.1140/epjc/s10052-013-2453-3}}.

\bibitem{Benayoun:2015gxa}
M.~Benayoun, P.~David, L.~DelBuono, F.~Jegerlehner, {Muon $g-2$ estimates: can
  one trust effective Lagrangians and global fits?}, Eur. Phys. J. C75~(12)
  (2015) 613.
\newblock \href {http://arxiv.org/abs/1507.02943} {\path{arXiv:1507.02943}},
  \href {http://dx.doi.org/10.1140/epjc/s10052-015-3830-x}
  {\path{doi:10.1140/epjc/s10052-015-3830-x}}.

\bibitem{Czarnecki:2001pv}
A.~Czarnecki, W.~J. Marciano, {The Muon anomalous magnetic moment: A Harbinger
  for 'new physics'}, Phys. Rev. D64 (2001) 013014.
\newblock \href {http://arxiv.org/abs/hep-ph/0102122}
  {\path{arXiv:hep-ph/0102122}}, \href
  {http://dx.doi.org/10.1103/PhysRevD.64.013014}
  {\path{doi:10.1103/PhysRevD.64.013014}}.

\bibitem{McKeen:2009ny}
D.~McKeen, {Contributions to the Muon's Anomalous Magnetic Moment from a Hidden
  Sector}, Annals Phys. 326 (2011) 1501--1514.
\newblock \href {http://arxiv.org/abs/0912.1076} {\path{arXiv:0912.1076}},
  \href {http://dx.doi.org/10.1016/j.aop.2010.12.015}
  {\path{doi:10.1016/j.aop.2010.12.015}}.

\bibitem{Jegerlehner:2009ry}
F.~Jegerlehner, A.~Nyffeler, {The Muon g-2}, Phys. Rept. 477 (2009) 1--110.
\newblock \href {http://arxiv.org/abs/0902.3360} {\path{arXiv:0902.3360}},
  \href {http://dx.doi.org/10.1016/j.physrep.2009.04.003}
  {\path{doi:10.1016/j.physrep.2009.04.003}}.

\bibitem{Branco:2011iw}
G.~C. Branco, P.~M. Ferreira, L.~Lavoura, M.~N. Rebelo, M.~Sher, J.~P. Silva,
  {Theory and phenomenology of two-Higgs-doublet models}, Phys. Rept. 516
  (2012) 1--102.
\newblock \href {http://arxiv.org/abs/1106.0034} {\path{arXiv:1106.0034}},
  \href {http://dx.doi.org/10.1016/j.physrep.2012.02.002}
  {\path{doi:10.1016/j.physrep.2012.02.002}}.

\bibitem{Cho:2011rk}
G.-C. Cho, K.~Hagiwara, Y.~Matsumoto, D.~Nomura, {The MSSM confronts the
  precision electroweak data and the muon $g-2$}, JHEP 11 (2011) 068.
\newblock \href {http://arxiv.org/abs/1104.1769} {\path{arXiv:1104.1769}},
  \href {http://dx.doi.org/10.1007/JHEP11(2011)068}
  {\path{doi:10.1007/JHEP11(2011)068}}.

\bibitem{Miller:2012opa}
J.~P. Miller, E.~de~Rafael, B.~L. Roberts, D.~Stöckinger, {Muon (g-2):
  Experiment and Theory}, Ann. Rev. Nucl. Part. Sci. 62 (2012) 237--264.
\newblock \href {http://dx.doi.org/10.1146/annurev-nucl-031312-120340}
  {\path{doi:10.1146/annurev-nucl-031312-120340}}.

\bibitem{Stockinger:2013rna}
D.~Stöckinger, {The muon magnetic moment and new physics}, Hyperfine Interact.
  214~(1-3) (2013) 13--19.
\newblock \href {http://dx.doi.org/10.1007/s10751-013-0804-y}
  {\path{doi:10.1007/s10751-013-0804-y}}.

\bibitem{Fargnoli:2013zia}
H.~Fargnoli, C.~Gnendiger, S.~Paßehr, D.~Stöckinger, H.~Stöckinger-Kim,
  {Two-loop corrections to the muon magnetic moment from fermion/sfermion loops
  in the MSSM: detailed results}, JHEP 02 (2014) 070.
\newblock \href {http://arxiv.org/abs/1311.1775} {\path{arXiv:1311.1775}},
  \href {http://dx.doi.org/10.1007/JHEP02(2014)070}
  {\path{doi:10.1007/JHEP02(2014)070}}.

\bibitem{Fargnoli:2013fra}
H.~G. Fargnoli, C.~Gnendiger, S.~Passehr, D.~Stöckinger, H.~Stöckinger-Kim,
  {The full electroweak Standard Model prediction for (g - 2) of the muon and
  improvements on the MSSM prediction}, Int. J. Mod. Phys. Conf. Ser. 35 (2014)
  1460419.
\newblock \href {http://arxiv.org/abs/1311.4890} {\path{arXiv:1311.4890}},
  \href {http://dx.doi.org/10.1142/S2010194514604190}
  {\path{doi:10.1142/S2010194514604190}}.

\bibitem{Agrawal:2014ufa}
P.~Agrawal, Z.~Chacko, C.~B. Verhaaren, {Leptophilic Dark Matter and the
  Anomalous Magnetic Moment of the Muon}, JHEP 08 (2014) 147.
\newblock \href {http://arxiv.org/abs/1402.7369} {\path{arXiv:1402.7369}},
  \href {http://dx.doi.org/10.1007/JHEP08(2014)147}
  {\path{doi:10.1007/JHEP08(2014)147}}.

\bibitem{Freitas:2014pua}
A.~Freitas, J.~Lykken, S.~Kell, S.~Westhoff, {Testing the Muon g-2 Anomaly at
  the LHC}, JHEP 05 (2014) 145, [Erratum: JHEP09,155(2014)].
\newblock \href {http://arxiv.org/abs/1402.7065} {\path{arXiv:1402.7065}},
  \href {http://dx.doi.org/10.1007/JHEP09(2014)155, 10.1007/JHEP05(2014)145}
  {\path{doi:10.1007/JHEP09(2014)155, 10.1007/JHEP05(2014)145}}.

\bibitem{Chiu:2014oma}
W.-C. Chiu, C.-Q. Geng, D.~Huang, {Correlation Between Muon $g-2$ and
  $\mu\rightarrow{e}{\gamma}$}, Phys. Rev. D91~(1) (2015) 013006.
\newblock \href {http://arxiv.org/abs/1409.4198} {\path{arXiv:1409.4198}},
  \href {http://dx.doi.org/10.1103/PhysRevD.91.013006}
  {\path{doi:10.1103/PhysRevD.91.013006}}.

\bibitem{Queiroz:2014zfa}
F.~S. Queiroz, W.~Shepherd, {New Physics Contributions to the Muon Anomalous
  Magnetic Moment: A Numerical Code}, Phys. Rev. D89~(9) (2014) 095024.
\newblock \href {http://arxiv.org/abs/1403.2309} {\path{arXiv:1403.2309}},
  \href {http://dx.doi.org/10.1103/PhysRevD.89.095024}
  {\path{doi:10.1103/PhysRevD.89.095024}}.

\bibitem{Gomes:2016ixw}
M.~Gomes, T.~Mariz, J.~R. Nascimento, A.~{\relax Yu}. Petrov, A.~J. da~Silva,
  {On the radiative corrections in the Horava–Lifshitz $z =$ 2 QED}, Phys.
  Lett. B764 (2017) 277--281.
\newblock \href {http://arxiv.org/abs/1607.01240} {\path{arXiv:1607.01240}},
  \href {http://dx.doi.org/10.1016/j.physletb.2016.11.042}
  {\path{doi:10.1016/j.physletb.2016.11.042}}.

\bibitem{Cheng:1977nv}
T.-P. Cheng, L.-F. Li, {Muon Number Nonconservation Effects in a Gauge Theory
  with V A Currents and Heavy Neutral Leptons}, Phys. Rev. D16 (1977) 1425.
\newblock \href {http://dx.doi.org/10.1103/PhysRevD.16.1425}
  {\path{doi:10.1103/PhysRevD.16.1425}}.

\bibitem{Meyer:1959zz}
P.~Meyer, G.~Salzman, {THE $\mu \rightarrow$ e gamma DECAY AND THE INTERMEDIATE
  CHARGED VECTOR BOSON THEORY}, Nuovo. Cim. 14 (1959) 1310.

\bibitem{Parker:1964zz}
S.~Parker, H.~L. Anderson, C.~Rey, {Search for the Decay $\mu^+ \rightarrow
  e^+$ + gamma}, Phys. Rev. 133 (1964) B768--B778.
\newblock \href {http://dx.doi.org/10.1103/PhysRev.133.B768}
  {\path{doi:10.1103/PhysRev.133.B768}}.

\bibitem{Depommier:1977yk}
P.~Depommier, et~al., {A New Limit on the $\mu^+ \rightarrow e^+$ gamma Decay},
  Phys. Rev. Lett. 39 (1977) 1113.
\newblock \href {http://dx.doi.org/10.1103/PhysRevLett.39.1113}
  {\path{doi:10.1103/PhysRevLett.39.1113}}.

\bibitem{Povel:1977sk}
H.~P. Povel, et~al., {A New Upper Limit for the Decay $\mu^+ \rightarrow e^+$
  gamma}, Phys. Lett. B72 (1977) 183--186.
\newblock \href {http://dx.doi.org/10.1016/0370-2693(77)90697-9}
  {\path{doi:10.1016/0370-2693(77)90697-9}}.

\bibitem{Bowman:1979bj}
J.~D. Bowman, et~al., {UPPER LIMIT FOR THE DECAY $\mu^+ \to e^+ \gamma$}, Phys.
  Rev. Lett. 42 (1979) 556--560.
\newblock \href {http://dx.doi.org/10.1103/PhysRevLett.42.556}
  {\path{doi:10.1103/PhysRevLett.42.556}}.

\bibitem{vanderSchaaf:1979hz}
A.~van~der Schaaf, R.~Engfer, H.~P. Povel, W.~Dey, H.~K. Walter, C.~Petitjean,
  {A Search for the Decay $\mu^+ \to e^+ \gamma$}, Nucl. Phys. A340 (1980)
  249--270.
\newblock \href {http://dx.doi.org/10.1016/0375-9474(80)90274-2}
  {\path{doi:10.1016/0375-9474(80)90274-2}}.

\bibitem{Kinnison:1981gc}
W.~W. Kinnison, et~al., {A Search for $\mu^+ \to e^+ \gamma$}, Phys. Rev. D25
  (1982) 2846.
\newblock \href {http://dx.doi.org/10.1103/PhysRevD.25.2846}
  {\path{doi:10.1103/PhysRevD.25.2846}}.

\bibitem{Azuelos:1983wx}
G.~Azuelos, et~al., {A New Upper Limit of the Decay $\mu \to e \gamma \gamma$},
  Phys. Rev. Lett. 51 (1983) 164.
\newblock \href {http://dx.doi.org/10.1103/PhysRevLett.51.164}
  {\path{doi:10.1103/PhysRevLett.51.164}}.

\bibitem{Bolton:1986tv}
R.~D. Bolton, et~al., {Search for the Decay $\mu^+ \to e^+ \gamma$}, Phys. Rev.
  Lett. 56 (1986) 2461--2464.
\newblock \href {http://dx.doi.org/10.1103/PhysRevLett.56.2461}
  {\path{doi:10.1103/PhysRevLett.56.2461}}.

\bibitem{Bolton:1988af}
R.~D. Bolton, et~al., {Search for Rare Muon Decays with the Crystal Box
  Detector}, Phys. Rev. D38 (1988) 2077.
\newblock \href {http://dx.doi.org/10.1103/PhysRevD.38.2077}
  {\path{doi:10.1103/PhysRevD.38.2077}}.

\bibitem{Brooks:1999pu}
M.~L. Brooks, et~al., {New limit for the family number nonconserving decay
  $\mu^+ \rightarrow e^+$ gamma}, Phys. Rev. Lett. 83 (1999) 1521--1524.
\newblock \href {http://arxiv.org/abs/hep-ex/9905013}
  {\path{arXiv:hep-ex/9905013}}, \href
  {http://dx.doi.org/10.1103/PhysRevLett.83.1521}
  {\path{doi:10.1103/PhysRevLett.83.1521}}.

\bibitem{Adam:2009ci}
J.~Adam, et~al., {A limit for the $\mu \rightarrow$ e gamma decay from the MEG
  experiment}, Nucl. Phys. B834 (2010) 1--12.
\newblock \href {http://arxiv.org/abs/0908.2594} {\path{arXiv:0908.2594}},
  \href {http://dx.doi.org/10.1016/j.nuclphysb.2010.03.030}
  {\path{doi:10.1016/j.nuclphysb.2010.03.030}}.

\bibitem{Golden:2011zz}
B.~Golden, {Results from the 2009 data of the MEG experiment in the search for
  $\mu^+ \rightarrow e^+$ gamma}, Nucl. Phys. Proc. Suppl. 218 (2011) 62--67.
\newblock \href {http://dx.doi.org/10.1016/j.nuclphysbps.2011.06.012}
  {\path{doi:10.1016/j.nuclphysbps.2011.06.012}}.

\bibitem{Adam:2011ch}
J.~Adam, et~al., {New limit on the lepton-flavour violating decay $\mu^{+} \to
  e^{+} \gamma$}, Phys. Rev. Lett. 107 (2011) 171801.
\newblock \href {http://arxiv.org/abs/1107.5547} {\path{arXiv:1107.5547}},
  \href {http://dx.doi.org/10.1103/PhysRevLett.107.171801}
  {\path{doi:10.1103/PhysRevLett.107.171801}}.

\bibitem{Adam:2013mnn}
J.~Adam, et~al., {New constraint on the existence of the $\mu^+ \rightarrow
  e^+\gamma$ decay}, Phys. Rev. Lett. 110 (2013) 201801.
\newblock \href {http://arxiv.org/abs/1303.0754} {\path{arXiv:1303.0754}},
  \href {http://dx.doi.org/10.1103/PhysRevLett.110.201801}
  {\path{doi:10.1103/PhysRevLett.110.201801}}.

\bibitem{TheMEG:2016wtm}
A.~M. Baldini, et~al., {Search for the lepton flavour violating decay $\mu ^+
  \rightarrow \mathrm {e}^+ \gamma $ with the full dataset of the MEG
  experiment}, Eur. Phys. J. C76~(8) (2016) 434.
\newblock \href {http://arxiv.org/abs/1605.05081} {\path{arXiv:1605.05081}},
  \href {http://dx.doi.org/10.1140/epjc/s10052-016-4271-x}
  {\path{doi:10.1140/epjc/s10052-016-4271-x}}.

\bibitem{Mori:2016vwi}
T.~Mori, {Final Results of the MEG Experiment}, Nuovo Cim. C39~(4) (2017) 325.
\newblock \href {http://arxiv.org/abs/1606.08168} {\path{arXiv:1606.08168}},
  \href {http://dx.doi.org/10.1393/ncc/i2016-16325-7}
  {\path{doi:10.1393/ncc/i2016-16325-7}}.

\bibitem{Hayes:1981bn}
K.~G. Hayes, et~al., {Experimental Upper Limits on Branching Fractions for
  Unexpected Decay Modes of the Tau Lepton}, Phys. Rev. D25 (1982) 2869.
\newblock \href {http://dx.doi.org/10.1103/PhysRevD.25.2869}
  {\path{doi:10.1103/PhysRevD.25.2869}}.

\bibitem{Hayes:1981ta}
K.~G. Hayes, {RECENT RESULTS ON THE TAU LEPTON}, in: {Proceedings, 16th
  Rencontres de Moriond : Session 2: New Flavours and Hadron Spectroscopy: Les
  Arcs, France, March 15-27, 1981}, 1981, pp. 145--154.

\bibitem{Keh:1988gs}
S.~Keh, et~al., {Search for Exotic Tau Decays}, Phys. Lett. B212 (1988)
  123--128.
\newblock \href {http://dx.doi.org/10.1016/0370-2693(88)91248-8}
  {\path{doi:10.1016/0370-2693(88)91248-8}}.

\bibitem{Albrecht:1992uba}
H.~Albrecht, et~al., {Search for neutrinoless tau decays}, Z. Phys. C55 (1992)
  179--190.

\bibitem{Bean:1992hh}
A.~Bean, et~al., {A Search for $\tau \rightarrow \gamma \mu^-$: A Test of
  lepton number conservation}, Phys. Rev. Lett. 70 (1993) 138--142.
\newblock \href {http://dx.doi.org/10.1103/PhysRevLett.70.138}
  {\path{doi:10.1103/PhysRevLett.70.138}}.

\bibitem{Abreu:1995gs}
P.~Abreu, et~al., {Upper limits on the branching ratios $\tau \rightarrow \mu
  \gamma$ and $\tau \rightarrow e \gamma$}, Phys. Lett. B359 (1995) 411--421.
\newblock \href {http://dx.doi.org/10.1016/0370-2693(95)01045-R}
  {\path{doi:10.1016/0370-2693(95)01045-R}}.

\bibitem{Edwards:1996te}
K.~W. Edwards, et~al., {Search for neutrinoless tau decays: $\tau \rightarrow e
  \gamma$ and $\tau \rightarrow \mu \gamma$}, Phys. Rev. D55 (1997) 3919--3923.
\newblock \href {http://dx.doi.org/10.1103/PhysRevD.55.3919}
  {\path{doi:10.1103/PhysRevD.55.3919}}.

\bibitem{Ahmed:1999gh}
S.~Ahmed, et~al., {Update of the search for the neutrinoless decay $\tau
  \rightarrow \mu \gamma$}, Phys. Rev. D61 (2000) 071101.
\newblock \href {http://arxiv.org/abs/hep-ex/9910060}
  {\path{arXiv:hep-ex/9910060}}, \href
  {http://dx.doi.org/10.1103/PhysRevD.61.071101}
  {\path{doi:10.1103/PhysRevD.61.071101}}.

\bibitem{Brown:2002mp}
C.~Brown, {Search for the lepton number violating decay $\tau \to \mu \gamma$},
  eConf C0209101 (2002) TU12, [,88(2002)].
\newblock \href {http://arxiv.org/abs/hep-ex/0212009}
  {\path{arXiv:hep-ex/0212009}}, \href
  {http://dx.doi.org/10.1016/S0920-5632(03)80311-0}
  {\path{doi:10.1016/S0920-5632(03)80311-0}}.

\bibitem{Inami:2002us}
K.~Inami, T.~Hokuue, T.~Ohshima, {Search for lepton flavor violating tau --->
  mu gamma decay}, eConf C0209101 (2002) TU11, [,82(2002)].
\newblock \href {http://arxiv.org/abs/hep-ex/0210036}
  {\path{arXiv:hep-ex/0210036}}, \href
  {http://dx.doi.org/10.1016/S0920-5632(03)80310-9}
  {\path{doi:10.1016/S0920-5632(03)80310-9}}.

\bibitem{Abe:2003sx}
K.~Abe, et~al., {An Upper bound on the decay $\tau \mu \gamma$ from Belle},
  Phys. Rev. Lett. 92 (2004) 171802.
\newblock \href {http://arxiv.org/abs/hep-ex/0310029}
  {\path{arXiv:hep-ex/0310029}}, \href
  {http://dx.doi.org/10.1103/PhysRevLett.92.171802}
  {\path{doi:10.1103/PhysRevLett.92.171802}}.

\bibitem{Hayasaka:2005xw}
K.~Hayasaka, et~al., {Search for $\tau \rightarrow e \gamma$ decay at BELLE},
  Phys. Lett. B613 (2005) 20--28.
\newblock \href {http://arxiv.org/abs/hep-ex/0501068}
  {\path{arXiv:hep-ex/0501068}}, \href
  {http://dx.doi.org/10.1016/j.physletb.2005.03.028}
  {\path{doi:10.1016/j.physletb.2005.03.028}}.

\bibitem{Aubert:2005ye}
B.~Aubert, et~al., {Search for lepton flavor violation in the decay $\tau \to
  \mu \gamma$}, Phys. Rev. Lett. 95 (2005) 041802.
\newblock \href {http://arxiv.org/abs/hep-ex/0502032}
  {\path{arXiv:hep-ex/0502032}}, \href
  {http://dx.doi.org/10.1103/PhysRevLett.95.041802}
  {\path{doi:10.1103/PhysRevLett.95.041802}}.

\bibitem{Aubert:2005wa}
B.~Aubert, et~al., {Search for lepton flavor violation in the decay $\tau^\pm
  \to e^\pm \gamma$}, Phys. Rev. Lett. 96 (2006) 041801.
\newblock \href {http://arxiv.org/abs/hep-ex/0508012}
  {\path{arXiv:hep-ex/0508012}}, \href
  {http://dx.doi.org/10.1103/PhysRevLett.96.041801}
  {\path{doi:10.1103/PhysRevLett.96.041801}}.

\bibitem{Hayasaka:2007vc}
K.~Hayasaka, et~al., {New search for $\tau \rightarrow \mu \gamma$ and $\tau
  \rightarrow e \gamma$ decays at Belle}, Phys. Lett. B666 (2008) 16--22.
\newblock \href {http://arxiv.org/abs/0705.0650} {\path{arXiv:0705.0650}},
  \href {http://dx.doi.org/10.1016/j.physletb.2008.06.056}
  {\path{doi:10.1016/j.physletb.2008.06.056}}.

\bibitem{Aubert:2009ag}
B.~Aubert, et~al., {Searches for Lepton Flavor Violation in the Decays
  $\tau^{+-} \rightarrow e^{+-} \gamma$ and $\tau^{+-} \rightarrow \mu^{+-}
  \gamma$}, Phys. Rev. Lett. 104 (2010) 021802.
\newblock \href {http://arxiv.org/abs/0908.2381} {\path{arXiv:0908.2381}},
  \href {http://dx.doi.org/10.1103/PhysRevLett.104.021802}
  {\path{doi:10.1103/PhysRevLett.104.021802}}.

\bibitem{Hayasaka:2012pj}
K.~Hayasaka, {Recent LFV results on tau lepton from Belle}, Nucl. Phys. Proc.
  Suppl. 225-227 (2012) 169--172.
\newblock \href {http://dx.doi.org/10.1016/j.nuclphysbps.2012.02.036}
  {\path{doi:10.1016/j.nuclphysbps.2012.02.036}}.

\bibitem{Hincks:1948vr}
E.~P. Hincks, B.~Pontecorvo, {Search for gamma-radiation in the 2.2-microsecond
  meson decay process}, Phys. Rev. 73 (1948) 257--258.
\newblock \href {http://dx.doi.org/10.1103/PhysRev.73.257}
  {\path{doi:10.1103/PhysRev.73.257}}.

\bibitem{Bernstein:2013hba}
R.~H. Bernstein, P.~S. Cooper, {Charged Lepton Flavor Violation: An
  Experimenter's Guide}, Phys. Rept. 532 (2013) 27--64.
\newblock \href {http://arxiv.org/abs/1307.5787} {\path{arXiv:1307.5787}},
  \href {http://dx.doi.org/10.1016/j.physrep.2013.07.002}
  {\path{doi:10.1016/j.physrep.2013.07.002}}.

\bibitem{Steinberger:1955hfk}
J.~Steinberger, H.~B. Wolfe, {Electrons from Muon Capture}, Phys. Rev. 100~(5)
  (1955) 1490.
\newblock \href {http://dx.doi.org/10.1103/PhysRev.100.1490}
  {\path{doi:10.1103/PhysRev.100.1490}}.

\bibitem{Marciano:1977wx}
W.~J. Marciano, A.~I. Sanda, {Exotic Decays of the Muon and Heavy Leptons in
  Gauge Theories}, Phys. Lett. 67B (1977) 303--305.
\newblock \href {http://dx.doi.org/10.1016/0370-2693(77)90377-X}
  {\path{doi:10.1016/0370-2693(77)90377-X}}.

\bibitem{Ilakovac:1994kj}
A.~Ilakovac, A.~Pilaftsis, {Flavor violating charged lepton decays in
  seesaw-type models}, Nucl. Phys. B437 (1995) 491.
\newblock \href {http://arxiv.org/abs/hep-ph/9403398}
  {\path{arXiv:hep-ph/9403398}}, \href
  {http://dx.doi.org/10.1016/0550-3213(94)00567-X}
  {\path{doi:10.1016/0550-3213(94)00567-X}}.

\bibitem{Gomez:1998wj}
M.~E. Gomez, G.~K. Leontaris, S.~Lola, J.~D. Vergados, {U(1) textures and
  lepton flavor violation}, Phys. Rev. D59 (1999) 116009.
\newblock \href {http://arxiv.org/abs/hep-ph/9810291}
  {\path{arXiv:hep-ph/9810291}}, \href
  {http://dx.doi.org/10.1103/PhysRevD.59.116009}
  {\path{doi:10.1103/PhysRevD.59.116009}}.

\bibitem{Feng:1999wt}
J.~L. Feng, Y.~Nir, Y.~Shadmi, {Neutrino parameters, Abelian flavor symmetries,
  and charged lepton flavor violation}, Phys. Rev. D61 (2000) 113005.
\newblock \href {http://arxiv.org/abs/hep-ph/9911370}
  {\path{arXiv:hep-ph/9911370}}, \href
  {http://dx.doi.org/10.1103/PhysRevD.61.113005}
  {\path{doi:10.1103/PhysRevD.61.113005}}.

\bibitem{Kuno:1999jp}
Y.~Kuno, Y.~Okada, {Muon decay and physics beyond the standard model}, Rev.
  Mod. Phys. 73 (2001) 151--202.
\newblock \href {http://arxiv.org/abs/hep-ph/9909265}
  {\path{arXiv:hep-ph/9909265}}, \href
  {http://dx.doi.org/10.1103/RevModPhys.73.151}
  {\path{doi:10.1103/RevModPhys.73.151}}.

\bibitem{Petcov:2003zb}
S.~T. Petcov, S.~Profumo, Y.~Takanishi, C.~E. Yaguna, {Charged lepton flavor
  violating decays: Leading logarithmic approximation versus full RG results},
  Nucl. Phys. B676 (2004) 453--480.
\newblock \href {http://arxiv.org/abs/hep-ph/0306195}
  {\path{arXiv:hep-ph/0306195}}, \href
  {http://dx.doi.org/10.1016/j.nuclphysb.2003.10.020}
  {\path{doi:10.1016/j.nuclphysb.2003.10.020}}.

\bibitem{Pascoli:2003uh}
S.~Pascoli, S.~T. Petcov, W.~Rodejohann, {On the connection of leptogenesis
  with low-energy CP violation and LFV charged lepton decays}, Phys. Rev. D68
  (2003) 093007.
\newblock \href {http://arxiv.org/abs/hep-ph/0302054}
  {\path{arXiv:hep-ph/0302054}}, \href
  {http://dx.doi.org/10.1103/PhysRevD.68.093007}
  {\path{doi:10.1103/PhysRevD.68.093007}}.

\bibitem{Cirigliano:2005ck}
V.~Cirigliano, B.~Grinstein, G.~Isidori, M.~B. Wise, {Minimal flavor violation
  in the lepton sector}, Nucl. Phys. B728 (2005) 121--134.
\newblock \href {http://arxiv.org/abs/hep-ph/0507001}
  {\path{arXiv:hep-ph/0507001}}, \href
  {http://dx.doi.org/10.1016/j.nuclphysb.2005.08.037}
  {\path{doi:10.1016/j.nuclphysb.2005.08.037}}.

\bibitem{Marciano:2008zz}
W.~J. Marciano, T.~Mori, J.~M. Roney, {Charged Lepton Flavor Violation
  Experiments}, Ann. Rev. Nucl. Part. Sci. 58 (2008) 315--341.
\newblock \href {http://dx.doi.org/10.1146/annurev.nucl.58.110707.171126}
  {\path{doi:10.1146/annurev.nucl.58.110707.171126}}.

\bibitem{Casas:2010wm}
J.~A. Casas, J.~M. Moreno, N.~Rius, R.~Ruiz~de Austri, B.~Zaldivar, {Fair scans
  of the seesaw. Consequences for predictions on LFV processes}, JHEP 03 (2011)
  034.
\newblock \href {http://arxiv.org/abs/1010.5751} {\path{arXiv:1010.5751}},
  \href {http://dx.doi.org/10.1007/JHEP03(2011)034}
  {\path{doi:10.1007/JHEP03(2011)034}}.

\bibitem{Deppisch:2012vj}
F.~F. Deppisch, {Lepton Flavour Violation and Flavour Symmetries}, Fortsch.
  Phys. 61 (2013) 622--644.
\newblock \href {http://arxiv.org/abs/1206.5212} {\path{arXiv:1206.5212}},
  \href {http://dx.doi.org/10.1002/prop.201200126}
  {\path{doi:10.1002/prop.201200126}}.

\bibitem{Davidson:2016edt}
S.~Davidson, {$\mu \rightarrow e \gamma $ and matching at $m_W$}, Eur. Phys. J.
  C76~(7) (2016) 370.
\newblock \href {http://arxiv.org/abs/1601.07166} {\path{arXiv:1601.07166}},
  \href {http://dx.doi.org/10.1140/epjc/s10052-016-4207-5}
  {\path{doi:10.1140/epjc/s10052-016-4207-5}}.

\bibitem{Davidson:2016utf}
S.~Davidson, {$\mu \rightarrow e \gamma $ in the 2HDM: an exercise in EFT},
  Eur. Phys. J. C76~(5) (2016) 258.
\newblock \href {http://arxiv.org/abs/1601.01949} {\path{arXiv:1601.01949}},
  \href {http://dx.doi.org/10.1140/epjc/s10052-016-4076-y}
  {\path{doi:10.1140/epjc/s10052-016-4076-y}}.

\bibitem{Cohen:1987fr}
E.~R. Cohen, B.~N. Taylor, {FUNDAMENTAL PHYSICAL CONSTANTS 1986 ADJUSTMENTS},
  Rev. Mod. Phys. 59 (1987) 1121, [Europhys. News18,65(1987)].
\newblock \href {http://dx.doi.org/10.1103/RevModPhys.59.1121}
  {\path{doi:10.1103/RevModPhys.59.1121}}.

\bibitem{Olive:2016xmw}
K.~A. Olive, {Review of Particle Physics}, Chin. Phys. C40~(10) (2016) 100001.
\newblock \href {http://dx.doi.org/10.1088/1674-1137/40/10/100001}
  {\path{doi:10.1088/1674-1137/40/10/100001}}.

\bibitem{Korenchenko:1975gf}
S.~M. Korenchenko, B.~F. Kostin, G.~Mitselmakher, K.~G. Nekrasov, V.~S.
  Smirnov, {Search for mu+ --> e+ e+ e- Decay}, Sov. Phys. JETP 43 (1976) 1,
  [Zh. Eksp. Teor. Fiz.70,3(1976)].

\bibitem{Bolton:1984qr}
R.~D. Bolton, et~al., {A Search for the Muon Number Violating Decay $\mu \to
  e^+ e^+ e^-$}, Phys. Rev. Lett. 53 (1984) 1415.
\newblock \href {http://dx.doi.org/10.1103/PhysRevLett.53.1415}
  {\path{doi:10.1103/PhysRevLett.53.1415}}.

\bibitem{Bertl:1984wi}
W.~H. Bertl, et~al., {A New Upper Limit for the Decay $\mu^+ \to e^+ e^+ e^-$},
  Phys. Lett. 140B (1984) 299.
\newblock \href {http://dx.doi.org/10.1016/0370-2693(84)90757-3}
  {\path{doi:10.1016/0370-2693(84)90757-3}}.

\bibitem{Bertl:1985mw}
W.~H. Bertl, et~al., {Search for the Decay $\mu^+ \to e^+ e^+ e^-$}, Nucl.
  Phys. B260 (1985) 1--31.
\newblock \href {http://dx.doi.org/10.1016/0550-3213(85)90308-6}
  {\path{doi:10.1016/0550-3213(85)90308-6}}.

\bibitem{Bellgardt:1987du}
U.~Bellgardt, et~al., {Search for the Decay $\mu^+ \to e^+ e^+ e^-$}, Nucl.
  Phys. B299 (1988) 1--6.
\newblock \href {http://dx.doi.org/10.1016/0550-3213(88)90462-2}
  {\path{doi:10.1016/0550-3213(88)90462-2}}.

\bibitem{Baranov:1990uh}
V.~A. Baranov, et~al., {Search for $\mu^+ \rightarrow e^+ e^+ e^-$ decay}, Sov.
  J. Nucl. Phys. 53 (1991) 802--807, [Yad. Fiz.53,1302(1991)].

\bibitem{Blondel:2013ia}
A.~Blondel, et~al., {Research Proposal for an Experiment to Search for the
  Decay $\mu \to eee$}\href {http://arxiv.org/abs/1301.6113}
  {\path{arXiv:1301.6113}}.

\bibitem{Bilenky:1977du}
S.~M. Bilenky, S.~T. Petcov, B.~Pontecorvo, {Lepton Mixing, $\mu \rightarrow e
  \gamma$ Decay and Neutrino Oscillations}, Phys. Lett. B67 (1977) 309.
\newblock \href {http://dx.doi.org/10.1016/0370-2693(77)90379-3}
  {\path{doi:10.1016/0370-2693(77)90379-3}}.

\bibitem{Cheng:1980tp}
T.~P. Cheng, L.-F. Li, {$\mu \to e \gamma$ in Theories With Dirac and Majorana
  Neutrino Mass Terms}, Phys. Rev. Lett. 45 (1980) 1908.
\newblock \href {http://dx.doi.org/10.1103/PhysRevLett.45.1908}
  {\path{doi:10.1103/PhysRevLett.45.1908}}.

\bibitem{Bartolotta:2017mff}
A.~Bartolotta, M.~J. Ramsey-Musolf, {Coherent $\mu-e$ Conversion at
  Next-to-Leading Order}\href {http://arxiv.org/abs/1710.02129}
  {\path{arXiv:1710.02129}}.

\bibitem{Bryman:1972rf}
D.~A. Bryman, M.~Blecher, K.~Gotow, R.~J. Powers, {Search for the reaction
  $\mu^- cu \rightarrow e^+ co$}, Phys. Rev. Lett. 28 (1972) 1469--1471.
\newblock \href {http://dx.doi.org/10.1103/PhysRevLett.28.1469}
  {\path{doi:10.1103/PhysRevLett.28.1469}}.

\bibitem{Badertscher:1981ay}
A.~Badertscher, et~al., {A Search for Muon - Electron and Muon - Positron
  Conversion in Sulfur}, Nucl. Phys. A377 (1982) 406--440.
\newblock \href {http://dx.doi.org/10.1016/0375-9474(82)90049-5}
  {\path{doi:10.1016/0375-9474(82)90049-5}}.

\bibitem{Bryman:1985fq}
D.~A. Bryman, et~al., {Search for $\mu e$ Conversion in Ti}, Phys. Rev. Lett.
  55 (1985) 465.
\newblock \href {http://dx.doi.org/10.1103/PhysRevLett.55.465}
  {\path{doi:10.1103/PhysRevLett.55.465}}.

\bibitem{Ahmad:1988ur}
S.~Ahmad, et~al., {Search for Muon - Electron and Muon - Positron Conversion},
  Phys. Rev. D38 (1988) 2102.
\newblock \href {http://dx.doi.org/10.1103/PhysRevD.38.2102}
  {\path{doi:10.1103/PhysRevD.38.2102}}.

\bibitem{Dohmen:1993mp}
C.~Dohmen, et~al., {Test of lepton flavor conservation in $\mu \to e$
  conversion on titanium}, Phys. Lett. B317 (1993) 631--636.
\newblock \href {http://dx.doi.org/10.1016/0370-2693(93)91383-X}
  {\path{doi:10.1016/0370-2693(93)91383-X}}.

\bibitem{Wintz:1998rp}
P.~Wintz, {Results of the SINDRUM-II experiment}, Conf. Proc. C980420 (1998)
  534--546.

\bibitem{Honecker:1996zf}
W.~Honecker, et~al., {Improved limit on the branching ratio of $\mu- e$
  conversion on lead}, Phys. Rev. Lett. 76 (1996) 200--203.
\newblock \href {http://dx.doi.org/10.1103/PhysRevLett.76.200}
  {\path{doi:10.1103/PhysRevLett.76.200}}.

\bibitem{Bertl:2006up}
W.~H. Bertl, et~al., {A Search for muon to electron conversion in muonic gold},
  Eur. Phys. J. C47 (2006) 337--346.
\newblock \href {http://dx.doi.org/10.1140/epjc/s2006-02582-x}
  {\path{doi:10.1140/epjc/s2006-02582-x}}.

\bibitem{Aysto:2001zs}
J.~Aysto, et~al., {Physics with low-energy muons at a neutrino factory complex}
  (2001) 259--306\href {http://arxiv.org/abs/hep-ph/0109217}
  {\path{arXiv:hep-ph/0109217}}, \href
  {http://dx.doi.org/10.5170/CERN-2004-002.259}
  {\path{doi:10.5170/CERN-2004-002.259}}.

\bibitem{deGouvea:2013zba}
A.~de~Gouvea, P.~Vogel, {Lepton Flavor and Number Conservation, and Physics
  Beyond the Standard Model}, Prog. Part. Nucl. Phys. 71 (2013) 75--92.
\newblock \href {http://arxiv.org/abs/1303.4097} {\path{arXiv:1303.4097}},
  \href {http://dx.doi.org/10.1016/j.ppnp.2013.03.006}
  {\path{doi:10.1016/j.ppnp.2013.03.006}}.

\bibitem{Geib:2015unm}
T.~Geib, A.~Merle, {Conversions of bound muons: Lepton flavor violation from
  doubly charged scalars}, Phys. Rev. D93~(5) (2016) 055039.
\newblock \href {http://arxiv.org/abs/1512.04225} {\path{arXiv:1512.04225}},
  \href {http://dx.doi.org/10.1103/PhysRevD.93.055039}
  {\path{doi:10.1103/PhysRevD.93.055039}}.

\bibitem{Melfo:2011nx}
A.~Melfo, M.~Nemevsek, F.~Nesti, G.~Senjanovic, Y.~Zhang, {Type II Seesaw at
  LHC: The Roadmap}, Phys. Rev. D85 (2012) 055018.
\newblock \href {http://arxiv.org/abs/1108.4416} {\path{arXiv:1108.4416}},
  \href {http://dx.doi.org/10.1103/PhysRevD.85.055018}
  {\path{doi:10.1103/PhysRevD.85.055018}}.

\bibitem{Alves:2011kc}
A.~Alves, E.~Ramirez~Barreto, A.~G. Dias, C.~A. de~S.~Pires, F.~S. Queiroz,
  P.~S. Rodrigues~da Silva, {Probing 3-3-1 Models in Diphoton Higgs Boson
  Decay}, Phys. Rev. D84 (2011) 115004.
\newblock \href {http://arxiv.org/abs/1109.0238} {\path{arXiv:1109.0238}},
  \href {http://dx.doi.org/10.1103/PhysRevD.84.115004}
  {\path{doi:10.1103/PhysRevD.84.115004}}.

\bibitem{Alves:2012yp}
A.~Alves, E.~Ramirez~Barreto, A.~G. Dias, C.~A. de~S.~Pires, F.~S. Queiroz,
  P.~S. Rodrigues~da Silva, {Explaining the Higgs Decays at the LHC with an
  Extended Electroweak Model}, Eur. Phys. J. C73~(2) (2013) 2288.
\newblock \href {http://arxiv.org/abs/1207.3699} {\path{arXiv:1207.3699}},
  \href {http://dx.doi.org/10.1140/epjc/s10052-013-2288-y}
  {\path{doi:10.1140/epjc/s10052-013-2288-y}}.

\bibitem{Akeroyd:2012ms}
A.~G. Akeroyd, S.~Moretti, {Enhancement of H to gamma gamma from doubly charged
  scalars in the Higgs Triplet Model}, Phys. Rev. D86 (2012) 035015.
\newblock \href {http://arxiv.org/abs/1206.0535} {\path{arXiv:1206.0535}},
  \href {http://dx.doi.org/10.1103/PhysRevD.86.035015}
  {\path{doi:10.1103/PhysRevD.86.035015}}.

\bibitem{Wang:2012ts}
L.~Wang, X.-F. Han, {130 GeV gamma-ray line and enhancement of
  $h\to\gamma\gamma$ in the Higgs triplet model plus a scalar dark matter},
  Phys. Rev. D87~(1) (2013) 015015.
\newblock \href {http://arxiv.org/abs/1209.0376} {\path{arXiv:1209.0376}},
  \href {http://dx.doi.org/10.1103/PhysRevD.87.015015}
  {\path{doi:10.1103/PhysRevD.87.015015}}.

\bibitem{Lavoura:2003xp}
L.~Lavoura, {General formulae for $f_1 \rightarrow f_2 \gamma$}, Eur. Phys. J.
  C29 (2003) 191--195.
\newblock \href {http://arxiv.org/abs/hep-ph/0302221}
  {\path{arXiv:hep-ph/0302221}}, \href
  {http://dx.doi.org/10.1140/epjc/s2003-01212-7}
  {\path{doi:10.1140/epjc/s2003-01212-7}}.

\bibitem{Cuypers:1996ia}
F.~Cuypers, S.~Davidson, {Bileptons: Present limits and future prospects}, Eur.
  Phys. J. C2 (1998) 503--528.
\newblock \href {http://arxiv.org/abs/hep-ph/9609487}
  {\path{arXiv:hep-ph/9609487}}, \href
  {http://dx.doi.org/10.1007/s100520050157} {\path{doi:10.1007/s100520050157}}.

\bibitem{Akeroyd:2009nu}
A.~G. Akeroyd, M.~Aoki, H.~Sugiyama, {Lepton Flavour Violating Decays $\tau
  \rightarrow \bar{l} ll$ and $\mu \rightarrow e \gamma$ in the Higgs Triplet
  Model}, Phys. Rev. D79 (2009) 113010.
\newblock \href {http://arxiv.org/abs/0904.3640} {\path{arXiv:0904.3640}},
  \href {http://dx.doi.org/10.1103/PhysRevD.79.113010}
  {\path{doi:10.1103/PhysRevD.79.113010}}.

\bibitem{Chakrabortty:2012vp}
J.~Chakrabortty, P.~Ghosh, W.~Rodejohann, {Lower Limits on $\mu \to e \gamma$
  from New Measurements on $U_{e3}$}, Phys. Rev. D86 (2012) 075020.
\newblock \href {http://arxiv.org/abs/1204.1000} {\path{arXiv:1204.1000}},
  \href {http://dx.doi.org/10.1103/PhysRevD.86.075020}
  {\path{doi:10.1103/PhysRevD.86.075020}}.

\bibitem{Chakrabortty:2015zpm}
J.~Chakrabortty, P.~Ghosh, S.~Mondal, T.~Srivastava, {Reconciling (g-2)$_μ$
  and charged lepton flavor violating processes through a doubly charged
  scalar}, Phys. Rev. D93~(11) (2016) 115004.
\newblock \href {http://arxiv.org/abs/1512.03581} {\path{arXiv:1512.03581}},
  \href {http://dx.doi.org/10.1103/PhysRevD.93.115004}
  {\path{doi:10.1103/PhysRevD.93.115004}}.

\bibitem{Leveille:1977rc}
J.~P. Leveille, {The Second Order Weak Correction to (G-2) of the Muon in
  Arbitrary Gauge Models}, Nucl. Phys. B137 (1978) 63--76.
\newblock \href {http://dx.doi.org/10.1016/0550-3213(78)90051-2}
  {\path{doi:10.1016/0550-3213(78)90051-2}}.

\bibitem{Abada:2014kba}
A.~Abada, M.~E. Krauss, W.~Porod, F.~Staub, A.~Vicente, C.~Weiland, {Lepton
  flavor violation in low-scale seesaw models: SUSY and non-SUSY
  contributions}, JHEP 11 (2014) 048.
\newblock \href {http://arxiv.org/abs/1408.0138} {\path{arXiv:1408.0138}},
  \href {http://dx.doi.org/10.1007/JHEP11(2014)048}
  {\path{doi:10.1007/JHEP11(2014)048}}.

\bibitem{Ponce:2006vw}
W.~A. Ponce, L.~A. Sanchez, {Systematic study of the SU(3)(c) X SU(4)(L) X
  U(1)(X) gauge symmetry}, Mod. Phys. Lett. A22 (2007) 435--448.
\newblock \href {http://arxiv.org/abs/hep-ph/0607175}
  {\path{arXiv:hep-ph/0607175}}, \href
  {http://dx.doi.org/10.1142/S0217732307021275}
  {\path{doi:10.1142/S0217732307021275}}.

\bibitem{Salazar:2007ym}
J.~C. Salazar, W.~A. Ponce, D.~A. Gutierrez, {Phenomenology of the SU(3)(c) x
  SU(3)(L) x U(1)(X) model with exotic charged leptons}, Phys. Rev. D75 (2007)
  075016.
\newblock \href {http://arxiv.org/abs/hep-ph/0703300}
  {\path{arXiv:hep-ph/0703300}}, \href
  {http://dx.doi.org/10.1103/PhysRevD.75.075016}
  {\path{doi:10.1103/PhysRevD.75.075016}}.

\bibitem{DePree:2008st}
E.~De~Pree, M.~Sher, I.~Turan, {Production of single heavy charged leptons at a
  linear collider}, Phys. Rev. D77 (2008) 093001.
\newblock \href {http://arxiv.org/abs/0803.0996} {\path{arXiv:0803.0996}},
  \href {http://dx.doi.org/10.1103/PhysRevD.77.093001}
  {\path{doi:10.1103/PhysRevD.77.093001}}.

\bibitem{Hewett:1988xc}
J.~L. Hewett, T.~G. Rizzo, {Low-Energy Phenomenology of Superstring Inspired
  E(6) Models}, Phys. Rept. 183 (1989) 193.
\newblock \href {http://dx.doi.org/10.1016/0370-1573(89)90071-9}
  {\path{doi:10.1016/0370-1573(89)90071-9}}.

\bibitem{Lee:2012xn}
H.-S. Lee, A.~Soni, {Fourth Generation Parity}, Phys. Rev. Lett. 110~(2) (2013)
  021802.
\newblock \href {http://arxiv.org/abs/1206.6110} {\path{arXiv:1206.6110}},
  \href {http://dx.doi.org/10.1103/PhysRevLett.110.021802}
  {\path{doi:10.1103/PhysRevLett.110.021802}}.

\bibitem{Ari:2013wda}
V.~Ari, O.~Çakir, S.~Kuday, {Pair Production of New Heavy Leptons with $U(1)'$
  Charge at Linear Colliders}, Int. J. Mod. Phys. A29 (2014) 1450055.
\newblock \href {http://arxiv.org/abs/1309.7444} {\path{arXiv:1309.7444}},
  \href {http://dx.doi.org/10.1142/S0217751X14500559}
  {\path{doi:10.1142/S0217751X14500559}}.

\bibitem{Ma:2013tda}
T.~Ma, B.~Zhang, G.~Cacciapaglia, {Triplet with a doubly-charged lepton at the
  LHC}, Phys. Rev. D89~(1) (2014) 015020.
\newblock \href {http://arxiv.org/abs/1309.7396} {\path{arXiv:1309.7396}},
  \href {http://dx.doi.org/10.1103/PhysRevD.89.015020}
  {\path{doi:10.1103/PhysRevD.89.015020}}.

\bibitem{Ma:2014zda}
T.~Ma, B.~Zhang, G.~Cacciapaglia, {Doubly Charged Lepton from an Exotic Doublet
  at the LHC}, Phys. Rev. D89~(9) (2014) 093022.
\newblock \href {http://arxiv.org/abs/1404.2375} {\path{arXiv:1404.2375}},
  \href {http://dx.doi.org/10.1103/PhysRevD.89.093022}
  {\path{doi:10.1103/PhysRevD.89.093022}}.

\bibitem{Altarelli:1977zu}
G.~Altarelli, L.~Baulieu, {Muon Number Nonconserving Processes in Models with
  Doubly Charged Leptons}, Lett. Nuovo Cim. 19 (1977) 463.
\newblock \href {http://dx.doi.org/10.1007/BF02745998}
  {\path{doi:10.1007/BF02745998}}.

\bibitem{Wilczek:1977wb}
F.~Wilczek, A.~Zee, {Rare Muon Decays, Natural Lepton Models, and Doubly
  Charged Leptons}, Phys. Rev. Lett. 38 (1977) 531.
\newblock \href {http://dx.doi.org/10.1103/PhysRevLett.38.531}
  {\path{doi:10.1103/PhysRevLett.38.531}}.

\bibitem{Biondini:2012ny}
S.~Biondini, O.~Panella, G.~Pancheri, Y.~N. Srivastava, L.~Fano, {Phenomenology
  of excited doubly charged heavy leptons at LHC}, Phys. Rev. D85 (2012)
  095018.
\newblock \href {http://arxiv.org/abs/1201.3764} {\path{arXiv:1201.3764}},
  \href {http://dx.doi.org/10.1103/PhysRevD.85.095018}
  {\path{doi:10.1103/PhysRevD.85.095018}}.

\bibitem{Li:2016ijq}
W.~Li, J.~N. Ng, {Doubly charged vector leptons and the Higgs portal}, Phys.
  Rev. D94~(9) (2016) 095012.
\newblock \href {http://arxiv.org/abs/1605.04885} {\path{arXiv:1605.04885}},
  \href {http://dx.doi.org/10.1103/PhysRevD.94.095012}
  {\path{doi:10.1103/PhysRevD.94.095012}}.

\bibitem{Pisano:1991ee}
F.~Pisano, V.~Pleitez, {An SU(3) x U(1) model for electroweak interactions},
  Phys. Rev. D46 (1992) 410--417.
\newblock \href {http://arxiv.org/abs/hep-ph/9206242}
  {\path{arXiv:hep-ph/9206242}}, \href
  {http://dx.doi.org/10.1103/PhysRevD.46.410}
  {\path{doi:10.1103/PhysRevD.46.410}}.

\bibitem{Kelso:2014qka}
C.~Kelso, H.~N. Long, R.~Martinez, F.~S. Queiroz, {Connection of $g-2_{\mu}$,
  electroweak, dark matter, and collider constraints on 331 models}, Phys. Rev.
  D90~(11) (2014) 113011.
\newblock \href {http://arxiv.org/abs/1408.6203} {\path{arXiv:1408.6203}},
  \href {http://dx.doi.org/10.1103/PhysRevD.90.113011}
  {\path{doi:10.1103/PhysRevD.90.113011}}.

\bibitem{Felipe:2014zka}
R.~González~Felipe, I.~P. Ivanov, C.~C. Nishi, H.~Serôdio, J.~P. Silva,
  {Constraining multi-Higgs flavour models}, Eur. Phys. J. C74~(7) (2014) 2953.
\newblock \href {http://arxiv.org/abs/1401.5807} {\path{arXiv:1401.5807}},
  \href {http://dx.doi.org/10.1140/epjc/s10052-014-2953-9}
  {\path{doi:10.1140/epjc/s10052-014-2953-9}}.

\bibitem{Aad:2014hja}
G.~Aad, et~al., {Search for new phenomena in events with three or more charged
  leptons in $pp$ collisions at $\sqrt{s}=8$ TeV with the ATLAS detector}, JHEP
  08 (2015) 138.
\newblock \href {http://arxiv.org/abs/1411.2921} {\path{arXiv:1411.2921}},
  \href {http://dx.doi.org/10.1007/JHEP08(2015)138}
  {\path{doi:10.1007/JHEP08(2015)138}}.

\bibitem{Khachatryan:2014dka}
V.~Khachatryan, et~al., {Search for heavy neutrinos and $\mathrm {W}$ bosons
  with right-handed couplings in proton-proton collisions at $\sqrt{s} =
  8\,\text {TeV} $}, Eur. Phys. J. C74~(11) (2014) 3149.
\newblock \href {http://arxiv.org/abs/1407.3683} {\path{arXiv:1407.3683}},
  \href {http://dx.doi.org/10.1140/epjc/s10052-014-3149-z}
  {\path{doi:10.1140/epjc/s10052-014-3149-z}}.

\bibitem{Lindner:2016lxq}
M.~Lindner, F.~S. Queiroz, W.~Rodejohann, C.~E. Yaguna, {Left-Right Symmetry
  and Lepton Number Violation at the Large Hadron Electron Collider}, JHEP 06
  (2016) 140.
\newblock \href {http://arxiv.org/abs/1604.08596} {\path{arXiv:1604.08596}},
  \href {http://dx.doi.org/10.1007/JHEP06(2016)140}
  {\path{doi:10.1007/JHEP06(2016)140}}.

\bibitem{Ruegg:2003ps}
H.~Ruegg, M.~Ruiz-Altaba, {The Stueckelberg field}, Int. J. Mod. Phys. A19
  (2004) 3265--3348.
\newblock \href {http://arxiv.org/abs/hep-th/0304245}
  {\path{arXiv:hep-th/0304245}}, \href
  {http://dx.doi.org/10.1142/S0217751X04019755}
  {\path{doi:10.1142/S0217751X04019755}}.

\bibitem{Feldman:2006wb}
D.~Feldman, Z.~Liu, P.~Nath, {The Stueckelberg $Z$ Prime at the LHC: Discovery
  Potential, Signature Spaces and Model Discrimination}, JHEP 11 (2006) 007.
\newblock \href {http://arxiv.org/abs/hep-ph/0606294}
  {\path{arXiv:hep-ph/0606294}}, \href
  {http://dx.doi.org/10.1088/1126-6708/2006/11/007}
  {\path{doi:10.1088/1126-6708/2006/11/007}}.

\bibitem{Feldman:2007wj}
D.~Feldman, Z.~Liu, P.~Nath, {The Stueckelberg Z-prime Extension with Kinetic
  Mixing and Milli-Charged Dark Matter From the Hidden Sector}, Phys. Rev. D75
  (2007) 115001.
\newblock \href {http://arxiv.org/abs/hep-ph/0702123}
  {\path{arXiv:hep-ph/0702123}}, \href
  {http://dx.doi.org/10.1103/PhysRevD.75.115001}
  {\path{doi:10.1103/PhysRevD.75.115001}}.

\bibitem{Feldman:2011ms}
D.~Feldman, P.~Fileviez~Perez, P.~Nath, {R-parity Conservation via the
  Stueckelberg Mechanism: LHC and Dark Matter Signals}, JHEP 01 (2012) 038.
\newblock \href {http://arxiv.org/abs/1109.2901} {\path{arXiv:1109.2901}},
  \href {http://dx.doi.org/10.1007/JHEP01(2012)038}
  {\path{doi:10.1007/JHEP01(2012)038}}.

\bibitem{Davidson:1978pm}
A.~Davidson, {$B^-$l as the Fourth Color, Quark - Lepton Correspondence, and
  Natural Masslessness of Neutrinos Within a Generalized Ws Model}, Phys. Rev.
  D20 (1979) 776.
\newblock \href {http://dx.doi.org/10.1103/PhysRevD.20.776}
  {\path{doi:10.1103/PhysRevD.20.776}}.

\bibitem{Mohapatra:1980qe}
R.~N. Mohapatra, R.~E. Marshak, {Local B-L Symmetry of Electroweak
  Interactions, Majorana Neutrinos and Neutron Oscillations}, Phys. Rev. Lett.
  44 (1980) 1316--1319, [Erratum: Phys. Rev. Lett.44,1643(1980)].
\newblock \href {http://dx.doi.org/10.1103/PhysRevLett.44.1316}
  {\path{doi:10.1103/PhysRevLett.44.1316}}.

\bibitem{Ismail:2016tod}
A.~Ismail, W.-Y. Keung, K.-H. Tsao, J.~Unwin, {Axial vector $Z′$ and anomaly
  cancellation}, Nucl. Phys. B918 (2017) 220--244.
\newblock \href {http://arxiv.org/abs/1609.02188} {\path{arXiv:1609.02188}},
  \href {http://dx.doi.org/10.1016/j.nuclphysb.2017.03.001}
  {\path{doi:10.1016/j.nuclphysb.2017.03.001}}.

\bibitem{Langacker:2000ju}
P.~Langacker, M.~Plumacher, {Flavor changing effects in theories with a heavy
  $Z^\prime$ boson with family nonuniversal couplings}, Phys. Rev. D62 (2000)
  013006.
\newblock \href {http://arxiv.org/abs/hep-ph/0001204}
  {\path{arXiv:hep-ph/0001204}}, \href
  {http://dx.doi.org/10.1103/PhysRevD.62.013006}
  {\path{doi:10.1103/PhysRevD.62.013006}}.

\bibitem{Aaboud:2016zkn}
M.~Aaboud, et~al., {Search for new resonances in events with one lepton and
  missing transverse momentum in $pp$ collisions at $\sqrt{s} = 13$ TeV with
  the ATLAS detector}, Phys. Lett. B762 (2016) 334--352.
\newblock \href {http://arxiv.org/abs/1606.03977} {\path{arXiv:1606.03977}},
  \href {http://dx.doi.org/10.1016/j.physletb.2016.09.040}
  {\path{doi:10.1016/j.physletb.2016.09.040}}.

\bibitem{Khachatryan:2014tva}
V.~Khachatryan, et~al., {Search for physics beyond the standard model in final
  states with a lepton and missing transverse energy in proton-proton
  collisions at sqrt(s) = 8 TeV}, Phys. Rev. D91~(9) (2015) 092005.
\newblock \href {http://arxiv.org/abs/1408.2745} {\path{arXiv:1408.2745}},
  \href {http://dx.doi.org/10.1103/PhysRevD.91.092005}
  {\path{doi:10.1103/PhysRevD.91.092005}}.

\bibitem{Wess:1974tw}
J.~Wess, B.~Zumino, {Supergauge Transformations in Four-Dimensions}, Nucl.
  Phys. B70 (1974) 39--50.
\newblock \href {http://dx.doi.org/10.1016/0550-3213(74)90355-1}
  {\path{doi:10.1016/0550-3213(74)90355-1}}.

\bibitem{Veltman:1980mj}
M.~J.~G. Veltman, {The Infrared - Ultraviolet Connection}, Acta Phys. Polon.
  B12 (1981) 437.

\bibitem{Amaldi:1991cn}
U.~Amaldi, W.~de~Boer, H.~Furstenau, {Comparison of grand unified theories with
  electroweak and strong coupling constants measured at LEP}, Phys. Lett. B260
  (1991) 447--455.
\newblock \href {http://dx.doi.org/10.1016/0370-2693(91)91641-8}
  {\path{doi:10.1016/0370-2693(91)91641-8}}.

\bibitem{Langacker:1991an}
P.~Langacker, M.-x. Luo, {Implications of precision electroweak experiments for
  $M_t$, $\rho_{0}$, $\sin^2\theta_W$ and grand unification}, Phys. Rev. D44
  (1991) 817--822.
\newblock \href {http://dx.doi.org/10.1103/PhysRevD.44.817}
  {\path{doi:10.1103/PhysRevD.44.817}}.

\bibitem{Chatrchyan:2011zy}
S.~Chatrchyan, et~al., {Search for Supersymmetry at the LHC in Events with Jets
  and Missing Transverse Energy}, Phys. Rev. Lett. 107 (2011) 221804.
\newblock \href {http://arxiv.org/abs/1109.2352} {\path{arXiv:1109.2352}},
  \href {http://dx.doi.org/10.1103/PhysRevLett.107.221804}
  {\path{doi:10.1103/PhysRevLett.107.221804}}.

\bibitem{ATLAS:2016kts}
T.~A. collaboration, {Further searches for squarks and gluinos in final states
  with jets and missing transverse momentum at $\sqrt{s}$ =13 TeV with the
  ATLAS detector}.

\bibitem{Khachatryan:2016xdt}
V.~Khachatryan, et~al., {Search for supersymmetry in events with one lepton and
  multiple jets in proton-proton collisions at $\sqrt{s} =$ 13 TeV}, Phys. Rev.
  D95~(1) (2017) 012011.
\newblock \href {http://arxiv.org/abs/1609.09386} {\path{arXiv:1609.09386}},
  \href {http://dx.doi.org/10.1103/PhysRevD.95.012011}
  {\path{doi:10.1103/PhysRevD.95.012011}}.

\bibitem{Cho:2000sf}
G.-C. Cho, K.~Hagiwara, M.~Hayakawa, {Muon g-2 and precision electroweak
  physics in the MSSM}, Phys. Lett. B478 (2000) 231--238.
\newblock \href {http://arxiv.org/abs/hep-ph/0001229}
  {\path{arXiv:hep-ph/0001229}}, \href
  {http://dx.doi.org/10.1016/S0370-2693(00)00288-4}
  {\path{doi:10.1016/S0370-2693(00)00288-4}}.

\bibitem{Choi:2001pz}
K.~Choi, K.~Hwang, S.~K. Kang, K.~Y. Lee, W.~Y. Song, {Probing the messenger of
  supersymmetry breaking by the muon anomalous magnetic moment}, Phys. Rev. D64
  (2001) 055001.
\newblock \href {http://arxiv.org/abs/hep-ph/0103048}
  {\path{arXiv:hep-ph/0103048}}, \href
  {http://dx.doi.org/10.1103/PhysRevD.64.055001}
  {\path{doi:10.1103/PhysRevD.64.055001}}.

\bibitem{Feng:2001tr}
J.~L. Feng, K.~T. Matchev, {Supersymmetry and the anomalous magnetic moment of
  the muon}, Phys. Rev. Lett. 86 (2001) 3480--3483.
\newblock \href {http://arxiv.org/abs/hep-ph/0102146}
  {\path{arXiv:hep-ph/0102146}}, \href
  {http://dx.doi.org/10.1103/PhysRevLett.86.3480}
  {\path{doi:10.1103/PhysRevLett.86.3480}}.

\bibitem{Cheung:2001hz}
K.-m. Cheung, C.-H. Chou, O.~C.~W. Kong, {Muon anomalous magnetic moment, two
  Higgs doublet model, and supersymmetry}, Phys. Rev. D64 (2001) 111301.
\newblock \href {http://arxiv.org/abs/hep-ph/0103183}
  {\path{arXiv:hep-ph/0103183}}, \href
  {http://dx.doi.org/10.1103/PhysRevD.64.111301}
  {\path{doi:10.1103/PhysRevD.64.111301}}.

\bibitem{Arnowitt:2001pm}
R.~L. Arnowitt, B.~Dutta, Y.~Santoso, {SUSY phases, the electron electric
  dipole moment and the muon magnetic moment}, Phys. Rev. D64 (2001) 113010.
\newblock \href {http://arxiv.org/abs/hep-ph/0106089}
  {\path{arXiv:hep-ph/0106089}}, \href
  {http://dx.doi.org/10.1103/PhysRevD.64.113010}
  {\path{doi:10.1103/PhysRevD.64.113010}}.

\bibitem{Belanger:2001am}
G.~Belanger, F.~Boudjema, A.~Cottrant, R.~M. Godbole, A.~Semenov, {The MSSM
  invisible Higgs in the light of dark matter and g-2}, Phys. Lett. B519 (2001)
  93--102.
\newblock \href {http://arxiv.org/abs/hep-ph/0106275}
  {\path{arXiv:hep-ph/0106275}}, \href
  {http://dx.doi.org/10.1016/S0370-2693(01)00976-5}
  {\path{doi:10.1016/S0370-2693(01)00976-5}}.

\bibitem{Byrne:2002cw}
M.~Byrne, C.~Kolda, J.~E. Lennon, {Updated implications of the muon anomalous
  magnetic moment for supersymmetry}, Phys. Rev. D67 (2003) 075004.
\newblock \href {http://arxiv.org/abs/hep-ph/0208067}
  {\path{arXiv:hep-ph/0208067}}, \href
  {http://dx.doi.org/10.1103/PhysRevD.67.075004}
  {\path{doi:10.1103/PhysRevD.67.075004}}.

\bibitem{Baek:2002cc}
S.~Baek, P.~Ko, J.-h. Park, {Muon anomalous magnetic moment from effective
  supersymmetry}, Eur. Phys. J. C24 (2002) 613--618.
\newblock \href {http://arxiv.org/abs/hep-ph/0203251}
  {\path{arXiv:hep-ph/0203251}}, \href
  {http://dx.doi.org/10.1007/s10052-002-0971-5}
  {\path{doi:10.1007/s10052-002-0971-5}}.

\bibitem{Stockinger:2006zn}
D.~Stockinger, {The Muon Magnetic Moment and Supersymmetry}, J. Phys. G34
  (2007) R45--R92.
\newblock \href {http://arxiv.org/abs/hep-ph/0609168}
  {\path{arXiv:hep-ph/0609168}}, \href
  {http://dx.doi.org/10.1088/0954-3899/34/2/R01}
  {\path{doi:10.1088/0954-3899/34/2/R01}}.

\bibitem{RamseyMusolf:2006vr}
M.~J. Ramsey-Musolf, S.~Su, {Low Energy Precision Test of Supersymmetry}, Phys.
  Rept. 456 (2008) 1--88.
\newblock \href {http://arxiv.org/abs/hep-ph/0612057}
  {\path{arXiv:hep-ph/0612057}}, \href
  {http://dx.doi.org/10.1016/j.physrep.2007.10.001}
  {\path{doi:10.1016/j.physrep.2007.10.001}}.

\bibitem{Heinemeyer:2003dq}
S.~Heinemeyer, D.~Stockinger, G.~Weiglein, {Two loop SUSY corrections to the
  anomalous magnetic moment of the muon}, Nucl. Phys. B690 (2004) 62--80.
\newblock \href {http://arxiv.org/abs/hep-ph/0312264}
  {\path{arXiv:hep-ph/0312264}}, \href
  {http://dx.doi.org/10.1016/j.nuclphysb.2004.04.017}
  {\path{doi:10.1016/j.nuclphysb.2004.04.017}}.

\bibitem{Babu:2014lwa}
K.~S. Babu, I.~Gogoladze, Q.~Shafi, C.~S. Ün, {Muon g-2, 125 GeV Higgs boson,
  and neutralino dark matter in a flavor symmetry-based MSSM}, Phys. Rev.
  D90~(11) (2014) 116002.
\newblock \href {http://arxiv.org/abs/1406.6965} {\path{arXiv:1406.6965}},
  \href {http://dx.doi.org/10.1103/PhysRevD.90.116002}
  {\path{doi:10.1103/PhysRevD.90.116002}}.

\bibitem{Gomez:2014uha}
M.~E. Gómez, T.~Hahn, S.~Heinemeyer, M.~Rehman, {Higgs masses and Electroweak
  Precision Observables in the Lepton-Flavor-Violating MSSM}, Phys. Rev.
  D90~(7) (2014) 074016.
\newblock \href {http://arxiv.org/abs/1408.0663} {\path{arXiv:1408.0663}},
  \href {http://dx.doi.org/10.1103/PhysRevD.90.074016}
  {\path{doi:10.1103/PhysRevD.90.074016}}.

\bibitem{Kobakhidze:2016mdx}
A.~Kobakhidze, M.~Talia, L.~Wu, {Probing the MSSM explanation of the muon g-2
  anomaly in dark matter experiments and at a 100 TeV $pp$ collider}, Phys.
  Rev. D95~(5) (2017) 055023.
\newblock \href {http://arxiv.org/abs/1608.03641} {\path{arXiv:1608.03641}},
  \href {http://dx.doi.org/10.1103/PhysRevD.95.055023}
  {\path{doi:10.1103/PhysRevD.95.055023}}.

\bibitem{Grifols:1982vx}
J.~A. Grifols, A.~Mendez, {Constraints on Supersymmetric Particle Masses From
  ($g-2$) $\mu$}, Phys. Rev. D26 (1982) 1809.
\newblock \href {http://dx.doi.org/10.1103/PhysRevD.26.1809}
  {\path{doi:10.1103/PhysRevD.26.1809}}.

\bibitem{Barbieri:1982aj}
R.~Barbieri, L.~Maiani, {The Muon Anomalous Magnetic Moment in Broken
  Supersymmetric Theories}, Phys. Lett. B117 (1982) 203--207.
\newblock \href {http://dx.doi.org/10.1016/0370-2693(82)90547-0}
  {\path{doi:10.1016/0370-2693(82)90547-0}}.

\bibitem{Ellis:1982by}
J.~R. Ellis, J.~S. Hagelin, D.~V. Nanopoulos, {Spin 0 Leptons and the Anomalous
  Magnetic Moment of the Muon}, Phys. Lett. B116 (1982) 283--286.
\newblock \href {http://dx.doi.org/10.1016/0370-2693(82)90343-4}
  {\path{doi:10.1016/0370-2693(82)90343-4}}.

\bibitem{Kosower:1983yw}
D.~A. Kosower, L.~M. Krauss, N.~Sakai, {Low-Energy Supergravity and the
  Anomalous Magnetic Moment of the Muon}, Phys. Lett. B133 (1983) 305--310.
\newblock \href {http://dx.doi.org/10.1016/0370-2693(83)90152-1}
  {\path{doi:10.1016/0370-2693(83)90152-1}}.

\bibitem{Yuan:1984ww}
T.~C. Yuan, R.~L. Arnowitt, A.~H. Chamseddine, P.~Nath, {Supersymmetric
  Electroweak Effects on G-2 (mu)}, Z. Phys. C26 (1984) 407.
\newblock \href {http://dx.doi.org/10.1007/BF01452567}
  {\path{doi:10.1007/BF01452567}}.

\bibitem{Romao:1984pn}
J.~C. Romao, A.~Barroso, M.~C. Bento, G.~C. Branco, {Flavor Violation in
  Supersymmetric Theories}, Nucl. Phys. B250 (1985) 295--311.
\newblock \href {http://dx.doi.org/10.1016/0550-3213(85)90483-3}
  {\path{doi:10.1016/0550-3213(85)90483-3}}.

\bibitem{Hisano:1995nq}
J.~Hisano, T.~Moroi, K.~Tobe, M.~Yamaguchi, T.~Yanagida, {Lepton flavor
  violation in the supersymmetric standard model with seesaw induced neutrino
  masses}, Phys. Lett. B357 (1995) 579--587.
\newblock \href {http://arxiv.org/abs/hep-ph/9501407}
  {\path{arXiv:hep-ph/9501407}}, \href
  {http://dx.doi.org/10.1016/0370-2693(95)00954-J}
  {\path{doi:10.1016/0370-2693(95)00954-J}}.

\bibitem{Barbieri:1995tw}
R.~Barbieri, L.~J. Hall, A.~Strumia, {Violations of lepton flavor and CP in
  supersymmetric unified theories}, Nucl. Phys. B445 (1995) 219--251.
\newblock \href {http://arxiv.org/abs/hep-ph/9501334}
  {\path{arXiv:hep-ph/9501334}}, \href
  {http://dx.doi.org/10.1016/0550-3213(95)00208-A}
  {\path{doi:10.1016/0550-3213(95)00208-A}}.

\bibitem{Ibrahim:1999aj}
T.~Ibrahim, P.~Nath, {Effects of large CP violating phases on g(muon) - 2 in
  MSSM}, Phys. Rev. D62 (2000) 015004.
\newblock \href {http://arxiv.org/abs/hep-ph/9908443}
  {\path{arXiv:hep-ph/9908443}}, \href
  {http://dx.doi.org/10.1103/PhysRevD.62.015004}
  {\path{doi:10.1103/PhysRevD.62.015004}}.

\bibitem{Deppisch:2002qw}
F.~Deppisch, H.~Pas, A.~Redelbach, R.~Ruckl, Y.~Shimizu, {Neutrino
  oscillations, SUSY seesaw mechanism and charged lepton flavor violation},
  Nucl. Phys. Proc. Suppl. 116 (2003) 316--320, [,316(2002)].
\newblock \href {http://arxiv.org/abs/hep-ph/0211138}
  {\path{arXiv:hep-ph/0211138}}, \href
  {http://dx.doi.org/10.1016/S0920-5632(03)80191-3}
  {\path{doi:10.1016/S0920-5632(03)80191-3}}.

\bibitem{Deppisch:2004xv}
F.~Deppisch, H.~Päs, A.~Redelbach, R.~Rückl, {Lepton flavor violation in the
  SUSY seesaw model: An Update}, Springer Proc. Phys. 98 (2005) 27--38.
\newblock \href {http://arxiv.org/abs/hep-ph/0403212}
  {\path{arXiv:hep-ph/0403212}}, \href
  {http://dx.doi.org/10.1007/3-540-26798-0_3}
  {\path{doi:10.1007/3-540-26798-0_3}}.

\bibitem{Martin:2002eu}
S.~P. Martin, J.~D. Wells, {Superconservative interpretation of muon g-2
  results applied to supersymmetry}, Phys. Rev. D67 (2003) 015002.
\newblock \href {http://arxiv.org/abs/hep-ph/0209309}
  {\path{arXiv:hep-ph/0209309}}, \href
  {http://dx.doi.org/10.1103/PhysRevD.67.015002}
  {\path{doi:10.1103/PhysRevD.67.015002}}.

\bibitem{Girrbach:2009uy}
J.~Girrbach, S.~Mertens, U.~Nierste, S.~Wiesenfeldt, {Lepton flavour violation
  in the MSSM}, JHEP 05 (2010) 026.
\newblock \href {http://arxiv.org/abs/0910.2663} {\path{arXiv:0910.2663}},
  \href {http://dx.doi.org/10.1007/JHEP05(2010)026}
  {\path{doi:10.1007/JHEP05(2010)026}}.

\bibitem{Heinemeyer:2004py}
S.~Heinemeyer, D.~Stockinger, G.~Weiglein,
  \href{http://weblib.cern.ch/abstract?CERN-PH-TH-2004-129}{{The Anomalous
  magnetic moment of the muon in the MSSM: Recent developments}}, in: {Linear
  colliders. Proceedings, International Conference, LCWS 2004, Paris, France,
  April 19-23, 2004}, 2004.
\newblock \href {http://arxiv.org/abs/hep-ph/0407171}
  {\path{arXiv:hep-ph/0407171}}.
\newline\urlprefix\url{http://weblib.cern.ch/abstract?CERN-PH-TH-2004-129}

\bibitem{Cirigliano:2004mv}
V.~Cirigliano, A.~Kurylov, M.~J. Ramsey-Musolf, P.~Vogel, {Lepton flavor
  violation without supersymmetry}, Phys. Rev. D70 (2004) 075007.
\newblock \href {http://arxiv.org/abs/hep-ph/0404233}
  {\path{arXiv:hep-ph/0404233}}, \href
  {http://dx.doi.org/10.1103/PhysRevD.70.075007}
  {\path{doi:10.1103/PhysRevD.70.075007}}.

\bibitem{Haestier:2006mg}
J.~Haestier, S.~Heinemeyer, W.~Hollik, D.~Stockinger, A.~M. Weber, G.~Weiglein,
  {Precision Observables in the MSSM: W mass and the muon magnetic moment}, AIP
  Conf. Proc. 903 (2007) 291--294, [,291(2006)].
\newblock \href {http://arxiv.org/abs/hep-ph/0610318}
  {\path{arXiv:hep-ph/0610318}}, \href {http://dx.doi.org/10.1063/1.2735182}
  {\path{doi:10.1063/1.2735182}}.

\bibitem{Cheung:2009fc}
K.~Cheung, O.~C.~W. Kong, J.~S. Lee, {Electric and anomalous magnetic dipole
  moments of the muon in the MSSM}, JHEP 06 (2009) 020.
\newblock \href {http://arxiv.org/abs/0904.4352} {\path{arXiv:0904.4352}},
  \href {http://dx.doi.org/10.1088/1126-6708/2009/06/020}
  {\path{doi:10.1088/1126-6708/2009/06/020}}.

\bibitem{Haber:1984rc}
H.~E. Haber, G.~L. Kane, {The Search for Supersymmetry: Probing Physics Beyond
  the Standard Model}, Phys. Rept. 117 (1985) 75--263.
\newblock \href {http://dx.doi.org/10.1016/0370-1573(85)90051-1}
  {\path{doi:10.1016/0370-1573(85)90051-1}}.

\bibitem{Martin:1997ns}
S.~P. Martin, {A Supersymmetry primer}[Adv. Ser. Direct. High Energy
  Phys.18,1(1998)].
\newblock \href {http://arxiv.org/abs/hep-ph/9709356}
  {\path{arXiv:hep-ph/9709356}}, \href
  {http://dx.doi.org/10.1142/9789812839657_0001, 10.1142/9789814307505_0001}
  {\path{doi:10.1142/9789812839657_0001, 10.1142/9789814307505_0001}}.

\bibitem{Moroi:1995yh}
T.~Moroi, {The Muon anomalous magnetic dipole moment in the minimal
  supersymmetric standard model}, Phys. Rev. D53 (1996) 6565--6575, [Erratum:
  Phys. Rev.D56,4424(1997)].
\newblock \href {http://arxiv.org/abs/hep-ph/9512396}
  {\path{arXiv:hep-ph/9512396}}, \href
  {http://dx.doi.org/10.1103/PhysRevD.53.6565, 10.1103/PhysRevD.56.4424}
  {\path{doi:10.1103/PhysRevD.53.6565, 10.1103/PhysRevD.56.4424}}.

\bibitem{Degrassi:1998es}
G.~Degrassi, G.~F. Giudice, {QED logarithms in the electroweak corrections to
  the muon anomalous magnetic moment}, Phys. Rev. D58 (1998) 053007.
\newblock \href {http://arxiv.org/abs/hep-ph/9803384}
  {\path{arXiv:hep-ph/9803384}}, \href
  {http://dx.doi.org/10.1103/PhysRevD.58.053007}
  {\path{doi:10.1103/PhysRevD.58.053007}}.

\bibitem{Athron:2015rva}
P.~Athron, M.~Bach, H.~G. Fargnoli, C.~Gnendiger, R.~Greifenhagen, J.-h. Park,
  S.~Paßehr, D.~Stöckinger, H.~Stöckinger-Kim, A.~Voigt, {GM2Calc: Precise
  MSSM prediction for $(g - 2)$ of the muon}, Eur. Phys. J. C76~(2) (2016) 62.
\newblock \href {http://arxiv.org/abs/1510.08071} {\path{arXiv:1510.08071}},
  \href {http://dx.doi.org/10.1140/epjc/s10052-015-3870-2}
  {\path{doi:10.1140/epjc/s10052-015-3870-2}}.

\bibitem{Bach:2015doa}
M.~Bach, J.-h. Park, D.~Stöckinger, H.~Stöckinger-Kim, {Large muon $(g-2)$
  with TeV-scale SUSY masses for $\tan\beta\to\infty$}, JHEP 10 (2015) 026.
\newblock \href {http://arxiv.org/abs/1504.05500} {\path{arXiv:1504.05500}},
  \href {http://dx.doi.org/10.1007/JHEP10(2015)026}
  {\path{doi:10.1007/JHEP10(2015)026}}.

\bibitem{Martin:2001st}
S.~P. Martin, J.~D. Wells, {Muon anomalous magnetic dipole moment in
  supersymmetric theories}, Phys. Rev. D64 (2001) 035003.
\newblock \href {http://arxiv.org/abs/hep-ph/0103067}
  {\path{arXiv:hep-ph/0103067}}, \href
  {http://dx.doi.org/10.1103/PhysRevD.64.035003}
  {\path{doi:10.1103/PhysRevD.64.035003}}.

\bibitem{Chacko:2001xd}
Z.~Chacko, G.~D. Kribs, {Constraints on lepton flavor violation in the MSSM
  from the muon anomalous magnetic moment measurement}, Phys. Rev. D64 (2001)
  075015.
\newblock \href {http://arxiv.org/abs/hep-ph/0104317}
  {\path{arXiv:hep-ph/0104317}}, \href
  {http://dx.doi.org/10.1103/PhysRevD.64.075015}
  {\path{doi:10.1103/PhysRevD.64.075015}}.

\bibitem{Bi:2002ra}
X.-J. Bi, Y.-p. Kuang, Y.-H. An, {Muon anomalous magnetic moment and lepton
  flavor violation in MSSM}, Eur. Phys. J. C30 (2003) 409--418.
\newblock \href {http://arxiv.org/abs/hep-ph/0211142}
  {\path{arXiv:hep-ph/0211142}}, \href
  {http://dx.doi.org/10.1140/epjc/s2003-01299-8}
  {\path{doi:10.1140/epjc/s2003-01299-8}}.

\bibitem{Ibrahim:2015hva}
T.~Ibrahim, A.~Itani, P.~Nath, {$\mu\to e \gamma$ decay in an MSSM extension},
  Phys. Rev. D92~(1) (2015) 015003.
\newblock \href {http://arxiv.org/abs/1503.01078} {\path{arXiv:1503.01078}},
  \href {http://dx.doi.org/10.1103/PhysRevD.92.015003}
  {\path{doi:10.1103/PhysRevD.92.015003}}.

\bibitem{Kersten:2014xaa}
J.~Kersten, J.-h. Park, D.~Stöckinger, L.~Velasco-Sevilla, {Understanding the
  correlation between $(g-2)_\mu$ and $\mu \rightarrow e \gamma$ in the MSSM},
  JHEP 08 (2014) 118.
\newblock \href {http://arxiv.org/abs/1405.2972} {\path{arXiv:1405.2972}},
  \href {http://dx.doi.org/10.1007/JHEP08(2014)118}
  {\path{doi:10.1007/JHEP08(2014)118}}.

\bibitem{Fargnoli:2013zda}
H.~G. Fargnoli, C.~Gnendiger, S.~Paßehr, D.~Stöckinger, H.~Stöckinger-Kim,
  {Non-decoupling two-loop corrections to $(g-2)_\mu$ from fermion/sfermion
  loops in the MSSM}, Phys. Lett. B726 (2013) 717--724.
\newblock \href {http://arxiv.org/abs/1309.0980} {\path{arXiv:1309.0980}},
  \href {http://dx.doi.org/10.1016/j.physletb.2013.09.034}
  {\path{doi:10.1016/j.physletb.2013.09.034}}.

\bibitem{Minkowski:1977sc}
P.~Minkowski, {$\mu \to e\gamma$ at a Rate of One Out of $10^{9}$ Muon
  Decays?}, Phys. Lett. B67 (1977) 421--428.
\newblock \href {http://dx.doi.org/10.1016/0370-2693(77)90435-X}
  {\path{doi:10.1016/0370-2693(77)90435-X}}.

\bibitem{Mohapatra:1979ia}
R.~N. Mohapatra, G.~Senjanovic, {Neutrino Mass and Spontaneous Parity
  Violation}, Phys. Rev. Lett. 44 (1980) 912.
\newblock \href {http://dx.doi.org/10.1103/PhysRevLett.44.912}
  {\path{doi:10.1103/PhysRevLett.44.912}}.

\bibitem{Lazarides:1980nt}
G.~Lazarides, Q.~Shafi, C.~Wetterich, {Proton Lifetime and Fermion Masses in an
  SO(10) Model}, Nucl. Phys. B181 (1981) 287--300.
\newblock \href {http://dx.doi.org/10.1016/0550-3213(81)90354-0}
  {\path{doi:10.1016/0550-3213(81)90354-0}}.

\bibitem{Mohapatra:1980yp}
R.~N. Mohapatra, G.~Senjanovic, {Neutrino Masses and Mixings in Gauge Models
  with Spontaneous Parity Violation}, Phys. Rev. D23 (1981) 165.
\newblock \href {http://dx.doi.org/10.1103/PhysRevD.23.165}
  {\path{doi:10.1103/PhysRevD.23.165}}.

\bibitem{Schechter:1980gr}
J.~Schechter, J.~W.~F. Valle, {Neutrino Masses in SU(2) x U(1) Theories}, Phys.
  Rev. D22 (1980) 2227.
\newblock \href {http://dx.doi.org/10.1103/PhysRevD.22.2227}
  {\path{doi:10.1103/PhysRevD.22.2227}}.

\bibitem{Senjanovic:1975rk}
G.~Senjanovic, R.~N. Mohapatra, {Exact Left-Right Symmetry and Spontaneous
  Violation of Parity}, Phys. Rev. D12 (1975) 1502.
\newblock \href {http://dx.doi.org/10.1103/PhysRevD.12.1502}
  {\path{doi:10.1103/PhysRevD.12.1502}}.

\bibitem{Senjanovic:1978ev}
G.~Senjanovic, {Spontaneous Breakdown of Parity in a Class of Gauge Theories},
  Nucl. Phys. B153 (1979) 334--364.
\newblock \href {http://dx.doi.org/10.1016/0550-3213(79)90604-7}
  {\path{doi:10.1016/0550-3213(79)90604-7}}.

\bibitem{Chang:1984uy}
D.~Chang, R.~N. Mohapatra, M.~K. Parida, {A New Approach to Left-Right Symmetry
  Breaking in Unified Gauge Theories}, Phys. Rev. D30 (1984) 1052.
\newblock \href {http://dx.doi.org/10.1103/PhysRevD.30.1052}
  {\path{doi:10.1103/PhysRevD.30.1052}}.

\bibitem{Das:2012ii}
S.~P. Das, F.~F. Deppisch, O.~Kittel, J.~W.~F. Valle, {Heavy Neutrinos and
  Lepton Flavour Violation in Left-Right Symmetric Models at the LHC}, Phys.
  Rev. D86 (2012) 055006.
\newblock \href {http://arxiv.org/abs/1206.0256} {\path{arXiv:1206.0256}},
  \href {http://dx.doi.org/10.1103/PhysRevD.86.055006}
  {\path{doi:10.1103/PhysRevD.86.055006}}.

\bibitem{AguilarSaavedra:2012fu}
J.~A. Aguilar-Saavedra, F.~Deppisch, O.~Kittel, J.~W.~F. Valle, {Flavour in
  heavy neutrino searches at the LHC}, Phys. Rev. D85 (2012) 091301.
\newblock \href {http://arxiv.org/abs/1203.5998} {\path{arXiv:1203.5998}},
  \href {http://dx.doi.org/10.1103/PhysRevD.85.091301}
  {\path{doi:10.1103/PhysRevD.85.091301}}.

\bibitem{Nemevsek:2011hz}
M.~Nemevsek, F.~Nesti, G.~Senjanovic, Y.~Zhang, {First Limits on Left-Right
  Symmetry Scale from LHC Data}, Phys. Rev. D83 (2011) 115014.
\newblock \href {http://arxiv.org/abs/1103.1627} {\path{arXiv:1103.1627}},
  \href {http://dx.doi.org/10.1103/PhysRevD.83.115014}
  {\path{doi:10.1103/PhysRevD.83.115014}}.

\bibitem{Helo:2013ika}
J.~C. Helo, M.~Hirsch, H.~Päs, S.~G. Kovalenko, {Short-range mechanisms of
  neutrinoless double beta decay at the LHC}, Phys. Rev. D88 (2013) 073011.
\newblock \href {http://arxiv.org/abs/1307.4849} {\path{arXiv:1307.4849}},
  \href {http://dx.doi.org/10.1103/PhysRevD.88.073011}
  {\path{doi:10.1103/PhysRevD.88.073011}}.

\bibitem{Fowlie:2014mza}
A.~Fowlie, L.~Marzola, {Testing quark mixing in minimal left–right symmetric
  models with b -tags at the LHC}, Nucl. Phys. B889 (2014) 36--45.
\newblock \href {http://arxiv.org/abs/1408.6699} {\path{arXiv:1408.6699}},
  \href {http://dx.doi.org/10.1016/j.nuclphysb.2014.10.009}
  {\path{doi:10.1016/j.nuclphysb.2014.10.009}}.

\bibitem{Aad:2014cka}
G.~Aad, et~al., {Search for high-mass dilepton resonances in pp collisions at
  $\sqrt{s}=8$  TeV with the ATLAS detector}, Phys. Rev. D90~(5) (2014)
  052005.
\newblock \href {http://arxiv.org/abs/1405.4123} {\path{arXiv:1405.4123}},
  \href {http://dx.doi.org/10.1103/PhysRevD.90.052005}
  {\path{doi:10.1103/PhysRevD.90.052005}}.

\bibitem{Parida:2014dla}
M.~K. Parida, B.~Sahoo, {Planck-scale induced left–right gauge theory at LHC
  and experimental tests}, Nucl. Phys. B906 (2016) 77--104.
\newblock \href {http://arxiv.org/abs/1411.6748} {\path{arXiv:1411.6748}},
  \href {http://dx.doi.org/10.1016/j.nuclphysb.2016.02.034}
  {\path{doi:10.1016/j.nuclphysb.2016.02.034}}.

\bibitem{Gao:2015irw}
Y.~Gao, T.~Ghosh, K.~Sinha, J.-H. Yu, {SU(2)×SU(2)×U(1) interpretations of
  the diboson and Wh excesses}, Phys. Rev. D92~(5) (2015) 055030.
\newblock \href {http://arxiv.org/abs/1506.07511} {\path{arXiv:1506.07511}},
  \href {http://dx.doi.org/10.1103/PhysRevD.92.055030}
  {\path{doi:10.1103/PhysRevD.92.055030}}.

\bibitem{Deppisch:2015cua}
F.~F. Deppisch, L.~Graf, S.~Kulkarni, S.~Patra, W.~Rodejohann, N.~Sahu,
  U.~Sarkar, {Reconciling the 2 TeV excesses at the LHC in a linear seesaw
  left-right model}, Phys. Rev. D93~(1) (2016) 013011.
\newblock \href {http://arxiv.org/abs/1508.05940} {\path{arXiv:1508.05940}},
  \href {http://dx.doi.org/10.1103/PhysRevD.93.013011}
  {\path{doi:10.1103/PhysRevD.93.013011}}.

\bibitem{Patra:2015bga}
S.~Patra, F.~S. Queiroz, W.~Rodejohann, {Stringent Dilepton Bounds on
  Left-Right Models using LHC data}, Phys. Lett. B752 (2016) 186--190.
\newblock \href {http://arxiv.org/abs/1506.03456} {\path{arXiv:1506.03456}},
  \href {http://dx.doi.org/10.1016/j.physletb.2015.11.009}
  {\path{doi:10.1016/j.physletb.2015.11.009}}.

\bibitem{Helo:2015ffa}
J.~C. Helo, M.~Hirsch, {LHC dijet constraints on double beta decay}, Phys. Rev.
  D92~(7) (2015) 073017.
\newblock \href {http://arxiv.org/abs/1509.00423} {\path{arXiv:1509.00423}},
  \href {http://dx.doi.org/10.1103/PhysRevD.92.073017}
  {\path{doi:10.1103/PhysRevD.92.073017}}.

\bibitem{Gluza:2015goa}
J.~Gluza, T.~Jeliński, {Heavy neutrinos and the pp→lljj CMS data}, Phys.
  Lett. B748 (2015) 125--131.
\newblock \href {http://arxiv.org/abs/1504.05568} {\path{arXiv:1504.05568}},
  \href {http://dx.doi.org/10.1016/j.physletb.2015.06.077}
  {\path{doi:10.1016/j.physletb.2015.06.077}}.

\bibitem{Brehmer:2015cia}
J.~Brehmer, J.~Hewett, J.~Kopp, T.~Rizzo, J.~Tattersall, {Symmetry Restored in
  Dibosons at the LHC?}, JHEP 10 (2015) 182.
\newblock \href {http://arxiv.org/abs/1507.00013} {\path{arXiv:1507.00013}},
  \href {http://dx.doi.org/10.1007/JHEP10(2015)182}
  {\path{doi:10.1007/JHEP10(2015)182}}.

\bibitem{Lindner:2016lpp}
M.~Lindner, F.~S. Queiroz, W.~Rodejohann, {Dilepton bounds on left–right
  symmetry at the LHC run II and neutrinoless double beta decay}, Phys. Lett.
  B762 (2016) 190--195.
\newblock \href {http://arxiv.org/abs/1604.07419} {\path{arXiv:1604.07419}},
  \href {http://dx.doi.org/10.1016/j.physletb.2016.08.068}
  {\path{doi:10.1016/j.physletb.2016.08.068}}.

\bibitem{Dev:2013oxa}
C.-H. Lee, P.~S. Bhupal~Dev, R.~N. Mohapatra, {Natural TeV-scale left-right
  seesaw mechanism for neutrinos and experimental tests}, Phys. Rev. D88~(9)
  (2013) 093010.
\newblock \href {http://arxiv.org/abs/1309.0774} {\path{arXiv:1309.0774}},
  \href {http://dx.doi.org/10.1103/PhysRevD.88.093010}
  {\path{doi:10.1103/PhysRevD.88.093010}}.

\bibitem{Gluza:2016qqv}
J.~Gluza, T.~Jelinski, R.~Szafron, {Lepton number violation and ‘Diracness’
  of massive neutrinos composed of Majorana states}, Phys. Rev. D93~(11) (2016)
  113017.
\newblock \href {http://arxiv.org/abs/1604.01388} {\path{arXiv:1604.01388}},
  \href {http://dx.doi.org/10.1103/PhysRevD.93.113017}
  {\path{doi:10.1103/PhysRevD.93.113017}}.

\bibitem{Mondal:2016kof}
S.~Mondal, S.~K. Rai, {Probing the Heavy Neutrinos of Inverse Seesaw Model at
  the LHeC}, Phys. Rev. D94~(3) (2016) 033008.
\newblock \href {http://arxiv.org/abs/1605.04508} {\path{arXiv:1605.04508}},
  \href {http://dx.doi.org/10.1103/PhysRevD.94.033008}
  {\path{doi:10.1103/PhysRevD.94.033008}}.

\bibitem{Queiroz:2016qmc}
F.~S. Queiroz, {Comment on “Polarized window for left-right symmetry and a
  right-handed neutrino at the Large Hadron-Electron Collider”}, Phys. Rev.
  D93~(11) (2016) 118701.
\newblock \href {http://dx.doi.org/10.1103/PhysRevD.93.118701}
  {\path{doi:10.1103/PhysRevD.93.118701}}.

\bibitem{Langacker:1980js}
P.~Langacker, {Grand Unified Theories and Proton Decay}, Phys. Rept. 72 (1981)
  185.
\newblock \href {http://dx.doi.org/10.1016/0370-1573(81)90059-4}
  {\path{doi:10.1016/0370-1573(81)90059-4}}.

\bibitem{Lee:1973iz}
T.~D. Lee, {A Theory of Spontaneous T Violation}, Phys. Rev. D8 (1973)
  1226--1239, [,516(1973)].
\newblock \href {http://dx.doi.org/10.1103/PhysRevD.8.1226}
  {\path{doi:10.1103/PhysRevD.8.1226}}.

\bibitem{Peccei:1977hh}
R.~D. Peccei, H.~R. Quinn, {CP Conservation in the Presence of Instantons},
  Phys. Rev. Lett. 38 (1977) 1440--1443.
\newblock \href {http://dx.doi.org/10.1103/PhysRevLett.38.1440}
  {\path{doi:10.1103/PhysRevLett.38.1440}}.

\bibitem{Kim:1986ax}
J.~E. Kim, {Light Pseudoscalars, Particle Physics and Cosmology}, Phys. Rept.
  150 (1987) 1--177.
\newblock \href {http://dx.doi.org/10.1016/0370-1573(87)90017-2}
  {\path{doi:10.1016/0370-1573(87)90017-2}}.

\bibitem{Dasgupta:2013cwa}
B.~Dasgupta, E.~Ma, K.~Tsumura, {Weakly interacting massive particle dark
  matter and radiative neutrino mass from Peccei-Quinn symmetry}, Phys. Rev.
  D89~(4) (2014) 041702.
\newblock \href {http://arxiv.org/abs/1308.4138} {\path{arXiv:1308.4138}},
  \href {http://dx.doi.org/10.1103/PhysRevD.89.041702}
  {\path{doi:10.1103/PhysRevD.89.041702}}.

\bibitem{Alves:2016bib}
A.~Alves, D.~A. Camargo, A.~G. Dias, R.~Longas, C.~C. Nishi, F.~S. Queiroz,
  {Collider and Dark Matter Searches in the Inert Doublet Model from
  Peccei-Quinn Symmetry}, JHEP 10 (2016) 015.
\newblock \href {http://arxiv.org/abs/1606.07086} {\path{arXiv:1606.07086}},
  \href {http://dx.doi.org/10.1007/JHEP10(2016)015}
  {\path{doi:10.1007/JHEP10(2016)015}}.

\bibitem{Trodden:1998qg}
M.~Trodden,
  \href{http://alice.cern.ch/format/showfull?sysnb=0278753}{{Electroweak
  baryogenesis: A Brief review}}, in: {Proceedings, 33rd Rencontres de Moriond
  98 electrowek interactions and unified theories: Les racs, France, Mar 14-21,
  1998}, 1998, pp. 471--480.
\newblock \href {http://arxiv.org/abs/hep-ph/9805252}
  {\path{arXiv:hep-ph/9805252}}.
\newline\urlprefix\url{http://alice.cern.ch/format/showfull?sysnb=0278753}

\bibitem{Joyce:1994zt}
M.~Joyce, T.~Prokopec, N.~Turok, {Nonlocal electroweak baryogenesis. Part 2:
  The Classical regime}, Phys. Rev. D53 (1996) 2958--2980.
\newblock \href {http://arxiv.org/abs/hep-ph/9410282}
  {\path{arXiv:hep-ph/9410282}}, \href
  {http://dx.doi.org/10.1103/PhysRevD.53.2958}
  {\path{doi:10.1103/PhysRevD.53.2958}}.

\bibitem{Funakubo:1993jg}
K.~Funakubo, A.~Kakuto, K.~Takenaga, {The Effective potential of electroweak
  theory with two massless Higgs doublets at finite temperature}, Prog. Theor.
  Phys. 91 (1994) 341--352.
\newblock \href {http://arxiv.org/abs/hep-ph/9310267}
  {\path{arXiv:hep-ph/9310267}}, \href {http://dx.doi.org/10.1143/PTP.91.341}
  {\path{doi:10.1143/PTP.91.341}}.

\bibitem{Davies:1994id}
A.~T. Davies, C.~D. froggatt, G.~Jenkins, R.~G. Moorhouse, {Baryogenesis
  constraints on two Higgs doublet models}, Phys. Lett. B336 (1994) 464--470.
\newblock \href {http://dx.doi.org/10.1016/0370-2693(94)90559-2}
  {\path{doi:10.1016/0370-2693(94)90559-2}}.

\bibitem{Cline:1995dg}
J.~M. Cline, K.~Kainulainen, A.~P. Vischer, {Dynamics of two Higgs doublet CP
  violation and baryogenesis at the electroweak phase transition}, Phys. Rev.
  D54 (1996) 2451--2472.
\newblock \href {http://arxiv.org/abs/hep-ph/9506284}
  {\path{arXiv:hep-ph/9506284}}, \href
  {http://dx.doi.org/10.1103/PhysRevD.54.2451}
  {\path{doi:10.1103/PhysRevD.54.2451}}.

\bibitem{Laine:2000rm}
M.~Laine, K.~Rummukainen, {Two Higgs doublet dynamics at the electroweak phase
  transition: A Nonperturbative study}, Nucl. Phys. B597 (2001) 23--69.
\newblock \href {http://arxiv.org/abs/hep-lat/0009025}
  {\path{arXiv:hep-lat/0009025}}, \href
  {http://dx.doi.org/10.1016/S0550-3213(00)00736-7}
  {\path{doi:10.1016/S0550-3213(00)00736-7}}.

\bibitem{Fromme:2006cm}
L.~Fromme, S.~J. Huber, M.~Seniuch, {Baryogenesis in the two-Higgs doublet
  model}, JHEP 11 (2006) 038.
\newblock \href {http://arxiv.org/abs/hep-ph/0605242}
  {\path{arXiv:hep-ph/0605242}}, \href
  {http://dx.doi.org/10.1088/1126-6708/2006/11/038}
  {\path{doi:10.1088/1126-6708/2006/11/038}}.

\bibitem{Kozhushko:2011ea}
A.~Kozhushko, V.~Skalozub, {The Parametric Space of the Two-Higgs-Doublet Model
  and Sakharov's Baryogenesis Conditions}, Ukr. J. Phys. 56 (2011) 431--442.
\newblock \href {http://arxiv.org/abs/1106.0790} {\path{arXiv:1106.0790}}.

\bibitem{Cline:2011mm}
J.~M. Cline, K.~Kainulainen, M.~Trott, {Electroweak Baryogenesis in Two Higgs
  Doublet Models and B meson anomalies}, JHEP 11 (2011) 089.
\newblock \href {http://arxiv.org/abs/1107.3559} {\path{arXiv:1107.3559}},
  \href {http://dx.doi.org/10.1007/JHEP11(2011)089}
  {\path{doi:10.1007/JHEP11(2011)089}}.

\bibitem{Tranberg:2012qu}
A.~Tranberg, B.~Wu, {On using Cold Baryogenesis to constrain the Two-Higgs
  Doublet Model}, JHEP 01 (2013) 046.
\newblock \href {http://arxiv.org/abs/1210.1779} {\path{arXiv:1210.1779}},
  \href {http://dx.doi.org/10.1007/JHEP01(2013)046}
  {\path{doi:10.1007/JHEP01(2013)046}}.

\bibitem{Ahmadvand:2013sna}
M.~Ahmadvand, {Baryogenesis within the two-Higgs-doublet model in the
  Electroweak scale}, Int. J. Mod. Phys. A29~(20) (2014) 1450090.
\newblock \href {http://arxiv.org/abs/1308.3767} {\path{arXiv:1308.3767}},
  \href {http://dx.doi.org/10.1142/S0217751X14500900}
  {\path{doi:10.1142/S0217751X14500900}}.

\bibitem{Mou:2015aia}
Z.-G. Mou, P.~M. Saffin, A.~Tranberg, {Cold Baryogenesis from first principles
  in the Two-Higgs Doublet model with Fermions}, JHEP 06 (2015) 163.
\newblock \href {http://arxiv.org/abs/1505.02692} {\path{arXiv:1505.02692}},
  \href {http://dx.doi.org/10.1007/JHEP06(2015)163}
  {\path{doi:10.1007/JHEP06(2015)163}}.

\bibitem{Bai:2012nv}
Y.~Bai, V.~Barger, L.~L. Everett, G.~Shaughnessy, {Two-Higgs-doublet-portal
  dark-matter model: LHC data and Fermi-LAT 135 GeV line}, Phys. Rev. D88
  (2013) 015008.
\newblock \href {http://arxiv.org/abs/1212.5604} {\path{arXiv:1212.5604}},
  \href {http://dx.doi.org/10.1103/PhysRevD.88.015008}
  {\path{doi:10.1103/PhysRevD.88.015008}}.

\bibitem{Drozd:2014yla}
A.~Drozd, B.~Grzadkowski, J.~F. Gunion, Y.~Jiang, {Extending two-Higgs-doublet
  models by a singlet scalar field - the Case for Dark Matter}, JHEP 11 (2014)
  105.
\newblock \href {http://arxiv.org/abs/1408.2106} {\path{arXiv:1408.2106}},
  \href {http://dx.doi.org/10.1007/JHEP11(2014)105}
  {\path{doi:10.1007/JHEP11(2014)105}}.

\bibitem{Wang:2014elb}
L.~Wang, X.-F. Han, {A simplified 2HDM with a scalar dark matter and the
  galactic center gamma-ray excess}, Phys. Lett. B739 (2014) 416--420.
\newblock \href {http://arxiv.org/abs/1406.3598} {\path{arXiv:1406.3598}},
  \href {http://dx.doi.org/10.1016/j.physletb.2014.11.016}
  {\path{doi:10.1016/j.physletb.2014.11.016}}.

\bibitem{Adulpravitchai:2015mna}
A.~Adulpravitchai, M.~A. Schmidt, {Sterile Neutrino Dark Matter Production in
  the Neutrino-phillic Two Higgs Doublet Model}, JHEP 12 (2015) 023.
\newblock \href {http://arxiv.org/abs/1507.05694} {\path{arXiv:1507.05694}},
  \href {http://dx.doi.org/10.1007/JHEP12(2015)023}
  {\path{doi:10.1007/JHEP12(2015)023}}.

\bibitem{Liu:2015oaa}
X.~Liu, L.~Bian, X.-Q. Li, J.~Shu, {Type-III two Higgs doublet model plus a
  pseudoscalar confronted with $h\rightarrow\mu\tau$, muon $g-2$ and dark
  matter}, Nucl. Phys. B909 (2016) 507--524.
\newblock \href {http://arxiv.org/abs/1508.05716} {\path{arXiv:1508.05716}},
  \href {http://dx.doi.org/10.1016/j.nuclphysb.2016.05.027}
  {\path{doi:10.1016/j.nuclphysb.2016.05.027}}.

\bibitem{Guo:2016ixx}
H.-K. Guo, Y.-Y. Li, T.~Liu, M.~Ramsey-Musolf, J.~Shu, {Lepton-Flavored
  Electroweak Baryogenesis}\href {http://arxiv.org/abs/1609.09849}
  {\path{arXiv:1609.09849}}.

\bibitem{Chiang:2016vgf}
C.-W. Chiang, K.~Fuyuto, E.~Senaha, {Electroweak Baryogenesis with Lepton
  Flavor Violation}, Phys. Lett. B762 (2016) 315--320.
\newblock \href {http://arxiv.org/abs/1607.07316} {\path{arXiv:1607.07316}},
  \href {http://dx.doi.org/10.1016/j.physletb.2016.09.052}
  {\path{doi:10.1016/j.physletb.2016.09.052}}.

\bibitem{Djouadi:2005gj}
A.~Djouadi, {The Anatomy of electro-weak symmetry breaking. II. The Higgs
  bosons in the minimal supersymmetric model}, Phys. Rept. 459 (2008) 1--241.
\newblock \href {http://arxiv.org/abs/hep-ph/0503173}
  {\path{arXiv:hep-ph/0503173}}, \href
  {http://dx.doi.org/10.1016/j.physrep.2007.10.005}
  {\path{doi:10.1016/j.physrep.2007.10.005}}.

\bibitem{Davidson:2010xv}
S.~Davidson, G.~J. Grenier, {Lepton flavour violating Higgs and tau to mu
  gamma}, Phys. Rev. D81 (2010) 095016.
\newblock \href {http://arxiv.org/abs/1001.0434} {\path{arXiv:1001.0434}},
  \href {http://dx.doi.org/10.1103/PhysRevD.81.095016}
  {\path{doi:10.1103/PhysRevD.81.095016}}.

\bibitem{Sierra:2014nqa}
D.~Aristizabal~Sierra, A.~Vicente, {Explaining the CMS Higgs flavor violating
  decay excess}, Phys. Rev. D90~(11) (2014) 115004.
\newblock \href {http://arxiv.org/abs/1409.7690} {\path{arXiv:1409.7690}},
  \href {http://dx.doi.org/10.1103/PhysRevD.90.115004}
  {\path{doi:10.1103/PhysRevD.90.115004}}.

\bibitem{Dorsner:2015mja}
I.~Dorsner, S.~Fajfer, A.~Greljo, J.~F. Kamenik, N.~Kosnik, I.~Nisandzic, {New
  Physics Models Facing Lepton Flavor Violating Higgs Decays at the Percent
  Level}, JHEP 06 (2015) 108.
\newblock \href {http://arxiv.org/abs/1502.07784} {\path{arXiv:1502.07784}},
  \href {http://dx.doi.org/10.1007/JHEP06(2015)108}
  {\path{doi:10.1007/JHEP06(2015)108}}.

\bibitem{Davidson:2005cw}
S.~Davidson, H.~E. Haber, {Basis-independent methods for the two-Higgs-doublet
  model}, Phys. Rev. D72 (2005) 035004, [Erratum: Phys. Rev.D72,099902(2005)].
\newblock \href {http://arxiv.org/abs/hep-ph/0504050}
  {\path{arXiv:hep-ph/0504050}}, \href
  {http://dx.doi.org/10.1103/PhysRevD.72.099902, 10.1103/PhysRevD.72.035004}
  {\path{doi:10.1103/PhysRevD.72.099902, 10.1103/PhysRevD.72.035004}}.

\bibitem{Haber:2015pua}
H.~E. Haber, O.~Stål, {New LHC benchmarks for the $\mathcal{CP}$ -conserving
  two-Higgs-doublet model}, Eur. Phys. J. C75~(10) (2015) 491, [Erratum: Eur.
  Phys. J.C76,no.6,312(2016)].
\newblock \href {http://arxiv.org/abs/1507.04281} {\path{arXiv:1507.04281}},
  \href {http://dx.doi.org/10.1140/epjc/s10052-015-3697-x,
  10.1140/epjc/s10052-016-4151-4} {\path{doi:10.1140/epjc/s10052-015-3697-x,
  10.1140/epjc/s10052-016-4151-4}}.

\bibitem{Diaz:2000cm}
R.~Diaz, R.~Martinez, J.~A. Rodriguez, {Lepton flavor violation in the two
  Higgs doublet model type III}, Phys. Rev. D63 (2001) 095007.
\newblock \href {http://arxiv.org/abs/hep-ph/0010149}
  {\path{arXiv:hep-ph/0010149}}, \href
  {http://dx.doi.org/10.1103/PhysRevD.63.095007}
  {\path{doi:10.1103/PhysRevD.63.095007}}.

\bibitem{Diaz:2002uk}
R.~A. Diaz, R.~Martinez, J.~A. Rodriguez, {Phenomenology of lepton flavor
  violation in 2HDM(3) from (g-2)(mu) and leptonic decays}, Phys. Rev. D67
  (2003) 075011.
\newblock \href {http://arxiv.org/abs/hep-ph/0208117}
  {\path{arXiv:hep-ph/0208117}}, \href
  {http://dx.doi.org/10.1103/PhysRevD.67.075011}
  {\path{doi:10.1103/PhysRevD.67.075011}}.

\bibitem{Bjorken:1977vt}
J.~D. Bjorken, S.~Weinberg, {A Mechanism for Nonconservation of Muon Number},
  Phys. Rev. Lett. 38 (1977) 622.
\newblock \href {http://dx.doi.org/10.1103/PhysRevLett.38.622}
  {\path{doi:10.1103/PhysRevLett.38.622}}.

\bibitem{Langacker:1988cm}
P.~Langacker, D.~London, {Analysis of Muon Decay With Lepton Number
  Nonconserving Interactions}, Phys. Rev. D39 (1989) 266.
\newblock \href {http://dx.doi.org/10.1103/PhysRevD.39.266}
  {\path{doi:10.1103/PhysRevD.39.266}}.

\bibitem{Leigh:1990kf}
R.~G. Leigh, S.~Paban, R.~M. Xu, {Electric dipole moment of electron}, Nucl.
  Phys. B352 (1991) 45--58.
\newblock \href {http://dx.doi.org/10.1016/0550-3213(91)90128-K}
  {\path{doi:10.1016/0550-3213(91)90128-K}}.

\bibitem{Broggio:2014mna}
A.~Broggio, E.~J. Chun, M.~Passera, K.~M. Patel, S.~K. Vempati, {Limiting
  two-Higgs-doublet models}, JHEP 11 (2014) 058.
\newblock \href {http://arxiv.org/abs/1409.3199} {\path{arXiv:1409.3199}},
  \href {http://dx.doi.org/10.1007/JHEP11(2014)058}
  {\path{doi:10.1007/JHEP11(2014)058}}.

\bibitem{Han:2015yys}
T.~Han, S.~K. Kang, J.~Sayre, {Muon $g-2$ in the aligned two Higgs doublet
  model}, JHEP 02 (2016) 097.
\newblock \href {http://arxiv.org/abs/1511.05162} {\path{arXiv:1511.05162}},
  \href {http://dx.doi.org/10.1007/JHEP02(2016)097}
  {\path{doi:10.1007/JHEP02(2016)097}}.

\bibitem{Cherchiglia:2016eui}
A.~Cherchiglia, P.~Kneschke, D.~Stöckinger, H.~Stöckinger-Kim, {The muon
  magnetic moment in the 2HDM: complete two-loop result}, JHEP 01 (2017) 007.
\newblock \href {http://arxiv.org/abs/1607.06292} {\path{arXiv:1607.06292}},
  \href {http://dx.doi.org/10.1007/JHEP01(2017)007}
  {\path{doi:10.1007/JHEP01(2017)007}}.

\bibitem{Ferreira:2009wh}
P.~M. Ferreira, H.~E. Haber, J.~P. Silva, {Generalized CP symmetries and
  special regions of parameter space in the two-Higgs-doublet model}, Phys.
  Rev. D79 (2009) 116004.
\newblock \href {http://arxiv.org/abs/0902.1537} {\path{arXiv:0902.1537}},
  \href {http://dx.doi.org/10.1103/PhysRevD.79.116004}
  {\path{doi:10.1103/PhysRevD.79.116004}}.

\bibitem{Haber:2010bw}
H.~E. Haber, D.~O'Neil, {Basis-independent methods for the two-Higgs-doublet
  model III: The CP-conserving limit, custodial symmetry, and the oblique
  parameters S, T, U}, Phys. Rev. D83 (2011) 055017.
\newblock \href {http://arxiv.org/abs/1011.6188} {\path{arXiv:1011.6188}},
  \href {http://dx.doi.org/10.1103/PhysRevD.83.055017}
  {\path{doi:10.1103/PhysRevD.83.055017}}.

\bibitem{Belanger:2012tt}
G.~Belanger, U.~Ellwanger, J.~F. Gunion, Y.~Jiang, S.~Kraml, J.~H. Schwarz,
  {Higgs Bosons at 98 and 125 GeV at LEP and the LHC}, JHEP 01 (2013) 069.
\newblock \href {http://arxiv.org/abs/1210.1976} {\path{arXiv:1210.1976}},
  \href {http://dx.doi.org/10.1007/JHEP01(2013)069}
  {\path{doi:10.1007/JHEP01(2013)069}}.

\bibitem{Dumont:2014wha}
B.~Dumont, J.~F. Gunion, Y.~Jiang, S.~Kraml, {Constraints on and future
  prospects for Two-Higgs-Doublet Models in light of the LHC Higgs signal},
  Phys. Rev. D90 (2014) 035021.
\newblock \href {http://arxiv.org/abs/1405.3584} {\path{arXiv:1405.3584}},
  \href {http://dx.doi.org/10.1103/PhysRevD.90.035021}
  {\path{doi:10.1103/PhysRevD.90.035021}}.

\bibitem{Ferreira:2015rha}
P.~Ferreira, H.~E. Haber, E.~Santos, {Preserving the validity of the Two-Higgs
  Doublet Model up to the Planck scale}, Phys. Rev. D92 (2015) 033003,
  [Erratum: Phys. Rev.D94,no.5,059903(2016)].
\newblock \href {http://arxiv.org/abs/1505.04001} {\path{arXiv:1505.04001}},
  \href {http://dx.doi.org/10.1103/PhysRevD.92.033003,
  10.1103/PhysRevD.94.059903} {\path{doi:10.1103/PhysRevD.92.033003,
  10.1103/PhysRevD.94.059903}}.

\bibitem{Bernon:2015wef}
J.~Bernon, J.~F. Gunion, H.~E. Haber, Y.~Jiang, S.~Kraml, {Scrutinizing the
  alignment limit in two-Higgs-doublet models. II. $m_H=125$  GeV}, Phys.
  Rev. D93~(3) (2016) 035027.
\newblock \href {http://arxiv.org/abs/1511.03682} {\path{arXiv:1511.03682}},
  \href {http://dx.doi.org/10.1103/PhysRevD.93.035027}
  {\path{doi:10.1103/PhysRevD.93.035027}}.

\bibitem{Cho:2017jym}
N.~Cho, X.-q. Li, F.~Su, X.~Zhang, {$K^0-\bar{K}^0$ mixing in the minimal
  flavor-violating two-Higgs-doublet models}, Adv. High Energy Physics 2017.
\newblock \href {http://arxiv.org/abs/1705.07638} {\path{arXiv:1705.07638}},
  \href {http://dx.doi.org/10.1155/2017/2863647}
  {\path{doi:10.1155/2017/2863647}}.

\bibitem{Primulando:2016eod}
R.~Primulando, P.~Uttayarat, {Probing Lepton Flavor Violation at the 13 TeV
  LHC}, JHEP 05 (2017) 055.
\newblock \href {http://arxiv.org/abs/1612.01644} {\path{arXiv:1612.01644}},
  \href {http://dx.doi.org/10.1007/JHEP05(2017)055}
  {\path{doi:10.1007/JHEP05(2017)055}}.

\bibitem{Ma:2006km}
E.~Ma, {Verifiable radiative seesaw mechanism of neutrino mass and dark
  matter}, Phys. Rev. D73 (2006) 077301.
\newblock \href {http://arxiv.org/abs/hep-ph/0601225}
  {\path{arXiv:hep-ph/0601225}}, \href
  {http://dx.doi.org/10.1103/PhysRevD.73.077301}
  {\path{doi:10.1103/PhysRevD.73.077301}}.

\bibitem{Ma:2012if}
E.~Ma, {Radiative Scaling Neutrino Mass and Warm Dark Matter}, Phys. Lett. B717
  (2012) 235--237.
\newblock \href {http://arxiv.org/abs/1206.1812} {\path{arXiv:1206.1812}},
  \href {http://dx.doi.org/10.1016/j.physletb.2012.09.046}
  {\path{doi:10.1016/j.physletb.2012.09.046}}.

\bibitem{Kashiwase:2013uy}
S.~Kashiwase, D.~Suematsu, {Leptogenesis and dark matter detection in a TeV
  scale neutrino mass model with inverted mass hierarchy}, Eur. Phys. J. C73
  (2013) 2484.
\newblock \href {http://arxiv.org/abs/1301.2087} {\path{arXiv:1301.2087}},
  \href {http://dx.doi.org/10.1140/epjc/s10052-013-2484-9}
  {\path{doi:10.1140/epjc/s10052-013-2484-9}}.

\bibitem{Klasen:2013jpa}
M.~Klasen, C.~E. Yaguna, J.~D. Ruiz-Alvarez, D.~Restrepo, O.~Zapata, {Scalar
  dark matter and fermion coannihilations in the radiative seesaw model}, JCAP
  1304 (2013) 044.
\newblock \href {http://arxiv.org/abs/1302.5298} {\path{arXiv:1302.5298}},
  \href {http://dx.doi.org/10.1088/1475-7516/2013/04/044}
  {\path{doi:10.1088/1475-7516/2013/04/044}}.

\bibitem{Toma:2013zsa}
T.~Toma, A.~Vicente, {Lepton Flavor Violation in the Scotogenic Model}, JHEP 01
  (2014) 160.
\newblock \href {http://arxiv.org/abs/1312.2840} {\path{arXiv:1312.2840}},
  \href {http://dx.doi.org/10.1007/JHEP01(2014)160}
  {\path{doi:10.1007/JHEP01(2014)160}}.

\bibitem{Vicente:2014wga}
A.~Vicente, C.~E. Yaguna, {Probing the scotogenic model with lepton flavor
  violating processes}, JHEP 02 (2015) 144.
\newblock \href {http://arxiv.org/abs/1412.2545} {\path{arXiv:1412.2545}},
  \href {http://dx.doi.org/10.1007/JHEP02(2015)144}
  {\path{doi:10.1007/JHEP02(2015)144}}.

\bibitem{Lindner:2016kqk}
M.~Lindner, M.~Platscher, C.~E. Yaguna, A.~Merle, {Fermionic WIMPs and vacuum
  stability in the scotogenic model}, Phys. Rev. D94~(11) (2016) 115027.
\newblock \href {http://arxiv.org/abs/1608.00577} {\path{arXiv:1608.00577}},
  \href {http://dx.doi.org/10.1103/PhysRevD.94.115027}
  {\path{doi:10.1103/PhysRevD.94.115027}}.

\bibitem{Merle:2015ica}
A.~Merle, M.~Platscher, {Running of radiative neutrino masses: the scotogenic
  model — revisited}, JHEP 11 (2015) 148.
\newblock \href {http://arxiv.org/abs/1507.06314} {\path{arXiv:1507.06314}},
  \href {http://dx.doi.org/10.1007/JHEP11(2015)148}
  {\path{doi:10.1007/JHEP11(2015)148}}.

\bibitem{Ma:2001mr}
E.~Ma, M.~Raidal, {Neutrino mass, muon anomalous magnetic moment, and lepton
  flavor nonconservation}, Phys. Rev. Lett. 87 (2001) 011802, [Erratum: Phys.
  Rev. Lett.87,159901(2001)].
\newblock \href {http://arxiv.org/abs/hep-ph/0102255}
  {\path{arXiv:hep-ph/0102255}}, \href
  {http://dx.doi.org/10.1103/PhysRevLett.87.159901,
  10.1103/PhysRevLett.87.011802} {\path{doi:10.1103/PhysRevLett.87.159901,
  10.1103/PhysRevLett.87.011802}}.

\bibitem{Arganda:2007jw}
E.~Arganda, M.~J. Herrero, A.~M. Teixeira, {mu-e conversion in nuclei within
  the CMSSM seesaw: Universality versus non-universality}, JHEP 10 (2007) 104.
\newblock \href {http://arxiv.org/abs/0707.2955} {\path{arXiv:0707.2955}},
  \href {http://dx.doi.org/10.1088/1126-6708/2007/10/104}
  {\path{doi:10.1088/1126-6708/2007/10/104}}.

\bibitem{Casas:2001sr}
J.~A. Casas, A.~Ibarra, {Oscillating neutrinos and muon ---> e, gamma}, Nucl.
  Phys. B618 (2001) 171--204.
\newblock \href {http://arxiv.org/abs/hep-ph/0103065}
  {\path{arXiv:hep-ph/0103065}}, \href
  {http://dx.doi.org/10.1016/S0550-3213(01)00475-8}
  {\path{doi:10.1016/S0550-3213(01)00475-8}}.

\bibitem{Zee:1985rj}
A.~Zee, {Charged Scalar Field and Quantum Number Violations}, Phys. Lett. B161
  (1985) 141--145.
\newblock \href {http://dx.doi.org/10.1016/0370-2693(85)90625-2}
  {\path{doi:10.1016/0370-2693(85)90625-2}}.

\bibitem{Zee:1985id}
A.~Zee, {Quantum Numbers of Majorana Neutrino Masses}, Nucl. Phys. B264 (1986)
  99--110.
\newblock \href {http://dx.doi.org/10.1016/0550-3213(86)90475-X}
  {\path{doi:10.1016/0550-3213(86)90475-X}}.

\bibitem{Babu:1988ki}
K.~S. Babu, {Model of 'Calculable' Majorana Neutrino Masses}, Phys. Lett. B203
  (1988) 132--136.
\newblock \href {http://dx.doi.org/10.1016/0370-2693(88)91584-5}
  {\path{doi:10.1016/0370-2693(88)91584-5}}.

\bibitem{McDonald:2003zj}
K.~L. McDonald, B.~H.~J. McKellar, {Evaluating the two loop diagram responsible
  for neutrino mass in Babu's model}\href {http://arxiv.org/abs/hep-ph/0309270}
  {\path{arXiv:hep-ph/0309270}}.

\bibitem{Babu:2002uu}
K.~S. Babu, C.~Macesanu, {Two loop neutrino mass generation and its
  experimental consequences}, Phys. Rev. D67 (2003) 073010.
\newblock \href {http://arxiv.org/abs/hep-ph/0212058}
  {\path{arXiv:hep-ph/0212058}}, \href
  {http://dx.doi.org/10.1103/PhysRevD.67.073010}
  {\path{doi:10.1103/PhysRevD.67.073010}}.

\bibitem{AristizabalSierra:2006gb}
D.~Aristizabal~Sierra, M.~Hirsch, {Experimental tests for the Babu-Zee two-loop
  model of Majorana neutrino masses}, JHEP 12 (2006) 052.
\newblock \href {http://arxiv.org/abs/hep-ph/0609307}
  {\path{arXiv:hep-ph/0609307}}, \href
  {http://dx.doi.org/10.1088/1126-6708/2006/12/052}
  {\path{doi:10.1088/1126-6708/2006/12/052}}.

\bibitem{Nebot:2007bc}
M.~Nebot, J.~F. Oliver, D.~Palao, A.~Santamaria, {Prospects for the Zee-Babu
  Model at the CERN LHC and low energy experiments}, Phys. Rev. D77 (2008)
  093013.
\newblock \href {http://arxiv.org/abs/0711.0483} {\path{arXiv:0711.0483}},
  \href {http://dx.doi.org/10.1103/PhysRevD.77.093013}
  {\path{doi:10.1103/PhysRevD.77.093013}}.

\bibitem{Ohlsson:2009vk}
T.~Ohlsson, T.~Schwetz, H.~Zhang, {Non-standard neutrino interactions in the
  Zee-Babu model}, Phys. Lett. B681 (2009) 269--275.
\newblock \href {http://arxiv.org/abs/0909.0455} {\path{arXiv:0909.0455}},
  \href {http://dx.doi.org/10.1016/j.physletb.2009.10.025}
  {\path{doi:10.1016/j.physletb.2009.10.025}}.

\bibitem{Araki:2010kq}
T.~Araki, C.~Q. Geng, {$\mu$-$\tau$ symmetry in Zee-Babu model}, Phys. Lett.
  B694 (2011) 113--118.
\newblock \href {http://arxiv.org/abs/1006.0629} {\path{arXiv:1006.0629}},
  \href {http://dx.doi.org/10.1016/j.physletb.2010.09.046}
  {\path{doi:10.1016/j.physletb.2010.09.046}}.

\bibitem{Schmidt:2014zoa}
D.~Schmidt, T.~Schwetz, H.~Zhang, {Status of the Zee–Babu model for neutrino
  mass and possible tests at a like-sign linear collider}, Nucl. Phys. B885
  (2014) 524--541.
\newblock \href {http://arxiv.org/abs/1402.2251} {\path{arXiv:1402.2251}},
  \href {http://dx.doi.org/10.1016/j.nuclphysb.2014.05.024}
  {\path{doi:10.1016/j.nuclphysb.2014.05.024}}.

\bibitem{Herrero-Garcia:2014hfa}
J.~Herrero-Garcia, M.~Nebot, N.~Rius, A.~Santamaria, {The Zee–Babu model
  revisited in the light of new data}, Nucl. Phys. B885 (2014) 542--570.
\newblock \href {http://arxiv.org/abs/1402.4491} {\path{arXiv:1402.4491}},
  \href {http://dx.doi.org/10.1016/j.nuclphysb.2014.06.001}
  {\path{doi:10.1016/j.nuclphysb.2014.06.001}}.

\bibitem{Okada:2014qsa}
H.~Okada, T.~Toma, K.~Yagyu, {Inert Extension of the Zee-Babu Model}, Phys.
  Rev. D90 (2014) 095005.
\newblock \href {http://arxiv.org/abs/1408.0961} {\path{arXiv:1408.0961}},
  \href {http://dx.doi.org/10.1103/PhysRevD.90.095005}
  {\path{doi:10.1103/PhysRevD.90.095005}}.

\bibitem{Herrero-Garcia:2014usa}
J.~Herrero-Garcia, M.~Nebot, N.~Rius, A.~Santamaria, {Testing the Zee-Babu
  model via neutrino data, lepton flavour violation and direct searches at the
  LHC}, in: {Proceedings, 37th International Conference on High Energy Physics
  (ICHEP 2014): Valencia, Spain, July 2-9, 2014}, 2016.
\newblock \href {http://arxiv.org/abs/1410.2299} {\path{arXiv:1410.2299}},
  \href {http://dx.doi.org/10.1016/j.nuclphysbps.2015.09.271}
  {\path{doi:10.1016/j.nuclphysbps.2015.09.271}}.

\bibitem{Nomura:2016rjf}
T.~Nomura, H.~Okada, {Generalized Zee–Babu model with 750 GeV diphoton
  resonance}, Phys. Lett. B756 (2016) 295--302.
\newblock \href {http://arxiv.org/abs/1601.07339} {\path{arXiv:1601.07339}},
  \href {http://dx.doi.org/10.1016/j.physletb.2016.03.034}
  {\path{doi:10.1016/j.physletb.2016.03.034}}.

\bibitem{Nomura:2016ask}
T.~Nomura, H.~Okada, {An Extended Colored Zee-Babu Model}, Phys. Rev. D94
  (2016) 075021.
\newblock \href {http://arxiv.org/abs/1607.04952} {\path{arXiv:1607.04952}},
  \href {http://dx.doi.org/10.1103/PhysRevD.94.075021}
  {\path{doi:10.1103/PhysRevD.94.075021}}.

\bibitem{Chang:2016zll}
W.-F. Chang, S.-C. Liou, C.-F. Wong, F.~Xu, {Charged Lepton Flavor Violating
  Processes and Scalar Leptoquark Decay Branching Ratios in the Colored
  Zee-Babu Model}, JHEP 10 (2016) 106.
\newblock \href {http://arxiv.org/abs/1608.05511} {\path{arXiv:1608.05511}},
  \href {http://dx.doi.org/10.1007/JHEP10(2016)106}
  {\path{doi:10.1007/JHEP10(2016)106}}.

\bibitem{Kitano:2002mt}
R.~Kitano, M.~Koike, Y.~Okada, {Detailed calculation of lepton flavor violating
  muon electron conversion rate for various nuclei}, Phys. Rev. D66 (2002)
  096002, [Erratum: Phys. Rev.D76,059902(2007)].
\newblock \href {http://arxiv.org/abs/hep-ph/0203110}
  {\path{arXiv:hep-ph/0203110}}, \href
  {http://dx.doi.org/10.1103/PhysRevD.76.059902, 10.1103/PhysRevD.66.096002}
  {\path{doi:10.1103/PhysRevD.76.059902, 10.1103/PhysRevD.66.096002}}.

\bibitem{Dinh:2012bp}
D.~N. Dinh, A.~Ibarra, E.~Molinaro, S.~T. Petcov, {The $\mu - e$ Conversion in
  Nuclei, $\mu \rightarrow e \gamma, \mu \rightarrow 3e$ Decays and TeV Scale
  See-Saw Scenarios of Neutrino Mass Generation}, JHEP 08 (2012) 125, [Erratum:
  JHEP09,023(2013)].
\newblock \href {http://arxiv.org/abs/1205.4671} {\path{arXiv:1205.4671}},
  \href {http://dx.doi.org/10.1007/JHEP09(2013)023, 10.1007/JHEP08(2012)125}
  {\path{doi:10.1007/JHEP09(2013)023, 10.1007/JHEP08(2012)125}}.

\bibitem{'tHooft:1976up}
G.~'t~Hooft, {Symmetry Breaking Through Bell-Jackiw Anomalies}, Phys. Rev.
  Lett. 37 (1976) 8--11.
\newblock \href {http://dx.doi.org/10.1103/PhysRevLett.37.8}
  {\path{doi:10.1103/PhysRevLett.37.8}}.

\bibitem{Klinkhamer:1984di}
F.~R. Klinkhamer, N.~S. Manton, {A Saddle Point Solution in the Weinberg-Salam
  Theory}, Phys. Rev. D30 (1984) 2212.
\newblock \href {http://dx.doi.org/10.1103/PhysRevD.30.2212}
  {\path{doi:10.1103/PhysRevD.30.2212}}.

\bibitem{Kuzmin:1985mm}
V.~A. Kuzmin, V.~A. Rubakov, M.~E. Shaposhnikov, {On the Anomalous Electroweak
  Baryon Number Nonconservation in the Early Universe}, Phys. Lett. B155 (1985)
  36.
\newblock \href {http://dx.doi.org/10.1016/0370-2693(85)91028-7}
  {\path{doi:10.1016/0370-2693(85)91028-7}}.

\bibitem{Fukugita:1986hr}
M.~Fukugita, T.~Yanagida, {Baryogenesis Without Grand Unification}, Phys. Lett.
  B174 (1986) 45--47.
\newblock \href {http://dx.doi.org/10.1016/0370-2693(86)91126-3}
  {\path{doi:10.1016/0370-2693(86)91126-3}}.

\bibitem{Arnold:1987mh}
P.~B. Arnold, L.~D. McLerran, {Sphalerons, Small Fluctuations and Baryon Number
  Violation in Electroweak Theory}, Phys. Rev. D36 (1987) 581.
\newblock \href {http://dx.doi.org/10.1103/PhysRevD.36.581}
  {\path{doi:10.1103/PhysRevD.36.581}}.

\bibitem{Alves:2015mua}
A.~Alves, A.~Berlin, S.~Profumo, F.~S. Queiroz, {Dirac-fermionic dark matter in
  U(1)$_{X}$ models}, JHEP 10 (2015) 076.
\newblock \href {http://arxiv.org/abs/1506.06767} {\path{arXiv:1506.06767}},
  \href {http://dx.doi.org/10.1007/JHEP10(2015)076}
  {\path{doi:10.1007/JHEP10(2015)076}}.

\bibitem{Rodejohann:2015lca}
W.~Rodejohann, C.~E. Yaguna, {Scalar dark matter in the B−L model}, JCAP
  1512~(12) (2015) 032.
\newblock \href {http://arxiv.org/abs/1509.04036} {\path{arXiv:1509.04036}},
  \href {http://dx.doi.org/10.1088/1475-7516/2015/12/032}
  {\path{doi:10.1088/1475-7516/2015/12/032}}.

\bibitem{Okada:2016gsh}
N.~Okada, S.~Okada, {$Z^\prime_{BL}$ portal dark matter and LHC Run-2 results},
  Phys. Rev. D93~(7) (2016) 075003.
\newblock \href {http://arxiv.org/abs/1601.07526} {\path{arXiv:1601.07526}},
  \href {http://dx.doi.org/10.1103/PhysRevD.93.075003}
  {\path{doi:10.1103/PhysRevD.93.075003}}.

\bibitem{Klasen:2016qux}
M.~Klasen, F.~Lyonnet, F.~S. Queiroz, {NLO+NLL collider bounds, Dirac fermion
  and scalar dark matter in the B–L model}, Eur. Phys. J. C77~(5) (2017) 348.
\newblock \href {http://arxiv.org/abs/1607.06468} {\path{arXiv:1607.06468}},
  \href {http://dx.doi.org/10.1140/epjc/s10052-017-4904-8}
  {\path{doi:10.1140/epjc/s10052-017-4904-8}}.

\bibitem{Batell:2014mga}
B.~Batell, R.~Essig, Z.~Surujon, {Strong Constraints on Sub-GeV Dark Sectors
  from SLAC Beam Dump E137}, Phys. Rev. Lett. 113~(17) (2014) 171802.
\newblock \href {http://arxiv.org/abs/1406.2698} {\path{arXiv:1406.2698}},
  \href {http://dx.doi.org/10.1103/PhysRevLett.113.171802}
  {\path{doi:10.1103/PhysRevLett.113.171802}}.

\bibitem{Abdallah:2011ew}
W.~Abdallah, A.~Awad, S.~Khalil, H.~Okada, {Muon Anomalous Magnetic Moment and
  $\mu \rightarrow e \gamma$ in B-L Model with Inverse Seesaw}, Eur. Phys. J.
  C72 (2012) 2108.
\newblock \href {http://arxiv.org/abs/1105.1047} {\path{arXiv:1105.1047}},
  \href {http://dx.doi.org/10.1140/epjc/s10052-012-2108-9}
  {\path{doi:10.1140/epjc/s10052-012-2108-9}}.

\bibitem{Khalil:2015wua}
S.~Khalil, C.~S. Un, {Muon Anomalous Magnetic Moment in SUSY B-L Model with
  Inverse Seesaw}, Phys. Lett. B763 (2016) 164--168.
\newblock \href {http://arxiv.org/abs/1509.05391} {\path{arXiv:1509.05391}},
  \href {http://dx.doi.org/10.1016/j.physletb.2016.10.035}
  {\path{doi:10.1016/j.physletb.2016.10.035}}.

\bibitem{'tHooft:1979bh}
G.~'t~Hooft, {Naturalness, chiral symmetry, and spontaneous chiral symmetry
  breaking}, NATO Sci. Ser. B 59 (1980) 135.

\bibitem{Dev:2009aw}
P.~S.~B. Dev, R.~N. Mohapatra, {TeV Scale Inverse Seesaw in SO(10) and Leptonic
  Non-Unitarity Effects}, Phys. Rev. D81 (2010) 013001.
\newblock \href {http://arxiv.org/abs/0910.3924} {\path{arXiv:0910.3924}},
  \href {http://dx.doi.org/10.1103/PhysRevD.81.013001}
  {\path{doi:10.1103/PhysRevD.81.013001}}.

\bibitem{Kanaya:1980cw}
K.~Kanaya, {Neutrino Mixing in the Minimal SO(10) Model}, Prog. Theor. Phys. 64
  (1980) 2278.
\newblock \href {http://dx.doi.org/10.1143/PTP.64.2278}
  {\path{doi:10.1143/PTP.64.2278}}.

\bibitem{Altarelli:2008yr}
G.~Altarelli, D.~Meloni, {CP violation in neutrino oscillations and new
  physics}, Nucl. Phys. B809 (2009) 158--182.
\newblock \href {http://arxiv.org/abs/0809.1041} {\path{arXiv:0809.1041}},
  \href {http://dx.doi.org/10.1016/j.nuclphysb.2008.09.044}
  {\path{doi:10.1016/j.nuclphysb.2008.09.044}}.

\bibitem{Khalil:2006yi}
S.~Khalil, {Low scale $B$ - L extension of the Standard Model at the LHC}, J.
  Phys. G35 (2008) 055001.
\newblock \href {http://arxiv.org/abs/hep-ph/0611205}
  {\path{arXiv:hep-ph/0611205}}, \href
  {http://dx.doi.org/10.1088/0954-3899/35/5/055001}
  {\path{doi:10.1088/0954-3899/35/5/055001}}.

\bibitem{Foot:1992rh}
R.~Foot, O.~F. Hernandez, F.~Pisano, V.~Pleitez, {Lepton masses in an SU(3)-L x
  U(1)-N gauge model}, Phys. Rev. D47 (1993) 4158--4161.
\newblock \href {http://arxiv.org/abs/hep-ph/9207264}
  {\path{arXiv:hep-ph/9207264}}, \href
  {http://dx.doi.org/10.1103/PhysRevD.47.4158}
  {\path{doi:10.1103/PhysRevD.47.4158}}.

\bibitem{Pleitez:1992xh}
V.~Pleitez, M.~D. Tonasse, {Heavy charged leptons in an SU(3)-L x U(1)-N
  model}, Phys. Rev. D48 (1993) 2353--2355.
\newblock \href {http://arxiv.org/abs/hep-ph/9301232}
  {\path{arXiv:hep-ph/9301232}}, \href
  {http://dx.doi.org/10.1103/PhysRevD.48.2353}
  {\path{doi:10.1103/PhysRevD.48.2353}}.

\bibitem{Foot:1994ym}
R.~Foot, H.~N. Long, T.~A. Tran, {$SU(3)_L \otimes U(1)_N$ and $SU(4)_L \otimes
  U(1)_N$ gauge models with right-handed neutrinos}, Phys. Rev. D50~(1) (1994)
  R34--R38.
\newblock \href {http://arxiv.org/abs/hep-ph/9402243}
  {\path{arXiv:hep-ph/9402243}}, \href
  {http://dx.doi.org/10.1103/PhysRevD.50.R34}
  {\path{doi:10.1103/PhysRevD.50.R34}}.

\bibitem{Hoang:1995vq}
H.~N. Long, {The 331 model with right handed neutrinos}, Phys. Rev. D53 (1996)
  437--445.
\newblock \href {http://arxiv.org/abs/hep-ph/9504274}
  {\path{arXiv:hep-ph/9504274}}, \href
  {http://dx.doi.org/10.1103/PhysRevD.53.437}
  {\path{doi:10.1103/PhysRevD.53.437}}.

\bibitem{Montero:2000ng}
J.~C. Montero, V.~Pleitez, M.~C. Rodriguez, {A Supersymmetric 3-3-1 model},
  Phys. Rev. D65 (2002) 035006.
\newblock \href {http://arxiv.org/abs/hep-ph/0012178}
  {\path{arXiv:hep-ph/0012178}}, \href
  {http://dx.doi.org/10.1103/PhysRevD.65.035006}
  {\path{doi:10.1103/PhysRevD.65.035006}}.

\bibitem{Dias:2003iq}
A.~G. Dias, C.~A. de~S.~Pires, P.~S. Rodrigues~da Silva, {Discrete symmetries,
  invisible axion and lepton number symmetry in an economic 3 3 1 model}, Phys.
  Rev. D68 (2003) 115009.
\newblock \href {http://arxiv.org/abs/hep-ph/0309058}
  {\path{arXiv:hep-ph/0309058}}, \href
  {http://dx.doi.org/10.1103/PhysRevD.68.115009}
  {\path{doi:10.1103/PhysRevD.68.115009}}.

\bibitem{VanDong:2005pi}
P.~V. Dong, H.~N. Long, {U(1)(Q) invariance and SU(3)(C) x SU(3)(L) x U(1)(X)
  models with beta arbitrary}, Eur. Phys. J. C42 (2005) 325--329.
\newblock \href {http://arxiv.org/abs/hep-ph/0506022}
  {\path{arXiv:hep-ph/0506022}}, \href
  {http://dx.doi.org/10.1140/epjc/s2005-02314-x}
  {\path{doi:10.1140/epjc/s2005-02314-x}}.

\bibitem{Dong:2006mg}
P.~V. Dong, H.~N. Long, D.~T. Nhung, D.~V. Soa, {SU(3)(C) x SU(3)(L) x U(1)(X)
  model with two Higgs triplets}, Phys. Rev. D73 (2006) 035004.
\newblock \href {http://arxiv.org/abs/hep-ph/0601046}
  {\path{arXiv:hep-ph/0601046}}, \href
  {http://dx.doi.org/10.1103/PhysRevD.73.035004}
  {\path{doi:10.1103/PhysRevD.73.035004}}.

\bibitem{Cogollo:2009yi}
D.~Cogollo, H.~Diniz, C.~A. de~S.~Pires, {KeV right-handed neutrinos from type
  II seesaw mechanism in a 3-3-1 model}, Phys. Lett. B677~(5) (2009) 338--342.
\newblock \href {http://arxiv.org/abs/0903.0370} {\path{arXiv:0903.0370}},
  \href {http://dx.doi.org/10.1016/j.physletb.2009.05.060}
  {\path{doi:10.1016/j.physletb.2009.05.060}}.

\bibitem{Mizukoshi:2010ky}
J.~K. Mizukoshi, C.~A. de~S.~Pires, F.~S. Queiroz, P.~S. Rodrigues~da Silva,
  {WIMPs in a 3-3-1 model with heavy Sterile neutrinos}, Phys. Rev. D83 (2011)
  065024.
\newblock \href {http://arxiv.org/abs/1010.4097} {\path{arXiv:1010.4097}},
  \href {http://dx.doi.org/10.1103/PhysRevD.83.065024}
  {\path{doi:10.1103/PhysRevD.83.065024}}.

\bibitem{Queiroz:2010rj}
F.~Queiroz, C.~A. de~S.~Pires, P.~S.~R. da~Silva, {A minimal 3-3-1 model with
  naturally sub-eV neutrinos}, Phys. Rev. D82 (2010) 065018.
\newblock \href {http://arxiv.org/abs/1003.1270} {\path{arXiv:1003.1270}},
  \href {http://dx.doi.org/10.1103/PhysRevD.82.065018}
  {\path{doi:10.1103/PhysRevD.82.065018}}.

\bibitem{Ferreira:2011hm}
J.~G. Ferreira, Jr, P.~R.~D. Pinheiro, C.~A. d.~S. Pires, P.~S.~R. da~Silva,
  {The Minimal 3-3-1 model with only two Higgs triplets}, Phys. Rev. D84 (2011)
  095019.
\newblock \href {http://arxiv.org/abs/1109.0031} {\path{arXiv:1109.0031}},
  \href {http://dx.doi.org/10.1103/PhysRevD.84.095019}
  {\path{doi:10.1103/PhysRevD.84.095019}}.

\bibitem{Dong:2011vb}
P.~V. Dong, H.~N. Long, C.~H. Nam, V.~V. Vien, {The $S_3$ flavor symmetry in
  3-3-1 models}, Phys. Rev. D85 (2012) 053001.
\newblock \href {http://arxiv.org/abs/1111.6360} {\path{arXiv:1111.6360}},
  \href {http://dx.doi.org/10.1103/PhysRevD.85.053001}
  {\path{doi:10.1103/PhysRevD.85.053001}}.

\bibitem{Huyen:2012uk}
V.~T.~N. Huyen, T.~T. Lam, H.~N. Long, V.~Q. Phong, {Neutral currents in
  reduced minimal 3-3-1 model}, Commun. Phys. 24 (2014) 97--112.
\newblock \href {http://arxiv.org/abs/1210.5833} {\path{arXiv:1210.5833}},
  \href {http://dx.doi.org/10.15625/0868-3166/24/2/3774}
  {\path{doi:10.15625/0868-3166/24/2/3774}}.

\bibitem{Huong:2012pg}
D.~T. Huong, L.~T. Hue, M.~C. Rodriguez, H.~N. Long, {Supersymmetric reduced
  minimal 3-3-1 model}, Nucl. Phys. B870 (2013) 293--322.
\newblock \href {http://arxiv.org/abs/1210.6776} {\path{arXiv:1210.6776}},
  \href {http://dx.doi.org/10.1016/j.nuclphysb.2013.01.016}
  {\path{doi:10.1016/j.nuclphysb.2013.01.016}}.

\bibitem{Alvares:2012qv}
J.~D. Ruiz-Alvarez, C.~A. de~S.~Pires, F.~S. Queiroz, D.~Restrepo, P.~S.
  Rodrigues~da Silva, {On the Connection of Gamma-Rays, Dark Matter and Higgs
  Searches at LHC}, Phys. Rev. D86 (2012) 075011.
\newblock \href {http://arxiv.org/abs/1206.5779} {\path{arXiv:1206.5779}},
  \href {http://dx.doi.org/10.1103/PhysRevD.86.075011}
  {\path{doi:10.1103/PhysRevD.86.075011}}.

\bibitem{Kelso:2013nwa}
C.~Kelso, C.~A. de~S.~Pires, S.~Profumo, F.~S. Queiroz, P.~S. Rodrigues~da
  Silva, {A 331 WIMPy Dark Radiation Model}, Eur. Phys. J. C74~(3) (2014) 2797.
\newblock \href {http://arxiv.org/abs/1308.6630} {\path{arXiv:1308.6630}},
  \href {http://dx.doi.org/10.1140/epjc/s10052-014-2797-3}
  {\path{doi:10.1140/epjc/s10052-014-2797-3}}.

\bibitem{Pires:2016vek}
C.~A. de~S.~Pires, P.~S. Rodrigues~da Silva, A.~C.~O. Santos, C.~Siqueira,
  {Higgs mass and right-handed sneutrino WIMP in a supersymmetric 3-3-1 model},
  Phys. Rev. D94~(5) (2016) 055014.
\newblock \href {http://arxiv.org/abs/1606.01853} {\path{arXiv:1606.01853}},
  \href {http://dx.doi.org/10.1103/PhysRevD.94.055014}
  {\path{doi:10.1103/PhysRevD.94.055014}}.

\bibitem{Pires:2016dqq}
C.~A. de~S.~Pires, J.~G. Rodrigues, P.~S. Rodrigues~da Silva, {Realizing the
  supersymmetric inverse seesaw model in the framework of R-parity violation},
  Phys. Lett. B759 (2016) 322--327.
\newblock \href {http://arxiv.org/abs/1602.08126} {\path{arXiv:1602.08126}},
  \href {http://dx.doi.org/10.1016/j.physletb.2016.05.089}
  {\path{doi:10.1016/j.physletb.2016.05.089}}.

\bibitem{Singer:1980sw}
M.~Singer, J.~W.~F. Valle, J.~Schechter, {Canonical Neutral Current Predictions
  From the Weak Electromagnetic Gauge Group SU(3) X $u$(1)}, Phys. Rev. D22
  (1980) 738.
\newblock \href {http://dx.doi.org/10.1103/PhysRevD.22.738}
  {\path{doi:10.1103/PhysRevD.22.738}}.

\bibitem{deS.Pires:2007gi}
C.~A. de~S.~Pires, P.~S. Rodrigues~da Silva, {Scalar Bilepton Dark Matter},
  JCAP 0712 (2007) 012.
\newblock \href {http://arxiv.org/abs/0710.2104} {\path{arXiv:0710.2104}},
  \href {http://dx.doi.org/10.1088/1475-7516/2007/12/012}
  {\path{doi:10.1088/1475-7516/2007/12/012}}.

\bibitem{Queiroz:2013lca}
F.~S. Queiroz, {Non-thermal WIMPs as Dark Radiation}, AIP Conf. Proc. 1604
  (2014) 83--90.
\newblock \href {http://arxiv.org/abs/1310.3026} {\path{arXiv:1310.3026}},
  \href {http://dx.doi.org/10.1063/1.4883415} {\path{doi:10.1063/1.4883415}}.

\bibitem{Profumo:2013sca}
S.~Profumo, F.~S. Queiroz, {Constraining the $Z'$ mass in 331 models using
  direct dark matter detection}, Eur. Phys. J. C74~(7) (2014) 2960.
\newblock \href {http://arxiv.org/abs/1307.7802} {\path{arXiv:1307.7802}},
  \href {http://dx.doi.org/10.1140/epjc/s10052-014-2960-x}
  {\path{doi:10.1140/epjc/s10052-014-2960-x}}.

\bibitem{Dong:2014esa}
P.~V. Dong, N.~T.~K. Ngan, D.~V. Soa, {Simple 3-3-1 model and implication for
  dark matter}, Phys. Rev. D90~(7) (2014) 075019.
\newblock \href {http://arxiv.org/abs/1407.3839} {\path{arXiv:1407.3839}},
  \href {http://dx.doi.org/10.1103/PhysRevD.90.075019}
  {\path{doi:10.1103/PhysRevD.90.075019}}.

\bibitem{Dong:2014wsa}
P.~V. Dong, D.~T. Huong, F.~S. Queiroz, N.~T. Thuy, {Phenomenology of the
  3-3-1-1 model}, Phys. Rev. D90~(7) (2014) 075021.
\newblock \href {http://arxiv.org/abs/1405.2591} {\path{arXiv:1405.2591}},
  \href {http://dx.doi.org/10.1103/PhysRevD.90.075021}
  {\path{doi:10.1103/PhysRevD.90.075021}}.

\bibitem{Cogollo:2014jia}
D.~Cogollo, A.~X. Gonzalez-Morales, F.~S. Queiroz, P.~R. Teles, {Excluding the
  Light Dark Matter Window of a 331 Model Using LHC and Direct Dark Matter
  Detection Data}, JCAP 1411~(11) (2014) 002.
\newblock \href {http://arxiv.org/abs/1402.3271} {\path{arXiv:1402.3271}},
  \href {http://dx.doi.org/10.1088/1475-7516/2014/11/002}
  {\path{doi:10.1088/1475-7516/2014/11/002}}.

\bibitem{Martinez:2014ova}
R.~Martínez, J.~Nisperuza, F.~Ochoa, J.~P. Rubio, {Scalar dark matter with
  CERN-LEP data and $Z′$ search at the LHC in an $U(1)′$ model}, Phys. Rev.
  D90~(9) (2014) 095004.
\newblock \href {http://arxiv.org/abs/1408.5153} {\path{arXiv:1408.5153}},
  \href {http://dx.doi.org/10.1103/PhysRevD.90.095004}
  {\path{doi:10.1103/PhysRevD.90.095004}}.

\bibitem{Martinez:2014rea}
R.~Martinez, J.~Nisperuza, F.~Ochoa, J.~P. Rubio, C.~F. Sierra, {Scalar
  coupling limits and diphoton Higgs decay from LHC in an U(1)′ model with
  scalar dark matter}, Phys. Rev. D92~(3) (2015) 035016.
\newblock \href {http://arxiv.org/abs/1411.1641} {\path{arXiv:1411.1641}},
  \href {http://dx.doi.org/10.1103/PhysRevD.92.035016}
  {\path{doi:10.1103/PhysRevD.92.035016}}.

\bibitem{daSilva:2014qba}
P.~S. Rodrigues~da Silva, {A Brief Review on WIMPs in 331 Electroweak Gauge
  Models}\href {http://arxiv.org/abs/1412.8633} {\path{arXiv:1412.8633}}.

\bibitem{Dong:2015rka}
P.~V. Dong, C.~S. Kim, D.~V. Soa, N.~T. Thuy, {Investigation of Dark Matter in
  Minimal 3-3-1 Models}, Phys. Rev. D91~(11) (2015) 115019.
\newblock \href {http://arxiv.org/abs/1501.04385} {\path{arXiv:1501.04385}},
  \href {http://dx.doi.org/10.1103/PhysRevD.91.115019}
  {\path{doi:10.1103/PhysRevD.91.115019}}.

\bibitem{Martinez:2015wrp}
R.~Martinez, F.~Ochoa, {Spin-independent interferences and spin-dependent
  interactions with scalar dark matter}, JHEP 05 (2016) 113.
\newblock \href {http://arxiv.org/abs/1512.04128} {\path{arXiv:1512.04128}},
  \href {http://dx.doi.org/10.1007/JHEP05(2016)113}
  {\path{doi:10.1007/JHEP05(2016)113}}.

\bibitem{Huong:2016ybt}
D.~T. Huong, P.~V. Dong, {Neutrino masses and superheavy dark matter in the
  3-3-1-1 model}, Eur. Phys. J. C77~(4) (2017) 204.
\newblock \href {http://arxiv.org/abs/1605.01216} {\path{arXiv:1605.01216}},
  \href {http://dx.doi.org/10.1140/epjc/s10052-017-4763-3}
  {\path{doi:10.1140/epjc/s10052-017-4763-3}}.

\bibitem{Schechter:1981cv}
J.~Schechter, J.~W.~F. Valle, {Neutrino Decay and Spontaneous Violation of
  Lepton Number}, Phys. Rev. D25 (1982) 774.
\newblock \href {http://dx.doi.org/10.1103/PhysRevD.25.774}
  {\path{doi:10.1103/PhysRevD.25.774}}.

\bibitem{Keung:1983uu}
W.-Y. Keung, G.~Senjanovic, {Majorana Neutrinos and the Production of the
  Right-handed Charged Gauge Boson}, Phys. Rev. Lett. 50 (1983) 1427.
\newblock \href {http://dx.doi.org/10.1103/PhysRevLett.50.1427}
  {\path{doi:10.1103/PhysRevLett.50.1427}}.

\bibitem{Dias:2002gg}
A.~G. Dias, V.~Pleitez, M.~D. Tonasse, {Naturally light invisible axion in
  models with large local discrete symmetries}, Phys. Rev. D67 (2003) 095008.
\newblock \href {http://arxiv.org/abs/hep-ph/0211107}
  {\path{arXiv:hep-ph/0211107}}, \href
  {http://dx.doi.org/10.1103/PhysRevD.67.095008}
  {\path{doi:10.1103/PhysRevD.67.095008}}.

\bibitem{Dias:2003zt}
A.~G. Dias, V.~Pleitez, {Stabilizing the invisible axion in 3-3-1 models},
  Phys. Rev. D69 (2004) 077702.
\newblock \href {http://arxiv.org/abs/hep-ph/0308037}
  {\path{arXiv:hep-ph/0308037}}, \href
  {http://dx.doi.org/10.1103/PhysRevD.69.077702}
  {\path{doi:10.1103/PhysRevD.69.077702}}.

\bibitem{Cogollo:2007qx}
D.~Cogollo, H.~Diniz, C.~A. de~S.~Pires, P.~S. Rodrigues~da Silva, {Fermion
  family number and the Z - Z-prime mixing in the 3-3-1 model with right-handed
  neutrinos}, Mod. Phys. Lett. A23 (2009) 3405--3410.
\newblock \href {http://arxiv.org/abs/0709.2913} {\path{arXiv:0709.2913}},
  \href {http://dx.doi.org/10.1142/S0217732308027382}
  {\path{doi:10.1142/S0217732308027382}}.

\bibitem{Phong:2014yca}
V.~Q. Phong, H.~N. Long, V.~T. Van, N.~C. Thanh, {Electroweak sphalerons in the
  reduced minimal 3-3-1 model}, Phys. Rev. D90~(8) (2014) 085019.
\newblock \href {http://arxiv.org/abs/1408.5657} {\path{arXiv:1408.5657}},
  \href {http://dx.doi.org/10.1103/PhysRevD.90.085019}
  {\path{doi:10.1103/PhysRevD.90.085019}}.

\bibitem{DeConto:2016osh}
G.~De~Conto, V.~Pleitez, {Neutron electric dipole moment in the minimal 3-3-1
  model}\href {http://arxiv.org/abs/1606.01747} {\path{arXiv:1606.01747}}.

\bibitem{Hernandez:2016eod}
A.~E. Carcamo~Hernandez, H.~N. Long, V.~V. Vien, {A 3-3-1 model with
  right-handed neutrinos based on the $\varDelta \left( 27\right) $ family
  symmetry}, Eur. Phys. J. C76~(5) (2016) 242.
\newblock \href {http://arxiv.org/abs/1601.05062} {\path{arXiv:1601.05062}},
  \href {http://dx.doi.org/10.1140/epjc/s10052-016-4074-0}
  {\path{doi:10.1140/epjc/s10052-016-4074-0}}.

\bibitem{Machado:2016jzb}
A.~C.~B. Machado, J.~Montaño, V.~Pleitez, {Lepton violating processes in the
  minimal 3-3-1 model with sterile neutrinos}\href
  {http://arxiv.org/abs/1604.08539} {\path{arXiv:1604.08539}}.

\bibitem{Fonseca:2016tbn}
R.~M. Fonseca, M.~Hirsch, {A flipped 331 model}, JHEP 08 (2016) 003.
\newblock \href {http://arxiv.org/abs/1606.01109} {\path{arXiv:1606.01109}},
  \href {http://dx.doi.org/10.1007/JHEP08(2016)003}
  {\path{doi:10.1007/JHEP08(2016)003}}.

\bibitem{Buras:2016dxz}
A.~J. Buras, F.~De~Fazio, {331 Models Facing the Tensions in $\Delta F=2$
  Processes with the Impact on $\varepsilon^\prime/\varepsilon$,
  $B_s\to\mu^+\mu^-$ and $B\to K^*\mu^+\mu^-$}, JHEP 08 (2016) 115.
\newblock \href {http://arxiv.org/abs/1604.02344} {\path{arXiv:1604.02344}},
  \href {http://dx.doi.org/10.1007/JHEP08(2016)115}
  {\path{doi:10.1007/JHEP08(2016)115}}.

\bibitem{Cabarcas:2013jba}
J.~M. Cabarcas, J.~Duarte, J.~A. Rodriguez, {Charged lepton mixing processes in
  331 Models}, Int. J. Mod. Phys. A29 (2014) 1450015.
\newblock \href {http://arxiv.org/abs/1310.1407} {\path{arXiv:1310.1407}},
  \href {http://dx.doi.org/10.1142/S0217751X14500158}
  {\path{doi:10.1142/S0217751X14500158}}.

\bibitem{Boucenna:2015zwa}
S.~M. Boucenna, J.~W.~F. Valle, A.~Vicente, {Predicting charged lepton flavor
  violation from 3-3-1 gauge symmetry}, Phys. Rev. D92~(5) (2015) 053001.
\newblock \href {http://arxiv.org/abs/1502.07546} {\path{arXiv:1502.07546}},
  \href {http://dx.doi.org/10.1103/PhysRevD.92.053001}
  {\path{doi:10.1103/PhysRevD.92.053001}}.

\bibitem{Hue:2015fbb}
L.~T. Hue, H.~N. Long, T.~T. Thuc, T.~Phong~Nguyen, {Lepton flavor violating
  decays of Standard-Model-like Higgs in 3-3-1 model with neutral lepton},
  Nucl. Phys. B907 (2016) 37--76.
\newblock \href {http://arxiv.org/abs/1512.03266} {\path{arXiv:1512.03266}},
  \href {http://dx.doi.org/10.1016/j.nuclphysb.2016.03.034}
  {\path{doi:10.1016/j.nuclphysb.2016.03.034}}.

\bibitem{Benavides:2009cn}
R.~H. Benavides, Y.~Giraldo, W.~A. Ponce, {FCNC in the 3-3-1 model with
  right-handed neutrinos}, Phys. Rev. D80 (2009) 113009.
\newblock \href {http://arxiv.org/abs/0911.3568} {\path{arXiv:0911.3568}},
  \href {http://dx.doi.org/10.1103/PhysRevD.80.113009}
  {\path{doi:10.1103/PhysRevD.80.113009}}.

\bibitem{Queiroz:2016gif}
F.~S. Queiroz, C.~Siqueira, J.~W.~F. Valle, {Constraining Flavor Changing
  Interactions from LHC Run-2 Dilepton Bounds with Vector Mediators}, Phys.
  Lett. B763 (2016) 269--274.
\newblock \href {http://arxiv.org/abs/1608.07295} {\path{arXiv:1608.07295}},
  \href {http://dx.doi.org/10.1016/j.physletb.2016.10.057}
  {\path{doi:10.1016/j.physletb.2016.10.057}}.

\bibitem{Alwall:2014hca}
J.~Alwall, R.~Frederix, S.~Frixione, V.~Hirschi, F.~Maltoni, O.~Mattelaer,
  H.~S. Shao, T.~Stelzer, P.~Torrielli, M.~Zaro, {The automated computation of
  tree-level and next-to-leading order differential cross sections, and their
  matching to parton shower simulations}, JHEP 07 (2014) 079.
\newblock \href {http://arxiv.org/abs/1405.0301} {\path{arXiv:1405.0301}},
  \href {http://dx.doi.org/10.1007/JHEP07(2014)079}
  {\path{doi:10.1007/JHEP07(2014)079}}.

\bibitem{Kretzer:2003it}
S.~Kretzer, H.~L. Lai, F.~I. Olness, W.~K. Tung, {Cteq6 parton distributions
  with heavy quark mass effects}, Phys. Rev. D69 (2004) 114005.
\newblock \href {http://arxiv.org/abs/hep-ph/0307022}
  {\path{arXiv:hep-ph/0307022}}, \href
  {http://dx.doi.org/10.1103/PhysRevD.69.114005}
  {\path{doi:10.1103/PhysRevD.69.114005}}.

\bibitem{deFavereau:2013fsa}
J.~de~Favereau, C.~Delaere, P.~Demin, A.~Giammanco, V.~Lemaître, A.~Mertens,
  M.~Selvaggi, {DELPHES 3, A modular framework for fast simulation of a generic
  collider experiment}, JHEP 02 (2014) 057.
\newblock \href {http://arxiv.org/abs/1307.6346} {\path{arXiv:1307.6346}},
  \href {http://dx.doi.org/10.1007/JHEP02(2014)057}
  {\path{doi:10.1007/JHEP02(2014)057}}.

\bibitem{Aaboud:2016ejt}
M.~Aaboud, et~al., {Search for new phenomena in events containing a
  same-flavour opposite-sign dilepton pair, jets, and large missing transverse
  momentum in $\sqrt{s}=$ 13 $pp$ collisions with the ATLAS detector}, Eur.
  Phys. J. C77~(3) (2017) 144.
\newblock \href {http://arxiv.org/abs/1611.05791} {\path{arXiv:1611.05791}},
  \href {http://dx.doi.org/10.1140/epjc/s10052-017-4700-5}
  {\path{doi:10.1140/epjc/s10052-017-4700-5}}.

\bibitem{Aaboud:2017buh}
M.~Aaboud, et~al., {Search for new high-mass phenomena in the dilepton final
  state using 36.1 fb$^{-1}$ of proton-proton collision data at $\sqrt{s}$ = 13
  TeV with the ATLAS detector}\href {http://arxiv.org/abs/1707.02424}
  {\path{arXiv:1707.02424}}.

\bibitem{ATLAS:2017wce}
T.~A. collaboration, {Search for new high-mass phenomena in the dilepton final
  state using 36.1 fb$^{-1}$ of proton-proton collision data at $\sqrt{s} =$ 13
  TeV with the ATLAS detector}.

\bibitem{Foot:1990mn}
R.~Foot, {New Physics From Electric Charge Quantization?}, Mod. Phys. Lett. A6
  (1991) 527--530.
\newblock \href {http://dx.doi.org/10.1142/S0217732391000543}
  {\path{doi:10.1142/S0217732391000543}}.

\bibitem{He:1991qd}
X.-G. He, G.~C. Joshi, H.~Lew, R.~R. Volkas, {Simplest Z-prime model}, Phys.
  Rev. D44 (1991) 2118--2132.
\newblock \href {http://dx.doi.org/10.1103/PhysRevD.44.2118}
  {\path{doi:10.1103/PhysRevD.44.2118}}.

\bibitem{Foot:1994vd}
R.~Foot, X.~G. He, H.~Lew, R.~R. Volkas, {Model for a light Z-prime boson},
  Phys. Rev. D50 (1994) 4571--4580.
\newblock \href {http://arxiv.org/abs/hep-ph/9401250}
  {\path{arXiv:hep-ph/9401250}}, \href
  {http://dx.doi.org/10.1103/PhysRevD.50.4571}
  {\path{doi:10.1103/PhysRevD.50.4571}}.

\bibitem{Bi:2009uj}
X.-J. Bi, X.-G. He, Q.~Yuan, {Parameters in a class of leptophilic models from
  PAMELA, ATIC and FERMI}, Phys. Lett. B678 (2009) 168--173.
\newblock \href {http://arxiv.org/abs/0903.0122} {\path{arXiv:0903.0122}},
  \href {http://dx.doi.org/10.1016/j.physletb.2009.06.009}
  {\path{doi:10.1016/j.physletb.2009.06.009}}.

\bibitem{Heeck:2010pg}
J.~Heeck, W.~Rodejohann, {Gauged $L_\mu - L_\tau$ and different Muon Neutrino
  and Anti-Neutrino Oscillations: MINOS and beyond}, J. Phys. G38 (2011)
  085005.
\newblock \href {http://arxiv.org/abs/1007.2655} {\path{arXiv:1007.2655}},
  \href {http://dx.doi.org/10.1088/0954-3899/38/8/085005}
  {\path{doi:10.1088/0954-3899/38/8/085005}}.

\bibitem{Heeck:2011wj}
J.~Heeck, W.~Rodejohann, {Gauged $L_\mu - L_\tau$ Symmetry at the Electroweak
  Scale}, Phys. Rev. D84 (2011) 075007.
\newblock \href {http://arxiv.org/abs/1107.5238} {\path{arXiv:1107.5238}},
  \href {http://dx.doi.org/10.1103/PhysRevD.84.075007}
  {\path{doi:10.1103/PhysRevD.84.075007}}.

\bibitem{Altmannshofer:2014cfa}
W.~Altmannshofer, S.~Gori, M.~Pospelov, I.~Yavin, {Quark flavor transitions in
  $L_\mu-L_\tau$ models}, Phys. Rev. D89 (2014) 095033.
\newblock \href {http://arxiv.org/abs/1403.1269} {\path{arXiv:1403.1269}},
  \href {http://dx.doi.org/10.1103/PhysRevD.89.095033}
  {\path{doi:10.1103/PhysRevD.89.095033}}.

\bibitem{Altmannshofer:2016brv}
W.~Altmannshofer, C.-Y. Chen, P.~S.~B. Dev, A.~Soni, {Lepton flavor violating
  $Z'$ explanation of the muon anomalous magnetic moment}, Phys. Lett. B762
  (2016) 389--398.
\newblock \href {http://arxiv.org/abs/1607.06832} {\path{arXiv:1607.06832}},
  \href {http://dx.doi.org/10.1016/j.physletb.2016.09.046}
  {\path{doi:10.1016/j.physletb.2016.09.046}}.

\bibitem{Crivellin:2015mga}
A.~Crivellin, G.~D'Ambrosio, J.~Heeck, {Explaining $h\to\mu^\pm\tau^\mp$, $B\to
  K^* \mu^+\mu^-$ and $B\to K \mu^+\mu^-/B\to K e^+e^-$ in a two-Higgs-doublet
  model with gauged $L_\mu-L_\tau$}, Phys. Rev. Lett. 114 (2015) 151801.
\newblock \href {http://arxiv.org/abs/1501.00993} {\path{arXiv:1501.00993}},
  \href {http://dx.doi.org/10.1103/PhysRevLett.114.151801}
  {\path{doi:10.1103/PhysRevLett.114.151801}}.

\bibitem{Crivellin:2015hha}
A.~Crivellin, J.~Heeck, P.~Stoffer, {A perturbed lepton-specific
  two-Higgs-doublet model facing experimental hints for physics beyond the
  Standard Model}, Phys. Rev. Lett. 116~(8) (2016) 081801.
\newblock \href {http://arxiv.org/abs/1507.07567} {\path{arXiv:1507.07567}},
  \href {http://dx.doi.org/10.1103/PhysRevLett.116.081801}
  {\path{doi:10.1103/PhysRevLett.116.081801}}.

\bibitem{Kim:2015fpa}
J.-C. Park, S.~C. Park, J.~Kim, {Galactic center GeV gamma-ray excess from dark
  matter with gauged lepton numbers}, Phys. Lett. B752 (2016) 59--65.
\newblock \href {http://arxiv.org/abs/1505.04620} {\path{arXiv:1505.04620}},
  \href {http://dx.doi.org/10.1016/j.physletb.2015.11.035}
  {\path{doi:10.1016/j.physletb.2015.11.035}}.

\bibitem{Baek:2015fea}
S.~Baek, {Dark matter and muon $(g-2)$ in local $U(1)_{L_\mu-L_\tau}$-extended
  Ma Model}, Phys. Lett. B756 (2016) 1--5.
\newblock \href {http://arxiv.org/abs/1510.02168} {\path{arXiv:1510.02168}},
  \href {http://dx.doi.org/10.1016/j.physletb.2016.02.062}
  {\path{doi:10.1016/j.physletb.2016.02.062}}.

\bibitem{Altmannshofer:2016jzy}
W.~Altmannshofer, S.~Gori, S.~Profumo, F.~S. Queiroz, {Explaining dark matter
  and B decay anomalies with an $L_\mu - L_\tau$ model}, JHEP 12 (2016) 106.
\newblock \href {http://arxiv.org/abs/1609.04026} {\path{arXiv:1609.04026}},
  \href {http://dx.doi.org/10.1007/JHEP12(2016)106}
  {\path{doi:10.1007/JHEP12(2016)106}}.

\bibitem{Biswas:2016yan}
A.~Biswas, S.~Choubey, S.~Khan, {Neutrino Mass, Dark Matter and Anomalous
  Magnetic Moment of Muon in a $U(1)_{L_{\mu}-L_{\tau}}$ Model}, JHEP 09 (2016)
  147.
\newblock \href {http://arxiv.org/abs/1608.04194} {\path{arXiv:1608.04194}},
  \href {http://dx.doi.org/10.1007/JHEP09(2016)147}
  {\path{doi:10.1007/JHEP09(2016)147}}.

\bibitem{Altmannshofer:2014pba}
W.~Altmannshofer, S.~Gori, M.~Pospelov, I.~Yavin, {Neutrino Trident Production:
  A Powerful Probe of New Physics with Neutrino Beams}, Phys. Rev. Lett. 113
  (2014) 091801.
\newblock \href {http://arxiv.org/abs/1406.2332} {\path{arXiv:1406.2332}},
  \href {http://dx.doi.org/10.1103/PhysRevLett.113.091801}
  {\path{doi:10.1103/PhysRevLett.113.091801}}.

\bibitem{Allanach:2015gkd}
B.~Allanach, F.~S. Queiroz, A.~Strumia, S.~Sun, {$Z^\prime$ models for the LHCb
  and $g-2$ muon anomalies}, Phys. Rev. D93~(5) (2016) 055045, [Erratum: Phys.
  Rev.D95,no.11,119902(2017)].
\newblock \href {http://arxiv.org/abs/1511.07447} {\path{arXiv:1511.07447}},
  \href {http://dx.doi.org/10.1103/PhysRevD.93.055045,
  10.1103/PhysRevD.95.119902} {\path{doi:10.1103/PhysRevD.93.055045,
  10.1103/PhysRevD.95.119902}}.

\bibitem{Holdom:1985ag}
B.~Holdom, {Two U(1)'s and Epsilon Charge Shifts}, Phys. Lett. B166 (1986)
  196--198.
\newblock \href {http://dx.doi.org/10.1016/0370-2693(86)91377-8}
  {\path{doi:10.1016/0370-2693(86)91377-8}}.

\bibitem{Galison:1983pa}
P.~Galison, A.~Manohar, {TWO Z's OR NOT TWO Z's?}, Phys. Lett. B136 (1984)
  279--283.
\newblock \href {http://dx.doi.org/10.1016/0370-2693(84)91161-4}
  {\path{doi:10.1016/0370-2693(84)91161-4}}.

\bibitem{Boehm:2003ha}
C.~Boehm, P.~Fayet, J.~Silk, {Light and heavy dark matter particles}, Phys.
  Rev. D69 (2004) 101302.
\newblock \href {http://arxiv.org/abs/hep-ph/0311143}
  {\path{arXiv:hep-ph/0311143}}, \href
  {http://dx.doi.org/10.1103/PhysRevD.69.101302}
  {\path{doi:10.1103/PhysRevD.69.101302}}.

\bibitem{Pospelov:2007mp}
M.~Pospelov, A.~Ritz, M.~B. Voloshin, {Secluded WIMP Dark Matter}, Phys. Lett.
  B662 (2008) 53--61.
\newblock \href {http://arxiv.org/abs/0711.4866} {\path{arXiv:0711.4866}},
  \href {http://dx.doi.org/10.1016/j.physletb.2008.02.052}
  {\path{doi:10.1016/j.physletb.2008.02.052}}.

\bibitem{ArkaniHamed:2008qn}
N.~Arkani-Hamed, D.~P. Finkbeiner, T.~R. Slatyer, N.~Weiner, {A Theory of Dark
  Matter}, Phys. Rev. D79 (2009) 015014.
\newblock \href {http://arxiv.org/abs/0810.0713} {\path{arXiv:0810.0713}},
  \href {http://dx.doi.org/10.1103/PhysRevD.79.015014}
  {\path{doi:10.1103/PhysRevD.79.015014}}.

\bibitem{Pospelov:2008zw}
M.~Pospelov, {Secluded U(1) below the weak scale}, Phys. Rev. D80 (2009)
  095002.
\newblock \href {http://arxiv.org/abs/0811.1030} {\path{arXiv:0811.1030}},
  \href {http://dx.doi.org/10.1103/PhysRevD.80.095002}
  {\path{doi:10.1103/PhysRevD.80.095002}}.

\bibitem{Campos:2017dgc}
M.~D. Campos, D.~Cogollo, M.~Lindner, T.~Melo, F.~S. Queiroz, W.~Rodejohann,
  {Neutrino Masses and Absence of Flavor Changing Interactions in the 2HDM from
  Gauge Principles}, JHEP 08 (2017) 092.
\newblock \href {http://arxiv.org/abs/1705.05388} {\path{arXiv:1705.05388}},
  \href {http://dx.doi.org/10.1007/JHEP08(2017)092}
  {\path{doi:10.1007/JHEP08(2017)092}}.

\bibitem{Alexander:2016aln}
J.~Alexander, et~al.,
  \href{http://inspirehep.net/record/1484628/files/arXiv:1608.08632.pdf}{{Dark
  Sectors 2016 Workshop: Community Report}}, 2016.
\newblock \href {http://arxiv.org/abs/1608.08632} {\path{arXiv:1608.08632}}.
\newline\urlprefix\url{http://inspirehep.net/record/1484628/files/arXiv:1608.08632.pdf}

\bibitem{Riordan:1987aw}
E.~M. Riordan, et~al., {A Search for Short Lived Axions in an Electron Beam
  Dump Experiment}, Phys. Rev. Lett. 59 (1987) 755.
\newblock \href {http://dx.doi.org/10.1103/PhysRevLett.59.755}
  {\path{doi:10.1103/PhysRevLett.59.755}}.

\bibitem{Batley:2015lha}
J.~R. Batley, et~al., {Search for the dark photon in $\pi^0$ decays}, Phys.
  Lett. B746 (2015) 178--185.
\newblock \href {http://arxiv.org/abs/1504.00607} {\path{arXiv:1504.00607}},
  \href {http://dx.doi.org/10.1016/j.physletb.2015.04.068}
  {\path{doi:10.1016/j.physletb.2015.04.068}}.

\bibitem{Davoudiasl:2014kua}
H.~Davoudiasl, H.-S. Lee, W.~J. Marciano, {Muon $g−2$, rare kaon decays, and
  parity violation from dark bosons}, Phys. Rev. D89~(9) (2014) 095006.
\newblock \href {http://arxiv.org/abs/1402.3620} {\path{arXiv:1402.3620}},
  \href {http://dx.doi.org/10.1103/PhysRevD.89.095006}
  {\path{doi:10.1103/PhysRevD.89.095006}}.

\bibitem{Suzuki:1987jf}
T.~Suzuki, D.~F. Measday, J.~P. Roalsvig, {Total Nuclear Capture Rates for
  Negative Muons}, Phys. Rev. C35 (1987) 2212.
\newblock \href {http://dx.doi.org/10.1103/PhysRevC.35.2212}
  {\path{doi:10.1103/PhysRevC.35.2212}}.

\bibitem{Alonso:2012ji}
R.~Alonso, M.~Dhen, M.~B. Gavela, T.~Hambye, {Muon conversion to electron in
  nuclei in type-I seesaw models}, JHEP 01 (2013) 118.
\newblock \href {http://arxiv.org/abs/1209.2679} {\path{arXiv:1209.2679}},
  \href {http://dx.doi.org/10.1007/JHEP01(2013)118}
  {\path{doi:10.1007/JHEP01(2013)118}}.

\bibitem{Deppisch:2015qwa}
F.~F. Deppisch, P.~S. Bhupal~Dev, A.~Pilaftsis, {Neutrinos and Collider
  Physics}, New J. Phys. 17~(7) (2015) 075019.
\newblock \href {http://arxiv.org/abs/1502.06541} {\path{arXiv:1502.06541}},
  \href {http://dx.doi.org/10.1088/1367-2630/17/7/075019}
  {\path{doi:10.1088/1367-2630/17/7/075019}}.

\bibitem{Mondal:2015zba}
S.~Mondal, S.~K. Rai, {Polarized window for left-right symmetry and a
  right-handed neutrino at the Large Hadron-Electron Collider}, Phys. Rev.
  D93~(1) (2016) 011702.
\newblock \href {http://arxiv.org/abs/1510.08632} {\path{arXiv:1510.08632}},
  \href {http://dx.doi.org/10.1103/PhysRevD.93.011702}
  {\path{doi:10.1103/PhysRevD.93.011702}}.

\bibitem{Biswal:2017nfl}
S.~S. Biswal, P.~S.~B. Dev, {Probing left-right seesaw models using beam
  polarization at an $e^+e^-$ collider}, Phys. Rev. D95~(11) (2017) 115031.
\newblock \href {http://arxiv.org/abs/1701.08751} {\path{arXiv:1701.08751}},
  \href {http://dx.doi.org/10.1103/PhysRevD.95.115031}
  {\path{doi:10.1103/PhysRevD.95.115031}}.

\bibitem{Banerjee:2015gca}
S.~Banerjee, P.~S.~B. Dev, A.~Ibarra, T.~Mandal, M.~Mitra, {Prospects of Heavy
  Neutrino Searches at Future Lepton Colliders}, Phys. Rev. D92 (2015) 075002.
\newblock \href {http://arxiv.org/abs/1503.05491} {\path{arXiv:1503.05491}},
  \href {http://dx.doi.org/10.1103/PhysRevD.92.075002}
  {\path{doi:10.1103/PhysRevD.92.075002}}.

\bibitem{Gorbunov:2007ak}
D.~Gorbunov, M.~Shaposhnikov, {How to find neutral leptons of the $\nu$MSM?},
  JHEP 10 (2007) 015, [Erratum: JHEP11,101(2013)].
\newblock \href {http://arxiv.org/abs/0705.1729} {\path{arXiv:0705.1729}},
  \href {http://dx.doi.org/10.1007/JHEP11(2013)101,
  10.1088/1126-6708/2007/10/015} {\path{doi:10.1007/JHEP11(2013)101,
  10.1088/1126-6708/2007/10/015}}.

\bibitem{Asakura:2014lma}
K.~Asakura, et~al., {Results from KamLAND-Zen}, AIP Conf. Proc. 1666 (2015)
  170003.
\newblock \href {http://arxiv.org/abs/1409.0077} {\path{arXiv:1409.0077}},
  \href {http://dx.doi.org/10.1063/1.4915593} {\path{doi:10.1063/1.4915593}}.

\bibitem{DellOro:2016tmg}
S.~Dell'Oro, S.~Marcocci, M.~Viel, F.~Vissani, {Neutrinoless double beta decay:
  2015 review}, Adv. High Energy Phys. 2016 (2016) 2162659.
\newblock \href {http://arxiv.org/abs/1601.07512} {\path{arXiv:1601.07512}},
  \href {http://dx.doi.org/10.1155/2016/2162659}
  {\path{doi:10.1155/2016/2162659}}.

\bibitem{Adams:2013qkq}
C.~Adams, et~al., {The Long-Baseline Neutrino Experiment: Exploring Fundamental
  Symmetries of the Universe}\href {http://arxiv.org/abs/1307.7335}
  {\path{arXiv:1307.7335}}.

\bibitem{Anelli:2015pba}
M.~Anelli, et~al., {A facility to Search for Hidden Particles (SHiP) at the
  CERN SPS}\href {http://arxiv.org/abs/1504.04956} {\path{arXiv:1504.04956}}.

\bibitem{Dittmar:1989yg}
M.~Dittmar, A.~Santamaria, M.~C. Gonzalez-Garcia, J.~W.~F. Valle, {Production
  Mechanisms and Signatures of Isosinglet Neutral Heavy Leptons in $Z^0$
  Decays}, Nucl. Phys. B332 (1990) 1--19.
\newblock \href {http://dx.doi.org/10.1016/0550-3213(90)90028-C}
  {\path{doi:10.1016/0550-3213(90)90028-C}}.

\bibitem{Blondel:2014bra}
A.~Blondel, E.~Graverini, N.~Serra, M.~Shaposhnikov, {Search for Heavy Right
  Handed Neutrinos at the FCC-ee}, Nucl. Part. Phys. Proc. 273-275 (2016)
  1883--1890.
\newblock \href {http://arxiv.org/abs/1411.5230} {\path{arXiv:1411.5230}},
  \href {http://dx.doi.org/10.1016/j.nuclphysbps.2015.09.304}
  {\path{doi:10.1016/j.nuclphysbps.2015.09.304}}.

\bibitem{Abada:2014cca}
A.~Abada, V.~De~Romeri, S.~Monteil, J.~Orloff, A.~M. Teixeira, {Indirect
  searches for sterile neutrinos at a high-luminosity Z-factory}, JHEP 04
  (2015) 051.
\newblock \href {http://arxiv.org/abs/1412.6322} {\path{arXiv:1412.6322}},
  \href {http://dx.doi.org/10.1007/JHEP04(2015)051}
  {\path{doi:10.1007/JHEP04(2015)051}}.

\bibitem{Magg:1980ut}
M.~Magg, C.~Wetterich, {Neutrino Mass Problem and Gauge Hierarchy}, Phys. Lett.
  B94 (1980) 61--64.
\newblock \href {http://dx.doi.org/10.1016/0370-2693(80)90825-4}
  {\path{doi:10.1016/0370-2693(80)90825-4}}.

\bibitem{Han:2006ip}
T.~Han, B.~Zhang, {Signatures for Majorana neutrinos at hadron colliders},
  Phys. Rev. Lett. 97 (2006) 171804.
\newblock \href {http://arxiv.org/abs/hep-ph/0604064}
  {\path{arXiv:hep-ph/0604064}}, \href
  {http://dx.doi.org/10.1103/PhysRevLett.97.171804}
  {\path{doi:10.1103/PhysRevLett.97.171804}}.

\bibitem{delAguila:2007qnc}
F.~del Aguila, J.~A. Aguilar-Saavedra, R.~Pittau, {Heavy neutrino signals at
  large hadron colliders}, JHEP 10 (2007) 047.
\newblock \href {http://arxiv.org/abs/hep-ph/0703261}
  {\path{arXiv:hep-ph/0703261}}, \href
  {http://dx.doi.org/10.1088/1126-6708/2007/10/047}
  {\path{doi:10.1088/1126-6708/2007/10/047}}.

\bibitem{Arhrib:2011uy}
A.~Arhrib, R.~Benbrik, M.~Chabab, G.~Moultaka, M.~C. Peyranere, L.~Rahili,
  J.~Ramadan, {The Higgs Potential in the Type II Seesaw Model}, Phys. Rev. D84
  (2011) 095005.
\newblock \href {http://arxiv.org/abs/1105.1925} {\path{arXiv:1105.1925}},
  \href {http://dx.doi.org/10.1103/PhysRevD.84.095005}
  {\path{doi:10.1103/PhysRevD.84.095005}}.

\bibitem{Akeroyd:2012nd}
A.~G. Akeroyd, S.~Moretti, H.~Sugiyama, {Five-lepton and six-lepton signatures
  from production of neutral triplet scalars in the Higgs Triplet Model}, Phys.
  Rev. D85 (2012) 055026.
\newblock \href {http://arxiv.org/abs/1201.5047} {\path{arXiv:1201.5047}},
  \href {http://dx.doi.org/10.1103/PhysRevD.85.055026}
  {\path{doi:10.1103/PhysRevD.85.055026}}.

\bibitem{Perez:2008ha}
P.~Fileviez~Perez, T.~Han, G.-y. Huang, T.~Li, K.~Wang, {Neutrino Masses and
  the CERN LHC: Testing Type II Seesaw}, Phys. Rev. D78 (2008) 015018.
\newblock \href {http://arxiv.org/abs/0805.3536} {\path{arXiv:0805.3536}},
  \href {http://dx.doi.org/10.1103/PhysRevD.78.015018}
  {\path{doi:10.1103/PhysRevD.78.015018}}.

\bibitem{ATLAS:2017iqw}
T.~A. collaboration, {Search for doubly-charged Higgs boson production in
  multi-lepton final states with the ATLAS detector using proton-proton
  collisions at $\sqrt{s}=13\,\mathrm{TeV}$}.

\bibitem{CMS:2017pet}
C.~Collaboration, {A search for doubly-charged Higgs boson production in three
  and four lepton final states at $\sqrt{s}=13~\mathrm{TeV}$}.

\bibitem{Foot:1988aq}
R.~Foot, H.~Lew, X.~G. He, G.~C. Joshi, {Seesaw Neutrino Masses Induced by a
  Triplet of Leptons}, Z. Phys. C44 (1989) 441.
\newblock \href {http://dx.doi.org/10.1007/BF01415558}
  {\path{doi:10.1007/BF01415558}}.

\bibitem{Bonilla:2015eha}
C.~Bonilla, R.~M. Fonseca, J.~W.~F. Valle, {Consistency of the triplet seesaw
  model revisited}, Phys. Rev. D92~(7) (2015) 075028.
\newblock \href {http://arxiv.org/abs/1508.02323} {\path{arXiv:1508.02323}},
  \href {http://dx.doi.org/10.1103/PhysRevD.92.075028}
  {\path{doi:10.1103/PhysRevD.92.075028}}.

\bibitem{Abada:2008ea}
A.~Abada, C.~Biggio, F.~Bonnet, M.~B. Gavela, T.~Hambye, {$\mu \rightarrow e
  \gamma$ and $\tau l \gamma$ decays in the fermion triplet seesaw model},
  Phys. Rev. D78 (2008) 033007.
\newblock \href {http://arxiv.org/abs/0803.0481} {\path{arXiv:0803.0481}},
  \href {http://dx.doi.org/10.1103/PhysRevD.78.033007}
  {\path{doi:10.1103/PhysRevD.78.033007}}.

\bibitem{Franceschini:2008pz}
R.~Franceschini, T.~Hambye, A.~Strumia, {Type-III see-saw at LHC}, Phys. Rev.
  D78 (2008) 033002.
\newblock \href {http://arxiv.org/abs/0805.1613} {\path{arXiv:0805.1613}},
  \href {http://dx.doi.org/10.1103/PhysRevD.78.033002}
  {\path{doi:10.1103/PhysRevD.78.033002}}.

\bibitem{Sirunyan:2017qkz}
A.~M. Sirunyan, et~al., {Search for evidence of the type-III seesaw mechanism
  in multilepton final states in proton-proton collisions at sqrt(s) = 13
  TeV}\href {http://arxiv.org/abs/1708.07962} {\path{arXiv:1708.07962}}.

\bibitem{Agostinelli:2002hh}
S.~Agostinelli, et~al., {GEANT4: A Simulation toolkit}, Nucl. Instrum. Meth.
  A506 (2003) 250--303.
\newblock \href {http://dx.doi.org/10.1016/S0168-9002(03)01368-8}
  {\path{doi:10.1016/S0168-9002(03)01368-8}}.

\bibitem{Mohapatra:1986bd}
R.~N. Mohapatra, J.~W.~F. Valle, {Neutrino Mass and Baryon Number
  Nonconservation in Superstring Models}, Phys. Rev. D34 (1986) 1642.
\newblock \href {http://dx.doi.org/10.1103/PhysRevD.34.1642}
  {\path{doi:10.1103/PhysRevD.34.1642}}.

\bibitem{Malinsky:2005bi}
M.~Malinsky, J.~C. Romao, J.~W.~F. Valle, {Novel supersymmetric SO(10) seesaw
  mechanism}, Phys. Rev. Lett. 95 (2005) 161801.
\newblock \href {http://arxiv.org/abs/hep-ph/0506296}
  {\path{arXiv:hep-ph/0506296}}, \href
  {http://dx.doi.org/10.1103/PhysRevLett.95.161801}
  {\path{doi:10.1103/PhysRevLett.95.161801}}.

\bibitem{Wyler:1982dd}
D.~Wyler, L.~Wolfenstein, {Massless Neutrinos in Left-Right Symmetric Models},
  Nucl. Phys. B218 (1983) 205--214.
\newblock \href {http://dx.doi.org/10.1016/0550-3213(83)90482-0}
  {\path{doi:10.1016/0550-3213(83)90482-0}}.

\bibitem{Witten:1985xc}
E.~Witten, {Symmetry Breaking Patterns in Superstring Models}, Nucl. Phys. B258
  (1985) 75.
\newblock \href {http://dx.doi.org/10.1016/0550-3213(85)90603-0}
  {\path{doi:10.1016/0550-3213(85)90603-0}}.

\bibitem{Abada:2014nwa}
A.~Abada, V.~De~Romeri, A.~M. Teixeira, {Effect of steriles states on lepton
  magnetic moments and neutrinoless double beta decay}, JHEP 09 (2014) 074.
\newblock \href {http://arxiv.org/abs/1406.6978} {\path{arXiv:1406.6978}},
  \href {http://dx.doi.org/10.1007/JHEP09(2014)074}
  {\path{doi:10.1007/JHEP09(2014)074}}.

\bibitem{Patel:2015tea}
H.~H. Patel, {Package-X: A Mathematica package for the analytic calculation of
  one-loop integrals}, Comput. Phys. Commun. 197 (2015) 276--290.
\newblock \href {http://arxiv.org/abs/1503.01469} {\path{arXiv:1503.01469}},
  \href {http://dx.doi.org/10.1016/j.cpc.2015.08.017}
  {\path{doi:10.1016/j.cpc.2015.08.017}}.

\end{thebibliography}

\end{document}